%% file: N2book-arXiv_withbbl.tex
\documentclass[12pt]{book}
\usepackage{amssymb,amsmath,latexsym,slashed}
\usepackage{amscd,nicefrac,float}
\usepackage{bbold}
\allowdisplaybreaks

\usepackage{makeidx}         
\usepackage{graphicx}        

   \usepackage[nosort]{cite}
     \usepackage[pdftex]{hyperref}
\usepackage{tcolorbox}
\newcommand{\ft}[2]{{\textstyle\frac{#1}{#2}}}
\def\Re{\mathop{\rm Re}\nolimits}
\def\Im{\mathop{\rm Im}\nolimits}
\DeclareMathOperator{\trace}{Tr} 
\def\rmi{{\rm i}}
\def\rmd{{\rm d}}
\def\rme{{\rm e}}
\newcommand{\hc}{{\rm h.c.}}
\def\unity{{\mathchoice {\mathrm{1\mskip-4mu l}} {\mathrm{ 1\mskip-4mu l}}
{\mathrm{ 1\mskip-4.5mu l}} {\mathrm{ 1\mskip-5mu l}}}}
\newcommand{\covder}{\nabla }
\newcommand{\bbox}{\lower.2ex\hbox{$\Box$}}
\def\IP{\relax{\rm I\kern-.18em P}}
\newcommand{\cmap}{{\bf c} map}
\newcommand{\rmap}{{\bf r} map}
\newcommand{\crmap}{{{\bf c}$\scriptstyle\circ${\bf r} map}}
\DeclareMathOperator{\SU}{SU}
\DeclareMathOperator{\SO}{SO}
\DeclareMathOperator{\Spin}{Spin}
\DeclareMathOperator{\U}{U}
\DeclareMathOperator{\USp}{USp}
\DeclareMathOperator{\OSp}{OSp}
\DeclareMathOperator{\Sp}{Sp}
\DeclareMathOperator{\Sl}{S\ell}
\DeclareMathOperator{\Gl}{G\ell }
\newcommand{\dr}{\raise.3ex\hbox{$\stackrel{\leftarrow}{\partial }$}{}}
\newcommand{\delr}{\raise.3ex\hbox{$\stackrel{\leftarrow}{\delta }$}{}}

\newcommand{\Cbar }{\mathbb{C}}

\newcommand{\Ae}{{\cal A}}
\newcommand{\Be}{{\cal B}}
\newcommand{\Ce}{{\cal C}}
\newcommand{\De}{{\cal D}}
\newcommand{\Ee}{{\cal E}}
\newcommand{\Fe}{{\cal F}}
\newcommand{\bAe}{{\bar {\cal A}}}
\newcommand{\bBe}{{\bar {\cal B}}}
\newcommand{\bCe}{{\bar {\cal C}}}
\newcommand{\bDe}{{\bar {\cal D}}}
\newcommand{\bEe}{{\bar {\cal E}}}
\newcommand{\Xu}{u}
\newcommand{\Yu}{v}
\newcommand{\Zu}{w}
\newcommand{\Wu}{x}
\newcommand{\Vu}{y}
\newcommand{\Ap}{A}
\newcommand{\Bp}{B}
\newcommand{\Cp}{C}

\newcommand{\bAp}{{\bar A}}
\newcommand{\bBp}{{\bar B}}
\newcommand{\bCp}{{\bar C}}
\newcommand{\bDp}{{\bar D}}
\newcommand{\hX}{X}  
\newcommand{\hY}{Y} 
\newcommand{\ralpha}{r} 
\newcommand{\rbeta}{s} 
\newcommand{\bi}{\bar\imath}
\newcommand{\bj}{\bar\jmath}
\newcommand{\goldstino}{\upsilon}
\newcommand{\nablaLC}{\nabla }

\newcommand{\cQ}{Q}
\newcommand{\poinc}{\boxdot}
\newcommand{\ZF}{{\cal F}}
\newcommand{\bZF}{\overline{{\cal F}}}

\numberwithin{equation}{chapter}
\numberwithin{table}{chapter}
\newcounter{exercise}[chapter]
\renewcommand{\theexercise}{\thechapter.\arabic{exercise}}
\newcommand{\bexer}{
   \begin{list}{\bf Exercise \theexercise : }{\refstepcounter{exercise}
           \setlength {\rightmargin}{\leftmargin}}\item }
\newcommand{\eexer}{\end{list}}
\begin{document}

\begin{titlepage}
\begin{center}
\baselineskip=16pt
{\LARGE   ${\cal N}=2$ Supergravity in $D=4,5,6$ Dimensions}
\vfill
{\large Edoardo Lauria$^{1,2}$  and Antoine Van Proeyen${}^{2}$ %
  } \\
\vfill
${}^{1}$
CPHT, Ecole Polytechnique\\
Palaiseau, France.

\vskip 5mm

${}^{2}$  KU Leuven, Institute for Theoretical Physics,\\
       Celestijnenlaan 200D B-3001 Leuven, Belgium.

\vskip 5mm

\texttt{edoardo.lauria@polytechnique.edu,~antoine.vanproeyen@fys.kuleuven.be} \\
\end{center}
\vfill
\begin{center}
{\bf Abstract}
\end{center}
{\small

An overview of matter-coupled ${\cal N}=2$ supergravity theories with 8 real supercharges, in 4,5
and 6 dimensions is given. The construction of the theories by
superconformal methods is explained from basic principles. Special geometry is obtained and
characterized. The relation between the theories in those dimensions is
discussed. This leads to the concepts of very special geometry and
quaternionic-K\"{a}hler manifolds, whose structures are
explained.}\vspace{2mm} \vfill
 \hrule width 3.cm {%
\noindent
This is the author version of the book with the same title, published by Spinger International Publishing, 2020  as \textbf{Lecture Notes in Physics 966},\\ see
\href{https://www.springer.com/gp/book/9783030337551}{https://www.springer.com/gp/book/9783030337551}.\\
\footnotesize eBook ISBN    978-3-030-33757-5,
DOI
    10.1007/978-3-030-33757-5,
Softcover ISBN
    978-3-030-33755-1,
Series ISSN
    0075-8450}
\end{titlepage}
\addtocounter{page}{1}
\include{preface}
\include{acknowledgement}

\tableofcontents{}
\include{ch1_BasicIngredients}

\include{ch2_Weylmultiplet}

\include{ch3_Mattermultiplets}
\include{ch4_Gaugefix}
\include{ch5_Specialgeometries}
\include{ch6_Finalresults}
\newpage
\appendix
\addcontentsline{toc}{chapter}{Appendices}
\renewcommand{\thesection}{\Alph{chapter}.\arabic{section}}
\include{appA_Notation}

\include{appB_Superalgebras}

\include{appC_Comparison}

\newpage
\providecommand{\href}[2]{#2}\begingroup\raggedright\endgroup

\begin{theindex}

  \item $D$-gauge, \hyperpage{112}
  \item $K$-gauge, \hyperpage{98}
  \item $S$-gauge, \hyperpage{112}, \hyperpage{129}
  \item $\SU(2)$-gauge, \hyperpage{129}
  \item $\SU(2)$-triplet, \hyperpage{196}
  \item $\U(1)$-gauge, \hyperpage{118}

  \indexspace

  \item affine special K\"{a}hler, \hyperpage{116}
  \item almost quaternionic structure, \hyperpage{74}, \hyperpage{165}
  \item attractor mechanism, \hyperpage{157}

  \indexspace

  \item black hole potential, \hyperpage{162}

  \indexspace

  \item c-map, \hyperpage{167}
  \item central charge, \hyperpage{187}
  \item charge conjugation, \hyperpage{198}
  \item charged black hole solution, \hyperpage{157}
  \item Chern--Simons term, \hyperpage{87}
  \item chiral density formula, \hyperpage{72}
  \item chiral multiplet, \hyperpage{69}
  \item chiral weight, \hyperpage{53}
  \item closed homothetic Killing vector, \hyperpage{22}
  \item conformal algebra, \hyperpage{20}
  \item conformal Killing equation, \hyperpage{19}
  \item conformal Killing vector, \hyperpage{24}
  \item conformal symmetry, \hyperpage{19}
  \item consistent constraints, \hyperpage{69}
  \item constrained chiral multiplet, \hyperpage{70}
  \item covariant curvature, \hyperpage{39}
  \item covariant derivative, \hyperpage{39}
  \item covariant general coordinate transformation, \hyperpage{34}

  \indexspace

  \item decomposition law, \hyperpage{98}, \hyperpage{186}
  \item degrees of freedom, \hyperpage{14}
  \item dilatation, \hyperpage{20}
  \item dilaton Weyl multiplet, \hyperpage{52}, \hyperpage{191}
  \item dual tensor, \hyperpage{195}
  \item duality symmetries, \hyperpage{140}

  \indexspace

  \item embedding space, \hyperpage{118}, \hyperpage{126}
  \item embedding tensor, \hyperpage{79}, \hyperpage{151}
  \item equivariance relation, \hyperpage{94}, \hyperpage{130},
		\hyperpage{152}
  \item exact homothetic Killing vector, \hyperpage{25}

  \indexspace

  \item Fayet--Iliopoulos term, \hyperpage{130}, \hyperpage{154},
		\hyperpage{176}
  \item fermion shift, \hyperpage{186}
  \item Fierz relation, \hyperpage{201}

  \indexspace

  \item gauge coupling, \hyperpage{136}
  \item general coordinate transformation (gct), \hyperpage{33}
  \item general scalar multiplet, \hyperpage{69}
  \item Goldstino, \hyperpage{182}
  \item graviphoton, \hyperpage{85}
  \item graviphoton field strength, \hyperpage{185}

  \indexspace

  \item higher derivatives, \hyperpage{191}
  \item homothetic Killing vector, \hyperpage{24}
  \item hyper-K\"{a}hler manifold, \hyperpage{135},
		\hyperpage{165, 166}

  \indexspace

  \item intrinsic parts, \hyperpage{21}

  \indexspace

  \item K\"{a}hler covariant derivatives, \hyperpage{138}
  \item K\"{a}hler curvature, \hyperpage{135}
  \item K\"{a}hler potential, \hyperpage{119}, \hyperpage{172}
  \item K\"{a}hler transformations, \hyperpage{118}, \hyperpage{123}

  \indexspace

  \item Levi-Civita connection, \hyperpage{193}
  \item Levi-Civita tensor, \hyperpage{194}
  \item Lie superalgebras, \hyperpage{205}

  \indexspace

  \item Majorana, \hyperpage{10}
  \item Majorana flip relation, \hyperpage{197}
  \item massive hypermultiplets, \hyperpage{72}
  \item matter fields, \hyperpage{35}
  \item matter multiplets, \hyperpage{9}
  \item moment map, \hyperpage{93}, \hyperpage{152}, \hyperpage{175}

  \indexspace

  \item Noether terms, \hyperpage{182}
  \item NW--SE convention, \hyperpage{196}

  \indexspace

  \item off-shell counting, \hyperpage{15}
  \item on-shell counting, \hyperpage{15}
  \item orbital parts, \hyperpage{21}

  \indexspace

  \item partial breaking, \hyperpage{69}, \hyperpage{147}
  \item Pauli term, \hyperpage{182}
  \item physical fermions, \hyperpage{178, 179}
  \item prepotential, \hyperpage{10}, \hyperpage{85}, \hyperpage{89}
  \item projective coordinates, \hyperpage{117}, \hyperpage{126}
  \item projective space, \hyperpage{118}, \hyperpage{126}
  \item projective special K\"{a}hler, \hyperpage{117}
  \item proper symmetry, \hyperpage{140}
  \item pseudo symmetry, \hyperpage{140}

  \indexspace

  \item quaternionic algebra, \hyperpage{74}
  \item quaternionic geometry, \hyperpage{165}
  \item quaternionic manifold, \hyperpage{135}
  \item quaternionic structures, \hyperpage{74}
  \item quaternionic-K\"{a}hler manifold, \hyperpage{166}

  \indexspace

  \item r-map, \hyperpage{167}
  \item Ricci tensor, \hyperpage{194}

  \indexspace

  \item scalar potential, \hyperpage{136}, \hyperpage{176},
		\hyperpage{186}, \hyperpage{190}
  \item Schouten identities, \hyperpage{195}
  \item self-dual, \hyperpage{195}
  \item signature of spacetime, \hyperpage{10}
  \item simple Lie superalgebra, \hyperpage{205}
  \item soft algebra, \hyperpage{54}
  \item special conformal transformation, \hyperpage{20},
		\hyperpage{22}
  \item special coordinates, \hyperpage{118}, \hyperpage{145}
  \item special geometries, \hyperpage{135}
  \item special K\"{a}hler, \hyperpage{10}, \hyperpage{121}
  \item special manifolds, \hyperpage{9}, \hyperpage{135}
  \item spin connection, \hyperpage{34}
  \item spinor indices, \hyperpage{14}
  \item standard gauge transformation, \hyperpage{33}
  \item standard Weyl multiplet, \hyperpage{52}, \hyperpage{191}
  \item superconformal tensor calculus, \hyperpage{11}
  \item symplectic inner product, \hyperpage{172}
  \item symplectic invariant, \hyperpage{138}
  \item symplectic Majorana, \hyperpage{10}, \hyperpage{14, 15},
		\hyperpage{199}
  \item symplectic Majorana--Weyl, \hyperpage{14}, \hyperpage{200}
  \item symplectic reparameterization, \hyperpage{140}
  \item symplectic transformation, \hyperpage{136, 137}
  \item symplectic vector, \hyperpage{138}, \hyperpage{172}

  \indexspace

  \item theta angles, \hyperpage{136}

  \indexspace

  \item vacuum expectation value, \hyperpage{136}
  \item very special K\"{a}hler manifolds, \hyperpage{122},
		\hyperpage{145}, \hyperpage{167}
  \item very special quaternionic manifolds, \hyperpage{167}
  \item very special real manifold, \hyperpage{167}
  \item Virasoro algebra, \hyperpage{20}

  \indexspace

  \item Weyl multiplet, \hyperpage{31}
  \item Weyl weight, \hyperpage{22}, \hyperpage{53}

\end{theindex}

\printindex
\end{document}

%% file: preface.tex
\section*{Preface}

\vspace{3cm}

Supergravity theories play an important role in present-day research on fundamental interactions. A profound knowledge can only be gained by understanding how supergravity theories are constructed. To understand the necessary techniques is not easy starting from the published literature. Despite the presence of a few books, it turns out that a detailed explanation is still lacking and this is particularly the case for supergravity theories with matter couplings in so-called ${\cal N}=2$ supergravity. In this book, we want to provide a detailed explanation of these theories and tools that researchers have used to construct supergravity theories with 8 real supercharges, in 4, 5 and 6 dimensions.

The basic technique is the superconformal calculus. We explain these methods from basic principles. Chapter \ref{ss:ingredients} will introduce the basic ingredients: the multiplets and the symmetries that will be instrumental in the rest of the book. In Chap.~\ref{ss:Weyl} it will be explained how these symmetry groups can be promoted to gauge groups. The matter multiplets will enter the game in Chap.~\ref{ss:matter}: first defining these multiplets in the superconformal context, and then actions will be constructed. These are reduced to super-Poincar\'{e} theories in Chap.~\ref{ss:gaugefix}. The beautiful geometric structure that these theories enjoy is discussed and characterized in Chap.~\ref{ss:specialgeom}. These geometries go together under the name of special geometries, which include special K\"{a}hler geometry and
quaternionic-K\"{a}hler manifolds.

For readers who want to progress soon to final results, Chap.~\ref{ch:finalresults}  gives a shorter account.

The text grew from lecture notes during the semester `Supergravity, superstrings and M-theory' at Institut Henri Poincar\'{e},
Paris, in  November 2000. Parts of it have been available on internet sites and were appreciated by many researchers.
We hope that this book will be useful as well for students as for researchers who want to know how these standard theories are constructed and what is their geometric structure.

\vspace{\baselineskip}
\begin{flushright}\noindent
Paris, France\hfill Edoardo Lauria \\
Leuven, Belgium\hfill Antoine Van Proeyen\\ September 2019\hfill\phantom{.}
\end{flushright}

%% file: acknowledgement.tex
\newpage

\section*{Acknowledgments}
\vspace{3cm}
\noindent
The notes for this book received input from many people in our field.
 We first of all thank Robin De Roover and Gabriele  Tartaglino-Mazzucchelli, who collaborated for some time to the writing of this book.
 \vspace{5mm}

\noindent The text is based on several fruitful collaborations on ${\cal N}=2$ supergravities, and we thank Eric Bergshoeff, Marco Bill\'{o}, Anna Ceresole, Piet Claus, Frederik Coomans, Ben Craps, Eug\`{e}ne Cremmer, Sorin Cucu, Riccardo D'Auria, Gianguido Dall'Agata, Mees de Roo, Bernard de Wit, Tim de~Wit, Jean-Pierre Derendinger, Martijn Derix, Sergio Ferrara, Pietro Fr\`{e}, Dan Freedman, Jos Gheerardyn, Luciano Girardello, Rein Halbersma, Renata Kallosh, Costas Kounnas,  Paul Lauwers, Mari\'{a}n Lled\'{o}, \'{O}scar Maci\'{a}, Roeland Philippe, Tullio Regge, Diederik Roest, Frederik Roose, Jan Rosseel, Ergin Sezgin, Paolo Soriani, S.-Q. Su, Walter Troost, Fran\c{c}ois Vanderseypen, Stefan Vandoren, Jan-Willem van Holten, Kor Van~Hoof, Veeravalli Varadarajan and Bert Vercnocke who contributed significantly to the content in this book.
 \vspace{5mm}

\noindent We thank the many colleagues whose remarks have been helpful to improve the text: Marco Baggio, Nikolay Bobev, Alessio Celi, Jan De Rydt, Alessandra Gnecchi, Dietmar Klemm, Joris Van den Bergh and Marco Zagermann. Special thanks also to Stefan Theisen who encouraged us to write this review.
 \vspace{5mm}

\noindent Most of this review has been prepared in Leuven in the Institute for Theoretical Physics. We thank the staff and the researchers for the ideal environment that was always present there. We have enjoyed the presence of inspiring and very friendly colleagues.
 \vspace{5mm}

\noindent EL is supported by the Simons Foundation grant $\#$488659 (Simons Collaboration on the non-perturbative bootstrap).
 \vspace{5mm}

\noindent Last but not least, we thank our partners Giuliana and Laura for their support during this work.

%% file: ch1_BasicIngredients.tex
\chapter{Basic Ingredients}
\label{ss:ingredients}
\begin{quote}
\textbf{Abstract.}
We give an introduction to the book, discussing the role of the ${\cal N}=2$ theories, its geometric structure, and the superconformal tensor calculus. We also refer to other treatments. We then set out the plan of the book.\\
In the second part of the chapter
we introduce tools that are useful for the
construction of superconformal gauge theory and multiplets. We first discuss the catalogue of supersymmetric theories with 8 supercharges (Sect. \ref{ss:susy8}) and
their multiplets (Sect. \ref{ss:multiplets}). After a short Sect. \ref{ss:strategy} with the strategy, we discuss the conformal (Sect. \ref{sec:conf_group}) and then superconformal (Sect. \ref{ss:scgroups}) groups. The transformations of the fields under the conformal symmetry is also given in Sect. \ref{sec:conf_group}, while for the fermionic symmetries, this is discussed in a short Sect.~\ref{ss:rigidsuperconf}.
\end{quote}
\section{Introduction}

Theories with 8 supercharges, i.e. ${\cal N}=2$ in $D=4$, 5 and 6, occupy a special place in the atlas of supersymmetric quantum field theories and supergravities. They lead to very interesting geometries called `special manifolds', \index{special manifolds} including special K\"{a}hler and quaternionic-K\"{a}hler geometries.

Why are the
theories with 8 supersymmetries so interesting? The maximal
supergravities\footnote{The restriction is due to interacting field
theory descriptions, which e.g. in 4 dimensions does not allow fields
with spin larger than~2.} contain 32 supersymmetries. These are the ${\cal N}=8$
theories in 4 dimensions, and exist in spaces of Lorentzian signature
with at most 11 dimensions, i.e. (10,1) spacetime dimensions. If one
allows more time directions, 32 supersymmetries are possible in 12
dimensions with (10,2) or (6,6) signature. However, these theories allow
no matter multiplets.\footnote{We distinguish the multiplet that contains
the graviton and gravitini, and is determined by specifying the dimension
and the number of supersymmetries, and other multiplets, which we call
`matter multiplets'.}\index{matter multiplets} For the geometry, determined by the kinetic terms
of the scalars, this means that the manifold is fixed once the dimension
is given. For all theories with 32 supersymmetries this is a symmetric
space.

Matter multiplets are possible if one limits the number of
supersymmetries to 16 (thus ${\cal N}=4$ in 4 dimensions). Theories with 16 real supersymmetries exist
up to 10 dimensions with Lorentzian signature. In this case, the geometry
is fixed to a particular coset geometry once the number of
matter multiplets that are coupled to supergravity is given.

The situation becomes more interesting if the number of supersymmetries
is~8. Now there are functions, which can be varied continuously, that
determine the geometry. This makes the geometries much more interesting.
Of course, if one further restricts to 4 supersymmetries, more geometries
would be possible. In 4 dimensions, e.g., general K\"{a}hler manifolds
appear. For 8 supersymmetries, these are restricted to `special K\"{a}hler
manifolds',\index{special K\"{a}hler} determined by a holomorphic prepotential~\cite{deWit:1984pk}. \index{prepotential}
This restriction makes the class of manifolds very interesting
and manageable. The holomorphicity is a useful ingredient, and was e.g.
essential to allow the solution of the theory in the Seiberg--Witten
model~\cite{Seiberg:1994rs,Seiberg:1994aj}. The theories with 8
supersymmetries are thus the maximally supersymmetric ones that are not
completely determined by the number of fields in the model, but allow
arbitrary functions in their definition, i.e. continuous deformations of
the metric of the manifolds.
\vspace{10pt}

A short account of the necessary techniques and results of this book has been given in Chaps. 20 and 21 of \cite{Freedman:2012zz}. The scope of this work is to elaborate further on these results, as well as allowing a more complete and detailed explanation of these techniques. Specifically, we will consider ${\cal N}=2$ supergravities in 4, 5 and 6 dimensions, and at the end pay special attention to the $D=4$ framework. We use the terminology ${\cal N}=2$ in 6
dimensions for what is also called $(1,0)$ in 6 dimensions. Indeed, in 6
dimensions one can have chiral and antichiral real supersymmetry generators, and the nomenclature can thus be done by giving multiples of 8 real chiral, namely the minimal spinor of
$\Spin(5,1)$. Also for ${\cal N}=2$ in 5 dimensions, the minimal spinor of
$\Spin(4,1)$ has 8 real components. We denote these theories still as ${\cal N}=2$
because for practical work we always use doublets of 4-component spinors. Note also
that for other signatures (2 or 3 time directions in $D=5$) one can impose Majorana conditions \index{Majorana}such that  only 4 of them survive (${\cal N}=1$). But for Minkowski
signature,\footnote{In this review, we only consider Minkowski signature of spacetime.\index{signature of spacetime} In the literature, also other signatures are discussed, e.g. Euclidean signature in a series of papers  \cite{Cortes:2003zd,Cortes:2005uq,Cortes:2009cs,Cortes:2015wca}, and special geometry has also been defined for other signatures \cite{Lledo:2006nr,Sabra:2017xvx,Gall:2018ogw,Cortes:2019mfa}.} one can only have (symplectic) Majorana conditions, \index{symplectic Majorana} which need
doublets of 4-component spinors. The basic properties of the spinors are
repeated in Appendix~\ref{app:gammaspinors}. As well for $D=4$, 5 and 6 we have an
$\SU(2)$ automorphism group of the spinors, and for 4 dimensions the
automorphism group has an extra $\U(1)$, important for the K\"{a}hler geometry that we will discuss.

The geometry described by the scalar fields in these theories is a main topic in this book and for these theories in general. This geometry is a direct product of two parts: the geometry of the scalars of vector multiplets and the one of the scalars of hypermultiplets.
Even the supergravity couplings do not mix these independent parts
\cite{deWit:1984rz}.
That hypermultiplets in ${\cal N}=2$ rigid supersymmetric theories lead to hyper-K\"{a}hler manifolds was already known since~\cite{Alvarez-Gaume:1981hm}.
In supergravity theories these geometries are upgraded to quaternionic \cite{Bagger:1983tt} geometries.\footnote{Strictly speaking: to quaternionic-K\"{a}hler geometry, which means that there is a metric. For the difference between these manifolds and the used terminology we refer to the review \cite{Bergshoeff:2005di}.} The fact that actions for vector multiplets can be constructed from an arbitrary holomorphic function in supersymmetry has been found first for rigid supersymmetry in \cite{Sierra:1983cc}. For the coupling in supergravity this appeared first in  \cite{deWit:1984pk}. In the following years it was called `K\"{a}hler manifolds of restricted type', see e.g.  \cite{Cremmer:1985hc}.
 Later, Strominger \cite{Strominger:1990pd} gave it the name `special K\"{a}hler geometry'. It was then also recognized that this is the geometry of the moduli of Calabi--Yau threefolds. In the first years only special coordinates were used. In  \cite{Castellani:1990zd} a start was made to formulate the geometry in arbitrary coordinates.

At the start of the second superstring revolution, people first considered models with ${\cal N}=2$ and special geometry  \cite{Seiberg:1994rs,Seiberg:1994aj} because of the restrictive nature (holomorphicity) of these theories. The concept of duality
became very important at that time. These dualities in ${\cal N}=2$, $D=4$ have
a natural formulation in the context of superconformal tensor calculus.

Superconformal tensor calculus\index{superconformal tensor calculus} is another main topic of this book.
It has been the basis of the first constructions of general matter
couplings. It was initiated in ${\cal N}=1$ due to the work of S. Ferrara, M.
Kaku, P.K. Townsend and P. van
Nieuwenhuizen~\cite{Ferrara:1977ij,Kaku:1978nz,Kaku:1978ea}. The extra
symmetries of the superconformal group give an advantage over the direct
super-Poincar\'{e} approach since many aspects of the theory get a clear
structure. In fact, the natural vectors in which the dualities have to be
formulated are the multiplets of the superconformal tensor calculus. In
this approach, the superconformal symmetry is used as a tool to obtain the
theories that have super-Poincar\'{e} symmetry. All the super-Poincar\'{e}
theories are constructed as broken superconformal theories.
Lately the conformal symmetry has gained importance. There is of course
its interest in AdS/CFT correspondence, although in that context one
mostly considers rigid conformal symmetry. But it may also be interesting
to consider which parts of the supergravity theory are explicitly
determined by the breaking of the superconformal invariance, and which
parts are generically determined by the superconformal structure. Thus the superconformal
approach may be more interesting than just as a tool to obtain
super-Poincar\'{e} results.

The tensor calculus clarifies also the off-shell structure, and as such has been used in the developments of `supersymmetric localization', where the auxiliary fields in the conformal calculus play an important role. We refer to \cite{Pestun:2016zxk} for an overview.

The conformal tensor calculus is not the only one to obtain the theories that we consider. There are several superspace methods\footnote{We thank G. Tartaglino-Mazzucchelli for his assistance in this overview.} developed for ${\cal N}=2$.
Standard superspace techniques for $D=4$, ${\cal N}=2$ conformal supergravity have been described long ago, see \cite{Howe:1981gz,Kuzenko:2009zu} and references therein. For $D=5$ and $D=6$ they have been developed in \cite{Kuzenko:2008wr,Linch:2012zh}. An approach that is closer to the superconformal methods is the `conformal superspace' method.
This has been developed in \cite{Butter:2009cp} first for ${\cal N}=1$, and extended to $D=4$, ${\cal N}=2$ in  \cite{Butter:2011sr,Butter:2012xg}, for $D=5$ in \cite{Butter:2014xxa}
and for $D=6$ in \cite{Butter:2016qkx,Butter:2017jqu}. In this method the entire superconformal symmetry is gauged as in the conformal tensor calculus, and the connection with the approach in components was developed in detail in
 \cite{Butter:2012xg,Butter:2014xxa,Butter:2017jqu}. Especially for hypermultiplets, the harmonic and projective superspace approaches are alternatives allowing off-shell matter couplings at the expense of an infinite number of auxiliary fields, packed in dependences on extra superspace coordinates. The harmonic superspace for rigid ${\cal N}=2$ supersymmetry has been developed in  \cite{Galperin:1984av} and reviewed in \cite{Galperin:2001uw}, while projective superspace was developed in \cite{Karlhede:1984vr,Lindstrom:1987ks,Lindstrom:1989ne} and is reviewed in \cite{Lindstrom:2008gs,Kuzenko:2010bd}.
For supergravity the harmonic and projective approaches have been developed in
 \cite{Galperin:1987em,Galperin:1987ek,Kuzenko:2007cj,Kuzenko:2007hu,Kuzenko:2008wr,Kuzenko:2008ep,Kuzenko:2009zu,Butter:2014gha,Butter:2014xua,Butter:2015nza}.

Furthermore there is the rheonomic
approach that has been used for another formulation of the general ${\cal N}=2$
theories in four dimensions~\cite{fresoriabook,Andrianopoli:1997cm}, and in five
dimensions~\cite{Ceresole:2000jd}. A relation with the conformal approach has been obtained in \cite{Cribiori:2018xdy}.
These methods (including the conformal tensor calculus) do not prove uniqueness of the ${\cal N}=2$ theories. Recently some uniqueness proofs (at least for the pure supergravities) have been given in \cite{Boulanger:2018fei} based on BRST-BV methods \cite{Batalin:1983jr,Henneaux:1990jq,Barnich:1993vg,Gomis:1995he}. They are consistent with the results found with other methods.

\bigskip

The plan of the book is as follows. In the rest of this chapter we will
introduce some basic concepts: the supersymmetries and multiplets and the superconformal groups.
In Chap. \ref{ss:Weyl} we gauge these groups by a gauge multiplet, which is called the Weyl multiplet.
This gauging of the spacetime symmetries is not straightforward and needs curvature constraints.

In Chap. \ref{ss:matter} the matter multiplets are introduced: the vector multiplets in 6, 5 and 4 dimensions in that order, and the hypermultiplets, whose structure is not dependent on the spacetime dimension. We have chosen to first present the multiplets independent of the actions. We thus give the definition of the fields, possible reality conditions, and how they represent the superconformal algebra. Then we present the construction of the superconformal-invariant actions for vector and for hypermultiplets. Another alternative order is possible in Chap. \ref{ss:matter}: those mainly interested in vector multiplets can go from Sect. \ref{vectmult} directly to Sect. \ref{ss:actvector4} for $D=4$ or Sect. \ref{ss:actVMD5} for $D=5$. On the other hand after discussing hypermultiplets in Sects. \ref{ss:hypermultiplets} and \ref{ss:hypermultipletConf} one can also directly go to the discussion of its actions in Sect. \ref{finsec_hypers}.

After the partial gauge fixing (i.e. gauge fixing of the symmetries that are not in the super-Poincar\'{e} algebra) the actions of vector- and hypermultiplets are connected. The total action is considered in Chap.~\ref{ss:gaugefix}. We first discuss in Sect.~\ref{ss:pureN2} pure supergravity in 4 dimensions, obtained by adding to the minimal field representation (Weyl multiplet + a compensating vector multiplet) a second compensating multiplet. Here we consider three possibilities. In the other parts of the book we always use the hypermultiplet as second compensating multiplet.
We also consider the reduction of these pure supergravity theories to ${\cal N}=1$ in Sect.~\ref{ss:N2toN1}. Then we go back to the complete matter-coupled supergravities. We discuss convenient coordinates for the scalars in these theories. The vector multiplets are discussed more thoroughly in preparation of the following chapter. The $D=5$ and $D=6$ theories are discussed shortly, referring to the techniques explained in detail for $D=4$.

The structure of the special geometries that result are studied in Chap.~\ref{ss:specialgeom}. This structure is also characterized independently of its supergravity construction. The isometries (gauged or non-gauged) are discussed. In Sect. \ref{ss:emcharges} the (charged) black holes are discussed as an application and the role of the symplectic transformations in this context becomes apparent. In Sect. \ref{ss:qKmanifolds} we
will also give an introduction to the properties of quaternionic-K\"{a}hler spaces independent of supergravity. The matter couplings with ${\cal N}=2$ in 5 and 6 dimensions are very much related to those in four dimensions as we discuss in Sect.~\ref{ss:HomSymSp}.
We discuss the differences and specific properties of the ${\cal N}=2$ theories in
these dimensions. Very special geometry will then show up in relations
between the scalar manifolds defined by these theories. This is further
clarified in relations between the homogeneous spaces.

Chapter \ref{ch:finalresults} gives the final super-Poincar\'{e} theory. In Sect. \ref{ss:finalresults} the actions, supersymmetry transformations and algebra are written explicitly for $D=4$. Section~\ref{ss:finalD5} is a bit similar for $D=5$, though we are less complete there, referring to \cite{Bergshoeff:2004kh} for the full results in the same notation.
In the final remarks in Sect.~\ref{ch:finalremarks} we stress the choices and restrictions that we made throughout the book. This includes then also references to papers that go beyond the present text, and the outlook to applications.

The notation in this book is the same as in \cite{Freedman:2012zz}. But in Appendix \ref{appendix_notations} this is explained more explicitly for the theories that we consider. We repeat the use of indices, spinor notations, $\SU(2)$ conventions, and the relations between spinors in $D=6$, $D=5$ and $D=4$. A short summary of simple superalgebras is given in Appendix \ref{app:groups}. Finally for convenience of the reader who also wants to use the original papers, we give in Appendix~\ref{ss:compareN2d4conv} the translations of our notation to those of these papers.

Readers may also choose to start immediately with Chap. \ref{ch:finalresults}. There we repeat the main ingredients and that chapter contain references to earlier parts when necessary.

\section{Supersymmetric Theories with 8 Real Supercharges}
\label{ss:susy8}
As mentioned in the introduction, we will be focusing on theories with $8$ supercharges in $D=4$, 5, 6, being the latter the maximal dimension for theories
with $8$ supercharges.\footnote{Of course one can introduce 8 supercharges in 1 or 2 dimensions,
where the elementary spinors have just one component. In 3 dimensions,
gamma matrices are $2\times 2$ matrices, and the theories with 8
supercharges are ${\cal N}=4$ theories where the spinors satisfy a reality
condition (a Majorana condition).}  In fact, one could first construct $D=6$ theories
and then derive several results for $D<6$ from dimensional reduction.
This programme has been started for the superconformal theories
in~\cite{Kugo:2000hn,Kugo:2000af}. There are many aspects that can be
treated at once for $D=4,5,6$. The treatment of the Weyl multiplets is
for a large part the same in these dimensions. Hypermultiplets (multiplets with scalars and
spinors only) do not feel the difference of dimension. This difference is relevant for vectors and tensors, which under dimensional reduction
decompose in several representations of the Lorentz algebra. However, only a
subset of the couplings of vectors multiplets in $D=4$ and $D=5$ can be
obtained from dimensional reduction, as we will explicitly demonstrate in
Sect.~\ref{ss:HomSymSp}. Spinors are also treated differently, and this has an immediate consequence on the supersymmetry algebra.

In 4 dimensions, the supersymmetries are represented by Majorana spinors, but in practice one can also use chiral spinors. We refer to \cite[Chap. 3]{Freedman:2012zz} for the notation and definitions of the
properties of the spinors. A supersymmetry operation is represented as
\begin{eqnarray}
 && \delta (\epsilon )=\bar \epsilon ^i \cQ_i +\bar \epsilon _i \cQ^i\,, \nonumber\\
 && \epsilon ^i=\gamma _*\epsilon ^i\,,\qquad  \cQ_i=\gamma _* \cQ_i\,,\qquad
  \epsilon _i=-\gamma _*\epsilon _i\,,\qquad  \cQ^i=-\gamma _* \cQ^i\,,
 \label{deltaEpsQ4}
\end{eqnarray}
where $\gamma _*=\rmi \gamma_0\gamma_1\gamma_2\gamma_{3}$ is Hermitian with $(\gamma_*)^2=\unity$  and
$\cQ$ is the supersymmetry operator that acts on the fields, e.g. if $\delta (\epsilon )X=\ft12\bar \epsilon ^i \lambda _i$, then $\cQ_i
X=\ft12\lambda _i$ and $\cQ^i X=0$.

In 5 dimensions, one uses symplectic Majorana spinors.\index{symplectic Majorana} The reality rules
of Appendix \ref{app:gamma5} imply that we have to insert a factor
$\rmi$ in $\delta(\epsilon)$ (see also the different meaning of the position of the indices
$i$, as explained in Appendix~\ref{app:gamma5}):
\begin{equation}
 \delta (\epsilon )=\rmi\bar \epsilon ^i \cQ_i\,.
 \label{deltaEpsQ5}
\end{equation}

In 6 dimensions, symplectic Majorana--Weyl spinors \index{symplectic Majorana--Weyl} can be used. In this
case we have
\begin{eqnarray}
&& \delta (\epsilon )=\bar \epsilon ^i \cQ_i\,,\nonumber\\
&& \gamma _*\epsilon ^i=\epsilon ^i\,,\qquad \gamma _* \cQ_i=- \cQ_i\,,
 \label{deltaEpsQ6}
\end{eqnarray}
where $\gamma _*$ is given in (\ref{gamma*D6}).

The properties of the spinors also imply how the translations appear in
the anticommutator of two supersymmetries (the overall real factor is a
matter of choice of normalization):
\begin{eqnarray}
 D=4 & : & \left\{ \cQ_\alpha ^i,\cQ_{\beta j}\right\}
  =-\ft12\left( \gamma _aP_L\right) _{\alpha \beta } \delta ^i_j P^a \nonumber\\
 D=5 \mbox{ and }D=6 & : & \left\{ \cQ_\alpha ^i,\cQ_\beta ^j\right\}=-\ft 12 \varepsilon ^{ij}
 \left( \gamma _a\right) _{\alpha \beta }  P^a \,.
 \label{QQP}
\end{eqnarray}
On the use of spinor indices, \index{spinor indices} see\cite[Sect. 3.2.2]{Freedman:2012zz}. The
normalization of the spinor generators is chosen to agree with much of the original literature in each case. See Appendix \ref{ss:compareN2d4conv}.
 \bexer
 Verify that, given (\ref{QQP}) one finds
\begin{equation}
  \left[ \delta (\epsilon _1),\delta (\epsilon _2)\right]
  =\xi ^a(\epsilon _1,\epsilon _2) P_a\,,
 \label{deltaEps12P}
\end{equation}
with
\begin{eqnarray}
 D=4 & :  & \xi ^a(\epsilon _1,\epsilon _2)=\ft12\left( \bar \epsilon_2 ^i\gamma ^a\epsilon _{1i}
+\bar \epsilon_{2i}\gamma^a\epsilon _1^i\right) \nonumber\\
 D=5 \mbox{ and }D=6 & : & \xi ^a(\epsilon _1,\epsilon _2)=\ft12\bar \epsilon _2\gamma^a\epsilon
 _1\,.
 \label{xiaeps12}
\end{eqnarray}
   \eexer
\subsection{Multiplets} \label{ss:multiplets}
On general grounds, in any supersymmetric theory, the bosonic and fermionic number of degrees of freedom \index{degrees of freedom} must be equal.\footnote{Different proofs of this statement can be given. A general argument has been given in \cite{Sohnius:1985qm}. This has been further discussed in \cite[App. 6B]{Freedman:2012zz}. For on-shell states in $D=4$, see also \cite[Sect. 6.4.1]{Freedman:2012zz}. For solutions of field equations that preserve supersymmetry, a detailed explanation is in Appendix~B of \cite{Binetruy:2004hh}.}
More precisely, this rule applies to transformations between states where the $Q$-transformation of a fermionic state gives a bosonic state, and vice versa, with symbolically $\{Q,Q\}=P$,
and $P$ invertible. The invertible $P$ can refer to translations in rigid supersymmetry or general coordinate transformations in supergravity. On the contrary, this rule does not apply to nonlinear realizations where some fields transform to constants.

Counting the number of degrees of freedom may be very subtle, and for this reason we need to be more precise about the correct procedure that we should adopt. Firstly, we will see that the anticommutator of supersymmetries often does not only contain the translations, but also gauge transformations. Therefore, the argument
about equal number of bosonic and fermionic states is only correct up to
gauge transformations. In other words, in general we should subtract the gauge degrees of
freedom in order to perform the counting.\footnote{General coordinate
transformations are also local gauge transformations that are more
general than fixed translations. So the general coordinate-equivalent states should also be subtracted.}

Secondly, there are two possible countings of degrees of freedom:  on-shell counting \index{on-shell counting} and off-shell counting. \index{off-shell counting} When the supersymmetry algebra is applied to the physical states determined by the equations of motion we denote this as `\textit{on-shell counting}'. In that case, the consistency requires that there are as many bosonic as fermionic physical states. On the other hand, in order to couple many different actions to the multiplets that we are constructing, we usually want the supersymmetry algebra to be satisfied independently of the equations of motion. In other words, we consider an \emph{off-shell closed} algebra and adopt \emph{off-shell counting} of degrees of freedom.
More details on the concepts of on-shell and off-shell degrees of freedom are explained in \cite{Freedman:2012zz} in the beginning of Chap. 4 and in Appendix 6B.

Let us first consider on-shell states. The massless on-shell multiplets for extended supersymmetry, on which we will later impose the superconformal algebra, are classified in the classical reference \cite{Strathdee:1987jr}. The multiplets of $\mathcal{N} = 2$ for the dimensions in which we are interested are given in Table \ref{tbl:onshellmult}.
The on-shell components for massless fields form representations of $\SO(D-2)$ with an internal $SU(2)$ group (which is the $R-$symmetry group of $\mathcal{N}=2$ theories and is labelled by $i$-type indices).
The representation content is indicated then as its multiplicity in $(\SO(D-2),\SU(2))$.

\begin{table}[!t]  
\caption{\it Massless on-shell representations.  The representation content is
  indicated as its multiplicity in $(\SO(D-2),\SU(2))$. The $\SU(2)$ representation can also be identified from the $i$-type indices.
   For 6 dimensions, $\SO(4)=\SU(2)\times \SU(2)$, and the
   corresponding decomposition is written. The non-symmetric
   representations are either chiral (chirality is indicated on the field)
   or they are antisymmetric tensors with
  self-dual (+) or anti-self-dual ($-$) field strengths. Of course, the
  two-form indicated here is not $\pm$ self dual, only its field strength is.}
  \label{tbl:onshellmult}
\begin{center}
$\begin{array}{|c|cccc|l|}\hline
D=4  & e_{\mu}^{a} & \psi _\mu ^i & V_\mu  &  & \mbox{Gravity} \\
\SO(2)& (2,1)          & (2,2)        & (2,1)  &  &  \\[2mm]
  & W_\mu  & \lambda_i  & X     & X^* & \mbox{Vector multiplet} \\
  & (2,1)  & (2,2)     & (1,1) & (1,1) &  \\[2mm]
  &  & \zeta  & q^i   &  & \mbox{Hypermultiplet} \\
  &  & (2,1)  & (1,2) &  &  \\
\hline
D=5  & e_{\mu}^{a} & \psi _\mu ^i & V_\mu  &  & \mbox{Gravity} \\
\SO(3)& (5,1)          & (4,2)        & (3,1)  &  &  \\[2mm]
  & W_\mu  & \lambda ^i & \sigma &  & \mbox{Vector multiplet}  \\
  & (3,1)  & (2,2)      & (1,1) &  & \\[2mm]
  &  & \zeta  & q^i  &  & \mbox{Hypermultiplet} \\
  &  & (2,1)  & (1,2)&  &  \\
\hline
D=6  & e_{\mu}^{a} & \psi _{\mu L} ^i & B^-_{\mu \nu }&  & \mbox{Gravity} \\
\SO(4)& (3,3;1)          & (2,3;2)      & (1,3;1)     &  &  \\[2mm]
  & B^+_{\mu \nu } & \psi_R ^i & \sigma  &  & \mbox{Tensor multiplet} \\
  & (3,1;1)      & (2,1;2) & (1,1;1)   &  &  \\[2mm]
  & W_\mu        & \lambda ^i &  &  & \mbox{Vector multiplet}  \\
  & (2,2;1)        & (1,2;2)   &  &  & \\[2mm]
  &  & \zeta  & q^i  &  & \mbox{Hypermultiplet} \\
  &  & (2,1;1)  & (1,1;2)&  &  \\
\hline
\end{array}$
\end{center}
\end{table}

The vectors ($V_{\mu}$ or $W_\mu $) are a fundamental representation of $\SO(D-2)$ and have therefore $D-2$ on-shell components (see the analysis in\cite[Sect.4.1.2]{Freedman:2012zz}). The gravitons $e_{\mu}^a$ are (after taking Lorentz symmetry into account) described as the symmetric $g_{\mu \nu }$. The on-shell components are symmetric traceless tensors so one counts $\ft12 (D-2+1)(D-2) - 1 = D(D-3)/2$ (see the analysis in\cite[Sect.8.2]{Freedman:2012zz}).

In order to count the degrees of freedom for spinors one must take into account that their characteristics differ in different dimensions. It is useful to first discuss the off-shell degrees of freedom and then go to the on-shell degrees of freedom which just differ by a factor $\ft12$. The number of complex components of a spinor in dimension $D$ is given by $2^{[D/2]}$. In $D=4$ we can furthermore impose consistent Majorana conditions, so that the minimal spinor has $4$ real components, or, equivalently consider one helicity with 2 complex components (and the other helicity is then related by complex conjugation). For $\mathcal{N}=2$, spinors in 4 dimensions will be taken to be part of an $\SU(2)$ doublet with index $i$, leading to 8 real components. For $D=5$ it is not possible to define the Majorana condition in a consistent manner and one must use the symplectic Majorana condition. \index{symplectic Majorana} This does not lower the number of components but rewrites them as a symplectic pair of Majorana spinors related through charge conjugation. Thus one finds that a spinor in 5 dimensions has 4 complex components or $2 \times 4$ real components. For $D = 6$ one still needs to use symplectic Majorana spinors but, moreover, one can impose the Weyl condition, which projects half of the components of the spinors, thus leaving us with $\ft12$ the number of complex components. So, one gets 8 complex components, which can be described by the symplectic condition as $2 \times 8$ real components, and due to the Weyl condition can be projected to $2 \times 4$ real components. Hence in all three cases this leads to 8 real off-shell degrees of freedom. Going from off-shell to on-shell degrees of freedom means multiplying the number of real components with $\ft12$ leading to 4 real on-shell degrees of freedom for the three cases that we consider.

The gravitini are described by massless vector--spinors. Their off-shell degrees of freedom are reduced by the supersymmetry, so that for $D=4$, 5, 6 we have $8(D-1)$ components. The on-shell degrees of freedom are $\gamma$-traceless in $D-2$ dimensions (see \cite[Sect.5.1]{Freedman:2012zz}). Therefore one has $(D-2-1)$ vector degrees of freedom where the $D-2$ comes from the vector representation and the $-1$ comes from the $\gamma$-traceless condition. One must then multiply these vector degrees of freedom with the spinor degrees of freedom, which leads to $4(D-3)$ degrees of freedom.

The vector multiplets in different dimensions are related by dimensional reduction. In the highest dimension ($D=6$) one has a vector $W_{\mu}$ and a symplectic pair of spinors $\lambda^i$. The scalars in lower dimensions can be considered as some of the vector components of $D=6$ after the reduction. For 5 dimensions one has an additional real scalar $\sigma=W_5$ and for 4 dimensions an additional complex scalar $X = \ft12 (W_4 - \rmi\sigma)$. The notion that there are scalar fields present in the vector multiplet for $D=4,5$ will give rise to special geometries described by these scalar fields. We will describe this in detail in Chap. \ref{ss:specialgeom}.

Looking at the hypermultiplets in Table~\ref{tbl:onshellmult} one notices that the scalars $q^i$ transform as a doublet under $\SU(2)$. Following from the fact that the $\SU(2)$-transformations are complex this means that the scalars cannot be real.
Taking them complex is equivalent to taking 4 real scalars $q^X$, which will due to supersymmetry define 3 complex structures. The scalars
naturally combine into quaternions as we will describe in Sect. \ref{ss:hypermultiplets}. The corresponding geometry will thus be of the type of a quaternionic manifold, which we will exhibit in Sect.~\ref{ss:qKmanifolds}.

Remark that here the on-shell massless multiplets have been mentioned. The
same multiplet may be represented by different off-shell multiplets. In 4
dimensions, physical scalars may be dualized to antisymmetric tensors.
E.g. when one of the scalars of a vector multiplet is replaced by an
antisymmetric tensor, we obtain the \textit{vector--tensor
multiplet}~\cite{Sohnius:1980it,deWit:1995zg,Claus:1998fk}
\begin{equation}
  \mbox{vector---tensor multiplet:}\  V_\mu ,\, \lambda ^i,\,  \phi ,\,
   B_{\mu \nu }\,.
 \label{vectortensormult}
\end{equation}
When one of the scalars of a hypermultiplet is replaced by an
antisymmetric tensor, then this gives the so-called \textit{linear
multiplet} (the name is due to its relation with a superfield that has a
linear constraint)
\begin{equation}
  \mbox{linear multiplet:}\ \varphi ^i,\, L^{(ij)}, \, E_a\,.
 \label{linearmult}
\end{equation}
We write the antisymmetric tensor here as a vector $E_a$ that satisfies a
constraint $\partial ^a E_a=0$, such that $E_a= \varepsilon
_{abcd}\partial ^b E^{cd}$. In this way, it can be generalized to $D=5$
and $D=6$ in a similar way. In these cases the $E_a$ is the field
strength of a 3-form or 4-form, respectively.

In 5 dimensions a vector is dual to a 2-index antisymmetric tensor.
Therefore the vector multiplets can be dualized to antisymmetric tensor
multiplets:
\begin{equation}
D=5  \mbox{ antisymmetric tensor multiplet:}\ H_{\mu \nu }, \,  \lambda
^i, \, \phi\,.
 \label{d5astensormult}
\end{equation}
For non-abelian multiplets, the two formulations are not equivalent. This
has been investigated in detail in a series of papers of G\"{u}naydin and
Zagermann~\cite{Gunaydin:1999zx,Gunaydin:2000xk,Gunaydin:2000ph} and further worked out in \cite{Ceresole:2000jd,Bergshoeff:2004kh}. We will come back to this in Sect. \ref{ss:finalD5}. The above overview is not exhaustive.

\subsection{The Strategy} \label{ss:strategy}
Our aim is to study the transformation laws and actions for the ${\cal N}=2$ multiplets, coupled to supergravity. As mentioned in the introduction,
there are several ways to accomplish this. One possible way is to use in intermediate steps symmetries that will not be present in the final action. Auxiliary enhancement of symmetry has the particular advantage that it facilitates the construction of the theories, clarifying a lot of their structures.

The extra symmetry will be the superconformal symmetry and the method we use goes under the name of \emph{superconformal tensor calculus}. In the classical work of Coleman and
Mandula~\cite{Coleman:1967ad} and its supersymmetric extension by Haag--{\L}opusza\'nski--Sohnius~\cite{Haag:1975qh}, it is proven that the superconformal groups is
 the largest spacetime symmetry allowed for a non-trivial quantum field theory.\footnote{Coleman and Mandula prove that the largest spacetime group
that is allowed without implying triviality of all scattering amplitudes is the conformal group.  This theorem is valid under some assumptions, like the analyticity of the elastic two-body scattering amplitudes. Haag, {\L}opusza\'nski and Sohnius base their analysis on the previous result and use group-theoretical arguments (essentially Jacobi identities). For an extension of the theorem to strongly coupled quantum field theories, see e.g. \cite{Maldacena:2011jn,Alba:2015upa}.} Although our motivation and use of the group is completely different, this gives an indication that the use of this group may be the most advantageous strategy.
Over the years we got more
convinced that indeed the use of conformal symmetries is a very useful and
clarifying method. Analyzing the steps that are taken in local
superspaces, we see that after using part of the superspace constraints,
the remaining part that leads to more insight is equivalent to the
structure that we use in the `superconformal tensor calculus'.

What we have in mind can be illustrated first for pure gravity. We show
how Poincar\'{e} supergravity is obtained after gauge fixing a conformal
invariant action. The details of this example will come back in
Sect.~\ref{ss:exampleBosComp}. We now just give the general idea. The
conformal invariant action for a scalar $\phi $ (in 4 dimensions)
is\footnote{Note that the scalar here has negative kinetic energy, and
the final gravity action has positive kinetic energy.}
\begin{eqnarray}
 {\cal L} & = & \sqrt{g}\left[  \ft12(\partial _\mu \phi )(\partial ^\mu \phi )+\ft1{12}R\phi
 ^2\right] \,,
 \nonumber\\
 \delta \phi  & = & \lambda _{\rm D}\phi \,,\qquad \delta g_{\mu \nu }=-2\lambda
 _{\rm D} g_{\mu \nu }\,,\label{confScalar}
 \end{eqnarray}
where the second line gives the local dilatation symmetry that leaves
this action invariant. Now, we can gauge fix this dilatation symmetry by
choosing\footnote{A gauge fixing can be interpreted as choosing better
coordinates such that only one field still transforms under the
corresponding transformations. Then, the invariance is expressed as the
absence of this field from the action. In this case we would use $g'_{\mu
\nu} =\frac{\kappa ^2}{6}g_{\mu \nu }\phi ^2$ as $D$-invariant metric.
One can check that this redefinition also leads to~(\ref{purePoincare})
in terms of the new field.} the gauge
\begin{equation}
  \phi =\frac{\sqrt{6}}{\kappa}\,.
 \label{gaugefixkappa}
\end{equation}
 This leads to the pure Poincar\'{e} action
\begin{equation}
  {\cal L}=\ft1{2\kappa ^2}\sqrt{g}R\,.
 \label{purePoincare}
\end{equation}
Pure Poincar\'{e} is in this way obtained from a conformal action of a scalar
after gauge fixing. This scalar, which we will denote further as
`compensating scalar', thus has no physical modes. Note also that the
mass scale of the Poincar\'{e} theory is introduced through the gauge
fixing~(\ref{gaugefixkappa}).

What we have seen is (1) the use of conformal symmetry, (2) construction of
a conformal invariant action, (3) gauge fixing of superfluous symmetries.

In the remaining part of this chapter, we will familiarize ourselves
first with the conformal group, and first of all as a rigid symmetry. We
will then take a look at possible superconformal groups, and repeat the
basis rules for gauging symmetries. In Chap.~\ref{ss:Weyl} we will
learn how to gauge the superconformal group. The step (2) of the above
overview involves the superconformal construction of multiplets and their
action. This will be the subject of Chap.~\ref{ss:matter}. The step (3)
will be taken in Chap.~\ref{ss:gaugefix}, which will allow us to obtain
the physical theories, and be the starting point for analysing their
physical and geometrical contents.
\subsection{Rigid Conformal Symmetry}
\label{sec:conf_group}

Conformal symmetry \index{conformal symmetry} is defined as the symmetry that preserves angles. Under a conformal transformation, the metric may change by a Weyl factor.
In terms of infinitesimal transformations $x'^\mu=x^\mu+\xi^\mu(x)$, for a given flat metric $\eta_{\mu\nu}$, the latter condition implies the following `conformal Killing equation' \index{conformal Killing equation} for $\xi^\mu\equiv \xi^\mu(x)$
\begin{equation}\label{Kill_eq}
\partial_{\mu}\xi_{\nu}+\partial_{\nu}\xi_{\mu}-\frac{2}{D} \eta_{\mu\nu}\partial_\rho \xi^\rho=0\,.
\end{equation}
 \bexer
 Get more insight in the meaning of the statement that these are the
 transformations that preserve `angles'.
 To consider angles, we should consider two variations of the same
spacetime point. Consider the vectors from $x$ to $x^\mu +(\Delta_1)^\mu
$ and another one to $x^\mu +(\Delta _2)^\mu $, where the deformations
are considered to be small. The angle between these two is
\begin{equation}
 \cos^2 \theta =\frac{\left(\Delta_1\cdot \Delta _2\right) ^2}
 {\left(\Delta_1\cdot \Delta _1\right) \left(\Delta_2\cdot \Delta _2\right)}
 \label{deftheta}
\end{equation}
Now we perform a spacetime transformation that takes a point $x$ to
$x'(x)=x+\xi (x)$. Then the first vector will be between $x' (x)$ and
$x'(x+\Delta_1)=x'(x)+\Delta _1\cdot \partial x'(x)$. The new vector is
thus
\begin{equation}
 \Delta' _1=\Delta _1\cdot \partial x'(x)=\Delta _1+ \Delta _1\cdot\partial \xi(x).
 \label{Delta1prime}
\end{equation}
We thus find that
\begin{equation}
 \Delta' _1 \cdot\Delta' _2= \Delta _1 \cdot\Delta _2 +\Delta _1^\rho  \left(\partial
 _\rho  \xi ^\mu (x)\right)\eta _{\mu \nu }\Delta _2^\nu
+\Delta _1^\mu\eta _{\mu \nu }\Delta _2^\rho  \left(\partial
 _\rho  \xi ^\mu (x)\right),
 \label{innerprodprime}
\end{equation}
where we added indices for clarity. If the last factor gives just a
scaling, i.e. if
\begin{equation}
 \Delta _1^\rho  \left(\partial
 _\rho  \xi ^\mu (x)\right)\eta _{\mu \nu }\Delta _2^\nu
+\Delta _1^\mu\eta _{\mu \nu }\Delta _2^\rho  \left(\partial
 _\rho  \xi ^\mu (x)\right) = 2\Lambda _{\rm D}(x)\Delta _1^\mu\eta _{\mu \nu }\Delta
 _2^\nu,
 \label{scaleDelta12}
\end{equation}
then it is easy to see that (\ref{deftheta}) is invariant. Indeed, all
factors scale with the same coefficient as the scale factor has to be
evaluated at the same spacetime point.

The requirement (\ref{scaleDelta12}) amounts to the scaling of the
metric. Indeed, if a metric scales under spacetime transformations,
\begin{equation}
  \delta (\rmd x^\mu\eta _{\mu \nu } \rmd x^\nu )\equiv 2\rmd x^\rho
  \left( \partial _\rho  \xi ^\mu(x)\right)  \eta _{\mu \nu } \rmd x^\nu =
   2\Lambda _{\rm D}(x) (\rmd x^\mu \eta _{\mu \nu } \rmd x^\nu  ),
 \label{scalemetric}
\end{equation}
then replace $\rmd x$ in the above by $(\Delta_1+\Delta _2)$ and subtract
the diagonal terms. This leads to (\ref{scaleDelta12}) and hence to the
invariance of the angle.

Angles are thus preserved by the transformations that scale the metric
and these are the conformal transformations.

 \eexer

In $D=2$ with as non-zero metric\footnote{In Minkowski space $z=\frac{1}{\sqrt{2}}\left( x^1+x^0\right) $ and $\bar z=\frac{1}{\sqrt{2}}\left(x^1-x^0\right) $ are not each other complex conjugates. In Euclidean signature they are complex conjugates.} elements $\eta_{z\bar z}=1$,  the
Killing equations are reduced to Cauchy--Riemann equations: $\partial_z  \xi _z = \partial_{\bar z}
\xi_{\bar z} = 0$. Therefore any holomorphic (and anti-holomorphic) function is locally a conformal transformation ($ \xi_{\bar z}(z)$ and  $ \xi _z(\bar z) $) and the conformal algebra becomes infinite dimensional (Virasoro algebra). \index{Virasoro algebra}
In dimensions $D>2$ the conformal algebra is
finite-dimensional and the general solutions of \eqref{Kill_eq} are
\begin{equation}
\xi^\mu(x)=a^\mu +\lambda^{\mu\nu}x_\nu+\lambda_{\rm D} x^\mu
+(x^2\lambda_{\rm K}^\mu-2x^\mu x\cdot \lambda_{\rm K}) \,. \label{ximu}
\end{equation}
They contain $D$ translations $P_\mu$ corresponding to the parameters $a^\mu$, one dilatation $D$ \index{dilatation} related to the parameter $\lambda_{\rm D}$, $D(D-1)/2$ (pseudo-)Lorentz rotations $M_{\mu\nu}$ corresponding to the parameters $\lambda^{\mu\nu}$ and a set of $D$ `special conformal transformations' \index{special conformal transformation} $K_\mu$ related to the parameters $\lambda_{\rm K}^\mu $.
One may notice that the full set of conformal transformations contains $(D+1)(D+2)/2$ parameters.

The most general conformal transformation, $\delta_{\rm C}(\xi )$, can be expressed in terms of generators as follows:
\begin{equation}
\delta_{\rm C}(\xi)= a^\mu  P_\mu + \ft12\lambda^{\mu\nu}M_{\mu\nu}+\lambda_{\rm D} D +
\lambda_{\rm K}^\mu K_\mu   \,.
\end{equation}
With these transformations, one can obtain the conformal algebra. \index{conformal algebra} The non-zero commutators are:
\begin{eqnarray}
&&[M_{\mu\nu} , M^{\rho\sigma}]=
-4\delta_{[\mu}^{[\rho} M_{\nu]}{}^{\sigma]} \,, \nonumber\\
&& [P_{\mu} , M_{\nu\rho} ] =2\eta_{\mu[\nu} P_{\rho]}\,, \qquad 
 [K_{\mu} , M_{\nu\rho} ] = 2\eta_{\mu[\nu}K_{\rho]} \,, \nonumber\\
&& [P_{\mu} , K_{\nu} ]= 2 (\eta_{\mu\nu} D +  M_{\mu\nu}) \,,
\nonumber\\
&& [D , P_{\mu}  ]= P_\mu \,,\qquad \quad \quad
\ [D , K_{\mu}  ]=-K_\mu \,.     \label{confalg}
\end{eqnarray}

The conformal algebra defined on a flat space with signature $(p,q)$, is isomorphic to the $SO(p+1,q+1)$ algebra. In what follows we will consider a $D$ dimensional Minkowski space (whose signature is $(D-1,1)$ in our conventions), such that the conformal algebra is the $\SO(D,2)$ algebra.\footnote{In the 2-dimensional case
$\SO(2,2)=\SU(1,1)\times \SU(1,1)$ is realized by the finite subgroup of
the infinite dimensional conformal group, and is well known in terms of
$L_{-1}= \ft12 (P_0 - P_1)$, $L_0=\ft12(D+M_{10})$, $L_1=\ft12 (K_0 +
K_1)$, $\bar L_{-1}= \ft12 (P_0 + P_1)$, $\bar L_0=\ft12(D-M_{10})$, $\bar
L_1=\ft12 (K_0 - K_1)$. Higher order $L_n, |n| \geq 2 $ have no analogs
in $D > 2$. } Indeed one can define
\begin{equation}
M^{\hat \mu\hat \nu}=\begin{pmatrix} M^{\mu\nu}&\ft12(P^\mu-K^\mu)&
\ft12(P^\mu+K^\mu)\cr -\ft12(P^\nu-K^\nu)&0&- D\cr
-\ft12(P^\nu+K^\nu) &D &0
\end{pmatrix}
\,,
\end{equation}
where indices are raised w.r.t. the rotation matrices $M^{\hat\mu}
{}_{\hat\nu} $ with the metric
\begin{equation}
  \hat\eta={\rm diag~}(-1,1,...,1,-1)\,.
 \label{etaSOd2}
\end{equation}

When applied on fields, conformal transformations have `orbital parts' \index{orbital parts} and may have `intrinsic parts'.\index{intrinsic parts} We absorb the orbital parts of all generators in the conformal Killing vector $\xi^\mu(x)$, given by \eqref{ximu}. The general form of the conformal transformations of fields $\phi^i(x)$  is\footnote{The expression $\xi$ always determines the parameters $\{a^\mu,\lambda^{\mu\nu},\lambda_{\rm D},\lambda_{\rm K}^\mu\}$ as in  (\ref{ximu}).}
\begin{eqnarray}
\delta_{\rm C}(\xi) \phi^i(x)&=& \xi^\mu(x)\partial_\mu \phi^i(x) -\ft12
\Lambda_M{}^{\mu\nu}(x)\, m_{\mu\nu}{}^i{}_{j}\phi^j(x) \nonumber\\ &&+
\Lambda_{\rm D}(x) \,k_{\rm D}{}^i(\phi )(x)+ \lambda_{\rm K}^\mu  k_\mu{}^i(\phi )(x)
\,, \label{deltaC}
\end{eqnarray}
where the $x$-dependent rotation $\Lambda_{M\,\mu\nu}(x)$ and
$x$-dependent dilatation $\Lambda_{\rm D}(x)$ are given by
\begin{eqnarray}
\Lambda_{M\,\mu\nu}(x)&=&\partial_{[\nu}\xi_{\mu]}=\lambda_{\mu\nu}
-4x_{[\mu} \lambda_{K\,\nu]} \,, \nonumber\\
\Lambda_{\rm D}(x)&=&\ft1D \partial_\rho \xi^\rho = \lambda_{\rm D} -2 x\cdot
\lambda_{\rm K}  \,,   \label{Lambdax}
\end{eqnarray}
and $m_{[\mu\nu]}{}^i{}_{j}\phi^j(x)$, $k_{\rm D}{}^i(\phi )(x)$ and $k_\mu^i (\phi)
(x)$ determine the intrinsic part of the conformal transformations on the fields.

To specify for each field $\phi^i$ its transformations under conformal
group one has to specify  the intrinsic part of the conformal transformation, namely:
\begin{description}
\item[i) Transformations under the Lorentz group,]
encoded into the matrix $(m_{\mu\nu})^i{}_j$. \\ The Lorentz transformation
matrix $m_{\mu\nu}$ should satisfy
\begin{equation}
m_{\mu\nu}{}^i{}_k m_{\rho\sigma}{}^k{}_j  - m_{\rho\sigma}{}^i{}_k
m_{\mu\nu}{}^k{}_j = 2\eta_{\mu[\rho}m_{\sigma]\nu}{} ^i{}_j
-2\eta_{\nu[\rho}m_{\sigma]\mu} {}^i{}_j \,. \label{algmatrlor}
\end{equation}
The explicit form for Lorentz transformation matrices is for vectors (the
indices $i$ and $j$ are of the same kind as $\mu$ and $\nu$)
\begin{equation}
m_{\mu\nu}{}^\rho{}_\sigma =2\delta^\rho_{[\mu}\eta_{\nu]\sigma} \,,\qquad \mbox{e.g. }\ \delta V^\mu=-\lambda^{\mu\nu}V_{\nu}\,,
\end{equation}
while for spinors, (where $i$ and $j$ are (unwritten) spinor indices)
\begin{equation}
m_{\mu\nu} =\ft12\gamma_{\mu\nu}\,,\qquad \mbox{e.g. }\ \delta \Psi=-\ft14\lambda^{\mu\nu}\gamma_{\mu\nu}\Psi \,.
\end{equation}
\item[ ii) Transformation under dilatation]  specified by $k_{\rm D}{}^i(\phi )$.
In most cases (and for all non-scalar fields), we just have
\begin{equation}
 k_{\rm D}{}^i(\phi)=w\phi ^i \,,
 \label{kDsimplew}
\end{equation}
where $w$ is a real number called the \emph{Weyl weight} \index{Weyl weight}of the field $\phi^i$ (in principle, different for each field). However, for scalars in a non-trivial manifold with affine
connection $\Gamma _{ij}{}^k$ (torsionless, i.e.~symmetric in $(ij)$),
these are closed homothetic Killing vectors, \index{closed homothetic Killing vector} see \cite[Sect. 15.7]{Freedman:2012zz}, i.e. solutions of
\begin{equation}
 \nabla  _i k_{\rm D}{}^j\equiv  \partial _i k_{\rm D}{}^j +\Gamma _{ik}{}^j k_{\rm D}{}^k=w\delta
  _i^j\,,
 \label{homothetic}
\end{equation}
where again $w$ is the `Weyl weight' and the derivatives $\partial_i$, $\nabla _i$ are with respect to the field $\phi^i$. For $\Gamma_{ij}{}^k=0$, this
reduces to the simple case~(\ref{kDsimplew}).
\item[ iii) Special conformal transformations] \index{special conformal transformation} can have extra parts apart from those
in~(\ref{ximu}) and~(\ref{Lambdax}), connected to translations, rotations
and dilatations. These are denoted as $k_\mu^i (\phi)$. From the commutator $[D,K_\mu]=-K_\mu$ and \eqref{kDsimplew} it is easy to see that\footnote{Remember that, for the intrinsic part, $\delta_K \delta_D \phi^i =\delta_K (k_D^i(\phi))=\partial_j k_D^i(\phi)\delta_K \phi^j$.}
\begin{equation}
  k_\mu{}^j\,\partial _jk_{\rm D}{}^i-k_{\rm D}{}^j\,\partial _jk_\mu{}^i=k_\mu{}^i\,.
 \label{commDKk}
\end{equation}
For the simple form of the dilatations~(\ref{kDsimplew}), this means that
$k_\mu{} ^i (\phi)$ must have Weyl weight one less than that of $\phi ^i$.
Also, for consistency of the $[K,K]$ commutators, the $k_\mu^i\partial _i $ should be
mutually commuting operators.
\end{description}
In this way, the algebra~(\ref{confalg}) is realized on the fields as
\begin{equation}
\left[\delta_{\rm C}(\xi_1),\delta_{\rm C}(\xi_2)\right]\phi=\delta_{\rm C}\left(
\xi^\mu=\xi_2^\nu\partial_\nu\xi_1^\mu-\xi_1^\nu\partial_\nu\xi_2^\mu\right)\phi\,,
\end{equation}
where $\xi^\mu  $ stands for the set $\{a^\mu ,\,\lambda ^{\mu \nu },\,\lambda _{\rm D},\, \lambda _{\rm K}^\mu \}$ building (\ref{ximu}).

To understand fully the meaning of the order of
the transformations, consider in detail the calculation of the commutator
of transformations of fields. See e.g.\  for a field of zero Weyl weight
($k_{\rm D}(\phi)=0$), and notice how the transformations act only on fields, not on
explicit spacetime points $x^\mu$:
\begin{eqnarray}
\lambda_{\rm D} a^\mu [D,P_\mu]\phi(x)&=& \left( \delta_{\rm D}(\lambda_{\rm D})
\delta_P(a^\mu)-\delta_P(a^\mu) \delta_{\rm D}(\lambda_{\rm D})\right) \phi(x)
\nonumber\\ &=& \delta_{\rm D}(\lambda_{\rm D})  a^\mu \partial_\mu \phi(x)   -
\delta_P(a^\mu) \lambda_{\rm D} x^\mu \partial_\mu \phi(x) \nonumber\\ &=&
a^\mu \partial_\mu  \left( \lambda_{\rm D} x^\nu \right)  \partial_\nu \phi(x)
\nonumber\\
&=& a^\mu \lambda_{\rm D} \partial_\mu  \phi(x)  = \lambda_{\rm D} a^\mu P_\mu\phi(x)
\,.
\end{eqnarray}
It is important to notice that the derivative of a field of Weyl weight
$w$ has weight $w+1$. E.g.\ for a scalar with dilatational transformation
determined by the vector $k_{\rm D}$ (and without extra special conformal
transformations) we obtain, using $\delta_{\rm C}(\xi) (\partial_{\mu}\phi(x))=\partial_{\mu}(\delta_{\rm C}(\xi) \phi(x))$
\begin{eqnarray}
\delta_{\rm C}(\xi)  \partial_\mu\phi(x)&=&\xi^\nu(x)\partial_\nu\partial_\mu \phi(x)
+\Lambda_{\rm D}(x)\,\partial_\mu k_{\rm D} (\phi)\nonumber\\ && -\Lambda_{M\mu}{}^\nu
(x)\partial_\nu \phi(x) +\Lambda_{\rm D}(x)
\partial_\mu \phi(x) -2\,\lambda_{K\mu}k_{\rm D} (\phi)\,.  \label{delCderphi}
\end{eqnarray}
The first term on the second line says that it behaves as a vector under
Lorentz transformations. Furthermore, this equation implies that the
dilatational transformation of $\partial _\mu \phi $ is determined by
$\partial _\mu (k_{\rm D} (\phi) + \phi)$. For the simple
transformation~(\ref{kDsimplew}), this means that the derivative also
satisfies such a simple transformation with weight $w+1$. Furthermore, the
derivative $\partial _\mu \phi $ has an extra part in the special
conformal transformation of the form $\lambda _{\rm K}^\nu k_\nu (\partial _\mu
\phi )=-2\lambda _{{\rm K}\mu }k_{\rm D} (\phi)$, as can be seen by looking at the special conformal transformation part in $\delta_C(\xi) \phi (x)$.
\par
With these rules the conformal algebra is satisfied. The question remains
when an action is conformally invariant. We consider local actions for fields $\phi ^i$ with at most first derivatives of the fields:
\begin{equation}
  S=\int \rmd^D x\,{\cal L}(\phi^i(x),\partial_\mu\phi^i(x))\,,
 \label{Slocal}
\end{equation}
The fields $\phi ^i$ are scalars or other representations of the Lorentz group, and have dilation transformation (\ref{kDsimplew}) with weight $w_i$ for each field (for an extension, see further Exercise~\ref{ex:scalarrigidconf}). For invariance of the action under $P_{\mu}$ and $M_{\mu\nu}$ one has the usual requirements of a covariant action. For invariance under dilatation one needs that the total weight of each term should be equal to the dimension $D$. Here $\partial_\mu$ counts also for 1, as can be seen from~(\ref{delCderphi}). Then
\begin{eqnarray}
\delta_{\rm D} {\cal L} &=& \xi^{\mu}\partial_{\mu}{\cal L} +  D \Lambda_{\rm D} (x) {\cal L} \nonumber\\
                &=& \partial_{\mu}(\xi^{\mu}{\cal L}) - (\partial_{\mu}\xi^{\mu}){\cal L} +  D\Lambda_{\rm D}(x){\cal L} = \partial_{\mu}(\xi^{\mu}{\cal L})\, ,
                \label{delDcalL}
\end{eqnarray}
so the remaining term is a total derivative which vanishes in the action.

Finally, we consider the condition for invariance of an action under special conformal transformation. The special conformal transformations sit in the last term of  (\ref{deltaC}), and hidden in $\Lambda_{\rm D}(x)$ and $\Lambda_{M\,\mu\nu}(x)$ in (\ref{Lambdax}). The contributions when the latter are not differentiated in the transformation of $\partial_\mu\phi^i$, are canceled by the steps mentioned above where the Lorentz transformations and dilatational transformations are considered (e.g. in (\ref{delDcalL})). When they are differentiated they give rise to extra terms
\begin{equation}
  \delta_{K,{\rm extra}}\partial _\mu\phi^i=w_i(\partial _\mu\Lambda_{\rm D})\phi^i-\ft12 (\partial _\mu\Lambda_{M}^{\rho\sigma})m_{\rho\sigma}{}^i{}_j \phi^j\,.
 \label{extraK}
\end{equation}
E.g. for scalars, the last term in  (\ref{delCderphi}) is the contribution due to the first term here. The contributions due to the last term of  (\ref{deltaC})
act, using the chain rule, as a field equation of $\phi ^i$ times the transformation $k_\mu{}^i(\phi)(x)$.
We thus remain with
\begin{equation}
\delta_{\rm K} S=2\lambda_{\rm K}^\mu  \int \rmd^Dx\, \frac{{\cal
L}\dr}{\partial(\partial_\nu\phi^i)} \left(- \eta_{\mu\nu}w_i\phi^i
-m_{\mu\nu}{}^i{}_j\phi^j \right) +\lambda_{\rm K}^\mu\frac{ S\delr}{\delta
\phi^i(x)}k_\mu{}^i(\phi)(x)\,. \label{deltaLK}
\end{equation}
where $\dr$ indicates a right derivative.  In many cases the first terms already cancel and no
$k_\mu(\phi)$ are necessary. In fact, the latter are often excluded
because of the requirement that they should have Weyl weight $w_i-1$, and
in many cases there are no such fields available.
 \bexer
There are typical cases in which the first two terms of (\ref{deltaLK}) cancel. Check the following ones
\begin{enumerate}
\item scalars with Weyl weight 0.
\item spinors appearing as $\slashed{\partial }\lambda$ if their Weyl weight is
$(D-1)/2$. This is also the appropriate weight for actions as $\bar
\lambda  \slashed{\partial }\lambda $.
\item \label{exampleconf}
Vectors or antisymmetric tensors whose derivatives appear only as field
strengths $ \partial_{[\mu_1}B_{\mu_2\ldots \mu_p]}$ if their Weyl weight
is $p-1$. This value of the Weyl weight is what we need also in order
that their gauge invariances and their zero modes commute with the
dilatations. Then scale invariance of the usual square of the field
strengths will fix $p=\ft d2$.
\item Scalars $X^i$ with Weyl weight $\ft d2-1$ and
\begin{equation}
{\cal L}=  (\partial_\mu X^i) A_{ij} (\partial^\mu X^j)  \,,
\label{scalarrigidaction}
\end{equation}
where $A_{ij}$ are constants.
\end{enumerate}
\eexer
\bexer \label{ex:scalarrigidconf}
When the scalars transform under dilatations and special conformal
transformations according to
\begin{equation}
  \delta \phi ^i= \xi ^\mu(x) \partial _\mu \phi ^i + \Lambda _{\rm D}(x) k^i_{\rm D}(\phi
  )\,,
 \label{conftrphi}
\end{equation}
with $k^i_{\rm D}(\phi )$ arbitrary, check that the conformal algebra is
satisfied. Consider now the action for scalars
\begin{equation}
  S_{\rm sc}=-\ft12\int \rmd^D x\, \partial _\mu \phi ^i\, g_{ij}(\phi )\,\partial
  ^\mu \phi ^j\,.
 \label{Ssc}
\end{equation}
It is invariant under translations and Lorentz rotations. Check that the
dilatational and special conformal transformations leave us with
\begin{align}
  \delta S_{\rm sc}= -\int \rmd^D x &\left\{ \Lambda _{\rm D}(x)\partial _\mu \phi ^i\,\partial
  ^\mu \phi ^j\left[ g_{k(i}\left( \partial _{j)} k_{\rm D}{}^k +\Gamma _{j)\ell
  }{}^kk_{\rm D}{}^\ell \right) -\ft12(D-2)g_{ij}\right] \right. \nonumber\\
  &\left.- 2\lambda _{\rm K}^\mu\partial _\mu \phi
  ^i\, g_{ij}k_{\rm D}{}^j\right\} ,
 \label{delconfSsc}
\end{align}
if one identifies the affine connection with the Levi-Civita connection
of the metric, similar to~(\ref{omegae}). The invariance under rigid
dilatations is thus obtained if
\begin{equation}
  \nabla _{(i}k_{j){\rm D}}=\ft12(D-2)g_{ij}\,,
 \label{confKilling}
\end{equation}
with the usual definition of a covariant derivative $\nabla _i$. Vectors
satisfying this equation are called `homothetic Killing
vectors'\footnote{The terminology reflects that the right-hand side is a
\textit{constant} times $g_{ij}$. For a function times $g_{ij}$ it is
just a `conformal Killing vector'.\index{conformal Killing vector}}. \index{homothetic Killing vector} However, to obtain special conformal
invariance, the last term of~(\ref{delconfSsc}), originating from a
contribution $\partial _\mu \Lambda _{\rm D}(x)$, should be a total derivative.
Thus one requires that
\begin{equation}
  k_{i{\rm D}}=\partial _i k\,,
 \label{kD=dk}
\end{equation}
for some $k$. Then $\nabla _i k_{j{\rm D}}$ is already symmetric, and thus with the requirement
(\ref{confKilling}) one obtains that (\ref{homothetic}) should be satisfied with $w=(D-2)/2$. The
vectors satisfying~(\ref{homothetic}) are `exact homothetic Killing vectors'.\index{exact homothetic Killing vector}
One can find that the scalar $k$ in (\ref{kD=dk}) is
\begin{equation}
   (D-2)k= k_{\rm D}{}^ig_{ij}k_{\rm D}{}^j\,.
 \label{k=kD2}
\end{equation}
A systematic investigation of conformal actions for scalars in gravity can
be found in~\cite{Sezgin:1995th}.
 \eexer
 \bexer
 Check that for a Lagrangian of the form
\begin{equation}
{\cal L}=(\partial_\mu\phi^1) (\partial^\mu\phi^2)\,,
\end{equation}
the conformal Killing equation has as solution
\begin{equation}
  k_{\rm D}{}^1= w_1 \phi ^1\,, \qquad k_{\rm D}{}^2=w_2 \phi ^2\,,\qquad w_1+w_2=D-2\,.
 \label{solKillflat}
\end{equation}
However, the Eq.~(\ref{homothetic}) gives also that
$w_1=w_2=\ft12(D-2)$. Check that this is necessary for special conformal
transformations.

Modifying the Lagrangian to
\begin{equation}
{\cal L}=\left( 1+\frac{\phi^1}{\phi^2}\right) (\partial_\mu\phi^1)
(\partial^\mu\phi^2)\,,
\end{equation}
one finds as only non-zero Levi-Civita connections
\begin{equation}
  \Gamma _{11}^1= \frac{1}{\phi ^1+\phi ^2}\,,\qquad
  \Gamma _{22}^2=-\frac{\phi ^1/\phi ^2}{\phi ^1+\phi ^2}\,.
 \label{Gammaex}
\end{equation}
The solutions of the conformal Killing equations~(\ref{confKilling})
already fix~(\ref{solKillflat}) with $w_1=w_2=\ft12(D-2)$. However, now
the Eq.~(\ref{homothetic}) gives no solution. Hence, this model can
only have rigid dilatations, but no rigid special conformal
transformations.
 \eexer
\subsection{Superconformal Groups} \label{ss:scgroups}
A classical work on the classification of superconformal groups is the
paper of Nahm~\cite{Nahm:1978tg}. In the groups that he classified, the
bosonic subgroup is a direct product of the conformal group and the
$R$-symmetry group, the automorphism group of the supersymmetries. The latter are bosonic symmetries that are not in
the conformal algebra, hence are spacetime scalars, as motivated by the
works of~\cite{Coleman:1967ad} and~\cite{Haag:1975qh}.\footnote{However,
with branes the assumptions of these papers may be too constrained. Other
examples have been considered first in 10 and 11 dimensions
in~\cite{vanHolten:1982mx}.}

Another classification has appeared
in~\cite{D'Auria:2000ec} from which we can extract\footnote{See Appendix~\ref{app:groups} for the
notations of groups and supergroups.}
Table~\ref{tbl:sconfalg} for dimensions from 3 to 11.
\begin{table}[!t] 
\caption{\it Superconformal algebras, with the two parts of the bosonic
subalgebra: one that contains the conformal algebra and  the other one is
the $R$-symmetry. In the cases $D=4$ and $D=8$, the U$(1)$ factor in the
$R$-symmetry group can be omitted for ${\cal N}\neq 4$ and ${\cal N}\neq 16$,
respectively.} \label{tbl:sconfalg}
\begin{tabular}{|c|lll|}
\hline
D & Supergroup & \multicolumn{2}{c|}{Bosonic group}  \\
\hline
 3 &$ \OSp({\cal N}|4) 
    $\hspace{2.7cm }\phantom{.} &$
    \Sp(4)=\SO(3,2)
 $\hspace{2.7cm }   \phantom{.}&$
  \SO({\cal N})       $ \\
 4 &$  \SU(2,2|{\cal N})  $&$    \SU(2,2)=\SO(4,2)   $&$   \SU({\cal N}) {\times} \U(1)   $ \\
 5 &$  \OSp(8^*|{\cal N}) $&$   \SO^*(8)\supset \SO(5,2)    $&$  \USp({\cal N})         $ \\
   &    F$(4)      $&$             \SO(5,2)  $&$    \SU(2) $ \\
 6 &$  \OSp(8^*|{\cal N}) $&$   \SO^*(8)= \SO(6,2)    $&$   \USp({\cal N})        $ \\
 7 &$  \OSp(16^*|{\cal N}) $&$  \SO^*(16)\supset \SO(7,2)    $&$ \USp({\cal N})          $ \\
 8 &$  \SU(8,8|{\cal N})  $&$    \SU(8,8) \supset \SO(8,2) $&$    \SU({\cal N}) {\times}\U(1)  $ \\
 9 &$  \OSp({\cal N}|32)  $&$   \Sp(32) \supset \SO(9,2)   $&$    \SO({\cal N})        $ \\
 10 &$\OSp({\cal N}|32)    $&$ \Sp(32) \supset \SO(10,2)    $&$ \SO({\cal N})           $ \\
 11 &$\OSp({\cal N}|64)    $&$ \Sp(64) \supset \SO(11,2)    $&$ \SO({\cal N})           $ \\
\hline
\end{tabular}
\end{table}
The bosonic subgroup contains always two
factors: the conformal group (one should consider its
covering that allows the spinor representation) and the $R-$symmetry group. When the first factor is just the
conformal group, then the algebra appears in Nahm's classification. Note
that 5 dimensions is a special case. There is a generic superconformal
algebra for any extension. But for the case ${\cal N}=2$ there exists a smaller
superconformal algebra that is in Nahm's list. Note that for $D=6$ or
$D=10$, where one can have chiral spinors, only the case that all
supersymmetries have the same chirality has been included. Non-chiral
supersymmetry can be obtained from the reduction in one more dimension,
so e.g. from the $D=7$ algebra $\OSp(16^*|{\cal N})$ we obtain $({\cal N},{\cal N})$ supersymmetry in
$D=6$~\cite{Lledo:2001ta}. So far, superconformal tensor calculus has
only been based on algebras of Nahm's type.\footnote{Note that the superalgebras that are relevant for quantum field theories, according to Coleman--Mandula~\cite{Coleman:1967ad} and Haag--{\L}opusza\'nski--Sohnius~\cite{Haag:1975qh}, only exist for $D\leq 6$. The maximal number of supercharges of the corresponding superconformal QFTs is bounded by the requirement that they admit a suitable stress-tensor multiplet, see e.g. \cite{Cordova:2016emh} for a recent discussion.}

In the case of our interest ${\cal N}=2$ in $D=4$, $D=5$ and $D=6$ the superconformal
algebras are, respectively, $\SU(2,2|2)$, F${}^2(4)$ and $\OSp(8^*|2)$ and with $R$-symmetry groups $\SU(2) \times \U(1)$, $\SU(2)$, $\SU(2)$. In the practical treatment, we will not see any fundamental difference in the structure of these supergroups.
The fermionic part of the superconformal group is generated by the supersymmetries $\cQ^i$ (superpartners of the translations $P_a$) and the special supersymmetries $S^i$ (superpartners of the special conformal transformation $K_a$).

The dilatations provide a 3-grading of the conformal algebra and a
5-grading of the superconformal algebra. In the conformal algebra, the
translations $P_a$ have weight~1, in the sense that $[D,P_a]=P_a$. The special
conformal transformations $K_a$ have weight $-1$, while
Lorentz generators $M_{ab}$ and dilatations have all weight~0. Of course, the $R$-symmetry algebra commutes with the conformal algebra and thus has weight~0. The supersymmetries have
weight~$+\ft12$ (see already that this is consistent with (\ref{QQP})),
and the special supersymmetries have weight $-\ft12$. Thus, there is a
clear structure in the superconformal algebra, ordering them according to
the Weyl (dilatational) weight:
\begin{eqnarray}
 1 & : & P_a  \nonumber\\
 \ft12 & : & \cQ \nonumber\\
 0 & : & D,\ M_{ab},\ U_i{}^j,\ T\nonumber\\
 -\ft12 & : & S \nonumber\\
 -1 & : & K_a    \,,
\label{weightsSCAlg}
\end{eqnarray}
This grading already determines the structure of many commutators.

Let us now present some details of the superconformal algebras.\footnote{\label{fn:normd6SKU}To prepare these formulae, we made
use of~\cite{thesisPietC} for $D=4$, apart from a sign change in the
choice of charge conjugation, such that the anticommutators of fermionic
generators have all opposite sign. For $D=5$, we made use
of~\cite{Bergshoeff:2001hc}, and for $D=6$ of~\cite{Bergshoeff:1986mz},
but replacing there $K$ by $-K$ and $U_{ij}$ by $-\ft12 U_{ij}$.} First, we have the conformal algebra, as given in (\ref{confalg}). The other part of the bosonic algebra is the $R$-symmetry. We denote $U_i{}^j$ as the anti-hermitian generators of $\SU(2)$ and $T$ as the real generator of $\U(1)$, the latter only for $D=4$.
The $\SU(2)$ algebra is
\begin{equation}
  [U_i{}^j, U_k{}^\ell ] \ =\ \delta_i{}^\ell  U_k{}^j - \delta_k{}^j
U_i{}^\ell \,.
 \label{SU2algebra}
\end{equation}

The antihermiticity and tracelessness properties of the $\SU(2)$ generators are
\begin{eqnarray}
 U_i{}^j&=&-\left( U_j{}^i\right) ^*\,,\qquad U_i{}^i=0\,,\nonumber\\
U_i{}^j&=&\vec{U}\cdot \vec{\tau }\,,
 \label{propUaHtr}
\end{eqnarray}
with $\vec{\tau }_i{}^j=\rmi \vec{\sigma }_i{}^j$ in terms of the three Pauli matrices $\vec{\sigma }_i{}^j$. The 3-vectors $\vec{U}$ have real components
$U_1$, $U_2$ and $U_3$ are real operators, for which the commutators (\ref{SU2algebra}) imply
\begin{equation}
  [U_1,\,U_2] = U_3\,,\qquad[U_2,\,U_3] = U_1\,,\qquad[U_3,\,U_1] = U_2\,.
 \label{U123algebra}
\end{equation}
More on the use of the $\SU(2)$ indices is collected in Appendix \ref{app:SU2Notation} and specifically for the triplet notation in Appendix \ref{SU2vectorNotation}.

The equations of the first line of (\ref{propUaHtr}) are valid as well in $D=4$, where raising and
lowering of $i$-indices is done by complex conjugation, and in $D=5,6$ where they are raised and lowered by the $\varepsilon^{ij}$ as in (\ref{NWSEconv}). In 4 dimensions, the first equation is written as
\begin{equation}
D=4\ :\  U_i{}^j=-U^j{}_i\,.
 \label{propU4}
\end{equation}
In 5 and 6 dimensions the one-index raised or lowered matrices are (e.g.
$\tau  _{1ij}=\tau  _{1i}{}^k\varepsilon _{kj}=\rmi\sigma _1\rmi\sigma
_2=-\rmi\sigma _3$)
\begin{equation}
D=5,6\ :\    U_{ij}=-\rmi U_1\sigma _3-U_2\unity +\rmi\sigma _1
U_3\,,\qquad U^{ij}=\rmi U_1\sigma _3-U_2\unity -\rmi\sigma _1 U_3\,.
 \label{Udu56}
\end{equation}
The first equation of (\ref{propUaHtr}) implies with the charge conjugation rule (see Appendices~\ref{app:gamma5} and~\ref{app:gamma6})
$M^C=\sigma _2   M^* \sigma _2$ that $U$ is a $C$-invariant matrix. The
tracelessness translates in symmetry of $U_{(ij)}$ and $U^{(ij)}$. Thus,
in conclusion, $U$ is $C$-invariant and symmetric (and thus also
$U_i{}^j=U^j{}_i$ here, different from (\ref{propU4})).

Then we include the fermionic generators. In 6
dimensions $\cQ^i$ and $S^i$ are symplectic Majorana--Weyl spinors with opposite chirality
\begin{equation}
\cQ^i=P_R \cQ^i=-\gamma_* \cQ^i,\,\qquad S^i=P_L S^i=\gamma _* S^i\,.
 \label{chiralSi6}
\end{equation}
In 4 dimensions, the $\cQ_i$ and $S_i$ have opposite chirality:
i.e. $\cQ_i$ and $S^i$ are left-handed (compare with (\ref{deltaEpsQ4}))
\begin{equation}
  S^i=P_L S^i=\gamma _*S^i\,,\qquad S_i=P_R S_i=-\gamma _* S_i\,.
 \label{chiralSi4}
\end{equation}
The commutators between the bosonic and fermionic generators are (we leave implicit the trivial ones)
\begin{equation}
\begin{array}{cll}
&[M_{ab} , \cQ_\alpha^i ] \ =\ - \ft12 ( \gamma_{ab} \cQ^i)_\alpha\,,\qquad
&[M_{ab} , S_\alpha^i ] \ =\ - \ft12 ( \gamma_{ab} S^i)_\alpha\,,
\\[2mm]
&[D,\cQ_\alpha{}^i] \ =\ \ft12 \cQ_\alpha{}^i\,,\qquad &[D,S_\alpha{}^i]=-
\ft12 S_\alpha^i\,,\\[2mm]
&[U_i{}^j, \cQ_\alpha{}^k] \ =\ \delta_i{}^k \cQ_\alpha^j - \ft 12
\delta_i{}^j \cQ_\alpha{}^k\,,\qquad &{}[U_i{}^j, S_\alpha{}^k]\ =\ \delta_i{}^k S_\alpha^j - \ft 12 \delta_i{}^j S_\alpha{}^k\,,\\[2mm]
&[U_i{}^j, \cQ_{\alpha k}] \ =\ -\delta_k{}^j \cQ_{\alpha i} + \ft 12
\delta_i{}^j \cQ_{\alpha k}\,,\qquad &{}[U_i{}^j, S_{\alpha k}]\ =\
-\delta_k{}^j S_{\alpha i} +\ft 12 \delta_i{}^j S_{\alpha k}\,,\\[2mm]
&[T, \cQ_\alpha{}^i] \ =\ \ft 12\rmi \cQ_\alpha{}^i\,,\qquad
&[T, S_\alpha{}^i] \ =\ \ft 12\rmi S_\alpha{}^i\,,\\
&&\hfill T \mbox{ exists only for }D=4\,,\\[2mm]
&[K_a, \cQ_{\alpha}^i] \ =\ s_D\, (\gamma_{a} S^i)_\alpha\,, \qquad & [P_a, S_\alpha{}^i] \ =\  \bar s_D( \gamma_{a} \cQ^i)_\alpha\,,\\[2mm]
&\phantom{.} \qquad \mbox{with}\qquad &\left\{ \begin{array}{ccc}
   D=4 & : &s_4=1,\\
   D=5 & : &s_5=\rmi ,\\
   D=6 & : &s_6=-1.
 \end{array}\right.
\label{ConfCommutators}
\end{array}
\end{equation}
and $\bar s_D$ is the complex conjugate of $s_D$ (only different for $D=5$).
Note that the commutators $[K,\cQ]$ and $[P,S]$ have a different form in different dimensions due to the difference in reality and chirality conditions.
The factors $\rmi$ are necessary in 5 dimensions if we use the same
symplectic Majorana condition for all the spinors. This can be seen easily
from the $C$-conjugation rules in Appendix~\ref{app:gamma5}.

The anticommutation relations between the fermionic generators are
\begin{eqnarray}
 D=4 & : & \{\cQ_\alpha{}^i, \cQ_j{}^\beta\} =\ -\ft12\delta_j{}^i (
\gamma^{a})_\alpha{}^\beta P_a \,,\qquad
\{\cQ_\alpha{}^i, \cQ^{j\beta}\}\ =\ 0\,,  \nonumber\\
   &   & \{S_\alpha{}^i, S_j{}^\beta\}\ =\
-\ft12\delta_j{}^i (\gamma^{a})_\alpha{}^\beta K_a\,,\qquad
\{S_\alpha{}^i,
S^{j\beta}\}\ =\ 0\,,\nonumber\\
&&\{\cQ_\alpha{}^i, S^{j\beta}\}\ =\ 0\,,\nonumber\\
&&\{\cQ_\alpha{}^i, S_j{}^\beta\}\  =\  -\ft12\delta_j{}^i\delta
_\alpha{}^\beta D - \ft14\delta_j{}^i (\gamma^{ab})_\alpha{}^\beta M_{ab}
-\ft12 \rmi \delta_j{}^i\delta _\alpha{}^\beta T +
\delta_\alpha{}^\beta U_j{}^i\,,
\nonumber\\
D=5,6 & : & \{ \cQ_{i\alpha }, \cQ^{j\beta }\}\ = \
 -\ft 12\delta _i{}^j (\gamma ^a)_\alpha{}^\beta  P_a \, ,\qquad
\{ S_{i\alpha }, S^{j\beta } \}\ = \
-\ft 12 \delta _i{}^j (\gamma ^a)_\alpha{}^\beta  K_a \, , \nonumber\\
D=5&:&\{ \cQ_{i\alpha }, S^{j\beta } \} \ = \ -\ft 12\rmi \left(\delta
_i{}^j \delta _\alpha {}^\beta  D +
     \ft12 \delta _i{}^j (\gamma ^{ab}) _\alpha {}^\beta  M_{ab}
      + 3\delta _\alpha {}^\beta  U_i{}^j \right)         \, ,\nonumber\\
D=6 &&\{ \cQ_{i\alpha }, S^{j\beta }\} \ = \ \ft 12 \left(\delta _i{}^j
\delta _\alpha {}^\beta  D +
    \ft12  \delta _i{}^j (\gamma ^{ab}) _\alpha {}^\beta  M_{ab}
      + 4\delta _\alpha {}^\beta  U_i{}^j \right)         \,.
 \label{anticomm}
\end{eqnarray}
Note that in 6 dimensions, the spinors have a chiral projection, and the
gamma matrices in the right-hand side should also be understood as their
chiral projection, thus e.g.\ in the last line $\delta_\alpha {}^\beta$
stands for $\ft12(\unity -\gamma _*)_\alpha {}^\beta$.

 \bexer
 Check that in 5 and 6 dimensions the $[U,\cQ]$ commutator can be written as
 $[U_{ij},\cQ^k]= \delta _{(i}{}^k \cQ_{j)}$.
 \eexer

\subsection{Rigid Superconformal Symmetry}
\label{ss:rigidsuperconf}
To upgrade a realization of supersymmetry to superconformal transformations, we thus have to define also the $S$-supersymmetry transformations on the fields on top of the $Q$-supersymmetry. At the bosonic side, we saw in  (\ref{deltaC}) that conformal transformations of fields have `orbital' parts and `intrinsic' parts. The same will be true for the supersymmetries. Though the word `orbital' is not really clear for supersymmetries, we will use the same terminology since the procedure is similar.
The orbital parts at the bosonic side were chosen such that the first term in (\ref{deltaC}) with the definition (\ref{ximu}) already satisfies the conformal algebra. We can do the same for supersymmetry: we replace the constant supersymmetry parameters $\epsilon $ of a rigid supersymmetric theory by
\begin{equation}
  \epsilon (x) = \epsilon + s_D \gamma _\mu x^\mu \eta \,,
 \label{epsilonx}
\end{equation}
where the $\epsilon $ and $\eta $ at the right-hand side are the constant parameters for rigid $Q$- and $S$-supersymmetry, respectively. The constants $s_D$ are those in  (\ref{ConfCommutators}), such that the algebra is satisfied.

Finally, in parallel with the special conformal transformations, we should expect intrinsic contributions from $S$-supersymmetry transformations. As follows from \eqref{weightsSCAlg}, these can only transform a field $A$ to a quantity $B$ that has dilatation weight $1/2$ less than $A$.

%% file: ch2_Weylmultiplet.tex
\chapter{Gauging Spacetime Symmetries: The Weyl Multiplet}\label{ss:Weyl}
\begin{quote}
\textbf{Abstract.}
In this chapter, we will discuss the gauge multiplets of the superconformal algebra, called Weyl multiplets.\index{Weyl multiplet}  We start by repeating the basic transformation rules for gauge fields and curvatures, and discuss then the modifications necessary for spacetime symmetries. We will see that one needs constraints on curvatures, and will learn how to deal with them. At the end of this chapter we obtain a Weyl multiplets for ${\cal N}=2$ theories in $D=4$, $D=5$ and $D=6$.
\end{quote}
\section{Rules of (Super)Gauge Theories, Gauge Fields and Curvatures}
 \label{ss:gaugerules}
Consider a general (super)algebra with commutators
\begin{equation}
[\delta( \epsilon_1),\delta(\epsilon_2)] =\delta\left(\epsilon^C_3=
\epsilon_2^B\epsilon_1^A f_{AB}{}^C\right) \,.
\end{equation}
In general $f_{AB}{}^C$ may be structure functions, i.e. depend on
the fields.
Moreover, the equality above may be satisfied only modulo equations of motion, as we will see below.
The $f_{AB}{}^C$ are related to the
abstract algebra introduced in Sect.~\ref{ss:scgroups}:
\begin{eqnarray}
 D=4,6 & : & [T_A,T_B\}=T_AT_B-(-)^{AB}T_BT_A= f_{AB}{}^C T_C \,,
 \nonumber\\
 D=5 & : & [T_A,T_B\}=T_AT_B-(-)^{AB}T_BT_A= (-)^{AB}f_{AB}{}^C T_C \,,
 \label{TTfT}
\end{eqnarray}
where $(-)^A$ is a minus sign if $T_A$ is fermionic. The extra sign
factor for $D=5$ is due to the factors $\rmi$ in (\ref{deltaEpsQ5}). It
is assumed that in all other cases the transformation is generated by
$\epsilon ^AT_A$.
\vspace{10pt}

The algebra above can be realized locally by introducing gauge fields and
curvatures. This means that for every generator there is a gauge field,
$B_\mu{}^A$, which transforms as follows:\footnote{Note that the order of the fields and parameters is relevant here. For
fermionic fields, the indices contain spinor indices and one may use the
conventions of \cite[Sect. 3.2.2]{Freedman:2012zz}. Although the objects may be
fermionic or bosonic, you do not see many sign factors. The trick to
avoid most sign factors is to keep objects with contracted indices
together. E.g., you see here the $B$ index of the gauge field next to the
$B$ index in the structure constants, and then the $C$ contracted indices
do not have other uncontracted indices between them.}
\begin{equation}
\delta(\epsilon) B_\mu{}^A=
\partial_\mu\epsilon^A +\epsilon^C B_\mu{}^B f_{BC}{}^A\,.
\label{delGaugef}
\end{equation}

Covariant derivatives have a term involving the gauge field for every
gauge transformation
\begin{equation}
{\cal D}_\mu =\partial_\mu -\delta(B_\mu)\,, \label{defnabla}
\end{equation}
and their commutators are new transformations with as parameters the
curvatures:
\begin{eqnarray}
[{\cal D}_\mu,{\cal D}_\nu]&=&-\delta(R_{\mu\nu})\,,\nonumber\\
 R_{\mu\nu}{}^A &=& 2\partial_{[\mu}B_{\nu]}{}^A+ B_\nu{}^C B_\mu{}^B f_{BC}{}^A\,,
 \label{defcurv}
\end{eqnarray}
which transform `covariantly':
\begin{equation}
\delta(\epsilon )  R_{\mu\nu}{}^A = \epsilon^C R_{\mu\nu}{}^B f_{BC}{}^A\,,
\label{deltacurv}
\end{equation}
and, with the definitions (\ref{defnabla}), satisfy Bianchi identities
\begin{equation}
  {\cal D}_{[\mu }R_{\nu \rho ]}{}^A=0\,.
 \label{Bianchi}
\end{equation}

Having in mind the motivations presented in Sect.~\ref{ss:strategy}, we now start gauging the superconformal group. In Table \ref{tablegaugefields} we give names to the gauge fields and parameters for the  generators that appeared in Sect.~\ref{ss:scgroups}.
\begin{table}[!t] 
  \caption{\it Superconformal gauge symmetries, their gauge fields and parameters }\label{tablegaugefields}
\begin{center}
$
  \begin{array}{c|cccccccc} \hline\noalign{\smallskip}
  T_A &  P_a & M_{ab} & D & K_a & U_i{}^j & T & Q & S \\ \hline
  B_\mu{} ^A &  e_\mu ^a & \omega _\mu {}^{ab}& b_\mu   & f_\mu {}^a & V_\mu {}_j{}^i & A_\mu  & \psi _\mu  & \phi _\mu  \\
  \epsilon ^A &  \xi ^a & \lambda ^{ab} & \lambda _{\rm D} & \lambda _{\rm K}^a & \lambda_j{}^i& \lambda_T & \epsilon  & \eta  \\
\noalign{\smallskip}\hline\noalign{\smallskip}
  \end{array}$
\end{center}
\end{table}
The relation between parameters (or gauge fields) and generators varies in different dimensions due to the different spinor properties and notations for raising and lowering $i$ indices.
This is encoded in the transformation rule $\delta =\epsilon ^AT_A$:
\begin{eqnarray}
\delta = \epsilon^{A}T_{A}&=& \ft12 \lambda ^{ab}M_{[ab]}+\lambda_{\rm D} D +
\lambda_{\rm K}^a K_a+ \nonumber\\
&&\left\{
\begin{array}{ll}
  + \lambda _i{}^jU_j{}^i+ \lambda_T\,T+\,\bar \epsilon ^i \cQ_i +\bar \epsilon _i\cQ^i+\bar \eta  ^i S_i +\bar \eta  _iS^i  & \mbox{for }D=4 \\
  + \lambda ^{ij}U_{ij}+\rmi \bar \epsilon ^i \cQ_i +\rmi \bar \eta  ^i S_i   & \mbox{for }D=5 \\
   + \lambda ^{ij}U_{ij}+ \bar \epsilon ^i \cQ_i + \bar \eta  ^i S_i   & \mbox{for }D=6\,.
\end{array}
\right.
 \label{paramgenerators}
\end{eqnarray}
The same correspondence holds for the gauge fields in expressions $B_\mu{} ^AT_A$. With this correspondence, the commutators can be written in terms of parameters in commutators of transformations, e.g. for $D=4$, we have
\begin{eqnarray}
  \left[\delta _S(\eta ),\delta _Q(\epsilon )\right]&=& \delta _{\rm D}(\lambda_{\rm D}(\epsilon ,\eta )) + \delta _{\rm M}(\lambda ^{ab}(\epsilon ,\eta )) + \delta _{\rm U}(\lambda _i{}^j(\epsilon ,\eta ))\,, \nonumber\\
 && \lambda_{\rm D}(\epsilon ,\eta )=\ft12(\epsilon ^i\eta _i+\hc)\,,\nonumber\\
 && \lambda ^{ab}(\epsilon ,\eta )= \ft14(\epsilon ^i\gamma ^{ab}\eta _i+\hc)\,,\nonumber\\
 &&\lambda _T(\epsilon ,\eta ) = \ft12\rmi (-\bar \epsilon ^i\eta _i+\bar \epsilon _i\eta ^i)\,,\nonumber\\
 && \lambda _i{}^j(\epsilon ,\eta )=\bar \epsilon ^j\eta _i-\bar \epsilon _i\eta ^j-\ft14\delta _i^j(\bar \epsilon^k\eta _k-\bar \epsilon _k\eta ^k)\,,
 \label{thirdcoeffQS}
\end{eqnarray}
and
\begin{align}
  \left[\delta _T(\lambda _T),\delta _Q(\epsilon )\right]= & \delta _Q(\epsilon '(\epsilon ,\lambda _T))\,,\qquad \epsilon '_i(\epsilon ,\lambda _T)=\ft12\rmi\lambda _T \epsilon _i\,, \nonumber\\
  \left[\delta _{\SU(2)}(\lambda),\delta _Q(\epsilon )\right]= & \delta _Q(\epsilon '(\epsilon ,\lambda))\,,\qquad \epsilon '_i(\epsilon ,\lambda )=\lambda _i{}^j\epsilon _j\,.
\label{commRQD4}
\end{align}

 \bexer Check that the first equation of (\ref{defR}) corresponds to the definition \eqref{defcurv} for $R_{\mu \nu }{}^{ab}$ if we just consider the
Lorentz group, i.e. the first equation of (\ref{confalg}), and
define $\omega _\mu {}^{ab}$ as the gauge field of $M_{ab}$.
 \eexer

\section{Gauge Theory of Spacetime Symmetries} \label{ss:gaugethspacetime}
The sole procedure explained in Sect.~\ref{ss:gaugerules} is not sufficient to define a suitable theory for local spacetime symmetries. In the following we will explain the necessary steps to improve the recipe given in Sect.~\ref{ss:gaugerules}.

\subsection{General Considerations}\label{ss:Gengaugethspacetime}
The main problem is that in general relativity one should have general coordinate transformations (gct) \index{general coordinate transformation (gct)} as a local symmetry rather than the local translations defined by rules of the previous section. The action of gct on spacetime scalar fields and gauge vectors is, by definition
\begin{align}
\delta_{\rm gct}(\xi)\phi(x)=&\,\,\xi^\mu(x)\partial_\mu \phi(x)\,, \label{defgct}\\
  \delta _{\rm gct}(\xi )B_\mu {}^A\equiv&\,\, \xi ^\nu \partial _\nu B_\mu {}^A +
  (\partial _\mu \xi ^\nu )B_\nu {}^A=\delta _B(\xi ^\nu B_\nu {}^B) B_\mu {}^{A}
  -\xi ^\nu R_{\mu \nu }{}^A\,,
 \label{gcttoR}
\end{align}
where in the last line we have rewritten the transformation using \eqref{delGaugef} and \eqref{defcurv}. On the other hand, the definition (\ref{delGaugef}) with the algebra (\ref{confalg}) would lead e.g. to $\delta _P b_\mu = 2 \xi _a f_\mu {}^a$, which is not yet a general coordinate transformation.
We will take a few steps to relate gct to local translations as defined by the algebra  (\ref{confalg}) and (\ref{ConfCommutators}).
\vspace{10pt}

\noindent \textbf{Step 1.} First of all we distinguish the translations from all the other transformations. All these others will be denoted as \emph{standard gauge transformations}. \index{standard gauge transformation} We therefore split the range of indices $A$ in the following way
\begin{align}
T_A=&\,(\underbrace{P_a}_{\text{Translations}},\underbrace{T_I}_{\text{Standard Gauge}}).
\end{align}
The gauge field of $P_a$, denoted as $e_\mu{} ^a$, is required to be invertible as a matrix, and it is interpreted as the frame field. \vspace{10pt}

\noindent \textbf{Step 2.} When we consider the parameter $a^\mu $ in (\ref{ximu}) as a local function, this absorbs all the other terms in  (\ref{ximu}), and we further denote it as $\xi ^\mu(x)$. In other words, a change of basis in the set of the gauge transformations is performed such that all the orbital parts of Lorentz rotations, dilatations and special conformal transformations are reabsorbed into the general coordinate transformation. In the same way the special conformal transformations in  (\ref{Lambdax}) are absorbed in the local parameters $\lambda^{ab}(x)$ and $\lambda _{\rm D}(x)$ and the $S$-supersymmetry part in (\ref{epsilonx}) is absorbed in $\epsilon (x)$.\vspace{10pt}

\noindent \textbf{Step 3.} A further basis change is performed using \emph{covariant general coordinate transformations} (cgct)~\cite{deWit:1975nq,Jackiw:1978ar}: \index{covariant general coordinate transformation}
\begin{equation}
\delta_{\rm cgct}(\xi)=\delta_{\rm gct}(\xi)-\delta_I\left(\xi^\mu
B_\mu{}^I\right)\,. \label{defcgct}
\end{equation}
Note that \eqref{defcgct} is a combination of general coordinate transformations and all the
non-translation transformations (standard gauge transformations) whose parameter $\epsilon ^I$ has been replaced
by $\xi^\mu B_\mu{}^I$. As we require that the final action is invariant both under
generic $\delta_{\rm gct}$ and $\delta_I$, it should be invariant also under cgct.
To summarize, we replace
\begin{equation}
  \delta (\epsilon )=\epsilon ^AT_A = \xi ^aP_a+ \epsilon ^I T_I\,,
 \label{deltaepstoPT}
\end{equation}
by a different transformation that leaves the action invariant:
\begin{equation}
  \bar \delta = \delta _{\rm cgct}(\xi )+ \epsilon ^I T_I = \delta _{\rm gct}(\xi )+ \delta _I\left(\epsilon ^I -\xi^\mu
B_\mu{}^I\right)\,.
 \label{bardelta}
\end{equation}
Below, we will discuss the action of cgct in more detail for the different types of fields.\vspace{10pt}

\noindent \textbf{Step 4.} In order to identify the cgct transformations with gauged translations we further need constraints on some curvatures. As we will see in Sect. \ref{ss:transfoframef}, the first one will be the curvature of translations that will be put to zero. This will imply that the gauge field of Lorentz rotations $\omega _\mu {}^{ab}$ will become a function of $e_\mu ^a$ as is common in general relativity, where it is then often called the `spin connection'. \index{spin connection} The expression in the supersymmetric theory will also involve other fields such as the gravitino. This is related to torsion, see e.g. \cite[Sect. 7.9]{Freedman:2012zz}. We will see in Sect.~\ref{ss:constraints} that more constraints will be imposed such that also other gauge fields of Table \ref{tablegaugefields}  are composite rather than independent fields.\vspace{10pt}

\noindent \textbf{Step 5.}  When general coordinate transformations are properly implemented, the numbers of bosonic and fermionic degrees of freedom in a supersymmetric theory should match, as we discussed in the beginning of Sect. \ref{ss:multiplets}. In the ${\cal N}=2$ theories, this will involve the introduction of  additional (auxiliary) degree of freedom, which we will discuss in Sect.~\ref{sect:Weyl}.
\vspace{10pt}

In summary the essential modifications of the procedure outlined in Sect.~\ref{ss:gaugerules} are
\begin{enumerate}
  \item Translations are replaced by general coordinate transformations, which are further combined with other symmetries to covariant general covariant transformations (cgct) \eqref{defcgct}.
  \item Some gauge fields will turn out to be `composite', i.e. functions of the other
  fields in the multiplet.
\item The multiplet that gauges the superconformal group will contain auxiliary fields.
\item The structure constants will be replaced by structure functions.
\end{enumerate}

In the following  we present the general form of cgct acting on frame fields, other gauge fields\footnote{The part on gauge fields could be generalized to $p$-form fields. We do not include a general setup for these, but the essential characteristics of cgct apply in the same way.} and matter fields, which will be useful for our considerations. At this point, it is important to keep in mind that the fundamental distinction between `gauge fields' and `matter fields' \index{matter fields} is that the former have a coordinate index $\mu $,\footnote{Of course one could change it to a frame index $a$ by multiplication with $e_a{}^\mu $, but we
consider it in the form $B_\mu ^A$ as the basic field and $B_a^A$ as the composite of $e_a{}^\mu B_\mu ^A$. This field $B_\mu ^A$ may
still be composite by itself, but that is not important at this point.} while matter fields have no coordinate indices. In other terms:
\vspace{10pt}
\begin{tcolorbox}[title=Matter vs Gauge fields]
The transformation of matter fields does not involve a derivative of a gauge parameter, while the transformation of gauge fields does have the $\partial _\mu \epsilon ^A$ term.
\end{tcolorbox}

\subsection{Transformations of the Frame Fields}
\label{ss:transfoframef}
Let us derive the explicit form of the transformations of the frame field. For a cgct the last term of (\ref{defcgct}) cancels the first term of the second expression in \eqref{gcttoR}, apart from the translation
part (where $B$ takes only the values corresponding to translations, i.e. $b$). As a result we obtain
\begin{equation}
 \delta _{\rm cgct}(\xi )e_\mu {}^a  =  \partial _\mu \xi ^a +
 \xi ^bB_\mu {}^Bf_{Bb}{}^a-\xi ^\nu R_{\mu \nu }(P^a) \,.
 \label{cgcte}
\end{equation}
The first two terms in the expression above are just (\ref{delGaugef}), with $\epsilon $ replaced by $\xi ^a$. In particular \emph{after imposing the constraint $R_{\mu \nu }(P^a)=0$, the cgct of the frame field is equal to its
$P_a$ transformation as it would directly follow from (\ref{delGaugef}).}

Let us now consider this first constraint using the explicit expressions of the commutators that are of the form
$[P,\cdot ]=P$ in the algebra (\ref{confalg}) and (\ref{ConfCommutators}).
\begin{equation}
  R_{\mu \nu }(P^a)=2\left(\partial_{[\mu}+b_{[\mu }\right)
e_{\nu]}^a+2\omega_{[\mu}{}^{ab}e_{\nu] b}+\xi ^a(\psi _\mu ,\psi _\nu
)=0\,,
 \label{RP0}
\end{equation}
where $\xi ^a$ is the function introduced in~(\ref{xiaeps12}). This constraint implies that $\omega _\mu {}^{ab}$ is the connection such that for pure gravity (hence $\psi _\mu =0$) the spacetime manifold
with metric $g_{\mu \nu }=e_\mu ^ae_{\nu a}$ is torsionless, while the gravitino terms define a torsion. We will come
back to this constraint and its consequences below, and explain its
solution for $\omega _\mu {}^{ab}$ in general. The expression \eqref{cgcte} then collapses to
\begin{equation}
 \delta _{\rm cgct}(\xi )e_\mu {}^a=
  (\partial _\mu+b_\mu ) \xi ^a+\omega _{\mu }{}^{ab}\xi _b\,.
 \label{cgcte2}
\end{equation}

The standard gauge transformations of the frame field follow straightforwardly from
the rule (\ref{delGaugef}). That is, using the notation $\xi ^a(\cdot ,\cdot
)$ from (\ref{xiaeps12}),
\begin{eqnarray}
  \delta _I(\epsilon ^I) e_\mu{}^a &=& \epsilon ^I B_\mu ^B f_{BI}{}^a
  = -\lambda _{\rm D} e_\mu{}^a -\lambda ^{ab}e_{\mu b}+\xi ^a(\psi _\mu ,\epsilon )\,,
  \nonumber\\
    D=4 &:& \xi ^a(\psi _\mu ,\epsilon )=
    \ft12\bar \epsilon^i\gamma ^a\psi _{\mu i}
    +\ft12\bar \epsilon _i\gamma ^a\psi _\mu ^i\,, \nonumber\\
    D=5,6 & : &\xi ^a(\psi _\mu ,\epsilon )=\ft12\bar \epsilon ^i\gamma ^a\psi _{\mu
    i}\,.
 \label{delIe}
\end{eqnarray}
\subsection{Transformations of the Other Gauge Fields}
Transformations (\ref{delGaugef}) of gauge fields other than $e_\mu{}^a$ are often deformed by the presence of matter fields in the multiplet. That is, the expression  (\ref{delGaugef}) is often not complete. This will be the case after other fields have been added as mentioned above under `Step 5', and will be discussed in full in Sect. \ref{sect:Weyl}. Here we will already discuss how this modifies some general rules that were presented in Sect. \ref{ss:gaugerules}. A simple example of this phenomenon appears in the supersymmetry transformation of the gauge field $W_\mu $ in the abelian vector multiplet, which contains a `gaugino', $\lambda ^i$:
\begin{equation}
  \delta_Q(\epsilon ) W_\mu =-\ft12\varepsilon _{ij}\bar \epsilon ^i\gamma _\mu \lambda ^j
  +\hc\,,\qquad (D=4)\,.
   \label{delWmu}
\end{equation}
Clearly the r.h.s. of the equation above cannot be seen as a part of (\ref{delGaugef}). Therefore, in order to account for these additional fields, we need to modify some of the general rules presented in Sect. \ref{ss:gaugerules}. We allow in general a modification of
(\ref{delGaugef}) by considering the following general form of standard gauge
transformations:
\begin{equation}
  \delta _J\left(\epsilon ^J\right)B_\mu ^I= \partial _\mu \epsilon ^I
  +\epsilon ^J B_\mu{}^Af_{AJ}{}^I+\epsilon ^JM_{\mu J}{} ^I \,.
 \label{delJhI1}
\end{equation}
The expression $M_{aJ}{}^I$ is a function of `matter fields', which should be a covariant
quantity (i.e. not transforming with a derivative of a parameter, see Sect.~\ref{ss:covquantities}). As an example, when the field $B_\mu ^I$ is the $\U(1)$ gauge
field $W_\mu $, the term written in (\ref{delWmu}) is of this form where the supersymmetry index is $J=(\alpha i)$ and $I$ refers to this gauge field, say $I=\centerdot$:
\begin{equation}
  M_{\mu \alpha i}{} ^{\centerdot}=  -\ft12\varepsilon _{ij}(\gamma _\mu \lambda ^j)_\alpha\,.
 \label{MexplicitW}
\end{equation}
As mentioned above, such terms will also appear for the gauge fields in Table \ref{tablegaugefields} after auxiliary fields have been added.

In (\ref{delJhI1}) there is still a sum over $A$, which we want to split in the standard gauge transformations, and the contribution when $B_\mu {}^A$ is the frame field $e_\mu {}^a$. Therefore we rewrite \eqref{delJhI1} as
\begin{eqnarray}
  \delta _J\left(\epsilon ^J\right)B_\mu ^I&=& \partial _\mu \epsilon ^I
  +\epsilon ^J B_\mu{}^Kf_{KJ}{}^I +\epsilon ^J {\cal M}_{\mu J}{} ^I, \nonumber\\
{\cal M}_{a J}{} ^I&=&f_{aJ}{}^I + M_{a J}{} ^I \,.
 \label{delJhI2}
\end{eqnarray}
While the terms containing $M_{a J}{} ^I$ are determined case by case, depending on the multiplet, those proportional to $f_{aJ}{}^I$ are fixed by the superconformal algebra (in particular by the commutators $[P_a,T_J]= f_{aJ}{^I}T_I$). Explicitly:
\begin{eqnarray}
 \delta _I\left(\epsilon ^I\right) \psi _\mu ^i & = & \ldots - s_D\gamma _\mu \eta ^i\,,\nonumber\\
 \delta _I\left(\epsilon ^I\right) b_\mu  & = & \ldots+ 2\lambda_{{\rm K}\mu }, \nonumber\\
 \delta _I\left(\epsilon ^I\right) \omega _\mu {}^{ab}&=&\ldots - 4\lambda _{\rm K}^{[a}e_\mu
 {}^{b]}\,.
 \label{delextra}
\end{eqnarray}
The constants $s_D$ are those that appear in the algebra (\ref{ConfCommutators}).
Including the matter terms as in (\ref{delJhI2}), the cgct (\ref{defcgct}) on the gauge fields $B_\mu^I$ are
\begin{eqnarray}
 \delta _{\rm cgct}(\xi )B_\mu ^I & = & -\xi ^\nu R_{\mu \nu }{}^I+
 \xi ^a B_\mu ^J f_{Ja}{}^I-\xi ^\nu B_\nu ^J M_{\mu J}{}^I \nonumber\\
   & = & -\xi ^\nu \widehat{R}_{\mu \nu }{}^I-\xi ^a B_\mu ^J{\cal
   M}_{aJ}{}^I\,.
 \label{cgcthI}
\end{eqnarray}
The second term in the first line only occurs for the transformation of
the  gauge fields of supersymmetry, dilatations and Lorentz rotations.
This is the original $P_a$ transformation of the gauge field. In the
second line appears a new covariant curvature, which takes the
transformations of the gauge fields to matter fields into account.
Indeed, the last term of (\ref{delJhI1}) implies that
$R_{\mu \nu }{}^I$ transforms in the derivative of a parameter, i.e.
there is a term $2\partial _{[\mu } \epsilon ^JM_{\nu ]J}{}^I$. In the last line we introduced the
modified curvature
\begin{equation}
  \widehat{R}_{\mu \nu }{}^I=R_{\mu \nu }{}^I-2B_{[\mu }{}^JM_{\nu ]J}{}^I=
  r_{\mu \nu }{}^I-2B_{[\mu }{}^J{\cal M}_{\nu ]J}{}^I\,,
 \label{hatR}
\end{equation}
which does not transform to a derivative of a parameter and we defined
\begin{equation}
  r_{\mu\nu}{}^I = 2\partial_{[\mu}B_{\nu]}{}^I+ B_\nu{}^K B_\mu{}^J f_{JK}
  {}^I\,,
 \label{defr}
\end{equation}
by stripping all contributions from translations out of
the curvatures.

 \bexer We saw already in \eqref{delJhI2}--\eqref{delextra} the explicit form of the terms that make the
difference between $M$ and ${\cal M}$. Using \eqref{hatR}, this should allow you to determine
that the only ones where these play a role are
\begin{eqnarray}
R_{\mu \nu }\left(M^{ab}\right) & = & r_{\mu \nu }\left(M^{ab}\right)+ 8f_{[\mu }{}^{[a}
 e_{\nu ]}{}^{b]} ,\nonumber\\
 R_{\mu \nu }(D) & = & r_{\mu \nu }(D)-4f_{[\mu }{}^a e_{\nu ]a},\nonumber\\
 R_{\mu \nu }(Q)&=& r_{\mu \nu }(Q)-2 s_D \gamma _{[\mu }\phi _{\nu ]}\,.
 \label{extraR}
\end{eqnarray}
 \eexer
\vspace{10pt}
The cgct of a standard gauge field, given by \eqref{cgcthI}, takes into account the matter fields as well as modifications of the gauge curvatures by the latter. On the frame field, upon imposing the constraint (\ref{RP0}), a cgct gives a local translation as dictated by \eqref{delGaugef}. In the case of standard gauge fields, the relation between local translations and these covariant general coordinate transformations is less obvious. First of all, we cannot
simply put all curvatures equal to zero to use (\ref{gcttoR}). This would
impose derivative constraints on all fields, and  thus restrict their dynamics. It could be done for $R(P)$ because in that case, this
constraint will just determine the spin connection in terms of the
frame field, without imposing further dynamical constraints.

We will only choose constraints that in the same way can be solved for gauge fields. In this respect,  (\ref{extraR}) will be useful, which already shows that with such constraints $f_\mu{}^a$ and $\phi _\mu $ may be determined in terms of other fields. How this is done exactly and what are the consequences will be the subject of Sect. \ref{ss:constraints}. Suppose that this is done, let us count the number of off-shell degrees of freedom associated to the independent fields of Table \ref{tablegaugefields}.
This is done in Table \ref{tbl:dofgauge}, where for comparison, we also included ${\cal N}=1$ in $D=4$.
\begin{table}[!t] 
  \caption{\it Off-shell degrees of freedom in gauge fields }\label{tbl:dofgauge}
$\begin{array}{l|cccc|cccccc|c|c|cc|c}
\hline
    & \multicolumn{11}{c|}{\mbox{Bosonic}} & \multicolumn{4}{c}{\mbox{Fermionic}}  \\ \cline{2-16}
 & \multicolumn{4}{ c|}{\mbox{Fields}} & \multicolumn{6}{ c|}{\mbox{Symm.}}& \# & \mbox{Field}& \multicolumn{2}{ c| }{\mbox{Symm.}}& \# \\
    & e_\mu ^a & b_\mu  & V_\mu {}_j{}^i & A_\mu & P_a & M_{ab} & D & K_a & U_i{}^j & T &\mbox{dof} & \psi _\mu    &  Q & S &\mbox{dof}\\ \hline
 {\cal  N}=1\ D=4   & 16 & 4 &   & 4 & -4 & -6 & -1 & -4 &  & -1 & 8 & 16 & -4 & -4 & 8 \\
 {\cal  N}=2\ D=4   & 16 & 4 &12 & 4 & -4 &-6 & -1 & -4 & -3 & -1 & 17 & 32 & -8 & -8 & 16 \\
 {\cal  N}=2\ D=5   & 25 & 5 & 15 &  & -5 & -10 & -1 & -5 & -3 &   & 21 & 40 & -8 & -8& 24 \\
 {\cal  N}=2\ D=6   & 36 & 6 & 18 &  & -6 & -15 & -1 & -6  &-3 &   & 29 & 48 & -8 & -8 & 32 \\ \hline\noalign{\smallskip}
\end{array}$
\end{table}
As explained in the beginning of Sect. \ref{ss:multiplets}, the number of off-shell degrees of freedom is given by total number of components minus the gauge degrees of freedom. From the components we have a factor $D$ for the index $\mu $ and $a$, a factor 3 for the $\SU(2)$ gauge fields, and a factor 8 for the spinors of the ${\cal N}=2$ theories (4 for ${\cal N}=1$). Upon subtracting the gauge degrees of freedom corresponding to the superconformal group, we see that for ${\cal N}=1$ there remain an equal number of bosonic and fermionic degrees of freedom. It turns out that indeed the gauged translations can be identified with general coordinate transformations. A detailed analysis can be found in  \cite{vanNieuwenhuizen:1985cy}. On the other hand, the numbers do not match for the ${\cal N}=2$ theories. A standard procedure to cure this mismatch introduces other (auxiliary) fields, which will be new `matter fields'. This solution will be presented in Sect. \ref{sect:Weyl}.

\subsection{Transformations of Matter Fields}
Matter fields transform by definition without derivatives on the parameters. Using this, we can rewrite the definition (\ref{defcgct}) of the cgct, extracting the $\xi ^\mu $, as
\begin{align}
  \delta_{\rm cgct}(\xi )\phi =&\xi ^\mu \partial _\mu \phi
   -\delta_I\left(\xi ^\mu B_\mu^I\right)\phi\nonumber\\ =&\xi ^\mu {\cal D} _\mu \phi\,,\qquad
 {\cal D} _\mu \phi\equiv\partial _\mu \phi
   -\delta_I\left(B_\mu^I\right)\phi\,.
 \label{defDmu}
\end{align}
Note that the new derivative ${\cal D}_\mu$ such defined, is not
a covariant quantity. Indeed, under general transformations its transformation includes $\partial _\mu \xi ^\nu {\cal D}_\nu \phi $, hence a derivative on a parameter. This can be avoided by defining ${\cal D}_a\phi$, as we will discuss below. We will then also prove that this is a covariant quantity, and then the action of translations on matter fields is $P_a\phi ={\cal D}_a\phi$.
\section{Covariant Quantities and Covariant Derivatives}
\label{ss:covquantities}
We reserve the present section for a more specific discussion on covariant quantities in gauge theories.
Several important steps for this were obtained in \cite[Sect. 3.2]{Bergshoeff:1986mz}. We start from the following definition:
\vspace{10pt}
\begin{tcolorbox}[title=Covariant quantity]
	A covariant quantity is a field whose transformation under {\em any} local
symmetry has no derivative on a
transformation parameter.
\end{tcolorbox}
The Lie derivative (general coordinate transformation) of a world scalar (namely an object with only frame indices) does not involve a derivative on the parameter, while all other fields that have coordinate indices (components of forms, world vectors, ...) transform under general coordinate transformations with derivatives of $\xi ^\mu $. Therefore, any covariant quantity must be a world
scalar.
Using this prescription, we have immediately two ways to build covariant quantities:
\begin{enumerate}
  \item A covariant derivative \index{covariant derivative} on a covariant quantity  with its index turned
  to a frame index:
\begin{equation}
  {\cal D}_a\phi =e_a{}^\mu {\cal D}_\mu \phi\,.
 \label{Daphi}
\end{equation}
   \item Covariant curvatures \index{covariant curvature} with their indices turned to
  frame indices:
\begin{equation}
  \widehat{R}_{ab}{}^I=e_a{}^\mu e_b{}^\nu \widehat{R}_{\mu \nu }{}^I\,.
 \label{Rab}
\end{equation}
\end{enumerate}

The elementary matter fields transform under the symmetries in other
covariant quantities. As will become clearer soon, this is not a general
property of covariant quantities. For quantities
that do not transform in covariant quantities, the expression of the
covariant derivative is not as in (\ref{defDmu}). Let us first look to
covariant quantities that do transform in covariant quantities. For those
we have the lemma\\[3mm]
\noindent \textbf{Lemma on Covariant Derivatives}. \textit{If a
covariant quantity $\phi $ transforms into covariant quantities
under standard gauge transformations, its covariant derivative ${\cal D}_a \phi$ given by (\ref{Daphi}),
with ${\cal D}_\mu \phi $ given by (\ref{defDmu}), is a covariant quantity. Moreover, if the algebra closes on the field $\phi$ then the standard gauge transformations of ${\cal D}_a\phi$ involve only covariant quantities.}\vspace{3mm}

An immediate consequence of the lemma is that under these conditions the cgct ${\cal D}_a\phi =P_a\phi $ gives a covariant quantity.

In this lemma a difference is made whether the algebra is `open' or `closed'
on the original field. We refer for more information to \cite[Sect. 11.1.3]{Freedman:2012zz}), but repeat here the essential statement. In general, the commutator of two transformations on a field $\phi $ is of the form
\begin{equation}
  \delta_I\left(\epsilon ^I\right) \phi =\epsilon ^I \chi _I\,,\qquad
  \left[ \delta _I\left(\epsilon ^I_1\right), \delta _J\left(\epsilon ^J_2\right)\right] \phi =
\epsilon ^J_2\epsilon ^I_1 \left( f_{IJ}{}^K \chi _K +f_{IJ}{}^a {\cal D}_a\phi
+\eta _{IJ}\right)\,.
 \label{commutatorphi}
\end{equation}
The last tensor, $\eta _{IJ}$ is the non-closure function for the field $\phi$. It is not always possible to close the algebra on all fields without using equations of motion. If these transformations occur for a theory based on an invariant action, the tensor $\eta _{IJ}$ is proportional to equations of motion, and the algebra is still `on-shell' closed, i.e. using trivial on-shell symmetries.\footnote{Note that in principle, $\eta _{IJ}$ may have contributions for all symmetries denoted by the indices $I,J$, but in practice (due to the engineering dimensions of the transformations) non-closure functions are related only to $[\delta_Q,\delta_Q]$.}
In conclusion if, for the field $\phi$, $\eta_{IJ}\neq 0$, then ${\cal D}_a\phi $ is still a covariant quantity, but its transformation $ \delta_I(\epsilon ^I){\cal D}_a\phi $ is not.
Vice versa, if $\eta_{IJ}=0$, then the algebra closes off-shell, and the lemma says that $\delta_I(\epsilon ^I){\cal D}_a\phi $  is again a covariant quantity.\vspace{3mm}

A similar lemma holds for curvatures, which are in fact the generalization of
the covariant derivatives for gauge fields:\\[3mm]
\noindent \textbf{Lemma on Covariant Curvatures}. \textit{For a gauge
field with transformation law as in (\ref{delJhI1}) where $M_{aJ}{}^I$ is a
covariant quantity, the covariant curvature (\ref{Rab}), with
(\ref{hatR}) is a covariant quantity.}\vspace{3mm}

The rest of this section is organized as follows. We first give a proof of the lemma of covariant derivatives in Sect. \ref{ss:proofcovder}. In Sects. \ref{ss:exd6VM} and \ref{ss:delFab} we
discuss the example of the $D=6$ abelian vector multiplet, in which we see in practice how the cgct indeed needs the seemingly non-covariant last term in (\ref{cgcthI}) and how the lemma
on the covariant curvature is realized: $\widehat{R}_{\mu \nu }$ does not
transform to a covariant quantity, but $\widehat{R}_{ab}$ does. As it will be clear soon, computing these variations is in general long and painful. Useful simplifications happen if we strategically make use the lemmas above, provided the algebra is closed, and this will be in fact the content of Sect. \ref{ss:easytransf},  the `easy way'. In Sect. \ref{ss:nonclosureD6VM} we will also consider an example with an open algebra.
\subsection{Proof of Lemma on Covariant Derivatives}
 \label{ss:proofcovder}
Let us start from computing $\delta _J(\epsilon ^J)  {\cal D}_\mu \phi$. Using the definitions \eqref{defDmu}, (\ref{delJhI2}) and (\ref{commutatorphi})
\begin{eqnarray}
\delta _J\left(\epsilon ^J\right)  {\cal D}_\mu \phi & = & \epsilon ^I\partial _\mu \chi _I
- \epsilon ^J B_\mu ^K
  f_{KJ}{}^I \chi _I-\epsilon ^J{\cal M}_{\mu J}{}^I\chi _I
  -B_\mu ^I\delta _J\left(\epsilon^J\right)\chi _I\nonumber\\
& = & \epsilon ^I\left({\cal D} _\mu + \delta _J\left(B_\mu ^J\right)\right)\chi _I -
\epsilon ^J B_\mu ^K
  f_{KJ}{}^I \chi _I\nonumber\\
&&-\epsilon ^J{\cal M}_{\mu J}{}^I\chi _I-\delta _J\left(\epsilon ^J\right)\delta
_I\left(B_\mu ^I\right)\phi \,. \label{delDmuphi1}
\end{eqnarray}
For the last term, we used the notation in the way it is used in
calculating a commutator. That means, the first $\delta (\epsilon )$
does not act on the $B_\mu ^I$ within the $\delta
_I(B_\mu ^I)\phi$, as it is done when one calculates a commutator. Then from this piece plus $\epsilon ^I\delta _J(B_\mu ^J)\chi _I $ in the second line one reconstructs a
commutator with $\epsilon _1^I$ replaced by $B_\mu^I $ and $\epsilon
_2^I$ by $\epsilon ^I$. Using then (\ref{commutatorphi}), we obtain
\begin{eqnarray}
 \delta _J\left(\epsilon^J\right) {\cal D}_\mu \phi & = & \epsilon ^I {\cal D} _\mu \chi _I
-\epsilon ^J{\cal M}_{\mu J}{}^I\chi _I +\epsilon ^JB_\mu ^I
\left( f_{IJ}{}^a {\cal D}_a\phi +\eta _{IJ}\right)\nonumber\\
   & = &\epsilon ^I {\cal D} _\mu \chi _I-\epsilon ^J{\cal M}_{\mu J}{}^I\chi _I +\epsilon ^JB_\mu ^I \eta _{IJ} \nonumber\\
&&+
\left( \delta _I(\epsilon ^I) e_\mu {}^a - \epsilon ^I e_\mu ^b
f_{bI}{}^a \right){\cal D}_a\phi \,.
 \label{delDmuphi2}
\end{eqnarray}
At the end, we made use of (\ref{delIe}). The first term in the bracket
has an explicit $\psi _\mu $, but is canceled when reverting to the
transformation of ${\cal D}_a\phi $:
\begin{equation}
\delta _J\left(\epsilon ^J\right) {\cal D}_a \phi=\epsilon ^I {\cal D} _a \chi _I -\epsilon
^J{\cal M}_{a J}{}^I\chi _I +\epsilon ^JB_a ^I \eta _{IJ} - \epsilon ^I
f_{aI}{}^b {\cal D}_b\phi \,,
 \label{finaldelDaphi}
\end{equation}
namely $ {\cal D}_a \phi$ is a covariant quantity, since $\delta _J(\epsilon ^J) {\cal D}_a \phi$ does not contain derivatives on the parameters. If $\eta _{IJ}=0$, the explicit gauge fields disappear from (\ref{finaldelDaphi}) and the transformation of the derivative is again a
covariant quantity. This proves the lemma on covariant derivatives. We will come back to the meaning of the last term of (\ref{finaldelDaphi}) in
Sect.~\ref{ss:easytransf}.


\subsection{Example: \texorpdfstring{$D=6$}{D=6} Abelian Vector Multiplet} \label{ss:exd6VM}
We will use an example to illustrate the closure of the algebra on
gauge fields. In Sect.~\ref{ss:delFab} this example is used to show
how the transformation of curvatures gives a covariant result. Finally,
in Sect.~\ref{ss:easytransf} it is used to illustrate how to facilitate
calculations using the lemmas.

The $D=6$ abelian vector multiplet consists of a vector $W_\mu $ and
a spinor $\lambda ^i$. The transformations\footnote{We apologize for the many occurrences of $\lambda $ here. $\lambda ^i$ is the spinor, while $\lambda ^{ab}$ and $\lambda ^{ij}$ are the parameters appearing in Table \ref{tablegaugefields}.} under standard gauge transformations are:
\begin{eqnarray}
  \delta_I\left(\epsilon ^I\right) W_\mu &=&\partial _\mu \theta +\ft12\bar \epsilon \gamma _\mu \lambda\,, \nonumber\\
  \delta_I\left(\epsilon ^I\right) \lambda ^i&=&
   \left(\ft32 \lambda _{\rm D} -\ft14 \gamma ^{ab}\lambda _{ab}\right)
   \lambda^i +\lambda ^{ij} \lambda _j-\ft14
    \gamma ^{ab}\widehat{F}_{ab} \epsilon^i\,,
 \label{TransfVM6onshell}
\end{eqnarray}
where $\theta $ is the parameter for the $\U(1)$ transformation that $W_\mu $
gauges. This $\U(1)$ commutes with all other symmetries. $W_\mu $ is thus
one of the $B_\mu ^I$ in the general treatment, and comparing with
(\ref{delJhI1}), we can identify
\begin{equation}
  M_{\mu (\alpha i) }{}^{\U(1)}=\ft12\left(  \gamma _\mu \lambda_i\right)_\alpha
  \,,
 \label{MinVM6}
\end{equation}
where $(\alpha i)$ stands for the combined index (spinor + extension) indicating a
supersymmetry. The $i$-index (and also the spinor index $\alpha $) is implicit in the first line of
(\ref{TransfVM6onshell}), as explained in (\ref{implicitIndBil}). The modified curvature, which is called $\widehat{F}_{\mu
\nu }$, is
\begin{equation}
  \widehat{F}_{\mu \nu }=F_{\mu \nu }-\bar \psi _{[\mu }\gamma _{\nu ]}
  \lambda, \qquad F_{\mu \nu }=2\partial _{[\mu }W_{\nu ]}.
 \label{hatFVM6}
\end{equation}
Instead of using (\ref{hatR}), just recognize that derivatives in $F_{\mu
\nu }$ have to be completed to covariant derivatives.

Let us now comment on $\delta_I(\epsilon ^I) \lambda ^i$.
The $\lambda _{\rm D}$-term states that the fermion $\lambda^i $ has `Weyl weight' $\ft32$. We will explain in
Sect.~\ref{ss:confpropMatter} how this weight can be easily obtained.
The Lorentz transformation is valid for all the spinors. Its form can
in fact already be seen from the first commutator in (\ref{ConfCommutators}).
Similarly, the $\SU(2)$ transformation is general for any doublet.
Finally, under supersymmetry transformations, the matter field $\lambda^i$ transforms to $\widehat{F}_{ab}$, which is a covariant quantity (as will be shown explicitly in Sect.~\ref{ss:delFab}). This should be the case, since $\lambda^i$ is a matter field and should thus transform to a covariant quantity.

Transformations under cgct follow from (\ref{cgcthI}):
\begin{eqnarray}
 \delta _{\rm cgct}(\xi )W_\mu  & = & -\xi ^\nu \widehat{F}_{\mu \nu }-\ft12
 \xi ^a\bar \psi _\mu \gamma _a \lambda \,\nonumber\\
 \delta _{\rm cgct}(\xi )\lambda ^i  & = & \xi ^a{\cal D}_a\lambda ^i \,,
\qquad {\cal D}_\mu \lambda ^i=D_\mu \lambda ^i+\ft14 \bar \psi _\mu^i
 \gamma ^{ab}\widehat{F}_{ab}\,, \nonumber\\
D_\mu \lambda ^i&=&\left( \partial _\mu -\ft32 b_\mu
  +\ft14 \gamma ^{ab}\omega  _{\mu ab}\right)
   \lambda^i -V_\mu {} ^{ij} \lambda_j
   \,.
 \label{delcgctVM6}
\end{eqnarray}
Let us first comment on the last expression. The $D_{\mu}$ introduced here only includes the $D$, $M_{ab}$ and $R$-symmetry, i.e. the linearly realized symmetries.\footnote{That turns out to be convenient in computations, as will be illustrated below.}

Note that the last term in the first line is necessary for
a correct result of the anticommutator of two supersymmetries. Indeed,
consider the exercise of calculating the commutator of two
supersymmetries on $W_\mu $. In order to close the algebra, this commutator should give a cgct.
\begin{equation}
  \delta_Q (\epsilon_1)\delta_Q (\epsilon _2)W_\mu =\ft12\bar \epsilon _2\gamma _\mu
\delta_Q (\epsilon_1)\lambda +\ft12\bar \epsilon _2\gamma _a\lambda\,\delta_Q
(\epsilon_1) e_\mu {}^a\,.
 \label{checkcommWd6}
\end{equation}
The first term gives clearly the covariant curvature, and the second one
leads to the second term in the last expression of (\ref{cgcthI}).  The $\{Q,Q\}$ algebra is thus closed on $W_\mu$.
 \bexer \label{ex:commWmu}
 Check further that one obtains indeed the right coefficient for the
 transformations, using (\ref{xiaeps12}), symmetries using \cite[(3.51)]{Freedman:2012zz},
 Fierz formulae (\ref{Fierzchiral6}), gamma manipulations using
 (\ref{sandwich}), and (\ref{as=epstrace}).
 \eexer

\subsection{Illustration of Full Calculation of the Transformation of a Curvature} \label{ss:delFab}
 We will now show that $\widehat{F}_{ab}$ is a covariant
quantity. We first calculate the supersymmetry transformation of
$\widehat{F}_{\mu \nu }$, defined in (\ref{hatFVM6}). Therefore we need $\delta_Q \psi _\mu $.
The matter terms are collected into the symbols $\Upsilon_\mu $. As we have seen, the latter are in principle determined by the gauge algebra. For the sake of simplicity we will leave the matter terms arbitrary in the following derivation.
The starting point is then
\begin{equation}
  \delta _Q\psi _\mu = D_\mu \epsilon + \Upsilon_\mu\,,\qquad
  D_\mu \epsilon^i \equiv \left( \partial _\mu +\ft12 b_\mu
  +\ft14 \gamma ^{ab}\omega  _{\mu ab}\right)
   \epsilon^i -V_\mu {} ^{ij} \epsilon _j\,,
 \label{delQpsi6}
\end{equation}
thus
\begin{equation}
 \delta_Q(\epsilon ) \widehat{F}_{\mu \nu }  =  \partial _{[\mu }\left( \bar \epsilon \gamma _{\nu]}
\lambda  \right) -\overline{(D_{[\mu } \epsilon)}\gamma _{\nu]}
\lambda -\bar \Upsilon_{[\mu }\gamma _{\nu]} \lambda -\bar \psi _{[\mu
}\delta _Q(\epsilon )\left( \gamma_{\nu]}\lambda  \right)\,.
 \label{delQFmunu}
\end{equation}
The first $\partial _\mu $ can be replaced by a covariant
derivative $D_\mu $ with only Yang--Mills gauge connections, as it acts on a $\SU(2)$ scalar, Lorentz
scalar, and D-invariant quantity. To say so, we consider for now the
parameter $\epsilon $ as an $\SU(2)$ doublet, Lorentz spinor, and of
dilatational weight $-\ft12$ (check that $\lambda $ got weight $\ft32$ and
the implicit $e_\nu {}^a$ has weight $-1$, see (\ref{delIe})). That is
implicit in the definition of $ D_\mu \epsilon^i$ in
(\ref{delQpsi6}). Then we can `distribute' this covariant derivative, and
terms with $D_\mu $ on $\epsilon $ cancel. This illustrates the
convenience to work with the $D_\mu $ derivatives. We are left with
\begin{equation}
 \delta_Q(\epsilon ) \widehat{F}_{\mu \nu }  =
 \bar \epsilon \gamma _{[\nu}
D _{\mu] }\lambda +\bar \epsilon\gamma _a \lambda\,D_{[\mu
}e_{\nu]}{}^a-\bar \Upsilon_{[\mu }\gamma _{\nu]} \lambda -2\delta_Q
(\epsilon )\delta_Q (\psi _{[\mu} )W_{\nu]} \,.
 \label{delQFmunu2}
\end{equation}
The writing of the last term is similar to what was done in
Sect.~\ref{ss:proofcovder}, and the following manipulations are
similar to those that we did there. Replacing the first derivative by
a full covariant derivative, this can be written as $D_\mu \lambda
={\cal D}_\mu \lambda +\delta _Q(\psi _\mu )\lambda $. The latter term nearly
leads to $\delta _Q(\psi _\mu )\delta _Q(\epsilon )W_\nu $, apart from
that we have to be careful that also the frame field transforms in the latter
expression. We obtain:
\begin{eqnarray}
 \delta_Q(\epsilon ) \widehat{F}_{\mu \nu } & =&
  \bar \epsilon \gamma _{[\nu}
{\cal D} _{\mu] }\lambda  +\bar \epsilon\gamma _a \lambda\,D_{[\mu
}e_{\nu]}{}^a-\bar \Upsilon_{[\mu }\gamma _{\nu]} \lambda \nonumber\\ &&
+2\left[ \delta_Q (\psi _{[\mu}),\delta_Q (\epsilon )\right] W_{\nu]}-
\bar \epsilon \gamma _a\lambda \delta _Q(\psi _{[\mu })e_{\nu]}{}^a \,.
 \label{delQFmunu3}
\end{eqnarray}
We already calculated the commutator on $W_\mu $, checking that it gives
the covariant general coordinate transformation. The parameter that we
have to use here is, see (\ref{xiaeps12}):
\begin{equation}
  \xi ^a(\psi _\mu ,\epsilon )=\ft12\bar \epsilon \gamma ^a\psi _\mu =
  \delta _Q(\epsilon )e_\mu {}^a\,.
 \label{xipsieps}
\end{equation}
That can be inserted in (\ref{delcgctVM6}) and one can use the constraint
(\ref{RP0}) to obtain
\begin{equation}
  0=2D_{[\mu }e_{\nu ]}{}^a + \xi ^a(\psi _\mu ,\psi _\nu )\,.
 \label{RP0calDxi}
\end{equation}
We leave to you that after further Fierz manipulations as in
Exercise~\ref{ex:commWmu}, one arrives at
\begin{equation}
   \delta_Q(\epsilon ) \widehat{F}_{\mu \nu }  =
 \bar \epsilon \gamma _{[\nu}
{\cal D} _{\mu] }\lambda -\bar \Upsilon_{[\mu }\gamma _{\nu]} \lambda
-2\widehat{F}_{a[\nu }\delta _Q(\epsilon )e_{\mu]}{}^a  \,.
 \label{delQFmunu4}
\end{equation}
The last term contains an explicit $\psi _\mu $. This is
canceled when calculating the transformation
\begin{equation}
  \delta _Q(\epsilon )\widehat{F}_{ab}=  \bar \epsilon \gamma _{[b}
{\cal D} _{a] }\lambda -\bar \Upsilon_{[a }\gamma _{b]} \lambda\,,
 \label{resultdelFab}
\end{equation}
which does not contain any explicit gauge fields. The gauge fields are
hidden in the covariant derivative.
\subsection{The Easy Way}  \label{ss:easytransf}

This was enough to show that the transformation works out as expected. But the calculation was complicated, and we were only
looking to a simple example! However, now comes the good news: once you
know some tricks, you never have to do all the computations. These tricks involve the knowledge
of some basic facts:
\begin{enumerate}
\item  ${\cal D}_a$ on a covariant quantity and $\widehat{R}_{ab}$ are covariant quantities;
  \item The transformation of a covariant quantity does not involve a
  derivative of a parameter (definition).
  \item \emph{If the algebra closes on the fields}, then the transformation
  of a covariant quantity is again a covariant quantity, i.e. gauge fields
  only appear either included in covariant derivatives or in curvatures.
\end{enumerate}

Let us consider again the calculation of $\delta_Q\widehat{F}_{ab}$ in the $D=6$ abelian vector multiplet, where from \eqref{checkcommWd6} we know that the algebra closes on the gauge fields.
The third principle implies that  $\delta_Q\widehat{F}_{ab}$ should be a covariant quantity. Let us start from the definition (\ref{hatFVM6})
\begin{equation}
  \widehat{F}_{\mu \nu }=F_{\mu \nu }-\bar \psi _{[\mu }\gamma _{\nu ]}
  \lambda, \qquad F_{\mu \nu }=2\partial _{[\mu }W_{\nu ]}\,.
 \label{defhatFD6}
\end{equation}
First of all, consider the transformation $\delta_Q{F}_{\mu\nu}\simeq 2\partial _{[\mu }\delta_QW_{\nu]}$, where $\delta_Q W_\mu= -\overline{\epsilon}\gamma_\mu \lambda$.  When taking the $\partial _\mu $
derivative, we delete the term where the derivative acts on the parameters $\alpha,\, \overline{\epsilon}$,
because these have to disappear in the transformation of a covariant quantity. Hence,
\begin{equation}
  \delta_Q F_{\mu\nu} \Rightarrow \overline{\epsilon}\gamma_{[\nu}\partial_{\mu]} \lambda\,.
 \label{delQFterm}
\end{equation}
Moreover we can forget about derivatives on a frame field. Indeed, a
derivative on any gauge field can only appear as its curvature, but we
know that the curvature for the frame field has been constrained to zero. Hence:
\begin{equation}
  \delta_Q F_{ab} \Rightarrow \overline{\epsilon}\gamma_{[b}\partial_{a]} \lambda \Rightarrow \overline{\epsilon}\gamma_{[b}{\cal D}_{a]} \lambda\,.
 \label{delQFDalambda}
\end{equation}
The last implication is just the `covariantization' of the term involving the partial derivative $\partial_a$, which is not covariant. The covariant quantity must involve ${\cal D}_a$, instead of $\partial_a$.

Let us now consider the second term, $\bar \psi _{[\mu}\gamma _{\nu ]}\lambda$. One can realize that the interesting term arises from
\begin{equation}
  \left(\delta _Q\bar \psi _{[\mu}\right)\gamma _{\nu ]}\lambda \,.
 \label{variedpart}
\end{equation}
Indeed, if we would act with $\delta _Q$ on the other factors, then a
$\psi _\mu $ remains and we know that these must be included in those covariant derivatives that we have already considered in the first term. For the same reason, regarding the variation of $\psi _\mu $, as for any $B_\mu
^I$ gauge field, one can neglect the first and second terms in the
first line of (\ref{delJhI2}):
\begin{equation}
  \delta_J\left(\epsilon^J\right)B_\mu^I \Rightarrow \epsilon^J \mathcal{M}_{\mu J}^I\,.
 \label{delBtermM}
\end{equation}
In our case this ${\cal M}$ is the $\Upsilon$-term in (\ref{delQpsi6}). These steps are quite general, leading to the conclusion that the
only relevant part of the transformation of gauge fields is ${\cal M}$.

\vspace{10pt}

In light of this smart strategy, we now review the result of the \textit{transformation of a covariant
derivative} on a covariant quantity, $\delta _J(\epsilon ^J) {\cal D}_a \phi$, as obtained in
(\ref{finaldelDaphi}). In order to apply this method we need to assume that the algebra closes on the field $\phi$, since otherwise $\delta_J(\epsilon^J){\cal D}_a \phi$ is not a covariant quantity and this `easy' method cannot be applied straightforwardly.

The first term in (\ref{finaldelDaphi}) is the covariantization of the $\partial _\mu \phi $ term in (\ref{delDmuphi1}).

The second term, the ${\cal M}$ term, is the only one that remains explicit from $\delta_J (\epsilon^J)B_\mu^I$, the rest being implicitly included in covariant derivatives. Thus, from a practical point of view, these non-gauge terms are  the most interesting terms in the transformation of gauge fields.
As shown in (\ref{delJhI2}), ${\cal M}$ terms are of two types. The first ones are those from the gauge algebra where the
gauge field was the frame field. These are the transformations explicitly given in (\ref{delextra}). The others are the matter terms, which we still have to find for each case. For the gauge field in the vector multiplet, that was the relevant term, which gave its supersymmetry transformation to the gaugino.

The third term appears only in case of non-closure. We will consider below the
fermionic field in the vector multiplet for which there is no closure.\footnote{
Closure could be obtained if we would have introduced auxiliary fields,
but for the didactical value of the example, it is good to consider the
situation without the auxiliary fields.}

Now consider the fourth term, which finds its origin in transformations
of the frame field to the frame field, (\ref{delIe}). This term refers to the first two terms in the
explicit expression of (\ref{delIe}):
\begin{equation}
  \delta_J \left(\epsilon^J\right){\cal D}_a \phi=...- \left(\lambda_{\rm D}\delta_a{}^b+\lambda_a{  }^{b}\right){\cal D}_b\phi\,,
 \label{delDaphifrome}
\end{equation}
which amount to the following. The first one implies that the Weyl weight of ${\cal D}_a\phi $ is one
higher than that of $\phi $. The second one implies that ${\cal D}_a\phi $ is a Lorentz vector.\vspace{3mm}

 \bexer
 Check that, whatever would be the $S$-transformation of $\lambda $ (here
 in 6 dimensions it is zero, however the corresponding fermion in 4
 and 5 dimensions has $S$-supersymmetry transformations), the
 $S$-variation of $\widehat{F}_{ab}$ is due to (\ref{delextra}):
\begin{equation}
  \delta _S(\eta )\widehat{F}_{ab}= \bar \eta \gamma _{ab}\lambda \,.
 \label{delShatF}
\end{equation}
  \eexer
 \bexer
One can even give a general formula for the transformation of curvatures,
correcting (\ref{deltacurv}) for the effects that gauge fields transform
with matter-like terms. To apply the methods explained earlier, the last
decomposition in (\ref{hatR}) is most useful. Indeed, explicit gauge
fields appear only quadratically in $r_{ab}{}^I$. You can then determine
that
\begin{equation}
  \delta _J\left(\epsilon^J\right)\widehat{R}_{ab}{}^I= \epsilon ^J\widehat{R}_{ab}{}^K
  f_{KJ}{}^I + 2\epsilon ^J{\cal D}_{[a}{\cal M}_{b]J}{}^I
  -2\epsilon ^K {\cal M}_{[aK}{}^J{\cal M}_{b]J}{}^I\,.
 \label{newdelR}
\end{equation}
Similarly, the Bianchi identity becomes
\begin{equation}
  {\cal D}_{[a}\widehat{R}_{bc]}{}^I + \widehat{R}_{[ab}{}^J{\cal M}_{c]J}{}^I=0\,.
 \label{newBianchi}
\end{equation}
 \eexer

\subsection{Non-closure Terms in \texorpdfstring{$D=6$}{D=6} Abelian Vector Multiplet}
\label{ss:nonclosureD6VM}
Finally, let us consider the fermionic field (gaugino) $\lambda ^i$ in the $D=6$ vector multiplet, on which the supersymmetry algebra does not close. As we have already calculated the supersymmetry transformation
of $\widehat{F}_{ab}$ in \eqref{resultdelFab}, it is easy to calculate the commutator of two
supersymmetries on the gaugino. We do not take the $\Upsilon$-term into
account, as we do not know its form yet, and it is independent of the
rest. The commutator is
\begin{align}
  \left[ \delta _Q(\epsilon _1),\delta _Q(\epsilon _2)\right] \lambda
  ^i=&-\ft12{\cal D}_c\lambda ^i\,\bar \epsilon _1\gamma ^c\epsilon _2\nonumber\\
&  +\ft3{16}\gamma _c\slashed{\cal D} \lambda^i\,\bar \epsilon _1\gamma ^c\epsilon _2
  +\ft1{96}\gamma _{cde}\slashed{\cal D} \lambda_j\,
  \bar \epsilon _1^{(i}\gamma^{cde}\epsilon ^{j)}  _2\,.
 \label{commOmega6}
\end{align}
The first term is the covariant general coordinate transformation. The
others are the non-closure terms, all proportional to
$\slashed{\cal D} \lambda$.  From these terms one finds that the supersymmetry variation of
${\cal D}_a\lambda $ is (neglecting again possible extra matter terms in
transformations of gauge fields)
\begin{equation}
  \delta _Q(\epsilon ){\cal D}_a\lambda^i =
  -\ft14 \gamma ^{bc}\epsilon^i {\cal D}_a\widehat{F}_{bc}
  +\ft3{16}\gamma _c\slashed{\cal D} \lambda^i\,\bar \psi _a\gamma ^c\epsilon
  +\ft1{96}\gamma _{cde}\slashed{\cal D} \lambda_j\,
  \bar\psi _a^{(i}\gamma^{cde}\epsilon ^{j)}  \,.
 \label{deltaQDOmega}
\end{equation}
To conclude, we see explicitly that the transformation of ${\cal D}_a\lambda $ is
not a covariant quantity. It is now not possible to define ${\cal D}_b{\cal D}_a\lambda
$ such that it does not transform to a derivative of a gauge field.
However, note that one can define it such that at least the antisymmetric
part in $[ab]$ does not transform in a derivative. That is analogous to a
curvature. Also on a gauge field we cannot define a covariant
generalization of $\partial _{(\mu }W_{\nu )}$. The covariant
${\cal D}_{[b}{\cal D}_{a]}$ should just have extra factors $\ft12$ for every term in
which there are two gauge fields. Also that is similar to a curvature:
\begin{align}
  {\cal D}_{[b}{\cal D}_{a]}\lambda ^i=&\partial _{[b}{\cal D}_{a]}\lambda ^i+
  \ft14 \gamma ^{cd}\psi _{[b}^i {\cal D}_{a]}\widehat{F}_{cd}\nonumber\\
&  -\ft3{32}\gamma _c\slashed{\cal D} \lambda^i\,\bar \psi _a\gamma ^c\psi _b
  -\ft1{192}\gamma _{cde}\slashed{\cal D} \lambda_j\,
  \bar\psi _a^{(i}\gamma^{cde}\psi _b ^{j)}  \,.
 \label{DbDaOmega}
\end{align}
But these are objects that one seldom needs.
\section{Curvature Constraints} \label{ss:constraints}
As we advocated in Sect. \ref{ss:gaugethspacetime}, we need to impose certain constraints in order that the translations on the gauge fields take the form of (covariant) general coordinate transformations. The kind of constraints we need are the so-called \emph{conventional constraints}, which are algebraic (and not differential) conditions on some fields. At the classical level, conventional constraints do not impose restrictions on the dynamic itself. Importantly, this discussion is not specific for conformal supergravity.\footnote{Indeed, one needs such constraints also in general relativity (as explained at the beginning of this chapter) where the connection $\omega_\mu{}^{ab}$ must be a function of $e^a_\mu$, and not an independent field.}

\subsection{Constraint on \texorpdfstring{$R(P)$}{R(P)}}
\label{ss:constraintRP}
The vanishing of the $P^a$ curvature has been already considered in (\ref{RP0}) and it is immediately clear why this leads to a conventional constraint. Concerning representation content, (\ref{RP0}) is a vector times an antisymmetric
tensor, and the same holds for the spin connection $\omega _\mu {}^{ab}$.
For this reason one can solve (\ref{RP0}) algebraically for $\omega _\mu {}^{ab}$, which from now on will become a dependent field\footnote{This is what we were
looking for (see the first item of the list of three shortcomings in Sect. \ref{ss:gaugethspacetime}).}:
\begin{align}
  \omega _\mu {}^{ab}=& \omega _\mu {}^{ab}(e,b)  -\xi ^{[a}(\psi _\mu ,\psi ^{b]})-\ft12\xi _\mu (\psi ^a,\psi ^b)\,,\nonumber\\
  &\omega _\mu {}^{ab}(e,b)= \omega _\mu {}^{ab}(e)+2e_\mu {}^{[a}b^{b]}\,,
 \label{solomega}
\end{align}
where $\omega _\mu {}^{ab}(e)$ is the usual expression of the spin
connection, see (\ref{omegae}), obtained by  the antisymmetric part of
(\ref{Nablae0}), which is a simplification of the above constraint
without $b_\mu $ and gravitini. The term with $b^b$ can be understood as necessary to reproduce
the special conformal invariance, see (\ref{delextra}). The gravitino
contribution is already known from pure Poincar\'{e} ${\cal N}=1$
supergravity~\cite{Freedman:1976xh,Deser:1976eh,VanNieuwenhuizen:1981ae}.
\vspace{3mm}

It is important to realize that the constraint (\ref{RP0}) is not invariant under all the symmetries if one assumes for $\omega_\mu{}^{ab}$ the transformation rule that follows from (\ref{delGaugef}), such that the curvatures would just transform as in (\ref{deltacurv}). Explicitly, the transformation $\delta _I(\epsilon ^I) R_{ab}(P^c)$ can be calculated from the definition (\ref{RP0}) using the `easy method'. The terms quadratic in the
gravitino do not play a role in the variation, as they will always leave
a naked gravitino behind. Thus we should only consider the variations of
the frame field in the derivative. According to (\ref{delIe}), this involves
only $\partial e$, i.e. $R_{ab}(P)$, and $\partial \psi $, i.e.
$\widehat{R}_{ab}(Q)$. The former are zero, so there remains only a
variation in $\widehat{R}_{ab}(Q)$:
\begin{equation}
  \delta _I\left(\epsilon^I\right) R_{ab}(P^c)=\xi ^c\left( \widehat{R}_{ab}(Q),\epsilon
   \right)\,.
 \label{delRP}
\end{equation}
One may therefore expect that there would be a new constraint $\widehat{R}_{ab}(Q)=0$. However, since in the expression for
$\widehat{R}_{ab}(Q)$ there is no field that is a spinor--antisymmetric tensor, this choice would be an unconventional constraint: it cannot be solved.
Instead, one should define the transformation laws of the dependent field from its definition as in (\ref{solomega}). As $\omega _\mu
{}^{ab}$ is now defined by the vanishing of $R_{ab}(P)$, its
transformations can be defined by its expression in terms of the other
fields.
Another way to say so, is that $\delta \omega _\mu {}^{ab}$ has
extra terms to compensate for the non-invariance of $R_{ab}(P)$, calculated above.

To see how this method works concretely, let us be slightly more general and consider a constraint $C(\omega(\phi),\phi)$, where $\omega(\phi)$ is a composite of the independent field $\phi$ and we assume that $C$ contains at least a curvature.
The transformation of $C$ consists of the pure gauge terms, from the gauge theory algebra and following the general rules, plus extra terms $\delta_{\cal M}\omega$ like in (\ref{delJhI2}):
\begin{equation}
\delta C = \delta_{\rm gauge}C + \delta_{\cal M}C = \delta_{\rm gauge}C + \frac{\partial C}{\partial \omega }\delta_{\cal M}\omega\,.
\label{deltaCgauge}
\end{equation}
Because $\frac{\partial C}{\partial \omega} $ is invertible we can calculate the extra terms in the transformation of $\omega $ that are necessary in order that the constraint is invariant, $\delta C=0$. The result for the constraint (\ref{RP0}) has the same structure as the solution (\ref{solomega}):
\begin{equation}
  \delta _{\cal M}\omega _\mu {}^{ab}=\xi ^{[a}\left(\widehat{R} _\mu{}^{b]}(Q) ,\epsilon\right)
  +\ft12\xi _\mu \left(\widehat{R}^{ab}(Q),\epsilon \right)\,,
 \label{delExtraOm}
\end{equation}
and can be thus interpret as the additional $M$ term in (\ref{delJhI2}) for the $\omega $ gauge field.

A final consequence of the  imposing (\ref{RP0}) can be seen from the Bianchi
identity (\ref{Bianchi}) applied to translations: ${\cal D}_{[\mu }R_{\nu \rho]}(P^a)=0$. Of course the covariant derivative used in this expression is not limited to the standard gauge transformations, and as such one will find covariantization terms corresponding to local translations, i.e. $e_\mu ^b f_{b J}{}^a R_{\nu \rho }{}^J$, where $f_{b J}{}^a$ are structure constants in the commutator $[P_b,T_J]$ proportional to $P^a$. This commutator is only non-zero for $T_J$ being dilatations and Lorentz rotations. According to the lemmas in Sect. \ref{ss:covquantities}, these covariantization terms (with all $\mu$ indices changed to tangent
spacetime indices) are the only ones that we have to consider when we upgrade the curvatures to covariant quantities.  Hence the Bianchi identity for translations immediately gives
\begin{equation}
  -\delta _{[a}{}^d \widehat{R}_{bc]}(D)+\widehat{R}_{[ab}\left(M_{c]}{}^d\right)=0\,.
 \label{BianchiP}
\end{equation}

So far, this is independent of supersymmetry or even of the conformal
group. In fact, the last identity, without $R(D)$ would then be the
well-known equation that the $M$-curvature antisymmetrized in 3 indices
is zero. Multiplying (\ref{BianchiP}) with $\delta _d{}^a $ gives that
the contracted $M$-curvature, i.e. the Ricci tensor, is symmetric if
there is no $R_{ab}(D)$:
\begin{equation}
(2-D)  \widehat{R}_{ab}(D)=2 \widehat{R}_{d[a}\left(M_{b]}{}^d\right)\,.
 \label{RD}
\end{equation}
\bexer The $M$-curvature can be calculated directly by taking the
derivative of~(\ref{solomega}). As the final result has to be covariant,
we can drop all other terms, and restrict to the $\partial \omega $ term,
where the derivative acts either on $\omega (e)$ or on $b_\mu $. This
leads to a direct relation expressing $\widehat{R}_{\mu \nu }(M^{ab})$ in
the covariantization of $\partial \omega (e)$, and $\widehat{R}_{\mu
\nu}$. Check that in this terminology, (\ref{BianchiP}) implies that the
former part satisfies the cyclic identity $R_{[\mu \nu \rho ]}{}^\sigma
=0$ and leads to a symmetric Ricci tensor.
 \eexer


\subsection{Other Conventional Curvature Constraints}\label{ss:otherConstr}

 Now that we have seen the procedure how $\omega _\mu {}^{ab}$ became
composite, we may consider whether this can be done for other gauge
fields in the same way. If we look for the smallest possible multiplet,
then we should use this procedure for all the gauge fields that we can
algebraically solve. The clue is given in (\ref{extraR}). These
equations show which gauge fields appear `linearly' in the curvatures, namely multiplied by a frame field. In particular, we may
choose constraints that determine $f_\mu {}^a$ and $\phi _\mu $. This
looks rather convenient. Indeed, without such a
constraint $f_\mu {}^a$ would already appear in bosonic conformal gravity, while we do not
know this field in the physical conformal gravity. At first it looks that
we can define it either from a constraint on the dilatational or on the
Lorentz curvature. But~(\ref{RD}) implies that $\widehat{R}(D)$ is a
function of $\widehat{R}(M)$, so that we can restrict ourselves to the
Lorentz curvature. To be able to eliminate $f_\mu {}^a$ completely, the
constraint should be a general $d\times d$ matrix:
\begin{equation}
  \widehat{R}_{ac}(M^{bc}) + C_a {}^b =0\,,
 \label{secondConstraint}
\end{equation}
where $C_a {}^c$ may be any covariant function that is independent of
$f_\mu {}^a$. Similarly, there can be a third constraint that determines
$\phi _\mu $ from a constraint of the form
\begin{equation}
   \gamma ^b\widehat{R}(Q)_{ba}+\rho_a=0\,,
 \label{ThirdConstraint}
\end{equation}
where $\rho _a$ is a covariant spinor-vector. One might impose the constraints with $C_a{}^b$ and $\rho _a$ vanishing, but we will see in Sect. \ref{sect:Weyl} that modifying the constraints in this way depending on other (auxiliary) fields can be convenient. In any case, the addition or not of such extra parts amounts at the end only to a field redefinition of the dependent gauge fields.

We go now again through the consequences of these constraints. Let us repeat the names of the
various covariantizations of curvatures (and introduce meanwhile another
one: $\widehat{R}'$):
\begin{eqnarray}
 r_{\mu\nu}{}^I &=& 2\partial_{[\mu}B_{\nu]}{}^I+ B_\nu{}^K B_\mu{}^J f_{JK}
  {}^I\,,\nonumber\\
 R_{\mu \nu }^I & = & r_{\mu \nu }{}^I -2B_{[\mu }{}^Jf_{\nu ]J}{}^I\,,\nonumber\\
  \widehat{R}_{\mu \nu }{}^I&=& R_{\mu \nu }{}^I-2B_{[\mu }{}^JM_{\nu ]J}{}^I=
  r_{\mu \nu }{}^I-2B_{[\mu }{}^J{\cal M}_{\nu ]J}{}^I\,,\nonumber\\
  \widehat{R}'_{\mu \nu }{}^I & = & r_{\mu \nu }{}^I-2B_{[\mu }{}^JM_{\nu
  ]J}{}^I= \widehat{R}_{\mu \nu }{}^I +2B_{[\mu }{}^Jf_{\nu ]J}{}^I\,.
 \label{allCurv}
\end{eqnarray}
Furthermore, as a completion of the Ricci tensor, see (\ref{defR}), we define
\begin{equation}
{\cal R}_{\mu \nu }=\widehat{R}'_{\rho \mu }\left(M^{ab}\right) e_a{}^\rho e_{\nu
b}\,,\qquad {\cal R}={\cal R}_\mu {}^\mu \,.
 \label{Riccicov}
\end{equation}
The solution of (\ref{secondConstraint}) and (\ref{ThirdConstraint}) is
\begin{eqnarray}
 2(D-2)f_\mu {}^a & =&-{\cal R}_\mu {}^a-C_\mu {}^a
 +\frac{1}{2(D-1)}e_\mu {}^a\left( {\cal R}+C_b{}^b\right) \,,    \nonumber\\
 -s_D(D-2)\phi _a & = & \gamma ^b \widehat{R}'_{ab}(Q)-\rho _a-\frac{1}{2(D-1)}\gamma
 _a\left( \gamma ^{bc}\widehat{R}'_{bc}(Q)-\gamma ^b\rho _b\right) \,.
 \label{solfphi}
\end{eqnarray}
This is thus the analogue of  (\ref{solomega}). Similar to what we found there, the constraints will have the consequence that the
transformation law of the fields $f_\mu {}^a$ and $\phi _\mu $ get extra
matter-like contributions. We assume that the extra $C_a{}^b$ and $\rho $ terms
in the constraints have the same Lorentz structure, $\SU(2)$
structure, dilatational and $\U(1)$ weight as the curvature term. As we explained at the beginning of Sect. \ref{ss:constraintRP}
(see e.g. around (\ref{deltaCgauge})), the
constraints do not break supersymmetry (in general, this can be as well $Q$
as $S$ supersymmetry) as long as one modifies the transformations of the composite fields. The exact rules depend on the transformations of
the matter terms in the constraint. The simplicity of the result, e.g.
invariance of the constraint under $S$-transformations may even be an
argument for the choice of the matter terms in the constraints.

The $K$ transformations of these constraints vanish due to the constraint
on $R_{\mu \nu }(P)$, at least when the extra matter fields do not
transform under $K$. Indeed, this will not be the case due to a
Weyl weight related argument that we will give in
Sect.~\ref{ss:matterinWeyl}.

Furthermore, there are relations between the curvatures as a consequence of Bianchi identities, similar to (\ref{BianchiP}). Those give
relations for the $S$ and $K$ curvatures. The explicit form depends on the matter sector.

\section{Example: Non-SUSY Sigma Model} \label{ss:actconf}
Before continuing with the full theory, let us illustrate the construction of a conformal invariant action by
looking at a scalar sigma model without supersymmetry. In this context, the scalars $\phi ^i$ are coordinates of a Riemannian manifold, with metric $g_{ij}$. The latter defines the
Levi-Civita connection $\Gamma _{ij}^k$. We assume that there is a closed
homothetic Killing vector as in (\ref{homothetic}), which allows to
define the conformal symmetry of the scalars. The gauge fields are $e_\mu {}^a$, $\omega_\mu {}^{ab}$,  $b_\mu $ and $f_\mu {}^a$ (with conventions given in Table \ref{tablegaugefields}) and their transformation under local dilatations and special conformal transformations follow directly from (\ref{delGaugef}):
\begin{eqnarray}
  \delta \phi ^i&=&k_{\rm D}{}^i\lambda _{\rm D}\,,\nonumber\\
  \delta e_\mu {}^a&=&-\lambda _{\rm D} e_\mu {}^a, \qquad \delta e^\mu {}_a=\lambda _{\rm D} e^\mu {}_a,\nonumber\\
  \delta \omega_\mu {}^{ab}&=&-4\lambda _{\rm K}^{[a}e_\mu ^{b]}\,,\nonumber\\
  \delta b_\mu &=&\partial _\mu \lambda _{\rm D} +2\lambda_{{\rm K}\mu }\,,\nonumber\\
  \delta f_\mu {}^a&=&\lambda _{\rm D} f_\mu {}^a+\partial _\mu \lambda
  _{\rm K}{}^a\,,
 \label{deltaDK}
\end{eqnarray}
where as usual, $e^\mu {}_a$ is the inverse of $e_\mu {}^a$. The scalar fields are thus invariant under special conformal transformations.\footnote{$k_\mu^i(\phi)=0$ in the terminology of (\ref{deltaC}), and this can be expressed by saying that $\phi ^i$ are \emph{primary fields}.}

The covariant derivative for the scalars follows from \eqref{defDmu}:
\begin{equation}
  {\cal D}_\mu \phi ^i= \partial _\mu \phi ^i-b_\mu k_{\rm D}{}^i\,, \qquad {\cal D}_a\phi ^i=e^\mu_a D_\mu \phi
  ^i\,,
 \label{Dmuphii}
\end{equation}
leading to
\begin{align}
  \delta {\cal D}_a\phi ^i&= \lambda _{\rm D}\left[ {\cal D}_a\phi^i+\left(\partial _jk_{\rm D}{}^i\right){\cal D}_a\phi
  ^j\right] -2\lambda_{{\rm K}a}k_{\rm D}{}^i\nonumber\\
  & = \lambda _{\rm D}\left[(w+1) {\cal D}_a\phi^i-k_{\rm D}{}^k\Gamma ^i_{jk}{\cal D}_a\phi
  ^j\right] -2\lambda_{{\rm K}a}k_{\rm D}{}^i\,.
 \label{delDaphi}
\end{align}
The second (contracted) covariant derivative is therefore
\begin{align}
  {\cal D}_a{\cal D}^a\phi ^i=&\partial _a{\cal D}^a\phi ^i+2f_a^a k_{\rm D}{}^i-b^a\left[(w+1) {\cal D}_a\phi^i-k^k\Gamma ^i_{jk}{\cal D}_a\phi
  ^j\right]\nonumber\\ &+e^{\mu a}\omega _{\mu a}{}^b{\cal D}_b\phi ^i\,.
 \label{DaDaphi}
\end{align}
The covariant box is
\begin{equation}
  \Box^c\phi ^i={\cal D}^a{\cal D}_a\phi ^i+\Gamma ^i_{jk}{\cal D}_a\phi ^j{\cal D}^a\phi ^k\,.
 \label{covbox}
\end{equation}
Since
\begin{equation}
  \delta _{\rm K}\Box^c\phi ^i= 2\lambda _{\rm K}^a(D-2-2w){\cal D}_a\phi ^i\,,
 \label{delKDDphi}
\end{equation}
 the covariant box is conformally invariant only for $w=(D-2)/2$.

We will show that the following action is conformal invariant:
\begin{equation}
  e^{-1}{\cal L}= -\frac12g_{ij}{\cal D}_a\phi ^i{\cal D}^a\phi ^j+\frac{1}{w}f^a_ag_{ij}k_{\rm D}{}^ik_{\rm D}{}^j\,.
 \label{Lscalar}
\end{equation}
Both terms on the right-hand side scale under dilatations separately with
a factor $2(w+1)$, while the left-hand side scales with weight $D$. So it
is consistent under the same condition
\begin{equation}
  w= \ft12 (D-2)\,,
 \label{wd}
\end{equation}
which keeps the conformal box also $K$-invariant. Now we will show the $K$-invariance, which is why we need the second term in (\ref{Lscalar}). The $K$ transformations
of the action give
\begin{equation}
  e^{-1}\delta _{\rm K}{\cal L}= 2\lambda_{\rm K}^a k_{\rm D}{}^i g_{ij}{\cal D}_a\phi ^j +\frac{1}{w}\left( \partial
  _a\lambda_{\rm K}^a\right) g_{ij}k_{\rm D}{}^ik_{\rm D}{}^j=\frac{1}{w}\partial _a\left( \Lambda_
  K^a g_{ij}k_{\rm D}{}^ik_{\rm D}{}^j\right),
 \label{Kinvaction}
\end{equation}
where one uses that
\begin{equation}
  \partial _a\left(g_{ij}k_{\rm D}{}^ik_{\rm D}{}^j\right)= 2wk_{\rm D}{}^ig_{ij}\partial _a\phi ^j\,.
 \label{auxdagkk}
\end{equation}
This finishes the proof of the invariance of the action (\ref{Lscalar}).

The equations of motion from \eqref{Lscalar} with \eqref{wd} read
\begin{equation}
  \frac{\delta }{\delta \phi ^i}\int \rmd^D {\cal L} =  g_{ij}\Box^c\phi ^j\,.
 \label{FEBox}
\end{equation}

We now consider the constraint (\ref{secondConstraint}) with $C_a{}^b=0$. Indeed, here we do not have any reason nor fields available to consider a different $C_a{}^b$.
Then (\ref{solfphi}) leads to
\begin{equation}
  f_a^a = -\frac{1}{4(D-1)}R\,.
 \label{faaR}
\end{equation}
We now just write $R$, rather than ${\cal R}$ to indicate that there are
no further corrections to the usual bosonic scalar curvature of the
metric. Therefore, the Lagrangian is
\begin{equation}
  e^{-1}{\cal L}=-\frac12g_{ij}{\cal D}_a\phi ^i{\cal D}^a\phi ^j-\frac{1}{2(D-1)(D-2)}Rg_{ij}k_{\rm D}{}^ik_{\rm D}{}^j\,.
 \label{Lfinal}
\end{equation}
In order to have positive energy for gravity, the metric $g_{ij}$ should have a negative signature in the direction of the homothetic Killing vector $k_{\rm D}{}^i$.
In flat space, $k_{\rm D}{}^i=w\phi ^i$ and we obtain
\begin{equation}
  e^{-1}{\cal L}=-\frac12g_{ij}{\cal D}_a\phi ^i{\cal D}^a\phi ^j-\frac{D-2}{8(D-1)}Rg_{ij}\phi ^i\phi ^j\,.
 \label{Lfinalflat}
\end{equation}
For $D=4$, in the case of a single scalar with negative kinetic energy, this reduces to (\ref{confScalar}).

\section{The Standard Weyl Multiplets}\label{sect:Weyl}

The aim of this section is to construct and describe the multiplet of pure superconformal gravity, called \emph{Weyl multiplet}. As long as we are not interested in higher-derivatives supergravity theories, we do not need to consider an action for the Weyl multiplet. Instead, in the spirit of the superconformal calculus, results given in the present section will be very useful to build up actions for Poincar\'{e} supergravity.

\subsection{Matter Fields Completing the Weyl Multiplet} \label{ss:matterinWeyl}
So far, the problem mentioned in the fifth step of Sect. \ref{ss:Gengaugethspacetime} has not been considered. We saw in Table \ref{tbl:dofgauge} that the number of bosonic and fermionic
components of the independent gauge fields do not match. As a consequence, the
supersymmetry algebra cannot give rise to invertible coordinate
transformations. We could expect this result, since given the general forms for the
covariant general coordinate transformations and for the other gauge
transformations, nothing guarantees that supersymmetries anticommute
to the covariant general coordinate transformations. The matter terms $M$
in the transformations of the gauge fields (\ref{delJhI1}) should be chosen appropriately
to obtain the right anticommutator, and they should be functions of new
matter fields.

The solution is not unique and the arguments to obtain it are not so
obvious. The first one of this nature was obtained in 4 dimensions from
splitting a linearized Poincar\'{e} multiplet~\cite{Fradkin:1979as,Fradkin:1979cw,deWit:1979pq} that was found earlier.
One way of constructing the complete set of fields is to make use of
supercurrents. One starts with a multiplet that has rigid superconformal
symmetry and considers the fields that couple to the Noether currents.
This method has been used in various cases, see~\cite{Kaku:1978nz,Ferrara:1975pz,Sohnius:1979pk,Bergshoeff:1981is,Bergshoeff:1983qk,Bergshoeff:1999db,Bergshoeff:2001hc}. We will not go into this subject any further here.

We present here the solutions with the fields mentioned in
Table~\ref{tbl:countWeyl}, which are called the `standard Weyl multiplets'. \index{standard Weyl multiplet} Alternative versions for Weyl
multiplets have been
constructed~\cite{Bergshoeff:1986mz,Bergshoeff:2001hc,Butter:2017pbp}, which differ from
the above ones in the choice of auxiliary fields (fields below the double
line in Table~\ref{tbl:countWeyl}). These multiplets contain a dilaton
auxiliary field and are therefore called the `dilaton Weyl multiplets'. \index{dilaton Weyl multiplet}
The existence of a $D=4$ version, recently found in \cite{Butter:2017pbp} was already noticed in \cite[Sect.~4.4]{Siegel:1996px}.
\begin{table}[!t] 
\caption{\it Number of off-shell components in the fields of the standard Weyl
  multiplet. The columns $D=4,5,6$ indicate the number of components of
  the fields in the first column with gauge transformations subtracted.
  The next columns
  contain the Weyl weight, the chiral weight (only for $D=4$, where it means
  that the transformation under $\U(1)$ is $\delta_T \phi =\rmi c\phi \lambda_T$),
  and the chirality for the fermions
  for even dimension. Note that for $D=4$, changing the position of the index changes the chirality,
  while in 6 dimensions the chirality is generic. Finally,
  we indicate the
symmetry for which it is a gauge field, and possibly other gauge
transformations that have been used to reduce its number of degrees of
freedom in this counting.}
  \label{tbl:countWeyl}
\begin{center}
$ \begin{array}{|l|rrr|rrc|cc|} \hline\hline
  & D=4 & D=5 & D=6 &w&c& \gamma _*& \mbox{Gauge transf.}&\mbox{Subtracted}\\
\hline\hline
e_\mu {}^a     & 5 & 9 & 14 &-1&0&&P^a & M_{ab},\,D\\
b_\mu    & \multicolumn{3}{c|}{\mbox{compensating } K^a} &0&0&&D&K^a\\
\omega _\mu {}^{ab}&\multicolumn{3}{c|}{\mbox{composite}} &0&0& &M^{ab}&\\
f_\mu {}^a &\multicolumn{3}{c|}{\mbox{composite}}&1&0& &K^a&\\
V_{\mu i}{}^j   & 9 & 12 & 15 &0&0& & \SU(2) & \\
A_\mu &3 &&&0&0& &\U(1)& \\
\psi _\mu {}^i  & 16 & 24 & 32 & -\ft12 &\ft12 &+&\cQ^i& S^i\\[1mm]
\phi _\mu {}^i & \multicolumn{3}{c|}{\mbox{composite}} &+\ft12&\ft12&-&S^i&\\[1mm] \hline\hline
T^-_{ab}, T_{abc}^-  & 6 & 10 & 10&1 & 1& &&\\
D                                  & 1 &  1 &  1&2&0&&&\\
\hline
&&&&&&\\[-2mm]
\chi ^i  & 8 & 8 & 8  &\ft32&\ft12&+&&\\[1mm]
\hline\hline
\mbox{Total}  & 24+24 & 32+32 & 40+40& &&&&\\
\hline\hline\noalign{\smallskip}
\end{array}$
\end{center}
\end{table}

The standard Weyl multiplet involves an
antisymmetric tensor $T$, with two indices in 4 dimensions ($T_{ab}$, you
may take it anti-self-dual, but then it is complex)~\cite{deWit:1980ug},
with two or three indices in 5 dimensions (these are dual to each
other)~\cite{Bergshoeff:2001hc}, or an anti-self-dual real tensor in 6
dimensions~\cite{Bergshoeff:1986mz}. Further there is a real scalar $D$
and a fermion doublet $\chi ^i$.

In order to obtain the transformation laws of the standard Weyl multiplet one uses the general procedure outlined in \cite{Bergshoeff:1986mz}.

(1) As a first step one must construct the linear $Q$-supersymmetry transformations of the matter fields and also the additional transformations for the gauge fields due to the presence of these matter fields. This is done by writing a general ansatz linear in the fields with yet undetermined coefficients. These coefficients will be fixed by demanding that the commutator of two supersymmetry transformations gives a local translation, i.e. $[\delta_Q(\epsilon_1), \delta_Q(\epsilon_2) ] = \xi^a(\epsilon_1,\epsilon_2 )P_a $ with $\xi^a(\epsilon_1,\epsilon_2 )$ given by (\ref{xiaeps12}). In writing the ansatz one makes use of the Lorentz structure of the fields and of course their spin. Recalling the tricks presented in Sect. \ref{ss:easytransf}, a derivative acting on a gauge field can be replaced with the corresponding (modified) curvature. Similarly, one must also covariantize all the derivatives using the covariant derivative. Both the explicit expressions of the covariant derivatives and the curvatures will change during the procedure that we outline here. More specifically one will encounter $M_{\mu J}{}^I$ terms, which will modify the transformation laws and curvatures as shown in (\ref{delJhI1}) and (\ref{extraR}) respectively.

(2) Next one would like to obtain the bosonic transformation rules of the matter fields. The Lorentz transformations follow straightforwardly from the index structures of the different fields. For the fields that we consider here, the action of the dilatation on a field is of the form
\begin{equation}
 \delta_{\rm D} \phi =w\, \lambda_{\rm D} \phi\,,
 \label{deltaDw}
\end{equation}
determined completely by the Weyl weight $w$ of the field. \index{Weyl weight} The Weyl weights of the gauge fields are easily found by looking at the algebra. For the matter fields one must use another consequence of the algebra, namely that the Weyl weight of the $Q$-supersymmetry transformation must be $\ft12$ higher than the original field on which it acts. This is equivalent to considering the Weyl weight of the parameter $\epsilon$ as $-\ft12$ and demand an equality of the Weyl weight of the left-hand side and right-hand side of the $Q$-supersymmetry transformations that we have just determined. In this one must also use that the Weyl weight of a covariant derivative is $1$ and that a curvature has Weyl weight $2$ higher than the corresponding gauge field. This is due to the presence of frame fields in their expressions. The determination of the Weyl weights is an important step because they restrict the other transformations severely. Especially for the $K$-transformation, which lowers the Weyl weight of the field by~$1$. In the same sense as for $\epsilon$ one can consider the parameter $\lambda_{\rm K}^a$ as having weight~$1$. So in order to construct these transformations, one must look whether there are fields present in the multiplet that obey this restriction.

One is then left with the $R$-symmetries. For the $T$ transformation in $D=4$ one uses similar rules as for dilatations, using a chiral weight $c$ \index{chiral weight} for every field with the meaning
\begin{equation}
  \delta_T \phi =\rmi c\phi \lambda_T\,.
 \label{deltaTc}
\end{equation}
The complex conjugate of a field has opposite chiral weight. In the same way, the right-projected field, e.g. $\chi _i$ has $c=-\ft12$ since $\chi ^i$ has $c=+\ft12$. To find the weights, the commutator $[T,Q^i]$ in (\ref{ConfCommutators}) implies that one can count $\epsilon ^i$ as having chiral weight $\ft12$, and thus $\epsilon _i$ as weight $-\ft12$. The $\SU(2)$ transformations are implicit in the position of the indices, see (\ref{lambdatransfSU2}).

(3) \label{it:Ssusy} The $S$-super\-symmetry transformation is determined by using the action of the commutator $[K, Q] \sim S$ on the fields. Since for the Weyl multiplet, $b_{\mu}$ is the only one of the independent fields that transforms under $K$-transformations (as in (\ref{delextra})), the $S$-transformation is easy to get in the following way. First consider all occurrences of $b_\mu $ in the transformation law (hidden in covariant derivatives and composite fields such as  (\ref{solomega})). This should lead to terms proportional to $b_\mu \gamma ^\mu \epsilon ^i$. Then replace  $b_\mu \gamma ^\mu \epsilon ^i$ by $ -2s_D \eta ^i$ where $s_D$ are the sign factors in  (\ref{ConfCommutators}).
\bexer
  An immediate example of how this procedure works consists in obtaining the first line in (\ref{delextra}) from the supersymmetry transformation of the gravitino
\begin{equation}
  \delta (\epsilon )\psi _\mu = D_\mu \epsilon + \ldots =  \left(\ft12 b_\mu +\ft14 \omega _\mu {}^{ab}(e,b)\gamma ^{ab}\right)\epsilon +\ldots
 \label{delepspsib}
\end{equation}
\eexer
\bigskip

At this point we have determined all the transformations for the fields of the standard Weyl multiplet. However, there are two complications which will further modify these transformations. A first modification applies only to the composite fields as discussed in Sect.~\ref{ss:constraints}. There we have shown that in order for the constraints to be invariant under the symmetries one must change the transformations of the composite fields. First, it is important that one uses the modified curvatures due to the extra matter terms in, for example, the $Q$-supersymmetry transformations of the gauge fields.

Further, as shown in Sect.~\ref{ss:otherConstr}, it is possible to change these constraints by adding additional covariant functions of fields. In some cases, these functions can be chosen in a way that they simplify the transformations of the composite fields. For $D=4$ and for $D=6$ (the (1,0) theory that we treat here) we will be able to choose the constraints invariant under $S$-supersymmetry, avoiding extra
$S$-transformations for the constrained fields. Observe, however, that
this is a choice for convenience, which is even not necessary possible
(it is not possible in $D=5$~\cite{Bergshoeff:2001hc} or for $(2,0)$ in
$D=6$~\cite{Bergshoeff:1999db}). In principle all constraints of the form
(\ref{secondConstraint}) and (\ref{ThirdConstraint}) are equivalent up to
field redefinitions.

A second, and more important, complication is the fact that one obtains additional transformations if one determines $\{Q,Q\}$. For example if one calculates this commutator on the frame field one will obtain a general coordinate transformation, as expected, but also an additional Lorentz transformation. These transformations will depend on the fields and thus change the algebra into a \emph{soft algebra}, \index{soft algebra} which has structure functions rather than constants, see \cite[Sect. 11.1.3]{Freedman:2012zz}. This change in the algebra is the start of an iterative process. First one imposes the newly found commutator $\{Q,Q\}$ on all the fields, i.e. one demands that the $Q$-transformations of all the fields are such that their commutator gives a Lorentz transformation acting on the field. In order to be able to do this one may need to use terms that are nonlinear in the fields, e.g. second order. If these nonlinear $Q$-transformations are found, one again calculates the commutator on all the fields. If one finds a new transformation as a result, one must impose this again on all the fields and thus change the $Q$-transformations. This iterative process stops if one finds no new transformations in calculating the commutator and thus obtains closure. A same procedure can be applied to $\{Q,S\}$ and $\{S,S\}$.

\subsection{\texorpdfstring{$D=4$}{D=4}}\label{WeylDeq4}
We present here the $Q$ and $S$ supersymmetry transformations in $D=4$. First for the independent gauge fields:
\begin{eqnarray}
       \delta_{Q,S}(\epsilon,\eta)e_\mu{}^a
      &=& \ft12\bar{\epsilon}^i\gamma^a\psi_{\mu i}+\hc\,, \nonumber\\
      \delta_{Q,S}(\epsilon,\eta)b_\mu &=&
      \ft12\bar{\epsilon}^i\phi_{\mu i} - \ft12\bar{\eta}^i\psi_{\mu i}\nonumber\\
      &&-\ft38\bar{\epsilon}^i\gamma_\mu\chi_i       +\hc
      \,,\nonumber\\
      \delta_{Q,S}(\epsilon,\eta)A_\mu &=&
     -\ft12\rmi \bar{\epsilon}^i\phi_{\mu i} -\ft12\rmi \bar{\eta}^i\psi_{\mu i} \nonumber\\
     &&  -\ft{3}8\rmi\bar{\epsilon}^i\gamma_\mu\chi_i  +\hc\,, \nonumber\\
      \delta_{Q,S}(\epsilon,\eta)V_{\mu\,i}{}^j &=&
      -\bar{\epsilon}_i\phi_\mu^j-\bar{\eta}_i\psi_\mu^j \nonumber\\
    &&+\ft34\bar{\epsilon}_i\gamma_\mu\chi^j     -(\hc \, ; \, {\rm traceless})\,, \nonumber\\
    \delta_{Q,S}(\epsilon,\eta)\psi_\mu^i &=& \left( \partial _\mu +\ft12 b_\mu
  +\ft14 \gamma ^{ab}\omega  _{\mu ab}-\ft12\rmi A_\mu \right)
   \epsilon^i-V_\mu{}_j {}^i \epsilon ^j  \nonumber\\
      &&-\ft1 {16} \gamma\cdot T^-\varepsilon^{ij}\gamma_\mu\epsilon_j -\gamma_\mu\eta^i \,,\nonumber\\
      \delta_{Q,S}(\epsilon,\eta)T_{ab}^-&=&
      2 \bar{\epsilon}^i \widehat{R}_{ab}(Q)^j\varepsilon_{ij}\,, \nonumber\\
     \delta_{Q,S}(\epsilon,\eta)\chi^i &=&
      -\ft1{24}\gamma\cdot\slashed{\mathcal{D}}  T^-\varepsilon^{ij}\epsilon_j
      -\ft{1}{6}\widehat{R}_j{}^{i}\cdot\gamma\epsilon^j
       +\ft{1}{6}\rmi \widehat{R}(T)\cdot\gamma\epsilon^i
      + \ft12 D\,\epsilon^i \nonumber\\
      &&+ \ft1{12} \gamma\cdot T^-\varepsilon^{ij}\eta_j
      \,,\nonumber\\
      \delta_{Q,S}(\epsilon,\eta)D &=&
       \ft12\bar\epsilon^i\slashed{\mathcal{D}}\chi_i+\hc\,,
  \label{transfo4}
  \end{eqnarray}
where $\widehat{R}_j{}^{i}$ is the $\SU(2)$ curvature, $R(T)$ the $\U(1)$
curvature, and $\gamma \cdot R=\gamma ^{ab}R_{ab}$. The notation $A^i{}_j
-(\hc \, ; \, {\rm traceless})$ stands for  $A^i{}_j- A_j{}^i-\ft12\delta
_j{}^i (A^k{}_k-A_k{}^k)$. For the gauge fields, the first line represents
the original gauge transformations, and the second line are the terms
that were symbolically represented by ${\cal M}$ in (\ref{cgcthI}). Observe the covariance
of these terms and the transformations of the matter terms.

The dependent fields are defined by the constraints
\begin{eqnarray}
0 &=& R^a_{\mu\nu}(P)\,,\nonumber\\
0 &=& \gamma^b \widehat R_{ba}(Q)^i +\ft32 \gamma_a\chi^i \,,\nonumber\\
0 &=& \widehat R_{a c}(M^{bc}) - \rmi \tilde {\widehat R}_{a}{}^b(T) +
\ft14 T^-_{ca} T^{+bc} + \ft32 \delta _a{}^b D\,. \label{covconstr}
\end{eqnarray}
The terms that could appear in these constraints are fixed by
compatibility with Weyl weights (thus that we do not want to modify the
dilatational transformations), and as mentioned above, the coefficients are chosen such that
they are invariant under $S$-supersymmetry.

The expressions of the dependent fields in terms of the physical fields are
\begin{eqnarray}
  \omega _\mu {}^{ab}&=&\omega _\mu {}^{ab}(e,b)
  +\ft12\left[ \bar \psi^i  _\mu \gamma ^{[a}\psi_i ^{b]}+\bar \psi_{\mu i} \gamma ^{[a}\psi ^{b]i}
  +\bar \psi_i ^{[a}\gamma _\mu \psi ^{b]i}\right] \,,\nonumber\\
  f_\mu {}^a &=&-\ft14 \hat{R}_\mu {}^a +\ft1{24}e_\mu {}^a
  \hat{R}+\ft14\rmi\widetilde {\widehat{R}}_\mu {}^a(T)-\ft1{16}T^-_{c\mu }T^{+ac}
  -\ft{1}{8}e_\mu {}^aD\,,\nonumber\\
  \phi _\mu^i &= &-\ft12 \gamma ^\nu  \widehat{R}'_{\mu \nu }(Q^i)+\ft1{12}\gamma _\mu \gamma
  ^{ab}\widehat{R}'_{ab}(Q^i)+\ft14\gamma _\mu \chi ^i\,.
\label{solfphiN2}
\end{eqnarray}
Here appears the dual of the $\U(1)$ curvature, and the Ricci tensor (\ref{Riccicov}) of
\begin{eqnarray}
 \widehat{R}'_{\mu \nu }\left(M^{ab}\right)&=&r_{\mu \nu }(M^{ab})
  +\left[ \bar \psi^i _{[\mu }\gamma _{\nu ]}\left( \ft34\gamma ^{ab}\chi _i+\widehat{R}^{ab}(Q_i)\right)\right.\nonumber\\
&&\left.  +\ft14\bar \psi _{\mu }^i\psi _\nu ^j\varepsilon _{ij}T^{+ab}+\hc\right]\,, \nonumber\\
  r_{\mu \nu }(M^{ab})&=&2\partial _{[\mu }\omega _{\nu ]}{}^{ab}+2\omega _{[\mu }{}^a{}_c\omega _{\nu ]}{}^{cb}
-\bar \psi^i_{[\mu }\gamma ^{ab}\phi _{\nu ]i}-\bar \psi_{i[\mu }\gamma ^{ab}\phi _{\nu ]}^i\,,
\label{hatRmunuab}
\end{eqnarray}
and
\begin{align}
 \widehat{R}'_{\mu \nu }(Q^i)=& 2 \left( \partial _{[\mu} +\ft12 b_{[\mu}
 +\ft14 \gamma _{ab}\omega  _{[\mu}{}^{ab}-\ft12\rmi A_{[\mu} \right)
   \psi _{\nu ]}^i +2V_{[\mu} {}^i{}_j \psi_{\nu ]} ^j\nonumber\\
&     -\ft1 {8} \gamma^{ab} T_{ab}^-\varepsilon^{ij}\gamma_{[\mu}\psi _{\nu ]j}\,.
 \label{RprimeQD4}
\end{align}

We present the transformations of the constrained
fields in 3 lines: first the original transformations, then the
$M$-transformations determined by the non-invariance of the constraints,
and finally terms that represent modified structure functions:
\begin{eqnarray}
     \delta_{Q,S}(\epsilon,\eta)\omega _\mu{}^{ab}
      &=& \ft12\bar{\epsilon}^i\gamma^{ab}\phi_{\mu i}+\ft12
      \bar{\eta}^i\gamma^{ab}\psi_{\mu i}
       \nonumber\\
      &&-\ft38\bar{\epsilon}^i\gamma_\mu\gamma^{ab}\chi_i
       -\ft12 \bar{\epsilon}^i\gamma_\mu\widehat{R}^{ab}(Q)_i  \nonumber\\
      &&-\ft14\bar{\epsilon}^iT^{+ab}\varepsilon _{ij}\psi_\mu^j +\hc  \,, \nonumber\\
      \delta_{Q,S}(\epsilon,\eta)f_\mu^a &=& \ft12\bar{\eta}^i\gamma^a\phi_{\mu i}
      \nonumber\\
      & & -\ft3{16}
      e_\mu{}^a\bar{\epsilon}^i\slashed{\mathcal{D}}\chi_i +\ft14 \bar{\epsilon}^i\gamma_\mu {\cal D}_b\widehat{R}^{ba}(Q)_i
     \nonumber\\
&& - \ft18\bar{\epsilon}^i\psi_\mu^j\, {\cal D}_b T^{+ba}\varepsilon _{ij}
-\ft3{16} \bar{\epsilon}^i\gamma^a\psi_{\mu i}\,D
    + \hc
      \,,\nonumber\\
       \delta_{Q,S}(\epsilon,\eta)\phi_\mu^i &=&\left( \partial _\mu -\ft12 b_\mu
  +\ft14 \gamma ^{ab}\omega  _{\mu ab}-\ft12\rmi A_\mu \right)
   \eta ^i +V_\mu {}^i{}_j \eta  ^j      - f_\mu^a\gamma_a\epsilon^i
 \nonumber\\ & &
  - \ft1{32}\slashed{\mathcal{D}} T^-\cdot\gamma \gamma_\mu\varepsilon^{ij}\epsilon_j
    -\ft{1}{8}\widehat{R}_j{}^i \cdot\gamma\gamma_\mu\epsilon^j -\ft
     18\rmi\widehat{R}(T)\cdot\gamma\gamma_\mu\epsilon^i \nonumber\\
&&        +\ft38\big[(\bar{\chi}_j\gamma^a\epsilon^j)\gamma_a\psi_\mu^i
      -(\bar{\chi}_j\gamma^a\psi_\mu^j)\gamma_a\epsilon^i\big]
      \,.
      \label{transfocompositeN4}
\end{eqnarray}
The third line of each of the \eqref{transfocompositeN4} show that the commutator between two
supersymmetries is modified as follows
\begin{equation}
[\delta_Q(\epsilon_1),\delta_Q(\epsilon_2)] = \delta_{\rm
cgct}\left(\xi_3^a\right)
+\delta_M\left(\lambda_3^{ab}\right)+\delta_{\rm K}\left(\lambda_{{\rm K}3}^a\right)
+ \delta_S\left(\eta_3^i\right)\,, \label{qqcomb}
\end{equation}
where the associated parameters are given by the following expressions:
\begin{eqnarray}
\xi_3^a& =  & \ft12\bar \epsilon_2^i\gamma^a\epsilon_{1i}+\hc\,, \nonumber\\
\lambda_3^{ab} &=&  \ft14 \bar\epsilon
^i_1\epsilon^j_2\,T^{+ab}\varepsilon_{ij}+\hc\,,
\nonumber\\
\Lambda_{{\rm K}3}^a &=& \ft18 \bar{\epsilon}^i_1\epsilon^j_2\,
{\cal D}_bT^{+ba}\varepsilon_{ij}
+\ft3{16}\bar{\epsilon}_2^i\gamma^a\epsilon_{1i}\,D+\hc\,, \nonumber\\
\eta_3^i &=& \ft34\,\bar{\epsilon}^i_{[1}\epsilon^j_{2]}\,\chi_j \,.
\label{qqparamsb}
\end{eqnarray}
These extra terms become important in applications where they can
give rise to central charges if the fields appearing in the structure
functions get non-zero vacuum expectation values. We will see that in the
presence of vector multiplets, there appear extra terms of a similar
nature involving the scalars of the vector multiplet.

\subsection{\texorpdfstring{$D=5$}{D=5}}
We report here the $Q$- and $S$-supersymmetry transformation laws of the independent
fields using again the same splitting in two lines for gauge terms and
matter terms\footnote{For $\delta \chi $ the split in two lines is
accidental due to the length of the expression.}
\begin{eqnarray}
\delta_{Q,S}(\epsilon,\eta) e_\mu{}^a   &=&  \ft 12\bar\epsilon \gamma^a
\psi_\mu
\,,\nonumber\\
\delta_{Q,S}(\epsilon,\eta) b_\mu       &=& \ft 12
\rmi\bar\epsilon\phi_\mu  + \ft12\rmi \bar\eta\psi_\mu \nonumber\\ &&
 -2 \bar\epsilon\gamma_\mu \chi\,,\nonumber\\
\delta_{Q,S}(\epsilon,\eta) V_\mu{}^{ij} &=&  -\ft32\rmi
\bar\epsilon^{(i} \phi_\mu^{j)}+ \ft32\rmi \bar\eta^{(i}\psi_\mu^{j)}
\nonumber\\ &&
 +4 \bar\epsilon^{(i}\gamma_\mu\chi^{j)}
  + \rmi \bar\epsilon^{(i} \gamma^{ab} T_{ab} \psi_\mu^{j)} \,,\nonumber\\
\delta_{Q,S}(\epsilon,\eta) \psi_\mu^i   &=& \partial_\mu \epsilon^i+\ft
12 b_\mu\epsilon^i +\ft 14 \omega_\mu^{ab} \gamma_{ab}\epsilon^i -
V^{ij}_\mu\epsilon _j -\rmi\gamma_\mu \eta^i\nonumber\\
&& + \rmi \gamma^{ab} T_{ab} \gamma_\mu \epsilon^i \,,\nonumber\\
 \delta_{Q,S}(\epsilon,\eta)
T_{ab}     &=& \ft12\rmi \bar\epsilon \gamma_{ab} \chi
-\ft {3}{32}\rmi \bar\epsilon\widehat R_{ab}(Q)    \,,\nonumber\\
\delta_{Q,S}(\epsilon,\eta) \chi^i     &=&  \ft 14 \epsilon^i D
-\ft{1}{64} \gamma^{ab} \widehat R_{ab}^{ij}(V) \epsilon_j
                   + \ft18\rmi \gamma^{ab} \slashed{\mathcal{D}} T_{ab} \epsilon^i
                   - \ft18\rmi \gamma^a {\cal D}^b T_{ab} \epsilon^i  \nonumber\\
               &&  -\ft 14 \gamma^{abcd} T_{ab}T_{cd} \epsilon^i + \ft 16 T^2 \epsilon^i
             +\ft 14 \gamma^{ab} T_{ab} \eta^i \,,\nonumber\\
\delta_{Q,S}(\epsilon,\eta) D         &=&  \bar\epsilon \slashed{\mathcal{D}} \chi
- \ft {5}{3} \rmi \bar\epsilon\gamma^{ab} T_{ab} \chi - \rmi \bar\eta\chi
\,. \label{modifiedtransf}
\end{eqnarray}
In 5 dimensions, not much simplifications are possible by taking
appropriate $C$ in~(\ref{secondConstraint}) or $\rho $
in~(\ref{ThirdConstraint}), so that we just took the
constraints~\cite{Bergshoeff:2001hc}
\begin{equation}
   R^a_{\mu\nu}(P)=0\,,\qquad
 \gamma^b \widehat R_{ba}(Q)^i =0 \,,\qquad
\widehat R_{a c}(M^{bc})=0\,.
 \label{constrD5}
\end{equation}
Following the methods in Sect. \ref{ss:constraints}, we find expressions for the dependent gauge fields:
gauge fields associated with $S$ and $K$ symmetries, respectively:
\begin{align}
\omega_\mu^{~ab}
=& \omega_\mu^{~ab}(e,b) -\ft12\bar{\psi}^{[b}\gamma ^{a]}\psi_{\mu}-\ft14\bar{\psi}^b\gamma _\mu \psi^a \,, \nonumber\\
\phi^i_\mu =&  \ft13\rmi \gamma^a \widehat{R}^\prime _{\mu a}{}^i(Q) - \ft1{24}\rmi
\gamma_\mu \gamma^{ab} \widehat{R}^\prime _{ab}{}^i(Q)\,,\nonumber\\
&\widehat{R}^\prime _{\mu \nu }{}^i(Q)= 2\partial _{[\mu}\psi^i_{\nu ]} +\ft12
\omega_{[\mu}{}^{ab} \gamma_{ab}\psi^i_{\nu ]} +b_{[\mu}\psi^i_{\nu ]}
-2V_{[\mu}{}^{ij}\psi_{\nu ]\,j} + 2\rmi \gamma \cdot T
\gamma_{[\mu} \psi_{\nu]}^i \, , \nonumber\\
f_a{}^{a}
=& \ft{1}{16} \left(- R(\omega) - \ft13 \bar{\psi}_\rho\gamma^{\rho\mu\nu} {\cal D}_\mu \psi_\nu \right.\nonumber\\
&+ \left.\ft13 \bar{\psi}_a^i\gamma ^{abc}\psi_b^jV_{cij} + 16 \bar{\psi}_a\gamma ^a\chi - 4 \rmi\bar{\psi}^a\psi^bT_{ab} + \ft43 \rmi\bar{\psi}^b\gamma _{abcd}\psi^aT^{cd} \right)\,.
\label{transfDepF}
\end{align}

The full commutator of two supersymmetry transformations is
\begin{equation}
 \left[\delta_Q(\epsilon_1),\delta_Q(\epsilon_2)\right] =  \delta_{\rm cgct}(\xi_3^\mu)+
\delta_M(\lambda^{ab}_3) + \delta_S(\eta_3)  + \delta_U(\lambda^{ij}_3)
+\delta_{\rm K}(\lambda^a_{{\rm K}3})  \,. \label{algebraQQ5}
\end{equation}
The covariant general coordinate transformations have been defined
in~(\ref{defcgct}). The parameters appearing in~(\ref{algebraQQ5}) are
 \begin{eqnarray}
\xi^\mu_3       &=& \ft 12 \bar\epsilon_2 \gamma_\mu \epsilon_1 \,,\nonumber\\
\lambda^{ab}_3      &=& - \rmi \bar \epsilon _2\gamma ^{[a}\gamma^{cd}T_{cd}
\gamma ^{b]}\epsilon _1 \,, \nonumber\\
\lambda^{ij}_3 &=& \rmi \bar\epsilon^{(i}_2 \gamma^{ab} T_{ab} \epsilon^{j)}_1\,, \nonumber\\
\eta^i_3       &=& - \ft {9}{4}\rmi\, \bar \epsilon_2 \epsilon_1 \chi^i
                   +\ft {7}{4}\rmi\, \bar \epsilon_2 \gamma_c \epsilon_1 \gamma^c \chi ^i \nonumber\\
               &&  + \ft1{4}\rmi\,  \bar \epsilon_2^{(i} \gamma_{cd} \epsilon_1^{j)}
\left( \gamma^{cd} \chi_j
                   + \ft 14\, \widehat R^{cd}{}_j(Q) \right) \, ,\nonumber\\
\lambda^a_{{\rm K}3} &=& -\ft 12 \bar\epsilon_2\gamma^a\epsilon_1 D + \ft{1}{96}
\bar\epsilon^i_2\gamma^{abc}\epsilon^j_1 \widehat R_{bcij}(V)  \nonumber\\
               && + \ft1{12}\rmi\bar\epsilon_2\left(-5\gamma^{abcd} {\cal D}_b T_{cd} +
9 {\cal D}_b T^{ba} \right)\epsilon_1  \nonumber\\
               && + \bar\epsilon_2\left(  \gamma^{abcde}T_{bc}T_{de}
                  - 4 \gamma^c T_{cd} T^{ad} +  \ft 23  \gamma^a T^2
                  \right)\epsilon_1 \,. \label{thirdpard5}
 \end{eqnarray}
For the $Q,S$ commutators we find the following algebra:
\begin{eqnarray}
 \left[\delta_S(\eta),
\delta_Q(\epsilon)\right] &=& \delta_{\rm D}( \ft12\rmi \bar\epsilon\eta ) + \delta_M( \ft12\rmi \bar\epsilon
\gamma^{ab} \eta) +      \delta_U(  -\ft32\rmi \bar\epsilon^{(i} \eta^{j)} )  +
\delta_{\rm K}(\lambda_{{\rm K}3}^a ) \,,\nonumber\\
\left[ \delta_S(\eta_1), \delta_S(\eta_2) \right] &=& \delta_{\rm K}( \ft 12 \bar\eta_2 \gamma^a
\eta_1 ) \,, \end{eqnarray}
 with
\begin{equation}
 \lambda_{{\rm K}3}^a= \ft16 \bar{\epsilon} \left(\gamma^{bc}T_{bc} \gamma_a - \ft12 \gamma_a \gamma^{bc}T_{bc}
\right) \eta \,. \label{algebraQSS}
\end{equation}
For practical purposes (see how to calculate transformations of covariant
derivatives), it is useful to give the extra parts of the transformation
laws of dependent gauge fields, i.e. the parts denoted by ${\cal M}_{\mu J}{}^I$
in~(\ref{delJhI2}). These are (for the gauge field of special conformal
transformations we suffice by giving the transformation of the trace, as
this is what one often needs)
\begin{eqnarray}
\delta_{Q,S}(\epsilon,\eta)\omega_\mu{}^{ab} &=& \dots  -\ft 12
\bar\epsilon\gamma^{[a} \widehat R_\mu{}^{b]}(Q)
 - \ft 14 \bar\epsilon\gamma_\mu \widehat R^{ab}(Q)  - 4 e_\mu{}^{[a} \bar\epsilon \gamma^{b]} \chi
\,, \nonumber\\
\delta_{Q,S}(\epsilon,\eta)\phi_\mu^i    &=& \dots -\ft1{12}\rmi \left\{
\gamma^{ab}\gamma_\mu - \ft 12 \gamma_\mu\gamma^{ab}
                    \right\} \widehat R_{ab}{}^i{}_j(V) \epsilon^j+  \nonumber\\
        &&  + \ft {1}{3}\left[ \slashed{\mathcal{D}} \gamma^{ab} T_{ab} \gamma_\mu - {\cal D}_\mu
\gamma^{ab} T_{ab}
            + \gamma_\mu\gamma^c{\cal D}^a T_{ac} \right] \epsilon^i+  \nonumber\\
        &&  + \rmi \left[ - \gamma _{\mu  abcd}T^{ab}T^{cd}
  +8 \gamma _\rho T^{\rho \sigma }T_{\mu \sigma } -2\gamma _\mu T^2 \right] \epsilon^i+  \nonumber\\
        &&  + \ft13\rmi \left( 8 \gamma^b T_{\mu b} - \gamma_\mu
\gamma\cdot T \right) \eta^i  \,, \nonumber\\
\delta_S(\eta ) f_a{}^a &=&  -5 \rmi \bar{\eta} \chi,
          \label{d5extraTransDep}
\end{eqnarray}
with $\gamma \cdot T=\gamma ^{ab}T_{ab}$ and $T^2=T^{ab}T_{ab}$. Note
that there are other terms, which however are proportional to gauge fields, determined by the algebra. For example, the $\lambda
_3^{ab}$ expression in~(\ref{thirdpard5}) implies that the supersymmetry
transformation of the spin connection contains a term
\begin{equation}
  \delta_{Q}(\epsilon)\omega_\mu{}^{ab} = \dots - \rmi \bar\epsilon \gamma^{[a}
\gamma^{cd} T_{cd} \gamma^{b]} \psi_\mu\,.
 \label{delAomegaTalg}
\end{equation}

\subsection{\texorpdfstring{$D=6$}{D=6}}
The transformation laws of the independent fields are
\begin{eqnarray}
       \delta_{Q,S}(\epsilon,\eta)e_\mu{}^a
      &=& \ft12\bar{\epsilon}\gamma^a\psi_{\mu }\,, \nonumber\\
      \delta_{Q,S}(\epsilon,\eta)b_\mu &=&
      -\ft12\bar{\epsilon}\phi_{\mu } + \ft12\bar{\eta}\psi_{\mu }\nonumber\\
      &&-\ft1{24}\bar{\epsilon}\gamma_\mu\chi
      \,,\nonumber\\
      \delta_{Q,S}(\epsilon,\eta)V_{\mu}{}^{ij} &=&
      2\bar{\epsilon}^{(i}\phi_\mu^{j)}+2\bar{\eta}^{(i}\psi_\mu^{j)} \nonumber\\
    && +\ft16\bar{\epsilon}^{(i}\gamma_\mu\chi^{j)} \,, \nonumber\\
    \delta_{Q,S}(\epsilon,\eta)\psi_\mu^i &=& \left( \partial _\mu +\ft12 b_\mu
  +\ft14 \gamma ^{ab}\omega  _{\mu ab}\right)
   \epsilon^i +V_\mu {}^i{}_j \epsilon ^j  \nonumber\\
      &&+\ft1 {24} \gamma\cdot T^-\gamma_\mu\epsilon^i +\gamma_\mu\eta^i \,,\nonumber\\
      \delta_{Q,S}(\epsilon,\eta)T_{abc}^-&=&-\ft1{32} \bar{\epsilon}\gamma ^{de}
      \gamma _{abc} \widehat{R}_{de}(Q)-\ft7{96}\bar \epsilon\gamma _{abc}\chi  \,, \nonumber\\
     \delta_{Q,S}(\epsilon,\eta)\chi^i &=&
      +\ft1{8}({\cal D}_\mu \gamma\cdot T^-)\gamma ^\mu \epsilon^i
      -\ft{3}{8}\widehat{R}^{ij}(V)\cdot\gamma\epsilon_j
      + \ft14 D\,\epsilon^i + \ft1{2} \gamma\cdot T^-\eta^ i
      \,,\nonumber\\
      \delta_{Q,S}(\epsilon,\eta)D &=&
       \bar\epsilon^i\slashed{\mathcal{D}}\chi_i-2\bar \eta \chi \,,
  \label{transfo6}
  \end{eqnarray}
  where $\gamma \cdot T\equiv \gamma ^{abc}T_{abc},\ldots $.

The constraints that we took in 6 dimensions are
\begin{eqnarray}
0 &=& R^a_{\mu\nu}(P)\,,\nonumber\\
0 &=& \gamma^b \widehat R_{ba}(Q)^i +\ft16 \gamma_a\chi^i \,,\nonumber\\
0 &=& \widehat R_{a c}(M^{bc}) - T^-_{acd} T^{-bcd} + \ft1{12} \delta
_a{}^b D\,. \label{covconstr6}
\end{eqnarray}
The last equation contains a sign correction\footnote{We thank T.~Kugo for
this correction.} to the equation in~\cite{Bergshoeff:1986mz}.

The transformations of the dependent gauge fields $\omega _\mu {}^{ab}$
and $\phi_\mu^i $ contain as covariant terms (terms not determined by the
algebra)
\begin{eqnarray}
 \delta _Q(\epsilon )\omega _\mu {}^{ab} & = &\dots
   -\ft 12 \bar\epsilon\gamma^{[a} \widehat R_\mu{}^{b]}(Q)
 - \ft 14 \bar\epsilon\gamma_\mu \widehat R^{ab}(Q)
  - \ft1{12} e_\mu{}^{[a} \bar\epsilon \gamma^{b]} \chi
\,, \nonumber\\
\delta_Q(\epsilon )\phi_\mu^i    &=& \dots +\ft1{32}\left\{
\gamma^{ab}\gamma_\mu - \ft 12 \gamma_\mu\gamma^{ab}
                    \right\} \widehat R_{ab}{}^i{}_j(V) \epsilon^j+  \nonumber\\
        &&  - \ft {1}{96}\left( \slashed{\mathcal{D}} \gamma^{abc} T^-_{abc} \gamma_\mu \right)
         \epsilon^i\,.
 \label{delQomegaphi}
\end{eqnarray}
In this case the constraints are also $S$-invariant, and thus there are no extra $S$-supersymmetry transformations.

The algebra is only modified in the anticommutator of two
$Q$-supersymmetries:
\begin{align}
  \left[\delta_Q(\epsilon_1),\delta_Q(\epsilon_2)\right] =&
    \delta_{\rm cgct}\left(\ft12\bar \epsilon _2\gamma _\mu \epsilon _1\right)+
\delta_M\left(\ft12\bar \epsilon _2\gamma _c\epsilon _1T^{-\,abc}\right)+ \delta_S\left(\ft1{24}\gamma _a\chi ^i\bar \epsilon _2\gamma
^a\epsilon _1\right) \nonumber\\
&+\delta_{\rm K}\left(-\ft18\bar \epsilon _2\gamma
_b\epsilon _1 \left( {\cal D}_cT^{-\,abc}+\ft1{12}\eta ^{ab}D\right) \right) \,.
\label{algebraQQ6}
\end{align}
Again, this implies that the transformations of gauge fields contain
extra terms with gauge fields as e.g. there is in the transformation of the spin connection
$+\ft12 \bar\epsilon \gamma_c \psi_\mu T^{-\,abc}$.

%% file: ch3_Mattermultiplets.tex
\chapter{Matter Multiplets} \label{ss:matter}

\begin{quote}
\textbf{Abstract.}
After the Weyl multiplet is introduced, we can now define matter multiplets whose transformations respect the algebra with structure functions that depend on the fields of the Weyl multiplet. We treat here vector multiplets and hypermultiplets. We define them for $D=4,5$ and 6, first for rigid supersymmetry and then for the superconformal theory. In the second part of this chapter we define actions for these multiplets, which will be the basis for the further chapters.
\end{quote}

\bigskip

The goal of this chapter is to construct local superconformal actions for the matter multiplets, exploiting our knowledge of the Weyl multiplet.
In principle there are many representations of the superconformal algebra that define matter multiplets. The physical theories for $D=4$ can all be obtained with vector multiplets and hypermultiplets. For $D=5$ and $D=6$ tensor multiplets can lead to inequivalent theories. For $D=5$, this has been included in the treatments of
\cite{Gunaydin:1999zx,Gunaydin:2000xk,Gunaydin:2000ph,Ceresole:2000jd,Bergshoeff:2004kh}, to which we will come back to this in Sect. \ref{ss:finalD5}.
 One  might also prefer formulations in terms of other multiplets to make connections with other descriptions, e.g. in string theory. We will briefly discuss the $D=4$ tensor multiplet in a superconformal background \cite{deWit:1981tn} in Sect. \ref{ss:tensormultiplet}. The action with one tensor multiplet was given in \cite{deWit:1982na} and extended to more multiplets in \cite{deWit:2006gn}. On-shell matter couplings using different formalisms have been given in \cite{Dall'Agata:2003yr,Sommovigo:2004vj,D'Auria:2004yi}. Recently \cite{Cribiori:2018xdy}, also the (off-shell) coupling of one tensor multiplet to an arbitrary number of vector multiplets has been obtained.

The main focus of this chapter will be on vector and hypermultiplets. Importantly, the latter will be used not only as physical multiplets, but also as compensating multiplets to describe super-Poincar\'{e} theories with matter couplings. This is in the spirit of the general strategy outlined in Sect.~\ref{ss:strategy} that we review in Sect. \ref{ss:exampleBosComp}.

The remainder of this chapter is split in two parts. In Sect. \ref{ss:confpropMatter} we explain the structure of first vector and then hypermultiplets and their embedding in the superconformal algebra. The construction of actions is postponed to the second part, Sect. \ref{ss:actions}. We explicitly construct the superconformal invariant actions for sets of these multiplets, which will be combined in Chap. \ref{ss:gaugefix} by the gauge fixing to Poincar\'{e} supergravity. Many parts of this chapter, especially for the case of $D=4$, have been obtained in the context of the master thesis of De Rydt and Vercnocke \cite{masterBertJan}.

\section{Review of the Strategy}\label{ss:exampleBosComp}
In Sect.~\ref{ss:strategy}, we already outlined the general idea of the
superconformal construction for actions with  super-Poincar\'{e}
invariance. At that time, we had not yet explained the gauging of the
conformal algebra. Now we can be more precise. For this example, we will
still restrict to the bosonic case. Consider a scalar field $\phi $ with
Weyl weight $w$ and no intrinsic special conformal transformations: $k_\mu(\phi)=0$. Its superconformal covariant derivative is
\begin{equation}
  {\cal D}_\mu\phi  =\left( \partial _\mu -wb_\mu \right)\phi \,.
 \label{Dmuphi}
\end{equation}
The transformation of the covariant derivative ${\cal D}_a\phi $ can be easily
obtained from the `easy method' (Sect. \ref{ss:easytransf}). One takes into account (\ref{delextra})
to find that there is a $K$ transformation. The transformation law of a
covariant derivative determines the covariant box
\begin{eqnarray}
\bbox^C \phi& \equiv &\eta^{ab}{\cal D}_b {\cal D}_a \phi=e^{a\mu}\left( \partial_\mu
{\cal D}_a \phi -(w+1) b_\mu {\cal D}_a \phi +\omega_{\mu\,ab}{\cal D}^b\phi+2wf_{\mu a}\phi \right) \nonumber\\
&=& e^{-1}\left( \partial_\mu-(w+2-D)b_\mu\right)e g^{\mu\nu}\left(
\partial_\nu-wb_\nu\right) \phi-\frac{w}{2(D-1)}R\,\phi . \label{bboxphi}
\end{eqnarray}
We use here the constraint (\ref{solfphi}) (without matter for the pure
bosonic case). The last term is the well-known $R/6$ term in
$D=4$. In fact, choosing $w=\frac{D}{2}-1$, one has a conformal invariant
scalar action
\begin{equation}
S=\int \rmd^D x\, e \phi\bbox^C \phi \,. \label{Iscalarconf}
\end{equation}
\bexer Show that $\int \rmd^D x\, e {\cal D}_a\phi\, {\cal D}^a \phi$ is not a special
conformal invariant, while $\bbox \phi $ is invariant under $K$ iff
$w=\frac{D}{2}-1$.
 \eexer

In order to obtain a Poincar\'{e} invariant action, we have to
break dilatations and special conformal transformations (as these are not
part of the Poincar\'{e} algebra). Considering (\ref{delextra}), it is clear
that the latter can be broken by a gauge choice
\begin{equation}
K-\mbox{gauge :}\qquad b_\mu=0\,.
\end{equation}
One could take as gauge choice for dilatations a fixed value of a scalar
$\phi$. As a consequence, the action (\ref{Iscalarconf})
reduces to the Poincar\'{e} gravity action: only the frame field of the `Weyl multiplet' (which was in the background) remains.

The lesson to learn is: once the gauge for the superfluous symmetries in the matter action is fixed, without considering any action for the Weyl multiplet, we find kinetic terms for the gravity sector.

We can schematically summarize this procedure in the following diagram:
\begin{eqnarray}
   & \mbox{Weyl multiplet: }e_\mu{} ^a,\, b_\mu  & \mbox{(Background)} \nonumber\\
   & + &  \nonumber\\
   & \mbox{matter field: }\phi & \nonumber\\
   &\downarrow & \mbox{gauge fixing }K_a,\, D \nonumber\\
   & \mbox{Poincar\'{e} gravity }e_\mu{} ^a\,,
 \label{schemecompGravity}
\end{eqnarray}
namely we introduce, in the background of the Weyl multiplet, the conformally invariant action of a matter field $\phi$ and we fix the gauge to get the action of Poincar\'{e} gravity. In the above scheme, the field $\phi $ provides the \emph{compensating field} degree of freedom that makes the combined field gauge equivalent to an irreducible multiplet of Poincar\'{e} gravity. We remark that, at the classical level, every gauge fixing is equivalent to redefinitions of the fields. In this case, defining (the conformal invariant)
\begin{equation}
  \tilde g_{\mu \nu }= g_{\mu \nu }\phi ^{4/(D-2)}\,,
 \label{tildeggf}
\end{equation}
and writing the action in terms of $R(\tilde g)$, the field $\phi $ disappears from the action
\begin{equation}
  S=-\frac{D-2}{4(D-1)} \int \rmd^Dx\,\sqrt{\tilde g}\,R(\tilde g)\,.
 \label{Stildeg}
\end{equation}
The absence of $\phi $ from the action above is just a consequence of dilatational invariance
\begin{equation}
  \int \rmd^Dx\,\left[\frac{\delta S(\tilde g,\phi )}{\delta \tilde g_{\mu \nu }(x)}\delta_{\rm D} \tilde g_{\mu \nu }(x)+ \frac{\delta S(\tilde g,\phi )}{\delta \phi (x) }\delta_{\rm D} \phi (x)\right]=0\,,
 \label{dilatationalinvS}
\end{equation}
which, together with $\delta_{\rm D} \tilde g_{\mu \nu }=0$, implies $S(\tilde g,\phi )\equiv S(\tilde g)$.

\section{Conformal Properties of the Multiplets}\label{ss:confpropMatter}

Having the Weyl multiplet, the further step now is to introduce other multiplets in the background of the Weyl multiplet. The resulting algebra, which depends for
part on the fields of the Weyl multiplet, is fixed for what concerns the
superconformal transformations. On the other hand, extra terms with
gauge transformations of extra vectors or antisymmetric tensors may still
appear in the algebra. As long as the fields of the Weyl multiplet are inert under these
transformations (as we will impose by hypothesis), these extra transformations do not modify our previous results.

 A first modification of this structure is obtained by the introduction of a gauge vector multiplet. The commutator of the supersymmetries can still be
modified by a gauge transformation that depends on fields of this vector
multiplet. For this structure to make sense, the algebra of the Weyl
multiplet had to close without using an equation of motion. Furthermore, as long as the vector multiplet is well defined off-shell, a matter
multiplet (in the background of both the vector and Weyl multiplet) may now be introduced whose algebra closes only modulo
equations of motion.

All fields in `matter multiplets' will now have to obey the same `soft' algebra defined by the Weyl multiplet. A first step is to define their transformations under the bosonic
symmetries. We assume the rules  (\ref{deltaDw}) and  (\ref{deltaTc}) under Weyl  and chiral transformations, where the weights will be given in Table \ref{tbl:matterMult}.
The $R$-symmetry  $\SU(2)$ transformation are implicit in the index structure of the fields.

\begin{table}[!t] 
\caption{\it Fields in some superconformal matter multiplets. We indicate
for each dimension the Weyl weight (and for $D=4$ chiral weight), the
number of real degrees of freedom,  the $\SU(2)$ representations, which
is the same in any dimension, and the chirality for $D=4$ and $D=6$. For
each multiplet we give first the bosonic fields, and then the fermionic
fields (below the line).} \label{tbl:matterMult}
\begin{center}
$\begin{array}{||cccc|ccc|ccc||cc||} 
\hline\noalign{\smallskip}\hline
 \multicolumn{4}{c|}{\mathbf{D=4}} & \multicolumn{3}{c|}{\mathbf{D=5}}  & \multicolumn{3}{c||}{\mathbf{D=6}}  &
& \\[2mm]
\hline
 \mbox{Field} & w & c & \mbox{\#} & \mbox{Field} & w  & \mbox{\#} & \mbox{Field} & w  & \mbox{\#} & \SU(2)&\gamma _*
 \\[2mm]
\hline\hline
 \multicolumn{12}{c}{\mbox{\bf Off-shell vector multiplet}}\\
\hline
 X & 1 & 1 & 2 & \sigma  & 1 & 1 &   &   &   & 1 &\\
 W_\mu  & 0 & 0 & 3 & W_\mu  & 0 & 4 & W_\mu  & 0 & 5 & 1& \\
 Y_{ij} & 2 & 0 &3 & Y_{ij} & 2 & 3 & Y_{ij} & 2 & 3 & 3&  \\
\hline
 \Omega  _i & 3/2 & 1/2 & 8 & \psi _i & 3/2 & 8 & \lambda _i & 3/2 & 8 & 2&+ \\
\hline\hline
 \multicolumn{4}{c|}{ }
  & \multicolumn{8}{c}{\mbox{\bf On-shell tensor multiplet}} \\
\cline{5-12}
 \multicolumn{4}{c|}{ } & B_{\mu \nu } & 0 & 3 & B_{\mu \nu }  & 0  & 3  &  1 &\\
\multicolumn{4}{c|}{ } & \phi  & 1 & 1 & \sigma   & 2  & 1  & 1 &\\
\cline{5-12}
 \multicolumn{4}{c|}{ }& \lambda ^i & 3/2 & 4 &  \psi ^i & 5/2  & 4  & 2 &\\
\hline\hline
 \multicolumn{12}{c}{\mbox{\bf On-shell hypermultiplet}} \\
\hline
 q^X & 1 & 0 & 4 & q^X & 3/2 & 4 & q^X &  2 & 4 & 2 &\\
 \hline
 \zeta ^A  & 3/2 & -1/2&4 & \zeta ^A & 2 &  4 & \zeta ^A & 5/2 & 4 & 1 & +\\
\hline\hline
  \multicolumn{4}{c|}{ \mbox{\bf Off-shell chiral multiplet}}&   &   &  &   &   &   &  & \\
\hline
 A & w & w   & 2 &   &   &  &   &   &   &  1&  \\
 B_{ij} & w+1 & w-1   & 6 &   &   &  &   &   &   & 3 & \\
 G^-_{ab} & w+1 & w-1   & 6 &   &   &  &   &   &   & 1&  \\
 C & w+2 & w-2   & 2 &   &   &  &   &   &   & 1 & \\
 \hline
 \Psi _i & w+\ft12 & w-\ft12 & 8 &   &   &  &   &   &   & 2  & + \\
 \Lambda _i & w+\ft32 & w-\ft32& 8 &   &   &  &   &   &   & 2  & - \\
\hline\hline
  \multicolumn{12}{c}{ \mbox{\bf Off-shell linear multiplet}}  \\
\hline
 L_{ij} & 2 & 0  & 3 & L_{ij} & 3 &3 & L_{ij}  & 4  & 3  & 3 &  \\
 E_a & 3 & 0  & 3 & E_a & 4 &4 & E_a  &  5 & 5  & 1&   \\
 G & 3 & -1& 2 & N & 4 & 1 &  &   &  & 1 &   \\
 \hline
 \varphi_i & 5/2 & 1/2 & 8 & \varphi ^i & 7/2 &8 & \varphi^i  & 9/2  & 8  & 2  & -\\
\hline\hline\noalign{\smallskip}
\end{array}  $
\end{center}
\end{table}

\subsection{Vector Multiplets}\label{vectmult}
Vector multiplets can first be defined in 6 dimensions, and then reduced to 5 or 4 dimensions.
\subsubsection{Vector Multiplet in 6 Dimensions (Abelian Case)}
Consider the vector multiplet in $D=6$, which has already been introduced in
Sect.~\ref{ss:exd6VM}. It has been shown in (\ref{commOmega6}) that
the supersymmetry transformations do not close. The solution to this issue is
well-known: the 5 bosonic components of the gauge vector, and the 8
components of the spinor, need an $\SU(2)$-triplet of real scalars,
$Y^{(ij)}$. The latter will appear in the transformation law of the fermion.

As an illustrative example, let us show how the transformation laws of the $D=6$ vector multiplet have been determined with methods
that can be used in general.
In general, it is useful to first consider the Weyl weights of the
fields. One useful principle is that gauge fields (beyond the Weyl multiplet) should have Weyl weight~0, as all transformations beyond the superconformal group must commute with the conformal generators. Equivalently, all the parameters beyond the superconformal group have to be considered\footnote{In principle parameters do not transform, but the
commutators of symmetries can be stated in these terms.}
as Weyl weight~0.

For the $\U(1)$ gauge vector $W_\mu$, whose abelian gauge transformation is $\delta_G W_\mu =\partial _\mu \theta $, the previous argument implies
that $W_\mu $ has Weyl weight~0.\footnote{We could straightforwardly have generalized to a non-abelian algebra. We will do this below for $D=5$ and $D=4$.} The same argument holds in fact for any
gauge field, or gauge two-form, \ldots~. Then the associated curvature $F_{ab}$ has
Weyl weight~2 (due to the frame fields involved in $F_{ab}=e_a{}^\mu
e_b{}^\nu F_{\mu \nu }$). As we have explained, these are the covariant
quantities that should appear in the transformations of other matter
fields. The supersymmetry parameter $\epsilon $ should be considered to
be of Weyl weight~$-\ft12$, identical to its gauge field $\psi _\mu $. Thus the
supersymmetry transformation of the gaugino to the field strength of the
gauge field determines that the conformal weight of $\lambda $ is indeed
$\ft32$.
 \bexer
 Determine the same result from the transformation of the gauge field to
 the gaugino.
 \eexer
The auxiliary field $Y^{ij}$ can appear in the transformation of the fermion via an extra term $\delta \lambda ^i= Y^{ij}\epsilon
_j$, hence the auxiliary field should be of Weyl weight~2. In its
supersymmetry transformation law can appear a covariant fermionic object
of Weyl weight~$\ft52$. This is consistent with a transformation to the
covariant derivative of the gaugino, in order to cancel
(\ref{commOmega6}). The full transformation laws are
\begin{eqnarray}
  \delta W_\mu &=&\partial _\mu \theta +\ft12\bar \epsilon \gamma _\mu \lambda\,, \nonumber\\
  \delta \lambda ^i&=&
   \left(\ft32 \lambda _{\rm D} -\ft14 \gamma ^{ab}\lambda _{ab}\right)
   \lambda^i +\lambda ^{ij} \lambda _j-\ft14
    \gamma ^{ab}\widehat{F}_{ab} \epsilon^i- Y^{ij}\epsilon _j\,,\nonumber\\
   \delta Y ^{ij}&=& 2\lambda _{\rm D} Y^{ij}+2\lambda^{k(i}Y^{j)}{}_k
   -\ft12 \bar \epsilon ^{(i}\slashed{\cal D} \lambda ^{j)}+\bar \eta ^{(i}\lambda ^{j)}\,.
 \label{TransfVM6offshell}
\end{eqnarray}
Starting from the rigid transformations, we replaced $F_{ab}$ by the covariant expression $\widehat{F}_{ab}$ and the derivative of $\lambda ^j$ has been replaced by a covariant derivative.
\begin{align}
  \widehat{F}_{\mu \nu }=&F_{\mu \nu }-\bar \psi _{[\mu }\gamma _{\nu ]}\,, \nonumber\\
  {\cal D}_\mu \lambda ^i=&\left( \partial _\mu -\ft32 b_\mu
  +\ft14 \gamma ^{ab}\omega  _{\mu ab}\right)
   \lambda^i -V_\mu {} ^{ij} \lambda_j
  +\ft14 \bar \psi _\mu^i
 \gamma ^{ab}\widehat{F}_{ab}+Y^{ij}\psi  _{\mu \,j}\,.
 \label{covderVMD6}
\end{align}

The consistency with Weyl weights does not leave place for other terms in the $Q$-transformations. Since the $S$-supersymmetry parameter $\eta $ has to be considered as having Weyl weight $\ft12$, the only $S$-transformation that can occur consistent
with Weyl weights is the last term in (\ref{TransfVM6offshell}). Its coefficient has to be fixed from calculating the
$[\delta _Q(\epsilon),\delta _Q(\eta )]$ commutator on the gaugino or from the method in item (3) in Sect. \ref{ss:matterinWeyl}. One
can check that the extra terms from $Y^{ij}$ cancel the non-closure terms
(\ref{commOmega6}).
 \bexer
 Check that all the transformation laws determine (and are consistent
 with) $\lambda $ to be a left-chiral spinor, in accordance with
 Table~\ref{tbl:onshellmult}.
 \eexer

\subsubsection{Vector Multiplet in 5 Dimensions}\label{vecmult5d}

The transformations of the vector multiplet in 5 dimensions can be obtained from dimensional reduction\footnote{The reader can easily find the linearized transformations from those in (\ref{TransfVM6offshell}) using the rules in Appendix \ref{app:spinors5to4}. It may be more difficult to find the nonlinear transformations, since there are redefinitions such as $W_\mu (D=6)= W_\mu (D=5)+e_\mu^5 \sigma $. It is easier to obtain the nonlinear transformations from directly imposing the supersymmetry algebra in $D=5$.} of the transformations for $D=6$. Note that one component of the $D=6$ vector is a real scalar $\sigma $ in $D=5$.

 We will introduce here the vector multiplet in a non-abelian
group, based on matrix representations with $[t_I, t_J] = f_{IJ}{}^K t_K$.  Note that we will use the index $I$ from now on to enumerate the
vector multiplets, and thus the generators of the non-abelian algebra
that can be gauged. We hope that this does not lead to confusion with the
index $I$ that was used so far to denote all standard gauge transformations as it was done in Chap.~\ref{ss:Weyl}.

 The full rules can be found in \cite{Bergshoeff:2002qk,Bergshoeff:2004kh} for a generalization containing also tensor multiplets. For simplicity, we give here the supersymmetry transformations for only vector multiplets:
\begin{eqnarray}
\delta W_\mu^I
&=& \partial _\mu \theta ^I -\theta^J W_\mu^Kf_{JK}{}^I+\ft12 \bar{\epsilon} {\gamma}_\mu {\psi}^I -\ft12 {\rmi} \sigma^I \bar\epsilon {\psi}_\mu , \nonumber\\
\delta Y^{ij  I}
&=& -\,\ft12 \bar{\epsilon}^{(i} \slashed{\mathcal{D}} {\psi}^{j)  I} +\ft12 \,{\rmi} \bar\epsilon^{(i}\gamma  \cdot T {\psi}^{j)  I} - 4 {\rmi} \sigma^{ I} \bar\epsilon^{(i} \chi^{j)} ,\nonumber\\
&&+\ft12 \rmi \bar{\epsilon}^{(i} f_{JK}{}^I\sigma^J \psi^{j)K}+ \ft12 {\rmi} \bar{\eta}^{(i} {\psi}^{j)  I} ,\nonumber\\
\delta {\psi}^{i  I} &=& - \ft14 \gamma  \cdot \widehat{{F}}{}^{ I} \epsilon^i -\ft12{\rmi} \slashed{\mathcal{D}} \sigma^{ I}
\epsilon^i - Y^{ij  I} \epsilon_j + \sigma^{ I} \gamma  \cdot T
\epsilon^i + \sigma^{ I} \eta^i , \nonumber\\
\delta \sigma^{ I} &=& \ft12 {\rmi} \bar{\epsilon}
{\psi}^{ I} . \label{tensorlocal}
\end{eqnarray}
The (superconformal) covariant derivatives are given by
\begin{eqnarray} \label{localderiv-tensor}
\mathcal{D}_\mu\, \sigma^{ I} &=& { D}_\mu \sigma^{ I} - \ft12{\rmi} \bar{\psi}_\mu \psi^{ I} ,\nonumber\\[2pt]
{ D}_\mu \sigma^{ I} &=& (\partial_\mu - b_\mu) \sigma^{ I}-f_{JK}{}^I W_\mu^K\sigma^J,\nonumber\\[2pt]
{\mathcal{D}}_\mu \psi^{i  I}
&=& { D}_\mu \psi^{i  I} + \ft14 \gamma  \cdot \widehat{{F}}^{ I} {\psi}_\mu^i + \ft12{\rmi} \slashed{\mathcal{D}} \sigma^{ I} {\psi}_\mu^i + Y^{ij  I} {\psi}_{\mu j} - \sigma^{ I} {\gamma}\cdot T {\psi}_\mu^i  ,\nonumber\\
&&+\ft12 f_{JK}{}^I \sigma^J \sigma^K \psi_\mu ^i- \sigma^{ I} \phi_\mu^i,\nonumber\\
{ D}_\mu \psi^{i  I} &=& \left(\partial_\mu - \ft32 b_\mu +
\ft14 {\gamma}_{ab} \hat{\omega}_\mu {}^{ab}\right) {\psi}^{i
I} - V_\mu^{ij} {\psi}_j^{ I} -f_{JK}{}^I W_\mu^K  \psi^{iJ}.
\end{eqnarray}
with $\widehat F^I_{\mu \nu}$ given by
\begin{equation}
\widehat{F}_{\mu\nu}^I ={F}_{\mu\nu}{}^I - \bar{\psi}_{[\mu} {\gamma}_{\nu]} {\psi}^I + \ft12 {\rmi} \sigma^I \bar{\psi}_{[\mu} {\psi}_{\nu]} \,,\qquad {F}_{\mu\nu}{}^I=2 \partial_{[\mu} W_{\nu]}^I + W_\mu^J W_\nu^K f_{JK}{}^I\,.
\label{widehatFD5}
\end{equation}

There is one more aspect in the dimensional reduction (whether the multiplet is abelian or not). Remember that the covariant general coordinate
transformations contain a linear combination of all gauge symmetries.
That involves also the gauge transformation of the vector. Thus in the
commutator of two supersymmetry transformations in $D=6$ is a term $\bar \epsilon
_2\gamma ^\mu \epsilon _1W_\mu $. When reduced to 5 dimensions (and below also to 4 dimensions), some
components of $W_\mu $ are replaced by the scalars $\sigma $. This is the
origin of a new term in the supersymmetry commutator involving structure
functions depending on the scalars, which is implicit in the form of the last term in $\widehat{F}_{\mu\nu}^I $, which is of the form of the last term in (\ref{defcurv}) for gravitini as gauge fields.

\subsubsection{Vector Multiplet in 4 Dimensions} Further dimensional reduction leads to the vector multiplet in 4 dimensions. As
mentioned already in Sect.~\ref{ss:multiplets}, it has then a complex
scalar, built from the fourth and fifth components of the vector of 6
dimensions. To get the right behaviour of gauge and general coordinate
transformations, one has to consider the reduction of the vector with tangent spacetime
indices (see~\cite{Kugo:2000hn,Kugo:2000af} and a useful general introduction to dimensional reduction is  \cite{Pope:2000}). In other words, the object from where the scalars originate in the dimensional reduction should  be a world scalar, $e_a{}^\mu W_\mu $, which has Weyl weight~1. Therefore the complex\footnote{To be in accordance with common practice here, we denote
the complex conjugates of the scalar fields by $\bar X$ rather than
$X^*$. } scalar $X$ of the $D=4$ vector multiplets has $w=1$.
\vspace{10pt}

Before giving the supersymmetry
transformations, we have to translate the reality of the triplet $Y_{ij}$
in appropriate notation for 4 dimensions. In 6 dimensions the reality is
$Y=Y^*=\sigma _2 Y^C \sigma _2$. It is in the form with $Y^C$ that we
have to translate it, thus giving rise to
\begin{equation}
  Y_{ij}=\varepsilon _{ik}\varepsilon _{j\ell }Y^{k\ell }\,,\qquad
  Y^{ij}=(Y_{ij})^*\,.
 \label{realYd4}
\end{equation}

As for $D=5$, we write the transformations for the non-abelian vector multiplet.
The transformations under dilatations and  chiral $\U(1)$ transformations
follow from Table~\ref{tbl:matterMult}, with the general
rules~(\ref{deltaDw}) and~(\ref{deltaTc}). The supersymmetry ($Q$ and $S$),
and the gauge transformations with parameter $\theta$
in 4 dimensions are\footnote{For the translation from $D=5$, we use $X^I=\ft12(W_4^I-\rmi \sigma ^I)$, and $\Omega $ has been defined with the opposite sign as would straightforwardly follow from Appendix \ref{app:spinors5to4}: $\psi ^{iI} = -\Omega ^{iI} -\Omega_j^I\varepsilon ^{ji}$. }
\begin{eqnarray}
 \delta X^I & = & 
 \ft12 \bar \epsilon ^i\Omega^I _i-\theta^J X^Kf_{JK}{}^I, \nonumber\\
  \delta \Omega _i^I & = & 
  \slashed{\cal D} X^I\epsilon _i +
  \ft14 \gamma^{ab} {\cal F}_{ab}^{I-}\varepsilon _{ij}\epsilon ^j
  + Y^I_{ij}\epsilon^j
  +X^J\bar X{}^Kf_{JK}{}^I\varepsilon _{ij}\epsilon ^j\nonumber\\
  &&+2X^I\eta _i-\theta^J\Omega _i^K f_{JK}{}^I , \nonumber\\
  \delta W_\mu^I &=&\ft12
 \varepsilon ^{ij}\bar \epsilon _{i}\gamma _\mu \Omega _j^I
  +\varepsilon ^{ij}\bar \epsilon _{i}\psi _{\mu j}X^I +\hc
  +\partial _\mu \theta^I -\theta^J W_\mu^Kf_{JK}{}^I, \nonumber\\
 \delta Y_{ij}^I&=&\ft12 \bar \epsilon _{(i}\slashed{\cal D} \Omega _{j)}^I
 +\ft12 \varepsilon _{ik}\varepsilon _{j\ell }\bar \epsilon ^{(k}\slashed{\cal D}
 \Omega^{\ell )I}
 +\varepsilon_{k(i}\left( \bar \epsilon _{j)}X^J\Omega ^{kK}
 -\bar \epsilon ^k \bar X^J\Omega _{j)K}\right)f_{JK}{}^I\nonumber\\
 &&-\theta^IY_{ij}^Kf_{JK}{}^I,
 \label{transfVM4}
\end{eqnarray}
where 
\begin{align}
 {\cal F}_{ab}^{I-}\equiv \widehat{F}_{ab}^{I-} -\ft12 \bar X^I T^-_{ab}.
 \label{calFVM4}
\end{align}
In the latter expression $\widehat{F}^{I-}_{ab}$ denotes the anti-self-dual part of $\widehat{F}_{ab}$, which is covariant with the new
structure functions, as dictated by definitions given in
Chap.~\ref{ss:Weyl} and reported here for convenience
 \begin{eqnarray}
 \widehat{F}_{\mu\nu}{}^I&=&{F}_{\mu\nu}{}^I+\left(-\varepsilon_{ij}\bar{\psi}^i_{[\mu}\gamma_{\nu]}\Omega^{Ij}-\varepsilon_{ij}\bar{\psi}^i_\mu \psi_\nu^j \bar{X}^I+\hc\right),\nonumber\\
 {F}_{\mu\nu}^I&=&\partial_\mu W_\nu^I -\partial_\nu W_\mu{}^I +W_\mu{}^J W_\nu{}^K f_{JK}{}^I.
 \label{FmunuD4}
 \end{eqnarray}

Indeed, the second term of the transformation of
the vector reflects the presence of the new term in the commutator of two
supersymmetries, as already discussed for $D=5$, and modifies (\ref{qqcomb}) to
\begin{eqnarray}
[\delta_Q(\epsilon_1),\delta_Q(\epsilon_2)]&=&
\delta_P\left(\xi_3^a\right)
+\delta_M\left(\lambda_3^{ab}\right)+\delta_{\rm K}\left(\lambda_{{\rm K}3}^a\right) +
\delta_S\left(\eta_3\right)\nonumber\\
&&+\delta _G\left(\theta_3^I(\epsilon _1,\epsilon _2) =\varepsilon
^{ij}\bar \epsilon _{2i}\epsilon _{1j}X^I +\hc \right) \,,
\label{QQcommG}
\end{eqnarray}
where $\delta _G$ is the (non-abelian) gauge transformation parameterized by $\theta ^I$.
 \bexer Check that this leads to the form of $\widehat F_{\mu\nu}{}^I$ as given in  (\ref{FmunuD4}).
 \eexer

 The covariant derivatives are
\begin{eqnarray}
  {\cal D}_{\mu}X^I&=& D_\mu X^I -\ft12 \bar \psi^i_\mu \Omega^I_i,\nonumber\\
  D_\mu X^I&=& \left(\partial_\mu-b_\mu-\rmi A_\mu\right)X^I +W_\mu^J X^K f_{JK}{}^I,\nonumber\\
  {\cal D}_{\mu}\Omega^I_i&=& D_\mu \Omega^I_i - \slashed{\cal D}X^I \psi_{\mu i}-\ft14 \gamma^{ab}{\cal F}^{I-}_{ab}\varepsilon_{ij}\psi_\mu^j\nonumber\\
  &&-Y_{ij}^I \psi_\mu^j-X^J\bar{X}^K f_{JK}{}^I \varepsilon_{ij}\psi_\mu^j-2X^I \phi_{\mu i}, \label{covD4N2part1}\\
  D_\mu \Omega^I_i&=& \left(\partial_\mu+\ft14 \omega_\mu{}^{ab}\gamma_{ab}-\ft32 b_\mu-\ft12 \rmi A_\mu\right)\Omega_i^I+V_{\mu i}{}^j \Omega_j{}^I+W_\mu^J\Omega^K_i f_{JK}{}^I\,.\nonumber
  \end{eqnarray}

As will become clear in the following section, the vector multiplet is a constrained chiral multiplet. This observation becomes relevant when constructing actions for the vector multiplet (Sect.~\ref{ss:actions}).
\subsection{Intermezzo: Chiral Multiplet}\label{ss:chiralmult}

A multiplet corresponds to a superfield in superspace. A multiplet or a superfield can be real or chiral, or carry a Lorentz representation, or be in a non-trivial representations of the $R$-symmetry, ...~. For the multiplet, this just reflects the property of its `lowest component'.\footnote{`Lowest' refers here to the Weyl weight in superconformal language (or to the engineering dimensions, if we do not discuss the superconformal properties).}  For example, a chiral multiplet \index{chiral multiplet} is characterized by the fact that its lowest component
transforms `chirally', i.e. only under the left-handed supersymmetry and not under the right-handed one. In superspace this means that one chiral superspace derivative vanishes on the field. Furthermore multiplets or superfields can be constrained. In this section we explain, first in the context of rigid supersymmetry, how further constraints on a chiral multiplet lead to the vector multiplet, which is smaller. A generalization to the rigid superconformal case follows.

\vspace{10pt}
Let us consider a general scalar multiplet, \index{general scalar multiplet} whose `lowest' component is a complex scalar $A$. In general, a complex scalar can  transform under $Q$ to arbitrary spinors
\begin{equation}
   \delta _Q(\epsilon )A=\ft12\bar \epsilon ^i\Psi _i + \ft12\bar \epsilon _i\Lambda ^i\,.
 \label{delQcomplexscalar}
\end{equation}
Then the transformations of
these arbitrary spinors $\Psi _i$ and $\Lambda ^i$ can have arbitrary expressions containing new
fields, as long as it is consistent with the algebra. See e.g. \cite[Sect. 14.1.1]{Freedman:2012zz}
for the example of ${\cal N}=1$ chiral
multiplets, and in Sect.~2.2 of~\cite{VanProeyen:1995sw} this is worked out for the chiral multiplet of ${\cal N}=2$, which we consider here.

If $\Lambda ^i=0$, then the lowest component only transforms under left supersymmetry:
\begin{equation}
  \delta _Q(\epsilon )A=\ft12\bar \epsilon ^i\Psi _i\,,
 \label{chiraltrA}
\end{equation}
and the multiplet is called \emph{chiral}.
Imposing the rigid supersymmetry algebra leads to the
following general expressions:
\begin{eqnarray}
 \delta _Q(\epsilon )A & = & \ft12\bar \epsilon ^i\Psi _i\,, \nonumber\\
   \delta _Q(\epsilon )\Psi_i & = & \slashed{\partial } A \epsilon_i +
 B_{ij}\epsilon^j+\ft14
\gamma_{ab}G^{-ab}\varepsilon_{ij}\epsilon^j\,,\nonumber\\
\delta_Q(\epsilon)B_{ij}&=&\ft12\bar \epsilon _{(i}\slashed{\partial } \Psi _{j)}
 -\ft12\bar \epsilon ^{k}\Lambda _{(i}\varepsilon _{j)k}\,, \nonumber\\
\delta_Q(\epsilon)G^-_{ab}&=&\ft14 \varepsilon ^{ij}\bar \epsilon _i\slashed{\partial }
\gamma_{ab}\Psi _i+\ft14\bar \epsilon ^i\gamma _{ab}\Lambda _i\,,\nonumber\\
\delta_Q(\epsilon)\Lambda _i&=&-\ft14\gamma^{ab}
G_{ab}^-\stackrel{\leftarrow} {\slashed{\partial }}\epsilon _i-\slashed{\partial }
B_{ij}\varepsilon ^{jk}\epsilon
_k+\ft12C\varepsilon _{ij}\epsilon ^j\,, \nonumber\\
\delta_Q(\epsilon)C&=&-\varepsilon ^{ij}\bar \epsilon_i\slashed{\partial } \Lambda _j\,.
 \label{lintransfchiral}
\end{eqnarray}
The reader can count that this is a $16+16$ multiplet counted as real components.
In fact it is reducible, since one can impose the following \emph{consistent constraints}\footnote{There is an extension possible that the first of these expressions is not zero \cite{Antoniadis:1995vb}  but a constant. This leads to magnetic couplings in rigid supersymmetry, and possibilities for partial breaking \index{partial breaking} to ${\cal N}=1$ supersymmetry. Recently \cite{Cribiori:2018jjh}, it has been shown how to generate these constants dynamically using multiplets with 3-form gauge fields, and in \cite{Antoniadis:2019gbd} this has been related to deformations in Dirac--Born--Infeld actions. It is not clear how to generalize this to supergravity, and hence we will not further discuss this.}: \index{consistent constraints}
\begin{eqnarray}
   &   & B_{ij}- \varepsilon_{ik}\varepsilon_{j\ell }B^{k\ell } =0\,, \nonumber\\
   &   & \slashed{\partial } \Psi^i-\varepsilon^{ij}\Lambda_j=0\,,\nonumber\\
   && \partial _b(G^{+ab}-G^{-ab})=0\,,\nonumber\\
   && C-2\partial _a\partial ^a \bar A=0\,,
 \label{constraintschiral}
\end{eqnarray}
where $B^{k\ell}$ is, as usual, defined by the complex conjugate of
$B_{k\ell }$, and similarly $G^+$ is the complex conjugate of $G^-$, and
thus self-dual as $G^-$ is anti-self-dual. These constraints are \emph{consistent}
in the sense that a supersymmetry variation of one of them leads to the
other equations, and this is a complete set in that sense.

The third equation is a Bianchi identity that can be solved by interpreting
$G_{ab}$ as the field strength of  a vector. To conclude, the independent components
are then those of the vector multiplet, with the following identifications:
\begin{equation}
  X=A\,,\qquad \Omega _i=\Psi _i\,, \qquad F_{ab}=G_{ab}\,, \qquad
  Y_{ij}=B_{ij}\,.
 \label{IdentVMchiral}
\end{equation}
Indeed the linear part of (\ref{transfVM4}) corresponds to (\ref{lintransfchiral}). We have thus identified the vector multiplet as a \emph{constrained} chiral multiplet.
\index{constrained chiral multiplet}

To define the chiral multiplet in the conformal algebra, one first allows
an arbitrary Weyl weight for $A$, say that this is $w$. Then consistency with Weyl weights imposes that a general $S$-supersymmetry transformation for $\Psi _i$ should be
proportional to $A$. Imposing the $\{Q,S\}$ anticommutator immediately shows
that the chiral $\U(1)$ weight of $A$ should be related to its Weyl
weight. In fact, to avoid the $\epsilon _i$ terms in this anticommutator,
one should impose that under dilatations and $\U(1)$,
\begin{equation}
  \delta_{{\rm D},T}(\lambda _{\rm D},\lambda_T) A= w\left( \lambda _{\rm D}+\rmi \lambda_T\right) A\,.
 \label{deltaDU1A}
\end{equation}
The same transformations for the other fields can be obtained by requiring compatibility with $Q$-transformations, to obtain
\begin{eqnarray}
   \delta _{{\rm D},T}(\lambda _{\rm D},\lambda_T)\Psi_i & = &\left((w+\ft12) \lambda _{\rm D}+\rmi (w-\ft12)\lambda_T\right)\Psi_i \,,\nonumber\\
\delta_{{\rm D},T}(\lambda _{\rm D},\lambda_T)B_{ij}&=&\left((w+1) \lambda _{\rm D}+\rmi(w-1) \lambda_T\right)B_{ij}\,, \nonumber\\
\delta_{{\rm D},T}(\lambda _{\rm D},\lambda_T)G^-_{ab}&=&\left((w+1) \lambda _{\rm D}+\rmi (w-1)\lambda_T\right)G^-_{ab}\,,\nonumber\\
\delta_{{\rm D},T}(\lambda _{\rm D},\lambda_T)\Lambda _i&=&\left((w+\ft32) \lambda _{\rm D}+\rmi(w-\ft32) \lambda_T\right)\Lambda _i\,, \nonumber\\
\delta_{{\rm D},T}(\lambda _{\rm D},\lambda_T)C&=&\left( (w+2)\lambda _{\rm D}+\rmi(w-2)
\lambda_T\right)C\,.
 \label{wcchiral}
\end{eqnarray}
To complete the superconformal multiplet, one has to add
$S$-transformations, and there are nonlinear transformations involving
the matter fields of the Weyl multiplet $\chi _i$ and $T_{ab}$, necessary
in order to represent the anticommutators (\ref{qqcomb}). The full result
was found in~\cite{deRoo:1980mm}:
\begin{eqnarray}
\delta_{Q,S}(\epsilon ,\eta )A &=& \ft12 \bar \epsilon^i  \Psi_i \,,\nonumber\\
\delta_{Q,S}(\epsilon ,\eta ) \Psi_i &=& \slashed{\cal D} A \epsilon_i +
B_{ij} \epsilon^j
+ \ft14 \gamma \cdot  G^- \varepsilon_{ij} \epsilon^j +2w A\eta_i \,,\nonumber\\
\delta_{Q,S}(\epsilon ,\eta ) B_{ij} &=&  \ft12\bar\epsilon_{(i} \slashed{\cal D}
\Psi_{j)} - \ft12\bar\epsilon^k
                     \Lambda_{(i} \varepsilon_{j)k}+(1-w)\bar \eta _{(i}\Psi _{j)} \,,\nonumber\\
\delta_{Q,S}(\epsilon ,\eta ) G_{ab}^-&=&\ft14\varepsilon^{ij}
\bar\epsilon_i \slashed{\cal D} \gamma _{ab}
 \Psi_j
                            +\ft14 \bar\epsilon^i \gamma _{ab}  \Lambda_i
 - \ft12\varepsilon^{ij}(1+w) \bar \eta_i \gamma _{ab}                              \Psi_j
                \,,\nonumber\\
\delta_{Q,S}(\epsilon ,\eta ) \Lambda_i &=&-\ft14 \gamma  \cdot G^-
\stackrel{\leftarrow} {\slashed{\cal D}}
                        \epsilon_i - \slashed{\cal D}  B_{ij} \epsilon_k \varepsilon^{jk} + \ft12
                         C \epsilon^j \varepsilon_{ij}\nonumber\\
                    & &
                    -\ft18 (\slashed{\cal D} A) T \cdot \gamma )\epsilon_i
    -\ft18 w A(\slashed{\cal D}  T) \cdot \gamma \epsilon_i        - \ft34(\bar
                        \chi_{[i}\gamma_a  \Psi_{j]}) \gamma^a \epsilon_k \varepsilon^{jk}
            \nonumber\\
                    & & - 2(1+w) B_{ij} \varepsilon^{jk} \eta_k
                    +\ft12(1-w)\gamma \cdot G^-\eta _i\,,\nonumber\\
\delta_{Q,S}(\epsilon ,\eta )  C &=& -  \varepsilon^{ij} \bar\epsilon_i
\slashed{\cal D} \Lambda_j
                -  6 \bar\epsilon_i \chi_j  B_{k\ell } \varepsilon^{ik}\varepsilon^{j\ell }
                \nonumber\\
&&   + \ft18 (w-1)\bar\epsilon_i \gamma \cdot T \stackrel{\leftarrow}
{\slashed{\cal D}} \Psi_j \varepsilon^{ij}
   + \ft18 \bar\epsilon_i \gamma \cdot T \slashed{\cal D}  \Psi_j \varepsilon^{ij}
                + 2 w\varepsilon^{ij} \bar\eta_i  \Lambda_j\,.\nonumber\\
\label{locTrChiralm}
\end{eqnarray}
This time, the set of consistent constraints is\footnote{For rigid supersymmetry, an imaginary constant in $B_{ij}$ would be possible, describing magnetic charges.}
\begin{eqnarray}
0&=& B_{ij} - \varepsilon_{ik}\varepsilon_{jl}  B^{kl}\,,\nonumber\\
0&=&\slashed{\cal D}  \Psi^i - \varepsilon^{ij}  \Lambda_j\,,\nonumber\\
0&=& {\cal D}^a( G_{ab}^+ -  G_{ab}^- + \ft12 A T_{ab} - \ft12 \bar A T_{ab}) -
\ft34 ( \varepsilon^{ij} \bar \chi_i\gamma_b
 \Psi_j
- \hc )\,,\nonumber\\
0&=&-2 \bbox \bar A - \ft 12  G^+_{\mu\nu} T^{\mu\nu} - 3 \bar \chi_i
\Psi^i -  C\,. \label{bian4}
\end{eqnarray}
Interestingly, the constraints above are consistent only for a specific choice of $w$. For example, the first constraint is a reality
condition, and it is easy to check that this is only consistent if the
chiral weight of $B_{ij}$ is zero. This fixes $w=1$, which in turn is the
appropriate value also to interpret $G_{ab}$ as a covariant field
strength. Note that the Bianchi identity in the third line of \eqref{bian4} shows the shift between the pure covariant
field strength and the $G$. Compare this with (\ref{calFVM4}).

The chiral multiplet plays an important role in the construction of the actions in rigid supersymmetry, as its highest component $C$ is a scalar transforming to a total
derivative. That action corresponds
in superspace to take the full chiral superspace integral of the chiral
superfield. However, in local supersymmetry, as in the superconformal transformations in (\ref{locTrChiralm}), the transformation of $C$ is not a pure derivative. Therefore in order to have an invariant action, one has to include more terms, i.e. something of the form
\begin{equation}
  I=\int \rmd^4 x\, e\, C+\cdots  +\hc\,.
 \label{I}
\end{equation}
The $+\cdots $ in (\ref{I}) are terms that should be such that the transformation of the integrand is a total
derivative.

But we can first make a few general observations. The integrand should be invariant under all superconformal transformations. Let us start with the Weyl transformations. The Weyl weight of the determinant of the frame field is $-4$, so $C$ should
have Weyl weight~4. It should also be invariant under $T$-transformations, which means that the chiral weight should be zero. We see from (\ref{wcchiral}) that these two requirements are consistent with a requirement that the chiral multiplet should have Weyl weight~2. Note that this implies that it will not be a constrained chiral (i.e. vector) multiplet. We found above that these have Weyl weight~1. But if we start from a vector multiplet, any holomorphic function of $X$ still transforms only under 1 chirality of $Q$. Hence any $F(X)$ is a chiral multiplet. If we take a homogeneous function of second degree in $X$, this gives us a chiral multiplet with $w=2$ on which we can use the action formula.

To determine the full expression in (\ref{I}) one considers other terms that have Weyl weight~4 and chiral weight~0, and imposes the condition of invariance of the action. In practice, imposing $S$-supersymmetry is easiest to determine all the coefficients of these terms. For local superconformal symmetry the result is~\cite{deRoo:1980mm}.
\begin{eqnarray}
 e^{-1}\mathcal{L}&=&C-\bar{\psi}_{i}\cdot\gamma\Lambda_j
\varepsilon^{ij}+\frac{1}{8}\bar{\psi}_{\mu i}\gamma\cdot T^+
\gamma^{\mu}\Psi_j\varepsilon^{ij}-\frac{1}{4}A T^+_{ab}T^{+ab}\nonumber\\&& -
\frac{1}{2}\bar{\psi}_{\mu i}\gamma^{\mu \nu}\psi_{\nu
j}B_{kl}\varepsilon^{ik}\varepsilon^{jl}+\bar{\psi}_{\mu i}\psi_{\nu j}\varepsilon^{ij}\left(G^{-\mu \nu}-AT^{+\mu \nu}\right)
\nonumber\\&&+\frac{1}{2}\rmi
\varepsilon^{ij}\varepsilon^{k\ell}e^{-1}\varepsilon^{\mu \nu \rho \sigma}\bar{\psi}_{\mu i}\psi_{\nu j}\bar{\psi}_{\rho k}\left(\gamma_{\sigma}\Psi_{\ell}+\psi_{\sigma\ell}A\right)
+\hc\,.
 \label{densityChirald4}
\end{eqnarray}
This is called the \emph{chiral density formula}. \index{chiral density formula}

\subsection{Rigid Hypermultiplets}
\label{ss:hypermultiplets}

Hypermultiplets are the analogues of the chiral multiplets of ${\cal N}=1$ supersymmetry. They contain four scalars and two spin- $1/2$ fields. In supergravity, they are defined in the background of the Weyl multiplet and
possibly also in the background of the vector multiplet (i.e. they can transform non-trivially
under the gauge transformations of the vector multiplets). One can further introduce auxiliary
fields to close the algebra for the simplest
quaternionic manifolds. The methods of harmonic or projective superspace mentioned in the introduction \cite{Galperin:1984av,Galperin:2001uw,Karlhede:1984vr,Lindstrom:1987ks,Lindstrom:1989ne,Lindstrom:2008gs,Kuzenko:2010bd}
are also equivalent to introducing an infinite number of auxiliary fields. However, we do not need auxiliary
fields any more at this point because the hypermultiplets are at
the end of the hierarchy line.\footnote{We are not going to introduce any further
multiplet in the background of the hypermultiplets, as these do not
introduce new gauge symmetries. This is to be confronted to when we considered the vector multiplets.
The construction of the latter had to take into account that the multiplets can be used
for various possible actions (including hypermultiplets or not).}

The closure of the supersymmetry algebra will impose equations that we will interpret as
equations of motion, even though we have not defined an action yet. Later
we will see how an action can be constructed that gives precisely these
equations as Euler--Lagrange equations.

Although our interest is the local case, the present section is mostly devoted to rigid super(conformal) symmetry. This choice has been made since the rigid case provides already simpler and explicative examples of the story. We remark that since dimensional reduction for scalars and spin-1/2 fermions leads to the same type of particles in lower dimension, the properties of the hypermultiplets do not depend on whether we consider $D=6$, $D=5$, or $D=4$ (or even $D=3$). There is a technical difference since the four on-shell (or eight off-shell) degrees of freedom are captured in symplectic Weyl, symplectic or Majorana spinors, respectively. In practice, we mostly report formulae in $D=5$. These can be translated to $D=6$ and $D=4$ by the rules in Appendix \ref{app:spinors5to4}.

Before starting the mathematical formulation, we still want to point out how massive hypermultiplets \index{massive hypermultiplets} can be described in this context, since the readers will mainly see equations of motion that describe only massless hypermultiplets. This is of course also related to the fact that we are mainly interested in conformal theories. Massive hypermultiplets in rigid supersymmetry are obtained in this setting by adding a coupling to a vector multiplet that has just a first scalar component equal to the mass, and all other components zero. The reader can glimpse at  (\ref{eqmozeta5d}) for $D=5$ with $\sigma ^I$ equal to a mass, or to (\ref{bewegingfermion1}) and (\ref{delQGamma}) for $X^I$ providing the mass to see that with a suitable choice of the Killing vectors these are massive field equations. In supergravity this will be natural for the $\sigma ^I$ or $X^I$ referring to the compensating multiplet.

\subsubsection{Rigid Supersymmetry}\label{hyp_rigid_notations}

We consider a set of $n_H$ hypermultiplets. The real scalars are denoted as $q^X$, with $X=1,...,4 n_H$, and the fermions are indicated by $\zeta^\Ae$, where the indices $\Ae=1,\ldots ,2n_H$ will indicate a fundamental representation of $\Sp(2n_H)$. Imposing the supersymmetry transformations
on the bosons lead to the identification of a hypercomplex manifold\footnote{In supergravity the scalars span a quaternionic manifold, see Sect. \ref{ss:qKmanifolds}.}
parameterized by these bosons $q^X$. The structure is determined by
frame fields $f^{i\Ae}{}_X$, connections $\omega _{X\Ae}{}^\Be$ and $\Gamma
_{XY}^Z$ (the latter symmetric in its lower indices) such that
\begin{equation}
  f^{i \Ae}{}_{Y}  f^{ X}{}_{i \Ae}
\,=\, \delta_{Y}^{ X} \,,\qquad  f^{i\Ae}{}_{ X}  f^{ X}{}_{j \Be} \,=\, \delta^i_j \delta^{ \Ae}_{ \Be}\,.
 \label{ff1}
\end{equation}
and
\begin{eqnarray}
\nabla _Y f^X{}_{i\Ae} \equiv
\partial_Y f^X{}_{i\Ae} - \omega_{Y\Ae}{}^\Be(q)   f^X{}_{i\Be} + \Gamma _{YZ}^X(q) f^Z{}_{i\Ae} =
0\,,\nonumber\\
\nabla _Y f^{i\Ae}{}_X \equiv
\partial_Y f^{i\Ae}{}_X +  f^{i\Be}{}_X\omega_{Y\Be}{}^\Ae(q)   - \Gamma _{YX}^Z(q) f^{i\Ae}{}_Z =
0\,,
 \label{Df0}
\end{eqnarray}
are satisfied. The frame field satisfies a reality condition, for which we will also introduce indices $\bAe$:
\begin{equation}
\left(f^{i\Ae}{}_X\right)^* =f^{j\Be}{}_X\varepsilon_{ji}\rho_{\Be \bAe}\,,\qquad \left(f^{X}{}_{i\Ae}\right)^* =\varepsilon^{ij}\rho^{\bAe \Be}f^{X}{}_{j\Be}\,,\label{introrho}
\end{equation}
in terms of a non-degenerate covariantly constant tensor $\rho_{\Ae \bBe}$ that satisfies
\begin{align}
\rho_{\Ae \bBe} \rho^{\bBe \Ce} =-\delta^{\Ce}_\Ae\,,\qquad \rho^{\bAe \Be}=\left(\rho_{\Ae \bBe}\right)^*\,.
\end{align}
By field redefinitions, we could bring it in the standard antisymmetric form
\begin{equation}
  \rho _{\Ae\bBe}=
  \begin{pmatrix}
  0&\unity _{n_H}\cr -\unity _{n_H} &0
  \end{pmatrix}= \rho ^{\bAe\Be}
  \,.
 \label{rhocanonical}
\end{equation}
We will not impose this basis choice in general. In Sect. \ref{ss:splithypers} we will show how such a basis could be implemented.

The complex conjugate of $\omega _{X\Ae}{}^\Be$ is
\begin{equation}
(\omega _{X\Ae}{}^\Be)^*\equiv \bar{\omega}_X{}^{\bAe}{}_\bBe=-\rho^{\bAe \Ce}\omega _{X\Ce}{}^\De\rho_{\De \bBe}.
\end{equation}
The above conditions lead to the identification of almost quaternionic structures \index{almost quaternionic structure}
\begin{align}
  2f^{i\Ae}{}_Xf^Y{}_{j\Ae}= \delta _X^Y\delta _j^i + J_X{}^Y{}_j{}^i\,,\qquad
&J_X{}^Y{}_j{}^i=   \vec{\tau }_j{}^i\cdot \vec{J}_X{}^Y\,,\nonumber\\
&\vec{J}_X{}^Y=\left(\vec{J}_X{}^Y\right)^*= -f^{i\Ae}{}_Xf^Y{}_{j\Ae}\vec{\tau }_i{}^j\,.
 \label{vpro-defJf}
\end{align}
We use here the 3-vectors notation and $\vec{\tau }_i{}^j=\rmi \vec{\sigma }_i{}^j$ in terms of the three Pauli-matrices $\vec{\sigma }_i{}^j$ as in (\ref{propUaHtr}), (\ref{propU4}). Related formulas are given in Appendix~\ref{SU2vectorNotation}.
The three matrices $\vec{J}$ satisfy the quaternionic algebra, \index{quaternionic algebra} i.e. for any vectors $\vec{A}$, $\vec{B}$
\begin{equation}\label{algebrJ}
\vec{A}\cdot  \vec{J}_X{}^Z \vec{B}\cdot  \vec{J}_Z{}^Y =-\delta_X{}^Y\vec{A}\cdot \vec{B}+(\vec{A}\times \vec{B})\cdot  \vec{J}_X{}^Y\,.
\end{equation}
In passing, we note that we can solve \eqref{Df0} for $\omega _{X\Ae}{}^\Be$, such that the independent connection is $\Gamma_{XY}{}^Z$. The latter is the unique connection on the scalar manifold respect to which
\begin{equation}\label{cov_const_J}
\nabla_Z \vec{J}_X{}^Y\equiv \partial_Z \vec{J}_X{}^Y-\Gamma_{ZX}{}^U \vec{J}_U{}^Y+\Gamma_{ZU}{}^Y \vec{J}_X{}^U=0.
\end{equation}
This last condition promotes $\vec{J}_X{}^Y$ to be quaternionic structures. \index{quaternionic structures}

The integrability condition of (\ref{Df0}) relates the curvatures defined by the two connections:
\begin{eqnarray}
 R_{XY}{}^W{}_Z & \equiv  & 2 \partial_{[X}\Gamma _{Y]Z}^W+2\Gamma^W _{V[X}\Gamma^V _{Y]Z}\,,\nonumber\\
  {\cal R}_{XY\Be}{}^\Ae &\equiv&  2\partial_{[X} \omega_{Y]\Be}{}^\Ae
+ 2\omega_{[X|\Ce|}{}^\Ae \omega_{Y]\Be}{}^\Ce\,,\nonumber\\
R_{XY}{}^W{}_Z&=&f^W{}_{i\Ae}f^{i\Be}{}_Z{\cal R}_{XY\Be} {}^\Ae \,,
\qquad {\cal R}_{XY\Be}{}^\Ae = \ft12 f^W{}_{i\Be} f^{i\Ae}{}_Z R_{XY}{}^Z{}_W\,.\nonumber\\
 \label{curvatureshyper}
\end{eqnarray}
In order to work with these tensors, it can be useful to introduce also tensors $L$ that are orthogonal to the complex structures:
\begin{align}
 L_Y{}^Z{}_\Ae{}^\Be \equiv f^Z{}_{i\Ae}f^{i\Be}{}_Y\,, \qquad & \vec{J}_Z{}^{Y } L_Y{}^Z{}_\Ae{}^\Be = 0 \,,\nonumber\\
 & L_X{}^Y{}_\Ae{}^\Be L_Y{}^Z{}_\Ce{}^\De = L_X{}^Z{}_\Ce{}^\Be \delta_\Ae{}^\De\,,\nonumber\\
 & L_X{}^X{}_\Ae{}^\Be = 2 \delta_\Ae{}^\Be\,,\qquad
  L_X{}^Y{}_\Ae{}^\Be L_Y{}^X{}_\Ce{}^\De = 2 \delta_\Ce{}^\Be \delta_\Ae{}^\De \,.\label{orthogJL}
\end{align}

If the affine connections is the Levi-Civita connection of a metric, then the curvatures satisfy the cyclicity properties $R_{(XY}{}^W{}_{Z)}=0$, and one can show that
\begin{equation}
  f^X{}_{i\Ce } f^Y{}_{j\De}{\cal R}_{XY\Be} {}^\Ae = - \ft12 \varepsilon_{ij} W_{\Ce\De\Be}{}^\Ae\,,\qquad W_{\Ae\Be\Ce}{}^\De   \equiv  -\varepsilon ^{ij} f^{X}{}_{i\Ae} f^Y{}_{j\Be} {\cal R}_{XY\Ce}{}^\De\,.
 \label{propWij}
\end{equation}
The tensor $W_{\Ae\Be\Ce}{}^\De$ is symmetric in its lower indices, and the other curvatures can be expressed in function of this one as
\begin{equation}
R_{XY}{}^W{}_Z=L_Z{}^W{}_\De{}^\Ce\,{\cal R}_{XY\Ce} {}^\De = -\varepsilon _{ij}\ft12 L_Z{}^W{}_\De{}^\Ce f^{i\Ae} {}_{X} f^{j\Be}{}_Y W_{\Ae\Be\Ce}{}^\De\,.
  \label{curvW}
\end{equation}
The Bianchi identity on ${\cal R}_{XY\Ae}{}^{\Be}$ implies also a symmetry of the covariant derivative of $W$:
\begin{equation}\label{DW}
f^X{}_{i\Ae}\nabla _X W_{\Be\Ce\De}{}^\Ee=f^X{}_{i(\Ae|}\nabla _XW_{|\Be\Ce\De)}{}^\Ee\,.
\end{equation}

When a metric will be defined on the manifold, the $W$-tensor will become symmetric in the 4 indices. As a consequence, the manifold will be \emph{Ricci flat}:
\begin{equation}
R_{YZ}=R_{XY}{}^X{}_Z=0\,.
\end{equation}

\subsubsection{Reparameterizations and Covariant Quantities}

The hypermultiplet is defined in terms of the scalars $q^X$, which form a parameterization of a $4n_H$-dimensional manifold, and the fermions $\zeta ^\Ae$, which are a parameterization of a $2n_H$-dimensional manifold of fermions. Both these basic parameterizations can be changed \cite{Bergshoeff:2002qk}. There are thus two kinds of
repara\-meterizations. The first ones are the target space diffeomorphisms,
$q^X\rightarrow {\widetilde q}^X(q)$, under which $f^X{}_{i\Ae}$ transforms
as a vector, $\omega_{X\Ae}{}^\Be$ as a one-form, and $\Gamma_{XY}{}^Z$ as a
connection. The second set are the reparameterizations, which act on the
tangent space indices $\Ae,\Be$ etc. On the fermions, they act as
\begin{equation}
\zeta^\Ae \rightarrow {\widetilde \zeta}^\Ae(q)=\zeta^\Be
U_\Be{}^\Ae(q)\,,\label{ferm-equiv}
\end{equation}
where $U_\Ae{}^\Be(q)$ is an invertible matrix, and the reality conditions
impose $U^*=\rho ^{-1}U\rho $, defining $\Gl(r,\mathbb H)$. In general, the right-hand side of (\ref{ferm-equiv}) depends on the $\zeta^\Ae$ and on the scalars. Thus the new basis ${\widetilde \zeta}^\Ae$ is a basis where the fermions depend on
the scalars $q^X$. In this sense, the hypermultiplet is written in a
special basis where $q^X$ and $\zeta^\Ae$ are independent fields. We will develop a covariant formalism which also takes into account these reparameterizations.

The supersymmetry transformations in $D=5$ are
\begin{eqnarray}
 \delta q^X &=&-\rmi\bar \epsilon ^i\zeta ^\Ae f^X{}_{i\Ae}\,,\nonumber\\
 \delta \zeta ^\Ae&=&\ft12\rmi f^{i\Ae}{}_X\slashed {\partial} q^X \epsilon _i-\zeta ^\Be\omega _{X\Be}{}^\Ae\delta q^X\,.
 \label{QhyperD5}
\end{eqnarray}

They are covariant under~(\ref{ferm-equiv}) if we transform $f^{i\Ae}{}_X(q)$ as a
vector and $\omega_{X\Ae}{}^\Be$ as a connection,
\begin{equation}
\omega_{X\Ae}{}^\Be\rightarrow {\widetilde \omega}_{X\Ae}{}^\Be=[(\partial_X
U^{-1})U+ U^{-1}\omega_X U]_\Ae{}^\Be\,.
\label{QhyperD5conn}
\end{equation}

These considerations lead us to define the covariant variation of vectors (see \cite[Appendix 14B]{Freedman:2012zz})
with indices in the tangent space, as $\zeta ^\Ae$, or a quantity $\Delta
^X$ with coordinate indices:
\begin{equation}
\widehat{\delta} \zeta^\Ae\equiv \delta \zeta^\Ae+\zeta^\Be\omega_{X\Be}{}^\Ae
\delta q^X\,, \qquad  \widehat{\delta} \Delta^X \equiv \delta \Delta^X +
\Delta^Y \Gamma_{YZ}{}^X \delta
  q^Z\,, \label{cov-var}
\end{equation}
for any transformation $\delta$ (as e.g. supersymmetry, conformal
transformations,\ldots).

Two models related by either target space diffeomorphisms or fermion
reparame\-terizations of the form~(\ref{ferm-equiv}) are equivalent; they
are different coordinate descriptions of the same system. We usually work in a basis where the fermions and the bosons are independent, i.e. $\partial _X\zeta ^\Ae=0$. But in  a
covariant formalism, after a transformation (\ref{ferm-equiv}), this is not anymore valid. This shows that the
expression $\partial _X\zeta ^\Ae$ has no basis-independent meaning. It makes only sense if one compares a transformed basis, like the $\tilde \zeta ^A$ with the original basis where $\partial _X\zeta ^\Ae=0$. But in the same way also the expression $\zeta^\Be
\omega_{X\Be}{}^\Ae$ makes only sense if one compares different bases, as the
connection has no absolute value. The only object that has a coordinate-invariant meaning is the
covariant derivative
\begin{equation}
  \covder_X \zeta ^\Ae\equiv \partial_X \zeta ^\Ae+\zeta ^\Be\omega
  _{X\Be}{}^\Ae\,.
 \label{calDXzeta}
\end{equation}
In the basis where the fermions $\zeta ^\Ae$ are considered independent of the bosons, i.e. $\partial _X\zeta ^\Ae=0$,  which is the basis used to write down the transformation rules (\ref{QhyperD5}), only the second term in the covariant derivative above remains, and thus \eqref{cov-var} becomes
\begin{equation}
 \widehat{\delta} \zeta
   ^\Ae= \delta \zeta ^\Ae +\covder_X \zeta ^\Ae\,\delta q^X\,.
 \label{hatdelzetacov}
\end{equation}
We will always consider independent bosons and fermions when we write variations. 

On any covariant coordinate quantity that depends only on the coordinates $q^X$, covariant transformations act by covariant derivatives, e.g. for some vectors $V^X(q)$, $W_\Ae$ or $W^\Ae$:
\begin{tcolorbox}
\begin{eqnarray}
\widehat{\delta} V^X(q) &=& \delta q^Y\nabla _Y V^X(q)=\delta q^Y\left(\partial _Y V^X(q) + \Gamma ^X_{YZ}V^Z(q) \right)\,,\nonumber\\
\widehat{\delta}  W^\Ae(q) &=& \delta q^Y\nabla _Y W^\Ae(q)=\delta q^Y\left(\partial _Y W^\Ae(q) +W^\Be(q) \omega _{Y\Be}{}^\Ae\right) \,,\nonumber\\
\widehat{\delta}  W_\Ae(q) &=& \delta q^Y\nabla _Y W_\Ae(q)=\delta q^Y\left(\partial _Y W^\Ae(q) - \omega _{Y\Ae}{}^\Be W_\Be(q)\right) \,.\nonumber\\
 \label{hatisnabla}
\end{eqnarray}
\end{tcolorbox}
\noindent In particular, $\widehat{\delta}$ of any covariantly constant object (like the frame fields $f^{i\Ae}{}_X$) is zero.

Note that we can exploit covariant transformations to calculate any
transformation on e.g.\ a quantity $W_\Ae(q)\zeta ^\Ae$:
\begin{equation}
  \delta \left(W_\Ae(q)\zeta ^\Ae\right)= \widehat{\delta}\left(W_\Ae(q)\zeta ^\Ae\right)= \delta q^X  \covder_XW_\Ae\,\zeta ^\Ae + W_\Ae\,\widehat{\delta} \zeta
   ^\Ae \,.
 \label{simpledelWzeta}
\end{equation}
Coordinates are not covariant, but their derivatives are, and e.g.
 the Laplacian\footnote{In the local (gravity) theory, the first term should be $(\sqrt{g})^{-1}\partial _\mu \sqrt{g}g^{\mu \nu }\partial _\nu $.}
\begin{equation}
\Box q^X =\nabla^\mu \partial  _\mu q^X = \partial^\mu \partial_\mu  q^X+ \Gamma_{YZ} {}^X\left( \partial_\mu  q^Y\right)
\left(\partial^\mu q^Z \right)\,,
\label{BoxqX}
\end{equation}
is covariant for target space transformations.

Another interesting relation is that the commutator of $\widehat{\delta}$ and $\nabla $ gives rise to curvature terms:
\begin{equation}
\widehat{\delta}\nabla _\mu V^X = \nabla _\mu \widehat{\delta}V^X+ R_{ZW}{}^X{}_YV^Y(\delta q ^Z)(\partial _\mu q^W)\,.
 \label{hatdnabla}
\end{equation}
Similarly the commutator gets adapted by curvature terms:
\begin{eqnarray}
  &&\left[\delta (\epsilon _1),\delta (\epsilon _2)\right]V^ X=\delta (\epsilon _3)V ^X\ \rightarrow \nonumber\\
  && \left[\hat{\delta} (\epsilon _1),\hat{\delta} (\epsilon _2)\right]V ^X=\widehat{\delta} (\epsilon _3)V ^X + R_{ZW}{}^X{}_YV^Y(\delta(\epsilon  _1)q^Z)( \delta(\epsilon  _2)q^W)\,,
 \label{commutcovdelta}
\end{eqnarray}
where $\epsilon _3$ is the function of $\epsilon _1$ and $\epsilon _2$ determined by the structure functions.
With these methods, it is easy to compute the commutator of two covariant derivatives. E.g. in $D=5$ with (\ref{QhyperD5}) for the fermions
\begin{align}
 \left[\widehat{\delta}(\epsilon _1),\widehat{\delta }(\epsilon _2)\right]\zeta ^\Ae  = & \ft12\gamma ^\mu \epsilon _{2i}f^{i\Ae}{}_X\bar \epsilon _1^j\nabla _\mu \zeta ^B f^X{}_{j\Be}- \ (1\leftrightarrow 2)\nonumber\\
    = & \ft14\gamma ^\mu \left[(\bar \epsilon _2\epsilon _1)+ \gamma ^\nu (\bar \epsilon _2\gamma _\nu \epsilon _1)\right]\nabla _\mu \zeta ^\Ae\nonumber\\
   =& \ft12\nabla _\mu \zeta ^\Ae(\bar \epsilon _2\gamma ^\mu  \epsilon _1)+\ft14\left[(\bar \epsilon _2\epsilon _1)- \gamma ^\nu (\bar \epsilon _2\gamma _\nu \epsilon _1)\right]\slashed{\nabla} \zeta ^\Ae\,,
 \label{calccommzetaA}
\end{align}
with the definition
\begin{align}
\nabla_\mu\zeta^\Ae\equiv 
\partial _\mu\zeta^\Ae + (\partial_\mu  q^X)
\zeta^\Be\omega_{X\Be} {}^\Ae.
\label{covDq5d}
\end{align}
Indices $i,j$ are raised and contracted as in Appendix \ref{app:gamma5}.
This result shows that the algebra does not close: we will interpret the extra parts as equations of motions of a putative action, see Sect. \ref{finsec_hypers}.

\subsubsection{Non-closure Relations for Fermions and Bosons}

From  the result (\ref{calccommzetaA}), using (\ref{commutcovdelta}), we can obtain the following commutator of transformations:
\begin{align}
 \left[\delta(\epsilon _1),\delta (\epsilon _2)\right]\zeta ^\Ae   = & \ft12\partial  _\mu \zeta ^\Ae(\bar \epsilon _2\gamma ^\mu  \epsilon _1)+\ft14\left[(\bar \epsilon _2\epsilon _1)- \gamma ^\nu (\bar \epsilon _2\gamma _\nu \epsilon _1)\right]\slashed{\nabla} \zeta ^\Ae \nonumber\\
 &+\zeta ^B {\cal R}_{XY\Be}{}^\Ae \bar \epsilon _1^i\zeta ^\Ce f^X{}_{i\Ce}\bar \epsilon _2^j\zeta^\De f^Y{}_{j\De}\,.
 \label{commzeta}
\end{align}
With (\ref{propWij}) and a Fierz transformation, we obtain that the non-closure terms (the last term on the first line and the second line) are
\begin{equation}
  +\ft14\left[(\bar \epsilon _2\epsilon _1)- \gamma ^\nu (\bar \epsilon _2\gamma _\nu \epsilon _1)\right]\slashed{\nabla} \zeta ^\Ae
   +\ft18 W_{\Ce\De\Be}{}^\Ae \zeta ^B  \bar \zeta ^\Ce\left[(\bar \epsilon _2\epsilon _1)+ \gamma ^\nu (\bar \epsilon _2\gamma _\nu \epsilon _1)\right] \zeta^\De \,.
 \label{nonclosureD5}
\end{equation}
Using some $D=5$ Fierz identities:
\begin{eqnarray}
 &   & 5\zeta^{(\Be}\bar{\zeta}^\Ce \zeta^{\De)} =
 -\gamma^\mu \zeta^{(\Be}\bar{\zeta}^\Ce\gamma_\mu  \zeta^{\De)}\,, \nonumber\\
   &   &\zeta^{(\Be}\bar{\zeta}^\Ce \gamma^\nu  \zeta^{\De)} =
-\gamma^\nu \zeta^{(\Be}\bar{\zeta}^\Ce \zeta^{\De)}\,,
 \label{Fierz3zeta}
\end{eqnarray}
we find
\begin{align}
 \left[\delta(\epsilon _1),\delta (\epsilon _2)\right] \zeta ^\Ae  & =  \xi ^\mu \partial  _\mu \zeta ^\Ae+\ft14\left[\left( {\bar  \epsilon} _2^i\epsilon _1^j\right)- \gamma ^\nu \left({\bar \epsilon} _2^i\gamma _\nu \epsilon _1^j\right)\right]\varepsilon _{ji} \rmi\Gamma ^\Ae\,,
 \label{nonclosurezeta}
\end{align}
i.e. the non-closure terms are proportional to\footnote{We inserted a factor $\rmi$ in order that $\Gamma ^A$ is symplectic Majorana.}
\begin{equation}
\rmi\Gamma^\Ae \equiv  \slashed{\nabla} \zeta^\Ae + \frac 12 W_{\Be\Ce\De} {}^\Ae
 \zeta^\Be\bar{\zeta}^\Ce \zeta^\De\,.
 \label{eqmozeta}
\end{equation}
The expression above must be interpreted as an equation of motion for the fermions.  The supersymmetry transformation of \eqref{eqmozeta} gives then
also an equation of motion for the scalar fields:
\begin{equation}
\widehat{\delta}(\epsilon ) \Gamma^\Ae =  \frac 12  f^{i\Ae}{}_{X}
\epsilon_i \Delta ^X \,,\label{delQGamma3}
\end{equation}
where
\begin{align}
\Delta^X=&\Box q^X -\frac 12 \bar{\zeta}^\Be \gamma_a \zeta^\De \partial^a q^Y f^{i\Ce}{}_Yf^X{}_{i\Ae} W_{\Be\Ce\De} {}^\Ae\nonumber\\ &-\frac{1}{4}\covder_Y W_{\Be\Ce\De}{}^\Ae \bar{\zeta}^\Ee \zeta^\De
\bar{\zeta}^\Ce \zeta^\Be f^{iY}{}_\Ee f^X{}_{i\Ae}\,.
\label{covscal}
\end{align}

The equations of motion given by~(\ref{eqmozeta}) and~(\ref{covscal}) form
a multiplet, since~(\ref{delQGamma3}) has the counterpart
\begin{equation}
  \widehat{\delta} (\epsilon ) \Delta^X= \bar{\epsilon}^i \slashed\covder \Gamma^\Ae f^X{}_{i\Ae}
  -\bar{\epsilon}^i \Gamma^\Be \bar{\zeta}^\Ce \zeta^\De f^X{}_{i\Ae}W_{\Be\Ce\De}{}^\Ae
\,, \label{susydelta}
\end{equation}
where the covariant derivative of $\Gamma^\Ae$ is defined similar to~(\ref{covDq5d}).
As announced before, we thus have already physical equations despite the absence of an action.

\subsubsection{Rigid Superconformal}

To allow the generalization to superconformal couplings, the essential
question is whether the manifold has dilatational symmetry. This means,
according to (\ref{homothetic}), that there is a `closed homothetic
Killing vector'~\cite{deWit:1998zg} (see also \cite[Sect. 15.7]{Freedman:2012zz}). The dilatations act as\footnote{Note
that we give here only the intrinsic part of the dilatations, i.e. the
$\lambda _{\rm D}$ term in (\ref{deltaC}), and not the `orbital' part included in the
general coordinate transformation $\xi^\mu  (x)$. Similarly for special
conformal transformations, we will write here only the intrinsic part
represented as $(k_\mu \phi )$ in that equation and also the `orbital' $S$-supersymmetry part (\ref{epsilonx}) is not mentioned explicitly.}
\begin{equation}
\delta _{\rm D}(\lambda _{\rm D})q^X=\lambda _{\rm D}k_{\rm D}{}^X(q),
 \label{delDphiA}
\end{equation}
where $k_{\rm D}{}^X$ satisfies (we generalize here already to $D$ dimensions, as
the modifications involve only a normalization factor)
\begin{equation}
  \nabla_Y k_{\rm D}{}^X \equiv \partial_Y k_{\rm D}{}^X + \Gamma _{YZ}^X k_{\rm D}{}^Z =
\frac{D-2}2 \delta_Y{}^X.
 \label{homotheticKilling}
\end{equation}
On a flat manifold, the fields $q^X$ have thus Weyl weight~$(D-2)/2$.
The presence of this vector allows one to extend the transformations of
rigid supersymmetry to the superconformal
group~\cite{deWit:1998zg,deWit:1999fp,Bergshoeff:2002qk}, with e.g.
transformations under the $\SU(2)$ $R$-symmetry group:
\begin{equation}
  \delta _{\SU(2)}(\vec\lambda )q^X= \mp2\,\vec{\lambda}\cdot \vec{k}^{X}\,,\qquad \vec{k}^{X}\equiv   \frac1{D-2}
  k_{\rm D}{}^Y\vec{J}_Y{}^X\,.
 \label{delDSU2q}
\end{equation}
Note the sign difference between $D=4$, upper sign, and $D=5,6$, lower sign, as in (\ref{convULambda}).

In general, one can introduce the sections
\begin{tcolorbox}
\begin{equation}
  A^{i\Ae} = k^X_{\rm D}f^{i\Ae}{}_X\,,
 \label{defAialpha}
\end{equation}
\end{tcolorbox}
and in terms of these
\begin{equation}
  \vec{k}^X = -\frac{1}{D-2}A^{i\Ae}\vec{\tau }_i{}^jf^X{}_{j\Ae}\,.
 \label{veckinA}
\end{equation}

Using the rules of covariant transformations (and in particular that $\nabla _Y f^{i\Ae}{}_X$ implies $\widehat{\delta}  f^{i\Ae}{}_X=0$), the $A^{i\Ae}$ transform as
\begin{align}
  \widehat{\delta} A^{i\Ae} = f^{i\Ae}{}_X \nabla _Y k^X_{\rm D}\delta q^Y =& \frac{D-2}{2}f^{i\Ae}{}_X\delta q^X\nonumber\\
  =& \frac{D-2}{2}\left(- \rmi{\bar
\epsilon}^i\zeta^\Ae +  \lambda _{\rm D} A^{i\Ae}\right)+ A^{j\Ae}\lambda_j{}^i \,,
 \label{widehatdelA}
\end{align}
using (\ref{homotheticKilling} and  (\ref{delDSU2q}). Note that the supersymmetry transformation in this equation is written for the symplectic spinors of $D=5,6$. Below, we will write them for $D=4$.
%

We can then derive the other (super)conformal transformations using the
algebra. The intrinsic special conformal transformations on $q^X$ and $\zeta ^\Ae$
vanish. They have only the `orbital' parts as follows from (\ref{deltaC}).
The latter imply e.g. that $\delta _{\rm K}(\lambda _{\rm K})
\slashed{\partial}q^X\neq0$. The algebra gives then for the intrinsic
$S$-supersymmetry
\begin{equation}
\delta_S(\eta^i)\zeta^\Ae=- A^{i\Ae} \eta_i\,.
\label{Stransfzeta}
\end{equation}
The (intrinsic) bosonic conformal symmetries act as
\begin{equation}
\widehat{\delta}_{\rm D} \zeta^\Ae = \frac{D-1}{2} \lambda _{\rm D} \zeta^\Ae \,, \qquad
\widehat{\delta}_{\SU(2)} \zeta^\Ae =0\,.
\end{equation}
The fermions are inert under $\SU(2)$ $R$-symmetry group.

\subsubsection{Isometries and Coupling to Vector Multiplets}

So far we considered the hypermultiplet with ungauged isometries. A more general situation includes couplings to vector multiplets and in this case one has to define the hypermultiplet in the
algebra including the vector multiplet with its gauge transformations. Let us consider general isometries (not necessarily gauged) of the hypermultiplet:
\begin{equation}\label{gen_iso_hyper}
\delta_{\rm G}(\theta ) q^X=\theta ^Ik_I{}^X(q)\,,
\end{equation}
where $\theta^I$ are constant parameters and the $k_I{}^X(q)$ represent the
transformations. The index $I$ identifies the different generators of the isometry group. Then a subgroup of these could be gauged, identified by an embedding tensor \index{embedding tensor} \cite{Cordaro:1998tx,Nicolai:2001sv,deWit:2005ub,Samtleben:2008pe} projecting from all the symmetries to those that are gauged.\footnote{However, we will here soon gauge the symmetries, and thus restrict the index $I$ to the gauged symmetries.} When we have a metric, $k_I{}^X(q)$ should be Killing
vectors in order to define symmetries of the action. As we have not discussed a metric yet, we could define here some
generalization of symmetries, but we just refer the interested reader to
\cite{Bergshoeff:2002qk}. The transformations \eqref{gen_iso_hyper} constitute an algebra with
structure constants $f_{IJ}{}^{K}$,
\begin{equation}
k_I{}^Y\partial_Yk_J{}^X-k_J{}^Y\partial_Yk_I{}^X =f_{IJ}{}^K\,
k_K{}^X\,. \label{killingclosure}
\end{equation}

We consider symmetries that respect the hypercomplex structure. This is
the requirement that $k_I{}^X(q)$ is tri-holomorphic:
\begin{equation}
  \left(\covder_X k^Y_I\right)\vec{J}_Y{}^Z= \vec{J}_X{}^Y\left(\covder_Y
  k^Z_I\right)\,.
 \label{commDkJ}
\end{equation}
Extracting affine connections from this equation, it can be written as
\begin{equation}
\label{intJ} \left( {\cal L}_{k_I} \vec{J}\right){} _X{}^Y\equiv  k_I{}^Z
\partial_Z \vec{J}_X{}^Y -\partial_Z k_I{}^Y \vec{J}_ X{}^Z+\partial_X
k_I{}^Z \vec{J}_Z{}^Y=0\,.
\end{equation}
This is the Lie derivative of the complex structure in the
direction of the vector $k_I$.

Multiplying  (\ref{commDkJ}) with $f^X{}_{i\Ae}f^{j\Be}{}_{Y}$  proves that $f^Y{}_{i\Ae}\covder_Yk_I{}^X f^{j\Be}{}_X$ should be proportional to $\delta _i^j$. This leads to the definition of the matrices
\begin{align} \label{def-t}
t_{I\Ae}{}^\Be=&\frac 12f^Y{}_{i\Ae}\covder_Yk_I{}^X f^{i\Be}{}_X,\qquad
f^Y{}_{i\Ae}\covder_Yk_I{}^X f^{j\Be}{}_X= \delta _i^j t_{I\Ae}{}^\Be \,.
\end{align}
These matrices satisfy a reality and an almost covariant constancy equation\footnote{Note that we defined $t_I{}^\bAe{}_\bBe$ using the common NW--SE convention for raising and lowering indices, and that the equation implies in this sense that $t_I$ is imaginary.}
\begin{equation}
  \left(t_{I\Ae}{}^\Be\right)^*=-\rho ^{\bAe\Ce}t_{I\Ce}{}^{\De}\rho _{\De\bBe}=-t_I{}^\bAe{}_\bBe\,,\qquad
  \nabla _X  t_{I\Ae}{}^\Be= k_I^Y R_{XYA}{}^B\,,
     \label{tIcc}
\end{equation}
as well as the commutation relations
\begin{equation}
[t_I,t_J]_\Be{}^\Ae=f_{IJ}{}^Kt_{K\Be}{}^\Ae-k^X_Ik^Y_J\,R_{XY\Be}{}^\Ae\,,
\end{equation}
which are consistent with (\ref{commutcovdelta}).

The transformation of the fermions under the gauge group follows from the requirement that the commutator of supersymmetry and Killing symmetries vanishes. It is given by the above-defined matrices:
\begin{equation}
 \widehat{\delta}_G(\theta )  \zeta^\Ae =  \theta^I t_{I\Be}{}^\Ae(q) \zeta^\Be\,.
 \label{delGferm}
\end{equation}

For the coupling of the hypermultiplet to the vector gauge multiplets in the
presence of the superconformal algebra, these isometries should be consistent with the conformal structure. The requirement that dilatations commute with the isometries is the equation
\begin{equation}
  0=k_{\rm D}{}^Y\partial  _{Y}k_I{}^X-k_I{}^Y\partial  _{Y}k_{\rm D}{}^X=
  k_{\rm D}{}^Y\nabla  _{Y}k_I{}^X- \frac{D-2}{2}k_I{}^X\,.
\label{WeylweightkI}
\end{equation}
This implies that the dilatations also commute with the $\SU(2)$ transformations
generated by $\vec{k}^{X}$, defined in (\ref{delDSU2q}). This equation can also be written as
\begin{equation}
  A^{i\Be} t_{I\Be}{}^\Ae= \frac{D-2}{2}f^{i\Ae}{}_X k_I{}^X\,.
 \label{Atiskf}
\end{equation}

One can also obtain the covariant transformation of $A^{i\Ae}$ (as for the other transformations in (\ref{widehatdelA})), using (\ref{commDkJ}) and (\ref{WeylweightkI})
\begin{equation}
  \widehat{\delta}_G(\theta ) A^{i\Ae}=A^{i\Be}\theta^I t_{I\Be}{}^\Ae\,.
 \label{delGAiA}
\end{equation}

\subsubsection{Non-closure Relations in $D=5$}
We now have all the ingredients to understand the case when the isometry with index $I$ is coupled to the
gauge symmetry of the vector multiplet (label by index $I$)---see Sect.~\ref{vecmult5d}. The full form
of~(\ref{QhyperD5}) is now
\begin{eqnarray}
\delta_Q (\epsilon) q^X &=&-\rmi\bar \epsilon ^i\zeta ^\Ae f^X{}_{i\Ae}\,,\nonumber\\
\widehat{\delta}_Q (\epsilon) \zeta^\Ae &=& \ft12 \rmi \slashed{D} q^X
f^{iA}{}_X\epsilon_i+\ft12 \sigma^I k_I{}^X f^{i\Ae}{}_X\epsilon _i\,,
 \label{delQzetagaugeConf}
\end{eqnarray}
with covariant derivatives defined as follows
\begin{align}
 D_\mu q^X &=\partial_\mu q^X-W_\mu^I k_I{}^X\,,\nonumber\\
 \nabla_\mu\zeta^\Ae&\equiv 
 \partial _\mu\zeta^\Ae + (\partial_\mu  q^X)
 \zeta^\Be\omega_{X\Be} {}^\Ae-W_\mu^I \zeta^\Be t_{I\Be}{}^\Ae.
 \label{CovDgauged5d}
\end{align}

Due to the gaugings, there are extra terms in the supersymmetry transformation of the fermions and the non-closure functions \eqref{eqmozeta} and (\ref{covscal}) are now modified to \cite{Bergshoeff:2002qk}
\begin{align}
\rmi\Gamma^\Ae =&  \slashed{\nabla} \zeta^\Ae + \frac 12 W_{\Be\Ce\De} {}^\Ae
 \zeta^\Be\bar{\zeta}^\Ce \zeta^\De-\rmi k_I{}^X f^{i\Ae}{}_{X}\psi_i{}^{I}+\rmi\zeta^\Be \sigma^I t_{I \Be}{}^{\Ae}\,,\nonumber\\
 \Delta^X=&\Box q^X -\frac 12 \bar{\zeta}^\Ae \gamma_a \zeta^\Be D^a q^Y W_Y{}^X{}_{\Ae\Be}
 -\frac{1}{4}f^X{}_{i\Ae}\varepsilon ^{ij}f^{Y}{}_{j\Ee}\covder_Y W_{\Be\Ce\De}{}^\Ae \bar{\zeta}^\Ee \zeta^\De\bar{\zeta}^\Ce \zeta^\Be\nonumber\\
&- k_I^Y \vec{J}_Y{}^X {}\cdot \vec{Y}^{I}+ \sigma^I \sigma^J k_J^Y \nabla _Y k_I^X\nonumber\\
&+ 2\rmi \bar{\psi}^{iI} \zeta^B t_{I\Be}{}^\Ae f^X{}_{i\Ae}
-\ft12\sigma ^I k_I^Y W_Y{}^X{}_{\Ae\Be}\bar{\zeta}^\Ae\zeta^\Be
\,,
\label{eqmozeta5d}
\end{align}
where $\Box q^X$ is now also covariant for gauge transformations:
\begin{equation}
\Box q^X=\partial_a D^a q^X- D_a q^Y \nabla _Y k_I^X W^{aI}+D_a q^Y D^a q^Z \Gamma_{YZ}^X\,,
\end{equation}
and we introduced the notation, using (\ref{orthogJL}),
\begin{equation}
  W_X{}^Y{}_{\Ae\Be}= L_X{}^Y{}_{\Ce}{}^\De W_{\Ae\Be\Ce}{}^\De\,.
 \label{defWXY}
\end{equation}

\subsubsection{Rigid Superconformal Case in $D=4$} \label{hyperD4conf}

To formulate the results in 4 dimensions, we consider the same bosonic
fields $q^X$. The fermionic formulae have to be translated using the rules explained in Appendix \ref{app:spinors5to4}. This leads again to $2n_H$ spinors, whose left-handed part is $\zeta^\Ae $, with $\Ae =1,\ldots, 2n_H$ and the left-handed ones
($C$-conjugates of the former) are $\zeta _\bAe$. Thus, in absence of an $\SU(2)$ index on
these spinors, the chirality is indicated by the fact that it has the
index $\Ae$ up or down. One can start again by allowing arbitrary
transformations for the scalars, and transformations of the spinors to
derivatives of the scalars and deduce again the conditions on quantities
that appear in these transformations. We would arrive again at
(\ref{ff1}) and (\ref{Df0}). But as we have already done all the work for
$D=5$ (for which in fact the formalism is easier) we can also translate
the results from what we already know.

This leads in 4 dimensions to the transformations \cite[(20.33)]{Freedman:2012zz}
\begin{eqnarray}
  \delta_Q(\epsilon ) q^X & = & -\rmi f^X{}_{i\Ae}\bar \epsilon ^i\zeta ^\Ae+\rmi f^{Xi\bAe}\bar \epsilon _i\zeta _\bAe\,,\nonumber\\
 \widehat{\delta}_Q(\epsilon ) \zeta ^\Ae  & = & \ft12\rmi f^{i\Ae}{}_X\slashed{D}q^X\epsilon _i+\rmi\bar X^Ik_I{}^X f^{i\Ae}{}_X\varepsilon _{ij}\epsilon ^j\,,
  \nonumber\\
 \delta_Q(\epsilon )\zeta _\bAe  & = & -\ft12\rmi f_{i\bAe X}\slashed{D }q^X\epsilon ^i-\rmi X^Ik_I{}^X f_{i\Ae X}\varepsilon ^{ij}\epsilon _j\,,
 \label{deltahyper}
\end{eqnarray}
where the complex conjugates of the frame fields are denoted as $f^{Xi\bAe}= (f^X{}_{i\Ae})^*$ and $f_{i\bAe X}=(f^{i\Ae}{}_X)^*$, see e.g. (\ref{fC}). $D_\mu q^X$ is given in (\ref{CovDgauged5d}).

The non-closure of the supersymmetries on the fermions is obtained in Appendix~\ref{app:spinors5to4} as an example of the translation rules from $D=5$ to $D=4$. The result is
\begin{eqnarray}
  \label{commzeta2}
\left[\delta_Q(\epsilon_1),\delta_Q(\epsilon_2)\right]
\zeta^\Ae&=&\xi^{\mu}\partial_{\mu}\zeta^\Ae - \ft12
\varepsilon^{ij}\rho^{\bBe\Ae}\Gamma_\bBe \bar\epsilon_{1i}\epsilon_{2j} - \ft12
\gamma_\mu \bar\epsilon_{[1}^i \gamma^\mu\epsilon_{2]i}\Gamma^\Ae\,,\nonumber\\
\end{eqnarray}
with $\xi ^\mu $ as in (\ref{xiaeps12}). The  non-closure functions are
\begin{eqnarray}
\Gamma^\Ae&\equiv&-\slashed{\nabla}
\zeta^\Ae + \ft12 W_{\Be\Ce}{}^{\bDe\Ae} \zeta_\bDe \bar{\zeta}^\Be\zeta^\Ce+2\bar X^I t_{I}{}^{\bBe\Ae}\zeta _{\bBe}+\rmi f^{i\Ae}{}_X k_I{}^X\varepsilon _{ij}\Omega ^{Ij}\,,\nonumber\\
\Gamma_\bAe&\equiv&-\slashed{\nabla}\zeta_\bAe+\ft12 W^{\bBe\bCe}{}_{\De\bAe}
\bar{\zeta}_\bBe\zeta_\bCe\zeta^\De+2 X^I t_{I\Be\bAe}\zeta ^\Be+\rmi k_I{}^Xf^{i\Be}{}_X\Omega ^I_i\rho _{\Be\bAe}\,,\nonumber\\
\label{bewegingfermion1}
\end{eqnarray}
where $W_{\Be\Ce}{}^{\bDe\Ae}=\rho ^{\bDe\Ee}W_{\Be\Ce\Ee}{}^\Ae$ and $W^{\bBe\bCe}{}_{\De\bAe}$ is
its complex conjugate. We will raise or lower indices changing the holomorphicity with the tensors $\rho _{\Ae\bBe}$ in NE--SW convention, e.g.
\begin{equation}
  t_I{}^{\bAe\Be}= \rho ^{\bAe\Ae}t_{I\Ae}{}^{\Be}\,,\qquad t_{I\Ae\bBe}= t_{I\Ae}{}^{\Be}\rho _{\Be\bBe}=\left(t_I{}^{\bAe\Be}\right)^*\,.
 \label{tIupup}
\end{equation}
These fermionic non-closure functions transform in real bosonic quantities $\Delta ^X$ as in~(\ref{delQGamma3})\footnote{Note that we use here the translation between $Y^{ij}$ and $\vec{Y}$ from (\ref{Yijupcc}), which will be used a lot further on.}:
\begin{align}
  \delta (\epsilon )\Gamma^\Ae=& -\ft12\rmi f^{i\Ae}{}_X\epsilon _i\Delta ^X\,,\nonumber\\
  \Delta ^X=& \Box q^ X +2(X^I\bar X^J+X^J\bar X^I)k_I^Y\nabla _Yk_J{}^X - 2 k_I{}^Y \vec{J}_Y{}^X \cdot \vec{Y}^I\nonumber\\
  &+ X^Ik_I{}^Y W_Y{}^X_{AB}\bar \zeta ^\Ae\zeta ^\Be
  +\bar X^I k_I{}^Y W_Y{}^{X\bAe\bBe}\bar \zeta_\bAe\zeta_\bAe+ \bar \zeta ^A \gamma _a \zeta _\bBe D_aq^YW_Y{}^{X}{}_\Ae{}^\bBe\nonumber\\
  &+\ft12 f^X{}_{i\Ae}\varepsilon ^{ij}f^Y{}_{j\Be}\nabla _Y W^{\bDe\bEe}{}_\Ce{}^A\bar \zeta ^\Be\zeta ^\Ce\bar \zeta _\bDe\zeta _\bEe\nonumber\\
  & -2\rmi f^X{}_{i\Ae}\bar \Omega ^{Ii}\zeta_\bBe t_I{}^{\bBe\Ae}+2\rmi f^{Xi\bAe}\bar \Omega^I _i\zeta ^\Be t_{I\Be\bAe}\,.
 \label{delQGamma}
\end{align}

Finally, for the remaining $\U(1)$ factor in the $R$-symmetry group we find
  \begin{eqnarray}
\widehat{\delta}_{\U(1)}q^X &=& 0,\nonumber \\
\widehat{\delta}_{\U(1)}\zeta^\Ae &=& \ft12 \rmi \lambda _T\zeta^\Ae.
\end{eqnarray}

\subsection{Hypermultiplets in Superconformal Gravity.}
\label{ss:hypermultipletConf}
The previous results (Sect. \ref{ss:hypermultiplets}) for rigid hypermultiplets can be generalized to local superconformal invariant theories by properly `covariantizing' the previous expressions with respect to the superconformal algebra.

\subsubsection{Case $D=5$}
The supersymmetry rules for the hypermultiplet coupled to the $D=5$ standard Weyl multiplet and the gauge symmetry of the vector multiplet were found to be \cite{Bergshoeff:2002qk}\footnote{A few changes of notation can be found in  (\ref{translationsoldD5}).}
\begin{eqnarray}
\delta q^X &=&-\rmi\bar\epsilon^i \zeta^\Ae f^X{}_{i\Ae}\,,\nonumber\\
\widehat{\delta} \zeta^\Ae &=& \ft12 \,{\rmi} \slashed{\cal D} q^X f^{i\Ae}{}_{X} \epsilon_i
+ \ft12 \sigma^I k_I{}^X f^{i\Ae}{}_{X} \epsilon_i -  A^{i\Ae} \eta_i\,.
 \label{delQzetagaugeLocalConf}
\end{eqnarray}
The new ingredients with respect to \eqref{delQzetagaugeConf} are the `matter terms' of the Weyl multiplets and the $S$-supersymmetry.
These transformations and the conformal and $R$-symmetry transformations determine the superconformal covariant derivatives
\begin{eqnarray}
{\mathcal D}_\mu  q^X
&=& D_\mu  q^X + {\rmi} \bar{\psi}_\mu^i \zeta^\Ae f^X{}_{i\Ae}, \nonumber\\
D_{\mu} q^X &=& \partial_\mu  q^X - b_\mu  k^X_{\rm D} -2 \vec{V}_\mu \cdot \vec{k}^X -
 W_{\mu}^I k_I{}^X , \nonumber\\
\widehat{{\mathcal D}}_\mu  \zeta^\Ae &=&\widehat{D}_\mu \zeta^\Ae -
\ft12 \,{\rmi} \slashed{D} q^X  f^{i\Ae}{}_{X} \psi_{\mu i} - \ft13 {\gamma}\cdot
T k_{\rm D}{}^X  f^{i\Ae}{}_{X} \psi_{\mu i} -  \ft12 \sigma^I k_I^X  f^{i\Ae}{}_{X} \psi_{\mu i}\nonumber\\
&&+ A^{i\Ae} \phi_{\mu i}\,, \label{hypCovD5}\\
\widehat{D}_\mu \zeta^\Ae &=& \partial _\mu  \zeta^\Ae  + \ft14 {\omega}_\mu {}^{bc} {\gamma}_{bc}
\zeta^\Ae - 2 b_\mu  \zeta^\Ae -  W_\mu^I \zeta^\Be t_{I\Be}{}^\Ae +\partial _\mu q^X \omega _{X\Be }{}^\Ae \zeta ^\Be \,.
\nonumber
\end{eqnarray}
The equations of motion for $\zeta^\Ae$ can be obtained by imposing the closure of the superconformal algebra
\begin{align}
\rmi\Gamma^\Ae \equiv & \slashed{\cal D} \zeta^\Ae + \frac 12 W_{\Be\Ce\De} {}^\Ae
 \zeta^\Be\bar{\zeta}^\Ce \zeta^\De+2\rmi \gamma^{ab} T_{ab}\zeta^\Ae\nonumber\\
 &-\rmi k_I{}^X f^{i\Ae}{}_{ X}\psi_i^{ I}+\rmi\zeta^\Be \sigma^I t_{I \Be}{}^{\Ae}+\ft83 \rmi k_{\rm D}{}^X f^{i\Ae}{}_{X}\chi_i\,.
 \label{eqmozeta5dlod}
\end{align}

\subsubsection{Case $D=4$}

The covariant supersymmetry transformations are those from (\ref{deltahyper}) with only a replacement of $D_\mu $ by the fully covariant ${\cal D}_\mu $, which are
\begin{eqnarray}\label{hypCovD4}
 {\cal D}_\mu q^X &= & D_\mu q^X+\rmi \bar{\psi }_\mu ^i\zeta^\Ae f^X{}_{i\Ae }
 -\rmi \varepsilon ^{ij}\rho ^{\bAe \Be } \bar{\psi }_{\mu i}\zeta_\bAe f^X{}_{j\Be }\,, \nonumber\\
D_\mu q^X&=& \partial _\mu q^X -b_\mu k_{\rm D}{}^X+2\vec{V}_\mu\cdot \vec{k}^{X}-
 W_\mu {}^Ik_I{}^X\,, \label{calDhyper}\\
\widehat{{\cal D}}_\mu \zeta ^\Ae  & = & \hat{D}_\mu \zeta ^\Ae
-\ft12\rmi f^{i\Ae }{}_X \slashed{\cal D}q^X
\psi _{\mu i}-\rmi \bar
X^Ik_I{}^Xf^{i\Ae }{}_X\varepsilon _{ij}\psi _\mu  ^j-\rmi A^{i\Ae }\phi _{\mu i}\,,\nonumber\\
\hat{D}_\mu \zeta ^\Ae &=& \left( \partial _\mu+\ft14\omega _\mu
{}^{ab}\gamma _{ab} -\ft32b_\mu +\ft12\rmi A_\mu\right) \zeta ^\Ae
-W_\mu ^It_{I\Be }{}^\Ae \zeta ^\Be +\partial _\mu q^X \omega _{X\Be }{}^\Ae \zeta ^\Be \,.
\nonumber
\end{eqnarray}
Note that the hatted covariant derivatives are covariant for target space transformations as well and that $\partial_\mu q^X$ in the last term should not be covariantized to obtain this covariant expression $\widehat{D}_\mu \zeta^{\Ae}$. Because of central-charge like terms, the algebra does not close on the spinors. The new non-closure functions $\Gamma ^A$ will be used to derive the action for the hypermultiplet, as we will explain in Sect.~\ref{finsec_hypers}.

In terms of $A^{i\Ae}$ (\ref{defAialpha}), the covariant transformations are
\begin{align}
  \widehat{\delta} A^{i\Ae} & \equiv \delta A^{i\Ae} + A^{i\Be}\omega _{X\Be}{}^\Ae\delta q^X= -\rmi\bar \epsilon ^i \zeta ^\Ae +\rmi\bar \epsilon_j \zeta _{\bBe}\,\varepsilon ^{ji}\rho ^{\bBe\Ae}\,, \nonumber\\
  \widehat{\delta}\zeta ^\Ae &= \ft12\rmi \widehat{\slashed{\cal D}}A^{i\Ae } \epsilon _i+\rmi \bar X^Ik_I{}^Xf^{i\Ae }{}_X\varepsilon _{ij}\epsilon   ^j+\rmi A^{i\Ae }\eta _i\,,
\label{covtranhyperD4A}
\end{align}
where we used
\begin{align}
  \widehat{{\cal D}}_\mu A^{i\Ae}
  =& f^{i\Ae}{}_X\nabla _Yk_{\rm D}{}^X {\cal D}_\mu q^Y= f^{i\Ae}{}_X {\cal D}_\mu q^X\nonumber\\
   =& f^{i\Ae}{}_X\partial _\mu q^X -b_\mu A^{i\Ae}-A^{j\Ae}V_{\mu j}{}^i -W_\mu ^IA^{i\Be} t_{I\Be}{}^\Ae
\nonumber\\
&  +\rmi\bar \psi _\mu ^i \zeta ^\Ae -\rmi\bar \psi _{\mu j}\zeta _{\bBe}\,\varepsilon ^{ji}\rho ^{\bBe\Ae}\,.
 \label{wideDA}
\end{align}
Note that the $\widehat{\delta}$ used in (\ref{covtranhyperD4A}) has no $\SU(2)$ connection, similar as in (\ref{Df0}).
\subsection{Tensor Multiplet in \texorpdfstring{$D=4$}{D=4} Local Superconformal Case.}
\label{ss:tensormultiplet}
The tensor multiplet in $D=4$ dimensions was obtained in
\cite{deWit:1982na}. It is in fact the multiplet of the constraints (\ref{bian4}). We can name these constraints, respectively, as $L_{ij}$, $\varphi^{i}$, $E_b$ (satisfying a differential constraint) and $G$. These transform in each other and thus form a multiplet. It starts from an $\SU(2)$ triplet $L_{ij}$ (hence satisfying the reality property as in (\ref{Yijupcc})). The constrained $E_a$ implies that the multiplet has a gauge tensor $E_{\mu \nu }$ (3 degrees of freedom) and a complex auxiliary $G$, to balance the 8 fermionic degrees of freedom in $\varphi _i$.
The transformation rules in the background of conformal supergravity are\footnote{Of course the tensor multiplet for rigid supersymmetry can be obtained from \eqref{eq:tensor-tr} by setting to zero the
fields of the Weyl multiplet ($T$ and $\psi _\mu $) and replacing the covariant derivatives by ordinary derivatives.}
\begin{equation}
  \label{eq:tensor-tr}
  \begin{split}
  \delta L_{ij} =& \bar\epsilon_{(i}\varphi_{j)} +
  \,\varepsilon_{ik}\varepsilon_{j\ell}\,
  \bar\epsilon^{(k}\varphi^{\ell)} +2\lambda_{\rm D}L_{ij} \,,\\
  \delta\varphi^{i} =& \,\ft12\slashed{\cal D} L^{ij} \,\epsilon_j +\ft12
  \varepsilon^{ij}\,\slashed{E}\,\epsilon_j - \ft12 G \,\epsilon^i
  + 2 L^{ij}\, \eta_j +\left(\ft52 \lambda _{\rm D}+\ft12\rmi \lambda_T\right)\varphi^{i}\,,\\
  \delta G =& -  \bar\epsilon_i \slashed{\cal D}  \varphi^{i}
  -  3 \bar\epsilon_i  L^{ij} \, \chi_j + \ft18\bar\epsilon_i
    \gamma^{ab}  T^+_{ab}   \varphi_j
    \varepsilon^{ij} + 2  \bar{\eta}_i\varphi^{i}+\left(3\lambda _{\rm D}-\rmi\lambda_T\right)G
    \, ,\\
  \delta E_{\mu\nu} =& \ft14 \rmi\bar\epsilon^i\gamma_{\mu\nu}
  \varphi^{j} \,\varepsilon_{ij} -\ft14 \rmi\bar\epsilon_i\gamma_{\mu\nu}
  \varphi_{j} \,\varepsilon^{ij} + \ft12\rmi L_{ij}
  \varepsilon^{jk}  \bar{\epsilon}^i \gamma_{[\mu} \psi_{\nu ]k}
  - \ft12 \rmi  L^{ij}  \varepsilon_{jk}  \bar{\epsilon}_i
    \gamma_{[\mu} \psi_{\nu ]}{}^k \, .
\end{split}
\end{equation}
where
\begin{equation}
  E^\mu= e^{-1}\varepsilon ^{\mu \nu \rho \sigma }\partial _\nu E_{\rho \sigma }-\ft12\left(\bar \psi _\nu ^i\gamma ^{\mu \nu }\varphi ^j\varepsilon _{ij} +\hc\right)
  -\ft12\rmi e^{-1}\varepsilon ^{\mu \nu \rho \sigma }L_{ij}\varepsilon ^{jk}\bar \psi _\nu ^i\gamma _\rho \psi _{\sigma k}\,.
 \label{defEmu}
\end{equation}

A first step in building actions from this multiplet has been set in
\cite{deWit:1982na}, but more applications can be found in
\cite{deWit:2006gn}.

\section{Construction of the Superconformal Actions}
\label{ss:actions}
This section is devoted to the construction of local superconformally invariant actions for the vector and the hypermultiplet. As we have shown in the example of Sect.~\ref{ss:exampleBosComp}, later one
gauge-fixes the extra symmetry such that the remaining theory has just the super-Poincar\'{e}
invariance. Crucially, as we will explain in Chap.~\ref{ss:gaugefix}, for these last steps one needs to include compensating multiplets. Besides interacting matter, the resulting action from the gauge fixing will contain also the pure gravity sector.

\subsection{Action for Vector Multiplets in \texorpdfstring{$D=4$}{D=4}}
 \label{ss:actvector4}

Let us consider the basic
supergravity multiplet coupled to $n$ vector multiplets. The physical content
that one should have (from representation theory of the super-Poincar\'{e}
group) can be represented in terms of particles with spin as follows:
\begin{equation}
\begin{array}{ccccccccc}
\multicolumn{3}{c}{\mbox{SUGRA}}&\qquad&&\multicolumn{3}{c}{\mbox{vector multiplet}}&\\
&2&                      &      &&&&&\\
\ft32&&\ft32             &      &&&&&\\
    &1&                  &      &&&1&&\rightarrow n+1\\
    & &                  &      &+ n\ *\quad&\ft12 & & \ft12 &\\
    & &                  &      &           &   0  & &  0&
\end{array}\label{schemeN2sugraVM}
\end{equation}
The supergravity sector contains the graviton, 2 gravitini and a
so-called graviphoton, \index{graviphoton} $W_\mu$ (that is a spin-1 field). When coupled to $n$ vector multiplets, $W_\mu$ gets part of a set of
$n+1$ vectors, which will be uniformly described by the special K\"{a}hler
geometry. The scalars inside these vector multiplets appear as $n$ complex fields $z^\alpha$, with
$\alpha=1,\ldots , n$.
\vspace{10pt}

In the framework of superconformal calculus, we consider $n+1$ superconformal vector
multiplets with scalars $X^I$ ($I=0,\ldots ,n$) in the background of the Weyl multiplet (main formulae can be found in Sect.~\ref{vectmult}). One of these
multiplets should contain the graviphoton, while we will use the missing fermions and scalars to fix
superfluous gauge symmetries of the superconformal algebra.

Exploiting the fact that vector multiplets are constrained chiral
multiplets (Sect. \ref{ss:chiralmult}), we can build an action for the vector multiplet from an action for a chiral multiplet. The lowest component of the chiral multiplet should be $A=\ft12\rmi F(X)$, \footnote{The overall
normalization is for later convenience to get a result with the
normalization that is most used in the literature} being then $A$ a new chiral superfield, given by an
arbitrary holomorphic function of the scalars in vector multiplets. This function $F(X)$ will determine the action, and is called the prepotential. \index{prepotential}
The further components are then defined by the transformation laws, which give,
comparing with (\ref{chiraltrA}), $\Psi_i= \ft12\rmi F_I \Omega^I_i$,
where we defined
\begin{eqnarray}
F_I(X) &=&\frac{\partial}{\partial X^I} F(X)\,,\qquad \bar F_I(\bar
X)\,=\,\frac{\partial}
{\partial \bar X^I}\bar F(\bar X)\,, \nonumber\\
F_{IJ}&=& \frac{\partial}{\partial X^I} \frac{\partial}{\partial X^J}
F(X)\qquad \ldots\,.
\end{eqnarray}
Calculating the transformation of $\Psi_i$ one finds $B_{ij}$, $G^-_{ab}$, \ldots
\begin{eqnarray}
A&=&\ft12\rmi F\nonumber\\
\Psi_i&=&\ft12\rmi F_I\Omega_i^I\nonumber\\
B_{ij}&=&\ft12\rmi F_I Y_{ij}^I-\ft18\rmi
F_{IJ}\bar{\Omega}_i^I\Omega_j^J\nonumber\\
G^-_{ab}&=&\ft12\rmi F_I \mathcal{F}^{- I}_{ab}-\ft{1}{16}\rmi
F_{IJ}\bar{\Omega}_i^I\gamma_{ab}\Omega_j^J\varepsilon^{ij}\nonumber\\
\Lambda_i&=&-\ft12\rmi F_I \slashed{D}
\Omega^{jI}\varepsilon_{ij}-\ft12\rmi F_I f^I_{JK}\bar{X}^J\Omega_i^K
-\ft18\rmi F_{IJ} \gamma^{ab}
\mathcal{F}^{-I}_{ab}\Omega_i^J\nonumber\\&&-\ft12\rmi
F_{IJ}\Omega_k^JY_{ij}^I\varepsilon^{jk}+\ft{1}{96}\rmi
F_{IJK}\gamma^{ab}\Omega_i^I
\bar{\Omega}_j^J\gamma_{ab}\Omega_k^K\varepsilon^{jk}\nonumber\\
C&=&-\rmi F_I D_aD^a \bar{X}^I - \ft14\rmi F_I
\mathcal{F}^{+I}_{ab}T^{+ab} -\ft32\rmi F_I
\bar{\chi}_i\Omega^{iI}+\ft12\rmi F_I
f_{JK}^I\bar{\Omega}^{iJ}\Omega^{jK}
\varepsilon_{ij}\nonumber\\
&&-\rmi  F_I f_{JK}^I f_{LM}^J \bar{X}^K \bar{X}^L
X^M-\ft12\rmi F_{IJ}Y^{ijI}Y_{ij}^J+\ft14\rmi F_{IJ}\mathcal{F}^{-I}_{ab}\mathcal{F}^{-abJ}\nonumber\\
&&+\ft12\rmi F_{IJ}\bar{\Omega}_i^{I} \slashed{D} \Omega^{iJ}-\ft12\rmi
F_{IJ} f^I_{KL}\bar{X}^K \bar{\Omega}_i^{J}\Omega_j^{L}\varepsilon^{ij}
+\ft14\rmi F_{IJK}Y^{ijI}\bar{\Omega}_i^{J}\Omega_j^{K}\nonumber\\&&-
\ft{1}{16}\rmi
F_{IJK}\varepsilon^{ij}\bar{\Omega}_i^{I}\gamma^{ab}\mathcal{F}^{-J}_{ab}\Omega_j^{K}+\ft{1}{48}\rmi
F_{IJKL}\bar{\Omega}^{I}_i\Omega^{J}_l
\bar{\Omega}^{K}_j\Omega^{L}_k\varepsilon^{ij}\varepsilon^{kl}.
\label{ChiralFromF}
\end{eqnarray}

This is the composite chiral multiplet that we discussed at the end of Sect. \ref{ss:chiralmult}, and on which we can apply the `density formula' (\ref{densityChirald4})
As mentioned, $F(X)$ must be homogeneous of weight~2, where the $X$ fields carry weight~1. This
implies the following relations for the derivatives of $F$:
\begin{equation}
 2F= F_IX^I\,,\qquad F_{IJ}X^J=F_I\,,\qquad F_{IJK}X^K=0.
 \label{homegeneqs}
\end{equation}
Inserting  (\ref{ChiralFromF}) in (\ref{densityChirald4}) leads to
\begin{eqnarray}
 \nonumber e^{-1}\mathcal{L}_g&=&-\rmi F_I D_aD^a \bar{X}^I
 +\ft14\rmi F_{IJ}\mathcal{F}^{-I}_{ab}\mathcal{F}^{-abJ}
  +\ft12\rmi F_{IJ}\bar{\Omega}_i^{I} \slashed{D} \Omega^{iJ}
  \nonumber\\
 &&
 -\ft12\rmi F_{IJ}Y^{ijI}Y_{ij}^J
 +\ft14\rmi F_{IJK}Y^{ijI}\bar{\Omega}_i^{J}\Omega_j^{K}
 \nonumber\\
&&
 -\ft1{16}\rmi F_{IJK}\varepsilon^{ij}\bar{\Omega}_i^{I}\gamma^{ab}\mathcal{F}^{-J}_{ab}\Omega_j^{K}
 +\ft1{48}\rmi F_{IJKL}\bar{\Omega}^{I}_i\Omega^{J}_\ell\bar{\Omega}^{K}_j\Omega^{L}_k\varepsilon^{ij}\varepsilon^{k\ell}
 \nonumber\\
&&
 +\ft12\rmi  F_I f_{JK}^I\bar{\Omega}^{iJ}\Omega^{jK} \varepsilon_{ij}
 -\ft12\rmi  F_{IJ} f^I_{KL}\bar{X}^K \bar{\Omega}_i^{J}\Omega_j^{L}\varepsilon^{ij} \nonumber\\
&& -\rmi  F_I f_{JK}^I f_{LM}^J \bar{X}^K \bar{X}^L X^M
 \nonumber\\
&&
 -\ft14\rmi F_I \mathcal{F}^{+I}_{ab}T^{+ab}
  -\ft32\rmi F_I \bar{\chi}_i\Omega^{iI}
 -\ft12\rmi F_{IJ}\bar{\psi}_i\cdot \gamma \Omega_j^I Y^{ijJ}
  \nonumber\\
&&
 +\ft12\rmi F_I f_{JK}^I\bar{X}^J \bar{\psi}_i\cdot\gamma\Omega_j^K\varepsilon^{ij}
 -\ft12\rmi F_I \bar{\psi}_i \cdot \gamma \slashed{D} \Omega^{iI}\nonumber\\
&& +\ft18\rmi F_{IJ} \mathcal{F}_{ab}^{-I}\bar{\psi}_i\cdot\gamma \gamma^{ab}\Omega_j^J \varepsilon^{ij}
 \nonumber\\
&&
 +\ft1{12}\rmi F_{IJK}\bar{\Omega}_\ell^J\Omega_j^K \bar{\psi}_i\cdot
\gamma \Omega_k^I \varepsilon^{ij}\varepsilon^{k\ell}
 +\ft1{16}\rmi F_I
\bar{\psi}_{\mu i}\gamma \cdot T^+\gamma^{\mu} \Omega_j^I
\varepsilon^{ij} \nonumber\\
&& -\ft{1}{8}\rmi F T^+_{ab}T^{+ab}
 \nonumber\\
&&
 -\ft12\rmi F_I \bar{\psi}_{\mu i}\gamma^{\mu \nu}\psi_{\nu j} Y^{ijI}
 -\ft12\rmi F T^{+ \mu\nu}\bar{\psi}_{\mu i}\psi_{\nu j}\varepsilon^{ij}
 +\ft12\rmi F_I\mathcal{F}^{-\mu \nu I}\bar{\psi}_{\mu i}\psi_{\nu j}\varepsilon^{ij}
 \nonumber\\
&&
 -\ft{1}{16}\rmi F_{IJ}\bar{\psi}_{\mu i}\psi_{\nu j}\bar{\Omega}_{k}^I\gamma^{\mu\nu}\Omega_\ell ^J\varepsilon^{ij}\varepsilon^{k\ell }
 +\ft18\rmi F_{IJ}\bar{\Omega}_k^I\Omega_l^J\bar{\psi}_{\mu i}\gamma^{\mu \nu}\psi_{\nu j}\varepsilon^{ik}\varepsilon^{jl}
 \nonumber\\
&&
 -\ft14\varepsilon^{ij}\varepsilon^{k\ell }e^{-1}\varepsilon^{\mu \nu \rho\sigma}\bar{\psi}_{\mu i}\psi_{\nu j}\bar{\psi}_{\rho k}\left(\gamma_{\sigma}F_I\Omega_\ell ^I+F\psi_{\sigma\ell}\right)
 \nonumber\\
  && + \hc
 \label{actVector1}
\end{eqnarray}
The first terms of the action (\ref{actVector1}) are kinetic terms for
the scalars $X$, the vectors, and the fermions $\Omega$. The following
term says that $Y_{ij}$ is an auxiliary field that can be eliminated by
its field equation. The first 5 lines are the ones that we would
encounter also in rigid supersymmetry, see \cite[(20.15)]{Freedman:2012zz}. For
these terms, the relations (\ref{homegeneqs}) have not been used, and
this part is thus the general result for rigid supersymmetry. The other
lines are due to the local superconformal symmetry. For those interested
in rigid symmetry, we repeat that in that case the covariant derivatives  (\ref{covD4N2part1})
reduce to e.g.
\begin{eqnarray}
 D_aX^I & = & \partial _a X^I-W_a^K  X^J f_{JK}{}^I, \nonumber\\
 D_a\Omega _i^I & = & \partial _a \Omega ^I-W_a^K  \Omega _i^J f_{JK}{}^I,\nonumber\\
 \mathcal{F}^{I}_{ab} & = & 2\partial _{[a}W_{b]}^I+W_b^K
 W_a^Jf_{JK}{}^I.
 \label{covderVectorm}
\end{eqnarray}
Note that the Lagrangian is a total derivative if $F(X)$ is a quadratic function of $X^I$ with real coefficients:
\begin{equation}
  F(X)= C_{IJ}X^I X^J\,,\qquad C_{IJ}\in\mathbb{R}\qquad \rightarrow \qquad S=\int \rmd^4 x\,{\cal L} _g =0\,.
 \label{FCS0}
\end{equation}

In deriving the above formulae, we assumed for simplicity that $F$ is a gauge-invariant function
such that $\mathcal{L}_g$ is invariant under gauge transformations.
However, the property (\ref{FCS0}) suggests that there is a more general situation~\cite{deWit:1985px,deWit:1987ph} in which $F$ transforms under the gauge transformations as
\begin{equation}
  \delta_G(\theta )  F\equiv F_I\theta ^K  X^J f_{JK}{}^I= - \theta ^I C_{I,JK}X^JX^K,
 \label{gdelFinC}
\end{equation}
where $C_{I,JK}$ are real constants. In fact, due to \eqref{FCS0}, the action is invariant for rigid transformations that satisfy  (\ref{gdelFinC}), transforming to
\begin{equation}
  \delta_G \mathcal{L}_g= \ft13 C_{I,JK}\varepsilon ^{\mu \nu \rho \sigma }\theta ^IF_{\mu \nu }^JF_{\rho \sigma }^K\,.
 \label{delGLg}
\end{equation}
In order to allow this extra possibility with local $\theta ^I$, one has to add to (\ref{actVector1}) a Chern--Simons term \index{Chern--Simons term}
\begin{equation}
  {\cal L}_{CS}=\ft23C_{I,JK}\varepsilon ^{\mu \nu \rho \sigma }W_\mu {}^IW_\nu {}^J\left(\partial _\rho W_\sigma {}^K+\ft38f_{LM}{}^KW_\rho {}^LW_\sigma {}^M \right)\,.
 \label{LCS}
\end{equation}
To prove the supersymmetry invariance of $\mathcal{L}_g+{\cal L}_{CS}$ one needs a few more relations that follow from (\ref{delGLg})
Replacing the arbitrary $\theta ^K$ by $X^K$ the variation vanishes, and thus for the consistency of (\ref{gdelFinC}) we should have
\begin{equation}
  C_{(I,JK)}X^IX^JX^K=0\,,
 \label{Cequationsymm}
\end{equation}
namely the completely symmetric part of $C_{I,JK}$ must vanish.

By taking two derivatives of (\ref{gdelFinC}) we obtain
\begin{equation}\label{defCijk}
C_{K,IJ}=f_{K(I}{}^L F_{J)L}-\ft12F_{IJL}X^Mf_{MK}{}^L=f_{K(I}{}^L \bar{F}_{J)L}-\ft12\bar F_{IJL}\bar X^Mf_{MK}{}^L\,.
\end{equation}
To prove the invariance of the sum of (\ref{actVector1}) and (\ref{LCS}) one needs an identity \cite{deWit:1985px}
\begin{equation}
  f_{KL}{}^M C_{M,IJ}= 2 f_{J[K}{}^M C_{L],IM}+2 f_{I[K}{}^MC_{L],JM}\,,
 \label{fCalgebra}
\end{equation}
which follows from the requirement that the gauge group closes on $F(X)$.
A simple example of the occurrence of a Chern--Simons term is given in \cite[(3.21)]{deWit:1985px}.

\subsubsection{Simplifications}

In order to get a more useful form of the action, one has to make the
conformal covariant derivatives explicit. The principle is explained for
the bosonic case in (\ref{bboxphi}). This leads here to
\begin{eqnarray}
 \bbox^C \bar{X}^I&=&\widehat{\partial}_{\mu}D^{\mu}\bar{X}^I-\omega_{\mu}^{\phantom{\mu} \mu \nu}D_{\nu}\bar{X}^I-\rmi W_{\mu}D^{\mu}\bar{X}^I
+ 2 f_{\mu}^{\phantom{\mu}\mu}\bar{X}^I-\ft{1}{2}\bar{\psi}_{\mu i}
D^{\mu}\Omega^{iI}\nonumber\\&&
+\ft{1}{32}\bar{\psi}^i_{\mu} \gamma^{\mu}
\gamma \cdot T^+\Omega^{jI}\varepsilon_{ij}
 -\ft{1}{2}\bar{\Omega}^{iI}\gamma\cdot \phi_i-\ft{3}{4}\bar{\psi}^i\cdot
\gamma\chi_i\bar{X}^I\nonumber\\ &&
-\ft{1}{2}\varepsilon^{ij}\bar{\psi}_i \cdot
\gamma\Omega_j^J\bar{X}^Kf^I_{JK}-\ft{1}{2}\varepsilon_{ij}\bar{\psi}^i\cdot
\gamma\Omega^{jJ}\bar{X}^Kf^I_{JK}.
 \label{bboxXI}
\end{eqnarray}
Hence the first term of (\ref{actVector1}) after adding a total derivative, is
\begin{align}
 -\rmi  F_I \square^C \bar{X}^I=&\rmi
 F_{IJ}\mathcal{D}_{\mu}X^I\left(\mathcal{D}^{\mu}\bar{X}^J-\ft12 \bar{\psi}_i^{\mu}\Omega^{iJ}\right) -2\rmi
F_If_{\mu}{}^\mu\bar{X}^I +\ft12\rmi  F_I \bar{\psi}_{\mu i} D^{\mu}\Omega^{iI}\nonumber\\
&-\ft{1}{32}\rmi  F_I \bar{\psi}_{\mu}^i\gamma^{\mu} \gamma
\cdot T^-\ft12\Omega^{jI}\varepsilon_{ij}+\ft12\rmi F_I
\bar{\Omega}^{iI}\gamma \cdot \phi_i+\ft34\rmi
F_I\bar{\psi}^i\cdot\gamma\chi_i\bar{X}^I \nonumber\\
&+\ft12\rmi
F_I\varepsilon^{ij}\bar{\psi}_i\cdot\gamma\Omega_j^J\bar{X}^Kf^I_{JK}
+\ft12\rmi
F_I\varepsilon_{ij}\bar{\psi}^i\cdot\gamma\Omega^{jJ}\bar{X}^Kf^I_{JK}
\nonumber\\
&+\rmi  F_I \bar{\psi}^i_{[\mu}\gamma^{\nu}\psi_{\nu]
i}\left(\mathcal{D}^{\mu}\bar{X}^I-\ft12\bar{\psi}^{\mu}_i\Omega^{iI}\right)\nonumber\\
&+\mbox{ total derivative.}
 \label{FboxX}
\end{align}
The other term that has to be written explicitly is the covariant
derivative of the fermions
\begin{align}
  \slashed{D} \Omega^{iJ}=&
\slashed{\mathcal{D}}\Omega^{iJ}  -\gamma^{\mu}\gamma^{\nu}\psi_{\mu}^i \left(\mathcal{D}_{\nu}\bar{X}^J-\ft12 \bar{\psi}_{\nu j}\Omega^{jJ}\right) \nonumber\\
&-\ft14\gamma_{\mu}\gamma \cdot
\mathcal{F}^{+J}\psi^{\mu}_j \varepsilon^{ij} -\gamma \cdot \psi_j \left(Y^{ijJ}
+\varepsilon^{ij}\bar{X}^KX^Lf^J_{KL}\right)
 -2\bar{X}^J\gamma \cdot \phi^i\,.
 \label{FermionderExplicit}
\end{align}

Deleting total derivatives, the action is at this point (adding also (\ref{LCS}))
\begin{align}
 e^{-1}\mathcal{L}_g=&
 \rmi F_{IJ}\mathcal{D}_{\mu}X^I\mathcal{D}^{\mu}\bar{X}^J
  -2\rmi F_If_{\mu}{}^\mu \bar{X}^I
 +\ft14\rmi F_{IJ}\mathcal{F}^{-I}_{ab}\mathcal{F}^{-abJ}
 -\ft{1}{8}\rmi F T^+_{ab}T^{+ab}
  \nonumber\\
&
 -\ft14\rmi F_I \mathcal{F}^{+I}_{ab}T^{+ab}
-\ft12\rmi F_{IJ}Y^{ijI}Y_{ij}^J
 -\rmi  F_I f_{JK}^I f_{LM}^J \bar{X}^K \bar{X}^L X^M+ e^{-1}{\cal L}_{CS}
  \nonumber\\
&
 +\rmi  F_I \bar{X}^I \bar{\psi}_{\mu i}\gamma^{\mu \nu}\phi^{i}_{\nu}
 +\ft12\rmi F_{IJ}\bar{\Omega}_i^{I}\slashed{\mathcal{D}}\Omega^{iJ}
 +\ft12\rmi  F_I \bar{\psi}_{\mu i}\gamma^{\mu \nu} \gamma^{\rho}
\mathcal{D}_{\rho} \bar{X}^I \psi_{\nu}^i
\nonumber\\
& +\ft12\rmi  F_I \bar{\Omega}^{iI}\gamma \cdot \phi_i
 \nonumber\\
 &
 +\ft18\rmi  F_I \bar{\psi}_{\mu i}\gamma^{\mu \nu} \gamma \cdot \mathcal{F}^{+I} \varepsilon^{ij}
\psi_{\nu j}
 -\ft1{32}\rmi  F_I \bar{\psi}_{\mu}^i\gamma^{\mu} \gamma \cdot T^+\Omega^{jI}\varepsilon_{ij}\nonumber\\
& +\ft34\rmi F_I\bar{\psi}^i\cdot\gamma\chi_i\bar{X}^I
 \nonumber\\
&
 +\ft12\rmi F_I\varepsilon^{ij}\bar{\psi}_i\cdot\gamma\Omega_j^J\bar{X}^Kf^I_{JK}
 +\ft12\rmi F_I\varepsilon_{ij}\bar{\psi}^i\cdot\gamma\Omega^{jJ}\bar{X}^Kf^I_{JK}
 \nonumber\\&
  +\rmi F_I \bar{\psi}^i_{[\mu}\gamma^{\nu}\psi_{\nu]i}\left(\mathcal{D}^{\mu}\bar{X}^I-\ft12
\bar{\psi}^{\mu}_i\Omega^{iI}\right)
 -\ft32\rmi F_I \bar{\chi}_i\Omega^{iI}\nonumber\\
& +\ft12\rmi  F_I f_{JK}^I\bar{\Omega}^{iJ}\Omega^{jK} \varepsilon_{ij}
 \nonumber\\
 &
 -\ft12\rmi F_{IJ}\mathcal{D}_{\mu}X^J \bar{\psi}_i^{\mu}\Omega^{iI}
  -\ft12 \rmi  F_I \bar{\psi}_{\mu i}\gamma^{\mu \nu} \mathcal{D}_{\nu}\Omega^{iI}
 \nonumber\\
&
 -\ft12\rmi F_{IJ}\bar{\Omega}_i^{I}\gamma^{\mu}\gamma^{\nu}\psi_{\mu}^i \mathcal{D}_{\nu}\bar{X}^J
 +\ft12\rmi F_I \mathcal{F}^{-\mu \nu I}\bar{\psi}_{\mu i}\psi_{\nu j}\varepsilon^{ij}
 \nonumber\\
&
 -\ft18\rmi F_{IJ}\bar{\Omega}_i^{I}\gamma_{\mu}\gamma \cdot \mathcal{F}^{+J}\psi^{\mu}_j \varepsilon^{ij}
 -\ft12\rmi F_{IJ}\bar{\Omega}_i^{I}\gamma \cdot \psi_j\varepsilon^{ij}\bar{X}^KX^Lf^J_{KL}
 \nonumber\\
 &
 -\rmi F_{IJ}\bar{X}^J\bar{\Omega}_i^{I}\gamma \cdot \phi^i
 -\ft12\rmi  F_{IJ} f^I_{KL}\bar{X}^K \bar{\Omega}_i^{J}\Omega_j^{L}\varepsilon^{ij}
 +\ft14\rmi F_{IJK}Y^{ijI}\bar{\Omega}_i^{J}\Omega_j^{K}
 \nonumber\\
&
 -\ft1{16}\rmi F_{IJK}\varepsilon^{ij}\bar{\Omega}_i^{I}\gamma^{ab}\mathcal{F}^{-J}_{ab}\Omega_j^{K}\nonumber\\
& -\ft12\rmi\varepsilon^{ij}\bar{\psi}_{\mu i}\left( F T^{+ \mu \nu}\psi_{\nu j}
 -\ft1{8}\gamma \cdot T^+\gamma^{\mu}F_I \Omega_j^I\right)
 \nonumber\\
&
 +\ft12\rmi  F_I f_{JK}^I\bar{X}^J \bar{\psi}_i\cdot \gamma \Omega_j^K \varepsilon^{ij}
 +\ft1{8}\rmi F_{IJ} \mathcal{F}_{ab}^{-I}\bar{\psi}_i\cdot \gamma \gamma^{ab}\Omega_j^J \varepsilon^{ij}
\nonumber\\
&
 +\ft{1}{12}\rmi F_{IJK}\bar{\Omega}_\ell ^J\Omega_j^K \bar{\psi}_i\cdot\gamma \Omega_k^I \varepsilon^{ij}\varepsilon^{k\ell }
 - \ft14 \rmi F_I \bar{\psi}_{\mu i}\gamma^{\mu \nu} \gamma^{\rho}\psi_{\nu}^i
\bar{\psi}_{\rho k}\Omega^{kI}
 \nonumber\\
&
 +\ft18\rmi F_{IJ}\bar{\Omega}_k^I\Omega_\ell ^J\bar{\psi}_{\mu i}\gamma^{\mu \nu}\psi_{\nu j}\varepsilon^{ik}\varepsilon^{j\ell }
  +\ft14\rmi F_{IJ}\bar{\Omega}_i^{I}\gamma^{\mu}\gamma^{\nu}\psi_{\mu}^i \bar{\psi}_{\nu j}\Omega^{jJ}
\nonumber\\
&
 +\ft1{48}\rmi F_{IJKL}\bar{\Omega}^{I}_i\Omega^{J}_\ell \bar{\Omega}^{K}_j\Omega^{L}_k\varepsilon^{ij}\varepsilon^{k\ell }
 -\ft1{16}\rmi F_{IJ}\bar{\psi}_{\mu i}\psi_{\nu j}\bar{\Omega}_{k}^I\gamma^{\mu \nu}\Omega_\ell ^J\varepsilon^{ij}\varepsilon^{k\ell }
 \nonumber\\
&
 -\ft14 \varepsilon^{ij}\varepsilon^{k\ell }e^{-1}\varepsilon^{\mu \nu \rho\sigma}\bar{\psi}_{\mu i}\psi_{\nu j}\bar{\psi}_{\rho k}\left(\gamma_{\sigma}F_I\Omega_\ell ^I+F\psi_{\sigma
\ell }\right)
 +\hc
 \label{actVMd4form2}
\end{align}
We use then (\ref{calFVM4}) and the values of the conformal gauge fields
that follow from the constraints:
\begin{align}
 f_{\mu}{}^\mu =&-\ft{1}{12}R-\ft{1}{2}D\nonumber\\
 &+\left\{\ft{1}{8}\bar{\psi}{}^i\cdot
\gamma \chi_i +\ft{1}{24}\rmi e^{-1}\varepsilon^{\mu \nu \rho
\sigma}\bar{\psi}{}^i_{\mu}\gamma_{\nu}\mathcal{D}_{\rho}\psi_{\sigma i}
+ \ft{1}{24} \bar{\psi}{}^i_{\mu}\psi_{\nu}^j\varepsilon_{ij}T^{+\mu
\nu}+\hc\right\}\,,\nonumber\\
\phi_{\mu}^i=&\ft{1}{4}\gamma_{\mu}\chi^i+\ft{1}{4}\left(\gamma^{\nu
\rho}\gamma_{\mu}-\ft{1}{3}\gamma_{\mu}\gamma^{\nu
\rho}\right)\left(\mathcal{D}_{\nu}\psi^i_{\rho}-\ft{1}{16}\gamma \cdot
T^- \varepsilon^{ij} \gamma_{\nu}\psi_{\rho j}\right).
 \label{d4valuefphi}
\end{align}
This leads to various simplifications, after which the vector action reduces to \cite[(20.89)]{Freedman:2012zz}:
\begin{eqnarray}
e^{-1}\mathcal{L}_g&=&-\ft16N\, R\,  -N\,D -N_{IJ}D_{\mu}X^ID^{\mu}\bar{X}^J+
N_{IJ}\vec{Y}^{I}\cdot \vec{Y}^J\nonumber\\
 && +\,N_{IJ}f_{KL}{}^I\bar{X}^K X^L
f_{MN}{}^J \bar{X}^M X^N + e^{-1}{\cal L}_{CS}\nonumber\\
 &&+ \left\{-\ft14\rmi\bar F_{IJ}\widehat F^{+I}_{\mu
\nu} \widehat F^{+\mu \nu J}
-\ft1{16}N_{IJ}X^I X^J T^+_{ab} T^{+ab}+\ft14 N_{IJ}X^I \widehat
F^{+J}_{ab}T^+{}^{ab}\right.\nonumber\\
&&-\,\ft14N_{IJ}\bar{\Omega}^{iI}\slashed{D}\Omega_i^J
+\ft16 N\, \bar\psi_{i\mu } \gamma^{\mu \nu \rho }D_\nu  \psi_\rho ^i \nonumber\\[3pt]&&
-\,\ft12N\, \bar{\psi}_{ia}  \gamma^a \chi^i
+N_{IJ}X^I \bar\Omega^{iJ}{\chi}_i
-\ft13N_{IJ}X^J\bar{\Omega}^{iI}\gamma^{\mu
\nu}D_{\mu}\psi_{\nu i}\nonumber\\[3pt]
&&+\,\ft12N_{IJ}\bar{\psi}^i_{\mu}\slashed{D}\bar{X}^I\gamma^{\mu}\Omega_i^J+\ft14
N_{IJ}\bar{X}^I\bar{\psi}_{ai}\gamma^{abc}\psi^i_bD_cX^J \nonumber\\[3pt]
&&+\,\ft12\rmi\bar F_{IJ}\varepsilon _{ij}\left(\bar \Omega ^{iI}\gamma _\mu-\bar X^I\bar \psi_\mu ^i\right)   \psi ^j_\nu \tilde {\hat{F}}{}^{\mu \nu J}
-\ft1{16}\rmi F_{IJK}\bar \Omega _i^I\gamma ^{\mu \nu }\Omega _j^J\varepsilon ^{ij} F_{\mu \nu }^{-\,K}\nonumber\\
&&+\,\ft12N_{IJ}\bar \Omega _i^If_{KL}{}^J\left( \Omega _j^L+\gamma ^a\psi _{aj}X^L\right)\bar X^K \varepsilon ^{ij}\nonumber\\
&& +\left(\ft1{12} N\,\bar\psi^a_{i}\psi^b_{j}  -\ft16 N_{IJ}\bar X^I\bar\Omega_i^J\gamma^a\psi^b_j
 +\ft1{32}\rmi F_{IJK} \bar\Omega_i^I \gamma^{ab}\Omega_j ^J\bar X^K\right) T^{-}_{ab}\varepsilon ^{ij}
  \nonumber\\&&\left.-\,\ft{1}{4}\rmi F_{IJK}D_{\mu}X^I \bar{\Omega}_i^J \gamma^{\mu}
\Omega^{iK}+\ft14\rmi
F_{IJK}Y^{ijI}\bar{\Omega}_i^{J}\Omega_j^{K}+\hc\right\}\nonumber\\
&&+\,\mbox{4-fermion terms}\,.\label{leadingLvectorsd4}
\end{eqnarray}
Here, important quantities are introduced, which will often be used below:
\begin{tcolorbox}
\begin{equation}
  N_{IJ}=N_{IJ}(X,\bar{X})\equiv 2\Im F_{IJ}= -\rmi F_{IJ}+\rmi \bar F_{IJ}\,,\qquad
N\equiv N_{I J}X^I\bar X^J\,.
 \label{NIJ}
\end{equation}
\end{tcolorbox}
Since $F_{IJ}$ is a function of $X^K$, we have the chain rule for the gauge transformations:
\begin{equation}
  \delta F_{IJ}=F_{IJK}\delta X^K= F_{IJK}\theta ^L X^M f_{ML}{}^K\,,
 \label{delGFIJ}
\end{equation}
and therefore the gauge transformation of $N_{IJ}$ is by \eqref{defCijk}
\begin{equation}\label{gaugeInvN}
\delta N_{IJ}= 2 \theta ^K f_{K(I}{}^L N_{J)L}\,.
\end{equation}

Covariant derivatives are presented in \eqref{covD4N2part1} together with
\begin{eqnarray}
  D_\mu \psi _{\nu i}&=&\left( \partial _\mu +\ft14\omega _\mu
{}^{ab}\gamma _{ab}+\ft12b_\mu +\ft12\rmi A_\mu \right)\psi _{\nu i}
 +V_{\mu i}{}^j\psi _{\nu j}\,.
\label{DpsiN2}
\end{eqnarray}

\subsection{Action for Vector Multiplets in \texorpdfstring{$D=5$}{D=5}}
\label{ss:actVMD5}
The scheme for $D=5$ is similar to (\ref{schemeN2sugraVM}) with the only exception that vector multiplets\footnote{We use here and below
freely the terminology `spin 1' for vectors, spin-$\frac{1}{2}$ for
spinors, ..., though of course only in 4 dimensions the representations
of the little group of the Lorentz group can be characterized by just one
number, which is called `spin'. In higher dimensions, the representations
should be characterized by more numbers, but often the same fields, like
graviton as a symmetric tensor, vectors, ... occur, and we denote them
freely with the terminology that is appropriate for the 4-dimensional
fields.} have only one real scalar, and thus at the end we will have $n$ real scalars, which are the $\sigma ^I$ that we saw already in Sect. \ref{vectmult}.
For $D=6$, which we will not treat in detail here, it is also be similar but without scalars in the vector multiplets.

The rigid superconformally invariant action $D=5$ is determined by a prepotential \index{prepotential} $C_{IJK}\sigma ^I\sigma ^J\sigma ^K$ cubic in the scalars $\sigma^I$ \cite{Gunaydin:1984bi,Seiberg:1996bd}. Since the vectors $W_\mu ^I$ can gauge a group, $I$ is also the index of the adjoint of the gauge group. As a consequence, the local superconformal action is determined by a gauge-invariant symmetric tensor $C_{IJK}$
\begin{equation}
f_{I(J}{}^MC_{KL)M}=0.
\end{equation}
Note that this tensor has no relation to the tensor $C_{I,JK}$ introduced for $D=4$ in  (\ref{gdelFinC}). The presence of this tensor allows Chern--Simons terms in the action. The kinetic term of the scalars defines the `very special real geometry' (see \cite[Sect.~20.3.2]{Freedman:2012zz}). The full superconformal invariant action takes the form \cite{Bergshoeff:2004kh}
{\allowdisplaybreaks
\begin{eqnarray}\label{conf-VTaction}
\lefteqn{e^{-1} {\cal L}_{g}
= C_{IJK} \left[ \left(- \ft14 {\widehat F}_{\mu\nu}^{ { I}} {\widehat F}^{\mu\nu { { J}}} - \ft12 \bar{\psi}^{ { I}} \slashed{\cal D} {\psi}^{ { J}} + \ft 13 \sigma^{ I} \Box^c \sigma^{ J}\right.\right.}
 \nonumber\\
&&\left. \left.+ \ft 16 {\cal D}_a \sigma^{ { I}} {\cal D}^a \sigma^{ { J}} +  2\vec{Y}^{I}\cdot \vec{Y}^J  \right) \sigma^{ { K}}\right.\nonumber\\
& & -\,\ft 43 \sigma^{ I} \sigma^{ J}
\sigma^{ K} \left(D + \ft{26}{3} T_{ab} T^{ab} \right)
+ 4 \sigma^{ I} \sigma^{ J} {\widehat  F}_{ab}^{ K} T^{ab}\nonumber\\
& & -\,\ft 18 {\rmi} \bar{\psi}^{ { I}} {\gamma}\cdot
{\widehat F}^{ { J}} {\psi}^{ { K}} -\ft12 {\rmi} \bar{\psi}^{i { { I}}} {\psi}^{j { { J}}}
Y_{ij}^{{ K}}  + {\rmi} \sigma^{ I}
\bar{\psi}^{ J}
{\gamma}\cdot T {\psi}^{ K} - 8 {\rmi} \sigma^{ I} \sigma^{ J} \bar{\psi}^{ K} \chi\nonumber\\
& & {+}\, \ft 16 \sigma^{ I} \bar {\psi}_\mu {\gamma}^\mu
\left({\rmi} \sigma^{ J} \slashed{\cal D} {\psi}^{ K} +
\ft12 {\rmi} (\slashed{\cal D} \sigma^{ J}) {\psi}^{ K} -
\ft 14 {\gamma}{\cdot} {\widehat  F}^{ J} {\psi}^{ K}
+ 2 \sigma^{ J} {\gamma}{\cdot} T {\psi}^{ K}\right.\nonumber\\
&&\left. - 8 \sigma^{ J} \sigma^{ K} \chi \right) \nonumber\\
&&
{-} \ft 16 \bar {\psi}_a {\gamma}_b {\psi}^{ I} \left(\sigma^J {\widehat  F}^{ab { K}} -8 \sigma^{ J} \sigma^{ K} T^{ab} \right) -\ft 1{12} \sigma^{ I} \bar {\psi}_\lambda  {\gamma}^{\mu\nu\lambda} {\psi}^{ J} {\widehat  F}_{\mu\nu}^{ K} \nonumber\\
&& {+}\,\ft 1{12} {\rmi} \sigma^{ I} \bar {\psi}_a {\psi}_b
\left(\sigma^{ J} {\widehat  F}^{ab { K}} -8
\sigma^{ J} \sigma^{ K} T^{ab} \right) +\ft 1{48}
{\rmi} \sigma^{ I} \sigma^{ J} \bar {\psi}_\lambda
{\gamma}^{\mu \nu \lambda \rho } {\psi}_\rho {\widehat
F}_{\mu\nu}^{ K}\nonumber\\&& {-}\, \ft12 \sigma^{ I}
\bar {\psi}_\mu^i {\gamma}^\mu {\psi}^{j  J} Y_{ij}^{
K} +\ft 1{6} {\rmi} \sigma^{ I} \sigma^{ J}
\bar{\psi}_\mu^i {\gamma}^{\mu\nu} {\psi}_\nu^j Y_{ij}^{ K}
-\ft{1}{24} {\rmi} \bar {\psi}_\mu {\gamma}_\nu {\psi}^{ I}
\bar {\psi}^{ J} {\gamma}^{\mu\nu} {\psi}^{
K}\nonumber\\&&  +\,\ft{1}{12} {\rmi} \bar {\psi}_\mu^i {\gamma}^\mu
{\psi}^{j { I}} \bar {\psi}_i^{ J}
{\psi}_j^{ K} -\ft{1}{48} \sigma^{ I} \bar {\psi}_\mu
{\psi}_\nu \bar {\psi}^{ J} {\gamma}^{\mu\nu}
{\psi}^{ K} +\ft {1}{24} \sigma^{ I} \bar{\psi}_\mu^i
{\gamma}^{\mu\nu} {\psi}_\nu^j \bar{\psi}_i^{ J}
{\psi}_j^{ K} \nonumber\\&&
 -\,\ft 1{12} \sigma^{ I} \bar {\psi}_\lambda
{\gamma}^{\mu\nu\lambda } {\psi}^{ J} \bar{\psi}_{\mu}
{\gamma}_{\nu} {\psi}^{ K} + \ft 1{24}\,{\rmi}
\sigma^{ I} \sigma^{ J} \bar {\psi}_\lambda  {\gamma}^{\mu\nu\lambda} {\psi}^{ K} \bar{\psi}_{\mu} {\psi}_{\nu}\nonumber\\
& &
 + \,\ft 1{48} {\rmi} \sigma^{ I} \sigma^{ J} \bar {\psi}_\lambda  {\gamma}^{\mu\nu \lambda \rho } {\psi}_\rho \bar{\psi}_{\mu} {\gamma}_{\nu} {\psi}^{ K} + \ft1{96}\sigma^{ I} \sigma^{ J} \sigma^{ K} \bar {\psi}_\lambda  {\gamma}^{\mu\nu \lambda \rho } {\psi}_\rho  \bar{\psi}_{\mu} {\psi}_{\nu}
  \nonumber\\
& &  -\,\ft 1{24} e^{-1} \varepsilon ^{\mu\nu \lambda \rho \sigma }
W_\mu^I\!
 \left(\!F_{\nu\lambda}^J F_{\rho\sigma}^K - f_{FG}{}^J \!W_\nu^F W_\lambda ^G   \left(\ft12  F_{\rho \sigma }^K - \ft1{10} f_{HL}{}^K W_\rho ^H W_\sigma^L  \right)\right)
\nonumber\\
 &&\left.+\, \ft14 {\rmi}\sigma ^I\sigma ^Jf_{LM}{}^K  \bar{\psi}^{ L}\psi^M\right]\,,
\label{Lform3}
\end{eqnarray}
}
\noindent where covariant derivatives and $\widehat{F}^I_{\mu \nu }$ are given in (\ref{localderiv-tensor}) and (\ref{widehatFD5}), and
the superconformal d'Alembertian is defined as
\begin{eqnarray}
 \Box^c
\sigma^I
&=& {\cal D}^a {\cal D}_a \sigma^{ I}\nonumber\\[2pt]
&=& \left( \partial^a  - 2 b^a + \omega_b^{~ba} \right) D_a
\sigma^{ I} + f_{J  K}{}^{ I} W_a^J {\cal D}^a
\sigma^{ K} - \ft{1}2{\rmi} \bar{\psi}_\mu D^\mu
{\psi}^{ I} - 2 \sigma^{ I} \bar{\psi}_\mu
{\gamma}^\mu \chi
\nonumber\\[2pt]
& & + \,\ft12 \bar{\psi}_\mu {\gamma}^\mu {\gamma}\cdot T
{\psi}^{ I} + \ft12 \bar{\phi }_\mu {\gamma}^\mu
{\psi}^{ I} + 2 f_\mu{}^\mu \sigma^{ I} -\ft12
\bar{\psi}_\mu {\gamma}^\mu f_{ J  K}{}^{ I}{\psi}^J \sigma^{ K}\,.
\end{eqnarray}
The dependent gauge fields are given in (\ref{transfDepF}).


\subsection{Action for Hypermultiplets}\label{finsec_hypers}
While the actions of vector multiplets were constructed using tensor
calculus manipulations, for the hypermultiplets we use another procedure. The main difference is that we have already the equations of motion from the non-closure relations, e.g. (\ref{eqmozeta5dlod}) in $D=5$, and we can therefore infer the action from the latter. To this end, we need a few ingredients that we are going to introduce in the following.

\subsubsection{Ingredients}

We first define a covariantly constant antisymmetric tensor $C_{\Ae\Be}(q)$ that describes the proportionality between the field equations for the fermions $\zeta^\Ae$.
and the non-closure functions. For example, in $D=5$,
\begin{equation}
\frac{\delta S_{\rm hyper}}{\delta \bar{\zeta}^\Ae}=2C_{\Ae\Be}\rmi\Gamma ^\Be\,.
\label{Casfieldeqn}
\end{equation}
Then, once the right-hand side of (\ref{Casfieldeqn}) is known, one can functionally integrate the above equation in order to obtain the action.
The properties of the tensor are (independent whether we consider $D=5$ or $D=4$):
\begin{align}
& \nabla _X C_{\Ae\Be}\equiv \partial_X C_{\Ae\Be}+2\omega _{X[\Ae}{}^\Ce C_{\Be]\Ce}=0\,,\nonumber\\
& C_{\Ae\Be}= -C_{\Be\Ae}\,,\nonumber\\
& C^{\bAe\bBe}\equiv (C_{\Ae\Be})^*= \rho^{\bAe\Ce}\rho^{\bBe\De}C_{\Ce\De}\,.
\label{Cq}
\end{align}

As will become clear below, the kinetic terms involve the Hermitian metric in tangent space
\begin{align}
  d^\bAe{}_\Be\equiv& -\rho ^{\bAe\Ce}C_{\Ce\Be}\,, \nonumber\\
 d^\bAe{}_\Be=&(d^\bBe{}_\Ae)^*=\rho ^{\bAe\Ce}d^{\bDe}{}_\Ce \rho _{\Be\bDe}\,,
\label{ddefine}
\end{align}
such that
\begin{equation}
 C_{\Ae\Be}= \rho _{\Ae\bCe}d^\bCe{}_\Be\,.
 \label{defCrhod}
\end{equation}
We also define an inverse
\begin{equation}
 C^{\Ae\Ce}C_{\Be\Ce}=\delta^{\Ae}{}_\Be \,,
 \label{Cinverse}
\end{equation}
so that we can use these matrices to  raise and lower $\Ae$ indices, using the common NE--SW convention.
\begin{equation}
  V_\Ae = V^{\Be}C_{\Be\Ae}\,,\qquad  V^\Ae = C^{\Ae\Be}V_{\Be}\,.
 \label{lowerAeindices}
\end{equation}
On the other hand, we raise or lower indices changing the holomorphicity as in (\ref{tIupup}). This is then consistent with changing the holomorphicity using $d^\bAe{}_\Be$. For example, for the gauge-transformation matrices in (\ref{tIupup}):
\begin{equation}
  t_{I\Ae\Be}=t_{I\Ae}{}^\Ce C_{\Ce\Be}= t_{I\Ae\bBe}d^\bBe{}_\Be\,.
 \label{tIbupup}
\end{equation}
Consistency of the transformations of the left- and right-hand side of (\ref{Casfieldeqn}) under the isometry group, determined by (\ref{delGferm}), implies that this matrix should be symmetric:
\begin{equation}
  t_{I\Ae\Be}=  t_{I\Be\Ae}\,.
 \label{tIantisymm}
\end{equation}
This equation is, using (\ref{tIcc}), equivalent to
\begin{equation}
  t_I{}^\bAe{}_\bBe d^\bBe{}_\Ce = t_{I\Ce}{}^\Be d^\bAe{}_\Be\,,
 \label{tId}
\end{equation}
which shows more clearly that it is related to the invariance of the action with signature matrix $d^\bAe{}_\Be$.

With the above conditions, $d^\bAe{}_\Be$ respects the quaternionic structure. It has been proven in \cite{deWit:1985px}, using the theorems of
\cite{Zumino:1962}, that at any point one can choose a basis such that $\rho$ is in the form (\ref{rhocanonical}) and at the same time
\begin{equation}
   d^\bAe{}_\Be=\begin{pmatrix}\eta &\phantom{\begin{pmatrix}a&b\cr c&d\end{pmatrix}}\cr \phantom{\begin{pmatrix}a&b\cr c&d\end{pmatrix}}&\eta \end{pmatrix}=
  \begin{pmatrix}
  -\unity _p&&&\cr &\unity_q &&\cr &&-\unity _p&\cr &&&\unity_q
  \end{pmatrix}
  \,,\qquad p+q=n_H\,.
\label{diagonalizedrhod}
\end{equation}
For rigid supersymmetry, positive kinetic terms will be obtained for $p=0$ and $q=n_H$. For supergravity we need one compensating multiplet and will use $p=1$.
These matrices should be covariantly constant. As we use a basis where
they are actually constant, this implies from \eqref{Cq} that (using the lowering of indices as in (\ref{lowerAeindices}))
\begin{equation}
2\omega _{X[\Ae}{}^\Ce C_{\Be]\Ce}=-\omega _{X\Ae\Be}+\omega _{X \Be \Ae}=0\,.
 \label{symmetryomegaAB}
\end{equation}
Thus the $\USp$-connection is symmetric in such bases. From
\begin{equation}
  \nabla _X d^\bAe{}_\Ce=-\bar{\omega}_X{}^\bAe{}_\bBe d^\bBe{}_\Ce-{\omega}_{X\Ce}{}^\De{} d^\bAe{}_\De=0\,,
 \label{nabladmatrix}
\end{equation}
one finds
\begin{equation}
   d^\bAe{}_\Ce\omega _{X\Be}{}^\Ce=-\bar{\omega} _X{}^\bAe{}_\bCe d^\bCe{}_\Be\,.
 \label{dCcondition}
\end{equation}
When $d=\unity $ the above condition is the antihermiticity of $\omega $.
In the preferred basis with (\ref{rhocanonical}) and (\ref{diagonalizedrhod}) we can also write
\begin{align}
  C_{\Ae\Be}=\begin{pmatrix}0&\eta \cr -\eta &0\end{pmatrix}\,,\qquad t_{I\Ae}{}^{\Be}=\begin{pmatrix}U_I&V_I\cr W_I&U_I{}^*  \end{pmatrix}\,,\qquad
  \begin{array}{ll}
  V_I{}^T =\eta  V_I\eta \,,\qquad    &W_I{}^T =\eta  W_I\eta \,,\\
    U^\dagger = -\eta U\eta \,, & V^*=-W \,.
  \end{array}
 \label{Candtcanonical}
\end{align}
This expresses that the transformations are in the subgroup of $\Gl (n_H,\mathbb{H})$ that preserves the antisymmetric metric $C_{\Ae\Be}$  and the metric $d^\bAe{}_\Be$, which is $\USp(2p,2q)$.

We define then the metric of the manifold to be
\begin{equation}
  g_{XY}=\left(f^{i\bAe}{}_X\right)^*d^\bAe{}_\Be f^{i\Be}{}_Y=f^{i\Ae}{}_X\varepsilon _{ij}C_{\Ae\Be}f^{j\Be}{}_Y\,,
 \label{defg}
\end{equation}
such that the holonomy associated to $g_{XY}$ is indeed $\USp(2p,2q)$.

The curvature tensor on the scalar manifold is determined in terms of a 4-index symmetric tensor in $\Sp(2n_H)$, denoted by $W_{\Ae\Be\Ce\De}$:
\begin{align}
W_{\Ae\Be\Ce\De}\equiv  W_{\Ae\Be\Ce}{}^{\Ee}C_{\Ee\De}=& -\varepsilon ^{ij} f^X{}_{i\Ae} f^Y{}_{j\Be} {\cal R}_{XY\Ce\De}\nonumber\\
 =& \ft12  f^{Xi}{}_{\Ae} f^Y{}_{i\Be} f^{Zk}{}_{\Ce} f^W{}_{k \De}
R_{XYZW}\,,
  \label{curvhypercomplex}
\end{align}
where we used the metric $g_{XY}$ \eqref{defg} to lower the indices.

\subsubsection{Remark on the Conformal Symmetry}
Due to the fact that we have now a metric available, we can invoke the
homothetic Killing equation (\ref{homotheticKilling}) and, similarly as in
 (\ref{k=kD2}), introduce a scalar function $\tilde{k}_{\rm D}$ such that
\begin{equation}
  k_{{\rm D}X}=g_{XY}k_{\rm D}{}^Y =\partial _X \tilde{k}_{\rm D}\,.
 \label{scalarkD}
\end{equation}
It is also possible to start from this scalar function, and generate the metric from
\begin{equation}
  g_{XY}=\frac{2}{D-2} \nabla_X \partial _Y \tilde{k}_{\rm D}\,.
 \label{gfromkD}
\end{equation}
We also define $k_{\rm D}{}^2$ using the metric \eqref{defg}
\begin{equation}
k_{\rm D}{}^2\equiv g_{XY}{k}_{\rm D}^X {k}_{\rm D}^Y \,.
 \label{defkd2}
\end{equation}
It will be also useful to express $k_{\rm D}{}^2$ in terms of the sections introduced in (\ref{defAialpha}) and their complex conjugates
\begin{equation}
  A_{i\bAe}=\left(A^{i\Ae}\right)^*= A^{j\Be}\rho _{\Be\bAe}\varepsilon _{ji}\,,\qquad  A^{i\Ae}= -\varepsilon ^{ij}\rho ^{\bBe\Ae}A_{j\bBe}\,.
 \label{Adown}
\end{equation}
To do so, we note that the matrix
\begin{equation}
  M^i{}_j \equiv A^{i\Ae}d^\bBe{}_\Ae A_{j\bBe}\,,
 \label{Mij}
\end{equation}
is Hermitian and equal to $\varepsilon ^{ik}\varepsilon _{j\ell}M^\ell{}_k$, i.e. $\sigma _2M^T\sigma ^2$. Therefore it should be proportional to the unit matrix. Indeed, using (\ref{defAialpha}) and \eqref{defg}
\begin{equation}
  M^i{}_j =\ft12\delta ^i_j A^{k\Ae}d^\bBe{}_\Ae A_{k\bBe}=\ft12\delta ^i_j k_{\rm D}{}^2\,.
 \label{A2iskD2}
\end{equation}

Another way in which $k_{\rm D}{}^2$ appears is in terms of an inner product of the $\SU(2)$ Killing vectors introduced in (\ref{delDSU2q}):
\begin{equation}
  k_{\rm D}{}^2= \ft13(D-2)^2 \vec{k}_X\cdot \vec{k}^X\,.
 \label{kD2fromveck}
\end{equation}
It is useful to record the relation between these quantities for arbitrary vectors $\vec{A}$ and $\vec{B}$:
\begin{equation}
   \vec{A}\cdot \vec{k}_X\,\vec{B}\cdot \vec{k}^X= \frac1{(D-2)^2}k_{\rm D}{}^2\vec{A}\cdot\vec{B}\,.
 \label{usevAvBk}
\end{equation}

\subsubsection{Moment Maps}

The isometries defined in (\ref{gen_iso_hyper}) can be expressed in terms of moment maps.\index{moment map} The definition of the latter depends on the theory. As we will discuss in Sect.~\ref{ss:isomKahler}, isometries for K\"{a}hler manifolds can be generically generated from a real moment map function using the complex structure and the metric. The hypermultiplet geometry has three complex structures, and as such have a triplet moment map for any isometry $\vec{P}_I$. They should satisfy
\begin{equation}\label{mommapshyper}
\partial_X \vec{P}_I=\vec{J}_X{}^Y k_{IY}\,.
\end{equation}
Furthermore, they satisfy an `equivariance relation'\index{equivariance relation}, which is necessary to build supersymmetric actions with these symmetries:
\begin{equation}
k_I{}^X \vec{J}_{XY}k_J{}^Y=f_{IJ}{}^K \vec{P}_K.
 \label{equivarianceglobal}
\end{equation}
With conformal symmetry, the solution of (\ref{mommapshyper}) is determined to be\footnote{It is a nice exercise to prove that  (\ref{mommapshyper}) is solved by (\ref{mom_mapPI}). You may replace the $\partial _X$ by covariant derivatives and use  (\ref{delDSU2q}), (\ref{cov_const_J}), (\ref{homotheticKilling}),  (\ref{commDkJ}) and  (\ref{WeylweightkI}).}
\begin{equation}
  \vec{P}_I = \vec{k}^Xk_{IX}= \frac{1}{D-2} k_{\rm D}{}^Y\vec{J}_Y{}^Xk_{IX}=\frac{2}{(D-2)^2}A^{i\Ae}t_{I\Ae\Be}\vec{\tau} _{ij}A^{j\Be}\,.
\label{mom_mapPI}
\end{equation}
In this context, it is also convenient to rewrite an expression that appears in the potential that occurs in these theories.
\begin{equation}
 k_I^X k_{JX} =\frac{4}{(D-2)^2} \varepsilon _{ij}A^{i\Ae}A^{j\Be}t_{I\Ae}{}^{\Ce}C_{\Ce\De}t_{J\Be}{}^{\De}\,.
 \label{kkastt}
\end{equation}

\subsubsection{Action for Hypermultiplets in $D=5$}
The resulting action is\cite{Bergshoeff:2004kh}
\begin{eqnarray}
e^{-1} \mathcal{L}_h &=& - \ft12 g_{XY}
D_a q^X {D^a} q^Y +\bar{\zeta}_A \slashed{\cal D} \zeta^A
+ \ft49 Dk_{\rm D}{}^2 + \ft{8}{27} T^{ab}T_{ab} k_{\rm D}{}^2
 \nonumber\\
&& +\, \ft{16}{3} \rmi \bar{\zeta}_A \chi_i k_{\rm D}{}^X f^{iA}{}_{X} + 2 \rmi
\bar{\zeta}_A {\gamma}\cdot T \zeta^A
-\ft14 W_{ABCD} \bar{\zeta}^A \zeta^B \bar{\zeta}^C \zeta^D  \nonumber\\
&& -\, \ft29 \bar{\psi}_a {\gamma}^a \chi k_{\rm D}{}^2 - \ft13 \bar{\zeta}_A
{\gamma}^a {\gamma}\cdot T \psi_{ai} k^X f^{iA}{}_{X}
- \ft12 \rmi \bar \zeta_A {\gamma}^a {\gamma}^b \psi_{ai} {D}_b q^X f^{iA}{}_{X}  \nonumber\\
&& + \, \ft23 f_a {}^a k_{\rm D}{}^2 - \ft16 \rmi \bar{\psi}_a {\gamma}^{ab}
\phi_b k_{\rm D}{}^2
+ \bar{\zeta}_A {\gamma}^a \phi_{ai} k_{\rm D}{}^X f^{iA}{}_{X} \nonumber\\
&& +\, \ft{1}{12} \bar{\psi}_a^i {\gamma}^{abc} \psi_b^j D_c
q^Y J_ {YX\,ij} k_{\rm D}{}^X - \ft19 \,\rmi k_{\rm D}{}^2\bar{\psi}^a \left(\psi^b T_{ab}
- \ft{1}{2} {\gamma}^{abcd} \psi_b T_{cd} \right)
\nonumber\\
&& + \rmi \sigma^I t_{IB} {}^A  \bar{\zeta}_A \zeta^B - 2
\rmi k_I^X f^{iA}{}_X \bar{\zeta}_A \psi_i^{I} -\ft12 \sigma^I k_I^X
f^{iA}{}_X \bar{\zeta}_A {\gamma}^a \psi_{ai}
 \nonumber\\
&& -\ft12 \bar{\psi}_a^i {\gamma}^a \psi^{jI}
P_{Iij}+\ft{1}{4} \rmi \bar{\psi}_a^i {\gamma}^{ab} \psi^j_b \sigma^I
P_{Iij}+ Y_{ij}^I P^{ij}_I  -\ft12 \sigma^I \sigma^J k_I^X k_{JX}\,,\nonumber\\
\label{conf-hyperaction}
\end{eqnarray}
with covariant derivatives given in \eqref{hypCovD5}.

\subsubsection{Action for Hypermultiplets in $D=4$}

When we discuss $D=4$, we can multiply (\ref{Casfieldeqn}) at both
sides with a chiral projection $P_R$. Using the rules (\ref{d54Gamma}) we
should now impose for the action $S_{\rm hyper}$
\begin{equation}
    \frac{\delta S_{\rm hyper}}{\delta \bar\zeta_\bAe} =  2 d^\bAe{}_\Be
  \Gamma^\Be\,. \label{dShypdzeta}
\end{equation}
We also want the action to generate the field equations for the scalars
that we have seen in (\ref{delQGamma}). This leads in rigid supersymmetry to
\begin{align}
 {\cal L}_h  =  & -\ft12 g_{XY}D_\mu q^X
  D^\mu q^Y-
\left(\bar\zeta_\bAe \slashed{\nabla }\zeta^\Be d^\bAe{}_\Be +\hc\right)
\nonumber\\
&  + \ft12 W_{\Ae\Be}{}^{\Ee \Fe }d^\bCe{}_\Ee d^\bDe{}_\Fe \bar\zeta_\bCe\zeta_\bDe
  \bar\zeta^\Ae \zeta^\Be  \nonumber\\
&  + \left(2 X^I
  t_{I\Ae\Be}\bar \zeta^\Ae\zeta^\Be  + 2 \rmi f^{i\Ae}_X k_I^X \bar\zeta
  _\bBe \Omega^{jI}\varepsilon _{ij}d^\bBe{}_\Ae +\hc\right)\nonumber\\
&  +\,2
\vec{P}_{I}\cdot \vec{Y}^{I}- 2 \bar X^I X^J k_I{}^X k_{JX}
\,,
\label{Lagrhypermgauged}
\end{align}
with the covariant derivatives in (\ref{CovDgauged5d}),
which satisfies (\ref{dShypdzeta}), and also
\begin{equation}
 \frac{\delta S_{\rm hyper}}{\delta q^X}=g_{XY}\Delta^Y-\left( 2
  \bar\zeta_\bAe\Gamma^\Be\omega_{X\Be}{}^{\Ce}d^\bAe{}_{\Ce}+\hc\right)\,.
  \label{v:varactie}
\end{equation}
See \cite[Exercises 20.8 and 20.9]{Freedman:2012zz} for a concrete example.

After gauge covariantization and using the values of the conformal gauge
fields as in (\ref{d4valuefphi}) and the covariant derivatives \eqref{hypCovD4}, the superconformal hypermultiplet action with gauged isometries in $D=4$ is \cite[(20.93)]{Freedman:2012zz}
\begin{eqnarray}
  e^{-1} {\cal L}_h&=& -\ft1{12}k_{\rm D}{}^2R +\ft14k_{\rm D}{}^2 D -\ft12 g_{XY}D_\mu  q^XD^\mu  q^Y
   -2  \bar X^I X^J k_I{}^X k_{JX}  +2\vec{P}_I\cdot \vec{Y}^I\nonumber\\
  &&+\Big{\{}-\bar\zeta_\bAe\widehat{\slashed{\cal D}} \zeta^\Be
  d^\bAe{}_\Be  +\ft1{12} k_{\rm D}{}^2\bar\psi_{i\mu } \gamma^{\mu \nu \rho }D_\nu  \psi_\rho ^i  \nonumber\\
                         && +\,\ft18k_{\rm D}{}^2\bar\psi_{ia}\gamma^a\chi^i- 2 \rmi d^\bAe{}_\Be  A^{i\Be } \bar\zeta_\bAe\chi_i \nonumber\\
                         &&+\,\ft12\rmi \bar\zeta_\bAe   \gamma^a \slashed{D} q^X \psi_{ai}f^{i\Be }{}_Xd^\bAe{}_\Be
  -\ft13 \rmi d^\bAe{}_\Be A^{i\Be } \bar\zeta_\bAe \gamma^{\mu \nu }D_\mu \psi_{\nu i } \nonumber\\&&
  +\left( \ft1{12}\rmi d^\bAe{}_\Be A^{i\Be } \bar\zeta_\bAe \gamma_a \psi_{b}^{j} -\ft1{48}k_{\rm D}{}^2\bar\psi_a^i\psi_b^j \right) T^{+ab} \varepsilon _{ij}\nonumber\\
                        &&  -\,\ft18 \bar\zeta_\bAe\gamma^{ab}
   T_{ab}^+ \zeta_\bBe C^{\bAe\bBe} +2\rmi \bar X^I
    k_I^X \bar\zeta_\bAe \gamma^a \psi_a^j \varepsilon _{ij}d^\bAe{}_\Be  f^{i\Be }{}_X\nonumber\\
    &&+\,2 X^I\bar\zeta ^\Ae \zeta ^\Be t_{I\Ae \Be }+ 2 \rmi k_I^X f^{i\Ae }{}_X
                          \bar\zeta_\bBe \Omega^{jI}\varepsilon _{ij}d^\bBe{}_\Ae \nonumber\\
                          &&+\,\ft12\bar\psi_{aj}\gamma^a
  \Omega^I_i
  P_I{}^{ij} +\ft12 \bar X^I \bar \psi_a^i \gamma^{ab} \psi_b^j
  P_{Iij}+\hc\Big{\}}\nonumber\\
&& +\, \ft12\bar\psi^i_a\gamma^{abc}\psi_{bj} D_c q^X
   \vec{k}_X\cdot \vec {\tau}_i{}^j +\mbox{4-fermion terms}
\,.\label{totaleLaghyp2}
\end{eqnarray}
We can rewrite the kinetic terms for the scalars $q^X$ in terms of the sections \eqref{defAialpha} using the bosonic part of (\ref{wideDA})
\begin{equation}
  g_{XY}D_\mu  q^XD^\mu  q^Y= \varepsilon _{ij}C_{\Ae\Be}\left(\widehat{D}_\mu A^{i\Ae}\right)\left(\widehat{D}^\mu A^{j\Be}\right)\,.
 \label{kinscalarsinA}
\end{equation}
\subsection{Splitting the Hypermultiplets and Example}
\label{ss:splithypers}

In general we did not use the basis (\ref{rhocanonical}). Sometimes, however, it will be convenient to use such basis in examples. To do this, we split the index $\Ae=1,\ldots ,2n_H$ into $\Ae=(\alpha a)$, with $\alpha =1,2$ and $a=1,\ldots ,n_H$.
The index $\bAe$ will then have the same form, but with $a$ in the opposite (up-down) position. We can then write the canonical basis with (\ref{rhocanonical}) and (\ref{diagonalizedrhod}) as
\begin{align}
  \rho _{\Ae\bBe}=& \varepsilon _{\alpha \beta }\delta _a^b\,,\qquad d^{\bAe}{}_{\Be}= \eta _{ab}\delta ^\alpha _\beta \,,\qquad C_{\Ae\Be}= \eta _{ab}\varepsilon _{\alpha \beta }\,.
\label{splitindex}
\end{align}
The components of $A^{i\Ae}$ can then be written as $A^{i\alpha a}$ and
 $(A^{i\alpha a})^*= A^{j\beta b}\varepsilon _{ji}\varepsilon _{\beta \alpha }$. Upon this splitting the action (\ref{totaleLaghyp2})  starts with
\begin{align}
  e^{-1} {\cal L}_h=& -\ft1{12}k_{\rm D}{}^2R +\ft14k_{\rm D}{}^2 D -\ft12 \widehat{D}_\mu A^{i\alpha a}\widehat{D}^\mu A^{j\beta b}\varepsilon _{ij}\varepsilon _{\alpha \beta }\eta _{ab}
+\cdots \,,\nonumber\\
 k_{\rm D}{}^2=& A^{i\alpha a}A^{j\beta b}\varepsilon _{ij}\varepsilon _{\alpha \beta }\eta _{ab}= A^{i\alpha a}(A^{i\beta b})^*\eta _{ab}\,.
 \label{Ssplithyper}
\end{align}
The conditions on the symmetry matrices $t_{I\Ae}{}^{\Be}$ (see \eqref{Candtcanonical}) are  such that they can be decomposed as
\begin{align}
  t_{I\alpha a}{}^{\beta b}=& t_{I0a}{}^b \delta _\alpha {}^\beta +\vec{t}_{Ia}{}^b\tau _\alpha {}^\beta\,, \qquad t_{I0a}{}^b,\,\vec{t}_{Ia}{}^b\in\mathbb{R}\,,\nonumber\\
  &t_{I0a}{}^b=-\eta _{ac}t_{I0d}{}^c\eta ^{db}\,,\qquad \vec{t}_{Ia}{}^b=\eta _{ac}\vec{t}_{Id}{}^c\eta ^{db}\,.
 \label{conditionstsplit}
\end{align}
As an example, we may consider
\begin{equation}
  t_{I\alpha a}{}^{\beta b}= \rmi Q_{Ia}{}^b (\sigma _3) _\alpha {}^\beta\,,\qquad Q_{Ia}{}^b\in\mathbb{R}\,,\qquad  Q_{Ia}{}^b=\eta ^{bc}Q_{Ic}{}^d\eta _{da}\,.
 \label{tabelian}
\end{equation}
Then from \eqref{Atiskf} and \eqref{mom_mapPI} we have
\begin{align}
  k_I{}^X =&\rmi f^X{}_{i(\alpha a)} A^{i\beta b} Q_{Ib}{}^a(\sigma _3) _\beta {}^\alpha\,,\nonumber\\
  \vec{P}_I =& \ft12\rmi A^{i\alpha a}Q_{Iab}(A^{j\beta b})^*\vec{\tau }_i{}^j (\sigma _3) _\alpha {}^\beta\,,
 \label{vecPexample}
\end{align}
with $Q_{Iab}=Q_{Ia}{}^c\eta _{cb}=Q_{Iba}$. 

%% file: ch4_Gaugefix.tex
\chapter{Gauge Fixing of Superconformal Symmetries} \label{ss:gaugefix}
\begin{quote}
\textbf{Abstract.}
In this chapter we combine the actions for vector multiplets and hypermultiplets, containing also compensating multiplets. We gauge fix the superconformal symmetries that are not necessary for the super-Poincar\'{e} theory, and in such way obtain matter-coupled Poincar\'{e} supergravity theories. We extensively discuss pure supergravity (using also other compensating multiplets leading to the off-shell theory) and discuss its reduction to ${\cal N}=1$. Then we discuss appropriate variables for the gauge fixing of general matter couplings.
\end{quote}

\section{General Considerations}

In the previous chapters, we constructed the local superconformal invariant actions involving vector multiplets and hypermultiplets.
The total action is then given by the sum of the vector multiplet and hypermultiplet actions, given respectively by  \eqref{leadingLvectorsd4} and \eqref{totaleLaghyp2} \cite{deWit:1985px}:\footnote{We concentrate here on 4 dimensions. The principles are the same for
$D=5$ and for a large part also for $D=6$.}
\begin{equation}
  {\cal L}= \mathcal{L}_g + \mathcal{L}_h\,,
 \label{Ltotalgh}
\end{equation}

It remains to break the superfluous symmetries in order to obtain a super-Poincar\'{e} invariant action.
This can be done by taking a `gauge choice' for the parts of the superconformal symmetry that are not in the super-Poincar\'{e} algebra:
dilatations, K-symmetries, S-supersymmetry and the $R-$symmetry.
As we illustrated at the end of Sect.~\ref{ss:exampleBosComp}, classical gauge invariance implies that some degree of freedom in the set of
fields $\{\phi ^i\}$ in fact disappears from the action $S[\phi^i]$. More precisely, from the gauge symmetry requirement one has
\begin{equation}
  0=\delta S[\phi]=\frac{\delta  S[\phi]}{\delta  \phi ^i}\delta\phi ^i \,,
 \label{gaugeinv}
\end{equation}
and if one can redefine the
basis of fields such that all $\delta \phi^i=0$ $\forall i$ except, for example, $i=0$, then \eqref{gaugeinv} clearly implies that $S$ does not depend on $\phi^0$. This argument immediately shows that, after writing out all covariant derivatives and dependent gauge fields, $b_\mu $ should disappear from the action. Indeed, as clear from Eq.~\eqref{deltaDK}, from our fundamental fields only $b_\mu $ transforms under $K$ and therefore
\begin{equation}
  \delta(\lambda _{\rm K}) S=\frac{\delta S}{\delta b_\mu} 2\lambda _{{\rm K}\mu }=0\,.
 \label{delKisnobmu}
\end{equation}
Since the action does not depend on $b_\mu$ in this basis, we can arbitrarily fix a gauge for $K$-invariance. Conventionally \index{$K$-gauge}
\begin{equation}\label{eq:K}
K\mbox{-gauge:}\quad b_\mu=0.
\end{equation}
We could repeat a similar procedure for the other gauged symmetries that we want to fix.\footnote{If there is one field that has the same number of degrees of
freedom than the gauge symmetry itself, one can often make redefinitions proportional to that field such that only that one field transforms.}
However, this procedure is often cumbersome (field redefinitions are often
non-local) while the
result is just the same as taking a `gauge choice' for this symmetry. The
example presented in Sect.~\ref{ss:strategy} can illustrate these
remarks. \vspace{10pt}

Note that if one imposes a gauge condition, it does not imply that one should forget about the transformations that are gauge fixed, e.g. the $K$ transformations after the choice (\ref{eq:K}). The correct conclusion is
that now the $K$ transformations are dependent on the other ones, in such
a way that the gauge condition is respected. In other words, the $K$ transformation must act in such a way on $b_{\mu}$ so that it eliminates all the other transformations that act on it. Only then will (\ref{eq:K}) be a consistent gauge condition that does maintain the invariance of the action, where the field truly vanishes.
In $D=4$, using (\ref{delextra}) and (\ref{transfo4}) while demanding that $\delta_I(\epsilon^I)b_{\mu}=0$, we obtain a \textit{decomposition law}:\index{decomposition law}
\begin{equation}
  \lambda_{\rm K}^a (\lambda _{\rm D},\epsilon ,\eta )= -\ft12e^{\mu a} \left[\partial _\mu \lambda _{\rm D}+ \ft12 \left( \bar \epsilon ^i \phi_{\mu i} -
   \ft34 \bar \epsilon ^i
\gamma _\mu \chi_i - \bar \eta^i \psi_{\mu i} + \hc\right)  \right]\,.
 \label{decompK}
\end{equation}
With the above expression for the parameter of $K$ transformations, (\ref{eq:K}) will be a consistent condition. The action of a $K$ transformation is now dependent on the other transformations, via the decomposition law (\ref{decompK}),  e.g. the first term in (\ref{decompK}) is a contribution to local dilatations of the
action where $b_\mu $ is omitted. \vspace{10pt}

The present chapter is structured as follows. In Sect. \ref{ss:pureN2} we explain the gauge fixing procedure for \emph{pure ${\cal N}=2$ supergravity in $D=4$}, i.e. where the physical fields are just the graviton, the doublet gravitino and the so-called graviphoton.
We will discuss 3 different ways of obtaining an off-shell representation using various sets of compensating multiplets.
In Sect.~\ref{ss:N2toN1} we consider the reduction of the ${\cal N}=2$, $D=4$ pure supergravity theory to ${\cal N}=1$.
In Sect.~\ref{ss:mattersugra} we will re-introduce vector multiplets and hypermultiplets (both as matter and compensators). We derive the equations of motion for the auxiliary fields of these actions in $D=4,5$ and choose appropriate gauge fixing conditions, focussing mostly on $D=4$ supergravity.
Then we will introduce appropriate (projective) coordinates, which will parameterize the Poincar\'{e} theory. In Sect. \ref{ss:projcoord} this will be done for the scalars of the vector multiplets, which will allow us to discuss already the main properties of special K\"{a}hler geometry. In Sect. \ref{coord_quatman} we will in the same way introduce appropriate coordinates to discuss the quaternionic-K\"{a}hler geometry for the scalars of the hypermultiplets.
We end this chapter in Sect.~\ref{ss:gfD56} with some remarks on the gauge fixing for $D=5$ and $D=6$ supergravity, which follows similar patterns as what we discussed for $D=4$.

All these ingredients will be intensively exploited in following chapters on special geometry and to define the final actions for ${\cal N}=2$ matter-coupled Poincar\'{e} supergravity in Chap. \ref{ch:finalresults}.

\section{Pure \texorpdfstring{${\cal N}=2$}{N=2} Supergravity}
\label{ss:pureN2}

This section is devoted to the construction of the pure ${\cal N}=2$ supergravity, i.e. there are no other physical fields than the graviton, the gravitino and the graviphoton. We will also pay attention to the off-shell structure of the theory, in particular discussing the auxiliary field structure. Though the principles that we discuss here are applicable in general, \emph{we will concentrate in this section on $D=4$.}

We start by introducing the so-called minimal field representations. As we will explain in Sect.~\ref{ss:minfieldrepr}, the latter is sufficient to fix the dilatation symmetry, but it leads to an inconsistent action. In order to overcome this problem we need a second compensating multiplet, for which there are several known choices: the hypermultiplet,  the so-called nonlinear multiplet \cite{deWit:1979pq,Fradkin:1979as,deWit:1980ug} and
the tensor multiplet. In each case one gets a different set of
auxiliary fields for the ${\cal N}=2$ super-Poincar\'{e} theory, with the same
number of field components ($40+40$ off-shell degrees of freedom) \cite{deWit:1982na}. We will discuss these in following subsections.

\subsection{The Minimal Field Representation}
\label{ss:minfieldrepr}

It is natural to perform the gauge fixing of dilatations and of the $\U(1)$ part of the $R$-symmetry group using the scalars of a vector multiplet. They are rather similar to the scalars of chiral multiplets in ${\cal N}=1$, which are used in that case to define the so-called old minimal sets of auxiliary fields, see  \cite{Kaku:1978ea,Ferrara:1978wj} and reviewed in \cite[Sect. 16.2]{Freedman:2012zz}.

The starting point is (\ref{leadingLvectorsd4}) for only one vector multiplet with complex scalar $X$:
\begin{equation}
  F(X)= -\ft14\rmi X^2\qquad \rightarrow\qquad  N_{00}=-1\,,\qquad N= -X\bar X\,.
 \label{Fpuresugra}
\end{equation}
The minus sign reflects that $X$ defines a compensating multiplet and not a physical multiplet, similar to the fact that the action (\ref{confScalar}) has negative kinetic terms for its scalar.
We fix the $\U(1)$ gauge by restricting $X$ to be real, and the dilatational gauge by taking $X=\kappa ^{-1}$, then
we are left with
\begin{align}
e^{-1}\mathcal{L}_g=&\ft16\kappa ^{-2}\, R\,  +\kappa ^{-2}\,D +\cdots \,.
\label{leadingpureD4VM}
\end{align}

The Weyl multiplet with the $24+24$ field components in Table \ref{tbl:countWeyl} and a compensating off-shell vector multiplet with $8+8$ components as in Table \ref{tbl:matterMult} forms a `minimal field representation' of ${\cal N}=2$ supergravity. However, when considering this minimal field representation for building an action, we are faced with two issues.
First, the action  (\ref{leadingpureD4VM}) contains a term linear in the $D$ field, which would impose $X\bar X=\kappa^{-2}=0$ as a consequence of the equation of motion for $D$, therefore removing  the Einstein--Hilbert term.
Second, there is still a remaining gauged $\SU(2)$ $R$-symmetry, because the scalars of the vector multiplets are invariant under these transformations. The super-Poincar\'{e} group does not contain this group, and thus a matter-coupled ${\cal N}=2$ supergravity does not necessarily have such an $\SU(2)$ gauge symmetry.

Both problems can be solved by introducing a second `compensating' multiplet
in the background of the vector multiplet, involving scalars that do transform under $\SU(2)$ and that can be therefore used to fix the $\SU(2)$.
In this book we concentrated a lot on the hypermultiplets, whose scalars indeed transform under $\SU(2)$ and whose action (\ref{totaleLaghyp2}) contains another term for the auxiliary field and for an Einstein--Hilbert term. This is indeed one of the possibilities, but in this section we will present also the other known solutions.
\bigskip

Off-shell fields form a massive representation of supersymmetry, which become propagating in a super-Weyl gravity \cite{deWit:1978pd}. Massive representations contain representations of $\USp(4)$ \cite{Ferrara:1980bh}. Let us recall the relevant massive representations of ${\cal N}=2$ supersymmetry\footnote{We first write the number of bosonic $+$ fermionic components ($2j+1$ components for spin $j$), and then the spin content of the fields with notation $(j)^{\mbox{\#}}$, separating $\USp(4)$ representations}:
\begin{eqnarray}
 \mbox{spin 2 multiplet}: &24+24 \   : & \{ 2^1, \, (\ft32)^4,\, 1^5,\, 1^1,\,(\ft12)^4,\,0^1\}\,,  \nonumber\\
 \mbox{spin 1 multiplet}: &8+8 \ : & \{  1^1,\,(\ft12)^4,\,0^5\}\,,  \nonumber\\
\mbox{spin }\ft12\mbox{ multiplet}:& 4+4 \ : &\{ (\ft12)^2,\,0^4\}\,.
 \label{massivereps}
\end{eqnarray}

The fields of the Weyl multiplet represent the components of a
massive spin 2 multiplet:
\begin{equation}
\begin{array}{|c|c|ccccc|}\hline
 \mbox{field}& \mbox{dof} & \mbox{spin } 2&\mbox{spin } \ft32&\mbox{spin } 1&\mbox{spin } \ft12&\mbox{spin } 0   \\
\hline
e_\mu {}^a & 5 & 1 &&&&\\
b_\mu & 0 &  &&&& \\
V_{\mu i}{}^j & 9 & &&3&&   \\
A_\mu & 3 & &&1&&  \\
T^-_{ab} & 6 && &2&&\\
D & 1 & &&&& 1 \\ \hline
\psi _\mu {}^i & 16 & &4&&&\\[1mm]
\chi ^i & 8 & &&&4& \\ \hline
\end{array}     \label{tblWeylN2}
\end{equation}
Note that in this and the following tables, the off-shell number of degrees of freedom (dof) of fields are given, subtracting all the superconformal gauge dof. The following columns order these according to the massive spin representations. We thus find indeed the multiplet as written in the first line of (\ref{massivereps}). Similarly, the
vector multiplet of the minimal field representation of ${\cal N}=2$ supergravity represents a spin 1 massive multiplet:
\begin{equation}
\begin{array}{|c|c|ccc|}\hline
 \mbox{field}& \mbox{dof} & \mbox{spin } 1&\mbox{spin } \ft12&\mbox{spin } 0 \\
\hline
X & 2 & & & 2\\
W_\mu & 3 & 1 &&  \\
Y_{ij}& 3 & & & 3\\ \hline
\Omega^i  & 8 &   & 4 &\\ \hline
\end{array}
\end{equation}

We can use some of these fields to fix superconformal symmetries. The first of these was already mentioned in (\ref{eq:K}). The complex scalar $X$ can be put to a constant to gauge fix the dilatation and the $\U(1)$ $R$-symmetry. Similar as in the example in Sect. \ref{ss:strategy} we can choose a convenient value to obtain the standard normalization of the Einstein--Hilbert action. But we will choose this only after we have considered the full action. Similarly, the fermion $\Omega ^i$ of the compensating multiplet can be fixed to a value to remove $S$-supersymmetry as an independent symmetry (similar to $K$ in (\ref{decompK})). The combination used for pure Poincar\'{e} supergravity is thus of the form (a spacetime index $\mu$ indicates that the field is a gauge field, while those with $a$ are not gauge, and thus have 4 components)
\begin{align}
   & \mbox{Weyl multiplet: }e_\mu{} ^a,\, b_\mu,\, \psi _\mu ^i,\,V_{\mu i}{}^j,\,A_\mu,
   \,T_{ab},\,\chi ^i,\,D     \nonumber\\
   & \phantom{\mbox{Weyl multiplet: }e_\mu{} ^a,}+  \nonumber\\
   & \mbox{vector multiplet: }X,\, \Omega ^i,\,W_\mu ,\,Y_{ij} \nonumber\\
   &\phantom{\mbox{Weyl multiplet: }}\downarrow 
   \begin{array}{rl}
 \mbox{gauge fixing symmetries }&K_a,\, D,\, \U(1),\, S^i  \\
 \mbox{ by choosing values for }&b_\mu ,\, X,\,\,\, \Omega ^i
                \end{array}\nonumber\\
&{\cal N}=2\ \mbox{super-Poincar\'{e} multiplet }(32+32)\nonumber\\
&\phantom{\mbox{Weyl}}\begin{array}{l}
  e_\mu{} ^a\,(6),\,W_\mu\, (3),\,V_{\mu   i}{}^j\,(9),\,
A_a\,(4),\,T_{ab}\,(6),\,D\,(1),\,Y_{ij}\,(3) \\
  \psi _\mu ^i\,(24),\ \chi ^i\, (8) \,.
\end{array}\nonumber\\
 \label{schemecompN2d4minimal}
\end{align}

The graviton, gravitino and graviphoton (the latter originating in the vector multiplet) are the physical fields. On the other hand, $V_{\mu i}{}^j$, $T_{ab}$, $\chi^i$ and $D$ are auxiliary fields in the Weyl multiplet, while $Y_{ij}$ is the $\SU(2)$ triplet auxiliary field originating in the vector multiplet.

We now consider the three possibilities for the second compensating multiplet.

\subsection{Version with Hypermultiplet Compensator}
\label{ss:hypercompensator}
We can use a hypermultiplet as a second compensating multiplet \cite{deWit:1981tn}. We use for this multiplet constant frame fields $f^{i\Ae}{}_X$, such that  (\ref{homotheticKilling}) is solved by $k_{\rm D}{}^X= q^X$. Then (\ref{defAialpha}) just defines another parameterization of the 4 physical scalars in the form $A^{i\Ae}$. They are thus written as a doublet of the $R$-symmetry group $\SU(2)$ and $\Ae=1,2$ are indices for a priori another $\USp(2)=\SU(2)$ group. To define the metric as in Sect. \ref{finsec_hypers} we take $d^\bAe{}_\Be= -\delta^\Ae{}_\Be$, where again the negative sign reflects that this is a compensating multiplet. With this choice, removing a bar from the index induces a minus sign and this is in accordance with Sect. \ref{ss:splithypers}, where $a$ now runs only over one value and $\eta _{11}=-1$. Normalizing $C_{\Ae\Be}=-\varepsilon _{\Ae\Be}$, we thus have e.g. $\rho_{\Ae\bBe}=\varepsilon _{\Ae\Be}$, such that $(A^{i\Ae})^*= A^{j\Be}\varepsilon _{ji}\varepsilon _{\Be\Ae}$. In this way the action (\ref{totaleLaghyp2}) starts with
\begin{align}
  e^{-1} {\cal L}_h=& -\ft1{12}k_{\rm D}{}^2R +\ft14k_{\rm D}{}^2 D +\ft12 \partial _\mu  A^{i\Ae}\partial ^\mu A^{j\Be}\varepsilon _{ij}\varepsilon _{\Ae\Be}
+\cdots \,,\nonumber\\
 k_{\rm D}{}^2=& -A^{i\Ae}A^{j\Be}\varepsilon _{ij}\varepsilon _{\Ae\Be}= -A^{i\Ae}(A^{i\Ae})^*\,.
 \label{SpureSGhyper}
\end{align}

In the case of a single hypermultiplet it is known how to close the supersymmetry algebra off-shell by including auxiliary fields $F^{i\Ae}$, appearing in the transformations of the fermions $\zeta^\Ae$ \cite{Fayet:1975yi,Fayet:1978ig,VanProeyen:1980na,deWit:1980gt,deWit:1985px}. In particular, one adds to the symmetry group transformations (\ref{gen_iso_hyper}) a non-compact generator $Z$, such that $k_Z{}^X f^{i\Ae}{}_X =F^{i\Ae}$. This generator is then gauged by the vector multiplet in the minimal field representation, leading to central charge transformations.  We do not discuss here the full procedure.\footnote{It is not clear whether such a procedure can also be done when the quaternionic geometry is not flat.}

The off-shell multiplet is then of the form
\begin{equation}
\begin{array}{|c|c|cc|}\hline
 \mbox{field}& \mbox{dof} & \mbox{spin } \ft12&\mbox{spin } 0   \\
\hline
A^{i\Ae} & 4 & & 4\\
F^{i\Ae} & 4 & & 4  \\ \hline
\zeta ^\Ae  & 8 &   4 &\\ \hline
\end{array}
\end{equation}

To gauge fix the $\SU(2)$, we restrict the quaternion to be a real number.\footnote{This is of course similar to the gauge fixing of $\U(1)$, which fixes the phase of the complex $X$ of the compensating multiplet.} We write
\begin{equation}
\SU(2)\mbox{ gauge: }\qquad   A^{i\Ae}= \sqrt{2}\kappa ^{-1}\varepsilon  ^{i\Ae}\rme^{u}\,,\qquad \rightarrow \qquad k_{\rm D}{}^2=- 4\kappa ^{-2}\rme^{2u}\,,
 \label{AiAgfu}
\end{equation}
being $\varepsilon  ^{i\Ae}$ the Levi-Civita antisymmetric symbol, which breaks the local $\SU(2)$ on the $i$ indices while leaving a global $\SU(2)$ group acting now on both indices.
Here $u$ is an auxiliary field, which in the sum of (\ref{leadingpureD4VM}) and (\ref{SpureSGhyper}) appears together with the auxiliary field $D$:
\begin{align}
  e^{-1} {\cal L}=&\ft16\kappa ^{-2}\, R\, \left(1 + 2\rme^{2u}\right)R + \kappa ^{-2}\left(1-\rme^{2u}\right)\,D +2\kappa ^{-2} \rme^{2u}\partial _\mu  u\,\partial ^\mu u+\cdots \,.
 \label{SpureSGtotal}
\end{align}
On-shell $u\approx 0$ and the Einstein--Hilbert term has the canonical normalization.

The scheme that extends (\ref{schemecompN2d4minimal}) is thus:
\begin{eqnarray}
   & \mbox{Minimal field representation}& \nonumber\\
   & + &  \nonumber\\
   & \mbox{second compensating multiplet: }A^{i\Ae} ,\, \zeta ^\Ae ,\,F^{i\Ae}  & \nonumber\\
   &\downarrow & \hspace{-45mm}
   \begin{array}{l}
 \mbox{gauge fixing}\ \SU(2) \mbox{ symmetry } \\
 \mbox{ by choosing }A^{i\Ae} \mbox{ in terms of real }u
                \end{array}
     \nonumber\\
   &{\cal N}=2\ \mbox{super-Poincar\'{e} gravity }(40+40)\nonumber\\
&\begin{array}{l}
  e_\mu{} ^a\,(6),\,W_\mu\, (3),\,V_{a i}{}^j\,(12),\,
A_a\,(4),\,T_{ab}\,(6),\,D\,(1),\,Y_{ij}\,(3),\,u\,(1),\,F^{i\Ae}\,(4) \\
  \psi _\mu ^i\,(24),\ \chi ^i\, (8),\ \zeta ^\Ae\,(8) \,.
\end{array}\nonumber\\
 \label{schemecompN2d4hyper}
\end{eqnarray}
In the remaining set of fields of the ${\cal N}=2$ super-Poincar\'{e} theory, $e_\mu {}^a$ and $W_\mu $ are the physical graviton
and the graviphoton, and $\psi _\mu ^i$ are the gravitini. All the others are auxiliary
fields, and the full set includes $40+40$ off-shell degrees of freedom.

One denotes by gauged ${\cal N}=2$ supergravity the situation where the
gravitinos transform under a local $\SU(2)$ and where this induces then
also a cosmological constant (anti-de Sitter supergravity). As an example, in the present formalism we can gauge a $\U(1) \subset \SU(2)$ if the scalar of the compensating multiplet
transforms under a $\U(1)$ subgroup of the $\USp(2)$ group acting on the indices $\Ae$.
We see from the first line in (\ref{totaleLaghyp2}) that such a gauging produces a potential. For example, since $g_{XY}$ for the compensating multiplet is negative definite, the term $2\bar X^IX^Jk_I{}^X g_{XY}k_J{}^Y$ produces a negative term in the potential. There will be another negative term after elimination $\vec{Y}^0$. Since the involved fields are now constants, this is a negative cosmological term, whose size is determined by the choice of the Killing vector. Thus we have anti-de Sitter supergravity. Due to the gauge condition (\ref{AiAgfu}), the gauge $\U(1)$ is then identified with a subgroup of the $\SU(2)$ $R$-symmetry group.

If more vector multiplets are included, the full automorphism group $\SU(2)$ can be
gauged by having the second compensating multiplet transforming under an
$\SU(2)$ group. Also in this case, the gauge fixing  (\ref{AiAgfu}) mixes the $\SU(2)$ factor of the superconformal group
with the group gauged by vector multiplets. The first compensating
multiplet and two physical vector multiplets can even gauge a non-compact $\SO(2,1)$ gauge
group~\cite{deWit:1984xe} leading to a massive vector multiplet.

\subsection{Version with Tensor Multiplet Compensator}
\label{ss:tensorcompensator}
The second choice is to use a tensor multiplet as second compensating multiplet~\cite{deWit:1984xe}. The tensor multiplet, shortly discussed in Sect.~\ref{ss:tensormultiplet}, contains a triplet of scalars $L_{ij}$, a gauge antisymmetric tensor $E_{\mu\nu}$, a doublet of Majorana spinors $\varphi^i$ and a complex scalar $G$. As massive fields they represent a spin 1 multiplet:
\begin{equation}
\begin{array}{|c|c|ccc|}\hline
 \mbox{field}& \mbox{dof} & \mbox{spin } 1&\mbox{spin } \ft12&\mbox{spin } 0  \\
\hline
L_{ij} & 3 & & & 3\\
E_{\mu \nu }& 3 & 1&&\\
G &   2 & & & 2  \\ \hline
\varphi ^i & 8 &  & 4 &\\ \hline
\end{array}
\end{equation}
The scalars $L_{ij}$ form a triplet of  the $\SU(2)$ part of the $R$-symmetry group. Therefore, a gauge choice on this field, e.g.
\begin{equation}
 \SU(2)/\U(1)\mbox{ gauge: }\qquad  L_{ij}=\delta _{ij}\rme^u
 \label{gaugeSU2L}
\end{equation}
breaks this group to a $\U(1)$ subgroup. This remaining subgroup is gauged by one of the components of $V_{\mu i}{}^j$: with the choice  (\ref{gaugeSU2L}) this is proportional to $V'_\mu =V_{\mu 1}{}^2-V_{\mu 2}{}^1$. On-shell, the latter is  identified with the graviphoton $W_\mu $. The other 2 components of $V_{\mu i}{}^j$ are then no more gauge fields, and indicated below as $V_{a i}{}^j$  (hence with $2\times4$ dof).
The scheme that extends (\ref{schemecompGravity}) is thus:
\begin{eqnarray}
   & \mbox{Minimal field representation}& \nonumber\\
   & + &  \nonumber\\
   & \mbox{second compensating multiplet: }L_{ij},\,  \varphi^i,\, G,\,E_{\mu\nu}  & \nonumber\\
    &\downarrow & \hspace{-42mm}
   \begin{array}{l}
 \mbox{gauge fixing}\ \SU(2)/\U(1) \mbox{ symmetry } \\
 \mbox{ by choosing }L_{ij} \mbox{ in terms of real }u
                \end{array}\nonumber\\
&{\cal N}=2\ \mbox{super-Poincar\'{e} gravity }(40+40)\nonumber\\
&\begin{array}{l}
  e_\mu {}^a\,(6),\,W_\mu\,(3),\,V'_{\mu }\,(3),\,V_{a i}{}^j\,(8),\,A_a\,(4),\,T_{ab}\,(6),\,D\,(1),\,Y_{ij}\,(3),\,\\ E_{\mu\nu}\,(3),\,u\,(1),\,G\,(2) \\
  \psi _\mu ^i\,(24),\ \chi ^i\, (8),\ \varphi^i\,(8) \,.
\end{array}\nonumber\\
\label{schemecompN2d4tensor}
\end{eqnarray}
Again the physical fields are $e_\mu{}^a,\, W_\mu,\,\psi _\mu ^i$ and the others are auxiliary.
With the gauge antisymmetric tensor and the remaining $\U(1)$ gauge group, this version has similar properties as the `new minimal' set of auxiliary fields of ${\cal N}=1$ supergravity  \cite{Sohnius:1981tp}.

Similarly to the previous case, the gauged ${\cal N}=2$ supergravity can be obtained by coupling the first compensating multiplet (vector multiplet) to the compensating tensor multiplet, which will include in the Lagrangian a term  $\varepsilon ^{\mu \nu \rho \sigma }E_{\mu\nu}F_{\rho \sigma }(W)$.

\subsection{Version with Nonlinear Multiplet Compensator}

The third choice (historically, it is the first version \cite{deWit:1979pq,deWit:1981tn}) is to use as a second compensating multiplet a \emph{nonlinear tensor multiplet}  \cite{Curtright:1979yz}, consisting of the fields $\Phi^\Ae_i,\,\lambda_i,\,M^{[ij]},\,V_a$. The scalar field $\Phi^\Ae_i$, $\Ae=1,2$ is a $2\times 2$ special unitary matrix, where the indices $\Ae$ are unrelated to the $R$-symmetry $\SU(2)$ indices $i$. This multiplet is called nonlinear because its transformation rules are nonlinear in the fields. The fields $\Phi^\alpha_i$ represent 3 degrees of freedom due to the nonlinear constraint. The linearized version of this multiplet is again the tensor multiplet. The vector field $V_a$ is constrained in the superconformal background due to the following (linearized) relation:
\begin{equation}\label{relationVD}
\partial_a V^a=D+\ft13 R,
\end{equation}
where $R$ is the Ricci scalar. One may view this equation as eliminating one of the 4 degrees of freedom of the vector in the nonlinear multiplet, which with the two real components in the complex $M^{[ij]}$ and $\Phi^\Ae_i $ complete the 8 bosonic degrees of freedom of this multiplet. However, in the superconformal framework we will use it to eliminate the auxiliary field $D$, which was the cause of the first problem of the minimal field representation mentioned at the start of Sect. \ref{ss:pureN2}.
The multiplet is also a spin 1 massive multiplet:
\begin{equation}
\begin{array}{|c|c|ccc|}\hline
 \mbox{field}& \mbox{dof} & \mbox{spin } 1&\mbox{spin } \ft12&\mbox{spin } 0  \\
\hline
\Phi^\Ae_i  & 3 & && 3\\
M^{[ij]} & 2 & & & 2\\
V_a &   3 & 1 & &  \\ \hline
\lambda  _i & 8 & & 4& \\ \hline
\end{array}
\end{equation}

To break the local $\SU(2)$ invariance one imposes on the 3 components of  $\Phi^\Ae_i$
\begin{equation}\label{gaugesu2nnlinear}
\SU(2)\mbox{ gauge: }\qquad\Phi^\Ae_i=\delta^\Ae_i,
\end{equation}
such that the distinction between indices $\Ae$ and $i,j,...$ is lost. The scheme that extends (\ref{schemecompGravity}) is thus:
\begin{eqnarray}
   & \mbox{Minimal field representation}& \nonumber\\
   & + &  \nonumber\\
   & \mbox{second compensating multiplet: } \Phi_i^\alpha, \,\lambda_i, \, M_{ij}, \,  V_{a}  & \nonumber\\
    &\downarrow & \hspace{-45mm}
   \begin{array}{l}
 \mbox{gauge fixing}\ \SU(2) \mbox{ symmetry } \\
 \mbox{ by choosing }\Phi^\Ae_i\mbox{ and solve for $D$}
                \end{array}
     \nonumber\\
   &{\cal N}=2\ \mbox{super-Poincar\'{e} gravity }(40+40)\nonumber\\
&\begin{array}{l}
  e_\mu{} ^a\,(6),\,W_\mu\, (3),\,V_{a i}{}^j\,(12),\,
A_a\,(4),\,T_{ab}\,(6),\,Y_{ij}\,(3),\,V_a\,(4),\,M_{[ij]}\,(2) \\
  \psi _\mu ^i\,(24),\ \chi ^i\, (8),\ \lambda_i\,(8) \,.
\end{array}\nonumber\\
 \label{schemecompN2d4nonlinear}
\end{eqnarray}
The theory remains invariant under rigid $\SU(2)$, corresponding to the residual group of \eqref{gaugesu2nnlinear}. Finally, a $\U(1)$ subgroup of this residual $\SU(2)$ (acting on the index $\Ae$) can be gauged via the first compensating multiplet. This produces gauged ${\cal N}=2$ supergravity.

\section{Reduction from \texorpdfstring{${\cal N}=2$}{N=2} to \texorpdfstring{${\cal N}=1$}{N=1}}
\label{ss:N2toN1}

We now discuss the reduction of the 3 sets of auxiliary fields to ${\cal N}=1$. This has been first considered in
the discussion section of \cite{deWit:1982na}. We will be more explicit here.
For any version of auxiliary fields, we have to start from the Weyl multiplet with the vector compensating multiplet.
We treat here just the pure supergravity theory. More general reductions in the Poincar\'{e} formulation have been discussed in \cite{Andrianopoli:2001zh,Andrianopoli:2001gm}. A conformal treatment has been given in \cite{Yamada:2019ttz}.

\subsection{Reduction of the \texorpdfstring{${\cal N}=2$}{N=2} Weyl Multiplet}
\label{ss:WeylN2N1}
To reduce the $D=4$, ${\cal N}=2$ Weyl multiplet, we consider (\ref{transfo4}) and  the Majorana spinors $\epsilon =\epsilon ^1+\epsilon _1$, $\eta  =\eta  ^1+\eta  _1$, and put $\epsilon ^2$ and $\eta _2$ to zero. We then compare with the transformations in \cite[Chap. 17]{Freedman:2012zz} for the reduction to ${\cal N}=1$.
Obviously this reduction breaks the $\SU(2)$ part of the $R$-symmetry group of ${\cal N}=2$ to a $\U(1)$ part, i.e. the part where
\begin{equation}
  \lambda_1{}^1=-\lambda _2{}^2\in \rmi \mathbb{R}\,,\qquad \lambda_1{}^2=\lambda _2{}^1=0\,.
 \label{remainingfromSU2}
\end{equation}

The frame field transforms to $\psi _{\mu 1}+\psi _\mu ^1$, which is thus the ${\cal N}=1$ gravitino. Its transformation is
\begin{equation}
  \delta\psi_\mu^1 = \left( \partial _\mu +\ft12 b_\mu   +\ft14 \gamma ^{ab}\omega  _{\mu ab}-\ft12\rmi A_\mu -V_\mu{}_1{}^1\right)    \epsilon^1 -\gamma _\mu \eta^1 \,.
 \label{delpsi1}
\end{equation}
This identifies the real combination
\begin{equation}
  A_\mu ^{(1)}= \ft13 A_\mu -\ft23\rmi V_\mu{}_1{}^1\,,
 \label{Amu1}
\end{equation}
as the $A_\mu $ field of ${\cal N}=1$, which transforms into
\begin{equation}
  \delta A_\mu ^{(1)}=-\ft12\rmi \epsilon ^1\phi _{\mu 1}+\ft18\rmi\epsilon ^1\gamma _\mu \chi _1+\ft12\rmi\eta _1\psi _\mu ^1+\hc
 \label{delAmu1}
\end{equation}
The composite field $\phi _\mu $ for  ${\cal N}=1$ is as in (\ref{solfphiN2}), but without the $\chi $-term, i.e.
\begin{equation}
  \phi _{\mu 1} = P_L \phi _\mu {}^{(1)}+\ft14\gamma _\mu \chi_1 \,,\qquad \phi _\mu {}^{(1)}=\phi _\mu -\ft14\gamma _\mu \chi \,,
 \label{phimudecomposed}
\end{equation}
so that the ($Q$-susy part of the) right-hand side of (\ref{delAmu1}) is indeed proportional to $\phi _\mu ^{(1)}$.

The transformation of $b_\mu $ in (\ref{transfo4}), rewritten using (\ref{phimudecomposed}), is
\begin{equation}
  \delta b_\mu =
      \ft12\bar{\epsilon}^1\phi_{\mu }{}^{(1)} -\ft14\bar{\epsilon}^1\gamma_\mu\chi_1   -\ft12\bar \eta ^1\psi _{\mu 1}    +\hc
 \label{delbmu}
\end{equation}
But $b_\mu $ transforms also under special conformal transformations, see (\ref{deltaDK}).
Therefore we identify the ${\cal N}=1$ supersymmetry as a linear combination of the $\epsilon ^1$ supersymmetry with a special conformal transformation:
\begin{equation}
 \delta ^{(1)}(\epsilon )= \delta (\epsilon ^1) + \delta _{\rm K}(\lambda _{{\rm K}\mu })\,,\qquad  \lambda _{{\rm K}\mu }=\ft18\bar \epsilon ^1\gamma _\mu\chi _1+\hc
 \label{Kcompensating}
\end{equation}
which eliminates the terms with $\chi _1$ in (\ref{delbmu}). The fields of the ${\cal N}=1$ Weyl multiplet are thus $\left\{e_\mu {}^a, \, b_\mu ,\, \psi _\mu ^1, ,\, A_\mu ^{(1)}\right\}$.
The combination (\ref{delAmu1}) implies that the superconformal $\U(1)$ of ${\cal N}=1$ is a diagonal subgroup of the $\U(1)\times \SU(2)$ of ${\cal N}=2$:
\begin{equation}
  \lambda _T^{(1)}= \ft13\left(\lambda _T -2\rmi \lambda _1{}^1\right)\,.
 \label{lambdaT1}
\end{equation}
Similarly, one can check that there is an ${\cal N}=1$ vector multiplet containing\footnote{The factors, e.g. $\rmi$ in the second component, are introduced such that the transformations agree with
\cite[(17.1)]{Freedman:2012zz}.}
\begin{equation}
  \left\{B_\mu ,\,\rmi(\chi ^1-\chi _1),\,D\right\}\,,\qquad B_\mu =-\ft23\left(A_\mu+\rmi V_\mu{}_1{}^1\right)\,.
 \label{N1multipletchi}
\end{equation}
The gauge field gauges a combination of $\lambda _T$ and $\rmi \lambda _1{}^1$, complementary to (\ref{lambdaT1}):
\begin{equation}
  \theta = -\ft23\left(\lambda _T+\rmi \lambda_1 {}^1\right)\,.
 \label{thetaN1}
\end{equation}

The remaining fields of the ${\cal N}=2$ Weyl multiplet form a double spin-3/2 multiplet:
\begin{equation}
\begin{array}{|c|cc|c|}\hline
{\cal N}=2 \mbox{ Weyl}& \mbox{Weyl multiplet} &\mbox{spin }1 \mbox{ multiplet}& 2\mbox{ spin }\ft32 \mbox{ multiplets}   \\
\hline
e_\mu {}^a & e_\mu {}^a & &\\
b_\mu & b_\mu &&\\
A_\mu ,\, V_{\mu 1}{}^1& A_\mu^{(1)} &B_\mu   &   \\
 V_{\mu 1}{}^2 &&&     V_{\mu 1}{}^2\in \mathbb{C} \\
T^-_{ab} & & & T_{ab}\\
D & & D & \\         \hline
\psi _\mu {}^i &\psi _\mu {}^1 & &\psi _\mu {}^2\\
\chi ^i &  &\chi ^1 &\chi ^2 \\ \hline
24+24&  8+8& 4+4 & 12+12\\ \hline
\end{array}
\label{N2N1Weyl}
\end{equation}
In conclusion, the $24+24$ fields, reduce to $8+8$ Weyl and a $4+4$ vector multiplet under the truncation. The remainder are two $6+6$ spin 3/2 multiplets. However, note that the spin 3/2 multiplets cannot be treated separately beyond the linear level, since the nonlinear transformations involve the frame field and the ${\cal N}=1$ gravitino.


\subsection{Reduction of the Compensating Vector Multiplet}
The superconformal transformations of the vector multiplet are in (\ref{transfVM4}). We now need the abelian version, i.e. we can omit the index $I$, and $f_{JK}{}^I=0$.
The complex scalar $X$ transforms chiral and is thus the basic field of a chiral multiplet:
\begin{equation}
  \left\{X,\, \frac{1}{\sqrt{2}}\Omega _1,\,Y_{11}\right\}\,.
 \label{chiralinvector}
\end{equation}
The scalar $X$ transforms under the $\theta $ transformation (\ref{thetaN1}). Indeed, considering the $\U(1)$ transformations:
\begin{equation}
  \delta X= \rmi \lambda _T X= \rmi \left(\lambda _T^{(1)}+\theta \right)X\,.
 \label{delXU1}
\end{equation}
Thus the chiral multiplet transforms under the gauge transformation of the vector multiplet (\ref{N1multipletchi}). This is also important to understand the following contribution to the $Q$-transformation of the auxiliary field $Y_{11}$. According to (\ref{covD4N2part1}), ${\cal D}_\mu \Omega _1$ contains a term $-2X\phi _{\mu 1}$, which, after use of (\ref{phimudecomposed}) implies that the transformation of $Y_{11}$ contains a term $\bar \epsilon _1 \chi _1\,X$. This term thus depends on the fermion of the gauge multiplet (\ref{N1multipletchi}). The latter is the transformation term that the auxiliary fields of chiral multiplets obtain in Wess--Zumino gauge in ${\cal N}=1$ supersymmetry due to the $\theta $ symmetry (\ref{delXU1}). For the normalization, see e.g.\cite[(17.3)]{Freedman:2012zz}.
\bigskip

The remaining fields of the ${\cal N}=2$ vector multiplet form the ${\cal N}=1$ vector multiplet:
\begin{equation}
  \left\{W_\mu ,\, -\Omega _2-\Omega ^2, \,2\rmi Y_{12}\right\}\,.
 \label{N1Wmultiplet}
\end{equation}
Thus the reduction goes as follows
\begin{equation}
\begin{array}{|c|c|c|}\hline
 {\cal N}=2\mbox{ vector}& \mbox{ spin } \ft12\mbox{ multiplet}& \mbox{spin }1\mbox{ multiplet}  \\
\hline
X & X\in\mathbb{C}& \\
W_\mu  &  &  W_\mu \\
Y_{ij} &Y_{11}\in\mathbb{C} &\rmi Y_{12}\in\mathbb{R} \\ \hline
\Omega _i & \Omega _1 & \Omega _2\\ \hline
8+8 &  4+4 & 4+4\\ \hline
\end{array}
\end{equation}
For clarity, we indicated the reality properties of the $Y_{11}$ and $Y_{12}$, see (\ref{Yijupcc}). The component $Y_{22}$ is the complex conjugate of $Y_{11}$.

One consistent reduction to ${\cal N}=1$ consists in keeping only the Weyl multiplet of (\ref{N2N1Weyl}) and the chiral multiplet (\ref{chiralinvector}), which provides then the compensating fields for dilatation, $\U(1)$ and S-supersymmetry. The conformal action of the chiral multiplet in the background of the Weyl multiplet gives the ${\cal N}=1$ pure supergravity with the `old-minimal' set of auxiliary fields.

If we keep also the spin-1 multiplet of (\ref{N2N1Weyl}) in the background, since the chiral multiplet transforms under $\theta $, see (\ref{delXU1}), this will induce a Fayet--Iliopoulos term $e\,X\,\bar X\,D$. However, to describe pure supergravity, there are no kinetic terms for the spin-1 multiplet. Therefore the field equation for $D$ is inconsistent with a gauge choice that gives a non-zero value to $X$. That is why we need the second compensating multiplet as viewed from the ${\cal N}=1$ perspective.

\subsection{Reduction of the Second Compensating Multiplet}

We comment here on the reduction of the two versions that have linear transformation laws.
When we use the hypermultiplet, there are two ${\cal N}=1$ chiral multiplets:
\begin{equation}
  \left\{A^{1\Ae},\, -\sqrt{2}\rmi \zeta ^\Ae,\, -\frac{1}{\sqrt{2}}\bar XF^{2\Ae}\right\}\,.
 \label{hyperN1}
\end{equation}
These are chiral multiplets for each value of $\Ae$. Remember that $A^{2\Ae}$ is dependent on the complex conjugate of $A^{1\Ae}$.
In the ${\cal N}=1$ reduction we can include one of these two chiral multiplets, e.g. $\Ae=1$.
Under the $\U(1)$ transformations:
\begin{equation}
  \delta A^{1\Ae}= A^{1\Ae}\lambda _1{}^1 =\rmi\left(\lambda _T^{(1)}+\ft12\theta \right) A^{1\Ae}\,.
 \label{delU1A}
\end{equation}
Hence, these multiplets transform also under $\theta $ and there is another Fayet--Iliopoulos term, such that the field equation of $D$ relates $X$ to $A^{1\Ae}$ and a consistent dilatational gauge (and gauge for $\lambda _T^{(1)}$ and $\theta $) can be chosen.

When we take for the second compensating multiplet of ${\cal N}=2$ the linear multiplet, we find that the ${\cal N}=1$ reduction leads to a chiral multiplet
\begin{equation}
  \left\{L^{11},\, \sqrt{2}\varphi ^1,\, -G\right\}\,,
 \label{chiralinlinear}
\end{equation}
and an ${\cal N}=1$ linear multiplet
\begin{equation}
  \left\{\rmi L^{12}, \, \phi ^2+\phi _2,\,E_{\mu \nu }\right\}\,.
 \label{linearN1inlinear}
\end{equation}
In Sect.~\ref{ss:tensorcompensator} we chose the gauge fixing (\ref{gaugeSU2L}). The proportionality with $\delta _{ij}$ was just an arbitrary choice in $\SU(2)$. Here we could choose the same (which means $L^{11}$ real, but as well $\rmi L^{12}$ to be the non-zero component. $L^{11}$ transforms under the remaining part of the $\SU(2)$, (\ref{remainingfromSU2}), and hence under the $\theta $ of (\ref{thetaN1}). On the other hand $L^{12}$ is invariant under that remaining $\U(1)$. Therefore, if we keep the chiral multiplet in the ${\cal N}=1$ reduction, it performs the same role as one of the multiplets (\ref{hyperN1}) to allow a consistent field equation for the auxiliary $D$. If we keep the linear multiplet in the reduction, then there is also a Fayet--Iliopoulos-like coupling of that field with $D$, but moreover a local $\U(1)$ remains. Hence we find the situation as in new minimal supergravity, having a preserved local $\U(1)$ (and gauge auxiliary tensor $E_{\mu \nu }$) coupled to the other matter multiplets of the reduction of the Weyl and first compensating multiplet.

\section{Matter-Coupled Supergravity}
\label{ss:mattersugra}
In this section we consider again the full action (\ref{Ltotalgh}), which includes vector multiplets and hypermultiplets, as well as the compensating multiplets. In light of the discussion in Sect.~\ref{ss:pureN2}, the latter is thus the extension of the action with an hypermultiplet as second compensating multiplet, considered in Sect.~\ref{ss:hypercompensator}. We choose the hypermultiplet as compensating multiplet such that the couplings of the physical hypermultiplets can be obtained as a projection from a conformal hyper-K\"{a}hler manifold, similar to the way in which the couplings of the vector multiplets are obtained as a projection from a conformal special K\"{a}hler manifold.
The disadvantage of this choice is that it is not known how to keep the structure with auxiliary fields $F^{i\Ae}$ mentioned in Sect.~\ref{ss:hypercompensator} for a non-trivial hyper-K\"{a}hler geometry without introducing an infinite number of fields as in harmonic or projective superspace. In total we therefore have $n+1$ vector multiplets and $n_H+1$ hypermultiplets, described by the action (\ref{Ltotalgh}), being $\mathcal{L}_g$, $\mathcal{L}_h$ the action for the vector multiplets and hypermultiplets, respectively. We thus redefine the range of the indices $X$ and $\Ae$ for the hypermultiplets to, respectively, $4(n+1)$ and $2(n_H+1)$ values.

Though in this section, as in the previous one, we will mostly focus on $D=4$ dimensions, some results for $D=5$ will also be presented along the way, exploiting the fact that the treatment is very similar. Full results for $D=5$ in the conformal setting are given in \cite{Kugo:2000af,Kugo:2002js,Bergshoeff:2002qk,Bergshoeff:2004kh} and with other methods in \cite{Gunaydin:1999zx,Ceresole:2000jd,Andrianopoli:2007xp}.

In Sect.~\ref{ss:pureN2} we took gauge choices, keeping the auxiliary fields. In the present section we first eliminate the auxiliary fields (Sect.~\ref{ss:eqom}) and then we consider the gauge fixing (Sect.~\ref{ss:gfwithmatter}). We conclude with the final formulae for the resulting action (Sect.~\ref{ss:fullD4}) and corresponding transformations (Sect.~\ref{ss:susygaugetrN2}).

\subsection{Elimination of Auxiliary Fields}
\label{ss:eqom}
We now present the equations of motion for the auxiliary fields of the matter-coupled supergravity in $D=4,5$. It has been expected for a long time, but proven only recently \cite{Vanhecke:2017chr}, that equations of motion can always be represented in the form of covariant equations. This means that gauge fields (in our case mostly the gravitino) appear only within covariant derivatives.\footnote{The proper field equation of a field $\phi ^i$ is in general not covariant if the derivative of that field $\partial _\mu \phi ^i$ appears in the transformation of another field $\delta \phi ^j$. However, using the field equation of $\phi ^j$ removes the non-covariant terms in the field equation of $\phi ^i$.\label{fn:coveom}}

\subsubsection{Case \texorpdfstring{$D=5$}{D=5}}

The vector multiplet and hypermultiplet actions are given by \eqref{Lform3} and \eqref{conf-hyperaction}. The theory then contains several auxiliary fields. In particular, the matter $D$, $T_{ab}$ and $\chi^i$, and gauge field $V_\mu^{ij}$ are inherited from the Weyl multiplet, while the gauge triplet $Y^I_{ij}$ belongs to the vector multiplet. Both $D$ and $\chi^i$ appear as Lagrange
multipliers in the action, leading to the following covariant equations of motion, where the symbol $\approx$ denotes the `on-shell' relation\footnote{Since the supersymmetry transformation of $D$ contains the spacetime derivative of $\chi $, see (\ref{modifiedtransf}), the proper field equation of $\chi $ contains a non-covariant term, which is however proportional to the field equation of $D$ and we have thus omitted here. Similar covariantizations have been done below and will not be indicated anymore.}:
\begin{eqnarray}
D &:& \qquad  C_{IJK} \sigma ^{I}
\sigma ^{J}\sigma ^{K} - \ft13 k_{\rm D}{}^2 \approx  0 \,,\nonumber\\
\chi^i &:& \qquad -8 \rmi \left(C_{ I J
K}\sigma ^{ I}\sigma ^{ J}\psi^{i K} -\ft{2}{3}\rmi \Ae^{i A} \zeta_{\Ae}\right)
\approx  0\,, \label{eq:Dchi-EOM}
\end{eqnarray}
where we used $A^{i \Ae}$ as defined by \eqref{defAialpha}.
The covariant equation of motion for the $\SU(2)$ gauge field $V_\mu^{ij}$ is
\begin{equation}
 0\approx k^{ij}_X {\cal D}_\mu q^{ X} - \ft 12
C_{ I J K} \sigma^{ K} \bar{\psi}^{i I}\gamma _\mu \psi^{j J}\,.
 \label{eomcovVmuD5}
\end{equation}
The first term, using (\ref{hypCovD5}) and  (\ref{usevAvBk}) contains, according to (\ref{hypCovD5}), $-\frac{2 }{9}k_{\rm D}{}^2\,V^{ij}_\mu $, so that this
equation can be read as
\begin{equation}
  \frac{2 }{9}k_{\rm D}{}^2\,V^{ij}_\mu \approx k^{ij}_X \partial _\mu q^{ X}-W_\mu ^IP_I^{ij}
  +\ft23\rmi \bar{\psi}_\mu^{(i} A^{j)\Ae}\zeta_\Ae - \ft 12
C_{ I J K} \sigma^{ K} \bar{\psi}^{i I}\gamma _\mu \psi^{j J}\,,
 \label{eomVmuD5}
\end{equation}
using (\ref{veckinA}) and the moment maps (\ref{mom_mapPI}).

The covariant equations of motion of the other auxiliary fields $Y^{ I}_{ij}$ and $T_{ab}$ are given by
\begin{align}
& Y^{ij\, J} C_{ I J K} \sigma^{ K} \approx - \ft12   P_I^{ij} + \ft14\,\rmi C_{ I J K} \bar\psi^{i J} \psi^{j K}\,,\nonumber\\
&\frac{64}{9} k_{\rm D}{}^2\,T_{ab} \approx  4 \sigma^{
I}\sigma^{ J}\hat{F}^{
K}_{ab}C_{ I\, J\, K} +\, \rmi\sigma^{ I}\,\bar{\psi}^{
J}{\gamma}_{ab} \psi^{ K}C_{ I J
K} + 2\rmi \bar{\zeta}_{ \Ae}{\gamma}_{ab}
\zeta^{\Ae}\,.\label{eq:T-EOM}
\end{align}

\subsubsection{Case \texorpdfstring{$D=4$}{D=4}} 

The vector multiplet and hypermultiplet actions are given, respectively, by  \eqref{leadingLvectorsd4} and \eqref{totaleLaghyp2}.
The fields $D$ and $\chi^i$ appear in the total action as Lagrange multipliers. Therefore, their (covariant) field equations imply the following conditions
\begin{eqnarray}
  \nonumber D: &&-N +\ft14k_{\rm D}{}^2 \approx  0,
  \nonumber\\
  \chi^i: && N_{IJ}\bar X^I \Omega_i^J +  \rmi d^{\bBe}{}_\Ae A_{i\bBe}\zeta^{\Ae} \approx 0\,.\label{v:veldvglD}
\end{eqnarray}

The covariant equations of motion for the $\U(1)$ gauge field $A_\mu $ and
$\SU(2)$ gauge field $V_{\mu i}{}^j$ are
\begin{align}
0\approx & \ft12\rmi N_{IJ}(X^I {\cal D}_\mu \bar{X}^J-\bar{X}^J {\cal D}_\mu X^I )+\ft18\rmi N_{IJ}\bar{\Omega}^{iI}\gamma_\mu \Omega_{i}^J -\ft12\rmi
 \bar{\zeta}_{\bar{\Ae}}\gamma_{\mu}\zeta^{\Be} d^{\bar{\Ae}}{}_\Be\,,\nonumber\\
0   \approx & g_{\hX \hY }{\cal D}_\mu  q^{\hX }\vec k^{\hY } - \ft14 N_{IJ} \bar\Omega^{I}_i\gamma_\mu \Omega^{Jj}\vec{\tau }_j{}^i \,,
\label{coveomD4AV}
\end{align}
which can be solved in the forms
\begin{align}
N\, A_\mu\approx & \ft12\rmi N_{IJ}(X^I{\partial}_\mu \bar{X}^J-\bar{X}^J {\partial}_\mu X^I )+W_\mu ^IP_I^0 +\ft14\rmi N_{IJ}\left(\bar X^I\bar \psi _\mu ^i\Omega _i^J-X^I\bar \psi _{\mu i}\Omega ^{iJ}\right)
\nonumber\\
&+\ft18\rmi N_{IJ}\bar{\Omega}^{iI}\gamma_\mu \Omega_{i}^J -\ft12\rmi  \bar{\zeta}_{\bar{\Ae}}\gamma_{\mu}\zeta^{\Be} d^{\bar{\Ae}}{}_\Be\,,\nonumber\\
-\ft12k_{\rm D}{}^2\,  \vec{V}_{\mu }   \approx & g_{\hX \hY }\partial _\mu  q^{\hX }\vec k^{\hY }-W_\mu ^I\vec{P}_I
+\ft12\rmi \vec{\tau }_i{}^j d^\bAe{}_\Be\left(A_{j\bAe}\bar \psi _\mu ^i\zeta ^A- A^{i\Be}\bar \psi _{\mu j}\zeta _\bAe\right)\nonumber\\
&- \ft14 N_{IJ} \bar\Omega^{I}_i\gamma_\mu \Omega^{Jj}\vec{\tau }_j{}^i\,,
\label{v:SU2veldvgl}
\end{align}
where for the $\SU(2)$ part again we used (\ref{veckinA}) and the moment maps (\ref{mom_mapPI}). For the $\U(1)$ part, the moment maps for the K\"{a}hler isometries $P^0_I$ appear. They are real functions, defined\footnote{For a general discussion of moment maps in K\"{a}hler manifolds, see e.g. \cite[Sect. 13.4.1]{Freedman:2012zz}, and also Sect.~\ref{ss:isomKahler} below.} such that
\begin{equation}
N_{IJ}  \delta(\theta ) X^J =-\rmi \theta ^I \partial _{\bar J}P^0_I\,.
 \label{defgoalPI0}
\end{equation}
The condition above is satisfied by
\begin{equation}
  P_I^0= \rmi N_{JL}f_{KI}{}^JX^K\bar X^L\,,
 \label{PI0def}
\end{equation}
due to the definition (\ref{NIJ}) and the homogeneity relations (\ref{homegeneqs}). Inserting again (\ref{NIJ}) and  (\ref{defCijk}) it can be rewritten in various ways:
\begin{eqnarray}
  P^0_I  &=& f_{IJ}{}^K(X^J\bar F_K+\bar X^J F_K)-2C_{I,JK}X^J\bar X^K\nonumber\\
   &=&-\rmi N_{IJ}f_{KL}{}^JX^K\bar X^L=\rmi N_{JL}f_{KI}{}^J X^K\bar X^L=-\rmi N_{JK}f_{LI}{}^JX^K\bar X^L \,, \label{P0N2}
\label{kalmomM}
\end{eqnarray}
and
\begin{equation}
  X^I P_I^0 =\bar X^IP_I^0 = 0\,.
 \label{XP0}
\end{equation}

The covariant field equations of the other auxiliary fields, $Y^{ij}$ and $T_{ab}$, are:
\begin{eqnarray}
   N_{IJ}\,Y^J_{ij} &\approx&
- P_{Iij}  -\ft14\rmi F_{IJK}\bar\Omega_i^J\Omega_j^K+\ft14\rmi \bar F_{IJK}\varepsilon _{ik}\varepsilon _{j\ell}\bar\Omega^{kJ}\Omega^{\ell K}\,,\nonumber\\
  \ft18 N_{IJ}X^I  X^J T^+_{ab} &\approx & \ft14 N_{IJ}X^I \widehat F^{+J}_{ab} -\ft1{32}\rmi\bar F_{IJK}\bar\Omega^{iI}\gamma_{ab}\Omega^{jJ}X^K\varepsilon _{ij}  -\ft18 \bar\zeta_{\bAe}\gamma_{ab}\zeta_{\bBe} C^{\bAe \bBe}\,.\nonumber\\
  \label{YTeom}
\end{eqnarray}

\subsection{Gauge Fixing for Matter-Coupled Supergravity}
\label{ss:gfwithmatter}
We now move to the gauge fixing of the total action \eqref{Ltotalgh}. Though the principles are the same for
 $D=5$ (and for a large part also for $D=6$) for simplicity here we concentrate on 4 dimensions, and comment on $D=5$ and $D=6$ in Sect.~\ref{ss:gfD56}.
\vspace{10pt}

As mentioned before, the conformal coupling of vector multiplets and hypermultiplets, has been chosen
such that there is in each sector one multiplet with negative kinetic
energy, used for gauge fixing, and all the others have positive kinetic energies.\footnote{The positivity of the kinetic energy is a further constraint on the choices of $N(X,\bar X)$ and the quaternionic metric, which we discuss further in Sect. \ref{ss:positivity}.} We recall that \eqref{Ltotalgh} contains the fields of the Weyl multiplet as a background. Dependent fields in this Weyl
multiplet have already been solved for in the final actions that we
presented. As mentioned in (\ref{eq:K}), the dilatation gauge field $b_\mu$ can be
eliminated by a gauge choice for the special conformal transformations
$K$. We will combine the field equation of the auxiliary field $D$, (\ref{v:veldvglD}), with a dilatation gauge
choice to eliminate the modulus of the scalars of the compensating
multiplets.
The $\U(1)$ gauge eliminates the phase of the scalar of the
compensating vector multiplet.\footnote{Here is the only difference with 5
dimensions. In that case the superconformal algebra does not contain a
$\U(1)$, but the scalar of the vector multiplet is also real, such that
this step is not necessary, see Sect.~\ref{ss:gfD56}.} Similarly the quaternionic phases of the
scalars of the compensating hypermultiplet are fixed by gauge choices of
the $\SU(2)$.

On the fermionic side the $\chi^i $ field equation will be combined  with a gauge for $S$-super\-symmetry to eliminate the fermions in the compensating multiplets.

The only remaining field of the compensating multiplets is
the vector field $W_\mu$, which becomes the graviphoton in the final
super-Poincar\'{e} action. The result will be the coupling of $n_H$
hypermultiplets and $n$ vector multiplets to $\mathcal{N} =2$, $D=4$ Poincar\'{e}
supergravity, see Sect. \ref{ss:fullD4}.

It is important to remark that the fixing of gauge symmetries leads to a change in the
definitions of covariant derivatives. Indeed, the fields that we use in
the Poincar\'{e} supergravity are chosen such that they do not transform any
more under the broken symmetries. Moreover, the parameters of the broken
symmetries will be expressed in terms of the independent parameters by
the decomposition laws, similar to (\ref{decompK}). 

We now give the explicit expressions for the gauge fixing of dilatations and $S$-super\-symmetry. For the dilatations, we have already
mentioned in the simple example (\ref{gaugefixkappa}) that our aim is to
have a standard kinetic term for gravity. Hence, we just collect the
terms with the scalar curvature and obtain\footnote{The dimensionful coupling constant $\kappa $ in fact appears here for the first time, thus breaking the dilatation invariance.} \index{$D$-gauge}
\begin{equation}
  D \mbox{-gauge:}\qquad -\ft16 N -\ft1{12}k_{\rm D}{}^2 = \ft12\kappa ^{-2}\,.
  \label{DDgauge}
\end{equation}
Combining this with the field equation for $D$, (\ref{v:veldvglD}), we
obtain
\begin{equation}\label{Dgaugegeneral}
  N = -\kappa ^{-2}\,, \qquad  k_{\rm D}{}^2 = -4\kappa ^{-2}\,.
\end{equation}
The condition that fixes the dilatation gauge is physically the requirement
that the kinetic terms of the scalars and the spin-2 field are not mixed.

The total action does contain terms of the form
$\gamma^{\mu\nu}\partial_\mu\psi^i_\nu$, multiplied with fermion fields
of the vector and hypermultiplet. These would imply a mixing of the
kinetic terms of spin $3/2$ and $1/2$. We choose, in analogy to the
bosonic sector, the $S$-gauge condition such that such a mixing does not
occur. Hence, we put to zero the coefficient of
$\gamma^{\mu\nu}\partial_\mu\psi^i_\nu$: \index{$S$-gauge}
\begin{equation}
S\mbox{-gauge:}\qquad  N_{IJ}\bar X^I \Omega^J_i -2\rmi
A_{i \bBe}d^{\bBe}{}_\Ae \zeta^{\Ae} =
0.\label{v:S-ijk}
\end{equation}
Combining  (\ref{v:S-ijk}) with the field equation of the auxiliary field $\chi_i$,
(\ref{v:veldvglD}), leads to:
\begin{equation}
N_{IJ}\bar X^I \Omega^J_i=0, \qquad  A_{i \bBe}d^{\bBe}{}_\Ae \zeta^{\Ae} = 0.\label{v:zeta_i02}
\end{equation}
Importantly, at this point the $R$-symmetry is still gauged. The gauge fixing of the latter will have some surprising consequences on the Nature of our ${\cal N}=2$ theories, leading to the so-called \emph{Special geometries}. This important passage will be therefore discussed separately, in Sect.~\ref{ss:projcoord}.

\subsection{Full Action for \texorpdfstring{$D=4$}{D=4}}
\label{ss:fullD4}
We now have all the ingredients to write down the full action. Using the gauge conditions and the definition (\ref{calFVM4}), we find:
 \begin{eqnarray}
e^{-1}\mathcal{L}&=&\kappa ^{-2}\, \left( \ft12R -\bar\psi_{i\mu } \gamma^{\mu \nu \rho }D_\nu  \psi_\rho ^i
\right)
   -N_{IJ}D_{\mu}X^ID^{\mu}\bar{X}^J -\ft12 g_{XY}D_\mu q^XD^\mu  q^Y\nonumber\\
 &&+ \left\{-\ft14\rmi\bar F_{IJ}F^{+I}_{\mu
\nu}  F^{+\mu \nu J}
+\ft1{16}N_{IJ}X^I X^J T^+_{ab} T^{+ab}+\hc\right\}\nonumber\\
 &&-\,
N_{IJ}\vec{Y}^{I}\cdot \vec{Y}^J 
                                 -N^{-1|IJ}P_I^0P_J^0
   -2  \bar X^I X^J k_I{}^X k_{JX} \nonumber\\
   &&+\,\ft23C_{I,JK}e^{-1}\varepsilon ^{\mu \nu \rho \sigma }W_\mu {}^IW_\nu {}^J\left(\partial _\rho W_\sigma {}^K+\ft38f_{LM}{}^KW_\rho {}^LW_\sigma {}^M \right) \nonumber\\
   &&+\,\ft12
\bar{\psi}_{ai}\gamma^{abc}\psi^j_b\left( \delta _j^i N_{IJ}\bar{X}^ID_cX^J + D_c q^X
   \vec{k}_X\cdot \vec {\tau}_j{}^i\right) \nonumber\\
&&+\left\{-\,\ft14N_{IJ}\bar{\Omega}^{iI}\hat{\slashed{D}}\Omega_i^J
 -\bar\zeta_\bAe\hat{\slashed{D}} \zeta^\Be
  d^\bAe{}_\Be\right.
 \nonumber\\
 &&+\,\ft12N_{IJ}\bar{\psi}_{ia}\slashed{D}X^I\gamma^{a}\Omega^{iJ}+  \rmi \bar\psi_{ia}
   \slashed{D} q^X\gamma^a\zeta_\bAe f^{i\Be }{}_Xd^\bAe{}_\Be
  \nonumber\\
&&  -\,\ft12N_{IJ}\varepsilon _{ij}\left(\bar \Omega ^{iI}\gamma _\mu-\bar X^I\bar \psi_\mu ^i\right)   \psi ^j_\nu F^{-\,\mu \nu J}
-\ft1{16}\rmi F_{IJK}\bar \Omega _i^I\gamma ^{\mu \nu }\Omega _j^J\varepsilon ^{ij} F_{\mu \nu }^{-\,K}\nonumber\\
&&+\,\ft12\bar\psi_{ai}\gamma^a\left[\Omega^I_j  P_I{}^{ij} +\Omega _j^IN_{IJ}f_{KL}{}^JX^L\bar X^K \varepsilon ^{ij}-4\rmi X^I k_I{}^X f^{i\Be }{}_X C_{\Be\Ae}  \zeta^\Ae \right]
 \nonumber\\
&&+\,\ft12N_{IJ}\bar \Omega _i^If_{KL}{}^J \Omega _j^L\bar X^K \varepsilon ^{ij}
+\ft12 \bar X^I \bar \psi_a^i \gamma^{ab} \psi_b^j
  P_{Iij}\nonumber\\
    &&+\,2 X^I\bar\zeta ^\Ae \zeta ^\Be t_{I\Ae \Be }
    + 2 \rmi k_I^X f^{i\Be }{}_X \varepsilon _{ij}d^\bAe{}_\Be  \bar\zeta_\bAe  \Omega^{jI}
+\hc\Big{\}}+\,\mbox{4-fermion terms.}\nonumber\\
\label{totaleLagN2D4}
\end{eqnarray}
We have used the field equations to determine the auxiliary fields, which are thus no longer to be
considered as independent fields, but stand for their values in  (\ref{v:SU2veldvgl})--(\ref{YTeom}).
For example, the term that was $+N_{IJ}\vec{Y}^{I}\cdot \vec{Y}^J +$ other $\vec{Y}$ terms in (\ref{leadingLvectorsd4}) and  (\ref{totaleLaghyp2}) are now written as $-N_{IJ}\vec{Y}^{I}\cdot \vec{Y}^J$.
The covariant derivatives are (omitting contributions that give
rise to 4-fermion terms)\footnote{In the previous chapter, the auxiliary fields were considered independent, and appeared as such in the `linear' part of the covariant derivatives $D_\mu $, see e.g. (\ref{covD4N2part1}) or (\ref{calDhyper}). Here we consider e.g. $A_\mu $ as its solution in (\ref{valueAVN2}), and the linear part written as $D_\mu $ contains only ${\cal A}_\mu $.}
\begin{eqnarray}
 D_\mu X^I&=& \left(\partial _\mu  -\rmi {\cal A}_\mu \right) X^I+W_\mu {}^J
X^Kf_{JK}{}^I \,,  \nonumber\\
  D_\mu q^X&=& \partial _\mu q^X+2\vec{{\cal V}}_\mu \cdot \vec{k}^{X}-
 W_\mu {}^Ik_I{}^X\,,\nonumber\\
 \hat{D}_\mu\Omega_i^I&=&D_\mu\Omega_i^I+\Gamma^I{}_{JK}\Omega^K D_\mu X^I\,,\nonumber\\
  D_\mu \Omega_i ^I&=&  \left( \partial _\mu +\ft14\omega _\mu
{}^{ab}(e)\gamma _{ab}-\ft12\rmi {\cal A}_\mu \right)\Omega^I _i
 +{\cal V}_{\mu i}{}^j\Omega _j{}^I
+W_\mu ^J\Omega _i^K f_{JK}{}^I\,, \nonumber\\
\hat{D}_\mu \zeta ^\Ae &=& \left( \partial _\mu+\ft14\omega _\mu
{}^{ab}(e)\gamma _{ab}  +\ft12\rmi {\cal A}_\mu\right) \zeta ^\Ae  -W_\mu
^It_{I\Be }{}^\Ae \zeta ^\Be +\partial _\mu q^X \omega _{X\Be }{}^\Ae \zeta ^\Be \,,\nonumber\\
  D_\mu \psi _{\nu i}&=&\left( \partial _\mu +\ft14\omega _\mu
{}^{ab}(e)\gamma _{ab}+\ft12\rmi {\cal A}_\mu \right)\psi _{\nu i}
 +{\cal V}_{\mu i}{}^j\psi _{\nu j}\,,
 \label{covderafterfieldeqn}
\end{eqnarray}
The connection $\Gamma ^{I}_{JK}$ refers to the completely holomorphic connection in a K\"{a}hler manifold with metric $N_{IJ}$. See below, (\ref{special_levicivita}), for more detail.
${\cal A}_\mu$ and ${\cal V}_{\mu i}{}^j$ are the bosonic parts of their expressions in \eqref{v:SU2veldvgl}:
\begin{eqnarray}
A_\mu &=& {\cal A}_\mu +A_\mu ^{\rm F}\,,\nonumber\\
&& {\cal A}_\mu   =  -\ft12\rmi\kappa ^2 N_{IJ}
(X^I\hat{\partial} _\mu\bar X^J-\bar X^J\hat{\partial}_\mu X^I)\nonumber\\
&&\phantom{ {\cal A}_\mu   }= {\cal A}_\mu ^0 -\kappa ^2W_\mu ^IP_I^0\,,\nonumber\\
&&{\cal A}_\mu ^0 = -\ft12\rmi\kappa ^2 N_{IJ}
(X^I{\partial} _\mu\bar X^J-\bar X^J{\partial}_\mu X^I)
\,,\nonumber\\
&& A_\mu^{\rm F}= -\ft18 \rmi\kappa ^2 N_{IJ}\bar \Omega ^{iI}\gamma _\mu \Omega _i^J+\ft12\rmi \kappa ^2 \bar \zeta _\bAe\gamma _\mu \zeta ^\Be  d^\bAe{}_\Be \,,
\nonumber\\
\vec{V}_\mu &=&\vec{{\cal V}}_\mu +\vec{V}_\mu^{\rm F},\nonumber\\
&&\vec{{\cal V}}_\mu  =  \ft12\kappa ^2\vec{k}_X\hat{\partial} _\mu q^X=\ft12\kappa ^2\left( \vec{k}_X\partial _\mu q^X -W_\mu ^I\vec{P}_I\right) \,,\nonumber\\
&&\vec{V}_\mu^{\rm F}= \ft18\kappa ^2N_{IJ}\bar \Omega ^{iI}\gamma _\mu
\Omega _j^J\vec{\tau }_i{}^j\,,
 \label{valueAVN2}
\end{eqnarray}
where we used
\begin{eqnarray}
  \hat{\partial }_\mu X^I&=&\partial _\mu  X^I+W_\mu {}^J
X^Kf_{JK}{}^I\,,\nonumber\\
 \hat{\partial }_\mu q^X&=&\partial _\mu q^X-W_\mu {}^Ik_I{}^X\,.
\label{hatpartialN2}
\end{eqnarray}
Another way to split the covariant derivative $D_\mu X^I$ in (\ref{covderafterfieldeqn}) splitting the geometric and the gauge parts is
\begin{align}
D_\mu X^I =& \nabla _\mu X^I
+ W_\mu ^K\left(\rmi\kappa ^2P_K^0X^I + f_{KJ}{}^IX^J\right)\,,\nonumber\\
  \nabla_\mu  X^I\equiv& \partial _\mu X^I- \rmi{\cal A}^0_\mu X^I\,.
 \label{nablaX}
\end{align}

\subsection{Supersymmetry Transformations}
 \label{ss:susygaugetrN2}
 The transformation laws for physical fields can be obtained by plugging into the superconformal transformations the values for the gauge fixing as well as the parameters as fixed by the decomposition laws.
The superconformal transformations from which we have to start are those of
the Weyl multiplet, (see Sect.~\ref{WeylDeq4}), the vector multiplets,
(\ref{transfVM4}), and the hypermultiplets (\ref{hypCovD4}). Restricting ourselves to the
physical fields, the $Q$- and $S$-supersymmetries are (using also (\ref{covtranhyperD4A}))
\begin{eqnarray}
       \delta e_\mu{}^a
      &=& \ft12\bar{\epsilon}^i\gamma^a\psi_{\mu i}+\hc\,, \nonumber\\
     \delta\psi_\mu^i &=& \left( \partial _\mu
  +\ft14 \gamma ^{ab}\omega  _{\mu ab}-\ft12\rmi A_\mu \right)
   \epsilon^i +V_\mu {}^i{}_j \epsilon ^j  -\ft1 {16} \gamma^{ab} T_{ab}^-\varepsilon^{ij}\gamma_\mu\epsilon_j -\gamma_\mu\eta^i \,,\nonumber\\
 \delta X^I & = &
 \ft12 \bar \epsilon ^i\Omega^I _i\,, \nonumber\\
  \delta \Omega _i^I & = &\slashed{\cal D} X^I\epsilon _i +
  \ft14 \gamma^{ab} {\cal F}_{ab}{}^I\varepsilon _{ij}\epsilon ^j
  +Y_{ij}{}^I\epsilon^j+X^J\bar X{}^Kf_{JK}{}^I\varepsilon _{ij}\epsilon ^j
  +2X^I\eta _i
  \,,\nonumber\\
  \delta W_\mu^I &=&\ft12
 \varepsilon ^{ij}\bar \epsilon _{i}\gamma _\mu \Omega _j{}^I+\varepsilon ^{ij}\bar \epsilon _{i}\psi _{\mu j}X^I
  +\hc
\,, \nonumber\\
 \delta q^X&=& -\rmi \bar{\epsilon}^i\zeta^\Ae f^X{}_{i\Ae }
+\rmi \varepsilon ^{ij}\rho ^{\bAe \Be } \bar{\epsilon}_i\zeta_\bAe f^X{}_{j\Be }\,,\nonumber\\
  \widehat{\delta} A^{i\Ae} & =& -\rmi\bar \epsilon ^i \zeta ^\Ae +\rmi\bar \epsilon_j \zeta _{\bBe}\,\varepsilon ^{ji}\rho ^{\bBe\Ae}\,, \nonumber\\
\widehat{\delta}\zeta^\Ae &=&\ft12\rmi \widehat{\slashed{\cal D}}A^{i\Ae }
\epsilon _i+\rmi \bar
X^Ik_I{}^Xf^{i\Ae }{}_X\varepsilon _{ij}\epsilon   ^j+\rmi A^{i\Ae }\eta _i \,,\label{transfoN4conf}
\end{eqnarray}
where we use the covariant transformations as defined in (\ref{hatisnabla}). Covariant derivatives and covariant gauge field strengths are given in (\ref{calFVM4}),  (\ref{covD4N2part1}), (\ref{wideDA}).
The relevant bosonic symmetry transformations are the $\U(1)$ and
$\SU(2)$ $R$-symmetries and the Yang--Mills gauge symmetries:
\begin{eqnarray}
\delta\psi_\mu^i &=&\ft12\rmi\lambda_T\psi_\mu^i+\psi_\mu^j\lambda_j{}^i \,,\nonumber\\
  \delta X^I &=&\rmi\lambda_T X^I+ \theta ^J X^K f_{KJ}{}^I\,,\nonumber\\
\delta \Omega _i^I&=& \ft12\rmi\lambda_T  \Omega _i^I-\lambda _i{}^j\Omega _j^I+\theta ^J \Omega _i^K f_{KJ}{}^I\,,\nonumber\\
\delta W_\mu^I&=& \partial _\mu \theta ^I +\theta ^J W_\mu^K f_{KJ}{}^I\,,\nonumber\\
\delta q^X &=&-2\vec{\lambda }\cdot\vec{k}^X +\theta ^I k_I{}^X\,,\nonumber\\
\hat{\delta}A^{i\Ae}&=&A^{j\Ae}\lambda_j {}^i+ A^{i\Be} \theta ^I t_{I\Be}{}^{\Ae}\,,\nonumber\\
\hat{\delta}\zeta^\Ae &=&-\ft12\rmi\lambda_T \zeta^\Ae +\theta ^I \zeta^\Be
t_{I\Be }{}^\Ae \,.
\label{delgaugeN2}
\end{eqnarray}

\subsubsection{Decomposition Laws}
The gauge choices from Sect.~\ref{ss:gfwithmatter} imply that the parameters of the gauge fixed
symmetries are functions of the remaining gauge symmetries. The decomposition law for the $K$ symmetry is given in (\ref{decompK}).
But since none of the fields in (\ref{transfoN4conf}) transforms under special conformal transformations, we do not need this decomposition law of $\lambda_{\rm K}^\mu$.

The $S$ gauge (\ref{v:S-ijk}) is the $Q$-supersymmetry transformed of the $D$-gauge (\ref{DDgauge}). Therefore, with this choice, the latter is invariant under
ordinary supersymmetry, and the invariance then implies that the dilatation parameter can be put to zero:
  \begin{equation}\label{dilchoic}
  \lambda _{\rm D} = 0\,.
  \end{equation}

We also need the decomposition law for the $S$-supersymmetry. We can
calculate the variation of any of the two equations in (\ref{v:zeta_i02}), which should lead to the same result. These equations are gauge invariant. We use the field equations of Sect. \ref{ss:eqom}.
When calculating the $Q$-supersymmetry of the first one,
a useful relation due to the
field equation (\ref{YTeom}) is
\begin{equation}
  N_{IJ}\bar X^J{\cal F}^{-I}_{ab}=
  -\ft18\rmi F_{IJK}\bar \Omega ^I_i\gamma _{ab}\Omega _j^J\bar X^K\varepsilon ^{ij}
  +\ft12C_{\Ap \Bp }\bar \zeta ^\Ap \gamma _{ab}\zeta ^\Bp \,.
\label{NcalF}
\end{equation}
In both ways we obtain
  \begin{align}
   \kappa ^{-2} \eta^i(\epsilon ) =&-\ft12 P_I^{ij}X^I\epsilon _j+\ft12\kappa ^{-2}\gamma ^a\epsilon ^j V^{\rm F}_{aj}{}^i\nonumber\\
   &+\ft14\gamma _a\epsilon ^i \bar \zeta _{\bAe}\gamma ^a\zeta ^\Be d^\bAe{}_\Be
    +\ft1{16}\gamma_{ab}\varepsilon ^{ij}\epsilon_jC^{\bAe\bBe} \bar \zeta_{\bAe}\gamma^{ab}\zeta_{\bBe}\,,
      \label{v:decomprule_eta}
  \end{align}
 where $V^{\rm F}_{ai}{}^j$ is the traceless expression in (\ref{valueAVN2}):
\begin{equation}
  V^{\rm F}_{ai}{}^j=-\ft14\kappa ^2\left(\bar \Omega _i^I \gamma _a\Omega ^{jJ} -\ft12 \delta _i^j \bar \Omega _k^I \gamma _a\Omega ^{kJ}\right)N_{IJ}\,.
 \label{VFij}
\end{equation}

Finally, the gauge fixing of the $R$-symmetries will be discussed in Sect. \ref{ss:projcoord}. The $\U(1)$ and $\SU(2)$ decomposition laws are presented (\ref{delzdecompcov}) and (\ref{veclambdagf}), respectively.

\section{Vector Multiplet Scalars:\\ Special K\"{a}hler Geometry}
\label{ss:projcoord}

The constraints (\ref{Dgaugegeneral}), obtained by combining the dilatational gauge fixing with the $D$-field equation, determine  a real condition on the manifold spanned by the scalars $\{X^I\}$ and $\{q^X\}$. As we will see, the gauge fixings of the $\U(1)$ and $\SU(2)$ $R$-symmetries remove another real degree of freedom of the $\{X^I\}$ and 3 real degrees of freedom of the $\{q^X\}$.
This procedure defines a complex $n$-dimensional submanifold of the manifold spanned by the $(n+1)$ variables $X^I$, as well as a quaternionic $n_H$-dimensional submanifold of the real $4(n_H+1)$-dimensional manifold spanned by the $q^X$. These submanifolds can be identified by choosing an appropriate system of (projective) coordinates. This identification, for what regards the scalars of the vector multiplets which defines special K\"{a}hler geometry, will be the subject of this section. Similarly, as we will present in Sect. \ref{coord_quatman}, the scalars of the hypermultiplets define quaternionic-K\"{a}hler manifolds.

As a further simplification, in this section we will focus on the geometric part for the scalars of the vector multiplet, hence reducing $D_\mu X^I$ to $\nabla _\mu  X^I$ as in (\ref{nablaX}), and leaving the connection with the gauge vectors $W_\mu ^I$ to the following chapters. Upon this reduction, the relevant term of (\ref{totaleLagN2D4}) are:
\begin{equation}
e^{-1}\mathcal{L}_{X}=-g^{\mu \nu }N_{I J}\nabla _\mu X^I\nabla _\nu \bar X^{ J}\,,\qquad \nabla_\mu  X^I\equiv \partial _\mu X^I- \rmi{\cal A}^0_\mu X^I\,,
\label{lagrangiancomplex}
\end{equation}
where $N_{IJ}$ is defined in (\ref{NIJ}). Due to the homogeneity of $F(X)$---see \eqref{homegeneqs} and \eqref{NIJ}---one also has
\begin{equation}
  N_{I J}=\frac{\partial }{\partial X^I}\frac{\partial }{\partial \bar X^J}N\equiv \partial_I\bar{\partial}_J N\,.
 \label{NKahlerhom}
\end{equation}

\subsection{Rigid Special K\"{a}hler Manifold}
\label{ss:rigidspecial}

We first look back to the vector multiplets in rigid supersymmetry with conformal symmetry. Hence, we just consider $\partial _\mu X^I$ in (\ref{lagrangiancomplex}). We also do not consider the constraints  (\ref{Dgaugegeneral}) and the fields $X^I$ are thus independent. Then (\ref{lagrangiancomplex}) defines $X^I$ as complex coordinates of a K\"{a}hler manifold with K\"{a}hler metric\footnote{To apply the general relations of K\"{a}hler manifolds we have to distinguish indices $I$ and $\bar I$, which cannot be done consistently for the relations that follow from the prepotential $F$, see e.g. (\ref{NIJ}) or the first of (\ref{GIJNIJ}) where at the right-hand side $N_{IJ}$ is symmetric and at the left-hand side there is a holomorphic and an anti-holomorphic index. In (\ref{special_levicivita}) we use the metric connection with all holomorphic indices, considering just $G_{I\bar J}$ as metric, ignoring these relations to a prepotential $F$.
\label{fn:Gammaholomoph}}:
\begin{equation}
  G_{I\bar J}=N_{IJ}\,,\qquad G_{IJ}= G_{\bar I\bar J}=0\,,
 \label{GIJNIJ}
\end{equation}
and K\"{a}hler potential $N$. Because the manifold is K\"{a}hler, the affine connection is given by \cite[(13.19)]{Freedman:2012zz}
\begin{equation}\label{special_levicivita}
\Gamma^I_{JK} =G^{I\bar L}\partial_J G_{K \bar L}= N^{-1|I L}\partial_J N_{K L}=-\rmi N^{-1|I L}F_{JKL}\,.
\end{equation}
The homogeneity equations also imply
\begin{equation}
 X^K {\partial_K} N_{I J}=-\rmi X^K F_{IJK}= 0\,,\qquad X^J\Gamma^I_{JK} =0\,,
\label{relNphi}
\end{equation}
and their complex conjugates.  Therefore $k_{\rm D}{}^I=X^I$ is a closed homothetic
Killing vector of this metric (look at (\ref{homothetic}) with $w=1$).

This K\"{a}hler geometry encountered in rigid supersymmetry is in the mathematics literature also indicated as affine special K\"{a}hler geometry \index{affine special K\"{a}hler}
 \cite{Sierra:1983cc,Craps:1997gp,Freed:1997dp,Alekseevsky:1999ts}.

\subsection{Coordinates in the Projective Special  K\"{a}hler Manifold}\label{proj_kahler_intro}

We now turn to the supergravity case, where a key ingredient is the gauged $\U(1)$ $T$-symmetry.
%
First, there is the contribution of the $T$-gauge field ${\cal A}^0$ in  (\ref{lagrangiancomplex}). Furthermore, we will impose the first relation of (\ref{Dgaugegeneral}), which was related to the gauge choice for dilatations, and a gauge choice for the  $T$-symmetry.\footnote{Note that such a $\U(1)$ symmetry, here included as part of the superconformal algebra, in fact follows from the presence of a complex structure and the dilatational symmetry, see e.g. \cite[Sect. 17.3.2]{Freedman:2012zz}. Similarly the $\SU(2)$ symmetry in the hypermultiplet sector follows from the hypercomplex structure and the dilatation.} We will see how this projection from the $n+1$-dimensional (complex) K\"{a}hler manifold of Sect. \ref{ss:rigidspecial} leads to $n$-dimensional special K\"{a}hler geometry~\cite{deWit:1984rz,deWit:1984pk,Strominger:1990pd}, which is sometimes also denoted as `projective' special K\"{a}hler, and has a non-trivial $\U(1)$ curvature.\footnote{The reader will find in Chap. \ref{ss:specialgeom} a more detailed discussion on special K\"{a}hler geometry independent of this construction procedure.} \index{projective special K\"{a}hler}

This mechanism for the emergence of projective special K\"{a}hler manifolds is similar to what appears in ${\cal N}=1$ and has as such been reviewed in \cite[Sect. 17.3]{Freedman:2012zz}.

To exhibit the projective nature of the manifold, we first consider the on-shell value of the auxiliary fields as it follows from (\ref{lagrangiancomplex}):
\begin{align}
  {\cal A}^0_\mu =&\frac{\rmi}{2N}\left[
X^IN_{I J}(\partial _\mu \bar X^{ J})-
  (\partial _\mu X^I)N_{I J} \bar X^{ J}\right]\nonumber\\ =&
  \frac{\rmi}{2N}\left[(\partial _\mu \bar X^{ J})\bar{\partial}_{J}N-(\partial _\mu X^I)\partial _IN \right] \,.
 \label{valueAKahler}
\end{align}
One can check that this agrees with  (\ref{valueAVN2}) upon using the gauge condition (\ref{Dgaugegeneral}).
The value of (\ref{lagrangiancomplex}) using  (\ref{valueAKahler}) is
\begin{eqnarray}
e^{-1}{\cal L}_{X}&=&
- N_{I J}\partial_\mu X^I\partial^\mu\bar X^{ J} +\frac
{1}{4N}\left[\partial_\mu\bar X^{ J}
\bar{\partial} _J N -\partial_\mu X^I\partial _I  N\right]^2\nonumber\\
&=& -\partial_\mu X^I\partial^\mu\bar X^{ J}\left[\partial _I\bar{\partial}_J N-\frac{1}{N}(\partial _I N)(\bar{\partial} _J N)\right]
-\frac {1}{4N}
\partial _\mu N\,\partial ^\mu N\nonumber\\
&=&-N\partial_\mu X^I\partial^\mu\bar X^{ J}\partial _I
\bar{\partial} _J\log |N|-\frac {1}{4N}
\partial _\mu N\,\partial ^\mu N\,.\label{lagranN}
\end{eqnarray}
On the surface defined by (\ref{Dgaugegeneral}), the last term vanishes. But the first one contains derivatives $N_I$ and $N_{\bar I}$ that are non-vanishing. This will lead to another K\"{a}hler manifold as compared to the rigid K\"{a}hler geometry.

\subsubsection{Projective Coordinates}\index{projective coordinates}
As we anticipated, the proper way to manifest this new structure is to select appropriate (projective) coordinates, in which we split between the direction orthogonal to the surface and along the surface. For this to be done, it is useful to split $(n+1)$ complex variables $\{X^I\}$ as $\{y,z^\alpha\}$, $\alpha=1,\ldots , n$ (and their complex conjugates $\{\bar y,\bar z^{\bar \alpha }\}$), where the variables $y$ will serve for defining the surface and $z^\alpha $ for coordinates on the $n$-dimensional surface.
We will therefore refer to the space spanned by the $\{X^I\}$ as the embedding space,\index{embedding space} while that spanned by the $\{z^\alpha \}$ as the projective space.  \index{projective space} The latter will describe the physical scalars.

The embedding in the $(n+1)$ dimensional manifold is given by functions $Z^I(z),\, \bar{Z}^I(\bar{z})$ defined as follows:
\begin{equation}
 X^I=y\,  Z^I(z)\,,\qquad  \bar X^I =\bar y\, \bar Z^I(\bar z) \,.
\label{Xrhoapp}
\end{equation}
As we require invariance under reparameterizations $z^\alpha\rightarrow z{'}{}^{\alpha}(z)$, the $Z^I(z)$ must be $n+1$ non-degenerate\footnote{The matrix $\partial
_\alpha Z^I$ has to be of rank $n$ and the matrix $(Z^I, \partial_\alpha
Z^I)$ has to be of rank $n+1$.} arbitrary holomorphic functions of the
$z^\alpha $.

We can now assign to $y$ the dilatation and $\U(1)$ charges carried by $X^I$ (according to Table \ref{tbl:matterMult}):
\begin{equation}
  \delta _{{\rm D},T}X^I= (\lambda _{\rm D}+ \rmi \lambda _T)X^I\ \rightarrow \qquad  \delta _{{\rm D},T}y= (\lambda _{\rm D}+ \rmi \lambda _T)y\,,\qquad \delta _{{\rm D},T}z^\alpha =0\,.
 \label{divideDTX}
\end{equation}

The non-uniqueness of the splitting \eqref{Xrhoapp} will be related to the K\"{a}hler transformations\index{K\"{a}hler transformations}, which will be discussed in Sect. \ref{ss:Kahlerrepar}. A practical choice for the functions $Z^I$, which is often used, is called `{\it special coordinates}' \index{special coordinates} and corresponds to
\begin{equation}
  Z^0(z)= 1\,,\qquad Z^\alpha(z) = z^\alpha \mbox{ for } \alpha =1,\ldots ,n\,.
 \label{specialcoord}
\end{equation}
For these `{\it special coordinates}' one can obviously write
\begin{equation}
z^\alpha = \frac{X^\alpha }{X^0}= \frac{Z^\alpha }{Z^0}\,. \label{defspcoor}
\end{equation}

\subsubsection{Gauge Fixing of Dilatations and $\U(1)$ $R$-symmetry}

The dilatation gauge fixing amounts in the vector multiplet to the first of the conditions (\ref{Dgaugegeneral}). We will rewrite it using (\ref{NIJ}), and use the homogeneity properties of $N_{IJ}$ (of degree 0 in $X^I$ and $\bar X^I$) to define $N_{IJ}(z,\bar z)$:
\begin{equation}
 N_{IJ}(X^I,\bar X^I)= N_{IJ}(Z^I(z),\bar Z^I(\bar z))= N_{IJ}(z,\bar z)\,.
 \label{NIJzbz}
\end{equation}
Therefore we can write the effective dilatational gauge condition as
\begin{equation}\label{subman_proj}
N(X,\bar{X})=|y|^2 Z^I(z) N_{IJ}(z,\bar z) \bar Z^I(\bar z)=-a\,, \qquad a= \left\{\begin{array}{cc}
                                                                                     3\kappa^{-2}&\mbox{ for }{\cal N}=1\,,\\
                                                                                     \kappa^{-2}&\mbox{ for }{\cal N}=2\,.
                                                                                   \end{array}
\right.
\end{equation}
We introduce here the arbitrary constant $a$, such that the treatment is the same for ${\cal N}=1$ and ${\cal N}=2$ supergravity, but for the purpose of this book $a=\kappa^{-2}$. The condition \eqref{subman_proj} fixes the modulus of $y$ in terms of $z^\alpha $ and their complex conjugates.

With these definitions and transformations (\ref{divideDTX}) a consistent $\U(1)$ gauge choice is \index{$\U(1)$-gauge}
\begin{equation}
T \mbox{-gauge:}\quad y=\bar y\,,
 \label{U1gauge}
\end{equation}
which removes the phase of $y$ as an independent variable and fixes completely $y$ (and $\bar{y}$)
\begin{equation}
  y= \left[-a^{-1} Z^I(z) N_{I J} (z,\bar z) \bar Z^{ J}(\bar z)\right]^{-1/2}\,.
 \label{valuey}
\end{equation}
As a result of the $D$, $\U(1)$ gauge fixing, we are left with only the $z^\alpha$ (and their complex conjugates) as unconstrained variables, which in turn will be related to the physical vector multiplet scalars in the Poincar\'{e} supergravity.

Before proceeding, we still want to make two remarks on the previous results:
\begin{itemize}
  \item At least as long as we consider only the bosonic part of the vector multiplet \eqref{lagrangiancomplex}, in practice we will not need to specify a $T$-gauge choice once in the projective basis \eqref{Xrhoapp}. Indeed, the phase of $y$ is the only bosonic quantity that transforms under $T$ and therefore---as shown at the beginning of Chap.~\ref{ss:gaugefix}---gauge invariance implies that the latter cannot appear in \eqref{lagrangiancomplex}. As a consequence, we will not need a $T$-decomposition law for the bosonic sector. On the other hand, fermions of the vector multiplet do transform under the $T$-symmetry.\footnote{We will relate these also to K\"{a}hler transformations in Sect. \ref{ss:Kahlerrepar}, and then this has a consequence on the global structure of the K\"{a}hler manifold, see footnote \ref{fn:fermionsHodge}.}
 \item The value (\ref{valuey}) for $y$ is a function of $z$ and $\bar z$. Therefore the transition from $X^I$ to $z^\alpha $ does not respect the
holomorphicity. In other words, the complex structure that is relevant in
the submanifold is not the same as the one in the embedding manifold.
\end{itemize}
\subsection{The K\"{a}hler Potential}
\label{ss:Kahlerpot}

It follows from our derivation that, after the projection, the resulting action for the scalars $z^\alpha $ is the first term of (\ref{lagranN}). We will now explicitly show that this defines again a K\"{a}hler manifold with a K\"{a}hler potential:
\index{K\"{a}hler potential}
\begin{tcolorbox}
\begin{equation}
   {\cal K}(z,\bar z)= 
   -a\ln\left[-a^{-1} Z^I(z) N_{I J}
 (z,\bar z) \bar Z^{ J}(\bar z)\right]\,.
 \label{Kahlerspecgeom}
\end{equation}
\end{tcolorbox}
\noindent To verify that \eqref{Kahlerspecgeom} is indeed correct, we can use the homogeneity properties of $N_{IJ}$ to obtain its derivatives. See e.g. (\ref{relNphi}), which implies that $\partial_\alpha(Z^I N_{I J}\bar Z^{ J})=(\partial_\alpha Z^I)(N_{I J}\bar Z^{ J})$, and hence we find the useful relations
\begin{eqnarray}
 \partial_{ \alpha } {\cal K} &=&-a\frac{N_{I J}\bar Z^{ J}
  \partial_{\alpha }  Z^I}{N_{KL}Z^K\bar Z^{L}}\,,\nonumber\\
\partial_\alpha \partial _{\bar \beta }{\cal K} &=&
a\partial_\alpha Z^I\partial _{\bar \beta }\bar Z^{ J}y\bar y
\left[ -\frac{N_{IJ}}{N}+\frac{N_{IK}\bar X^{K}X^LN_{LJ}}{N^2}\right]\nonumber\\ &=&-a\partial _\alpha Z^I\partial _{\bar \beta
}\bar Z^{ J}\,y\bar y\,\partial _I\bar{\partial}_{J}\ln
N\,.\label{partialK}
\end{eqnarray}
Hence we can write
\begin{equation}
  e^{-1}{\cal L}_X = -\partial _\mu z^\alpha  \partial ^\mu \bar z^{\bar \beta }\partial
  _\alpha \partial _{\bar \beta }{\cal K}\,,
 \label{finalkinScVMd4}
\end{equation}
confirming that ${\cal K}$ is the K\"{a}hler potential of the projective manifold.

It will also be useful to write the on-shell value of the $T$-gauge field in terms of the K\"{a}hler potential
\begin{equation}
{\cal A}_\mu^0 =\ft1{2}\rmi a^{-1}\left( \partial_\alpha {\cal K}\,
  \partial _\mu z^\alpha   -
  \partial _{\bar \alpha }{\cal K}\,\partial _\mu \bar z^{\bar \alpha }\right) -\ft12{\rmi}\partial _\mu
  \ln \frac{y }{\bar y }\,.
 \label{valuetilAmu}
\end{equation}
Note that we have not implemented yet the gauge condition (\ref{U1gauge}), which would eliminate the last term, which is anyway a pure gauge term.
Note also that the $\partial_\mu y, \partial^\mu \bar{y},\partial_\mu y\partial^\mu \bar{y}$ terms cancel as a consequence of the relations \eqref{relNphi}.
Furthermore, we can now rewrite  (\ref{valuey}) in terms of $\cal{K}$
\begin{equation}
  y=\bar y=\rme^{{\cal K}/(2a)}\,,
 \label{Ychoice}
\end{equation}
and thus, with  (\ref{Xrhoapp}), we have
\begin{equation}
  X^I =\rme^{{\cal K}/(2a)}Z^I (z)\ ,\qquad\bar X^I = \rme^{{\cal K}/(2a)}\bar Z^I
(\bar z)\,.
 \label{XinKz}
\end{equation}

In view of the importance of the K\"{a}hler potential, we still give an alternative way to present (\ref{Kahlerspecgeom}). First we remark that due to the homogeneity, $F_I(X)$ is first order in $X$, and we can thus write
\begin{equation}
  F_I (X)=y \ZF_I(Z(z))\,,\qquad \bar F_I (\bar X)=y \bZF_I(\bar Z(\bar z))\,,
 \label{defZF}
\end{equation}
where $\ZF_I(Z)$ is the same functional dependence as $F_I(X)$ and can thus also be written as\footnote{Note that $\bZF_I$ is obtained as $\bar F_I(\bar Z(\bar z))$ using for $\bar F_I(\bar Z)$ the same functional dependence as $\bar F_I(\bar X)$.}
\begin{align}\label{ZFIasdZF}
  \ZF_I (z)=   \frac{\partial }{\partial Z^I}\ZF (Z(z)) \,,
\end{align}
where $\ZF (Z)$ is the same function as $F(X)$. Finally
\begin{tcolorbox}
\begin{align}
{\cal K}(z,\bar z)=& -\kappa^{-2}\ln\left[\rmi \kappa^2 \bar Z^I (\bar z)\,\ZF_I(z) -\rmi\kappa^2 Z^I (z)\, \bar \ZF_I (\bar z)\right] \,,
\label{KP}
\end{align}
\end{tcolorbox}
\noindent and thus the K\"{a}hler potential is written in terms of derivatives of the holomorphic function $\ZF(Z(z))$. Note that the last equations, starting from  (\ref{defZF}), are only valid for ${\cal N}=2$, and thus we used $a=\kappa ^{-2}$ in (\ref{KP}).

\subsubsection{Interpretation as Sasakian Cone}
\label{ss:Sasakian}
There is an alternative interpretation of the metric \eqref{finalkinScVMd4}, which gives another geometrical view on the projective manifold.\footnote{This will not be important for what follows, and the reader may thus skip this part.} Before any gauge fixing we can write the action \eqref{lagrangiancomplex} as follows
\begin{equation}
e^{-1}{\cal L}_X= -\ft14 N^{-1}\left( \partial _\mu
N\right) ^2
-N \left( A_\mu-{\cal A}_\mu^0 \right) ^2
  +\frac{N}{a}\left(\partial_{\bar \beta }\partial _\alpha  {\cal K}\right)
  \partial  _\mu z^\alpha \, \partial ^\mu z^{\bar \beta }\,,
\label{rewrittenL}
\end{equation}
where ${\cal A}_\mu^0 $ is the geometric part of $A_\mu$ that we wrote in
(\ref{valueAKahler}).

The obtained metric in (\ref{rewrittenL}) is a cone~\cite{Gibbons:1998xa,deWit:1999fp}. Indeed, starting from the $n+1$ complex variables $\{X^I\}$ we have defined $\{y , z^\alpha \}$, where we can choose $y=|y|\rme^{\rmi\theta }$. As follows from (\ref{subman_proj}), the modulus $|y|$ appears in $N$ (in order to obtain a canonical parameterization of a cone, it is useful to define $r^2=-N$). The resulting coordinates are thus:
\begin{itemize}
  \item
a radius $r$ (gauge degree of freedom for dilatations)
  \item
an angle $\theta $ ($\U(1)$ $R-$ symmetry degree of freedom);
  \item $n$ complex variables $z^\alpha $.
\end{itemize}
In terms of $(r,\theta, z^\alpha, \bar{z}^{\bar \alpha})$ the metric (\ref{rewrittenL}) takes the form
\begin{align}
  \rmd s^2=&\rmd r^2 +r^2 \left[A+ \rmd\theta + \ft12\rmi
  \left(\partial _\alpha {\cal K}(z,\bar z)\, \rmd z^\alpha-   \partial _{\bar \alpha }{\cal K}(z,\bar z)\,\rmd\bar z^{\bar \alpha }\right)\right]   ^2\nonumber\\
  & -   r^2\partial _\alpha \partial _{\bar \alpha }{\cal K}(z,\bar z)\, \rmd z^\alpha \rmd\bar z^{\bar \alpha   }\,,
\label{dsSasakian}
\end{align}
where $A$ is the one-form gauging of the $\U(1)$ group, and ${\cal K}(z,\bar
z)$ is a function of the holomorphic prepotential $F(X)$. When $\U(1)$ is
not gauged ($A_\mu =0$), the base of the cone
 (the manifold with fixed $N$ or $r$) is a Sasakian
manifold with a $\U(1)$ invariance.\footnote{This has been remarked first
in a similar situation with hypermultiplets in ${\cal N}=2$
in~\cite{deWit:1999fp}, and has been looked at systematically
in~\cite{Gibbons:1998xa}.} When the $\U(1)$ is gauged, the auxiliary field $A_\mu $ can be redefined such
that the whole expression in square brackets is the field equation of $A_\mu $ itself, and it drops out on-shell. In that case, with fixed $r$ (gauge fixing the superfluous
dilatations), the remaining manifold is K\"{a}hler, with the K\"{a}hler potential
${\cal K}$ determined by $F(X)$. Therefore in the situation where $\U(1)$ is gauged, the geometry is further constrained to special K\"{a}hler,  \index{special K\"{a}hler} living on a K\"{a}hler submanifold of the
$(n+1)$-complex-dimensional manifold defined by a constant value of $r$. This is a real condition, but the $\U(1)$ invariance implies
that the other real variable $\theta $ has disappeared.

\subsection{Positivity Requirements}
\label{ss:positivity}
A standard requirement for a suitable Lagrangian of a physical system is that its kinetic terms should define positive kinetic energy.
The kinetic terms of the spin 2 particle, the graviton, are included in the Einstein--Hilbert term, the first term in (\ref{totaleLagN2D4}). This is positive due to the gauge choices in (\ref{DDgauge}), and for the vector multiplet part this is the choice of the sign in (\ref{subman_proj}). In fact, according to \eqref{subman_proj}, this choice requires that the prepotential $F(X)$ and the range of the scalars allow a domain where
\begin{equation}
  Z^I N_{IJ} \bar Z^J <0 \,.
\label{signN}
\end{equation}
Thus $N_{IJ}(z,\bar z)$ should have at least one negative
eigenvalue for all values of the scalars in the domain. The positivity of the kinetic energy of the scalars in $\mathcal{L}_0$ requires $g_{\alpha \bar \beta }$ to be a positive definite matrix. Using \eqref{partialK} with $N$ negative, this implies that $N_{IJ}$ should be a matrix with $n$ positive eigenvalues and 1 negative one. The negative one has the significance of the direction of the `compensating multiplet' in the $(n+1)$-dimensional embedding space $\{X^I\}$. The separation of the positive and negative definite parts will become explicit in  the next chapter, see (\ref{MatrixPositiveProduct}). These conditions define a so-called positivity domain for the scalars $\{ z^\alpha \}$. The requirement that the positivity domain is non-empty, restricts the space of prepotentials $F$ that can be used.
Finally, these two conditions also imply the positivity of the kinetic terms of the spin-1 part, namely that $\Im {\cal N}_{IJ}$ is a negative-definite matrix \cite{Cremmer:1985hj}. The mentioned matrix equation (\ref{MatrixPositiveProduct}) will make this property explicit. Since a symmetric matrix with a negative-definite imaginary part is invertible, these remarks prove that the inverse of ${\cal N}_{IJ}$ is well defined. Several theorems on these matrices are collected in \cite{Craps:1997gp}.

\subsection{Examples}
\label{ss:examplesSpecK}

We give here some examples of functions $F(X)$ and their corresponding
target spaces:
\begin{eqnarray}
F=-\rmi\,X^0X^1  &\quad& \frac{\SU(1,1)}{\U(1)}  \label{FiX0X1}\\
 F=(X^1)^3/X^0  &\qquad&  \frac{\SU(1,1)}{\U(1)}  \label{FX13}\\
 F=-2\sqrt{X^0(X^1)^3}   &&  \frac{\SU(1,1)}{\U(1)}  \label{Fsqrt} \\
F=\ft14\rmi X^I\eta_{IJ}X^J &&
\frac{\SU(1,n)}{\SU(n)\otimes \U(1)}\label{Feta}\\
 F=\frac{d_{ABC} X^A X^B X^C}{X^0} &&  \mbox{`very special'} \label{Fvsp}
\end{eqnarray}
The first three functions give rise to the manifold $\SU(1, 1)/\U(1)$.
However, the first one is not equivalent to the other two as the
manifolds have a different value of the curvature~\cite{Cremmer:1985hc}.
The latter two are, however, equivalent by means of a symplectic
transformation, as we will show in Sect.~\ref{ss:symplSpecial}. In the
fourth example $\eta$ is a constant non-degenerate real symmetric matrix.
In order that the manifold has a non-empty positivity domain, the
signature of this matrix should be $(-+\cdots +)$.

The last example, defined by a real symmetric tensor $d_{ABC}$, with $A,B,C=1,\ldots n$,
defines a class of special K\"{a}hler manifolds, which are denoted as `very
special' K\"{a}hler manifolds. \index{very special K\"{a}hler manifolds} These can be obtained by a dimensional reduction of $D=5$ supergravity--vector multiplet couplings~\cite{Gunaydin:1984bi} where the tensor $d_{ABC}$ is identified with the tensor $C_{IJK}$ introduced in Sect. \ref{ss:actVMD5}. We will say more on these in Sect. \ref{ss:HomSymSp}.

 \bexer \label{ex:Feta}
We go through the example (\ref{Feta}) in some detail. We get easily the
second derivative
\begin{equation}
  F_{IJ}=\ft12 \rmi \eta _{IJ}\,,\qquad  N_{IJ}=\eta _{IJ}\,.
 \label{FIJNIJFeta}
\end{equation}
We will now specify to the case
\begin{equation}
  \eta _{IJ}=\begin{pmatrix}-1&0\cr 0&1\end{pmatrix}\,.
 \label{choiceeta}
\end{equation}
We thus write the formulae for $n=1$, but they can be easily generalized
for arbitrary $n$ by taking the lower-right entry to be a unit $n\times
n$ matrix. From (\ref{XinKz}) we get
\begin{equation}
  \rme^{-\kappa ^2{\cal K}(z,\bar z)}=Z^0(z) \bar Z^0(\bar z) -Z^1(z) \bar Z^1(\bar z)\,.
 \label{e-KFeta}
\end{equation}
The right-hand side should be positive. A convenient parameterization consists in using the special coordinates (\ref{specialcoord}), i.e. $Z^0=1$, $Z^1=z$. Indeed, then $\partial _z Z^I$ is of rank 1, while
the $2\times 2$ matrix $(Z^I, \partial _z ^I)$ is of rank~2. The domain
for $z$ is then $|z|^2<1$.

This leads to
\begin{equation}
\kappa ^2\partial _z {\cal K}= \frac{\bar z}{1-z\bar z}\,,\qquad
\kappa ^2  g_{z\bar z}=\kappa ^2\partial _z\partial _{\bar z} {\cal K}=\frac{1}{\left( 1-z\bar z\right)
  ^2}\,.
 \label{metricFeta}
\end{equation}
You may check that the indefinite signature of (\ref{choiceeta}) was
necessary to have a positive metric.
 \eexer

\subsection{K\"{a}hler Reparameterizations}
\label{ss:Kahlerrepar}

The careful reader may have noticed that the splitting \eqref{Xrhoapp}  is not unique. The projected coordinates are defined up to K\"{a}hler repara\-meterizations \index{K\"{a}hler transformations} that leave invariant $X$ and $\bar{X}$:
\newcommand{\Ktransfo}{f(z)}
\newcommand{\Ktransfob}{\bar f(\bar z)}
\begin{equation}
y'= y \, \rme^{\Ktransfo/a}, \qquad
Z'^I=Z^I\,\rme^{-\Ktransfo/a}\,, \label{Kahlertransf}
\end{equation}
where $f(z)$ is an arbitrary holomorphic function.\footnote{We stress the fact that K\"{a}hler repara\-meterizations are transformations of the target space functions, like $Z^I(z)$. They do not act on the coordinates $\{z,\,\bar{z}\}$.}  The K\"{a}hler potential is not invariant under these reparameterizations. From \eqref{Kahlerspecgeom} we see that
\begin{equation}
{\cal K}'(z,\bar z)= {\cal K}(z,\bar z)+\Ktransfo + \Ktransfob\,, \label{KahlertransfK}
\end{equation}
where we consider $\Ktransfo$ and $ \Ktransfob$ as independent transformations.
If we choose the $\U(1)$ gauge \eqref{U1gauge}, then this leaves a
combination of the $\U(1)$ and the K\"{a}hler transformation. The decomposition law for K\"{a}hler transformations is thus\footnote{The symbol $\poinc$ for Point-carr\'{e} is used to indicate transformations in the Poincar\'{e} theory.}
\begin{equation}
\delta_K ^\poinc [\Ktransfo,\Ktransfob] = \delta_K [\Ktransfo,\Ktransfob] + \delta _T [\lambda_T = \ft12a^{-1}(\Ktransfo -\Ktransfob)]\,, \label{decompU1K}
\end{equation}
where $\delta_K [\Ktransfo,\Ktransfob]$ are the transformations in the conformal setting, i.e. induced by (\ref{Kahlertransf}) for small $f$.
The remaining K\"{a}hler transformation can e.g. be
used to choose one of the $Z^I$, say $Z^0$, equal to~1. In any case, one
can choose the parameterization of the $n$ physical scalars $z^\alpha$
(with $\alpha=1,\ldots ,n$) at random, as stressed
in~\cite{Castellani:1990tp,Castellani:1990zd,D'Auria:1991fj}.\footnote{Note that, as it is clear from Sect. \ref{ss:susygaugetrN2}, the fermions transform under the
superconformal $\U(1)$ factor, and hence, by (\ref{decompU1K}), under the
(finite) K\"{a}hler transformations. This implies that the K\"{a}hler form should be of even integer cohomology ({\em
K\"{a}hler--Hodge manifold})~\cite{Witten:1982hu,Bagger:1984ge}. This is similar to what is needed in ${\cal N}=1$ and has been explained in several steps in Appendix 17.A of \cite{Freedman:2012zz}.
\label{fn:fermionsHodge}}

\subsection{The K\"{a}hler Covariant Derivatives}
\label{ss:Kacovder}

Let us consider the $\U(1)$-covariant derivative $\nabla_\mu X^I$ as defined in \eqref{nablaX}. This object is of course invariant under K\"{a}hler transformations since it was defined before splitting the variables $\{X^I\}$ as in (\ref{Xrhoapp}). However, after defining $y$ and $Z^I$, we have to split  (\ref{nablaX}) into derivatives of these new variables, which are not invariant under K\"{a}hler transformations. To do so in a K\"{a}hler-covariant way, it is convenient to introduce a connection for the K\"{a}hler transformations \cite[(17.71)]{Freedman:2012zz} (see also more details in \cite{Freedman:2017obq}) as follows:
\begin{equation}
  \nabla _\mu y \equiv   \partial _\mu y- \rmi{\cal A}^0_\mu y -a^{-1}\omega _\mu y\,,\qquad \nabla _\mu Z^I\equiv \partial _\mu Z^I +a^{-1}\omega _\mu Z^I\,,
 \label{nablayZ}
\end{equation}
where
\begin{equation}
  \omega _\mu = \omega _\alpha \partial _\mu z^\alpha \,,
 \label{omegamu}
\end{equation}
and $\omega _\alpha $ is a K\"{a}hler connection, which means that under small holomorphic K\"{a}hler transformations it should transform as
\begin{equation}
  \delta \omega _\alpha =\partial _\alpha f\,.
 \label{Kahleromega}
\end{equation}
In this way $\nabla _\mu y$ and $\partial _\mu Z^I$ transform as $y$ and $Z^I$. Since the K\"{a}hler potential transforms as in (\ref{KahlertransfK}), we can identify
\begin{equation}
  \omega _\alpha =\partial _\alpha{\cal K} \,.
 \label{omaldef}
\end{equation}

We can then also write
\begin{equation}
   \nabla _\mu Z^I  =  \nabla _\alpha Z^I \partial _\mu z^\alpha \,,\qquad \nabla _\alpha Z^I \equiv \left(\partial _\alpha +a^{-1}\omega _\alpha\right)Z^I\,.
 \label{nablaalZ}
\end{equation}
Using the expression of ${\cal A}_\mu^0 $ in \eqref{valuetilAmu}, the covariant derivative $\nabla _\mu y$ is given by
\begin{equation}
 \nabla _\mu  y  =  \frac12 y \partial _\mu (\ln y\bar y)-\ft12 a^{-1} y\partial _\mu {\cal K}\,.
\label{nablamuyZAfilled}
\end{equation}
So far we have only redefined the fields $X^I$. If we now perform the gauge fixing for dilatations and $T$-symmetry: \eqref{Ychoice} we find
\begin{equation}
  \nabla _\mu  y = 0\,.
 \label{nablay0}
\end{equation}
We can then also write the value of \eqref{valuetilAmu} as a pullback of derivatives on the projective manifold:
\begin{align}
  {\cal A}^0_\mu  =& {\cal A}_\alpha \partial _\mu z^\alpha +{\cal A}_{\bar \alpha }\partial _\mu \bar z^{\bar \alpha }\,,\nonumber\\
&  {\cal A}_\alpha = \rmi\partial _\alpha\ln  y=\frac{1}{2a}\rmi\partial _\alpha {\cal K}=\frac{1}{2a}\rmi\omega _\alpha  \,,\nonumber\\
&  {\cal A}_{\bar \alpha }= -\rmi \partial _{\bar \alpha }\ln y=-\frac{1}{2a}\rmi\partial _{\bar \alpha} {\cal K}=-\frac{1}{2a}\rmi\bar \omega  _{\bar \alpha} \,.
 \label{calAmualpha}
\end{align}
\vspace{10pt}

\newcommand{\fG}{G}
The K\"{a}hler covariant derivatives on a scalar quantity $\fG(z,\bar z)$ on the projective space can be defined in general as
\begin{equation}
  \nabla _\mu \fG(z,\bar z)= \nabla _\alpha \fG \,\partial _\mu z^\alpha + \nabla _{\bar \alpha} \fG\partial _\mu\bar z^{\bar \alpha }\,,
 \label{nablaVsplit}
\end{equation}
with\footnote{From \eqref{calAmualpha}, ${\cal A}_\alpha$ and $\omega _\alpha$ are both related on $\partial_\alpha\cal{K}$ , hence \eqref{nablaVsplit} should contain only combinations of weight $w_{\pm}$ and $c$.}
\vspace{10pt}
\begin{tcolorbox}
\begin{equation}
  \nabla _\alpha \fG =\partial _\alpha \fG +a^{-1} \hat{w}_+\fG\left(\partial _\alpha {\cal K}\right)\,,\qquad
  {\nabla} _{\bar \alpha} \fG =\partial _{\bar \alpha} \fG +a^{-1} \hat{w}_- \fG\left( \partial _{\bar \alpha} {\cal K}\right)\,,
 \label{defnablaV}
\end{equation}
\end{tcolorbox}
\noindent and
\begin{equation}
\hat{w}_+={w}_+ +\ft12 c\,,\qquad \hat{w}_-={w}_- -\ft12 c\,,
\end{equation}
 where $c$ is the chiral weight, and $w_+$ and $w_-$ are K\"{a}hler weights for a transformation with infinitesimal $f(z)$ and $\bar f(\bar z)$, according to
\begin{equation}
  \left(\delta _T[\lambda _T] + \delta _K[f,\bar f]\right)\fG(z,\bar z) =\left(\rmi c\lambda _T -w_+ a^{-1}f(z)-w_- a^{-1}\bar f(\bar z)\right)\fG(z,\bar z)\,.
 \label{defweightscw}
\end{equation}
In the Poincar\'{e} frame,  (\ref{decompU1K})  leads thus to
\begin{equation}
  \delta^\poinc _K[f,\bar f]\fG(z,\bar z) =-a^{-1}\left( \hat{w}_+f(z)+\hat{w}_-\bar f(\bar z)\right)\fG(z,\bar z)\,.
 \label{delpoincKG}
\end{equation}
For example,
\begin{equation}
  \begin{array}{c|ccc|cc}
     \fG  & c & w_+ & w_- & {\hat w}_+ & {\hat w}_-\\  \hline
     X^I,\,F_I & 1 & 0 & 0 & \nicefrac12 & - \nicefrac12 \\
     \bar X^I,\,\bar F_I & -1 & 0 & 0 & - \nicefrac12 & \nicefrac12 \\
     y & 1 & -1 & 0 & - \nicefrac12 & - \nicefrac12 \\
     \bar y & -1 & 0 & -1 & - \nicefrac12 & - \nicefrac12  \\
     Z^I,\,\ZF_I  & 0 & 1 & 0 & 1 & 0 \\
     \bar Z^{\bar I },\,\bZF_{\bar I} & 0 & 0 & 1 & 0 & 1 \,.
        \end{array}
 \label{weightscw}
\end{equation}
The weights of $\nabla _\alpha \fG$ are equal to the weights of $\fG$. Furthermore,
\begin{equation}
  \left[\nabla _\alpha ,\,\nabla _{\bar \beta }\right] \fG = a^{-1}(\hat{w}_--\hat{w}_+)g_{\alpha \bar \beta }\fG\,.
 \label{commnablas}
\end{equation}

Note that \eqref{nablay0} together with \eqref{nablaVsplit} imply that
\begin{equation}
  \nabla _\alpha  y(z,\bar z)= \nabla _{\bar \alpha }y (z,\bar z)=0\,,
 \label{nablay}
\end{equation}
and hence
\begin{align}\label{nablalphaZ}
\nabla_\alpha X^I&=y\nabla_\alpha Z^I\,,\nonumber\\
\nabla_{\bar{\alpha}} \bar{X}^I&=y\nabla_{\bar{\alpha}} \bar{Z}^I\,,\nonumber\\
\nabla_\alpha \bar{X}^I&=\nabla_{\bar{\alpha}} X^I=0\,.
\end{align}
The covariant derivative of $N=-a$, using  (\ref{relNphi}), implies
\begin{equation}
  (\nabla _\alpha X^I)N_{IJ}\bar X^J=0\,.
 \label{nablaN}
\end{equation}

These relations will be very useful when rewriting the conformal theory in Poincar\'{e} language.
%
%
In particular we will use `covariant transformations'  \cite{Freedman:2017obq} for all ordinary transformations that are defined on functions of $z$ and $\bar z$ by
\begin{tcolorbox}
\begin{equation}
  \widehat{\delta} \fG(z,\bar z)= \left(\delta z^\alpha \nabla  _\alpha + \delta \bar z^{\bar \alpha }\nabla_{\bar{\alpha}}\right)\fG(z,\bar z)\,.
 \label{hatdeltaf}
\end{equation}
\end{tcolorbox}
See the similarity with (\ref{hatisnabla}). Note that this differs from the ordinary transformation in Poincar\'{e} frame:
\begin{align}
  \delta ^ \poinc \fG(z,\bar z)= &\left(\delta z^\alpha \partial   _\alpha + \delta \bar z^{\bar \alpha }\partial _{\bar{\alpha}}\right)\fG(z,\bar z)\,,\nonumber\\
  \widehat{\delta} \fG(z,\bar z)=&\delta^ \poinc  \fG(z,\bar z)+a^{-1} \fG(z,\bar z)\left(\hat{w}_+ \delta z^\alpha \partial   _\alpha + \hat{w}_-\delta \bar z^{\bar \alpha }\partial _{\bar{\alpha}}\right){\cal K}\nonumber\\
 =&\delta^ \poinc  \fG(z,\bar z)- \delta_K^\poinc\left[f= \delta z^\alpha \partial _\alpha {\cal K},\, \bar f= \delta \bar z^{\bar \alpha }\partial _{\bar \alpha }{\cal K}\right] \,.
   \label{deltaf}
\end{align}
Since both terms in the right-hand side are symmetries, also this covariant transformation is a symmetry.
Note that  (\ref{nablay}) implies that for whatever symmetry transformation
\begin{equation}
  \widehat{\delta} y =0\,.
 \label{hatdely0}
\end{equation}

\section{Coordinates in the Quaternionic-K\"{a}hler Manifold.}\label{coord_quatman}
We will now introduce also convenient coordinates for the hypermultiplet side. This concerns  the scalars $q^\hX$, where the index $\hX$ runs over $4(n_H+1)$ values. In this case, we want to take coordinates that take into account the second parts of  (\ref{Dgaugegeneral}) and  (\ref{v:zeta_i02}). We want to project out the four directions defined by the homothetic and $\SU(2)$ Killing vectors in the hyper-K\"{a}hler manifold (they are, respectively, $k_{\rm D}{}^{\hX }$ and $\vec{k}^\hX$, see \eqref{delDphiA} and \eqref{delDSU2q}).

\subsection{Projective Coordinates}\index{projective coordinates}
We denote the direction of the dilatation generator with the coordinate $q^0$ and the directions of the $\SU(2)$ $R$-symmetry generators with $q^\ralpha$ ($\ralpha =1,2,3$). On these coordinates we will apply gauge fixing conditions for, respectively, dilatations and the $\SU(2)$ transformations (\ref{delDSU2q}). The remaining $4n_H$ real variables are indicated as $q^\Xu$.
Our new basis for the scalars of the hypermultiplet is therefore
\begin{equation}
  \{q^{\hX }\} = \{q^0,\,q^\ralpha,\,q^\Xu\}\,,\label{v:basisikeuze_hyp}
\end{equation}
and
\begin{equation}
  k_{\rm D}{}^{\hX } = \{2  q^0,\,0,\,0\}\,, \qquad  {\vec k}^{\hX } =
  \{0,\,\vec k^\ralpha,\,0\}
  \,.\label{v:basiskeuze_kX}
\end{equation}
In analogy with the special K\"{a}hler manifolds, we will refer to the $\{q^{\hX }\}$ as the coordinates of the embedding space\index{embedding space}, and to the $\{q^\Xu\}$ as those of the projective space. \index{projective space} Note that we use  the index $\ralpha $ for the choice of coordinates
and the vector sign for the 3 directions of the $\SU(2)$ vectors. The vector ${\vec k}^{\ralpha}$ connects these as a 3-bein and can be a function of $q^\ralpha$ and $q^\Xu$.

This choice of coordinates leads to equations for the complex structures in this basis.
Using \eqref{v:basiskeuze_kX}, from \eqref{delDSU2q} it follows that $\vec{J}_0{}^X$ has only components in the $X=\ralpha$ directions. Similarly, after some algebraic manipulation one can obtain the following expressions for coordinates of the complex structures \cite{Bergshoeff:2004nf}:
\begin{align}\label{JprojQuat}
&\vec{J}_0{}^0=0\,,\qquad \vec{J}_0{}^\ralpha = (q^0)^{-1}\vec{k}^\ralpha\,,\qquad \vec{J}_0{}^u=0\,,\nonumber\\
&\vec{J}_\ralpha{}^0=\vec{k}_\ralpha\,, \qquad \vec{J}_\ralpha{}^\rbeta = (q^0)^{-1}\vec{k}_\ralpha \times \vec{k}^\rbeta\,,\qquad \vec{J}_\ralpha{}^u=0\,,\nonumber\\
& \vec{J}_u{}^0 = \vec{k}_u  \,,\qquad \vec{J}_u{}^\rbeta = (q^0)^{-1}\left(\vec{k}_u\times \vec{k}^\rbeta+ \vec{J}_u{}^v \vec{k}_v\cdot \vec{k}^\rbeta\right)\,,
\end{align}
where
\begin{equation}
  \vec{k}^\ralpha \cdot \vec{k}_\rbeta =-q^0\delta^{\ralpha}_\rbeta\,,
 \label{veckorthogonal}
\end{equation}
and $\vec{J}_u{}^v$ separately satisfy the quaternionic algebra \eqref{algebrJ}. Using the constraints of dilatation invariance and general properties of the
quaternionic frame fields, one can derive relations between the connection coefficients \cite{Bergshoeff:2004nf}. The quaternionic metric \eqref{gfromkD} takes the form\footnote{We normalize here $h_{XY}$ with a factor $q^0$ different from  \cite{Bergshoeff:2004nf}.}
\begin{eqnarray}
  g_{\hX  \hY } \rmd q^{\hX } \rmd q^{\hY }&=& -\frac{(\rmd q^0)^2}{q^0} +h_{\Xu \Yu }(q)\rmd q^\Xu \rmd q^\Yu
-\frac{1}{q^0}\vec{k}_\hX \cdot \vec{k}_\hY\rmd q^{\hX } \rmd q^{\hY }\,,\nonumber\\
&&\vec{k}_\hX\rmd q^{\hX }= \vec{k}_\ralpha \rmd q^\ralpha + \vec{k}_\Xu \rmd q^\Xu \,.
\label{v:hypermetriek}
\end{eqnarray}
Note that although $\vec{k}^{X} $ has only 3 non-zero components according to (\ref{v:basiskeuze_kX}), the vector $\vec{k}_{X} $ has
components in the $\ralpha $- and $\Xu $-directions. One can now first check from \eqref{kD2fromveck} that
\begin{equation}
  k_{\rm D}{}^2= -4q^0\,,
 \label{kDq0}
\end{equation}
and  (\ref{usevAvBk}) then implies for arbitrary vectors $\vec{A}$ and $\vec{B}$:
\begin{equation}
  \vec{A}\cdot \vec{k}^X \vec{k}_X \cdot \vec{B}=\vec{A}\cdot \vec{k}^\ralpha \vec{k}_\ralpha \cdot \vec{B}=-q^0 \vec{A}\cdot \vec{B}\,.
 \label{innerprodveck}
\end{equation}
With the present notation we have that $\vec{k}_u = g_{u\ralpha}\vec{k}^\ralpha$ and $\vec{k}_\ralpha = g_{\ralpha\rbeta}\vec{k}^\rbeta$.
\vspace{10pt}

For the fermionic side of the hypermultiplet, we denote by ${\cal A}$ the coordinates on the target space enumerating the $2(n_H+1)$ fermions, and these can be split in $2+2n_H$. This is consistent, since the distinction of $q^0$ and $q^\ralpha $ splits the structure group $\Gl(n_H+1,\mathbb{H})$ to $\SU(2)\times\Gl(n_H,\mathbb{H})$. We thus write
\begin{equation}
\{{\cal A}\} = \{i,A\}\,,
  \label{fermcoordhyper}
\end{equation}
where $i=1,2$ is an $\SU(2)$-index. We will use this split also for distinguishing compensating and physical fermions in Sect.~\ref{phys_ferm}
\begin{equation}
  \{\zeta^{{\cal A}}\} = \{\zeta^{i},\zeta^A\}\,.
  \label{fermcoordhyperzeta}
\end{equation}
It has been shown in detail in \cite{Bergshoeff:2004nf} that the coordinates can be chosen such that some components of $f^{i{\cal A}}{}_\hX$ and $f^{\hX}{}_{i{\cal A}}$ vanish and one obtains, e.g.
\begin{align}
  &f^{iA}{}_0 = f^{iA}{}_\ralpha = f^u{}_{ij}= f^0{}_{iA} =0\,,\qquad  f^\ralpha {}_{i\Ap}=(q^0)^{-1} f^\Xu{}_{i\Ap}\vec{k}^\ralpha\cdot \vec{k}_u\,,\nonumber\\
  &f^{ij}{}_0 =\rmi \varepsilon ^{ij}\sqrt{\frac{1}{2q^0}}\,,\qquad f^0{}_{ij} =-\rmi \varepsilon_{ij}\sqrt{\frac{q^0}{2}}\,.
 \label{fcomponents0}
\end{align}
In these coordinates, the $\rho _{\Ae \bBe}$ introduced in  (\ref{introrho}) as well as the $C_{{\cal A}{\cal B}}$ and $d^\bAe{}_\Be$  from Sect.~\ref{finsec_hypers} are block-diagonal, e.g.
\begin{align}
  &C_{ij}= \varepsilon _{ij}\,,\qquad C_{iA}=0\,,\qquad d^{\bi}{}_j = -\delta ^i{}_j\,,\qquad d^{\bi}{}_A=0\,,\nonumber\\
   &\rho _{i\bj}=-\varepsilon _{ij}\,,\qquad \rho _{i\bar A}=0\,.
 \label{blockdiagCrhod}
\end{align}
The $A^{i\Ae}$ introduced in (\ref{defAialpha}) has only components for $\Ae$ in the doublet range:
\begin{equation}
  A^{ij}= \rmi\sqrt{2q^0}\varepsilon ^{ij}\,,\qquad A^{iA}=0\,,\qquad A_{ij}= -\rmi\sqrt{2q^0}\varepsilon _{ij}\,,\qquad A_{iA}=0\,.
 \label{AiAcomponents}
\end{equation}

We can now reduce the fundamental relations of the quaternionic structures in the embedding space to relations in the projective space, making use of (\ref{fcomponents0}):
\begin{align}
  \begin{array}{ll}
  f^{iA}{}_v f^{u}{}_{iA}=\delta^{u}_v\,,\qquad & f^{iA}{}_u f^{u}{}_{jB}=\delta^{i}_j\delta^{A}_B,\\
\left(f^{iA}{}_u\right)^* =f^{jB}{}_u\varepsilon_{ji}\rho_{B \bar{A}}\,,& \left(f^{u}{}_{iA}\right)^* =\varepsilon^{ij}\rho^{\bar{A} B}f^{u}{}_{jB},\\
\rho_{A \bar{B}} \rho^{\bar{B} C} =-\delta^{C}_A\,,& \rho^{\bar{A} B}=\left(\rho_{A \bar{B}}\right)^*,\\
  2f^{iA}{}_uf^v{}_{jA}= \delta _u^v\delta _j^i + \vec{\tau }_j{}^i\cdot \vec{J}_u{}^v\,,\qquad&
\vec{J}_u{}^v=\left(\vec{J}_u{}^v\right)^*= -f^{iA}{}_uf^v{}_{jA}\vec{\tau }_i{}^j\,,\\
  d^\bAp{}_\Bp=(d^\bBp{}_\Ap)^*=\rho ^{\bAp\Cp}d^{\bDp}{}_\Cp \rho _{\Bp\bDp}\,, \\
  C_{\Ap\Bp}= - C_{\Bp\Ap}= \rho _{\Ap\bCp}d^{\bCp}{}_\Bp\,,& C^{\bAp\bBp}=(C_{\Ap\Bp})^*= \rho ^{\bAp\Cp}d^{\bBp}{}_\Cp\,,
\\
h_{uv}=f^{iA}{}_u \varepsilon_{ij} C_{AB}f^{jB}{}_v= \left(f^{i\Ap}{}_u\right)^* d^\bAp{}_\Bp f^{iB}{}_v\,.
   \end{array}
 \label{projecrealf}
\end{align}
Using \eqref{algebrJ} and \eqref{JprojQuat} the authors of \cite{Bergshoeff:2004nf} found that the condition \eqref{cov_const_J} is projected to\footnote{For these projections one needs the Levi-Civita connection of the embedding metric $g_{XY}$ expressed in the quantities of the projective manifold. They are all given in  \cite{Bergshoeff:2004nf}. We repeat here a few components useful for the projections:
\begin{align*}
 & \Gamma _{00}{}^X= -\ft12(q^0)^{-1}\delta _0^X\,, \qquad \Gamma _{0r}{}^0=\Gamma _{0u}{}^0=0 \,,\qquad \Gamma _{0r}{}^s= \ft12(q^0)^{-1}\delta _r^s\,,\qquad  \Gamma _{0u}{}^v= \ft12(q^0)^{-1}\delta _u^v\,,\nonumber\\
   & \Gamma _{uv}{}^0=\ft12g_{uv}\,,\qquad \Gamma _{ur}{}^0=-\ft12(q^0)^{-1}\vec{k}_u\cdot \vec{k}_r\,.
\end{align*}\label{fn:Gammaquat} }
\begin{equation}
\widetilde \nabla _w \vec{J}_u{}^v\equiv  \nabla _w \vec{J}_u{}^v+2\vec{\omega}_w \times \vec{J}_u{}^v=\partial_w \vec{J}_u{}^v-\Gamma _{wu}^z\vec{J}_z{}^v+\Gamma _{wz}^v\vec{J}_u{}^z+2\vec{\omega}_w \times \vec{J}_u{}^v=0\,,
\label{nablaJproj}
\end{equation}
where $\Gamma_{\Xu \Yu }^\Zu =\Gamma _{\Yu \Xu }^\Zu $ is the torsionless Levi-Civita connection of the metric $h_{\Xu\Yu }$.
There is a new term, originating in the embedding space from $\Gamma _{wr}{}^v\vec{J}_v{}^r$, containing
\begin{equation}
  \vec{\omega }_u\equiv -\frac{1}{2q^0}\vec{k}_u\,.
 \label{defvecomega}
\end{equation}
The triplet $\vec{\omega}_\Xu$ is an $\SU(2)$ connection on the projected manifold, characteristic of the quaternionic-K\"{a}hler geometry. Another important projection from (\ref{cov_const_J}), namely from the antisymmetric part of $\nabla _{[u}J_{v]}{}^0=0$ is\footnote{We write in the right-hand side explicitly $\vec{J}_{\Xu} {}^\Zu h_{\Zu\Yu }$ since so far all raising and lowering of indices was done with $g_{XY}$, e.g.
$\vec{J}_{\Xu\Yu}= \vec{J}_{\Xu}{}^rg_{r\Yu}+\vec{J}_{\Xu}{}^\Zu g_{\Zu\Yu}= \vec{J}_{\Xu} {}^\Zu h_{\Zu\Yu }+(q^0)^{-1}\vec{k}_{\Xu}\times \vec{k}_{\Yu}$.
}
\begin{equation}
  \vec{R}_{\Xu \Yu }\equiv 2\partial_{[\Xu }\vec \omega _{\Yu ]}
+2\vec \omega _\Xu \times \vec \omega _\Yu =-\ft12(q^0)^{-1}\vec{J}_{\Xu} {}^\Zu h_{\Zu\Yu }.
 \label{SU2curvJ}
\end{equation}
Thus, the $\SU(2)$ curvature becomes proportional to the complex structure. This is a main property of a quaternionic-K\"{a}hler manifold, as will be further discussed in Sect. \ref{ss:qKmanifolds}.

Finally, from \eqref{projecrealf} the condition \eqref{Df0} is projected to
\begin{equation}
\widetilde \nabla _v f^{iA}{}_u \equiv
\partial _v  f^{iA}{}_u-\Gamma _{vu}^w f^{iA}{}_w+f^{jA}{}_u\omega _{vj}{}^i +f^{iB}{}_u\omega
_{v B}{}^A=0\,.
 \label{covconstf}
\end{equation}

\subsection{S-Supersymmetry, Dilatations and \texorpdfstring{$\SU(2)$}{SU(2)} Gauge Fixing}
Having chosen a convenient set of coordinates, it is now easy to proceed with the gauge fixings to the Poincar\'{e} theory.
We start from the dilatations: using (\ref{v:basiskeuze_kX}) and (\ref{v:hypermetriek}), the second condition in (\ref{Dgaugegeneral}) becomes
\begin{equation}
q^0=\kappa ^{-2}\equiv -\nu^{-1} \,.
 \label{v:q^0constant}
\end{equation}
The parameter $\nu$ will play a role in the characterisation of Quater\-nionic-K\"{a}hler manifolds, see Sect.~\ref{ss:qKmanifolds}.
Similarly, we gauge fix the $\SU(2)$ by choosing the phases of the compensating quaternion in the hypermultiplet to be some constants $q^\ralpha_0$: \index{$\SU(2)$-gauge}
\begin{align}
  \SU(2)\mbox{-gauge:}\qquad q^\ralpha = q^\ralpha_0\,.
  \label{SU2gauge}
\end{align}
With the above conditions, the bosonic part of $\vec{V}_\mu $ in (\ref{valueAVN2}) becomes
\begin{equation}
 \vec{{\cal V}}_\mu =
  \left.\vec{V}_\mu\right|_{\rm bos}=\ft12\kappa ^2\left(\vec{k}_u \partial _\mu q^u -W_\mu ^I\vec{P}_I\right)=-\vec{\omega }_u \partial _\mu q^u-\ft12\kappa ^2W_\mu ^I\vec{P}_I \,.
 \label{Vmubos_nongauge}
\end{equation}
Inserting the above expression into the covariant derivatives \eqref{covderafterfieldeqn} and plugging in the $\SU(2)$-gauge \eqref{SU2gauge} we obtain
\begin{equation}
  D_\mu q^X= \nabla  _\mu q^X-W_\mu ^I\left(k_I^X + \kappa ^2\vec{P}_I\cdot \vec{k}^X\right)\,,\qquad
  \nabla  _\mu q^X\equiv \partial _\mu q^X-2(\vec{\omega }_u \partial _\mu q^u) \cdot \vec{k}^X\,.
 \label{DqXnabla}
\end{equation}
Furthermore, due to (\ref{v:basiskeuze_kX}), the second term in $\nabla  _\mu q^X$ only contributes for $X=\ralpha$.
The various cases give
\begin{equation}
  \nabla _\mu q^0 =0\,, \qquad \nabla _\mu q^\ralpha = -2(\partial_\mu q^u)\vec{\omega }_u\cdot \vec{k}^\ralpha\,,\qquad \nabla _\mu q^u = \partial _\mu q^u\,.
 \label{DmuqXsplit}
\end{equation}
If we now restrict to the non-gauge related part in $D_\mu q^X$, upon using the above relations together with (\ref{innerprodveck}), the kinetic terms of the $q^X$ in (\ref{totaleLagN2D4}) reduce to
\begin{align}
{\cal L}_{q}=  -\ft12 g_{XY}\nabla _\mu q^X\,\nabla ^\mu q^Y = -\ft12\partial _\mu q^u \partial _\mu q^v\left(g_{uv}+\kappa ^2\vec{k}_u\cdot \vec{k}_v\right)\,.
 \label{calcqkmetric}
\end{align}
A quick comparison with \eqref{v:hypermetriek} shows therefore that the final metric on the $4n_H$-dimensional space is  $h_{\Xu \Yu }$, used in (\ref{v:hypermetriek}):
\begin{equation}
  h_{\Xu\Yu}= g_{\Xu\Yu}+4\kappa ^{-2}\vec{\omega }_\Xu\cdot\vec{\omega }_\Yu\,.
 \label{hquat}
\end{equation}

The $S$-supersymmetry gauge fixing together with the $\chi ^i$ field equation \eqref{v:veldvglD} lead to (\ref{v:zeta_i02}). Upon using (\ref{AiAcomponents}), the second of these implies the $S$-gauge \index{$S$-gauge}
 \begin{equation}\label{SgaugeHyper}
 S\mbox{-gauge:}\qquad \zeta ^i=0\,.
\end{equation}

\subsection{Isometries in the Projective Space}
Let us now explain how the isometries of the target space (defined in (\ref{gen_iso_hyper})) get projected onto the spaces of the $\{q^u\}$. Upon requiring the corresponding Killing vectors in the target manifold $k_I{}^X$ to commute with the closed homothetic Killing vector (this is the condition \eqref{WeylweightkI}), one can show that there exists a frame in which the components of these $k_I{}^X$ in the projection are \cite{Bergshoeff:2004nf}
  \begin{equation}\label{KillProjSymm}
k_{I}{}^{\hX }(q) = \{k_{I}{}^{0}=0,\,k_{I}{}^{\ralpha}= \vec{k}^\ralpha \cdot \vec{r}_I,\, k_{I}{}^{u}\}\,,
  \end{equation}
where $\vec{r}_I (q)$ are arbitrary holomorphic vectors.\footnote{We can make this  choice since the $\vec{k}^\ralpha $ introduced in (\ref{veckorthogonal}) can be considered as invertible $3\times 3$ matrices. The first 0 in (\ref{KillProjSymm}) is then the statement that the isometries commute with dilatations, which according to (\ref{v:basiskeuze_kX}) act only on the 0-component.} The moment maps in the embedding space satisfy the condition (\ref{mommapshyper}), which can be rewritten in the projective space using \eqref{KillProjSymm} and (\ref{JprojQuat}) as
\begin{equation}
  \widetilde{\nabla }_u \vec{P}_I \equiv  \partial _u\vec{P}_I +2\vec{\omega }_u\times \vec{P}_I = \vec{J}_u{}^vh_{vw}k_I{}^w\,.
 \label{nabPproj}
\end{equation}
The value of the moment map $\vec{P}_I$ in (\ref{mom_mapPI}) can be written using (\ref{innerprodveck}) and (\ref{defvecomega}) as
\begin{equation}
  \vec{P}_I = k_I{}^X \vec{k}_X= k_I{}^\ralpha \vec{k}_\ralpha + k_I{}^u\vec{k}_u= -\kappa ^{-2} \left( \vec{r}_I+2 k_I{}^u\vec{\omega }_u\right)\,.
 \label{kIalphakalpha}
\end{equation}
A similar relation for the special K\"{a}hler geometry will appear in Sect.~\ref{ss:gaugeisom}, where also a $r_\Lambda $ (or $r_I $ for the gauged symmetries) will describe the transformation of the compensating fields, see (\ref{deltaGY}).
Using the metric in (\ref{v:hypermetriek}) and (\ref{innerprodveck}), the relation \eqref{kIalphakalpha} implies
\begin{equation}
  k_{Ir}= -\kappa ^2 \vec{k}_r\cdot \vec{P}_I\,.
 \label{kIr}
\end{equation}
The equivariance relation\index{equivariance relation}  (\ref{equivarianceglobal}) gets also an extra part when we split the index $X$ and use the metric $h_{\Xu\Yu}$:
\begin{equation}
  k_I{}^\Xu \vec{J}_{\Xu}{}^{\Zu}h_{\Zu\Yu}k_J{}^\Yu + \kappa ^2 \vec{P}_I\times \vec{P}_J=f_{IJ}{}^K \vec{P}_K.
 \label{equivariancelocal}
\end{equation}
This equation admits unique solutions for $\vec{P}_I$ if the quaternionic-K\"{a}hler manifold is non-trivial ($n_H\neq 0$):
\begin{equation}
  2n_H\kappa ^2 \vec{P}_I=-\vec{J}_\Xu {}^\Yu \covder_\Yu k_I{}^\Xu .
 \label{solnPI}
\end{equation}
On the other hand, when $n_H=0$ the first term in \eqref{equivariancelocal} drops and there are two possible solutions for the
moment maps, which are then called Fayet--Iliopoulos (FI) terms. \index{Fayet--Iliopoulos term} First, in the
case where the gauge group contains an $\SU(2)$ factor, we can have
\begin{equation}
  \vec P_I= \vec e_I \xi ,
 \label{SU2FI}
\end{equation}
for any arbitrary constant $\xi$, and $\vec e_I$ being non-zero constants only for $I$ in the range of the $\SU(2)$ factor and satisfying
\begin{equation}
 \kappa ^2\xi  \vec e_I\times \vec e_J= f_{IJ}{}^K \vec e_K,
 \label{vece}
\end{equation}
in order that~(\ref{equivariancelocal}) is verified.

The second case are the $\U(1)$ FI terms. In that case the only remaining term in (\ref{equivariancelocal}) tells that $\vec{P}_I$ and $\vec{P}_J$ should be in the same direction in $\SU(2)$ space. Hence
\begin{equation}
   \vec P_I= \vec e\, \xi_I ,
 \label{U1FI}
\end{equation}
where $\vec e$ is an arbitrary vector in $\SU(2)$ space and $\xi_I$ are
constants for the $I$ corresponding to $\U(1)$ factors in the gauge
group.

\subsection{Decomposition Rules}
To find the decomposition laws, we start from the complete transformations of the hypermultiplet scalars in the conformal setting (from
\eqref{deltahyper}, \eqref{delDphiA}, \eqref{delDSU2q}, \eqref{gen_iso_hyper}):
\begin{align}
  \delta q^X =& \left(\delta _Q[\epsilon ]+ \delta _D[\lambda _{\rm D}]+ \delta _{\SU(2)}[\vec{\lambda }]+\delta _{G}[\theta]\right)q^X\nonumber\\
=& \left[  -\rmi \bar{\epsilon}^i\zeta^{{\cal A}} f^X{}_{i{{\cal A}} }+\hc\right] +\lambda _{\rm D}k_{\rm D}{} ^X-2\vec{\lambda }\cdot \vec{k} ^X + \theta ^I k_I^X\,.
 \label{conftransfqX}
\end{align}
We now consider what remains from \eqref{conftransfqX} when we restrict to Poincar\'{e} transformations (denoted as $\delta ^\poinc$), namely those combinations of (\ref{conftransfqX}) that preserve the gauge conditions. As we explained previously, under projection and gauge fixing, among the $q^X$, the only remaining physical fields are $q^\Xu$. Thus $\delta ^\poinc$ are the transformations induced by the dependence on $q^\Xu$, times the transformations of the latter.

First, for $X=0$, since $q^0$ is a constant in the Poincar\'{e} frame and $\vec{k}^0=\zeta^{{\cal A}}f^X{}_{i{{\cal A}} }=k_I{}^0=0$ (see \eqref{v:basiskeuze_kX}, \eqref{fcomponents0}, \eqref{SgaugeHyper},  \eqref{KillProjSymm}), only the dilatation term remains, and we thus find $0=\delta_D(\lambda_{\rm D})q^0$, i.e. implying that we should put $\lambda _{\rm D}=0$ in order to obtain Poincar\'{e} transformations.
Also $q^\ralpha $ is a constant by (\ref{SU2gauge}), and thus we get for $X=\ralpha$ the decomposition law for the $\SU(2)$ symmetry:
\begin{align}
0=\delta^\poinc q^\ralpha =& \left[  -\rmi \bar{\epsilon}^i\zeta^\Ap f^\ralpha {}_{i\Ap }
+\hc\right] -2\vec{\lambda }\cdot \vec{k} ^\ralpha + \theta ^I k_I^\ralpha \,.
 \label{0delqalpha}
\end{align}
Using (\ref{fcomponents0}) and (\ref{KillProjSymm}) all terms are proportional to $\vec{k}^r$, which is invertible as a $3\times 3$ matrix, and we find
\begin{equation}
  \vec{\lambda }= -\vec{\omega }_\Xu \left[  -\rmi \bar{\epsilon}^i\zeta^\Ap f^\Xu {}_{i\Ap }+\hc\right]+\ft12\theta ^I\vec{r}_I\,.
 \label{veclambda}
\end{equation}
The last case of (\ref{conftransfqX}) is
\begin{align}
  \delta^\poinc q^\Xu =\delta q^\Xu =&  \left[  -\rmi \bar{\epsilon}^i\zeta^{{\cal A}} f^\Xu{}_{i{{\cal A}} }+\hc\right] + \theta ^I k_I^\Xu\,.
 \label{conftransfqu}
\end{align}
Inserting this in (\ref{veclambda}), we obtain
\begin{equation}
  \vec{\lambda }= -\vec{\omega }_u\delta q^\Xu -\ft12\kappa ^2 \theta ^I\vec{P}_I \,.
 \label{veclambdagf}
\end{equation}
The second term determines that any gauge symmetry in the Poincar\'{e} theory has a contribution from the $\SU(2)$ $R$-symmetry in the conformal theory.

\section{\texorpdfstring{$D=5$ and $D=6$, ${\cal N}=2$}{D=5 and D=6, N=2}
Supergravities}
\label{ss:gfD56}

So far we discussed $D=4$ theories, but the main procedure is very similar for $D=5$ and $D=6$ theories. We review in the present section the main result for $D=5$ and $D=6$, ${\cal N}=2$
supergravities.

\subsection{\texorpdfstring{$D=5$}{D=5}}

 For $D=5$ one can again use a vector multiplet and a hypermultiplet as compensating multiplets.\footnote{For an overview of constructions with linear multiplets and other Weyl multiplets, see \cite{Ozkan:2016csy}.}  The difference from $D=4$ is that now the vector multiplet has only a real scalar. This fits remarkably well
with the fact that the $R$-symmetry group contains only $\SU(2)$, and no
$\U(1)$, which we used in $D=4$ to fix the phase of the scalar field
of the vector multiplet. This is schematically represented in
Table~\ref{tbl:fieldsConf}.
\begin{table}[!t] 
  \caption{\it Multiplets and fields in the superconformal construction for $D=5$}
  \label{tbl:fieldsConf}
\begin{tabular}{p{8mm}|p{1.5cm}p{1.5cm}p{1.5cm}|p{3.5cm}p{2cm}}
\hline
Spin& Weyl  & Vector  & Hyper & Gauge fix &Auxiliary \\
\hline
2&  $e_\mu ^a$    &               &         & &\\
$\ft32$&  $\psi _\mu ^i$ &              &         &  &\\
&  $V_{\mu i}{}^j$, $T_{ab}$ &   &         & &Auxiliary \\
 1&                & $n+1$      &           &  &\\
$\ft12$&$\chi ^i$ &    $2(n+1)$ & $2(n_H+1)$ & 2: $S$ & $\chi ^i$ with 2
others  \\
0& $D$ &            $n+1$ & $4(n_H+1)$ & 1: dilatations, 3: $\SU(2)$& $D$ and
 1 other   \\
\hline\noalign{\smallskip}
\end{tabular}
\end{table}
We already gave the field equations in Sect.~\ref{ss:eqom}. The field equations in (\ref{eq:Dchi-EOM}) constrain a doublet of spinors and a real scalar together with the gauge fixings of dilatation and $S$-supersymmetry, similar to how this happens in $D=4$, see Sect.~\ref{ss:gfwithmatter}. Vice versa, the field equations of these components of the vector and hypermultiplet eliminate $\chi ^i$ and $D$.
We give here the bosonic gravity sector in more detail. The Riemann scalar appears in $f_a{}^a =-R/16+\ldots $ in (\ref{transfDepF}), which in its turn appears in the sum of (\ref{conf-VTaction}) and  (\ref{conf-hyperaction})
\begin{align}
  {\cal L} = & {\cal L}_g +{\cal L}_g =-\frac{1}{24} e\, R \, \left(C_{IJK}\sigma ^I\sigma ^J\sigma ^K + k_{\rm D}{}^2\right)+\cdots \,.
\end{align}
The dilatational gauge fixing is chosen such that this gives the canonical normalization of the Einstein--Hilbert action, i.e. $eR/(2\kappa ^2)$. Combining with  (\ref{eq:Dchi-EOM}) leads to the analogue of (\ref{Dgaugegeneral}):
\begin{equation}
    C_{IJK}\sigma ^I\sigma ^J\sigma ^K = -3\kappa ^{-2}\,, \qquad  k_{\rm D}{}^2 = -9\kappa ^{-2}\,.
 \label{DgaugegeneralD5}
\end{equation}
This is used to obtain the Poincar\'{e} theory as we will further illustrate in Sect. \ref{ss:finalD5}.

\subsection{\texorpdfstring{$D=6$}{D=6}}

As we discussed in Sect \ref{sect:Weyl}, the $D=6$ Weyl multiplet contains an antisymmetric tensor $T^-_{abc}$, which is anti-self-dual. If one wants to build an action with manifest Lorentz-invariance, this has to be combined with a self-dual tensor $F^+_{abc}$, which sits in a tensor multiplet. The sum of these two can then be considered as the field strength of a physical antisymmetric tensor $B_{\mu \nu }$. The added tensor multiplet acts also as first compensating multiplet. There are constructions with as second compensating multiplet a hypermultiplet or with a linear multiplet, see e.g.
\cite{Bergshoeff:1986mz,VanProeyen:1985ib,Bergshoeff:1985mv,Coomans:2011ih}.

Importantly, the construction of $D=6$ minimal supergravity differs slightly from that for $D=4$ and $D=5$, since the tensor multiplet involves constraints that define the field $D$ of the Weyl multiplet in terms of fields of the tensor multiplet. This has to be contrasted with the cases of $D=4,5$, in which the field equation of the field $D$ in  (\ref{v:veldvglD}) combines with a dilatation gauge condition (\ref{DDgauge}) to give a condition on a scalar of the first compensating multiplet and another condition on a scalar of the second multiplet, see (\ref{Dgaugegeneral}). Indeed, in $D=6$, the field $D$ of the Weyl multiplet is not anymore independent from the tensor multiplet, and we thus have only one condition, which imply that there remains a physical scalar. The latter is in the Poincar\'{e} theory part of the tensor multiplet. This results matches with the expectation that, also in the Poincar\'{e} theory, we need one tensor multiplet in order to be able to construct Lorentz-invariant actions.\footnote{Clearly the fermionic part is analogous: the field $\chi ^i$ of the Weyl multiplet is determined by a constraint of the tensor multiplet. The $S$-gauge determines the fermion field of the second compensating multiplet, leaving the fermion of the tensor multiplet as physical field.}

Adding an arbitrary number of vector and hypermultiplets, one obtains super-Poincar\'{e} theories which all contain one tensor multiplet.\footnote{If one is only interested in field equations, and not in Lorenz-invariant actions, one can add an arbitrary number of tensor multiplets, or even no tensor multiplet  \cite{Marcus:1982yu}.}
Such actions were also constructed independently in  \cite{Nishino:1984gk,Nishino:1986dc}. The general super-Poincar\'{e} theory is given in \cite{Riccioni:2001bg}. It
builds on earlier work, e.g. \cite{Romans:1986er,Nishino:1997ff,Ferrara:1997gh}.

Famously, theories in $4n+2$ dimensions can suffer from gravitational (and gauge) anomalies. These can be calculated with the methods of Alvarez-Gaum\'{e} and Witten  \cite{AlvarezGaume:1983ig}. Anomaly-free theories, using the Green--Schwarz mechanism, have been identified in \cite{Green:1984bx,RandjbarDaemi:1985wc,Nishino:1986dc,Sagnotti:1992qw}.

%% file: ch5_Specialgeometries.tex
\chapter{Special Geometries}\label{ss:specialgeom}
\begin{quote}
\textbf{Abstract.}
In this chapter we will discuss the geometric concepts behind the special geometries. In terms of the supergravity theories, this means that we will restrict to the bosonic sector of these theories. In the first part of this chapter, we will examine the scalars of the vector multiplets, and we will see that the special K\"{a}hler geometry for the scalars is very much related to the symplectic transformations defined by the dualities of the vector fields. We will also discuss the attractor mechanism for black holes in this context. Then we will study the geometry of quaternionic-K\"{a}hler manifolds, related to the scalars of the hypermultiplets. Finally, we will present the relations between these manifolds and their versions in the different dimensions.
\end{quote}
\bigskip

The \emph{special geometries} \index{special geometries}are defined as the geometries of the scalars in ${\cal N}=2$ supergravities in $D=4,5,6$. As we will discuss, a very important role in these geometries is played by the $R$-symmetry group, namely $\SU(2)$ for $D=5,\,6$, and $\SU(2)\times \U(1)$
for $D=4$. In particular:
\begin{itemize}
\item
The $\SU(2)$ subgroup acts on the scalars of the
hypermultiplets, leading to the three complex structures of the
corresponding manifolds. The gauge connection of $\SU(2)$ promotes the hyper-K\"{a}hler manifold of
  hypermultiplets to a quaternionic manifold.\index{quaternionic manifold} \index{hyper-K\"{a}hler manifold}
\item
 In $D=4$, the $\U(1)$ factor acts on the complex scalars
of the vector multiplet, whose manifold therefore inherits one
complex structure. The gauge connection of this $\U(1)$ will be the K\"{a}hler curvature. \index{K\"{a}hler curvature}
\end{itemize}
There are three types of \emph{special manifolds}\index{special manifolds}. These are either associated to geometry of the real scalars of vector multiplets in $D=5$, to the complex
scalars of $D=4$ vector multiplets or to the quaternionic scalars of
hypermultiplets.\footnote{Since there are no scalars in the vector multiplets of
$D=6$ (see Table \ref{tbl:matterMult}), there is no geometry in that case.} However, these spaces are not completely
independent. An additional relation among them, called \cmap~and \rmap~(see Sect. \ref{ss:HomSymSp}), defines a precise correspondence.

Most of this chapter will be concerned with the special K\"{a}hler geometry of the scalars in the $D=4$ vector multiplet, which was obtained in the previous chapter.  We will see how properties of this geometry, due to supersymmetry, are inherited from symplectic transformations \index{symplectic transformation} acting on the gauge vectors in the multiplet.  The latter generates dualities between $D=4$, ${\cal N}=2$ theories \cite{Ferrara:1977iq,deWit:1979sh,Cremmer:1979up,Gaillard:1981rj}, which can be thought as generalizations of the well-known electric-magnetic dualities of Maxwell equations.
Afterwards, we will discuss quaternionic-K\"{a}hler geometries (Sect. \ref{ss:qKmanifolds}), and finally relations between all the manifolds (Sect. \ref{ss:HomSymSp}), where we will also consider special situations when the manifolds are homogeneous or symmetric.

\section{\texorpdfstring{$D=4$, ${\cal N}=2$}{D=4, N=2} Bosonic Action}\label{ss:resultact}

Before we start, let us rewrite the bosonic part of the action associated to the vector multiplets.
 The bosonic part of the matter-coupled supergravity action appears in the first 4 lines of (\ref{totaleLagN2D4}). Omitting for now the hypermultiplet sector ($q^X$ terms) and the Chern--Simons term in the fourth line (this will be included in the full action later, Sect. \ref{ss:finalresults}) we find
\begin{align}
e^{-1}\,{\cal L}_{\rm bos}=&\ft12 \kappa^{-2}R +{\cal L}_0+{\cal L}_1\,,\nonumber\\
{\cal L}_0 =& -g_{\alpha \bar \beta}\nabla _\mu z^\alpha
\nabla ^\mu \bar z^\beta\ -\ V(z,\bar z)\,, \qquad g_{\alpha \bar \beta
}=\partial_\alpha
\partial_{\bar \beta } {\cal K}(z,\bar z)\,,\nonumber\\
{\cal L}_1=& \ft 12 \Im \left( {\cal N}_{IJ}\,F_{\mu\nu}^{+I}
 F^{+\mu\nu\,J }\right)\nonumber\\
  =&\ft14(\Im {\cal N}_{IJ})F_{\mu\nu}^I F^{\mu\nu J} -\ft 18 e^{-1} (\Re {\cal
N}_{IJ}) \varepsilon^{\mu\nu\rho\sigma}F_{\mu\nu}^I F_{\rho\sigma}^J \,,\label{genL01}
\end{align}
 with $\nabla_\mu z^\alpha $ defined in (\ref{nablaX}), and we have used the results of Sect. \ref{ss:projcoord} to write the kinetic terms in ${\cal L}_0$.
This is in fact the general form for a theory with scalar fields
$z^\alpha$ in a K\"{a}hler manifold, and $n+1$ non-abelian vector fields labelled by an index $I$. The scalar potential $V(z,\bar z)$ \index{scalar potential} is very important for physical applications. Here we want to emphasize that $V(z,\bar z)$ is determined by the gauging as in the third line of (\ref{totaleLagN2D4}) (see Sect. \ref{ss:potentialN2} for more details).

\vspace{10pt}
Crucially, the kinetic terms in ${\cal L}_1$ are controlled by the complex functions ${\cal N}_{IJ}$. The latter are defined by the second line of (\ref{totaleLagN2D4}), upon using the equation of motion of $T_{ab}$ (\ref{YTeom}),
\begin{equation}
{\cal N}_{IJ}(z,\bar z) =\bar F_{IJ} +\rmi \frac{(N_{IN}X^N )\,(N_{JK}
X^K)}{N_{LM}\ X^L X^M}\,. \label{Ndef}
\end{equation}
These functions become the effective (complexified) gauge couplings, whenever the scalars have a non-zero vacuum expectation value (v.e.v.)\index{vacuum expectation value} on some classical background. In particular the v.e.v. of $\Im {\cal N}$ gives the effective gauge coupling\index{gauge coupling}, while that of $\Re {\cal N}$ gives the so-called theta angles. \index{theta angles}
 \bexer \label{ex:FetaVectors}
Consider the model of Exercise~\ref{ex:Feta}. Check that in this case, the kinetic terms of the two vectors are determined by the matrix
(\ref{Ndef})
\begin{equation}
  {\cal N}_{IJ}=\frac12\rmi\begin{pmatrix}1&0\cr 0&-1\end{pmatrix}
  +\frac{\rmi}{z^2-1}\begin{pmatrix}1&-z\cr -z&z^2\end{pmatrix}=-\frac{\rmi}{1-z^2}
  \begin{pmatrix}\ft12(1+z^2)&-z\cr -z&\ft12(1+z^2)\end{pmatrix}\,.
 \label{calNFeta}
\end{equation}
The imaginary part is negative definite as it should be for positive
kinetic terms of the vectors.
 \eexer

\section{Symplectic Transformations}
 A prerequisite to understand the structure of special K\"{a}hler geometry is a study of symplectic transformations. As we will explain in this section, the latter first appear as generalized electric-magnetic dualities in the context of pure $D=4$ abelian gauge theories. We will argue that symplectic transformations and consequent dualities can be extended to the full ${\cal N}=2$ vector multiplets by means of supersymmetry.

\subsection{Electric-Magnetic Dualities of Vector Fields in \texorpdfstring{$D=4$}{D=4}}
\label{ss:dualities}
A striking property of pure abelian gauge theories is that their equation of motions and Bianchi identities are left invariant by a
group of symplectic transformations, which act on the (complexified) gauge couplings. In other words, symplectic transformations generate dualities of these theories.

Let us recall some key ingredients to understand symplectic transformations, following the main ideas discussed in  \cite[Sect. 4.2.4]{Freedman:2012zz}.\footnote{There the kinetic matrix ${\cal N}_{IJ}$ is written as ${\cal N}_{IJ}=-\rmi\bar f_{IJ}$.} The starting point is to consider the kinetic terms of the vector fields, ${\cal L}_1$ in
(\ref{genL01}), which can be written as
\begin{equation}
{\cal L}_1= \ft 12 \Im \left( {\cal N}_{IJ}\,F_{\mu\nu}^{+I}
 F^{+\mu\nu\,J }\right) = \ft 12 \Im \left(
 F_{\mu\nu}^{+I}   G_{+I}^{\mu \nu } \right) \,,
 \label{L1FG}
\end{equation}
with, viewing ${\cal L}_1$ as function of $F_{\mu\nu}^{+I}$ and $F_{\mu\nu}^{-I}$,
\begin{equation}
  G_{+I }^{\mu\nu}\equiv 2\rmi\frac{\partial{\mathcal L}_1}
  {\partial { F}^{+I }_{\mu\nu}}=
{\mathcal N}_{I J }{ F}^{+J \,\mu\nu}\,, \label{defG}
\end{equation}
and the tensor $\cal N$ defined in \eqref{Ndef}. Importantly, in selecting the kinetic terms of \eqref{genL01}, we are setting $f_{IJ}{}^{K}=0$, such that \eqref{genL01} becomes effectively a $\U(1)^{n+1}$ gauge theory. Non-abelian parts will be discussed later, in Sect. \ref{ss:gaugeisom}, when the symmetries of the kinetic terms will be identified and gauged.

The field equations and Bianchi identities are invariant under $\Sp(2(n+1),\mathbb R)$ transformations\footnote{We
use here and below the tilde to indicate the transformed fields. We hope
that the reader does not confuse these with the duals, (\ref{dual46}),
which are contained in the (anti)self-dual $F^{\pm }$.}
\begin{equation}
\begin{pmatrix}\widetilde{ F}^+\cr \widetilde
G_+\end{pmatrix}={\mathcal S} \begin{pmatrix}{ F}^+\cr G_+
\end{pmatrix}\,,\qquad {\mathcal S} \in \Sp(2(n+1),\mathbb R)\,.
\label{FGsympl}
\end{equation}
Specifically ${\mathcal S}$ is a symplectic transformation\index{symplectic transformation}, namely a real matrix\footnote{Quantization of electric and magnetic charges breaks $\Sp(2(n+1),\mathbb R)$ to $\Sp(2(n+1),\mathbb Z)$. We do not discuss quantum effects in this section.} satisfying
\begin{equation} {\mathcal S}= \begin{pmatrix}A&B\cr C&D\end{pmatrix}\,,\qquad  {\mathcal S}^T  \Omega   {\mathcal S}   =  \Omega  \qquad\mbox{where}\qquad
\Omega= \begin{pmatrix}0&\unity \cr -\unity &0\end{pmatrix}\,.
\label{Scond}
\end{equation}
This is equivalent to the requirements
\begin{equation}
A^T C-C^T A=0\,,\qquad  B^T D- D^T B=0\,,\qquad   A^T D-C^T B=\unity\,.
\label{symplABCD}
\end{equation}
It is not hard to verify that under \eqref{FGsympl} the kinetic matrix ${\cal N}$ transforms as
\begin{equation}
  \widetilde{\mathcal N} = (C + D{\mathcal N})(A+B{\mathcal
N})^{-1}\,.
 \label{sympltrN}
\end{equation}
Importantly, \eqref{sympltrN} preserves the condition $\Im{\mathcal N}<0$, namely the positivity of the kinetic
terms for $\mathcal{L}_1$ (see Sect. \ref{ss:positivity}). In the language of symplectic geometry, a $2(n+1)$-component column $V$ transforming as $\widetilde V = {\cal S}V$. The prime example is \eqref{FGsympl}. The invariant inner product of two symplectic vectors\index{symplectic invariant}
$V$ and $W$ is
\begin{equation}
 \langle V , W \rangle \equiv V^T \Omega W \ .
\label{innerProdSymplV}
\end{equation}

Finally, observe that if we apply  (\ref{FGsympl}) to the last of (\ref{L1FG}), the action is invariant only if $B=0$ (see that $\Im F^{+I}F^{+J}$ is a total derivative). For this reason the subgroup of $\Sp(2(n+1),\mathbb R)$ with $B=0$ are sometimes called `perturbative symmetries'. It is possible to construct an action invariant under the more general transformations with $B\neq 0$, at the price of introducing magnetic duals of the vector fields $W_\mu ^I$.

\subsection{Symplectic Transformations in \texorpdfstring{${\cal N}=2$}{N=2}}
\label{ss:symplSpecial}

Let us apply the arguments of the previous section to ${\cal N}=2$ supergravities. Our starting point is the tensor ${\cal N}$, which is determined as a function of $F$, as explained in Sect.~\ref{ss:resultact}. From its definition (\ref{Ndef}) we can immediately derive the following suggestive relations
\begin{align}
  {\cal N}_{IJ}   X^J =&(\bar F_{IJ}+\rmi N_{IJ})X^J = F_{IJ}X^J = F_I\,, \nonumber\\
  {\cal N}_{IJ} \nabla _{\bar\alpha} \bar X^J =& \bar F_{IJ}\nabla _{\bar\alpha} \bar X^J =\nabla _{\bar\alpha}\bar F_I \,,
\label{calNXbarF}
\end{align}
where in the last line we used also (the complex conjugate of)  (\ref{nablaN})  and we introduced the K\"{a}hler covariant\index{K\"{a}hler covariant derivatives} derivatives of Sect.~\ref{ss:Kacovder})
\begin{equation}
  \nabla_{\bar \alpha} \bar F_I=\partial_{\bar \alpha} \bar F_I+\ft12
  a^{-1}(\partial_{\bar   \alpha}{\cal K})\bar F_I\,,\qquad
\nabla_{\bar \alpha} \bar X^J=\partial_{\bar \alpha} \bar X^J+\ft12
  a^{-1}(\partial_{\bar   \alpha}{\cal K})\bar X^J\,.
\label{Dbarbar}
\end{equation}
The identities (\ref{calNXbarF}) are of a similar form as (\ref{defG}) relating the lower components of a symplectic vector to its upper components by multiplication with ${\cal N}_{IJ}$. Hence we can identify two $2(n+1)$-component symplectic vectors (and their complex conjugates)\index{symplectic vector}
\begin{equation}
 V=\begin{pmatrix}X^I\cr F_I\end{pmatrix}  \,,\qquad
U_\alpha =\begin{pmatrix}\nabla _\alpha X^I\cr \nabla _\alpha F_I\end{pmatrix}\,,
 \label{defVU}
\end{equation}
and check that they indeed transform as in (\ref{FGsympl}). With this identification in mind, we can reconsider
the kinetic terms of the scalars and note that the K\"{a}hler
potential (\ref{KP}), and the constraint (\ref{subman_proj}) are
symplectic invariants. This is the starting point of the symplectic formulation of the special
geometry, which we will discuss in Sect.~\ref{ss:defSpK}. There we will see that  (\ref{calNXbarF}) can be seen as an alternative definition of the matrix ${\cal N}$.

An immediate consequence of this construction is that a symplectic transformation on $V$ induces a change of coordinates $X^I \mapsto \tilde X^I$ such that
\begin{equation}
  \tilde X^I= A^I{}_J X^J + B^{IJ} F_J(X)\,,\qquad \tilde F_I= C_{IJ} X^J + D_I{}^J F_J(X)\,.
 \label{tildeXX}
\end{equation}
If the first equation is invertible,\footnote{The full
symplectic matrix is always invertible due to (\ref{Scond}), but this part may not be. In
rigid supersymmetry, the invertibility of this transformation is
necessary for the invertibility of ${\cal N}$ (due to the positive
definiteness of the full metric), but in supergravity we may have that
the $\tilde X^I$ do not form an independent set, and then $\tilde F$
cannot be defined. See below.} i.e. defines a relation $X^I(\tilde X)$, then
\begin{equation}
  \frac{\partial }{\partial \tilde X^J}\tilde F_I = \frac{\partial X^K}{\partial \tilde X^J}\,\frac{\partial\tilde  F_I}{\partial X^K}\,.
 \label{integrabFI}
\end{equation}
The inverse of the first factor and the last factor can be obtained from (\ref{tildeXX}) and then  (\ref{symplABCD}) implies that (\ref{integrabFI}) is symmetric in $(IJ)$. This is the integrability condition for the (local) existence of a new \emph{prepotential} $\tilde F(\tilde X)$, such that:
\begin{equation}
\tilde F_I(\tilde X)=\frac{\partial \tilde F(\tilde X)} {\partial \tilde X^I}\,.
\end{equation}
Invertibility of \eqref{tildeXX} thus guarantees the existence of a new formulation of the theory, and thus of the
target space manifold, in terms of the prepotential $\tilde F(\tilde X)$.

\vspace{10pt}
The attentive reader may have noticed that the prepotential has entered automatically our superconformal construction in Sect.~\ref{ss:actvector4}. This is not surprising due to the homogeneity conditions \eqref{homegeneqs}, also valid for $\tilde F $, 
which allow one to determine $\tilde F$ from \eqref{tildeXX} as follows:
\begin{equation}
\tilde F(\tilde X(X))= \frac{1}{2} V^T \begin{pmatrix}C^T A& C^T B \cr
D^T A & D^T B\end{pmatrix}V\,. \label{tilFexpl}
\end{equation}
 However, as we will show in the following, there are also symplectic transformations such that the prepotential is not defined in the new symplectic basis, i.e. that the new $\tilde F_I$ cannot be obtained as a derivative of a function $\tilde F(\tilde X)$. A concrete example exhibiting this feature will be given in Sect. \ref{ss:defSpK}.
Therefore, one may expect new formulations without a prepotential to be generically available. We will show in the following that this is indeed the case.
\vspace{10pt}

Let us first consider an example  \cite{deWit:1996jd}, starting from the prepotential (\ref{FX13}).
If we apply the symplectic transformation
\begin{equation}
{\cal S}=\begin{pmatrix}A&B\cr C&D\end{pmatrix}=\begin{pmatrix}1&0&0&0\cr
0&0&0&1/3\cr 0&0&1&0\cr 0&-3&0&0\end{pmatrix}\,,
\end{equation}
using (\ref{tilFexpl}) one arrives at $\tilde F(\tilde X(X))= -2 (X^1)^3X^0$, which is just (\ref{Fsqrt}) after the trivial field redefinition $X\rightarrow \tilde X$. Thus the two theories based on (\ref{FX13}) and  (\ref{Fsqrt}) are equivalent. We will call this a \emph{pseudo symmetry}\index{pseudo symmetry} between the two formulations, and the transformation is called  \emph{symplectic reparameterization} \index{symplectic reparameterization}.

On the other hand consider
\begin{equation}\label{cosetS11}
{\cal S}=\begin{pmatrix}1+3\epsilon&\mu    &0&0\cr
           \lambda& 1+\epsilon &0&2\mu/9\cr
            0&0           &1-3\epsilon&-\lambda\cr
            0& -6\lambda &-\mu&1-\epsilon \end{pmatrix}\,,
\end{equation}
 for infinitesimal $\epsilon,\mu,\lambda$. Since $F$ is invariant under this transformation, the latter defines a symmetry of the model. On the scalar field $z=X^1/X^0$, the transformations act as
\begin{equation}
\delta z=\lambda-2\epsilon z-\mu z^2/3 \,.
\end{equation}
All in all, the set of transformations \eqref{cosetS11} form an $\SU(1,1)$ isometry group of the scalar manifold. The domain
were the metric is positive definite is $\Im z >0$, hence proving the
identification of the manifold as the coset space in (\ref{FX13}), or equivalently (\ref{Fsqrt}).

In general, we have to distinguish two  situations: \cite{Hull:1995gk}
\begin{enumerate}
  \item The function $\tilde F(\tilde
X)$ is different from $F(\tilde X)$. In that case $\tilde F$ provides a dual description of the classical field theory with prepotential $F$. The two functions
describe equivalent physics and we have a {\em pseudo symmetry}. The transformations are called symplectic reparameterizations~\cite{Cecotti:1989qn}. Hence we may find a variety of
descriptions of the same theory in terms of different functions  $F$.
\item If a symplectic transformation leads to the same function $F$,
then we are dealing with a {\em proper symmetry}. \index{proper symmetry}
This invariance reflects itself in an isometry of the target space manifold.

Note that a symplectic transformation with
\begin{eqnarray}
{\cal S}=\begin{pmatrix}\unity & 0 \cr C &\unity \end{pmatrix}\,,
\label{10C0}
\end{eqnarray}
does not change the $X^I$ and gives $\tilde F=F+\ft12 C_{IJ}X^IX^J$. The matrix $C$ must be symmetric as required for ${\cal S}$ to be symplectic. Indeed, the difference between $\tilde F$ and $F$ is then of the form (\ref{FCS0}), and the action is invariant. Hence these are proper symmetries.
\end{enumerate}
The two cases above are called `duality
symmetries',\index{duality
symmetries} as they are generically accompanied by duality
transformations on the field equations and the Bianchi identities. The
question remains whether the duality symmetries comprise all the
isometries of the target space, i.e. whether
\begin{equation}
Iso(\mbox{scalar manifold})\subset \Sp(2(n+1),\mathbb{R})\,.
\label{isosubSp}
\end{equation}
This question was investigated in~\cite{deWit:1992cr} for the very special
K\"{a}hler manifolds, and it was found that in that case one does obtain the
complete set of isometries from the symplectic transformations. For
generic special K\"{a}hler manifolds no isometries have been found that are
not induced by symplectic transformations, but on the other hand we do not know a proof that these do not exist.

That the full supersymmetric theory allows such symplectic
transformations can also be seen in another way. We mentioned before that
the vector multiplets are chiral multiplets, with $w=1$ that satisfy
constraints (\ref{bian4}). One of these constraints is the Bianchi
identity for the vector. The functions $F_I(X)$ transform also in a
chiral way under supersymmetry, and thus define also chiral multiplets,
with Weyl weight $w=1$. Now it turns out that the same constraints on these chiral multiplets are in fact the field equations. One of these is the
field equation of the vectors. Thus the symplectic vectors $V$
(\ref{defVU}) are the lowest components of a symplectic vector of chiral
multiplets. If the vector of chiral multiplets satisfy the mentioned constraints, then these imply as well that the multiplets are vector multiplets rather than general chiral multiplets, and that the fields of these vector multiplets satisfies the field equations. This is thus a supersymmetric generalization of the symplectic
set-up of Sect.~\ref{ss:dualities}. For the supergravity case this has
been worked out in~\cite{Claus:1999np}. In this way, even the equations
for models in a parameterization without prepotential can be obtained. Such
situations will be explained in the next section.

\section{Characteristics of a Special Geometry} \label{ss:defSpK}
In Chap. \ref{ss:specialgeom} we defined the Special K\"{a}hler geometry~\cite{deWit:1984pk} as a property of the scalar couplings in $\mathcal{N}=2$ supergravities.
However, one may wonder what is the
definition of special geometry, independent of supersymmetry. A first step
in that direction has been taken by Strominger~\cite{Strominger:1990pd}.
He had in mind the moduli spaces of Calabi--Yau spaces. His definition is
already based on the symplectic structure, which we also have emphasized.
However, being already in the context of Calabi--Yau moduli spaces, his
definition of special K\"{a}hler geometry omitted some ingredients that are
automatically present in any Calabi--Yau moduli space, but have to be
included as necessary ingredients in a generic definition. Another
important step was made in~\cite{Ceresole:1995jg}. Before, special
geometry was necessarily connected to the existence of a holomorphic prepotential $F(X)$ or $\ZF(Z(z))$: it was recognized as those K\"{a}hler manifolds where the K\"{a}hler potential is of the form
(\ref{KP}) with (\ref{ZFIasdZF}).

However, the authors of~\cite{Ceresole:1995jg} found examples of ${\cal N}=2$ supergravities coupled to Maxwell
multiplets where the $\ZF_I$ in (\ref{KP}) is not a derivative of a prepotential as in  (\ref{ZFIasdZF}). Crucially the latter were
obtained from a symplectic transformation of a model admitting a prepotential. This surprising fact raised some natural questions: are all the models without
prepotential symplectic dual to models with a prepotential? Can one still
define special K\"{a}hler geometry as always symplectic dual to a formulation with a
prepotential? And, of course, is there a more convenient definition that does not
involve this prepotential? These questions have been answered
in~\cite{Craps:1997gp}, and are reviewed here. A lot of this is based on formalism that has been developed using other methods in \cite{Castellani:1990tp,Ceresole:1995ca,Andrianopoli:1996vr,Andrianopoli:1997cm}.

\subsection{Symplectic Formulation of the Projective K\"{a}hler Geometry}
\label{ss:symplformul}

The scope of this section is to reformulate the special K\"{a}hler geometry of Sect.~\ref{ss:resultact}  using the symplectic formalism. The basic building block is the symplectic vector $V$, which transforms as in (\ref{FGsympl}),
\begin{equation}
  V=\begin{pmatrix}X^I\cr F_I\end{pmatrix}\,.
\label{Vlocal}
\end{equation}
It is important to realize that $F_I$ is not necessarily a function of $X^I$, as is the case in the formulation with the prepotential. The
dilatational gauge fixing condition (\ref{subman_proj}) is chosen in order to decouple kinetic terms of the
graviton from those of the scalars.
It can be written as a condition on the symplectic inner product \eqref{innerProdSymplV}:
\begin{equation}
  \langle V,{\bar V}\rangle=X^I\bar F_I-F_I\bar X^I=\rmi a=\rmi\kappa ^{-2}\,.
\label{VbarVisi}
\end{equation}
To solve this condition, we define
\begin{equation}
  V=y(z,\bar z)v(z)\,,\qquad y(z,\bar z)=\rme^{\kappa ^2{\cal K}(z,\bar z)/2}\,,
\label{Vtov}
\end{equation}
where $v(z)$ is a holomorphic symplectic vector
\begin{equation}
v(z)= \begin{pmatrix}Z^I(z)\cr \ZF_I(z)
\end{pmatrix}\,, \label{vz}
\end{equation}
 and ${\cal K}(z,\bar z)$ will be the K\"{a}hler potential. $Z^I(z)$ and $\ZF_I(z)$ are so far $2(n+1)$ arbitrary holomorphic functions in the $n$ complex variables $z^\alpha $, reflecting the freedom of choice of the coordinates $z^\alpha $, up to a non-dege\-nera\-cy condition that we will soon discover.
In terms of \eqref{Vtov} the condition (\ref{VbarVisi}) implies a value for the K\"{a}hler potential in terms of a symplectic invariant:
\begin{equation}
 y^{-2}= \rme^{-\kappa ^2{\cal K}(z,\bar z)}=-\rmi \kappa ^2\langle {v},{\bar v}\rangle\,.
\label{Kahlerv}
\end{equation}

We now act on (\ref{VbarVisi}) and (\ref{Kahlerv}) with the following K\"{a}hler covariant derivatives (see Sect.~\ref{ss:Kacovder}):
\begin{align}
 \nabla_\alpha v (z)\equiv&  \partial _\alpha v(z)+\kappa ^2 v(z)(\partial _\alpha {\cal K})\,,\quad &\nabla _\alpha \bar v=0\,,\nonumber\\
 \nabla_\alpha V\equiv&  \partial _\alpha V+\ft12\kappa ^2 V(\partial _\alpha {\cal K})\,,\quad &\nabla _\alpha \bar V=0\,.
\label{KahcovD22}
\end{align}
This implies
\begin{equation}
  \langle \nabla_\alpha {v},{\bar   v}\rangle=0\,,\qquad \langle \nabla_\alpha V,{\bar   V}\rangle=0\,.
 \label{Dalphavbarv}
\end{equation}
A further derivation implies, using (\ref{commnablas})
\begin{equation}
\langle\nabla _\alpha V,\,\nabla _{\bar \beta }{\bar V}\rangle  =-\langle \nabla _{\bar \beta }\nabla _\alpha V,{\bar V}\rangle= -\rmi g_{\alpha \bar \beta }\,.
 \label{stepVbarVg}
\end{equation}
 Hence the K\"{a}hler metric can be obtained as
\begin{equation}
  g_{\alpha \bar \beta
    }=\rmi\langle \nabla_{ \alpha}  V ,\nabla_{\bar \beta}\bar  V\rangle\,.
 \label{gassymplprod}
\end{equation}

There is an important extra condition on the inner product
\begin{equation}
  \langle\nabla _\alpha V,\,V\rangle=0\,.
 \label{nabVV0}
\end{equation}
In most examples\footnote{It has been proven \cite{Craps:1997gp} that there always exist a symplectic frame where this is the case.} the matrix (\ref{matXDX}) is invertible and we will prove at the end of Sect. \ref{ss:specKahlerdef} that then this condition implies the existence of a prepotential. Here we can already see that the condition is always satisfied for any $F_I = \partial _I F$.
Indeed, in that case, due to the homogeneity of $F_I$:
\begin{equation}
  \langle\nabla _\alpha V,\,V\rangle=  (\nabla _\alpha X^I)F_I - (\nabla _\alpha F_I )X^I= (\nabla _\alpha X^I)F_I - F_{IJ}(\nabla _\alpha X^J)X^I = 0\,.
 \label{proofnabVVF}
\end{equation}
However, in full generality and without requiring a prepotential, we will impose (\ref{nabVV0}) in a definition of special K\"{a}hler geometry, see Sect.~\ref{ss:specKahlerdef}.
The conditions (\ref{VbarVisi}), (\ref{gassymplprod}) and (\ref{nabVV0}) can be combined in the $(n+1)\times (n+1)$ matrix relation
\begin{tcolorbox}
\begin{equation}
G \equiv \rmi \langle\begin{pmatrix} \bar V \cr\nabla _\alpha V\end{pmatrix}, \,\begin{pmatrix}V& \nabla_{\bar \beta}\bar  V\end{pmatrix}\rangle =\begin{pmatrix}\kappa ^{-2}& 0\\ 0&g_{\alpha \bar \beta }\end{pmatrix}\,.
 \label{matrixeqnspecK}
\end{equation}
\end{tcolorbox}
If we further require positivity of $G$, namely of the kinetic matrices $g_{\alpha \bar \beta }$ and $\kappa ^{-2}>0$, \footnote{These are thus the positivity requirements of the kinetic terms of the spin~0 and spin~2 fields. Recall from Sect.~\ref{ss:positivity} that this in turn implies positivity for the kinetic terms of the spin 1 fields.}
this in turn implies that the $(n+1)\times (n+1)$ matrix
\begin{equation}
 \begin{pmatrix}\bar X^I &\nabla _{\alpha}X^I\end{pmatrix}= y \begin{pmatrix}\bar Z^I &\nabla _{\alpha}Z^I\end{pmatrix}\,,
 \label{invertible}
\end{equation}
which is the upper part of $\begin{pmatrix}\bar V&\nabla _\alpha V\end{pmatrix}$, is invertible.
For clarity: the matrix  (\ref{invertible}) is in full
\begin{equation}
  \begin{pmatrix}{\bar X}^0\quad \nabla_{1}{ X}^0\quad\dots\quad\nabla_{ n} { X}^0\cr
    {\bar  X}^1\quad\nabla_{ 1}{ X}^1\quad\dots\quad\nabla_{ n} { X}^1\cr
    \cdots\cr{\bar X}^n\quad\nabla_{ 1}{ X}^n\quad\dots\quad\nabla_{ n} { X}^n\cr \end{pmatrix}\,.
 \label{explicituppermatrix}
\end{equation}

Such theorems on positivity and invertibility of symplectic matrices are collected in  \cite[Appendix B]{Craps:1997gp}. To prove the above statement, suppose that (\ref{invertible}) is not invertible. Then there is a zero mode $a\bar X^I+b_\alpha \nabla _\alpha X^I=0$. Also its complex conjugate is then zero, which leads to
\begin{align}
  0= &\langle a\bar V+b_\alpha \nabla _\alpha V,\, \bar a V+\bar b_\alpha \nabla _\alpha \bar V \rangle= \begin{pmatrix}a &b^\alpha \end{pmatrix} \langle\begin{pmatrix} \bar V \cr\nabla _\alpha V\end{pmatrix}, \,\begin{pmatrix}V& \nabla_{\bar \beta}\bar  V\end{pmatrix}\rangle
 \begin{pmatrix}\bar a \cr\bar b^\alpha \end{pmatrix} \nonumber\\
  &=\begin{pmatrix}a &b^\alpha \end{pmatrix} (-\rmi G)\begin{pmatrix}\bar a \cr\bar b^\alpha \end{pmatrix} \,.
\label{proofinvertible}
\end{align}
 This is contradictory to the positivity of $G$.
Invertibility of (\ref{invertible}) is the non-dege\-nera\-cy condition on the choice of (\ref{vz}) that we mentioned below that equation.

In terms of these quantities, and using the invertibility of (\ref{invertible}), the kinetic matrix for the vectors is defined by
\begin{equation} {\mathcal N}_{IJ}= \begin{pmatrix}F_I&\nabla_{\bar \alpha} \bar F_I \end{pmatrix}
\begin{pmatrix}X^J& \nabla_{\bar \alpha} \bar X^J\end{pmatrix}^{-1}\,, \label{Nsugra}
\end{equation}
which is the product of two $(n+1)\times (n+1)$ matrices. This agrees with the definition of ${\cal N}_{IJ}$ starting from a prepotential, since the latter implied (\ref{calNXbarF}).

\subsubsection{Models without a Prepotential and Examples}
In the previous sections, all the requirements listed above were satisfied with $F_I$ defined via a prepotential, i.e. $F_I =\frac{\partial }{\partial X^I}F(X)$. The conclusion of the current analysis is that we can get rid completely of this last requirement. We can now appreciate how a symplectic transformation may relate symplectic vectors of a special K\"{a}hler manifold determined by a prepotential to new symplectic vectors where $\tilde F_I$ cannot be written as the derivative of a function $\tilde F(\tilde X)$.
A famous example \cite{Ceresole:1995jg} comes from the reduction to ${\cal N}=2$ of two versions of ${\cal N}=4$ supergravity, known,
respectively, as the `$\SO(4)$ formulation'~\cite{Das:1977uy,Cremmer:1977zt, Cremmer:1977tc} and the
`$\SU(4)$ formulation' of pure ${\cal N}=4$ supergravity~\cite{Cremmer:1978tt}.\footnote{This was revisited in exercises 20.18 and 20.19 in \cite{Freedman:2012zz}.}  In the initial duality frame the prepotential is
\begin{equation}
 F=-\rmi \kappa ^{-2}X^0X^1\,.
 \label{FexampleO1}
\end{equation}
After a symplectic transformation, one finds a new model based on the symplectic vector
\begin{equation}
v=\begin{pmatrix}1\cr \rmi\cr -\kappa ^{-2} z\cr
-\rmi \kappa ^{-2}z\end{pmatrix}\,. \label{tildevExample}
\end{equation}
After this mapping, $z$ does not appear anymore in $(\widetilde Z^0,\widetilde Z^1)$, the upper two components of the symplectic vector.
It is then clear that from (\ref{vz}) we cannot define the function $\widetilde F(\widetilde Z^0,\widetilde
Z^1)$, which would play the role of a prepotential.
Another surprising phenomenon showing up in this model,\footnote{This is explicitly worked out for the above example in exercise 20.20 in \cite{Freedman:2012zz}.} but being in fact very general, is that the set of perturbative symmetries (those with $B=0$) is different before or after the transformation. In particular in the formulation with a prepotential there is only one `perturbative' symmetry, while in the formulation without prepotential there are three.
\bigskip

The existence of a prepotential is intimately related to the invertibility of the matrix
\begin{equation}\begin{pmatrix}X^I&\nabla_\alpha  X^I\end{pmatrix}.
\label{matXDX}
\end{equation}
One can easily check that, as opposed to (\ref{invertible}), the above is not always invertible---for example in the formulation (\ref{tildevExample}). \textbf{If}
that matrix is invertible, then a prepotential exists~\cite{Craps:1997gp}. This will be proven below, related to (\ref{existenceFproof}). It turns out that the following 3 conditions are equivalent:
\begin{enumerate}\label{equivCondF}
  \item The matrix (\ref{matXDX}) invertible;
  \item \label{condspeccoord} Special coordinates \index{special coordinates} are possible; these are coordinates defined in (\ref{specialcoord}).
  \item  A prepotential $F(X)$ exists.
\end{enumerate}

\bigskip
Another important example of the absence of a prepotential occurs when describing the manifold
\begin{equation}
 \frac{\SU(1,1)}{\U(1)}\otimes
\frac{\SO(r,2)}{\SO(r)\otimes \SO(2)}\ .\label{manSTr}
\end{equation}
This is the only special K\"{a}hler manifold that is a product of two
factors~\cite{Ferrara:1989py}. In physics, these manifold emerge in the classical limit of the compactified heterotic string, where the
dilaton does not mix with the scalars of the other vector multiplets and thus the target space should be a product of two factors. The first formulation of these spaces
used a prepotential $F$ of the form (\ref{Fvsp}) \cite{Cremmer:1985hj}. As mentioned in Sect. \ref{ss:examplesSpecK} this is therefore an example of a `very special K\"{a}hler manifold'.\footnote{This example will reappear in Sect. \ref{ss:HomSymSpaces}: see the third line of Table \ref{tbl:symvs}.}\index{very special K\"{a}hler manifolds}
In this case
\begin{equation}
  d_{ABC}X^A X^B X^C= X^1X^a\eta _{ab}X^b\,,\qquad a,b=2,\ldots ,r+1\,,
 \label{dsymbolssplit}
\end{equation}
with $\eta _{ab}$ a constant metric of signature $(+,-,\ldots ,-)$.

In this description of (\ref{manSTr}) by a prepotential, only part of $\SO(r,2)$ belongs to the set of perturbative symmetries. In other words, one needs $B\neq 0$ in the duality group to generate the full $\SO(2,r)$. This is somehow unexpected from string theory point of view, as from the superstring compactification one expects the full $\SO(2,r;\mathbb{Z} )$ to be explicitly realized as a perturbative ($T$-duality) symmetry group. The key point is that there exists another symplectic frame where the full $\SO(r,2)$ is realized as  a perturbative symmetry~\cite{Ceresole:1995jg}. After that symplectic transformation one finds a symplectic vector $(X^I,F_I)$ satisfying
\begin{equation}
 X^I\,\eta_{IJ}\,X^J =0\ ;\qquad
F_I =S\,\eta_{IJ}\,X^J \ ,
\end{equation}
where $\eta_{IJ}$ is a metric for $\SO(2,r)$ and the dilaton $S$ is one of the variables $\{z^\alpha \}$. The first constraint comes
on top of (\ref{VbarVisi}), and thus implies that the
variables $\{z^\alpha \}$ cannot be chosen between the $X^I$ only. Indeed, $S$ occurs only in $F_I$. Therefore, in this new formulation, which realizes the full $\SO(r,2)$ as a symmetry of the action, special coordinates are not possible and we do not have a prepotential.

\subsection{Definitions}
\label{ss:specKahlerdef}
After this extension of the formulation, the reader may wonder what is then really special K\"{a}hler geometry. This question has been addressed in \cite{Craps:1997gp} and leads to  a few equivalent formulations of a definition. We will first give a definition using the prepotential, and then a second one using
only the symplectic vectors. We will then show the equivalence. There is
also a more mathematically inspired definition \cite{Freed:1997dp}, which was summarized in \cite[Appendix 20C]{Freedman:2012zz} and we will not repeat here.
\vspace{20pt}

\noindent \textbf{Definition 1 of (Local) Special K\"{a}hler Geometry.}

\noindent A special K\"{a}hler manifold is an $n$-dimensional K\"{a}hler--Hodge manifold
with on any chart $n+1$ holomorphic functions $Z^I(z)$ and a holomorphic
function $F(Z)$, homogeneous of second degree, such that, with
(\ref{vz}), the K\"{a}hler potential is given by  \eqref{Kahlerv} and the $v(z)$ are connected by symplectic
transformations $\Sp(2(n+1),\mathbb R)$ and/or  K\"{a}hler transformations.
\begin{equation} v(z) \rightarrow e^{f(z)}{\mathcal S} v(z)\,,
\end{equation}
on overlap of charts. The transition functions should satisfy the cocycle condition on overlap
of regions of three charts.

We mentioned that there are formulations without a prepotential, e.g. the one based on the symplectic vector (\ref{tildevExample}).
In this case, the above definition turns
out not to be applicable in an arbitrary symplectic frame. Therefore we
will now give a second definition, and will comment about their equivalence.

\vspace{10pt}
\noindent \textbf{Definition 2 of (Local) Special K\"{a}hler Geometry.}

\noindent A special K\"{a}hler manifold is an $n$-dimensional K\"{a}hler--Hodge manifold,
that is the base manifold of a  $\Sp(2(n+1))\times \U(1)$ bundle. There
should exist a holomorphic section $v(z)$ such that the K\"{a}hler potential
can be written as (\ref{Kahlerv}) and it should satisfy the condition
\begin{equation}
\langle{\nabla_\alpha v},{\nabla_\beta v}\rangle =0\,.
\label{DavDbv0}
\end{equation}

Note that the latter condition guarantees the symmetry of ${\mathcal
N}_{IJ}$. This condition did not appear in~\cite{Strominger:1990pd},
where the author had in mind Calabi--Yau manifolds. As we will see below,
in those applications, this condition is automatically fulfilled. For
$n>1$ the condition can be replaced by the equivalent condition
\begin{equation}
\langle {\nabla_\alpha v},{v}\rangle  =0\,. \label{Davv0}
\end{equation}
For $n=1$, the condition (\ref{DavDbv0}) is empty, while (\ref{Davv0}) is
not. In~\cite{Claus:1999np} it has been shown that models with $n=1$ not
satisfying (\ref{Davv0}) can be formulated.

\vspace{10pt}
\noindent \textbf{Equivalence of the Two Definitions.}

\noindent It is thus legitimate to ask about the equivalence of the two
definitions. Indeed, we saw that in some cases definition 2 is satisfied,
but one cannot obtain a prepotential $F$. However, that example, as
others in~\cite{Ceresole:1995jg}, was obtained by performing a
symplectic transformation from a formulation where the prepotential does
exist. In~\cite{Craps:1997gp} it was shown that this is true in general.
If definition 2 is applicable, then there exists a symplectic
transformation to a basis such that $F(Z)$ exists. However, in the way
physical problems are handled, the existence of formulations without
prepotentials can play an important role. For example, these formulations
were used to prove that one can break ${\cal N}=2$ supersymmetry  partially to
${\cal N}=1$ \cite{Ferrara:1995gu} and not necessarily to ${\cal N}=0$, as it was
thought before. This is an  extremely important property for
phenomenological applications. Note that by introducing magnetic vectors the partial breaking \index{partial breaking} can be obtained in the context of the theory with a prepotential, since this is just a dual formulation of the same theory. This is understood well using the embedding tensor formalism \cite{Cordaro:1998tx,Nicolai:2001sv,deWit:2005ub,Samtleben:2008pe,Trigiante:2016mnt}, as clarified in \cite{Louis:2010ui}. Recent investigations of the possibilities of partial breaking have been performed in \cite{Antoniadis:2018blk,Abe:2019svc}.\footnote{Breaking by nonlinear terms has been investigated in \cite{Farakos:2018aml}.}

On page \pageref{equivCondF}, we gave  equivalent conditions on the symplectic basis for the existence of a prepotential.
The first and second conditions are clearly equivalent. Indeed, the invertibility of (\ref{matXDX}) is equivalent with the invertibility of $\begin{pmatrix}X^I&\partial _\alpha  X^I\end{pmatrix}$, i.e. with ordinary derivatives, since the difference is proportional to the first column. Then one can define the coordinates as in  (\ref{specialcoord}), which clearly satisfy the invertibility condition.
In these coordinates it can be shown that a prepotential exists.
One first notices that $(X^I/X^0)$ is independent of $\bar z$. Then also $F_I/X^0$ is by the homogeneity only function of $z$ and one can define holomorphic functions $F_I(X)$, by replacing those $z^\alpha $ by their value in (\ref{specialcoord}):
\begin{equation}
F_I(X)\equiv X^0 \frac{F_I}{X^0} \left( z\left(\frac{X^\alpha }{X^0} \right)
\right) \,.
\end{equation}
The constraints (\ref{DavDbv0})  then imply
\begin{equation}
\begin{pmatrix}X^I\cr \partial_\alpha X^I\end{pmatrix}\partial_{[I}F_{J]} \ \begin{pmatrix}
X^J&\partial_\alpha X^J\end{pmatrix}=0\,,
\label{existenceFproof}
\end{equation}
and since the first and last factor is invertible, it follows that the middle factor should vanish, which is the integrability condition that in any patch $F_J=\frac{\partial}{\partial
X^J}F(X)$ for some $F(X)$.

\subsection{Symplectic Equations and the Curvature Tensor}
Let us first summarize the symplectic inner products that we found. They
can be simply written in terms of the $2(n+1)\times 2(n+1)$ matrix
\begin{align}
  {\cal V}= &\begin{pmatrix}\bar V & U_\alpha &   V & \kappa ^{-2}   \bar U^\alpha\end{pmatrix}\,,\nonumber\\
   &U_\alpha \equiv \nabla_\alpha V=\left[\partial _\alpha +\ft12\kappa ^2 \left(\partial _\alpha {\cal K}\right)\right] V
 \,,  \nonumber\\
  &\bar U_{\bar \alpha} \equiv \nabla_{\bar \alpha}\bar  V
  = \left[\partial _{\bar \alpha }+\ft12\kappa ^2 \left(\partial _{\bar \alpha } {\cal K}\right)\right]\bar V\,,\qquad \bar U^\alpha =g^{\alpha \bar \beta }\bar U_{\bar\beta}\,,
 \label{defcalV}
\end{align}
as
\begin{equation}
  {\cal V}^T \Omega {\cal V}=-\rmi \kappa ^{-2}\Omega\,,
 \label{allinnerV}
\end{equation}
where $\Omega $ is the standard antisymmetric matrix in (\ref{Scond}).

Thus, the matrix ${\cal V}$ is also invertible. Covariant derivatives on
these equations lead to new ones, like
\begin{eqnarray}
  &&\langle V,\nabla_\alpha U_\beta \rangle = \langle \bar V,\nabla_\alpha
  U_\beta\rangle= \langle \bar U_{\bar \gamma },\nabla_\alpha
  U_\beta\rangle=0,\nonumber\\
  &&\langle \nabla_\alpha  U_\beta ,U_\gamma \rangle +\langle U_\beta ,\nabla_\alpha U_\gamma
  \rangle=0.
 \label{Dsymplinner}
\end{eqnarray}
Note that $\nabla_\alpha U_\beta $ contains also Levi-Civita
connection.
\begin{equation}
 \nabla_\alpha U_\beta = \left[\partial _\alpha +\ft12\kappa ^2 \left(\partial _\alpha {\cal K}\right)\right]\left[\partial _\beta +\ft12\kappa ^2 \left(\partial _\beta {\cal K}\right)\right]V -\Gamma _{\alpha \beta }^\gamma U_\gamma \,.
 \label{nabU}
\end{equation}

Due to these equations (\ref{Dsymplinner}) and the invertibility of
(\ref{defcalV}),  $\nabla_\alpha U_\beta $ must be proportional to a
$\bar U^\gamma$, defining a third rank tensor $C_{\alpha \beta \gamma }$:
\begin{equation}
  \nabla_\alpha U_\beta=C_{\alpha \beta \gamma }\bar U^\gamma , \qquad
   C_{\alpha \beta \gamma }\equiv -\rmi\langle \nabla_\alpha U_\beta ,U_\gamma
  \rangle.
 \label{defC}
\end{equation}
Since (\ref{nabU}) is already symmetric in
$(\alpha \beta)$,  (\ref{Dsymplinner}) implies that $C_{\alpha \beta \gamma }$ is completely symmetric.
\bexer
 Check that if there is a prepotential, then one can write
\begin{equation}
  C_{\alpha \beta \gamma }=\rmi F_{IJK}\nabla_\alpha X^I\nabla_\beta
  X^J\nabla_\gamma X^K,
 \label{CFIJK}
\end{equation}
where the covariant derivatives may be replaced by ordinary derivatives
due to (\ref{homegeneqs}).
 \eexer

The curvature of the projective manifold can be obtained from the commutator of
covariant derivatives.
The connections contain
Levi-Civita and K\"{a}hler connections, see (\ref{nabU}), so the commutator will lead to a sum
of the curvature of the K\"{a}hler manifold as function of the Levi-Civita
connection of $g_{\alpha \bar \beta }$ and the curvature of the K\"{a}hler
connection. The latter is the original $\U(1)$ from the superconformal
algebra (on K\"{a}hler-invariant quantities) when it is pulled back to the spacetime, see (\ref{calAmualpha}).
\begin{equation}
  \left[ \nabla_\alpha ,\nabla_{\bar \beta }\right] V =-\rmi R^{\rm K}_{\alpha \bar \beta   } V\,, \qquad R^{\rm K}_{\alpha \bar \beta   }=\partial _\alpha {\cal A}_{\bar \beta }-\partial _{\bar \beta } {\cal A}_\alpha = -\rmi\kappa ^2 g_{\alpha \bar \beta}\,.
\end{equation}
Hence we have
\begin{equation}
\left[ \nabla_\alpha ,\nabla_{\bar \beta }\right] V = -\kappa ^2 g_{\alpha \bar \beta}V,\qquad \rightarrow \qquad  \nabla_{\bar \beta   }\nabla _\alpha V =\nabla_{\bar \beta   } U_\alpha =\kappa ^2 g_{\alpha \bar \beta}V.
 \label{commDDKaRV}
\end{equation}
On $U_\gamma  $ we get
\begin{equation}
  \left[ \nabla_\alpha ,\,\nabla_{\bar \beta }\right] U_\gamma =-\rmi R^{\rm K}_{\alpha \bar \beta   } U_\gamma-R_{\alpha \bar \beta}{}^\delta {}_\gamma U_\delta\,.
 \label{commDDKaRU}
\end{equation}

To make the calculation of the left-hand side, one lemma that we still
need is that $C_{\alpha \beta \gamma }$ is covariantly holomorphic. From the definition (\ref{defC})
\begin{equation}
\rmi  \nabla_{\bar\delta} C_{\alpha \beta \gamma }=\langle \nabla_{\bar\delta} \nabla_\alpha U_\beta ,\,U_\gamma \rangle +\langle
\nabla_\alpha U_\beta ,\, \nabla_{\bar\delta} U_\gamma\rangle =0.
 \label{covholC}
\end{equation}
Both terms vanish separately. For the first term,  use (\ref{commDDKaRU})
and (\ref{commDDKaRV}) and $\langle V ,U_\alpha  \rangle =0$ and $\langle U_\alpha ,U_\beta \rangle =0$ (both part of  (\ref{allinnerV})).
The second term is by  (\ref{commDDKaRV}) and (\ref{Dsymplinner}) immediately zero.

Now we can calculate the left-hand side of (\ref{commDDKaRU}):
\begin{align}
 \nabla_\alpha \nabla_{\bar\beta }  U_\gamma  =& \kappa ^2 g_{\gamma \bar\beta
  }  U_\alpha\,,\nonumber\\
 \nabla_{\bar\beta} \nabla_\alpha U_\gamma  =& C_{\alpha \gamma  \epsilon }g^{\epsilon \bar \epsilon }\bar
  C_{\bar \epsilon \bar \beta  \bar \delta }g^{\bar \delta \delta }U_\delta\,.
\label{calculationCurvature}
\end{align}
Subtracting the two  and insert the result in (\ref{commDDKaRU}) gives
\begin{equation}
\left(\kappa ^2 g _{\gamma \bar\beta } \delta_\alpha ^\delta
  - C_{\alpha\gamma \epsilon }g^{\epsilon \bar \epsilon }\bar
  C_{\bar\epsilon \bar \beta\bar \delta }g^{\bar \delta \delta }\right)U_\delta =  \left(-\kappa ^2g _{\alpha \bar\beta }  \delta
  _\gamma ^\delta-R_{\alpha\bar \beta  }{}^\delta{}_\gamma\right) U_\delta\,.
 \label{CommRelU}
\end{equation}
We can drop the $U_\delta $, e.g. by taking a symplectic product of this
relation with $\bar U^\phi$. This establishes the form of the curvature
tensor~\cite{Cremmer:1985hj}
\begin{equation}
  R_{\alpha \bar \beta\gamma \bar \delta }=\kappa ^2\left(g _{\alpha \bar\beta}g_{\gamma \bar \delta }+g _{\alpha \bar\delta }g_{\gamma\bar\beta } \right)
  - C_{\alpha\gamma \epsilon }g^{\epsilon \bar \epsilon }\bar   C_{\bar\epsilon \bar \beta \bar \delta }\,.
 \label{RspecKahl}
\end{equation}
Hence, this has a contribution from the gauged K\"{a}hler symmetry and one from the K\"{a}hler curvature.

\bigskip

Having all this machinery, we want to derive a few more relations that
are often used in special K\"{a}hler geometry.
Define the $2(n+1)\times (n+1)$ matrix ${\cal U}$ as the left part of
${\cal V}$, (\ref{defcalV}),
\begin{equation}
  {\cal U}= \begin{pmatrix}\bar V &U_\alpha\end{pmatrix} =\begin{pmatrix}\bar X^I&\nabla_\alpha X^I\cr \bar
  F_I&\nabla_\alpha F_I\end{pmatrix}=\begin{pmatrix}x^I\cr f_I\end{pmatrix},
 \label{defcalUxf}
\end{equation}
where the last equation defines $(n+1)\times (n+1)$ matrices. The matrix
$x^I$ (columns as in (\ref{defcalUxf}) and rows defined by $I$) is invertible. With  these definitions, the matrix ${\cal N}$ in (\ref{Nsugra})
is written as
\begin{equation}
  {\cal N}_{IJ}= \bar f_I(\bar x)^{-1}_J\,,\qquad \mbox{or}\qquad  {\cal U}=\begin{pmatrix}1\cr \bar {\cal N}\end{pmatrix} x.
 \label{calU1Nx}
\end{equation}
The $(n+1)\times (n+1)$ matrix $G$, introduced in (\ref{matrixeqnspecK}), can then also be written as
\begin{equation}
  G=\begin{pmatrix}\kappa ^{-2}&0\cr 0&g_{\alpha \bar \beta }\end{pmatrix}= \rmi\, {\cal
  U}^T\Omega \bar {\cal U}= -2 \begin{pmatrix}\bar X^I\cr\nabla_\alpha X^I\end{pmatrix}\Im {\cal N}_{IJ}
  \begin{pmatrix}X^J&\nabla_{\bar \beta }\bar X^J\end{pmatrix}\,.
 \label{MatrixPositiveProduct}
\end{equation}
The above relation implies that $\Im {\cal N}_{IJ}$ is negative definite if the metric $g_{\alpha\bar \beta}$ is positive definite.
The following consequences are often used:\footnote{The first is the inverse of (\ref{MatrixPositiveProduct}), while for the second, one proves first from the definition (\ref{Ndef}) that $N^{-1|IJ} + \kappa ^2(X^I\bar X^J+\bar X^IX^J)$ is the inverse of $-2\Im{\cal N}_{IJ}$.}
\begin{align}
  -\ft12(\Im{\cal N})^{-1|IJ}=&\nabla_\alpha X^I g^{\alpha \bar \beta}\nabla_{\bar \beta }\bar
  X^J+\kappa ^2 \bar X^IX^J\,,\nonumber\\
 N^{-1|IJ}=& -\ft12(\Im{{\cal N}})^{-1|IJ}-\kappa ^2\left( X^I\bar X^J+\bar X^IX^J\right)\nonumber\\
 =&\nabla_\alpha X^I g^{\alpha \bar \beta}\nabla_{\bar \beta }\bar
    X^J -\kappa ^2  X^I \bar X^J\,.\label{InverseImN}
\end{align}

Another way to write (\ref{MatrixPositiveProduct}) is
\begin{equation}
  \bar {\cal U} G^{-1}{\cal U}^T =\kappa ^2V\bar V^T+\bar U_{\bar \beta }g^{\bar \beta \alpha }U_\alpha ^T= -\frac{1}{2}\begin{pmatrix}1\cr{\cal
  N}\end{pmatrix}(\Im{\cal N})^{-1}\begin{pmatrix}1&\bar {\cal
  N}\end{pmatrix}= -\frac12{\cal M}+\frac12\rmi\Omega \,,
  \label{UGU}
\end{equation}
in terms of the real matrix ${\cal M}$ introduced in \cite{Ceresole:1995ca}, that is often used for describing black hole solutions:
\begin{equation}
{\cal M}\equiv \begin{pmatrix}\Im{\cal N}^{-1}&\Im{\cal N}^{-1}\Re{\cal
N}\cr\Re{\cal N}\Im{\cal N}^{-1} &\Im{\cal N}+\Re{\cal N}\Im{\cal
N}^{-1}\Re{\cal
  N}\end{pmatrix}\,.
 \label{defcalM}
\end{equation}

  \bexer
 Check from the definition (\ref{Ndef}) and the gauge fixing  (\ref{Dgaugegeneral}) the following useful relations:
\begin{align}
  \Im{\cal N}_{IJ}\bar X^J=& -\frac{1}{2\kappa ^2}\frac{N_{IJ}X^J}{N_{LM}\ X^L   X^M}\,,\nonumber\\
X^I(\Im {\cal N}_{IJ})X^J =& \frac{1}{2\kappa  ^4\bar X^IN_{IJ}\bar X^J}\,,\nonumber\\
  C_{\alpha \beta \gamma }g^{\gamma \bar  \delta }{\nabla }_{\bar \delta }\bar X^I(\Im {\cal N}_{IJ})X^J =&
   -\frac{\rmi\kappa ^{-2}}{2\bar X^IN_{IJ}\bar X^J}
  F_{KLM}\nabla _\alpha X^K\nabla _\beta X^L \bar X^M\,.
 \label{ImcNX}
\end{align}
Check also that  (\ref{MatrixPositiveProduct}) is consistent with the matrix equation
\begin{align}
 \begin{pmatrix}X^I\cr\nabla_\alpha X^I\end{pmatrix}N_{IJ}
  \begin{pmatrix}\bar X^J&\nabla_{\bar \beta }\bar X^J\end{pmatrix} =
\begin{pmatrix}-\kappa ^{-2}&0\cr 0&g_{\alpha \bar \beta }\end{pmatrix}\,.
\label{Nmatrixeqn}
\end{align}
The latter is similar to the one in ${\cal N}=1$ supergravity, and illustrates that $N_{IJ}$ has to be of indefinite signature.
 \eexer

Finally, we discuss the relation between the curvature in the projective manifold (\ref{RspecKahl}) and the geometry in the embedding manifold.
The embedding manifold of the $X^I$ in the formulation with a prepotential and metric (\ref{GIJNIJ}) has affine connections (\ref{special_levicivita}) and therefore (see e.g. \cite[(13.22)]{Freedman:2012zz})
\begin{equation}
  R_{I\bar JK\bar L}= G_{L\bar L}\bar \partial _{\bar J}\Gamma ^L_{IK}=  -F_{IKM}N^{-1|M N}\bar F_{\bar N\bar J\bar L}\,.
 \label{curvembedding}
\end{equation}
 With (\ref{CFIJK}) and the relation (\ref{InverseImN}) for $N^{-1|MN}$ we can relate  (\ref{RspecKahl}) to this curvature:
\begin{equation}
  R_{\alpha \bar \beta \gamma\bar \delta }=\kappa ^2\left(g _{\alpha \bar\beta}g_{\gamma \bar \delta }+g _{\alpha \bar\delta }g_{\gamma \bar\beta} \right)
 + \nabla _\alpha X^I\nabla _{\bar\beta }\bar  X^J \nabla _\gamma X^K \nabla _{\bar \delta } \bar X^L R_{I\bar JK\bar L}\,.
 \label{curvsplit}
\end{equation}
This relation is also true in K\"{a}hler geometry in ${\cal N}=1$ supergravity and can elegantly be shown when including $\Gamma ^I_{JK}$ connections  (\ref{relNphi}) in a covariant derivative $\widehat{\nabla }$. First note that
\begin{align}
\widehat{\nabla} _\alpha X^I \equiv  & (\partial _\alpha + \ft12\kappa ^{-2}(\partial _\alpha {\cal K}))X^I+ \Gamma ^I{}_{JK}X^K(\partial _\alpha X^J)= \nabla _\alpha X^I \,,
\label{nablaXGamma}
\end{align}
since the $\Gamma $ term is vanishing due to  (\ref{relNphi}). However, we then have
\begin{align}
\widehat{\nabla }_\alpha \nabla _\beta X^I\equiv     & (\partial _\alpha + \ft12\kappa ^{-2}(\partial _\alpha {\cal K}))\nabla _\beta X^I -\Gamma _{\alpha \beta }^\gamma \nabla _\gamma X^I +\Gamma ^I{}_{JK}\nabla _\beta X^K(\partial _\alpha X^J)=0\,.
\label{nabnabXdef}
\end{align}
Indeed from (\ref{special_levicivita}) with  (\ref{InverseImN}) and the homogeneity relation $X^IF_{IJK}=0$ we find
\begin{equation}
\Gamma ^I{}_{JK}\nabla _\beta X^K\,\nabla _\alpha X^J=- \nabla _{\bar \gamma  }X^I g^{\gamma  \bar \gamma }C_{\alpha \beta \gamma }=-\rmi \nabla _\alpha \nabla _\beta  X^I\,,
 \label{GammaItoC}
\end{equation}
using (\ref{defC}). This vanishing of  (\ref{nabnabXdef}) can also be shown directly  \cite{Freedman:2017obq} from applying a $\widehat{\nabla }_\alpha $ derivative to the second line of (\ref{Nmatrixeqn}) in the formulation where the last factor in that equation is invertible (existence of prepotential). Since $N_{IJ}=G_{I\bar J}$ it is inert under $\widehat{\nabla }_\alpha $.

The curvature relation  (\ref{curvsplit}) can then be obtained by acting with $\left[\widehat{\nabla }_\alpha ,\widehat{\nabla }_{\bar \beta }\right]$ on $\nabla _\gamma X^I$ using the vanishing of (\ref{nabnabXdef}). We summarize below the relevant equations for $\widehat{\nabla }$ on $X^I$:
\begin{align}
  &\widehat{\nabla }_\alpha X^I = \nabla_\alpha  X^I\,,\qquad  \widehat{\nabla }_{\bar \alpha} X^I = \nabla_{\bar \alpha}  X^I=0\,,  \nonumber\\
  &\widehat{\nabla }_\alpha \nabla _\beta X^I =0 \,,\qquad \widehat{\nabla }_{\bar\beta}\nabla _\alpha X^I= {\nabla }_{\bar \beta}\nabla _\alpha X^I= \kappa ^2g_{\alpha \bar \beta }X^I\,.
\label{widehatnabX}
\end{align}

However, we will only use $\widehat{\nabla }$ on objects where it is clear whether the $I$ index is holomorphic or anti-holomorphic. The mixing of these two is in fact due to the constraints in chiral multiplets, in  (\ref{constraintschiral}), which, applied to the gauge multiplet, connect $Y_{ij}^I$ with its complex conjugate, and thus does not preserve a distinction of indices $I$ for fields of the chiral multiplets and $\bar I$ for those of the antichiral multiplet. The same holds then also for the gauge field strengths.

\section{Isometries and Symplectic Geometry}
 \label{ss:gaugeisom}

In considering the symmetries of special K\"{a}hler manifolds, there are a priori three different concepts, which we will relate:
\begin{itemize}
  \item Isometries of the K\"{a}hler manifold,
  \item Symplectic transformations,
  \item Gauge group of the vector multiplets.
\end{itemize}
We will show that the group of transformations gauged by the vector multiplet belongs to the isometry group of the K\"{a}hler manifold and can be embedded in the group of symplectic transformations. On the other hand, not all isometries of the K\"{a}hler manifold are gauged. The gauged isometries can be selected from the general set of isometries by an embedding tensor formalism.\index{embedding tensor}   We do not treat this here, and refer for the embedding tensor approach to \cite{Cordaro:1998tx,Nicolai:2001sv,deWit:2005ub} and to \cite{Samtleben:2008pe,Trigiante:2016mnt} for reviews.

\subsection{Isometries of a K\"{a}hler Metric}
\label{ss:isomKahler}

Isometries of a K\"{a}hler metric are defined as symmetries that preserve both the Hermitian metric and the covariantly constant complex structure. In the present section we will explain how isometries of the embedding manifold, with coordinates $\{X^I\}$, descend to isometries of the projective manifold. These steps apply to both ${\cal N}=1$ and ${\cal N}=2$ supergravity.  For the convenience of the reader, we will first review the main results on isometries of a K\"{a}hler metric (more details can be found in \cite[Sect. 13.4.1]{Freedman:2012zz}).

\vspace{10pt}
We start with a generic K\"{a}hler metric $g_{\alpha\bar{\beta}}=\partial_\alpha\partial_{\bar{\beta}}{\cal K}(z,\bar{z})$ parameterized by scalars $z^\alpha$. Let us consider general (not gauged) isometries labelled by an index $\Lambda $. Such isometries act on the scalars, defining a Killing vector $k_\Lambda{}^\alpha $\footnote{The `embedding tensor', mentioned above would be a tensor relating the index $\Lambda $ to the index for gauge vectors, electric or magnetic. We will restrict here for gauging to identifying the index $\Lambda $, or part of its range, to $I$ index of the gauge vectors $W_\mu ^I$.}:
\begin{equation}
  \delta(\theta ) z^\alpha =  \theta^\Lambda k _\Lambda{}^\alpha (z)\,,
 \label{delGzalpha}
\end{equation}
and $ k_\Lambda{}^\alpha$ must be holomorphic in $z$ in order to preserve the complex structure. Killing vectors are generated by a \emph{moment map}\index{moment map}, $\mathcal{P}_\Lambda (z,\bar{z})$, which is a real function defined as
\begin{equation}\label{killMomentMap}
k_\Lambda{}^\alpha (z)=-\rmi g^{\alpha\bar{\beta}}\partial_{\bar{\beta}}\mathcal{P}_\Lambda (z,\bar{z})\,,
\end{equation}
and satisfying
\begin{equation}
\nabla_\alpha\partial_{\beta}\mathcal{P}_\Lambda (z,\bar{z})=0\,.
\end{equation}

The isometries \eqref{delGzalpha} are not required to be symmetries of the K\"{a}hler potential. The latter can transform with a
K\"{a}hler transformation depending on a holomorphic
function $r_\Lambda(z)$:
\begin{equation}\label{deltaKrrbar}
\delta(\theta)\mathcal{K}(z,\bar {z})=\theta^\Lambda [r_\Lambda(z)+\bar{r}_\Lambda (\bar{z})]\,,
\end{equation}
since the new potential then leads to the same K\"{a}hler metric. Due to this fact, one can find a general solution of \eqref{killMomentMap} as
\begin{equation}\label{solkillMomentMap}
\mathcal{P}_\Lambda (z,\bar{z})=\rmi [k_\Lambda{}^\alpha (z) \partial_\alpha \mathcal{K}(z,\bar{z})-r_\Lambda (z)]
= -\rmi [k_\Lambda{}^{\bar \alpha} (\bar z) \partial_{\bar \alpha} \mathcal{K}(z,\bar{z})-\bar r_\Lambda (z)]\,.
\end{equation}

Furthermore, when the isometries are non-abelian, the Killing vectors generate a Lie algebra
\begin{equation}
 k_\Lambda{}^\beta\partial _\beta k_\Sigma{}^\alpha -
  k_\Sigma{}^\beta\partial _\beta k_\Lambda{}^\alpha=
  f_{\Lambda \Sigma }{}^\Gamma k_\Gamma {}^\alpha\,.
 \label{Galgebraz}
\end{equation}
One can restrict the moment maps by a so-called `equivariance relation' \index{equivariance relation} \cite{Andrianopoli:1997cm}
\begin{equation}
k_\Lambda{}^\alpha g_{\alpha\bar{\beta}}k_\Sigma{}^{\bar{\beta}}-k_\Sigma{}^\alpha g_{\alpha\bar{\beta}}k_\Lambda{}^{\bar{\beta}}=\rmi f_{\Lambda\Sigma}{}^\Gamma \mathcal{P}_\Gamma\,.
\label{equivarianceP}
\end{equation}
This relation is satisfied by the moment maps that appear in a supersymmetric Lagrangian.
\vspace{5pt}

The properties discussed above are general requirements that isometries for a general K\"{a}hler manifold must fulfill. In particular, they must apply for the isometries of the embedding manifold, which are further constrained by the conditions that the dilatational and $R$-symmetry $\U(1)$ structure have to be preserved.  We can start with the general results written above using the following substitutions:
\begin{equation}
\begin{array}{lll}
\displaystyle{z^\alpha}&\displaystyle{\longrightarrow}&\displaystyle{X^I}\,,\\
\displaystyle{\mathcal{K}}&\displaystyle{\longrightarrow}&\displaystyle{N}\,,\\
\displaystyle{g_{\alpha\bar\beta}}&\displaystyle{\longrightarrow}&\displaystyle{N_{IJ}}\,,\\
\displaystyle{r_\Lambda,\bar{r}_\Lambda}&\displaystyle{\longrightarrow}&\displaystyle{0}\,.
\end{array}
\label{substitutionsembedding}
\end{equation}
The first three rules follow directly from what was discussed in Sect. \ref{proj_kahler_intro}. The last rule in the previous equation follows from the fact that a transformation of the form (\ref{deltaKrrbar}) would now imply
\begin{equation}
  \delta(\theta)N(X,\bar {X})=\theta^\Lambda [r_\Lambda(X)+\bar{r}_\Lambda (\bar{X})]\,,
 \label{zerorargument}
\end{equation}
which would be inconsistent with the (Weyl,chiral) weight (2,0) of the left-hand side. All together, the isometries of the embedding K\"{a}hler manifold act on the scalars $X^I$ as follows:
\begin{equation} \label{delXI}
  {\delta}(\theta) X^I =  \theta^\Lambda k _\Lambda{}^I (X)\,.
\end{equation}
The moment map (\ref{solkillMomentMap}) then becomes
\begin{equation}\label{solkillMomentMap2}
\mathcal{P}_\Lambda (X,\bar{X})=\rmi k_\Lambda{}^I N_{IJ}\bar{X}^J.
\end{equation}

The goal is now to interpret these isometries of the embedding manifold as isometries in the projective manifold. We go to the projective manifold by splitting the $X^I$:
\begin{equation}\label{splittsplit}
X^I=\{y,z^\alpha\},
\end{equation}
In these variables, only $y$ carries the dilatational and $\U(1)$ weight. The requirement to preserve the dilatational and $\U(1)$ structure thus implies that $y$ should transform linear in $y$ and the transformation of $z^\alpha $ should not contain $y$. We write this as\footnote{We keep here the notation $a=\kappa ^{-2}$ since this part is also valid for ${\cal N}=1$ with $a=3\kappa ^{-2}$.}
\begin{equation}\label{deltaGY}
\delta(\theta)y=a^{-1}\,y\, r_\Lambda(z)\,\theta^\Lambda\,.
\end{equation}
Note that we have introduced here a new holomorphic function $r_\Lambda(z)$. This is not to be confused with the function that is set to zero in (\ref{substitutionsembedding}). It will turn out that the latter $r_\Lambda(z)$ is the function that describes the non-invariance of the K\"{a}hler potential in the projective manifold. Imposing the commutation relations \eqref{Galgebraz} for (\ref{deltaGY}) leads to the condition
\begin{equation}
  k_\Lambda {}^\alpha \partial _\alpha r_\Sigma -
  k_\Sigma{} ^\alpha \partial _\alpha r_\Lambda=
 f_{\Lambda \Sigma }{}^\Gamma r_\Gamma\,.
 \label{GalgebraY}
\end{equation}

The gauge fixing, i.e. restriction to the projective manifold, is performed by selecting the section (\ref{Ychoice}).
Since $y$ transforms both under (\ref{divideDTX}) and (\ref{deltaGY}), there is a decomposition law for the invariance of the condition $y=\bar y$. This determines a dependence of $\lambda _T$ on the parameters of the isometry $\theta^\Lambda $. We indicate the remaining transformation  as $\delta ^\poinc(\theta )$:
\begin{align}
\delta ^\poinc(\theta )=&{\delta}(\theta)+\delta_T (\lambda_T(\theta))\,,\nonumber\\
  \lambda_T(\theta)=&\frac{1}{2a}\rmi
  \,\theta ^\Lambda\left(r_\Lambda (z)-\bar r_\Lambda (\bar z)\right)=\frac{1}{a} \theta ^\Lambda\left(-{\cal P}_\Lambda +\ft12\rmi (k_\Lambda {}^\alpha \partial _\alpha {\cal K}-k_\Lambda {}^{\bar \alpha }\partial _{\bar \alpha } {\cal K})\right)\,.
 \label{deltahatdelta}
\end{align}
This implies for $y$
\begin{equation}
  \delta ^\poinc(\theta )y= \frac{1}{2a}y\,\theta ^\Lambda \left(r_\Lambda +\bar r_\Lambda \right)=\frac{1}{2a}y\,\theta ^\Lambda \left(k_\Lambda ^\alpha \partial _\alpha {\cal K}+k_\Lambda ^{\bar \alpha }\partial _{\bar \alpha } {\cal K}\right) \,.
 \label{hatdeltay}
\end{equation}
Note that, together with (\ref{Ychoice}), \eqref{hatdeltay} implies that ${\cal K}$ indeed transforms as in (\ref{deltaKrrbar}), by an amount precisely controlled by $r_\Lambda(z)$.\footnote{Note that this is consistent with (\ref{hatdely0}).}
Equation \eqref{deltahatdelta} shows that, for any field that carries a non-trivial $\U(1)$ weight in the embedding manifold, a $T$ transformation must be included in its effective symmetry transformation in the projective manifold, $\delta ^\poinc(\theta )$. The new terms from $\lambda_T$ have no effect on the scalars $z^\alpha$ since they are invariant under $T$ transformations, while fermions do transform.  In other words the scalars $z^\alpha$ transform as
\begin{equation} \label{delzalpha}
  \delta ^\poinc(\theta )z^\alpha =\delta(\theta)z^\alpha =  \theta^\Lambda k _\Lambda{}^\alpha (z)\,,
\end{equation}
and the Killing vectors in the projective manifold are related to those in the embedding manifold via
\begin{equation}
k_\Lambda{}^I=y(k_\Lambda{}^\alpha \nabla_\alpha Z^I+\rmi a^{-1} \mathcal{P}_\Lambda Z^I)\,,
\label{Killingembeddingtoprojective}
\end{equation}
using the K\"{a}hler covariant derivative \eqref{nablaalZ}. The moment map ${\cal P}_\Lambda $ is the same in the embedded and projected manifold. In the former it is written as  (\ref{solkillMomentMap2}). In the projected manifold an imaginary constant in ${\cal P}_\Lambda$ is undetermined from the transformations of the scalars. We can still define it as in (\ref{solkillMomentMap}), satisfying (\ref{killMomentMap}). The undetermined constant imaginary parts of $r_\Lambda $ or real parts of ${\cal P}_\Lambda $ are the so-called Fayet--Iliopoulos terms.\index{Fayet--Iliopoulos term} For a non-abelian theory these arbitrary constants are eliminated by the equivariance condition (\ref{equivarianceP}). The curious reader may have noticed already the similarity between (\ref{solkillMomentMap}), i.e.
\begin{equation}
  k_\Lambda{} ^\alpha \partial _\alpha {\cal K}= r_\Lambda -\rmi{\cal P}_\Lambda \,,\qquad k_\Lambda {}^{\bar \alpha }\partial _{\bar \alpha } {\cal K}= \bar r_\Lambda +\rmi{\cal P}_\Lambda \,,
 \label{kdK}
\end{equation}
and \eqref{kIalphakalpha} for the projected quaternionic geometry, keeping in mind the identification of $\partial _\alpha {\cal K}$ as $\omega_\alpha $ in (\ref{omaldef}). Note that the derivative of the first of (\ref{kdK}) w.r.t. $\bar z^{\bar \alpha }$  brings us back to the definition of the moment map in (\ref{killMomentMap}).

\subsection{Isometries in Symplectic Formulation }
\label{ss:isomsympl}
In this section we will show how the general isometries discussed in Sect. \ref{ss:isomKahler} can be embedded in the symplectic group. It will be useful to adopt the symplectic formulation (see Sect.~\ref{ss:symplformul}) and work with covariant transformations (\ref{hatdeltaf}) under the isometries:
\begin{align}
  \widehat{\delta}(\theta )v=& \theta  ^\Lambda  k_\Lambda{} ^\alpha \nabla_\alpha v\,,\nonumber\\
   \widehat{\delta}(\theta )V=& \theta  ^\Lambda  k_\Lambda{} ^\alpha \nabla_\alpha V= y \widehat{\delta}(\theta )v\,.
 \label{covtransftheta}
\end{align}
Covariant transformations $\widehat{\delta}$ are related to the Poincar\'{e} transformations $\delta ^{\poinc}$ (namely those induced by the dependence on $z^\alpha $ and $\bar z^{\bar \alpha }$) by  (\ref{deltaf})
\begin{align}
   \widehat{\delta}(\theta )v=&\delta ^{\poinc}(\theta )v + \kappa ^2\theta ^\Lambda (r_\Lambda - \rmi{\cal P}_\Lambda) v\,,\nonumber\\
     \widehat{\delta}(\theta )V=&\delta ^{\poinc}(\theta )V + \ft12\kappa ^2\theta ^\Lambda (r_\Lambda -\bar r_\Lambda)V -\rmi\kappa ^2\theta ^\Lambda{\cal P}_\Lambda V\,,
 \label{hatdeldelpoincv}
\end{align}
where we used the weights in the first and one but last row in (\ref{weightscw}) and (\ref{kdK}). Furthermore, the Poincar\'{e} transformations are related to the transformations in the conformal setting $\delta$ by  (\ref{deltahatdelta}), and thus
\begin{align}
   \widehat{\delta}(\theta )v=&\delta(\theta )v + \kappa ^2\theta ^\Lambda (r_\Lambda - \rmi{\cal P}_\Lambda) v\,,\nonumber\\
    \widehat{\delta}(\theta )V=&\delta (\theta )V -\rmi \kappa ^2\theta ^\Lambda{\cal P}_\Lambda V\,.
 \label{hatdeldelconfv}
\end{align}
The correspondence between the two expressions above agrees with $V=yv$ and $\widehat{\delta}V=  y \widehat{\delta}v$ using  (\ref{deltaGY}). We can re-express the second of \eqref{hatdeldelconfv} using (\ref{covtransftheta}) as
\begin{equation}
  \delta (\theta )V= \theta  ^\Lambda\left(k_\Lambda{} ^\alpha \nabla_\alpha V+ \rmi \kappa ^2{\cal P}_\Lambda V\right)\,.
 \label{delV}
\end{equation}
Therefore, upon contracting with $\bar V$ and using \eqref{Dalphavbarv}, the last equation implies a symplectic expression for the moment map associated to the isometries in the embedding manifold
\begin{equation}
\theta ^\Lambda  {\cal P}_\Lambda = -\langle \delta (\theta ) V,\bar V\rangle\,.
\label{PLambdainsympl}
\end{equation}
From the expression above it is clear that $\delta (\theta ) V$ must be part of the symplectic group. Indeed, being $\delta (\theta )V$ a transformation in the conformal setting, it must be holomorphic in $X$ and consistent with dilatational symmetry. Moreover, since \eqref{PLambdainsympl} is symplectic invariant, the only possibility is\footnote{We use here the opposite sign convention for $T_\Lambda $ as in \cite[(20.147)]{Freedman:2012zz}.}
\begin{equation}
  {\delta} (\theta) V= -\theta ^\Lambda  T_\Lambda V\,, \qquad T_\Lambda\in \Sp(2(n+1),\mathbb{R})\,,
 \label{deltaGvgeneral}
\end{equation}
where $\unity + T_\Lambda$ is a symplectic matrix, such that the symplectic inner products discussed in Sect.~\ref{ss:defSpK} are preserved. The most general form of $T_\Lambda$, according to (\ref{symplABCD}), is then
\begin{equation}
  T_\Lambda =\begin{pmatrix}a_\Lambda{}^I{}_J &b_\Lambda{}^{IJ}\cr c_{\Lambda IJ} &- a_\Lambda {}^J{}_I \end{pmatrix}\,,\qquad b_\Lambda{}^{IJ}= b_\Lambda{}^{JI}\,,\qquad c_{\Lambda IJ}= c_{\Lambda JI}\,.
 \label{TLambdaform}
\end{equation}
In particular for isometries we require the transformation \eqref{TLambdaform} to be a symmetry of the action, which implies $b_\Lambda{}^{IJ}=0$, according to the discussion at the end of \ref{ss:dualities}. Isometries of the embedding manifold are therefore part of the symplectic group, as simply dictated by the following identity
\begin{equation}
 {\cal P}_\Lambda = \langle T_\Lambda V,\bar V\rangle=  \rme^{\kappa ^2{\cal K}} \langle T_\Lambda v,\bar v\rangle\,.
\label{PLambdainT}
\end{equation}

\bexer
Here is an exercise leading to a formula that will be important in the context of the potential.
First combine equations of this section to
\begin{equation}
  k_\Lambda{} ^\alpha \nabla_\alpha V= -T_\Lambda V-\rmi \kappa ^2\theta ^\Lambda{\cal P}_\Lambda V\,.
 \label{knabV}
\end{equation}
Then obtain
\begin{equation}
  k_\Lambda {}^\alpha g_{\alpha \bar \beta }k_\Sigma  {}^{\bar \beta }
  =\rmi k_\Lambda {}^\alpha\langle \nabla _\alpha V,\nabla _{\bar \beta }\bar V\rangle k_\Sigma {} ^{\bar \beta }
  =\rmi\langle T_\Lambda  V,T_\Sigma \bar V \rangle + \kappa ^2{\cal P}_\Lambda {\cal P}_\Sigma \,.
  \label{kgbark}
\end{equation}
\eexer

\subsection{Gauged Isometries as Symplectic Transformations}
The symmetries gauged by the vector multiplet are of course a subgroup of the generic isometries $\Lambda$ considered so far, and as such are expected to be represented by symplectic transformations. To specify such transformations we fix the index $\Lambda$ to be in the range of $I$ (which label the vector multiplets) as each symmetry is gauged by a vector. The moment map ${\cal P}_\Lambda$ is then identified with the $P_I^0$, introduced in (\ref{PI0def}).

The gauge transformations $\delta_G(\theta )$ of the scalars in the vector multiplet were written in (\ref{transfVM4})
\begin{equation}
  \delta_G(\theta ) X^I= -\theta ^K f_{KJ}{}^I X^J \,,
 \label{delXIalpha}
\end{equation}
where $\theta^K$ are the parameters. Notice that $\delta_G$ in \eqref{delXIalpha} is understood to be a symmetry \emph{before} the gauge fixing to the Poincar\'{e} theory. The transformations \eqref{delXIalpha} fit in the general scheme of isometries defined by Killing vectors
\begin{equation}
  k _J {}^I(X)=X^K f_{KJ}{}^I\,.
  \label{kJIinf}
\end{equation}
The associated moment map, from \eqref{solkillMomentMap2}, is called the `K\"{a}hler moment map' and it is identified with (\ref{PI0def}). The matrix $T_\Lambda $ of  (\ref{deltaGvgeneral}) is thus of the form (\ref{TLambdaform}) and comparing with (\ref{delXIalpha}), we write
\begin{equation}
T_K V =\begin{pmatrix}f_{KJ}{}^I&0\cr 2C_{K,IJ}&-f_{KI}{}^J\end{pmatrix}
   \begin{pmatrix}X^J \cr F_J\end{pmatrix},
 \label{defTKV}
\end{equation}
where $C_{K,IJ}$ is symmetric in the last two indices according to (\ref{TLambdaform}). The transformation (\ref{deltaGvgeneral}) with this matrix leads to (\ref{delXIalpha}) and
\begin{eqnarray}
 \delta_G (\theta) F_I & = & -2\theta^K C_{K,IJ}X^J+\theta^K f_{KI}{}^J F_J \,.
  \label{delGXF}
\end{eqnarray}
In particular
\begin{equation}
  \delta_G(\theta) [X^IF_I]= -2\theta^KC_{K,IJ}X^IX^J\,.
 \label{deltaGXF}
\end{equation}
In the tensor calculus, where $F_I$ are functions of $X^I$ and $X^IF_I=2\,F(X)$, Eq.~\eqref{deltaGXF} tells us that the prepotential transforms in a real quadratic function, which does not contribute to the action. In fact, this corresponds to (\ref{gdelFinC}).\footnote{This is the reason why we put the factor 2 in (\ref{defTKV}).}
Furthermore, the matrices $T_K$ should satisfy the algebra of gauge transformations \eqref{delXIalpha}:
\begin{equation}\label{algYMIJK}
[T_I,\,T_J]=f_{IJ}{}^K T_K\,,
\end{equation}
which imposes again the condition (\ref{fCalgebra}).

In the more general case without a prepotential, we can proceed from comparing (\ref{delV}) and  (\ref{deltaGvgeneral}) replacing the $\theta ^\Lambda $ by $X^J$, leading to
\begin{align}
  \delta_G (X )X^I=& X ^J k_J{}^\alpha \nabla_\alpha X^I
  +\rmi \kappa ^2{P}_J^0 X ^JX^I =0\,, \nonumber\\
  \delta_G (X )F_I= & X^J k_J{}^\alpha \nabla_\alpha F_I
  +\rmi \kappa ^2 { P}_J^0X ^JF_I=-2X^K C_{K,IJ} X^J+X^Kf_{KI}{}^JF_J\,.
 \label{delGF2expressions}
\end{align}
Multiplying the latter with $X^I$ and using the first one gives
\begin{equation}
  -2X^IX^J C_{J,IK} X^K = X^IX^J k_J^\alpha \nabla_\alpha F_I
  -X ^Jk_J^\alpha F_I \nabla_\alpha X^I=-X^J k_J^\alpha\langle V,\nabla _\alpha V\rangle =0\,.
 \label{Cmultiplied3X}
\end{equation}
We thus re-obtain (\ref{Cequationsymm}) in this more general setting. This is the equation that says that if all the $X^I$ are independent, $C_{I,JK}$ does not have a completely symmetric part.

For the gauged symmetries, we already found other relations for the moment maps in (\ref{PI0def})--(\ref{XP0}). The relation between the embedding and projective gauge transformations (\ref{Killingembeddingtoprojective}) is now written as
\begin{equation}
  k_J{}^I=y(k_J{}^\alpha \nabla_\alpha Z^I+\rmi\kappa ^2 P^0_J Z^I)=k_J{}^\alpha \nabla_\alpha X^I+\rmi\kappa ^2 P^0_J X^I= X^K f_{KJ}{}^I \,,
 \label{KJIkalpha}
\end{equation}
where we repeat at the end also the expression (\ref{kJIinf}) for easy reference below.
\bexer
Prove the following interesting relation from the definition (\ref{killMomentMap}) and (\ref{XP0}):
\begin{equation}
  X^I k_I{}^\alpha = -\rmi g^{\alpha \bar \beta }X^I\partial _{\bar \beta }P_I^0= \rmi g^{\alpha \bar \beta }P_I^0\nabla  _{\bar \beta }X^I\,.
 \label{Xkalpha}
\end{equation}
Further, with (\ref{KJIkalpha}) rewrite the gauge terms in the covariant derivative (\ref{nablaX})  as
\begin{align}
D_\mu X^I =& \nabla _\mu X^I - W_\mu ^Kk_K{}^\alpha \nabla _\alpha X^I= \nabla _\alpha X^I\left(\partial _\mu z^\alpha - W_\mu ^Kk_K{}^\alpha\right) \,.
 \label{nablaXshort}
\end{align}
\eexer


\section{Electric-Magnetic Charges; Attractor Phenomenon}
 \label{ss:emcharges}
 In this section we put all the ingredients of special geometry together to study some universal property of charged black hole solutions\index{charged black hole solution}  in ${\cal N}=2$ supergravity.\footnote{A recent review on black hole solutions in these theories can be found in \cite{Gallerati:2019mzs}.} It turns out that scalars in the background of charged black holes take fixed values at the horizon. These values are universally determined by (conserved) electric and magnetic charges, and independent of the initial configuration of the scalars (i.e. their value at infinite
distance from the black hole). This phenomenon is called `attractor mechanism'\index{attractor mechanism} in special geometry
\cite{Ferrara:1995ih,Strominger:1996kf}. Even though we will restrict here to $D=4$ for simplicity, the phenomenon has a $D=5,6$ counterpart in ${\cal N}=2$ theories\cite{Kraus:2005gh,Ferrara:2006xx,Larsen:2006xm}.

This section is structured as follows.
First, we obtain from the supergravity action \eqref{genL01} an effective action for the extremal black holes. In the usual way to proceed, at least in the case of static and spherically symmetric black holes, one eliminates the vectors through their equation of motion and then integrates out the angular dependence of the supergravity action to obtain an effective theory that governs the dynamics of the scalars. Crucially, the integration introduces a charge-dependent black hole potential along the way. For completeness, a Hamiltonian constraint has to be imposed on the system. Instead, we will employ an alternative derivation of such an effective action using only Einstein's equations, together with the equation of motion for the scalars and those on the vectors \cite{VanProeyen:2007pe}. The Einstein equations that cannot be derived from this effective action, will become constraints that have to be imposed on the solution. Along the way we will show that the black hole potential governing the effective action has a simple expression in the symplectic language. We will conclude this section by showing the universal solution corresponding to the attractor mechanism for the scalars.

\subsection{The Spacetime Ansatz and an Effective Action}

Consider a static spacetime metric\footnote{`Static' means that it admits a
global, nowhere zero, timelike hypersurface orthogonal Killing vector
field. A generalization are the `stationary' spacetimes, which admit a
global, nowhere zero timelike Killing vector field. In that case the
components $g_{0m}$ could be non-zero. For simplicity we look here to the
static spacetimes.}:
\begin{eqnarray}
 && \rmd s^2 =- \rme^{2U}\rmd t\rmd t + \rme^{-2U}\gamma _{mn}\rmd x^m\rmd
  x^n,\nonumber\\&&\qquad \mbox{i.e. }g_{00}=- \rme^{2U},\quad g_{mn}=\rme^{-2U}\gamma
  _{mn}, \qquad \sqrt{g}=\rme^{-2U}\sqrt{\gamma }\,, \nonumber\\  &&\partial _tU=\partial _t\gamma _{mn}=0\,.
 \label{stationaryMetric}
\end{eqnarray}
We are interested in the theory of a complex scalar $z$ coupled to Maxwell fields and gravity.\footnote{The generalization to non-trivial K\"{a}hler manifolds is obvious.} The relevant action can be recovered from the more general \eqref{genL01} and takes the schematic form
\begin{align}
S=&S_{\rm Einstein}+S^{(0)}+S^{(1)}\,,\nonumber\\
& S_{\rm Einstein}= \int \rmd^4 x \ft12 \sqrt{g}R(g)\,,\nonumber\\
&  S^{(0)}=-\int \rmd^4 x \sqrt{g}g^{\mu \nu }\partial
_\mu z\partial _\nu \bar z\,,\nonumber\\
&  S^{(1)} =   \int \rmd^4 x\left[\ft14\sqrt{g}(\Im {\cal
N}_{IJ})F_{\mu\nu}^I F^{\mu\nu J} -\ft 18 (\Re {\cal N}_{IJ})
\varepsilon^{\mu\nu\rho\sigma}F_{\mu\nu}^I F_{\rho\sigma}^J   \right]\,,
 \label{S1}
\end{align}
where $F^I_{\mu\nu}=2\partial_{[\mu} W_{\nu]}^I$. Note that the metric appears in $S^{(1)}$ only in the term with $\Im{\cal N}$.
\vspace{10pt}

As explained at the beginning of this section, we want to obtain an effective action starting from the Einstein equations. Let us consider the Einstein tensor:
\begin{equation}
 G_{\mu \nu }=2(\sqrt{g})^{-1}\frac{\delta S_{\rm
Einstein}}{\delta g^{\mu \nu }}=R_{\mu \nu }-\ft12g_{\mu \nu }R\,.
 \label{EinsteinTensor}
\end{equation}
For the static metric \eqref{stationaryMetric}, one obtains
\begin{eqnarray}
G_{00}&=&\rme^{4U}
\left( \ft12R(\gamma )+2D_m\partial ^mU-\partial ^mU\partial _mU\right) ,\nonumber\\
G_{mn}&=&\left(\gamma _{mn}\gamma ^{rs}-2\delta _m^r\delta _n^s\right)
\left(-\ft12R_{rs}(\gamma )+\partial_rU\partial _sU\right)\,.
 \label{GEinstein}
\end{eqnarray}
Here $\gamma _{mn}$ is used to raise and lower indices, and to define the
covariant derivative $D_m$. For the bosonic sector, the Einstein equations are
\begin{equation}
  G_{\mu \nu }=T_{\mu \nu }\,,
 \label{Einsteineqs}
\end{equation}
where $T_{\mu \nu }$ is the energy--momentum tensor. We split it in the
scalar part and the spin 1 part:
\begin{eqnarray}
  T_{\mu \nu }&=&T^{(0)}_{\mu \nu }+T^{(1)}_{\mu \nu }, \nonumber\\
&&T^{(0)}_{\mu \nu }=-2(\sqrt{g})^{-1}\frac{\delta S^{(0)}}{\delta g^{\mu
\nu }},\qquad T^{(1)}_{\mu \nu }=-2(\sqrt{g})^{-1}\frac{\delta
S^{(1)}}{\delta g^{\mu \nu }}\,.
 \label{Tmunusplit}
\end{eqnarray}
The energy--momentum tensor induced from $S^{(0)}$ is
\begin{equation}
  T^{(0)}_{\mu \nu }=-g_{\mu \nu }g^{\rho \sigma }\partial_\rho
 z\partial _\sigma  \bar z+2\partial _\mu z\partial _\nu \bar z\,.
 \label{FEg0}
\end{equation}
For our metric ansatz \eqref{stationaryMetric}, this gives
\begin{equation}
  T^{(0)}_{00}= \rme^{4U}\gamma ^{mn}\partial _mz\partial _n\bar z\,, \qquad
  T^{(0)}_{mn}=\left(-\gamma _{mn}\gamma ^{rs}+2\delta _{(m}^r\delta _{n)}^s\right)
  \partial _rz\partial _s\bar z\,.
 \label{T0}
\end{equation}
The energy--momentum tensor induced from $S^{(1)}$ is
\begin{equation}
T^{(1)}_{\mu \nu }= -\Im{\cal N}_{IJ}\left( -\ft14 g_{\mu \nu }F^I_{\rho
\sigma }F^{J\rho \sigma }+F_{\mu \rho }^IF_{\nu \sigma }^Jg^{\rho \sigma
}\right) .
 \label{T1general}
\end{equation}
If we now use the spacetime metric (\ref{stationaryMetric}), the non-zero
terms are
\begin{eqnarray}
  T^{(1)}_{00}&=&-\Im{\cal N}_{IJ}\left( \ft12\rme^{2U}F_{0m}^I\gamma
  ^{mn}F^J_{0n}+\ft14\rme^{6U}F^I_{mn}\gamma ^{mp}\gamma
  ^{nq}F^J_{pq}\right)\,,\nonumber\\
  T^{(1)}_{mn}&=&-\Im{\cal N}_{IJ}\left( \ft12\rme^{-2U}\gamma _{mn}F_{0p}^I\gamma
  ^{pq}F^J_{0q}-\ft14\rme^{2U}\gamma _{mn}F^I_{pq}\gamma ^{pp'}\gamma
  ^{qq'}F^J_{p'q'}\right.\nonumber\\&&\phantom{\Im{\cal N}_{IJ}}\left.-\rme^{-2U}F_{0m}^IF_{0n}^J+\rme^{2U}F_{mp}^I\gamma ^{pq}F_{nq}^J\right)\,.
 \label{T1}
\end{eqnarray}
We now introduce the magnetic vectors
\begin{equation}
  F^I_m=\ft12\gamma _{mn}(\sqrt{\gamma })^{-1}\varepsilon ^{npq}F_{pq}^I,
 \label{magnvectors}
\end{equation}
such that
\begin{eqnarray}
  &&F_{mn}^I=\sqrt{\gamma }\varepsilon _{mnp}\gamma ^{pq}F_q^I,\nonumber\\
&&  F_{mn}^I\gamma ^{mp}\gamma ^{nq}F_{pq}^J=2F^I_r\gamma
^{rs}F_s^J,\qquad F_{mp}^I\gamma
^{pq}F_{nq}^J=\gamma_{mn}F^I_r\gamma^{rs}F_s^J-F_n^IF_m^J.\nonumber\\
 \label{identitiesFmn}
\end{eqnarray}
Using these, we can again write the energy--momentum tensor in a similar
form as for the gravity field and for the scalars. We find
\begin{equation}
 T^{(1)}_{00}  =  \rme^{6U}\gamma ^{mn}V_{mn}\,, \qquad 
 T^{(1)}_{mn}  =  \rme^{2U} \left(\gamma _{mn}\gamma ^{rs}-2\delta _m^r\delta
_n^s\right)V_{rs}\,,
 \label{T0100mn}
\end{equation}
where
\begin{equation}
  V_{mn}=-\ft12\Im{\cal N}_{IJ}\left(
\rme^{-4U}F^I_{0m}F^J_{0n}+F_m^IF_n^J\right).
 \label{gammaVBH}
\end{equation}

The Einstein equations thus reduce to the following two equations
\begin{eqnarray}
 && -\ft12R_{mn}(\gamma )+\partial _mU\,\partial _n U + \partial _{(m}z\partial _{n)}\bar
  z-\rme^{2U}V_{mn}=0\,,\label{Einsteinconstraint}\\
  && D_m\partial ^m U -\rme^{2U}\gamma ^{mn}V_{mn}=0\,.\label{EinsteinFE}
\end{eqnarray}
We can obtain \eqref{EinsteinFE} using the following effective action
\begin{equation}
  S_{\rm eff}=\int \rmd^3 x \sqrt{\gamma }\left[
  -\partial_mU\,\gamma ^{mn}\partial _n U -\partial
  _mz\,\gamma ^{mn}\partial _n\bar z-\rme^{2U}\gamma ^{mn}V_{mn}\right]
  \,.
 \label{EffectiveSgamma}
\end{equation}
Indeed, the field equation of this action for $U$ is (\ref{EinsteinFE}) if we
keep $V_{mn}$ fixed during the variation. Of course, as $U$ is never involved in the $z$-dependent part of the action, the latter is not determined by
this requirement. We will prove that the field equation for the scalar $z$
can also be derived from this action for a specific parameterization of
$V_{mn}$. This will be clarified in Sect.~\ref{ss:MaxwBHpot}. Only then
it will be clear how to use this effective action. We remark here that $\gamma^{mn}$ should not be seen as a dynamical variable in this action. Instead of its field equations as following from this effective action, we impose the extra
constraint (\ref{Einsteinconstraint}), which is not derivable from \eqref{EffectiveSgamma}.

\subsection{Maxwell  Equations and the Black Hole Potential}
\label{ss:MaxwBHpot}

The quantity (\ref{gammaVBH}) is expressed in components of the field strengths
$F_{\mu \nu }^I$. However, we can write it in terms of the symplectic
vectors of field strengths and field equations. To do so, we start from the real
form of (\ref{defG}), which can be rewritten using the Hodge duality relation (\ref{dual46}) as follows:
\begin{equation}
  G_{I\mu \nu }= \Re{\cal N}_{IJ}\,F^J_{\mu \nu }+\ft12\Im{\cal N}_{IJ}\,g_{\mu \mu
  '}g_{\nu \nu '}(\sqrt{g})^{-1}\varepsilon ^{\mu' \nu' \rho \sigma }F^J_{\rho \sigma}\,.
 \label{defGF}
\end{equation}
Then $F^I_{\mu \nu }$, $G_{I\mu \nu }$ are components of a symplectic vector. Introducing now the three-dimensional duals as a generalization of
(\ref{magnvectors}):
\begin{equation}
  F^I_m=\ft12\gamma _{mn}(\sqrt{\gamma })^{-1}\varepsilon ^{npq}F_{pq}^I,\qquad
  G_{Im}=\ft12\gamma _{mn}(\sqrt{\gamma })^{-1}\varepsilon ^{npq}G_{Ipq},
 \label{3ddualF}
\end{equation}
we obtain from (\ref{defGF}) with $\varepsilon^{0npq}=-\varepsilon
^{npq}$ (as we use $\varepsilon^{0123}=-1$, see (\ref{valueLeCi}))
\begin{equation}
  \begin{pmatrix}F_{0m}^I\cr G_{I0m}\end{pmatrix}=-\rme^{2U}{\cal M}\Omega \begin{pmatrix}F^J_m\cr
  G_{Jm}\end{pmatrix},
 \label{F0mFpq}
\end{equation}
where ${\cal M}$ was given in (\ref{defcalM}) and $\Omega $ is the
symplectic metric (\ref{Scond}). These matrices contain indices $I$ and
$J$ at appropriate positions automatically for (\ref{F0mFpq}) to make
sense. Equation (\ref{gammaVBH}) then becomes
\begin{equation}
  V_{mn}=\ft{1}{2}\begin{pmatrix}F^I_m & G_{Im}\end{pmatrix}
  \Omega {\cal M}\Omega\begin{pmatrix}F^J_n\cr G_{Jn}\end{pmatrix}\,,
 \label{Vmnsympl}
\end{equation}
where
\begin{equation}
  \Omega {\cal M}\Omega=\begin{pmatrix}-I-RI^{-1}R&RI^{-1}\cr
  I^{-1}R&-I^{-1}\end{pmatrix}\,, \qquad R=\Re{\cal N}\,,\quad I=\Im{\cal
  N}\,.
 \label{OmMOm}
\end{equation}
Note that $U$ does not appear in this expression for $V_{mn}$. This implies that if
we consider $V_{mn}$ as a function of $F^I_m$, $G_{Im}$ and the scalars
implicitly present in (\ref{OmMOm}) and we insert it as such in the
effective action (\ref{EffectiveSgamma}) then this
action still generates the field equation (\ref{EinsteinFE}) for $U$. We now check that in
this way it also generates the same scalar field equations as those
obtained from the original action $S^{(0)}$ and $S^{(1)}$, where the
vector fields $W_\mu ^I$ were the other independent variables. Hence
these field equations that should be reproduced are
\begin{equation}
 0= \partial _\mu \sqrt{g}g^{\mu \nu }\partial _\nu \bar z+\ft14\sqrt{g}\partial _z(\Im {\cal
N}_{IJ})F_{\mu\nu}^I F^{\mu\nu J} -\ft 18 \partial _z(\Re {\cal N}_{IJ})
\varepsilon^{\mu\nu\rho\sigma}F_{\mu\nu}^I F_{\rho\sigma}^J\,.
 \label{scalarfieldeqnorig}
\end{equation}
Specifying the metric (\ref{stationaryMetric}) and the expressions for
the field strengths in terms of $F^I_m$ and $G_{Im}$, this becomes
\begin{equation}
 0= \partial _m\sqrt{\gamma }\gamma ^{mn}\partial _n\bar z+\ft{1}{2}\sqrt{\gamma
  }\rme^{2U}\gamma ^{mn}\begin{pmatrix}F^I_m&G_{mI}\end{pmatrix}
\Omega \partial_z{\cal M}\Omega  \begin{pmatrix}F^J_n\cr
G_{nJ}\end{pmatrix}\,,
 \label{scalarfieldeqn2}
\end{equation}
where the indices $I$ and $J$ appear again in appropriate positions on
the submatrices of $\Omega \partial_z{\cal M}\Omega$. The latter is
indeed the field equation obtained from the effective action
\begin{align}
   S_{\rm eff}(U,z)=\int \rmd^3 x \sqrt{\gamma }\gamma ^{mn}&\left[
  -\partial_mU\,\partial _n U -\partial
  _mz\,\partial _n\bar z\phantom{\begin{pmatrix}F^J_n\cr G_{Jn}\end{pmatrix}}\right.\nonumber\\
 & \left.-\ft{1}{2}\rme^{2U}\begin{pmatrix}F^I_m & G_{Im}\end{pmatrix}
  \Omega {\cal M}\Omega\begin{pmatrix}F^J_n\cr G_{Jn}\end{pmatrix}\right]   \,.
 \label{EffectiveSgammaUz}
\end{align}
The $(U,z)$ in the left-hand side indicates that $S_{\rm eff}$ should be considered
as an effective action for varying with respect to $\{U,z\}$, while
$\gamma ^{mn}$, $F^I_m$ and $G_{Im}$ should be considered as background.
We saw already that the field equations of the original action for
$\gamma ^{mn}$ lead to the constraint (\ref{Einsteinconstraint}). We will now
check what the field equations of the vector sector impose.

The field equations from $S^{(1)}$ in (\ref{S1}) with independent vectors $W_\mu ^I$
are equivalent to the field equations and Bianchi identities
\begin{equation}
  \varepsilon ^{\mu \nu \rho \sigma }\partial _\nu \begin{pmatrix}F^I_{\rho \sigma }\cr G_{I\rho \sigma
  }\end{pmatrix}=0\,.
 \label{FEBianchi}
\end{equation}
Using our preferred variables, this gives
\begin{equation}
 \partial _m\sqrt{\gamma
}\gamma ^{mn} \begin{pmatrix}F^I_{n}\cr G_{In}\end{pmatrix}=0, \qquad
\partial _{[m}\rme^{2U}{\cal M}\Omega\begin{pmatrix}F^I_{n]}\cr
G_{In]}\end{pmatrix}=0.
 \label{FEBianchi0m}
\end{equation}

One way of solving these equations is to define $F^I_m=\partial _mH^I$ and $G_{Im}=\partial _mH_I$, such that the first of  \eqref{FEBianchi0m} becomes
\begin{eqnarray}
&&\partial _m  \sqrt{\gamma }\gamma ^{mn}\partial _n H^I=\partial _m
\sqrt{\gamma }\gamma ^{mn}\partial _n
  H_I=0\,.
 \label{harmonicH}
\end{eqnarray}
We remain then with Bianchi identities of the form
\begin{equation}
  \left( \partial _{[m}\rme^{2U}{\cal M}\right)\Omega\partial _{n]} \begin{pmatrix}H^I\cr
H_I\end{pmatrix}=0\,,
 \label{RestBianchi}
\end{equation}
which can be solved by assuming that all fields ($U$, the scalars $z$ and
the harmonic $H^I$ and $H_I$) depend only on one coordinate. This ensures that the
$\partial _m$ and $\partial _n$ for $m\neq n$ in the above equation
cannot both be nonvanishing. We denote this one coordinate as $\tau $.
Thus $U(\tau )$, $z(\tau )$, $H^I(\tau )$ and $H_I(\tau )$.

A convenient metric is, e.g., \cite{Ferrara:1997tw}
\begin{equation}
\gamma _{mn}\rmd x^m\rmd x^n = \frac{c^4}{\sinh^4 c\tau}\rmd \tau^2 +
\frac{c^2}{\sinh^2 c\tau}(\rmd \theta ^2 + \sin^2\theta \rmd \phi ^2)\,.
 \label{sphericalcoord}
\end{equation}
Details on this parameterization are given in an appendix of
\cite{Kallosh:2006bt}. This parameterization has the property
$\sqrt{\gamma }\gamma ^{\tau \tau }=\sin\theta $, which will be useful. In this basis, condition \eqref{harmonicH} is just $H''=0$ (where a prime is now a derivative w.r.t. $\tau $), so we can take
\begin{equation}
H=  \begin{pmatrix}H^I\cr H_I\end{pmatrix}= \Gamma \tau +h,\qquad
h=\begin{pmatrix}h^I\cr h_I\end{pmatrix}, \qquad \Gamma
=\begin{pmatrix}p^I\cr q_I\end{pmatrix}.
 \label{Hpqh}
\end{equation}
We have here introduced the magnetic and electric charges in the
symplectic vector $\Gamma $. We come back to their meaning in Sect.~\ref{ss:charges}. The effective Lagrangian for the scalars is obtained upon plugging these solutions into \eqref{EffectiveSgammaUz}. Up to a constant we get
\begin{equation}
  {\cal L}_{\rm eff}=U'^2+\rme^{2U}V_{\rm BH} +z'\bar z'\,,
 \label{SeffU}
\end{equation}
where the `black hole potential' \index{black hole potential} is now
\begin{equation}
  V_{\rm BH}=V_{\tau \tau }=\ft12\Gamma ^T\Omega {\cal M}\Omega \Gamma.
 \label{VBH}
\end{equation}

The one-dimensional effective Lagrangian \eqref{SeffU} does not reproduce all the Einstein equations. Indeed the Einstein equations
(\ref{Einsteinconstraint})-(\ref{EinsteinFE}) lead to 2 independent equations:
\begin{equation}
c^2-U'^2-z'\bar z'+\rme^{2U}V_{\rm BH}=0\,, \qquad -U''+\rme^{2U}V_{\rm
BH}=0\,.
 \label{IndepEinstein}
\end{equation}
The second one is the one that can be obtained from the effective action, while the first one cannot. It must be considered as an extra constraint.

\subsection{Field Strengths and Charges}
\label{ss:charges}

In (\ref{Hpqh}) we already wrote $p^I$ and $q_I$ for the components of $\Gamma$, suggesting that they are charges. Indeed, when we consider field configurations with electric and/or magnetic charges in
4 dimensions, this means that there are 2-cycles $S^2$ surrounding the
sources such that
\begin{equation}
  \int _{S^2}F^I_{\mu \nu }\rmd x^\mu \wedge \rmd x^\nu = 8\pi p^I, \qquad
\int _{S^2}G_{I\mu \nu }\rmd x^\mu \wedge \rmd x^\nu = 8\pi q_I.
 \label{chargesqp}
\end{equation}
 \bexer  Check that the solution that we gave above, leads indeed to the
 identification of the charges here and in (\ref{Hpqh}).
 \eexer

There is also the field strength that occurs in the gravitino
transformation, see e.g. (\ref{transfo4}), which is the value of the
auxiliary field $T_{\mu \nu }$ of the Weyl multiplet. When we restrict
to the bosonic part of its value, determined in (\ref{YTeom}),
we obtain
\begin{eqnarray}
  T^-_{\mu \nu }&=&2T_IF^{-I}_{\mu \nu}\,, \label{bosvalueT}
 \\
  T_I&=& \frac{ N_{IJ}\bar X^J}{\bar X^LN_{LM}\bar X^M}
  = -2 \kappa ^2\Im{\cal N}_{IJ}X^J=\rmi \kappa ^2\left( F_I-\bar {\cal N}_{IJ}X^J\right), \qquad
  \bar X^IT_I=1,\nonumber
\end{eqnarray}
where use has been made of (\ref{ImcNX}).

Assuming that $F_I$ and $X^I$ are sufficiently constant in the integration region
(such that they can be taken in and out of the integral), the integral of this quantity gives
\cite{Billo:1995ge}
\begin{align}
  {\cal Z}\equiv \frac{\rmi}{16\pi \kappa ^2 }\int _{S^2}T^-_{\mu \nu }\rmd x^\mu \wedge \rmd x^\nu =&\frac{1}{8\pi }\int _{S^2}\left(X^IG^-_{I\mu \nu }-F_IF^{-I}_{\mu \nu}\right) \rmd x^\mu \wedge \rmd x^\nu \nonumber\\
=&\frac{1}{8\pi }\int _{S^2}\left(X^IG_{I\mu \nu }-F_IF^I_{\mu
\nu}\right) \rmd x^\mu \wedge \rmd x^\nu\nonumber\\
=& X^Iq_I- F_Ip^I\,. \label{intTS2}
\end{align}
Between the first and the second line we used that the combination with the self-dual field strengths vanishes due
to $F_I={\cal N}_{IJ}X^J$. The object ${\cal Z}$ is called the central charge,
because its value appears in the commutator of two supersymmetries, as
can be seen from (\ref{qqcomb})-(\ref{qqparamsb}).

When we take the holomorphic covariant derivatives of the final expression,
then we have to use $\nabla_\alpha F_I=\bar {\cal N}_{IJ}\nabla _\alpha X^J$, and therefore only the self-dual parts remain. This gives
thus
\begin{eqnarray}
  \nabla_\alpha {\cal Z}&=& \nabla_\alpha X^Iq_I- \nabla_\alpha F_Ip^I\nonumber\\
  &=&\frac{1}{8\pi }\int _{S^2}\left(\nabla_\alpha X^IG^+_{I\mu \nu }-\nabla_\alpha F_IF^{+I}_{\mu \nu}\right) \rmd
x^\mu \wedge \rmd x^\nu\nonumber\\ &=&\frac{2\rmi}{8\pi }\int _{S^2}{\cal
D}_\alpha X^I\Im{\cal N}_{IJ}F^{+I}_{\mu \nu}\rmd x^\mu \wedge \rmd
x^\nu\,.\label{DalphaZ}
\end{eqnarray}
The latter quantities $\nabla_\alpha {\cal Z}$ are the objects that appear also in the
transformation laws of the physical gauginos. Indeed, the fermions of the
conformal multiplets transform according to (\ref{transfVM4}) to
quantities ${\cal F}^-_{\mu \nu }$, whose bosonic part is
\begin{equation}
  {\cal F}^{-I}_{\mu \nu }= \left( \delta ^I_J-\bar X^IT_J\right) F^{-J}_{\mu \nu}\,.
 \label{calFbos}
\end{equation}
The physical fermions are the ones that satisfy the $S$-gauge condition
(\ref{v:zeta_i02}), which means that they vanish under projection with
$T_I$. We find indeed $T_I{\cal F}^{-I}_{\mu \nu }= 0$.

Using the vector $\Gamma $ of (\ref{Hpqh}), we can write in a symplectic notation
\begin{equation}
  {\cal Z}=\langle V,\Gamma \rangle=V^T\Omega \Gamma\,,\qquad \nabla _\alpha {\cal Z}= \langle U_\alpha ,\Gamma \rangle\,,
 \label{Zsymplectic}
\end{equation}
and can derive from (\ref{UGU}) a simple expression for the `black hole potential'
\cite{Ceresole:1995ca,Ferrara:1996dd,Ferrara:1997tw}
\begin{equation}
  V_{\rm BH}\equiv  {\cal Z}\bar {\cal Z}+ \nabla_\alpha {\cal Z} g^{\alpha \bar \beta }\nabla_{\bar \beta }\bar
  {\cal Z}= \ft12 \Gamma ^T\Omega {\cal M}\Omega \Gamma\,.
 \label{VBHZ}
\end{equation}
Similarly, by using the same identity, one derives:
\begin{equation}
  V\bar {\cal Z} + \bar U_{\bar \beta }g^{\bar \beta \alpha }\nabla_\alpha {\cal Z}=
  -\ft12 \left( {\cal M}\Omega +\rmi\unity \right)  \Gamma\,.
 \label{VZ+UDZ}
\end{equation}

\subsection{Attractors}

The attractor solution
\cite{Ferrara:1995ih,Strominger:1996kf,Ferrara:1996dd} is the solution
near the horizon. This is the large $\tau $ behaviour. In that case supersymmetry is preserved, which is
expressed as $\nabla_\alpha {\cal Z}=0$. This extremizes the black hole
potential. So it is consistent with constant $z^\alpha $ as solution of
the field equation for the scalars. In this case (\ref{VZ+UDZ})
simplifies. The imaginary part is
\begin{equation}
-2  \Im (V\bar {\cal Z})=\Gamma\,.
 \label{GammaBPS}
\end{equation}
These are the attractor equations. The BH potential reduces to
\begin{equation}
  V_{\rm BH,BPS}= |{\cal Z}|^2\,,
 \label{VBHBPS}
\end{equation}
where the indication `BPS' is used because we started from an extremal black hole solution. Then we determine $U$ by the constraint
\begin{equation}
  {\dot U}^2=\rme^{2U}V_{\rm BH,BPS},\qquad \mbox{i.e.}\qquad \dot
  U=\pm\rme^U V_{\rm BH,BPS}\,.
 \label{dotU2}
\end{equation}
The $V_{\rm BH,BPS}$ being constant, this automatically implies the other
field equation
\begin{equation}
  \ddot U=\rme^{2U}V_{\rm BH,BPS}\,.
 \label{ddotUBPS}
\end{equation}
Finally, the solution is
\begin{equation}
  \rme^{-U}=\mp|{\cal Z}|\tau +\mbox{constant}\,.
 \label{solnU}
\end{equation}
Since ${\cal Z}$ is completely determined by the charges via \eqref{intTS2}, the solution above shows that the near-horizon ($\tau\rightarrow \infty$) value of the scalars is universal and independent of the initial conditions.

\section{Quaternionic-K\"{a}hler Manifolds}
 \label{ss:qKmanifolds}

Quaternionic-K\"{a}hler manifolds entered supergravity research first in the seminal work~\cite{Bagger:1983tt}. In \cite{Galicki:1987ja} a lot of interesting properties were already discussed. Workshops on quaternionic geometry have been organized where mathematics and physics
results were brought together~\cite{QuatWorksh2}. Other important papers that reviewed the properties of quaternionic manifolds
are~\cite{Fre:2001jd,D'Auria:2001kv,Bergshoeff:2002qk}.

As a result of Swann~\cite{Swann}, every
quaternionic manifold can be obtained as a cone with an $\SU(2)$ gauging. The condition that a
hyper-K\"{a}hler manifold can be formulated in a conformal way is equivalent
to the condition that there is a cone. Therefore, those hyper-K\"{a}hler that
satisfy this condition are one-to-one related with the hyper-K\"{a}hler
manifolds that can be made quaternionic by an $\SU(2)$ gauging.
The result of Swann has been made explicit in \cite{Bergshoeff:2004nf} by the construction with the embedding of the manifold in a conformal manifold, as written in Sect. \ref{coord_quatman}.
There it has been shown how any
quaternionic-K\"{a}hler metric can be obtained in this way. The procedure is the same whether
applied for $D=4$, $D=5$ or $D=6$.

\subsection{Supersymmetry and Quaternionic Geometry.}
The quaternionic geometry\index{quaternionic geometry} is a bi-product of the supersymmetry algebra, which leads to the definition of the almost quaternionic structures $\vec{J}_X{}^Y$\index{almost quaternionic structure}, see \eqref{vpro-defJf}. Furthermore, in order to build an action, we require this manifold to possess an invariant metric $g_{XY}$, which was defined in \eqref{defg}. If the almost quaternionic structures are covariantly constant (up to a rotation among them) with the Levi-Civita connection associated to $g_{XY}$ (see \eqref{cov_const_J}), the manifold is promoted to a hyper-K\"{a}hler manifold\index{hyper-K\"{a}hler manifold}. The requirement of conformal
symmetry further restricts the manifold. The gauge fixing then leads from a $4(n_H+1)$-dimensional embedding manifold to a
projection on a $4n_H$-dimensional submanifold, which turns out to be quaternionic-K\"{a}hler (see Sect. \ref{coord_quatman}). The latter is parameterized by scalar fields $q^\Xu $, while its tangent space
contains vectors labelled by indices $(i,A)$. Using the basis (\ref{v:basisikeuze_hyp}), a quaternionic-K\"{a}hler metric $h_{\Xu \Yu }(q)$, is obtained from the embedding metric in  (\ref{hquat}). Similarly all other geometrical quantities are obtained by a projection from the embedding to the quaternionic-K\"{a}hler space.\footnote{The terminology `quaternionic' and `quaternionic-K\"{a}hler' is used in mathematics in this sense: the first requires a quaternionic structure on the manifold, and the second demands moreover compatibility with a metric structure.}

Considering the differential equation on the frame fields in the quaternionic-K\"{a}hler space, (\ref{covconstf}), we see that it contains $\SU(2)$ connections $\omega _{uj}{}^i$ as well as $\USp(2n_H)$ connections\footnote{We demanded the preservation of the metric $g_{XY}$ (see \eqref{defg}), therefore the $\Gl(n_H,\mathbb{H})$ is restricted to $\USp(2n_H)$.} $\omega_{u B}{}^A$. The latter is similar to the condition in the embedding manifold,  (\ref{Df0}), but the $\SU(2)$ connection was absent in the hyper-K\"{a}hler manifold.
This $\SU(2)$ connection $\vec{\omega }_u$, promotes the almost quaternionic structure to a quaternionic
structure, such that the resulting manifolds are `quaternionic'. If the $\SU(2)$
connection is zero, they are called `hypercomplex', which is the case
before the gauge fixing (and in rigid supersymmetry).
Note that the same $\SU(2)$ connection also appears  in the differential equation on the hypercomplex structures in the quaternionic-K\"{a}hler manifold,  (\ref{nablaJproj}), while it was absent in that of the hyper-K\"{a}hler structure (\ref{cov_const_J}). For physics applications we further require a positive definite energy in the Einstein--Hilbert action, which results in additional constraints on the geometry.

\subsection{Quaternionic Manifolds}
As mentioned above, a main ingredient is the equation that states the covariant constancy of the frame fields \eqref{covconstf}, which can be considered in terms of a connection on the tangent space
\begin{equation}
\Omega _{\Xu \,jB}{}^{iA}\equiv   f^\Yu {}_{jB}\left( \partial _\Xu  f^{iA}_\Yu-\Gamma
_{\Xu \Yu }^\Zu f^{iA}{}_\Zu\right)=-\omega _{\Xu j}{}^i \delta _B{}^A-\omega
_{\Xu B}{}^A\delta _j{}^i\,.
 \label{vpro-Omega}
\end{equation}
If this $\Omega _{\Xu \,jB}{}^{iA}$,
for each $\Xu $, would be a general $4n_H\times 4n_H$ matrix, then we would say
that the holonomy is not restricted (or sits in $\Gl (4n_H,\mathbb{R})$). The splitting as in the right-hand side of this equation implies that the
holonomy group is restricted to $\SU(2)\times \Gl (n_H,\mathbb{H})$.

The integrability condition of (\ref{covconstf}) leads to an expression for the curvature of quaternionic manifolds:
\begin{equation}
R^\Zu {}_{\Wu \Xu \Yu }= f^\Zu {}_{iA}f^{jA}{}_\Wu{\cal R} _{\Xu \Yu j}{}^i +f^\Zu {}_{iA}f^{iB}{}_\Wu{\cal
R}_{\Xu \Yu B}{}^A= -\vec{J}_\Wu {}^{\Zu}\vec{{\cal R}} _{\Xu \Yu }
+f^\Zu{} _{iA}f^{iB}{}_\Wu{\cal R}_{\Xu \Yu B}{}^A \,,
 \label{vpro-Rdecomp}
\end{equation}
where
\begin{align}
  R^\Zu {}_{\Wu \Xu \Yu }\equiv& 2\partial_{[\Xu } \Gamma _{\Yu ]\Wu }^\Zu +2\Gamma ^\Zu _{\Vu [\Xu }\Gamma ^\Vu _{\Yu ]\Wu }\,,\nonumber\\
  \vec{\cal R}_{\Xu \Yu }\equiv&  2\partial _{[\Xu }\vec{\omega} _{\Yu ]}  +2\vec{\omega} _{[\Xu  }\times \vec{\omega} _{\Yu ]}\,,\nonumber\\
  {\cal R}_{\Xu \Yu B}{}^A\equiv & 2\partial_{[\Xu} \omega_{\Yu]B}{}^A
+ 2\omega_{[\Xu|C|}{}^A \omega_{\Yu]B}{}^C\,.
 \label{allcurvatures}
\end{align}
Thus the curvature of the affine connection is split in an $\SU(2)$ part and a $\Gl (n_H,\mathbb{H})$ part.

\subsection{Quaternionic-K\"{a}hler Manifolds}
Quaternionic-K\"{a}hler manifolds \index{quaternionic-K\"{a}hler manifold}(which include `hyper-K\"{a}hler manifolds' \index{hyper-K\"{a}hler manifold} in
the limiting case that the $\SU(2)$ curvature vanishes) by definition have a
metric, $h_{\Xu\Yu}$. This corresponds to the requirement in the matter couplings that there is an action.
There is the request that the connections preserve the metric, which restricts the holonomy group to $\SU(2)\times \USp(2n_H)$.
The affine connection $\Gamma _{\Xu \Yu }^\Zu$ is now the metric connection, and the first line of (\ref{allcurvatures}) is the metric curvature, while the last line is the $\USp(2n_H)$ curvature.

For $n_H>1$ one can prove that these manifolds are Einstein, and that the
$\SU(2)$ curvatures are proportional to the complex structures\footnote{In the conventions of this book, such proofs are given in \cite[Appendix B]{Bergshoeff:2002qk}.}
\begin{equation}
 R_{\Xu \Yu }\equiv R^\Zu {}_{\Xu \Zu \Yu }=\frac{1}{4n_H} h_{\Xu \Yu } R\,,\qquad    \vec{{\cal R}}_{\Xu \Yu } =\ft12\nu  \vec{J}_{\Xu \Yu }\,,\qquad
\nu  =\frac{1}{4n_H(n_H+2)}R\,.
 \label{RlambdaJ}
\end{equation}
For $n_H=1$ this is part of the definition of
quaternionic-K\"{a}hler manifolds. Hyper-K\"{a}hler manifolds are those where the
$\SU(2)$ curvature is zero, and these are thus also Ricci-flat.

\subsection{Quaternionic-K\"{a}hler Manifolds in Supergravity}

In supergravity we find all these constraints from requiring a
supersymmetric action. Moreover, we need for the invariance of the action
that the last equation of (\ref{RlambdaJ}) is satisfied with $\nu =-\kappa^2$, see \eqref{v:q^0constant}.
This implies that the scalar curvature is $R=-4n_H(n_H+2)\kappa^2$. The fact that this is negative excludes e.g. the compact symmetric spaces. In Sect. \ref{ss:HomSymSpaces} all the symmetric spaces in this class will be mentioned, and these are thus all noncompact.

\section{Relations Between Special Manifolds} \label{ss:HomSymSp}

So far we discussed separately the K\"{a}hler manifolds (associated to the vector multiplets) and the quaternionic-K\"{a}hler manifolds (hypermultiplets), even though we paid attention, along the way, to various similarities among these spaces. Although our discussion has been mostly limited to $D=4$, as we pointed out at the beginning of Chap.~\ref{ss:specialgeom}, there exists an analogous special geometry associated to the real scalars in the $D=5$ vector multiplets. Perhaps not surprisingly, all these `special manifolds' are related by maps, which are denoted as \cmap~and \rmap.

\subsection{c-Map and r-Map}
\label{ss:crmap}

The \cmap~\cite{Cecotti:1989qn} \index{c-map} connects a special
K\"{a}hler to a quaternionic-K\"{a}hler manifold. It is induced by dimensional reduction of an ${\cal N}=2$
supergravity theory from $D=4$ to $D=3$, by suppressing the dependence on one of the
(spatial) coordinates. The resulting $D=3$ supergravity theory can be
written in terms of $D=3$ fields and this rearranges the original fields
such that the number of scalar fields increases from $2n$ to $4(n+1)$. This is shown in Table~\ref{tbl:cmap}.
\begin{table}
[!t]
\caption{\it The \cmap\ as dimensional reduction from $D=4$
to $D=3$ supergravity. The number of fields of various spins is indicated.
} \label{tbl:cmap}.
\begin{center}\begin{tabular}{||r|ccc||}\hline
$D=4$ spins& 2 & 1 & 0 \\  \cline{2-4} numbers &  1  & $n+1$ & 2$n$ \\
\hline
$D=3$ spins & & & \\
2  & 1 &  &  \\
0  & 2 & 2$(n+1)$ & 2$n$ \\
\hline
\end{tabular}
\end{center}
%
\end{table}
Essential in this map is that $D=4$ vectors leave first a scalar component in $D=3$, but also the other part, vectors in $D=3$, are dual to scalars.
This map is also obtained in string theory context by changing from the reduction of a type IIA theory to the reduction of the $T$-dual type IIB theory or vice versa.

This leads to the notion of `{\it special quaternionic manifolds}', which
are those manifolds appearing in the image of the \cmap. They are a
subclass of the quaternionic manifolds.

The \rmap \cite{deWit:1992cr} \index{r-map} is determined by the reduction of vector multiplets in $D=5$ to vector multiplets in $D=4$. Starting with $n-1$ physical vector multiplets in $D=5$, one ends up with $n$ vector multiplets in $D=4$ ($2n$ real scalars) as schematically shown in
Table~\ref{tbl:rmap}.
\begin{table}[!t] 
\caption{\it The \rmap\ induced by dimensional reduction from $D=5$ to $D=4$ supergravity. The number of fields of integer spins is indicated.} \label{tbl:rmap}
\begin{center}\begin{tabular}{||r|ccc||}\hline
$D=5$ spins& 2 & 1 & 0 \\  \cline{2-4} numbers &  1  & $n$ & $n-1$ \\
\hline
$D=4$ spins & & & \\
2  & 1 &  &  \\
1  & 1 & $n$ &  \\
0  & 1 & $n$ & $n-1$ \\ \hline
\end{tabular}
\end{center}
%
%
\end{table}
The manifolds defined by coupling (real) scalars to vector multiplets in 5 dimensions are called `very special real
manifolds'. \index{very special real manifold}
They are determined by the symmetric tensor $C_{IJK}$ in Sect.~\ref{ss:actVMD5}. The corresponding $D=4$ couplings will then also be determined by such a tensor, and one parameterization is in terms of the prepotential (\ref{Fvsp}). As already indicated there, the special K\"{a}hler manifolds that are in the image of this \rmap\ are then denoted as  `very special K\"{a}hler manifolds' \index{very special K\"{a}hler manifolds}. Their further image under the \crmap\ are then called `very special
quaternionic manifolds'.\index{very special quaternionic manifolds}

The \cmap\ has been studied in superspace in \cite{Gates:1999zv,Rocek:2005ij} and the mathematical structure has been clarified in terms of the so-called Swann bundle in \cite{Rocek:2006xb}. An off-shell extension has been found \cite{Banerjee:2015uee} in view of applications with higher derivatives. It is then useful to formulate the hypermultiplets in the form of tensor multiplets \cite{deWit:2006gn}.

\subsection{Homogeneous and Symmetric Spaces}
\label{ss:HomSymSpaces}

Homogeneous and symmetric spaces are the most known manifolds. These are
spaces of the form $G/H$, where $G$ is the isometry group and $H$ is its
isotropy subgroup. The group $G$ is not necessarily a semi-simple group,
and thus not all the homogeneous spaces have a clear name. The symmetric
spaces are those for which the algebra splits as $g=h+k$ and all
commutators $[k,k]\subset h$. The homogeneous special manifolds are
classified in~\cite{deWit:1992nm}.

It turns out that homogeneous special manifolds are in one-to-one
correspondence to realizations of real Clifford algebras with signature
$(q+1,1)$ for real, $(q+2, 2)$ for K\"{a}hler, and $(q+3,3)$ for quaternionic
manifolds. Thus, the spaces are identified by giving the number $q$,
which specifies the Clifford algebra, and by specifying its
representation. If $q$ is not a multiple of 4, then these Clifford algebras have only one
irreducible representations, and we thus just have to mention the
multiplicity $P$ of this representation. The spaces are denoted as
$L(q,P)$. If $q=4m$ then there are two inequivalent representations,
chiral and antichiral, and the spaces are denoted as $L(q,P,\dot P)$. The fact that the chiral and the antichiral representations are conjugate implies $L(4m,P,\dot
P)=L(4m,\dot P,P)$. A special case is $q=0$ for which $L(0,n)=L(n,0)$. These manifolds are listed in Table \ref{tbl:homsp}.
\begin{table}[!t]
\caption{\it Homogeneous manifolds.
In this table, $q$, $P$, $\dot P$ and $m$ denote positive integers or
zero, and $q\neq 4m$. SG denotes an empty space, which corresponds to
supergravity models without scalars. The horizontal lines separate
spaces of different rank. The first non-empty space in each column has
rank~1. Going to the right or down a line increases the rank by~1. The
manifolds indicated by a $\star$ did not get a name. The number $n$ is
the complex dimension of the K\"{a}hler space as in (\ref{nformula}), which simplifies in many cases.} \label{tbl:homsp}
\begin{center}
\begin{tabular}{|l|r|ccc|}
\hline\noalign{\smallskip}
&$n$ &Real & K\"{a}hler & Quaternionic \\
\hline &&&&\\[-3mm]
$L(-3,P)$&$P$&&   & $\frac{\USp(2P+2,2)}{\USp(2P+2)\otimes \SU(2)} $
     \\[2mm]
$SG_4$&0&&  SG       &$\frac{\U(1,2)}{\U(1)\otimes \U(2)} $
    \\[2mm]
\hline&&&&\\[-3mm]
$L(-2,P)$&$1+P$&&$\frac{\U(P+1,1)}{\U(P+1)\otimes \U(1)}$
    &$\frac{\SU(P+2,2)}{\SU(P+2)\otimes \SU(2)\otimes \U(1)} $
        \\[2mm]
$SG_5$&1&SG  & $\frac{\SU(1,1)}{\U(1)}$
    &$\frac{G_2}{\SU(2)\otimes \SU(2)}$  \\[2mm]
\hline&&&&\\[-3mm]
$L(-1,P)$&$2+P$&$\frac{\SO(P+1,1)}{\SO(P+1)}$& $\star$ & $\star$ \\[2mm]
\hline&&&&\\[-3mm]
$L(4m,P,\dot P)$&& $\star$ & $\star$& $\star$ \\[2mm]
$L(q,P)$&&$X(P,q)$&$H(P,q)$&$V(P,q)$\\ 
\end{tabular}
\end{center}
\end{table}
If we use $n$ as the complex dimension of the special K\"{a}hler space, the
dimension of these manifolds is ($\dot P=0$ if $q\neq 4m$)
\begin{equation}
  n=3+q+(P+\dot P){\cal D}_{q+1}\,,\qquad\left\{
  \begin{array}{l}
    \mathop{\rm dim}\nolimits_\mathbb{R}[\mbox{very special real }L(q,P,\dot P)] =n-1 \\
    \mathop{\rm dim}\nolimits_\mathbb{R}[\mbox{special K\"{a}hler }L(q,P,\dot P)]=2n \\
    \mathop{\rm dim}\nolimits_\mathbb{R}[\mbox{quaternionic-K\"{a}hler }L(q,P,\dot P)]=4(n+1). \
  \end{array}\right.
 \label{nformula}
\end{equation}
where ${\cal D}_{q+1}$ is the dimension of the irreducible representation
of the Clifford algebra in $q+1$ dimensions with positive signature, i.e.
\begin{align}
& {\cal D}_{q+1}=1 \ \ \mbox{for }q=-1,0\,,\qquad {\cal D}_{q+1}=2 \ \
\mbox{for }q=1\,,\qquad {\cal D}_{q+1}=4 \ \
\mbox{for }q=2\,,\nonumber\\
&{\cal D}_{q+1}=8 \ \mbox{for }q=3,4\,,\qquad{\cal D}_{q+1}=16 \
\mbox{for }q=5,6,7,8\,,\qquad {\cal D}_{q+8}=16\, {\cal D}_q\,.
 \label{dimcalDq}
\end{align}
The very special manifolds are defined by coefficients ${\cal C}_{IJK}$. For the homogeneous ones, we can write them as
\begin{equation}
{\cal C}_{IJK}\,h^Ih^Jh^K = 3\left\{ h^1\,\big(h^2\big)^2 -h^1\,
\big(h^\mu\big)^2 -h^2\,\big(h^i\big)^2 +\gamma_{\mu ij}\,h^\mu\,
h^i\,h^j \right\}.\label{soldhom}
\end{equation}
We decomposed the indices $I=1,\ldots ,n$ into $I= 1, 2, \mu, i$, with
$\mu=1,\ldots, q+1$ and $i=1,\ldots, (P+\dot P){\cal D}_{q+1}$. Here,
$\gamma_{\mu ij}$ is the $(q+1,0)$ Clifford algebra representation that
we mentioned. Note that these models have predecessors in 6 dimensions,
with $q+1$ tensor multiplets and $(P+\dot P){\cal D}_{q+1}$ vector
multiplets. The gamma matrices are then the corresponding coupling
constants between the vector and tensor multiplets.

Considering further Table~\ref{tbl:homsp}, we find in the
quaternionic spaces the homogeneous ones that were found
in~\cite{Alekseevsky1975}, together with those that were discovered
in~\cite{deWit:1992nm} (the ones with a $\star$ except for the series
$L(0,P,\dot P)$, which were already in~\cite{Alekseevsky1975}, and
denoted there as $W(P,\dot P)$).
A new overview of the properties of these homogeneous manifolds can be found in \cite{Anastasiou:2017nsz}, where these (apart from the pure $D=4$ and pure $D=5$ theories)  are also constructed as double copies of Yang--Mills theories.

Observe that the classification of homogeneous spaces exhibits that the
quaternionic projective spaces have no predecessor in special geometry,
and that the complex projective spaces have no predecessor in very
special real manifolds. Similarly, only those with $q\geq -1$ can be
obtained from 6 dimensions. $L(-1,0)$ corresponds to pure
supergravity in 6 dimensions. In general, the scalars of the tensor multiplets in $D=6$ describe a SO$(1,q+1)/$SO$(q+1)$ manifold.

In the range $q\geq -1$, some of these manifolds are in fact symmetric manifolds. These are collected
in Table~\ref{tbl:symvs}.
\begin{table}[!t]
\caption{\it Symmetric very special manifolds. Note that
the very special real manifolds $L(-1,P)$ are symmetric, but not their
images under the {\bf r} map. The number $n$ is the dimension as in table
\ref{tbl:homsp}.} \label{tbl:symvs}
\begin{center}
\begin{tabular}{|l|r|ccc|}
\hline 
&$n$ &Real & K\"{a}hler & Quaternionic \\
\hline
$L(-1,0)$&2&$\SO(1,1)$&$\left[\frac{\SU(1,1)}{\U(1)}\right]^2$&$\frac{\SO(3,4)}{(
\SU(2)) ^ 3 } $ \\
$L(-1,P)$&$2+P$&$\frac{\SO(P+1,1)}{\SO(P+1)}$& & \\
 $L(0,P)$&$3+P$&$ \SO(1,1)\otimes \frac{\SO(P+1,1)}{\SO(P+1)}$&
$\frac{\SU(1,1)}{\U(1)}\otimes\frac{\SO(P+2,2)}{\SO(P+2)\otimes \SO(2)}$ &
$\frac{\SO(P+4,4)}{\SO(P+4)\otimes \SO(4)} $\\
$L(1,1)$&6& $\frac{{\mathop{\rm
S}}\ell(3,\mathbb{R})}{\SO(3)}$&$\frac{\Sp(6)}{\U(3)
}$&$\frac{F_4}{\USp(6)\otimes \SU(2)}$\\
$L(2,1)$&9& $\frac{{\mathop{\rm
S}}\ell(3,C)}{\SU(3)}$&$\frac{\SU(3,3)}{\SU(3)\otimes
\SU(3)\otimes \U(1)}$&$\frac{E_6}{\SU(6)\otimes \SU(2)}$\\
$L(4,1)$&15& $\frac{\SU^*(6)}{\USp(6)}$&$\frac{\SO^*(12)}{\SU(6)\otimes
\U(1)}$&$\frac{E_7}{\overline{\SO(12)}\otimes \SU(2)}$\\
$L(8,1)$&27& $\frac{E_6}{F_4}$&$\frac{E_7}{E_6\otimes
 \U(1)}$&$\frac{E_8}{E_7\otimes \SU(2)}$\\
\hline
\end{tabular}
\end{center}
\end{table}
For the symmetric special K\"{a}hler spaces, this reproduces the
classification obtained in~\cite{Cremmer:1985hc}. A study of the full set
of isometries could be done systematically in these models. All this has
been summarised in~\cite{deWit:1995tf}.

%% file: ch6_Finalresults.tex
\chapter{Final Results}
\label{ch:finalresults}
\begin{quote}
\textbf{Abstract.}
The previous part of this book first considered conformal couplings, and finally these were projected to a subspace after gauge choices. We saw how then special and quaternionic-K\"{a}hler geometries emerge.
In this chapter we obtain the final action and transformation
laws for $\mathcal{N} =2$, $D=4$ (Sect. \ref{ss:finalresults}) and $D=5$ (Sect. \ref{ss:finalD5}) Poincar\'{e} supergravity coupled to $n_V$ vector multiplets and $n_H$ hypermultiplets
after elimination of auxiliary fields and in terms of the variables of Poincar\'e supergravity.
We end with final remarks indicating also future directions.
\end{quote}

\section{Final \texorpdfstring{$D=4$}{D=4} Poincar\'e Supergravity Results}
\label{ss:finalresults}

This chapter repeats all definitions and is readable by itself for what concerns the final results. For the origin of the equations and more detailed explanations we refer to the previous chapters.
Here we will first collect the already obtained relevant results for the bosonic sector, and then complete it for the fermionic sector. These final results could also have been obtained in other ways. In particular a geometric and rheonomic approach has been used in \cite{Andrianopoli:1997cm}, where the complete results found in this chapter have also been obtained.

\vspace{10pt}
We consider the theory with
\begin{itemize}
  \item Supergravity including the frame field $e_\mu ^a$, the doublet gravitino $\psi_\mu ^i=P_L\psi  _\mu ^i$ and the graviphoton. The description of the graviphoton is unified with the description of the vector  multiplets.
  \item $n=n_V$ vector multiplets with $n_V$ complex scalars $z^\alpha $, $n_V$ physical fermions, described as $\chi ^{\alpha}_i=P_L \chi ^{\alpha}_i$ (and their right-handed components $\chi ^{i\bar{\alpha}}$) and $n_V$ vectors. The unified description of these vectors and the graviphoton is in $W_\mu ^I$, $I=0, \ldots ,
    n_V$, with the field strengths $F_{\mu \nu }^I$
\begin{equation}
F_{\mu\nu}^I = 2 \partial_{[\mu} W_{\nu]}^I + f_{JK}{}^I W_\mu^J W_\nu^K\,.
 \label{FdefD4}
 \end{equation}
The gauge algebra has structure constants $f_{JK}{}^I$. We will say more about the  definition of the physical fermions in Sect. \ref{phys_ferm}. Here we concentrate on the description of the scalars. The latter are described as independent variables in a holomorphic $2(n_V+1)$ component
symplectic vector,\index{symplectic vector}
\begin{equation}
v(z)= \begin{pmatrix}Z^I(z)\cr \ZF_I(z)
\end{pmatrix}\,.
\label{vzresult}
\end{equation}
This symplectic vector satisfies constraints such that the independent physical scalars  are $z^\alpha $ with $\alpha =1,\ldots ,n_V$. The constraint is expressed using a symplectic inner product\index{symplectic inner product} defined by (\ref{innerProdSymplV}):
\begin{equation}
\langle\nabla_\alpha v,\nabla_\beta v\rangle =2\left(
\nabla_{[\alpha} Z^I\right) \left(  \nabla_{\beta]}
\ZF_I\right) =0,\qquad \nabla_\alpha v\equiv
\partial _\alpha v +\kappa^2(\partial _\alpha {\cal K})v,
\label{DavDbv0Result}
\end{equation}
where the K\"{a}hler potential\index{K\"{a}hler potential} ${\cal K}$ is determined by
\begin{equation}
  \rme^{-\kappa^2{\cal K}(z,\bar z)}=-\rmi \kappa^2\langle {v},\,{\bar
  v}\rangle=-\rmi \kappa^2 \left( Z^I\bZF_I-\ZF_I\bar Z^I\right) \,.
\label{KahlervResult}
\end{equation}
This K\"{a}hler potential is related to the metric of the scalar manifold via:
\begin{equation}
  g_{\alpha \bar \beta }=\partial _\alpha \partial _{\bar \beta }{\cal
K}=\rmi\rme^{\kappa^2{\cal K}}\langle{\nabla_\alpha v},{\nabla_{\bar
\beta} \bar v}\rangle .
 \label{metricspKah}
\end{equation}
We require this metric\footnote{There are further cohomological
restrictions concerning the global structure of the metric, i.e. it
should be K\"{a}hler manifolds of restricted type or `Hodge manifolds', but
these global restrictions are not discussed here. See footnote \ref{fn:fermionsHodge} page \pageref{fn:fermionsHodge}.} to be positive
definite in the physical domain of the scalars $z$.

The geometric structure of the manifold is most visible in the fundamental equation (\ref{matrixeqnspecK}):
\begin{equation}
 \rmi \left<
 \begin{pmatrix} \bar V &\nabla _\alpha V\end{pmatrix}, \,\begin{pmatrix}V\\\nabla_{\bar \beta}\bar  V\end{pmatrix}
 \right>
 = G\equiv \begin{pmatrix}\kappa ^{-2}& 0\\ 0&g_{\alpha \bar \beta }\end{pmatrix}\,,
 \label{matrixeqnspecKrepeat}
\end{equation}
where we use the symplectic sections
\begin{equation}
  V(z,\bar z)= y(z,\bar z) v(z)=\begin{pmatrix}X^I\cr F_I\end{pmatrix}
\,,\qquad y= \rme^{\kappa ^2{\cal K}/2}\,.
 \label{defVfromv}
\end{equation}
Demanding positive kinetic terms, the matrix $G$ should be positive definite, and thus the vector
\begin{equation}
  \begin{pmatrix}Z^I& \nabla_{\bar \alpha} \bar Z^I\end{pmatrix}\qquad \mbox{or}\qquad \begin{pmatrix}\bar Z^I& \nabla_{ \alpha} Z^I\end{pmatrix}
 \label{invertiblealways}
\end{equation}
should be invertible (see proof in (\ref{proofinvertible})).
This then allows to define the kinetic matrix for the vectors (see the action below in (\ref{4dtruncated})):
\begin{equation}
{\mathcal N}_{IJ}=  \begin{pmatrix}\ZF_I& \nabla_{\bar \alpha} \bZF_I\end{pmatrix}
\begin{pmatrix}Z^J& \nabla_{\bar \alpha} \bar Z^J\end{pmatrix}^{-1}\,. \label{sympldefcN}
\end{equation}
The positivity properties of $G$ imply that the imaginary part of ${\cal N}_{IJ}$ is
negative definite.

The `usual case' is
when the $(n_V+1)\times (n_V+1)$ matrix
\begin{equation}
  \begin{pmatrix} Z^I & \nabla_\alpha Z^I \end{pmatrix}
 \label{invertiblematrixF}
\end{equation}
is invertible. This can always be obtained by symplectic
transformations (but one might sometimes
prefer not to use such a symplectic basis). In this `usual case' the condition (\ref{DavDbv0}) implies the existence of a
holomorphic function $\ZF(Z)$, homogeneous of second order in $Z$, such that
\begin{equation}
  \ZF_I(z)=\frac{\partial }{\partial Z^I}\ZF(Z(z)).
 \label{FIfromF}
\end{equation}
In the basis where a prepotential $\ZF(Z)$ exists,  (\ref{sympldefcN}) can be expressed as
in \eqref{Ndef}
\begin{align}
&{\cal N}_{IJ}(z,\bar z) =\bZF_{IJ} +\rmi
\frac{N_{IN}N_{JK}\,Z^N Z^K}{N_{LM}\ Z^L Z^M}\,,\nonumber\\
&N_{IJ}\equiv 2\Im \ZF_{IJ}= -\rmi \ZF_{IJ}+\rmi \bZF_{IJ}\,,\qquad \ZF_{IJ}=\partial _I\partial _J \ZF\,,
 \label{NdefResult}
\end{align}
where $\partial _I$ are derivatives w.r.t. $Z^I$. Due to the homogeneity of $F(X)$, we have $\ZF_{IJ}=F_{IJ}$, where the latter is the second derivative of $F(X)$ w.r.t. $X^I$.
\item $n_H$ hypermultiplets with scalars $q^\Xu $, $u=1,\ldots ,4n_H$ and spinors $\zeta ^A=P_L \zeta ^A$ (their right components are $\zeta _{\bar{A}}$) with $A,\bar{A}=1,\ldots ,2n_H$. We introduced the fermions already in (\ref{fermcoordhyperzeta}) and give more details in Sect. \ref{phys_ferm}. The scalar manifold has a quaternionic-K\"{a}hler property, which has also been obtained from a projection in (\ref{hquat}).
    The main ingredients are the   metric $h_{\Xu \Yu } $, the triplet of hypercomplex structures  $\vec{J}_u{}^v$ and a triplet connection $\vec{\omega}_\Xu(q)$. They are related by (\ref{nablaJproj}), which extends the Levi-Civita covariant derivatives $\nabla _\Xu$, defined from the metric $h_{\Xu \Yu } $, to covariant derivatives $\widetilde \nabla_\Xu$ that preserve the complex structures:
\begin{equation}
\widetilde \nabla _\Zu \vec{J}_\Xu{}^\Yu\equiv  \nablaLC _\Zu \vec{J}_\Xu{}^\Yu+2\,\vec{\omega }_\Zu\times \vec{J}_\Xu{}^\Yu=0\,.
\label{covconstquatphys}
\end{equation}
The interactions are determined by the frame fields, $f^{iA}{}_\Xu $, invertible as $4n_H\times 4n_H$ matrices, which satisfy a covariant constancy condition that contains also a $\USp(2n_H)$ connection $\omega_{\Xu\Ap}{}^\Bp$
\begin{equation}
  \widetilde{\nabla }_\Yu f^{i\Ap}{}_\Xu \equiv
   \partial_\Yu f^{i\Ap}{}_\Xu  + f^{j\Ap}{}_\Xu\omega_{\Yu j}{}^i +  f^{i\Bp}{}_\Xu\omega_{\Yu\Bp}{}^\Ap  - \Gamma _{\Yu\Xu}^\Zu f^{i\Ap}{}_\Zu   = 0\,.
\label{fprojcovconst}
\end{equation}
Notice that the connection $\vec{\omega} _\Xu$ is written here as a traceless anti-hermitian $\omega_{\Xu i}{}^j$. We often switch between one and the other notation using (\ref{tripletvecij}).
There are also reality conditions on these objects, for which we refer to (\ref{projecrealf}).
\end{itemize}

Summarizing, Table~\ref{tbl:superPoincD4} gives the physical fields in the Poincar\'{e} theory, multiplets in Poincar\'{e} language, and the corresponding range of indices.
\begin{table}[!t]
\caption{\it Multiplets and fields of the super-Poincar\'{e} theories for $D=4$}\label{tbl:superPoincD4}
\begin{center}
\begin{tabular}
{|l|c|ccc|l|}
\hline
Field & Spin & Pure SG & Vector mult. & Hypermult. & Indices \\
\hline
$e_\mu ^a$  & 2 & 1 &   &   & $\mu ,a=0,\ldots ,3$ \\
$\psi _\mu ^i$ & $\ft32$ &   2    &  & & $i=1,2$  \\
$W_\mu ^I$ & 1 & 1 & $n_V$ &   & $I=0,\ldots ,n_V$ \\
 $\chi ^i_\alpha $, $\zeta ^A$ & $\ft12$ &   & $2n_V$ & $2n_H$ & $A=1,\ldots ,2n_H$ \\
$z^\alpha $, $q^\Xu $ & 0 &   & $2n_V$ & $4n_H$ & $\alpha =1,\ldots ,n_V$; $\Xu =1,\ldots ,4n_H$ \\
\hline
\end{tabular}
\end{center}\end{table}

\newpage 

\subsection{The Bosonic Action}
\label{ss:D4bosonicaction}

We start by summarizing the results for the bosonic sector. The bosonic action is
\begin{eqnarray}
e^{-1}\mathcal{L}_{\rm bos}&=&
\ft12\kappa^{-2} R-g_{\alpha \bar \beta}\hat{\partial}_\mu z^\alpha \hat{\partial}^\mu \bar
z^\beta -\ft1{2}h_{\Xu \Yu }\hat{\partial}_\mu q^{\Xu }\hat{\partial}^\mu q^{\Yu }- V(z,\bar z,q)\nonumber\\
&&
+\ft14(\Im {\cal N}_{IJ})F_{\mu\nu}^I F^{\mu\nu J} -\ft
18 (\Re {\cal N}_{IJ}) e^{-1}\varepsilon^{\mu\nu\rho\sigma}F_{\mu\nu}^I
F_{\rho\sigma}^J
\nonumber\\
&&+ \ft23 C_{I,JK}e^{-1}\varepsilon^{\mu\nu\rho\sigma}W_\mu ^IW_\nu
^J\left( \partial _\rho W_\sigma ^K+\ft38f_{LM}{}^KW_\rho ^LW_\sigma
^M\right).
 \label{4dtruncated}
\end{eqnarray}
The first line starts with the Einstein--Hilbert term, which is the kinetic term for the graviton. It has been normalized to this form by our choice of dilatational gauge (\ref{DDgauge}).
Then appear the kinetic terms for the scalars related to the geometric structure. The main feature visible in these kinetic terms is that the scalar field space is divided in a special K\"{a}hler manifold, parameterized by $\{z^\alpha ,\,\bar z^\alpha \}$ (and was obtained in (\ref{genL01})) and a quaternionic-K\"{a}hler manifold parameterized by $\{q^\Xu\}$, with a metric obtained in (\ref{calcqkmetric}).
The hats on the kinetic terms are related to the gauging, see Sect. \ref{ss:gauging}.
The potential $V(z,\bar z,q)$ will be discussed in Sect. \ref{ss:potentialN2}, since it is determined by this gauging. The second line contains the kinetic terms of the vectors using the matrix (\ref{sympldefcN}) or (\ref{NdefResult}). The last line contains a Chern--Simons term that is also related to the embedding of the gauge group in the symplectic group.

\subsubsection{The Gauging}
\label{ss:gauging}
The scalars have covariant kinetic terms for possible gaugings by the gauge vectors $W_\mu ^I$, with $I=0,\ldots ,n$:
\begin{equation}
   \hat{\partial }_\mu z^\alpha =\partial_\mu z^\alpha -W_\mu {}^Ik_I{}^\alpha\,,\qquad
 \hat{\partial }_\mu q^\Xu=\partial _\mu q^\Xu-W_\mu {}^Ik_I{}^\Xu,\label{hatcovderzq}
\end{equation}
First note that in these covariant derivatives the $\U(1)$, respectively $\SU(2)$, connections are absent, since the physical
scalars $z^\alpha $ do not transform under the $\U(1)$ transformation and $q^\Xu$ do not transform under the $\SU(2)$ (see \eqref{v:basisikeuze_hyp}).

The covariant derivatives are covariant for the gauge transformations, which are for the bosons:
\begin{eqnarray}
 \delta_G(\theta ) W_\mu^I &=& \partial_\mu\theta^I -\theta ^Jf_{JK}{}^IW_\mu^K\,,   \nonumber\\
\delta _G(\theta )z^\alpha &=&\theta ^Ik_I{}^\alpha\,, \nonumber\\
\delta _G(\theta )q^\Xu &=&\theta ^Ik_I{}^\Xu \,.
 \label{gaugetransfod4}
\end{eqnarray}
The Killing vectors $k_I{}^\alpha(z)$ and $k_I{}^\Xu(q)$ separately satisfy the algebra with structure constants $f_{JK}{}^I$. They are projections of the Killing vectors of the embedding manifolds, see  (\ref{KJIkalpha}) and (\ref{KillProjSymm}). They, respectively, preserve the complex and quaternionic structures, which can be seen in that they can be related to a real moment map $P_I^0$ and triplet real moment map $\vec{P}_I$, see  (\ref{killMomentMap}) and (\ref{nabPproj}):
\begin{align}
  g_{\alpha \bar \beta }k_I^{\bar\beta} =& \rmi \partial _\alpha P_I^0\,,\nonumber\\
  h_{uv}k_I^v =& -\ft13 \vec{J}_u{}^v\cdot \widetilde{\nabla }_v\vec{P}_I\,.
\label{tedP}
\end{align}
The gauge transformations do not necessarily leave the K\"{a}hler potential invariant
\begin{equation}
  \delta _G(\theta ) {\cal K}(z,\bar z)= \theta ^I\left(r_I(z)+ \bar r_I(\bar z)\right)\,,\qquad r_I(z)=k_I^\alpha \partial _\alpha {\cal K}+\rmi P_I^0\,.
 \label{delGK}
\end{equation}
Supersymmetry implies that the gauge transformations of the scalars are related to those of the vectors, and thus related to the structure constants.

This can best be seen in the symplectic formulation, where the vectors $V$ transform in the conformal setting with a matrix
\begin{equation}
  T_I
=\begin{pmatrix} f_{IK}{}^J& 0\cr 2C_{I,JK}& -f_{IJ}{}^K\end{pmatrix}
\,,
 \label{TISympl}
\end{equation}
where $C_{I,JK}$ are real coefficients, symmetric in the last two indices and\\ $Z^IZ^JZ^KC_{I,JK}=0$. They appear in the last line of (\ref{4dtruncated}).

Following Sect. \ref{ss:isomsympl} the gauge transformations in the Poincar\'{e} frame and the covariant transformations are of the form
\begin{align}
 \delta^\poinc _G(\theta ) v(z) = \theta ^I(-T_I-\kappa^2 r_I)v(z)\,,\qquad  \widehat{\delta}_G(\theta )v(z) = \theta ^I(-T_I-\kappa^2 P_I^0)v(z)\,.
\label{delGv}
\end{align}
The moment map \index{moment map} can be written as a symplectic inner product
\begin{equation}
  P_I^0=\rme^{\kappa ^2\cal K}\langle T_I v(z),\,\bar v(\bar z) \rangle= -\rme^{\kappa ^2\cal K}\langle  v(z),\, T_I \bar v(\bar z) \rangle\,.
 \label{PI0sympl}
\end{equation}
Other ways to write $P_I^0$ are given in (\ref{kalmomM}).

On the quaternionic side, the moment maps $\vec{P}_I$ are defined from the Killing vectors in (\ref{nabPproj}). We use here $h_{uv}$ to raise and lower indices, and thus the main equations are
\begin{align}
  &\widetilde{\nabla }_u \vec{P}_I \equiv  \partial _u\vec{P}_I +2\vec{\omega }_u\times \vec{P}_I = \vec{J}_{uv}k_I{}^v\,,\nonumber\\
  & k_I{}^\Xu \vec{J}_{\Xu\Yu}k_J{}^\Yu + \kappa ^2 \vec{P}_I\times \vec{P}_J=f_{IJ}{}^K \vec{P}_K\,.
  \label{momentmapvectorsumm}
\end{align}
As we discussed around  (\ref{solnPI}) , the solution for $\vec{P}_I$ is unique when $n_H\neq 0$, and for $n_H=0$ there are two types of FI terms possible: for $\SU(2)$ and for $U(1)$ factors.

From the derivatives of the quaternionic Killing vectors, one can define matrices
\begin{equation} \label{def-tp}
   t_{I\Ap}{}^\Bp\equiv \ft12f^\Yu{}_{i\Ap}\nabla _\Yu k_I{}^\Xu f^{i\Bp}{}_\Xu\,,
\end{equation}
which satisfy the gauge algebra and are relevant for the gauge transformation of the fermions of the hypermultiplets.

\subsubsection{The Potential}
\label{ss:potentialN2}
The remaining bosonic part of the action (\ref{4dtruncated}) is the scalar potential. It originates in the conformal setting from the third line in (\ref{totaleLagN2D4}). We can use the bosonic part of the field equations for $\vec{Y}$ \eqref{YTeom} to obtain \index{scalar potential}
\begin{equation}
  V= N^{-1|IJ}\vec{P}_I\cdot \vec{P}_J +N^{-1|IJ}P_I^0P_J^0
   +2  \bar X^I X^J k_I{}^{X} k_J{}^{Y}g_{XY}\,.
\label{VhomogN2}
\end{equation}
We rewrite this now in the Poincar\'{e} frame. For the first two terms we use (\ref{InverseImN}) and (\ref{XP0}). For the last term, we use the decomposition of the metric \eqref{v:hypermetriek} and Killing vector (\ref{KillProjSymm}) with the conditions \eqref{v:q^0constant},  (\ref{kIalphakalpha}), (\ref{innerprodveck}):
\begin{equation}
  k_I{}^{X} k_J{}^{Y}g_{XY}=k_I{}^\Xu k_J{}^\Yu h_{\Xu\Yu}-\kappa ^2\vec{P}_I\cdot \vec{P}_J\,.
 \label{Vkkg}
\end{equation}
We obtain
\begin{tcolorbox}
\begin{equation}
  V=-4\kappa ^2X^I\bar X^J \vec{P}_I\cdot \vec{P}_J  -\ft12(\Im{{\cal N}})^{-1|IJ}\left[\vec{P}_I\cdot \vec{P}_J+P_I^0P_J^0\right]
  +2\bar X^I X^Jk_I{}^\Xu k_J{}^\Yu h_{\Xu\Yu}.
\label{VN2construct}
\end{equation}
\end{tcolorbox}
Since $\Im{{\cal N}}$ is negative definite, only the first term gives a negative contribution to the potential. If there are no physical hypermultiplets ($n_H=0$) this term can still be there for constant vectors ${\vec P}_I$, which are then the `Fayet--Iliopoulos terms'.\index{Fayet--Iliopoulos term}

Since the potential is important for applications, it is worthwhile to give alternative expressions. Therefore, let us first introduce \cite{Ceresole:1995ca,Andrianopoli:1996vr,Andrianopoli:1997cm}
\begin{equation}
  {U}^{IJ} \equiv g^{\alpha \bar \beta}\nabla_\alpha X^I
  \nabla_{\bar \beta} \bar X^J
=  -\ft{1}{2} \left( \mbox{Im}\mathcal{N}\right) ^{-1 \vert IJ}
-\kappa ^2 \overline{X}^I \, X^J. \label{formulaU}
\end{equation}
and using (\ref{XP0}), we can write \index{scalar potential}
\begin{equation}
V=-3\kappa ^2\bar X^I X^J \vec{P}_I \cdot \vec{P}_J + U^{IJ}(\vec{P}_I \cdot \vec{P}_J + P_I^0P_J^0)
+2 h_{\Xu \Yu }k^\Xu _I k^\Yu _J \bar X^I X^J\,.
\label{Vd4total}
\end{equation}
Only the first term is negative. After we have discussed the supersymmetry transformations, see (\ref{WardidVN2}), we will understand the three terms in this expression as squares of supersymmetry transformations of, respectively, the gravitino, the gauginos
(split in the $\SU(2)$ triplet and $\SU(2)$ singlet part) and the hyperinos. The fact that the scalar potential can be written in such a way is a general feature in supergravity (see, e.g., \cite{D'Auria:2001kv}).

Finally, let us still give alternative expressions for the part
\begin{equation}
  V_1= U^{IJ}P_I^0P_J^0=g^{\alpha \bar \beta}\nabla_\alpha X^I
  \nabla_{\bar \beta} \bar X^J P_I^0P_J^0 =
  g_{\alpha \bar \beta}k_I{}^\alpha k_J{}^{\bar \beta }\,,
\end{equation}
where for the last expression we used (\ref{Xkalpha}).
Another way uses (\ref{kgbark}) on the last expression (remember that for gauged symmetries ${\cal P}_\Lambda $ is $P_I^0$), leading to
\begin{equation}
  V_1= \rmi \bar{X}^I X^J\langle T_I  V,T_J \bar V \rangle\,.
 \label{V1intprod}
\end{equation}
Using the explicit form of the matrices in (\ref{defTKV}) we get
 \begin{align}
  V_1=&-\rmi \bar X^I X^J f_{IJ}{}^K
 X^L \left(  f_{LK}{}^M\bar F_M+2C_{L,KM}\bar X^M\right)  +\hc
 \label{V1alternative}
\end{align}

 \bexer
As a simple example, consider the special K\"{a}hler manifold that we have
discussed in example \ref{ex:Feta}, which has $n=1$, i.e. 2 vectors. The
algebra should be abelian to leave $F$ invariant. We consider no physical
hypermultiplet, i.e. in the superconformal setup there is just a trivial
one that is needed for compensation. Then the potential can only
originate in the gauging of the compensating hypermultiplet, which is
equivalent to having constant moment maps $\vec{P}_I$. In the
equivariance condition, second of (\ref{momentmapvectorsumm}), only the second term survives, which says that $\vec{P}_1$ and $\vec{P}_2$
should point in the same direction in $\SU(2)$ space. Hence as in (\ref{U1FI}) we take
\begin{equation}
  \vec{P}_I = \xi _I \vec{e}\,, \qquad  \vec e\cdot \vec e = 1\,.
 \label{PIgI}
\end{equation}
To calculate the potential we first obtain $U^{IJ}$, which can most
easily be determined using $\nabla_z Z^I$. This gives, using
(\ref{metricFeta}),
\begin{align}
 \nabla_zZ^I = & \partial _z Z^I + \kappa ^2 Z^I\partial _z{\cal K}=\frac{1}{1-z\bar z}
 \begin{pmatrix}\bar z\cr 1\end{pmatrix}\,,\qquad \nabla_zX^I= \rme^{\kappa ^2{\cal K}/2}\nabla_zZ^I \,,\nonumber\\
 \rme^{\kappa ^2{\cal K}}=& \frac{1}{1-z\bar z}\,, \qquad g^{z\bar z}= \kappa ^2(1-z\bar z)^2\,,\nonumber\\
 U^{IJ}  = & \frac{\kappa ^2}{1-z\bar z}
 \begin{pmatrix}z\bar z&\bar z\cr z &1\end{pmatrix}\,,\qquad
 \bar X^IX^J=\frac{1}{1-z\bar z} \begin{pmatrix}1&\bar z\cr z &z\bar z\end{pmatrix}\,.
\label{calcVFeta}
\end{align}
Thus the potential, originating from the first two terms of (\ref{Vd4total}) leads to
\begin{equation}
   V=\frac{\kappa ^2}{1-z\bar z}\left[ \xi _0^2(z\bar z-3)-2\xi _0\xi _1(z+\bar
 z)+\xi _1^2(1-3z\bar z)\right] \,.
 \label{Vexample}
\end{equation}

Before the gauging the model has an $\SU(1,1)$ rigid symmetry. The
properties of the potential depend on the $\SU(1,1)$-invariant
$\xi _0^2-\xi _1^2$. There are thus 3 relevant cases, whether this invariant is
positive, negative or zero \cite{deWit:1984pk}. In each case, we can take a standard choice by reparameterization, and we find, respectively, the
following extrema
\begin{itemize}
  \item Take $\xi _0=\xi $ and $\xi _1=0$. There is an extremum at $z=0$ with
  negative $V(z=0)$, i.e. anti-de Sitter. $\xi _IU^{IJ}\xi _J$, which
  is the contribution from the supersymmetry of the gaugini, vanishes. Therefore
  this vacuum preserves supersymmetry.
  \item $\xi _0=\xi _1$. In this case there is no extremum in the positivity
  domain $|z|<1$.
  \item $\xi _0=0$ and $\xi _1=\xi $. There is an extremum with positive $V$, i.e.
  de Sitter, with non-vanishing $\xi _IU^{IJ}\xi _J$, i.e. (spontaneously) broken supersymmetry.
\end{itemize}
Note that in the first case we can omit the scalars, and this case thus occurs in
pure ${\cal N}=2$ supergravity with possible gauging leading to the
anti-de Sitter ${\cal N}=2$ supergravity.
 \eexer

\subsection{Physical Fermions}\label{phys_ferm}
In the bosonic sector we have defined appropriate coordinates for implementing the conditions  (\ref{Dgaugegeneral}). We saw in the vector multiplet sector that it was often useful to use the invertible matrix (\ref{invertiblealways}) to split equations between the compensating part and physical part, as is most clear from (\ref{matrixeqnspecKrepeat}). For usual symplectic bases where a prepotential is defined, we can simplify it to (\ref{invertiblematrixF}). We will use such a basis in this part, since the tensor calculus on which we base ourselves is developed in this setting.\footnote{For a more general formulations where the fermions are also in a symplectic vectors, see \cite{Claus:1999np}.} For the hypermultiplets we defined an appropriate basis (\ref{v:basisikeuze_hyp}) where the first quaternion $\{q^0,q^r\}$ is put to a constant using the conformal gauge conditions, and the physical fields are $q^\Xu$.

Similarly, we will now split the fermions in the parts that are put equal to zero by (\ref{v:zeta_i02}) and the physical fermions \index{physical fermions} of the matter-coupled Poincar\'e supergravity theory.

\subsubsection{Fermions of the \texorpdfstring{${\cal N}=2$}{N=2} Vector Multiplets.}\label{FermionsVector}
For the fermions of the conformal vector multiplet we consider the following splitting in 1 + $n$ doublets:
\begin{equation}
\Omega_i^I=\chi_i^0 X^I+\chi_i^\alpha \nabla_\alpha X^I\,,\qquad \Omega^{iI}=\chi^{i0} \bar{X}^I+\chi^{i\bar{\alpha}} \nabla_{\bar{\alpha}} \bar{X}^I\,.
\label{deffermionsPoinc}
\end{equation}
Since the first condition in (\ref{Dgaugegeneral}) leads to (\ref{Nmatrixeqn}), the first of \eqref{v:zeta_i02} imposes
\begin{equation}
  \chi_i^{0}=0\,,\qquad \rightarrow\qquad \Omega_i^I=\chi_i^\alpha \nabla_\alpha X^I\,.
 \label{SgaugefixedOmegachi}
\end{equation}
The inverse can be obtained using (\ref{Nmatrixeqn}) or (\ref{MatrixPositiveProduct}):
\begin{equation}
  \chi _i^\alpha  = g^{\alpha \bar \beta }\nabla _{\bar \beta }\bar X^I N_{IJ}\Omega _i^J= -2 g^{\alpha \bar \beta }\nabla _{\bar \beta }\bar X^I \Im{\cal N}_{IJ}\Omega _i^J\,.
 \label{chifromOmega}
\end{equation}
Thus $\chi ^\alpha _i$ will be the $n$ physical spin 1/2 fields, and their conjugates are defined as
\begin{equation}
  \Omega^{iI}=\chi^{i\bar \alpha} \bar \nabla_{\bar \alpha}\bar  X^I\,.
 \label{Omegachibar}
\end{equation}

In the conformal setting the transformations of $X^I$ are
\begin{equation}
  \delta X^I = \ft12 \bar \epsilon ^i\Omega_i ^I+ \left(\lambda _{\rm D}+\rmi \lambda _T\right)X^I+\theta ^Kk_K{}^I\,.
 \label{delXIconform}
\end{equation}
If we use  (\ref{SgaugefixedOmegachi}) and  (\ref{KJIkalpha}) for the gauged isometries, we can write this as
\begin{equation}
 \delta X^I = \ft12 \bar \epsilon ^i \chi_i^\alpha \nabla_\alpha X^I + \left(\lambda _{\rm D}+\rmi \lambda _T\right)X^I + \theta ^K\left(
k_K{}^\alpha \nabla_\alpha X^I+\rmi \kappa ^2{P}_K^0 X^I\right)\,.
 \label{sconfsymmX}
\end{equation}

In the Poincar\'{e} theory, only supersymmetry and the gauge symmetry survive, and the other parameters $\lambda _{\rm D}$ and $\lambda _T$ become functions of $\epsilon $ and $\theta ^K$ by the decomposition laws.

On the other hand, after these gauge transformations  $X^I$ is only function of $z^\alpha $ and $\bar z^{\bar \alpha }$, and thus these transformations should also be obtainable as
\begin{align}
  \delta ^\poinc X^I =& \delta z^\alpha \partial _\alpha X^I + \delta \bar z^{\bar \alpha }\partial _{\bar \alpha} X^I\nonumber\\
  =&\delta z^\alpha \left(\nabla  _\alpha  -\ft12\kappa ^2 \partial _\alpha {\cal K}\right) X^I + \ft12\kappa ^2 \delta \bar z^{\bar \alpha }\partial_{\bar \alpha}{\cal K} X^I\,,
 \label{delpoincXI}
\end{align}
where we used  (\ref{nablalphaZ}) and the definition of the covariant derivatives in that section.
Identifying this with (\ref{sconfsymmX}) we find
\begin{align}
\delta(\epsilon,\theta ) z^\alpha =& \ft12 \bar \epsilon^i\chi_i^\alpha+\theta ^Kk_K{}^\alpha \,,\nonumber\\
 \lambda _{\rm D}(\epsilon,\theta )=&0\,,\qquad \lambda _T(\epsilon,\theta )=
\ft12\rmi\kappa ^2(\delta z^\alpha  \partial _\alpha {\cal K}- \delta \bar z^{\bar \alpha }\partial_{\bar \alpha}{\cal K})
-\kappa ^2\theta ^K P_K^0\,.
\label{delzdecompcov}
\end{align}
Thus our parameterization  (\ref{deffermionsPoinc}) is convenient for defining the physical fermions $\chi_i^\alpha$ as supersymmetry partners of the $z^\alpha $.

On the other hand, when we use covariant transformations (\ref{hatdeltaf}), the $\U(1)$ compensation is much simpler. Indeed, the covariant transformation is
\begin{equation}
  \hat{\delta}X^I= \nabla  _\alpha X^I\, \delta z^\alpha\,,
 \label{hatdeltaX}
\end{equation}
which is then identified with (\ref{sconfsymmX}) by the same $z^\alpha $ transformation and
\begin{equation}
  \lambda _{\rm D}(\epsilon,\theta )=0\,,\qquad \lambda _T(\epsilon,\theta )=-\kappa ^2\theta ^K P_K^0\,.
 \label{covdecompU1}
\end{equation}
One can also check that the difference between (\ref{hatdeltaX}) and (\ref{delpoincXI}) is in agreement with (\ref{deltaf}).

\subsubsection{Fermions of the \texorpdfstring{${\cal N}=2$}{N=2} Hypermultiplets.}\label{FermionsHyper}
For the hypermultiplet we use the coordinates introduced in Sect. \ref{coord_quatman}:
\begin{equation}
  \{\zeta^{{\cal A}}\} = \{\zeta^{i},\zeta^A\}\,.
  \label{fermcoordhyperzetasec7}
\end{equation}
In these coordinates the second part of (\ref{v:zeta_i02}) with (\ref{AiAcomponents}) reduces to
\begin{equation}
  \zeta^i = 0\,,
 \end{equation}
and the physical fermions \index{physical fermions} are $\zeta ^A$. The reduction of (\ref{transfoN4conf}), including (\ref{gaugetransfod4}), is
\begin{equation}
  \delta q^u = -\rmi \bar{\epsilon}^i\zeta^A f^u{}_{iA }
+\rmi \varepsilon ^{ij}\rho ^{\bAp \Bp } \bar{\epsilon}_i\zeta_\bAp f^u{}_{j\Bp }+ \theta ^I k_I{}^u\,.
 \label{delquproj}
\end{equation}

The Poincar\'{e} supersymmetry transformations have a contribution from the $\SU(2)$ symmetry as in (\ref{veclambdagf}):
\begin{equation}
  \vec{\lambda }(\epsilon ,\theta )= -\vec{\omega }_u\delta q^\Xu -\ft12\kappa ^2 \theta ^I\vec{P}_I \,.
 \label{veclambdagfrepeat}
\end{equation}
One may notice the similarity with (\ref{delzdecompcov}).
 However, with the coordinate choice (\ref{v:basiskeuze_kX}), these transformations  (\ref{delDSU2q}) do not act on the physical scalars $q^u$.

For objects that transform under the $\SU(2)$, we can define covariant transformations that contain the geometric part of the gauge field (\ref{Vmubos_nongauge}), which then, similarly to the case for the $T$-transformations above, reduces the remaining compensation of $\SU(2)$ transformations to a gauge transformation:
\begin{equation}
  \vec{\lambda }(\epsilon ,\theta )= -\ft12\kappa ^2 \theta ^I\vec{P}_I \,,
 \label{veclambdaCovariant}
\end{equation}
similar to (\ref{covdecompU1}).

\subsection{The Fermionic Part of the \texorpdfstring{Poincar\'{e}}{Poincare} Action}
All this allows us to obtain the action for the physical fields. In the language of conformal fields, this action was given in (\ref{totaleLagN2D4}). The first four lines gave rise to the bosonic action (\ref{4dtruncated}), and the gravitino kinetic term
\begin{align}
  {\cal L}_{\rm kin,gravitino}=& -e\bar\psi_{i\mu } \gamma^{\mu \nu \rho }D_\nu  \psi_\rho ^i\,,\nonumber\\
  &D_\mu \psi _{\nu i}=\left( \partial _\mu +\ft14\omega _\mu
{}^{ab}(e)\gamma _{ab}+\ft12\rmi {\cal A}_\mu \right)\psi _{\nu i}
 +{\cal V}_{\mu i}{}^j\psi _{\nu j},.
 \label{Lgravitino}
\end{align}
Here and below appear the (bosonic part of the) effective $\U(1)$ and $\SU(2)$ composite gauge fields:
\begin{align}
{\cal A}_\mu =& \ft12\kappa ^{2}\rmi\left[(\partial  _\alpha{\cal K}) \partial _\mu z^\alpha
-(\partial  _{\bar \alpha }{\cal K})\partial _\mu \bar z^{\bar \alpha }\right]-\kappa ^{2}W_\mu ^I P^0_I\nonumber\\
=& \ft12\kappa ^{2}\rmi\left[(\partial  _\alpha{\cal K}) \hat{\partial} _\mu z^\alpha
-(\partial  _{\bar \alpha }{\cal K})\hat{\partial}_\mu \bar z^{\bar \alpha }\right]+\ft12\rmi W_\mu  ^I (r_I-\bar r_I),\nonumber\\
 \vec{{\cal V}} _\mu=& -\vec{\omega }_\Xu\partial _\mu  q^\Xu -\ft12\kappa ^2 W_\mu  ^I \vec{P}_I\nonumber\\
 =&  -\vec{\omega }_\Xu\hat{\partial} _\mu  q^\Xu +\ft12 W_\mu  ^I \vec{r}_I\,,
\label{valueAVmu}
\end{align}
where $\vec{\omega }_\Xu$, which is in the conformal variables $-\ft12\kappa ^2\vec{k}_u$, is in the Poincar\'{e} variables the connection defined by  (\ref{covconstquatphys}).

Moreover these first four lines of (\ref{totaleLagN2D4}) contain auxiliary fields, whose expressions in the conformal terminology have been given in  (\ref{YTeom}) and  (\ref{valueAVN2}) and contain also fermionic terms.

To translate to Poincar\'{e} language, it is first useful to use the expression of $D_\mu X^I$ obtained in (\ref{nablaXshort})
\begin{align}
  D_\mu X^I =& \nabla _\alpha X^I\, \widehat{\partial }_\mu z^\alpha \,,
\label{DmuXinnabdz}
\end{align}
with $\widehat{\partial }_\mu z^\alpha$ defined in (\ref{hatcovderzq}).
It is then convenient to repeatedly use  (\ref{Nmatrixeqn}), which then e.g. shows that the first term on the fifth line of (\ref{totaleLagN2D4}) vanishes. For the second term, we note that the bosonic part of (\ref{coveomD4AV}) is $D_\mu q^X\vec{k}_X$, which thus vanishes when we use field equations.

The sixth line of (\ref{totaleLagN2D4}) contains the kinetic terms of the gauginos and hyperinos. We obtain for the covariant derivatives of the gauginos
\begin{align}
  \widehat{D}_\mu \Omega _i^I =& (\widehat{D}_\mu \chi _i^\alpha )\nabla _\alpha X^I + \chi _i^\alpha \kappa ^2 g_{\alpha \bar \beta } X^I \hat\partial _\mu \bar z^{\bar \beta }\,,\label{DmuOmegachi} \\
  \widehat{D}_\mu \chi _i^\alpha =&  \left( \partial _\mu +\ft14\omega _\mu
{}^{ab}(e)\gamma _{ab}+\ft12\rmi {\cal A}_\mu \right)\chi  _i^\alpha
 +{\cal V}_{\mu i}{}^j\chi  _j^\alpha
-W_\mu ^I\chi _i^\beta \nabla    _\beta k_I{}^\alpha \nonumber\\
&+ \Gamma ^\alpha _{\beta \gamma } \chi_i^\gamma \partial _\mu z^\beta\,.
 \label{hatDmuchi}
\end{align}
Let us take some time to explain this in more detail, since this is also relevant to understand the gauge transformations of the $\chi ^i_\alpha$ in Sect. \ref{ss:remaindersusy}.
The covariant derivative that is used here is given in (\ref{covderafterfieldeqn}). Then  (\ref{SgaugefixedOmegachi}) should be used. The chiral weight of $\Omega _i^I$ is 1/2. The chiral weight of $\nabla _\alpha X^I$ is the same of that of $X^I$, i.e. +1. Thus the chiral weight of $\chi _i^\alpha $ is $-1/2$. We will have to care then also for the difference between the term $-\ft12\rmi{\cal A}_\mu \Omega ^i_I$ in ${D}_\mu \Omega _i^I$ and the one included in the term with $D_\mu \chi _i^\alpha $, i.e. this remaining term is
\begin{equation}
  ({D}_\mu \Omega _i^I - {D}_\mu \chi _i^\alpha )\nabla _\alpha X^I = \ldots -\rmi {\cal A}_\mu \chi  _i^\alpha\nabla _\alpha X^I\,.
 \label{missingAconnection}
\end{equation}
For the geometric part, we write
\begin{align}
 \widehat{D}_\mu \Omega _i^I = &  \widehat{D}_\mu\left(\chi _i^\alpha\nabla _\alpha X^I\right)=\left(\widehat{D}_\mu\chi _i^\alpha\right)\nabla _\alpha X^I +\chi _i^\alpha\widehat{\nabla} _\beta \nabla _\alpha X^I \partial _\mu z^\beta + \chi _i^\alpha\widehat{\nabla} _{\bar \beta} \nabla _\alpha X^I \partial _\mu \bar z^{\bar \beta} \,,
\label{splitDmuOmega}
\end{align}
and with (\ref{widehatnabX}) this leads to 
 (\ref{DmuOmegachi})
(and the covariant derivative  $\widehat{\nabla} _\beta \nabla _\alpha X^I$ takes into account  (\ref{missingAconnection}) for what concerns the non-gauge part). The remaining question concerns the gauge part: how it reduces to the $W_\mu ^I$ terms mentioned in 
 (\ref{hatDmuchi}).
For that, we see that the explicit $W_\mu ^I$ terms in (\ref{covderafterfieldeqn}) lead to the terms
\begin{equation}
  \widehat{D}_\mu \Omega _i^I =\ldots  + \chi _i^\alpha \widehat{\nabla} _\alpha (W_\mu ^JX^Kf_{JK}{}^I)\,.
 \label{gaugepartDOmega0}
\end{equation}
Note that $W_\mu ^J$ appears here inside the bracket in order that the covariantization is respected. In this equation we then also treat the index $J$ as holomorphic. The interpretation is similar to the discussion after  (\ref{ferm-equiv}): we treat the $W_\mu ^J$ as independent of the $z^\alpha $, but then $\widehat{\nabla} _\alpha W_\mu ^J = \Gamma ^J_{KL}W_\mu ^L\partial _\alpha X^K$. Then we insert
(\ref{KJIkalpha}) and obtain
\begin{align}
 \widehat{D}_\mu \Omega _i^I =&\ldots  - \chi _i^\alpha \widehat{\nabla} _\alpha (W_\mu ^Jk_{J}{}^I)\nonumber\\
 =&\ldots  - \chi _i^\alpha \nabla _\alpha (W_\mu ^Jk_J{}^\beta) \nabla _\beta X^I-\rmi \kappa ^2\chi _i^\alpha \nabla _\alpha(W_\mu ^JP_J^0) X^I -\chi _i^\alpha W_\mu ^JP_J^0\nabla _\alpha X^I \,,
\label{gaugepartDOmega}
\end{align}
using once more  (\ref{widehatnabX}). Note that in the last line we removed the hats on the covariant derivatives either because the $J$ index is anyway contracted or the fact  that $\widehat{\nabla} _\alpha X^I= \nabla _\alpha X^I$. This then allows to extract the $W_\mu ^J$ out of the covariant derivative.  In the last term we use that the gauge part of ${\cal A}_\mu $ is $-\kappa ^2W_\mu ^JP_J^0$ and this term thus cancels the gauge part in (\ref{missingAconnection}). The second term can be identified with the gauge term included in $\hat\partial _\mu \bar z^{\bar \beta }$ in (\ref{DmuOmegachi}) using  (\ref{killMomentMap}). This finally leaves the explicit $W_\mu ^I$ term in (\ref{hatDmuchi}).

Using (\ref{Nmatrixeqn}), the parts proportional to $X^I$ in $\widehat{D}_\mu \Omega _i^I$ do not contribute to the kinetic terms, which reduce to
\begin{equation}
  e^{-1}{\cal L}_{\rm kin, matter ferm}= -\,\ft14g_{\alpha \bar \beta }\bar{\chi }_i^{\alpha }\widehat{\slashed{D}}\chi ^{i\bar \beta }
 -\bar\zeta_\bAp\hat{\slashed{D}} \zeta^\Bp d^\bAp{}_\Bp+\hc\,,
 \label{Lkinfermmatter}
\end{equation}
where our task at the hypermultiplets side was simple since the gauge choices just restrict $\zeta ^\Ae$ to $\zeta^\Ap$.
The covariant derivative is
\begin{eqnarray}
 \hat{D}_\mu  \zeta^\Ap&=&\left( \partial _\mu +\ft14\omega _\mu
{}^{ab}(e)\gamma _{ab}+\ft12\rmi {\cal A}_\mu \right)\zeta^\Ap-W_\mu
^It_{I\Bp }{}^\Ap \zeta ^\Bp +\partial _\mu q^\Xu \omega _{\Xu\Bp }{}^\Ap \zeta ^\Bp\,,\nonumber\\
\label{DmufermN2}
\end{eqnarray}
where $\omega_{\Xu\Bp }{}^\Ap $  appears in (\ref{fprojcovconst}).
Note that since we now are restricted to the physical fermions we could replace $d^\bAe{}_\Be$ by $\delta^\Ae{}_\Be $, thus reducing also $\zeta ^\bAe$ to $\zeta_\Ap$. We will keep here the covariant way of writing.

For the terms bilinear in fermions and derivatives of bosons, which are Noether terms (seventh line), we use  (\ref{DmuXinnabdz}) for the gaugino part, and for the hyperini the restriction of $\zeta ^\Ae$ to $\zeta^\Ap$ implies that we can use
\begin{equation}
 D_\mu q^X f^{iA}{}_X =D_\mu q^u f^{iA}{}_u = \widehat{\partial }_\mu q^u f^{iA}{}_u\,,\qquad \hat\partial _\mu q^u= \partial _\mu q^u-W_\mu{} {}^Ik_I^u\,.
  \label{Dmuqu}
\end{equation}
Therefore, these Noether terms \index{Noether terms} are
\begin{equation}
 e^{-1} {\cal L}_{\rm Noether}=\bar \psi _{i\mu}\left[\ft12g_{\alpha \bar \beta } (\widehat{\slashed{\partial }} z^{\alpha} )\gamma ^\mu \chi ^{i\bar \beta
  }+\rmi (\widehat{\slashed{\partial
 } }q^\Xu )\gamma ^\mu f^{iA}{}_{\Xu} \zeta _A\right]+\hc
 \label{Noether}
\end{equation}

For the coupling of the gauge field strengths to the fermions, we first have to take into account that the graviphoton
field strength (last line of \eqref{YTeom}) contains also a fermionic part, which is thus included in the $TT$ term in the second line of (\ref{totaleLagN2D4}). We can rewrite the expression for $T_{ab}$ using the special geometry relations \eqref{ImcNX} and \eqref{defC}  as follows:
\begin{align}
   T^-_{ab} =-\kappa ^2\Im{\cal N}_{IJ}X^J&\left[ 4\widehat F^{-I}_{ab} + \ft12  C_{\alpha \beta \gamma }g^{\gamma \bar \delta }{\nabla }_{\bar \delta }\bar X^I
 \bar \chi _i^\alpha \gamma _{ab}\chi _j^\beta \varepsilon ^{ij}\right.\nonumber\\
&\left.
 +2\kappa ^2 X^I\bar\zeta^\Ap\gamma_{ab}\zeta^{\Bp} C_{\Ap\Bp}\right]\,.
\label{Tfermions}
\end{align}
The square of this expression contains 2-fermion terms, which combine with the terms on the eighth line in (\ref{totaleLagN2D4}) to the so called `Pauli terms'\footnote{In the expression of $Q^{ab\,-J} $ we did not explicitly write that one should take the anti-self-dual combination in $[ab]$ since anyway this is multiplied by an anti-self-dual field strength in the Pauli terms.}\index{Pauli term}
\begin{eqnarray}
 e^{-1}{\cal L}_{\rm Pauli} & =  & F_{ab}^{-I}\Im{\cal N}_{IJ}Q^{ab\,-J}+\hc,\nonumber\\
Q^{ab\,-J} &\equiv & {\nabla }_{\bar \alpha  }\bar X^J\left(
\ft18g^{\beta  \bar \alpha } C_{ \beta \gamma\delta  }
\bar \chi _i^\gamma \gamma ^{ab}\chi _j^\delta  \varepsilon ^{ij}+\bar \chi^{\bar \alpha i}\gamma ^a\psi ^{bj}\varepsilon _{ij}\right)\nonumber\\
&&+\,X^J\left( \bar \psi ^a_i\psi^b_j\varepsilon ^{ij}+\ft12\kappa ^2\bar
\zeta ^\Ap\gamma ^{ab}\zeta ^\Bp C_{\Ap\Bp}\right) \,.
\label{LPauliN2construct}
\end{eqnarray}
Notice that the above expression contains the coupling between fermions and gauge field strengths. The precise form of this term is governed by symplectic geometry. In particular, by similarity with \eqref{Nsugra} we can identify
   \begin{equation}
   \begin{pmatrix}Q^{ab-J}\cr \mathcal{N}_{IJ}Q^{ab-J}
   \end{pmatrix},
   \label{Qzresult}
   \end{equation}
   as a symplectic vector \cite{Ceresole:1995jg}. The Pauli term is then $\rmi/2$ times the symplectic inner product of this with the symplectic vector of $F_{ab}^{-I}$ and $G_{ab\,I}^{-}$  defined in Sect. \ref{ss:dualities}.

The ninth line in (\ref{totaleLagN2D4}) are Goldstino \index{Goldstino} terms:
\begin{align}
 e^{-1} {\cal L}_{\rm goldstino}=& -(\bar \psi _i\cdot \gamma \goldstino^i +\hc)\,,\qquad \goldstino^i=\ft12 W_\alpha {}^{ij}\chi_j^\alpha +2 N^i{}_\Ap \zeta^\Ap\,,\nonumber\\
&W_\alpha {}^{ij}\equiv
 \left( \rmi\varepsilon ^{ij}P_I^0- P_I{}^{ij}\right)\nabla _{ \alpha }X^I= -\varepsilon ^{ij}g_{\alpha \bar \beta }k_I{}^{\bar \beta }X^I -
P_I{}^{ij}\nabla _{ \alpha }X^I,\nonumber\\
& N^i{}_\Ap\equiv  \rmi X^Ik_I{}^\Xu f^{i\Bp}{}_{\Xu}C_{\Bp\Ap}\,,
 \label{Lgoldstino}
\end{align}
where we used (\ref{kalmomM}).

The remaining terms in (\ref{totaleLagN2D4}) are in the last two lines, and fermionic terms originating from the auxiliary field term $-N_{IJ}\vec{Y}^I\cdot \vec{Y}^J$, using the expression in (\ref{YTeom}).
The latter we can write using (\ref{InverseImN}) and (\ref{CFIJK}) as
\begin{equation}
  Y^I_{ij} =- N^{-1|IJ}P_{Jij}-\ft14\left[\bar \chi _i^\alpha \chi _j^\beta  C_{\alpha \beta \gamma }g^{\gamma \bar \gamma }\nabla _{\bar \gamma }\bar X^I +
  \bar \chi ^{i\bar \alpha} \chi ^{j\bar \beta} \bar  C_{\bar \alpha \bar \beta \bar \gamma }g^{\gamma \bar \gamma }\nabla _{\gamma } X^I\right]\,.
  \label{Yeomrewritten}
\end{equation}
These are thus fermion mass terms, which we can parameterize as
\begin{align}
 {\cal L}_m  =&\ft12S_{ij}\bar \psi _\mu ^i\gamma ^{\mu \nu }\psi^j _\nu-\ft12m^{ij}{}_{\alpha \beta }\bar \chi _i^\alpha \chi _j^\beta
   -m^i_{ \alpha \Ap}\bar \chi _{i}^\alpha\zeta ^\Ap
   -\ft12m_{\Ap\Bp}\bar\zeta ^\Ap\zeta ^\Bp+\hc
 \label{LmN2}
\end{align}
All these expressions  are determined by the gauging. We obtain
\begin{eqnarray}
 S_{ij}&=&P_{Iij}\bar X^I,\nonumber\\
 m^{ij}{}_{\alpha \beta } & = & \ft12 P_{I}^{ij} C_{\alpha \beta \gamma }g^{\gamma \bar \delta }{\nabla }_{\bar \delta }\bar X^I
 + \varepsilon ^{ij}\nabla _\alpha X^I k_I{}^{\bar \gamma }g_{\beta \bar \gamma }\,,\nonumber\\
m_{i\bar \alpha}^\bAp & = & 2 \rmi k_I^\Xu \varepsilon _{ij} f^{j\Ap}{}_\Xu 
{\nabla }_{\bar \alpha }\bar X^I\,,\nonumber\\
m^i_{ \alpha \Ap} & = &- 2 \rmi k_I^\Xu  f^{i\Bp}{}_\Xu C_{\Bp\Ap}\nabla _\alpha X^I\,,\nonumber\\
m_{\Ap\Bp}&=&-4 X^It_{I\Ap\Bp}\,.
\label{mMatrN2}
\end{eqnarray}
For the $\chi \chi $ mass term we used  (\ref{KJIkalpha}).

\subsection{Total Action}

We have rewritten the action in Poincar\'{e} variables. The result is
\begin{align}
\mathcal{L}=& \mathcal{L}_{\rm bos} &  (\ref{4dtruncated})\nonumber\\
&+ {\cal L}_{\rm kin,gravitino}  & (\ref{Lgravitino})\nonumber\\
&+ {\cal L}_{\rm kin, matter ferm}& (\ref{Lkinfermmatter})  \nonumber\\
& + {\cal L}_{\rm Noether}& (\ref{Noether})\nonumber\\
&+ {\cal L}_{\rm Pauli}&  (\ref{LPauliN2construct})\nonumber\\
& +  {\cal L}_{\rm goldstino}&  (\ref{Lgoldstino})\nonumber\\
& +{\cal L}_m &  (\ref{LmN2})\nonumber\\
&+\mbox{4-fermion terms}, \label{totPoincN2D4}
\end{align}
where we indicate where these parts can be found.

\subsection{Supersymmetry and Gauge Transformations}
\label{ss:remaindersusy}

We still have to obtain the transformation rules under which this action is invariant. After all the gauge choices, the remaining local symmetries are (apart from the obvious general coordinate transformations and Lorentz symmetry): supersymmetry and the gauge transformations parameterized by $\theta ^I$.

In Sect. \ref{phys_ferm} we already obtained the transformations of the scalar fields in the Poincar\'{e} theory\footnote{All the transformations in this section are the Poincar\'{e} transformations, indicated at some places in this book with $\poinc$. We will use this indication only at the end of this section when we make the connection with the superconformal theory.}:
\begin{align}
\delta(\epsilon,\theta ) z^\alpha =& \ft12 \bar \epsilon^i\chi_i^\alpha+\theta ^Kk_K{}^\alpha\,,\nonumber\\     \delta(\epsilon,\theta ) q^u =&-\rmi \bar{\epsilon}^i\zeta^A f^u{}_{iA }
+\rmi \varepsilon ^{ij}\rho ^{\bAp \Bp } \bar{\epsilon}_i\zeta_\bAp f^u{}_{j\Bp }+ \theta^I k_I{}^u\,.
 \label{delscalars}
\end{align}

The other bosonic physical fields are the gravitons and the gauge fields $W_\mu ^I$, for which we can immediately obtain from (\ref{transfoN4conf}) and  (\ref{delgaugeN2})
\begin{align}
   \delta(\epsilon,\theta ) e_\mu{}^a
      =& \ft12\bar{\epsilon}^i\gamma^a\psi_{\mu i}+\hc, \nonumber\\
 \delta(\epsilon,\theta ) W_\mu {}^I =&\ft12
 \varepsilon ^{ij}\bar \epsilon _{i}\gamma _\mu \chi _j^\alpha \nabla _\alpha  X^I+\varepsilon ^{ij}\bar \epsilon _{i}\psi _{\mu j}X^I
  +\hc + \partial _\mu \theta ^I +\theta ^J W_\mu^K f_{KJ}{}^I\,.
  \label{susyeW}
\end{align}

We can read off the gauge transformations of the fermions from their covariant derivatives (\ref{hatDmuchi}) and (\ref{DmufermN2}). We explained the derivation there in full. The explanation of the gauge transformations is identical by replacing $W_\mu ^I$ by $\theta ^I$. In that derivation it was clear that the covariant transformations are a useful tool. The contributions of  (\ref{covdecompU1}) and (\ref{veclambdaCovariant}) are identical to the $W_\mu ^I$ parts in the expressions of the auxiliary fields ${\cal A}_\mu $ and $\vec{{\cal V}}_\mu$ in  (\ref{valueAVmu}). Thus we get
\begin{align}
  \widehat{\delta }(\theta )\chi_i^\alpha   = &\theta  ^I\chi _i^\beta \nabla    _\beta k_I{}^\alpha + \ft12\kappa ^2\theta ^I \left(\rmi P_I^0\chi  _i^\alpha
 +P_i{}^j\chi  _j^\alpha\right)\,,\nonumber\\
  \widehat{\delta }(\theta ) \zeta^\Ap=&\theta ^It_{I\Bp }{}^\Ap \zeta ^\Bp + \ft12\kappa ^2\theta ^I \rmi P_I^0    \zeta^\Ap \,.
\label{thetatransfFermions}
\end{align}
Due to the compensating $\U(1)$ and $\SU(2)$ transformations, also the gravitino transforms under the gauge transformations.
The covariant transformation is
\begin{equation}
  \widehat{\delta }(\theta )\psi _\mu ^i = -\ft12\kappa ^2\theta ^I \left(\rmi P_I^0 \psi _\mu ^i +  {P} _j{}^i \psi _\mu ^j\right)\,.
 \label{thetatransfGravitino}
\end{equation}

For the supersymmetry transformations of the fermions we need the values of the auxiliary fields, which we rewrite here in the conformal context.
We obtain from (\ref{valueAVN2})
\begin{align}
  A_\mu^{\rm F}=& -\ft18 \rmi\kappa ^2 g_{\alpha \bar \beta }\bar \chi ^{i\bar \beta  }\gamma _\mu \chi  _i^\alpha +\ft12\rmi \kappa ^2 \bar \zeta _\bAp\gamma _\mu \zeta ^\Bp  d^\bAp{}_\Bp \,,
\nonumber\\
\vec{V}_\mu^{\rm F}=& \ft18\kappa ^2 g_{\alpha \bar \beta }\bar \chi ^{i\bar \beta  }\gamma _\mu \chi  _j^\alpha
\vec{\tau }_i{}^j\,.
 \label{projvaluesAFVF}
\end{align}
These appear in the covariant derivatives.
 The linear part of ${\cal D}_\mu X^I$ in (\ref{covD4N2part1}) was simplified in (\ref{nablaXshort}), and we obtain the full result:
\begin{align}
  {\cal D}_\mu X^I =& \nabla _\alpha  X^I \widehat{\partial }_\mu z^\alpha  -\rmi A_\mu ^{\rm F} X^I -\ft12 \bar \psi^i_\mu \Omega^I_i\nonumber\\
=& \nabla _\alpha  X^I \left(\widehat{\partial }_\mu z^\alpha-\ft12 \bar \psi^i_\mu \chi ^\alpha _i\right)+\ft12 \kappa ^2 X^I\left( -\ft14 g_{\alpha \bar \beta }\bar \chi ^{i\bar \beta  }\gamma _\mu \chi  _i^\alpha
 + \bar \zeta _\bAp\gamma _\mu \zeta ^\Bp  d^\bAp{}_\Bp \right)
\,.
\label{calDPoincare}
\end{align}

The covariant derivative of the hyperscalars in (\ref{wideDA}) are
\begin{align}
  \widehat{{\cal D}}_\mu A^{i\Ap}
  =& f^{i\Ap}{}_u {\cal D}_\mu q^u\,,\nonumber\\
 {\cal D}_\mu q^u  =& \widehat{\partial} _\mu q^X +\rmi\bar \psi _\mu ^i \zeta ^\Ap -\rmi\bar \psi _{\mu j}\zeta _{\bBp}\,\varepsilon ^{ji}\rho ^{\bBp\Ap}\,.
 \label{wideDAproj}
\end{align}

With  (\ref{projvaluesAFVF}) we can also rewrite the contribution of $S$-supersymmetry from (\ref{v:decomprule_eta}) as
\begin{eqnarray}
  \kappa ^{-2}\eta ^i(\epsilon )&=& -\ft12 X^I P_{I}{}^{ij}\epsilon _j
  +\ft18\gamma ^a\epsilon ^j g_{\alpha \bar \beta }\bar\chi^{i\bar \beta }\gamma _a\chi _j^\alpha
  +\ft1{16}\gamma ^{ab}\varepsilon ^{ij}\epsilon _jC^{\bAp \bBp }\bar \zeta _\bAp \gamma _{ab}\zeta _\bBp \nonumber\\
&&  +\, \gamma ^a\epsilon^i\left[ \ft1{16}g_{\alpha \bar \beta }\bar\chi_j ^\alpha \gamma _a\chi ^{j\bar \beta }+
  \ft14\bar \zeta_\bAp \gamma _a\zeta ^\Bp d^\bAp{}_\Bp \right]\,.
\label{SdecompN2}
\end{eqnarray}

The value of the auxiliary field $T_{ab}$ was already given in (\ref{Tfermions}). It appears in the transformation of the gravitino and is therefore called the graviphoton field strength. \index{graviphoton field strength}
Its bosonic part can also be written as
\begin{equation}
\left.T^-_{ab}\right|_{\rm bos} =-4\kappa ^2 X^I \Im{\cal N}_{IJ} F^{-J}_{ab}=-2\rmi \kappa^2 \left(X^I G^-_{I\,ab}-F_I F^{-I}_{ab} \right),
\label{Tbos}
\end{equation}
where $G_{I\,ab}$ was defined in (\ref{defG}). The graviphoton is thus a scalar-field dependent combination
of the field strengths of the $(n_V+1)$ gauge fields. The last expression in (\ref{Tbos}) is symplectic invariant and can
thus also be used in the absence of a prepotential.

Note that this graviphoton field strength is the `projection' of $\Im{\cal N}_{IJ} F^{-J}_{ab}$ on $X^I$. The gaugino field strengths are the projection of these quantities on $\nabla_{\bar\alpha }\bar X^I$.
\begin{equation}
  G^{-\,\alpha }_{ab}= g^{\alpha \bar \beta } {\nabla }_{\bar \beta }\bar X^I\Im{\cal N}_{IJ}F_{ab}^{-J}.
\label{defGalpha}
\end{equation}

We are ready to give the supersymmetry transformations of the fermions. The most relevant part for applications is the bosonic part in the transformations, and we will here restrict to these. For the full transformations, again covariant transformations would be helpful. But the difference between the two is quadratic in fermions.

For the gravitino, we obtain from  (\ref{transfoN4conf})
\begin{align}
  \delta(\epsilon )\psi_\mu^i =& \left( \partial _\mu+\ft14\omega _\mu  {}^{ab}\gamma _{ab}-\ft12\rmi {\cal A}_\mu \right) \epsilon^i -{\cal V}_{\mu\,j}{}^i\epsilon ^j-\ft1{16}\gamma^{ab}T_{ab}^-\varepsilon^{ij}\gamma_\mu \epsilon_j\nonumber\\
  &+\ft12 \kappa^2 \gamma_\mu S^{ij}\epsilon_j+\ldots\,,
 \label{susygravitino}
\end{align}
where $\ldots$ refers to terms of higher order in the fermions; ${\cal A}_\mu $,  ${\cal V}_\mu{}^i{}_j$ and $T_{ab}^-$ are found in (\ref{valueAVmu}) and (\ref{Tbos}). In (\ref{SdecompN2}) only the first term is bosonic, and is proportional to the triplet $S^{ij}$ defined as mass matrix of the gravitini in (\ref{mMatrN2}).

For the gauginos, we insert the transformation (\ref{transfoN4conf}) in (\ref{chifromOmega}) and obtain
\begin{align}
  \delta (\epsilon )\chi _i^\alpha  =& g^{\alpha \bar \beta }\nabla _{\bar \beta }\bar X^I N_{IJ}\left[\slashed{\cal D} X^J\epsilon _i +
  \ft14 \gamma^{ab} {\cal F}_{ab}{}^J\varepsilon _{ij}\epsilon ^j
  +Y_{ij}{}^J\epsilon^j\right.\nonumber\\
  &\qquad \qquad \qquad \left.+X^K\bar X{}^Lf_{KL}{}^J\varepsilon _{ij}\epsilon ^j
  +2X^J\eta _i\right]+\ldots \nonumber\\
  =&\hat{\slashed\partial} z^\alpha \epsilon_i -\ft12 G_{ab}^{-\alpha}\gamma^{ab}\varepsilon_{ij}\epsilon^j  +g^{\alpha\bar{\beta}}\epsilon^j\overline{W}_{\bar{\beta}ji}+\ldots
 \label{susygaugino}
\end{align}
 The $\eta $ term does not contribute here due to one of the expressions in (\ref{Nmatrixeqn}). In fact, the part of $\Omega _i^I$ that transforms under $\eta _i$ is the orthogonal one to $\chi _i^\alpha$, which has been gauge-fixed to zero. The first term follows directly from the bosonic part of (\ref{calDPoincare}). For the second term, one uses that (\ref{defGalpha}) can also be written as
\begin{equation}
 G^{-\,\alpha }_{ab}=-\ft12 g^{\alpha \bar \beta } \nabla_{\bar\beta }\bar X^I N_{IJ}{\cal F}_{ab}^{-J}+\mbox{fermionic terms}
\,,
 \label{nabXNcalF}
\end{equation}
which follows from the definition (\ref{calFVM4}), the expression for $T^-$ in (\ref{Tbos}) and the difference between the two formulas in (\ref{InverseImN}). For the following terms we use  (\ref{Yeomrewritten}) and  (\ref{kalmomM}), and these combine in the expression $\overline{W}_{\bar{\beta}ji}$, the complex conjugate of the expression in (\ref{Lgoldstino})
\begin{equation}
  \overline{W}_{\bar \alpha  ij}=\left( -\rmi\varepsilon _{ij}P_I^0- P_{Iij}\right){\nabla} _{ \bar \alpha }\bar X^I
  = -\varepsilon _{ij}g_{\beta  \bar\alpha }k_I{}^{\beta }\bar X^I -
P_{Iij}{\nabla} _{ \bar \alpha }\bar X^I\,.
 \label{barW}
\end{equation}

For the hyperini, we project from (\ref{transfoN4conf}), use (\ref{wideDAproj}) and the $S$-supersymmetry again does not contribute due to (\ref{AiAcomponents}) (again because the physical fermions have been chosen like this). Also (\ref{fcomponents0}) implies that only one term contributes from the Killing vectors (the index $X$ reduces to $u$) and thus we get
\begin{align}
{\delta}(\epsilon )\zeta^\Ap =&\ft12\rmi f^{i\Ap}{}_\Xu \hat{\slashed{\partial}}q^\Xu \epsilon_i+\rmi \bar
X^Ik_I{}^Xf^{i\Ap }{}_X\varepsilon _{ij}\epsilon   ^j+ \bar{N}_i{}^\bAp \epsilon^i+\ldots \,,
\label{susyhyperino}
\end{align}
with again the complex
conjugate of the expressions in (\ref{Lgoldstino}) (using (\ref{projecrealf}))
\begin{equation}
 \bar N_i{}^\bAp=(N^i{}_\Ap)^*=   -\rmi \varepsilon _{ij} d^\bAp{}_\Bp f^{j\Bp }{}_{\Xu}  k_I{}^\Xu \bar  X^I.
\label{barWN}
\end{equation}

Let us introduce the name `fermion shifts'\index{fermion shift}  \cite{Cecotti:1986wn,D'Auria:2001kv} for the scalar, non-derivative parts of the supersymmetry transformations of the fermions. Thus these are the last terms in (\ref{susygravitino}),  (\ref{susygaugino}) and  (\ref{susyhyperino}).
The fact that these matrices appear in the mass terms of the gravitino (\ref{LmN2}) and in the Goldstino (\ref{Lgoldstino}) is a general fact, or Ward equation of all supergravities.\footnote{This can be proven by looking just at the variation of the action linear in matter fermions and with one spacetime derivative.} They also lead to another interpretation of the potential. A general formula in supergravity says that the square of the  fermion shifts, using the kinetic matrix of the fermions, obeys \index{scalar potential}
\begin{tcolorbox}
\begin{align}
 \delta ^i_j V=& -3\kappa ^2S^{ik}S_{jk}+ W_\alpha {}^{ik}g^{\alpha \bar\beta  }\overline{W}_{\bar \beta jk}+4N^i{}_\Ap  (d^{-1})^\Ap {}_\bBp   \bar N_j{}^\bBp \,,\nonumber\\
 2V=& -3\kappa ^2S^{ik}S_{ik}+ W_\alpha {}^{ik}g^{\alpha \bar\beta  }\overline{W}_{\bar \beta ik}+4N^i{}_\Ap  (d^{-1})^\Ap {}_\bBp   \bar N_i{}^\bBp \,.
\label{WardidVN2}
\end{align}
\end{tcolorbox}
To prove this explicitly for the non-diagonal part, one needs the equivariance relation (\ref{equivariancelocal}). For the diagonal part, inserting the definitions and e.g.  (\ref{productSU2}) one re-obtains the form of the potential as in (\ref{Vd4total}).

\subsubsection{Relation Between Symmetries}
A few final remarks are in order for the relation between the supersymmetry and gauge transformations. Let us first summarize how the symmetries of the Poincar\'{e} theory are related to those of the conformal theory (decomposition law): \index{decomposition law}
\begin{align}
  \delta ^{\poinc}[\epsilon ,\theta ]= & \delta _Q[\epsilon ] + \delta _G[\theta ] + \delta _S[\eta (\epsilon )] + \delta _T[\lambda _T(\epsilon,\theta )] +\delta _{\SU(2)}[\vec{\lambda }(\epsilon ,\theta )]\nonumber\\
  &+ \delta_K[\lambda_{\rm K}^a (0,\epsilon ,\eta )]\,.
\label{fulldecomp}
\end{align}
At the right-hand side of this equation are the superconformal transformations. In the third term appears  (\ref{SdecompN2}). For the $T$ and $\SU(2)$ transformations, the parameters are given by  (\ref{delzdecompcov}) and  (\ref{veclambdagfrepeat}). When using covariant transformations, they can be reduced to (\ref{covdecompU1}) and (\ref{veclambdaCovariant}). The final term uses (\ref{decompK}), in which we can now put $\lambda _{\rm D}=0$. This last term is not often important since none of the independent fields transforms under $K$-symmetry.

The commutators between symmetries in the Poincar\'{e} theories are then obtained from the conformal commutators (\ref{QQcommG}),  (\ref{thirdcoeffQS}),  (\ref{commRQD4}) combined by (\ref{fulldecomp}). We obtain (omitting terms of higher order in fermions)
\begin{align}
[\delta^\poinc(\epsilon_1),\delta^\poinc(\epsilon_2)]=&
\delta_{\rm cgct}\left(\xi_3^a(\epsilon _1,\epsilon _2)\right)
+\delta_M\left(\lambda_3^{ab}(\epsilon _1,\epsilon
_2)\right)+\delta^\poinc _G\left(\theta_3^I(\epsilon _1,\epsilon _2)\right)\,,\nonumber\\
&\xi_3^a =   \ft12\bar \epsilon_2^i\gamma^a\epsilon_{1i}+\hc\,, \nonumber\\
&\lambda_3^{ab} (\epsilon _1,\epsilon _2)=  \ft14\bar\epsilon^i_1\left(\,T^{+ab}\varepsilon_{ij}+\kappa ^2 X^I \gamma ^{ab}P_{I\,ij}\right)\epsilon^j_2+\hc\,,\nonumber\\
&\theta_3^I(\epsilon _1,\epsilon _2)=\varepsilon ^{ij}\bar \epsilon _{2i}\epsilon _{1j}X^I +\hc\,,
\label{QQcommD4}
\end{align}
and
\begin{align}
  [\delta^\poinc _G(\theta),\delta^\poinc _Q(\epsilon )]=&\delta _Q\left[\epsilon_3 (\epsilon ,\theta )\right]\,,\nonumber\\
&\epsilon_3 ^i(\epsilon ,\theta )  =\ft12\kappa ^2 \theta ^I
  \left( \rmi P_I^0\epsilon ^i + P_{Ij}{}^i \right)\epsilon ^j .
 \label{commGQ}
\end{align}
For example, the second term in $\lambda_3^{ab} (\epsilon _1,\epsilon _2)$ originates from the $[Q,S]$ conformal commutator where the bosonic part of the parameter  (\ref{SdecompN2}) is used.

One can also see that the appearance of the last term in (\ref{QQcommD4}) is consistent with the transformation of the gauge vectors proportional to the gravitino in  (\ref{susyeW}) following general rules as in (\ref{delGaugef}). Similarly, the gauge transformation of the gravitino  (\ref{thetatransfGravitino}) or its supersymmetry transformation proportional to $W_\mu ^I$ included in (\ref{susygravitino}) implies that these gauge transformations do not commute with the supersymmetry and also lead to (\ref{commGQ}). This non-vanishing is the contribution of gauge transformations to $R$-symmetry and is an interpretation of the moment maps.

The result (\ref{QQcommD4}) is important for interpreting solutions of the field equations. For example, for solutions with non-vanishing value of $X^I$ and a gauged isometry group or non-vanishing value of $T^{ab}$ (charged black hole solutions), this implies that there is a central charge \index{central charge} in the preserved algebra.


\section{Final Results for \texorpdfstring{$D=5$ Poincar\'{e}}{D=5 Poincare} Supergravity}
\label{ss:finalD5}

As in the previous chapter for $D=4$, we will collect here the main results of $D=5$ for general couplings with vector multiplets and hypermultiplets, though we will be much shorter since the results in our conventions have been collected in \cite{Bergshoeff:2004kh}, for a large part based on previous work, e.g.
\cite{Gunaydin:1984bi,Gunaydin:1999zx,Ceresole:2000jd}.  Actually, in giving these results, we will extend the vector multiplets, which we discussed in Chaps.~\ref{ss:matter} and \ref{ss:gaugefix}, to vector-tensor multiplets, where the vector multiplets are off-shell, but the tensor multiplets are on-shell.

\vspace{10pt}
We consider the theory with
\begin{itemize}
  \item Supergravity, including the frame field $e_\mu ^a$ and gravitino $\psi
  _\mu ^i$. The graviphoton is included in the vector
  multiplets.
  \item Vector-tensor multiplets enumerated by $\tilde I=0, \ldots ,
  n_V+n_T$ where $n_V$ is the number of vector multiplets and $n_T$ is
  the number of tensor multiplets. The index is further split as $\tilde
  I=(I,M)$, where $I=0,\ldots ,n_V$ and $M=n_V+1,\ldots ,n_V+n_T$.
  The vector or tensor fields are grouped in
\begin{equation}
  H_{\mu \nu }^{\widetilde{I}} \equiv
\left(F_{\mu \nu }^I,\widetilde{B}_{\mu \nu }^M\right)\,, \qquad
F_{\mu\nu}^I \equiv 2
\partial_{[\mu} W_{\nu]}^I + f_{JK}{}^I W_\mu^J W_\nu^K\,.
 \label{defHtildeI}
\end{equation}
where $\widetilde{B}_{\mu \nu }^M$ are the fundamental tensor fields and
$W_\mu ^I$ are the fundamental vector fields, gauging an algebra with
structure constants $f_{JK}{}^I$. The fermions of these multiplets are denoted as $\lambda ^{xi}$ and the real scalars as $\phi ^x$ where $x=1,\ldots ,n_V+n_T$.
In Sect. \ref{ss:actVMD5}, we constructed actions for the vector multiplets based on a symmetric constant tensor $C_{IJK}$. For vector-tensor multiplets this is extended to a tensor
$C_{\tilde I\tilde J\tilde K}$. In order to get a
standard normalization, we rescale the $C_{\widetilde I\,\widetilde
J\,\widetilde K}$ symbol and the vector multiplet scalars $\sigma ^{\widetilde I}$ (generalizations of the $\sigma ^I$ introduced in Sect. \ref{vectmult}) as follows:
\begin{equation}
\sigma^{\widetilde I} \equiv \sqrt{\frac{3}{2\kappa^2}} h^{\widetilde I},
\qquad C_{\widetilde I\,\widetilde J\,\widetilde K} \equiv -
2\sqrt{\frac{2\kappa^2}{3}} {\cal C}_{\widetilde I\,\widetilde
J\,\widetilde K}\,.
\end{equation}
Furthermore, for the tensor multiplets we need an antisymmetric and invertible metric
$\Omega_{MN}$ and the structure constants $f_{IJ}{}^K$ are extended to transformation matrices $t_{I\tilde J}{}^{\tilde
K}$ related by
\begin{eqnarray}
   &   &{\cal C}_{M\widetilde J \widetilde K} =  \sqrt{\frac{3}{8\kappa^2}} t_{(\widetilde{J} \widetilde{K})}{}^P
\Omega_{ PM}, \qquad  
   t_{I[M}{}^P\Omega _{N]P}=0,\qquad  
  t_{I(\widetilde J}{}^{\widetilde M} {\cal C}_{\widetilde K\widetilde
L)\widetilde M} = 0 ,\nonumber\\
&& (t_M){}_{\widetilde J}{}^{\widetilde K}  = 0,\qquad 
(t_I){}_{\widetilde J}{}^{\widetilde K} =
\begin{pmatrix}
f_{IJ}{}^K & (t_I)_J{}^N \\
0 & (t_I)_M{}^N
\end{pmatrix}.
 \label{COmegarelations}
\end{eqnarray}
Note that this implies that at least one index of a non-zero ${\cal
C}_{\tilde I\tilde J\tilde K}$ should correspond to a vector multiplet.
The above-mentioned physical scalars $\phi ^x$ are a parameterization of the manifold defined as solution of the constraint (\ref{DgaugegeneralD5}), which is now written as
\begin{equation}
{\cal C}_{\widetilde I\,\widetilde J\,\widetilde K} h^{\widetilde
I}(\phi)h^{\widetilde J}(\phi) h^{\widetilde K}(\phi) = 1.
\label{eq:hypersurface_vector}
\end{equation}
\item $n_H$ hypermultiplets with scalars $q^\Xu $ and spinors $\zeta ^A$,
  where $\Xu =1,\ldots ,4n_H$ and $A=1,\ldots ,2n_H$. Their interactions are
  determined by the frame fields, $f^{iA}{}_\Xu $, invertible as $4n_H\times
  4n_H$ matrices, identical to what we saw for $D=4$.
\end{itemize}
This leads to the independent physical fields in Table~\ref{tbl:superPoincD5} .
\begin{table}[!t]
\caption{\it Multiplets and fields of the super-Poincar\'{e} theories for $D=5$}\label{tbl:superPoincD5}
\begin{center}
\begin{tabular}
{|l|c|ccc|l|}
\hline
Field  & Pure SG & Vector/tensor mult. & Hypermult. & Indices \\
\hline
$e_\mu ^a$   & 1 &   &   & $\mu ,a=0,\ldots ,4$ \\
$\psi _\mu ^i$  &   2    &  & & $i=1,2$  \\
$W_\mu ^I$& 1 & $n_V$ &   & $I=0,\ldots ,n_V$ \\
$\tilde B_{\mu \nu }^M$ & & $ n_T $&  & $M= n_V+1,\ldots ,n_V+ n_T=n $,\\
  & & & & $\tilde I=0,\ldots , n_V+n_T=n$\\
 $\lambda ^x_i$, $\zeta ^A$  &   & $2n$ & $2n_H$ & $A=1,\ldots ,2n_H$ \\
$\phi ^x$, $q^\Xu $ &   & $n$ & $4n_H$ & $x=1,\ldots ,n $; $\Xu =1,\ldots ,4n_H$ \\
\noalign{\smallskip}\hline\noalign{\smallskip}
\end{tabular}
\end{center}\end{table}

As mentioned, the full results are in \cite{Bergshoeff:2004kh}.\footnote{A few changes of notation are mentioned in  (\ref{translationsoldD5}).} We repeat the main results.
The bosonic action is
\begin{align}
e^{-1}\mathcal{L}_{\rm bos}=&
\ft1{2\kappa^2} R -\ft14a_{\widetilde{I}\widetilde{J}}
H^{\widetilde{I}}_{\mu\nu}H^{\widetilde{J}\mu \nu} -\ft12
g_{xy}\hat{D}_\mu \phi^x\hat{D}^\mu \phi^y
-\ft1{2}g_{\Xu \Yu }\hat{D}_\mu q^{\Xu }\hat{D}^\mu q^{\Yu }-V\nonumber\\&
+\, \frac {1}{16 g} e^{-1} \varepsilon ^{\mu \nu \rho \sigma \tau}
\Omega_{MN} \widetilde{B}_{\mu \nu}^M \Big(\partial_\rho
\widetilde{B}_{\sigma \tau}^N+2g t_{IJ}{}^NW_\rho ^I F_{\sigma \tau}^J
   + t_{IP}{}^N W_\rho^I \widetilde{B}_{\sigma\tau}^P\Big) \nonumber\\&
+\ft{\kappa}{12}\sqrt{\ft23} e^{-1} \varepsilon ^{\mu \nu \lambda \rho \sigma } {\cal C}_{IJK} W_\mu^I \left[ F_{\nu \lambda }^J F_{\rho \sigma }^K \right.\nonumber\\
& \phantom{\ft{\kappa}{12}\sqrt{\ft23} e^{-1} \varepsilon ^{\mu \nu \lambda \rho \sigma } {\cal C}_{IJK} W_\mu^I}\left.+ f_{FG}{}^J W_\nu^F W_\lambda ^G \left(- \ft12 F_{\rho \sigma }^K + \ft1{10}   f_{HL}{}^K W_\rho ^H W_\sigma^L \right)\right]\nonumber\\
& - \ft{1}{8} e^{-1} \varepsilon^{\mu\nu\lambda\rho\sigma} \Omega_{MN}
t_{IK} {}^M t_{FG}{}^N W_{\mu}^I W_\nu^F W_\lambda^G \left(-\ft 12
F_{\rho\sigma}^K + \ft{1}{10}   f_{HL}{}^K W_\rho^H W_\sigma^L \right).
\label{5dtruncated}
\end{align}
The metrics for the vectors and the vector--scalars are defined by
\begin{eqnarray}
 a_{\widetilde I\,\widetilde J} &=& -2 {\cal C}_{\widetilde I\,\widetilde
J\widetilde K} h^{\widetilde K} + 3 h_{\widetilde I} h_{\widetilde J},\qquad
 h_{\widetilde I} \equiv {\cal C}_{\widetilde I\,\widetilde J\,\widetilde K} h^{\widetilde J} h^{\widetilde K}
 =a_{\widetilde I\,\widetilde J} h^{\widetilde J}\,,\nonumber\\
g_{xy}& =& h^{\widetilde I}_x h^{\widetilde J}_y a_{\widetilde
I\widetilde J},\qquad h_x^{\widetilde I} \equiv
-\sqrt{\frac{3}{2\kappa^2}} \partial _xh^{\widetilde I}(\phi).
 \label{metricsrealspecial}
\end{eqnarray}
Many useful relations are given in Appendix C of \cite{Bergshoeff:2004kh}.

The domain of the variables should be limited to $h^I(\phi )\neq 0$ and
the metrics $a_{IJ}$ and $g_{xy}$ should be positive definite. Similar to (\ref{Nmatrixeqn}) for $D=4$, one can block-diagonalize the metric $a_{IJ}$ in a singlet corresponding to the compensating multiplet, and the part corresponding to the physical scalars:
\begin{equation}
  \begin{pmatrix}h^I\cr h^I_x\end{pmatrix} a_{IJ}\begin{pmatrix}h^J&h^J_y\end{pmatrix}
  =\begin{pmatrix}1&0\cr 0&g_{xy}\end{pmatrix}\,.
 \label{matrixag}
\end{equation}

The metric for the hyperscalars is the same as discussed for $D=4$ in Sect. \ref{ss:D4bosonicaction} with curvature  as in (\ref{RlambdaJ}).

All other quantities in (\ref{5dtruncated}) are related to gauged
symmetries. The gauge symmetry transformations (with parameters
$\theta^I$) of the bosons are
\begin{eqnarray}
\delta_GW_\mu^I &=& \partial_\mu\theta^I -
\theta^Jf_{JK}{}^IW_\mu^K,\qquad
 \delta _G\widetilde{B}_{\mu \nu }^M=
 -\theta ^Jt_{J\widetilde K}{}^M H_{\mu \nu }^{\widetilde{K}},
  \nonumber\\
 \delta _G h^{\widetilde{I}}(\phi ) & = & -\theta ^Jt_{J\widetilde K}{}^{\widetilde{I}}h^{\widetilde{K}}\,,
 \nonumber\\
 \delta _G\phi ^x&=&\theta ^Ik_I{}^x,\qquad k_I{}^x \equiv  \frac1\kappa
\sqrt\frac32 t_{I{\widetilde J}}{}^{\widetilde K} h^{{\widetilde J}}
h_{\widetilde K}{}^x,\nonumber\\
\delta _Gq^\Xu &=&\theta ^Ik_I{}^\Xu ,
 \label{GaugeVTtransfo}
\end{eqnarray}
where $k_I{}^\Xu $ should be isometries of the quaternionic-K\"{a}hler metric
$h_{\Xu \Yu }$, whose commutators are determined by the structure constants
$f_{IJ}{}^K$:
\begin{equation}
  2k_{[I}{}^\Yu \partial _\Yu k_{J]}{}^\Xu=f_{IJ}{}^Kk_K{}^\Xu \,.
 \label{algebraquatKilling}
\end{equation}
These transformations determine immediately the covariant derivatives in
this bosonic truncation:
\begin{equation}
\widehat{D}_\mu \phi^x =\partial_\mu\phi^x-W_\mu^Ik_I{}^x  ,\qquad
\widehat{D}_\mu q^{\Xu } = \partial_\mu q^{\Xu }- W_\mu^I k_I{}^{\Xu }.
\end{equation}
The gauge transformations are also in one-to-one relation with triplet moment maps. For example, the moment map is obtained from the Killing vectors in
 (\ref{solnPI}), derived in the treatment of $D=4$ and also valid here. The potential $V$ will appear below in (\ref{Vd5}) after discussing the supersymmetry transformations.

The ${\cal N}=2$ supersymmetry rules of the fermionic fields, up to bilinears in
the fermions, are given by\footnote{Remember that indices are lowered with the symplectic metrics $\varepsilon _{ij}$
and $C_{AB}$ using the NW-SE convention as in (\ref{NWSEconv}), while the index $x$ is lowered with $g_{xy}$. The translation from triplet to doublet notation in (\ref{tripletvecij}) holds for all vectors such as $\vec{P}$ or $\vec{P}_x$.}
\begin{eqnarray}
\delta(\epsilon ) \psi_\mu^i&=&  D_\mu(\omega)\epsilon^i + \frac{\rmi
}{4\sqrt6}\left[\kappa h_{\widetilde I} H^{\widetilde I\nu \rho} \left(
{\gamma}_{\mu\nu \rho } - 4 g_{\mu \nu } {\gamma}_\rho\right) \epsilon^i
+ 2 \kappa^{-1} P_j{}^i\gamma_\mu  \epsilon^j\right] ,\nonumber\\
&&D_\mu(\omega) \epsilon^i =
 \left( \partial_\mu + \ft14 \omega_\mu{}^{ab} \gamma _{ab}\right) \epsilon ^i
 +\partial_\mu q^\Xu  \omega_{\Xu j}{}^i\epsilon^j + \ft12\kappa^2 W_\mu^I P_{Ij}{}^i\epsilon^j,
 \nonumber\\
\delta(\epsilon )  \lambda^{xi} &=& - \ft{1}{2} \rmi\widehat{\slashed{D}} \phi^x
\epsilon^i + \ft14 \gamma\cdot H^{\widetilde I}
 h^x_{\widetilde I} \epsilon ^i  + \kappa^{-2} P^x{}_j{}^i \epsilon^j
 + \ft{1}{2}\kappa^{-2} W^x \epsilon^i ,\nonumber\\
\delta(\epsilon )  \zeta^A&=& \ft12 \rmi \gamma^\mu \hat{D}_\mu q^\Xu  f^{iA}{}_\Xu
\epsilon _i - \kappa^{-1} {\mathcal N}^{iA} \epsilon_i  .
 \label{Dmuepsilond5}
\end{eqnarray}
Here appear quantities with some similarity between the vector and hypermultiplet sector\footnote{One could even improve the similarity by e.g. defining $W_{xi}{}^j =\ft12 W_x\delta _i{}^j + P_{xi}{}^j$ to those in (\ref{Lgoldstino}) and (\ref{mMatrN2}).}
\begin{eqnarray}
&&W ^x \equiv -\frac{\sqrt6}{4}\kappa h^I k_I{}^x = -\frac{3}{4}
t_{J{\widetilde I}}{}^{\widetilde P} h^J h^{\widetilde I} h^x_{\widetilde
P}\,,\qquad {\mathcal N}^{iA} \equiv
-\frac{\sqrt6}{4}\kappa h^I k_I{}^\Xu  f^{iA}{}_\Xu\,,\nonumber\\
&& \vec{P} \equiv \kappa ^2 h^I \vec{P}_I , \qquad  \vec{P}_x \equiv
\kappa^2h_x^I \,.
 \label{fermionshiftsd5}
\end{eqnarray}
See e.g. that $\vec{P}$ is the real analogue of $S_{ij}$ in (\ref{mMatrN2}). As in (\ref{WardidVN2}) the supersymmetry transformations determine the potential, which is \index{scalar potential}
\begin{tcolorbox}
\begin{equation}
V=\kappa^{-4}\left( \vec{P}\cdot \vec{P}
-\ft12\vec{P}^x\cdot\vec{P}_x- 2 W_x W^x
-2\mathcal{N}^{iA}\varepsilon _{ij}C_{AB}\mathcal{N}^{jB}\right).
  \label{Vd5}
\end{equation}
\end{tcolorbox}

Let us finally note that the remarks about the gauge transformation of the gravitino  (\ref{thetatransfGravitino}) and the non-vanishing commutator between gauge and supersymmetry transformations (\ref{commGQ}) that we had for $D=4$ are also applicable for $D=5$ by just omitting $P_I^0$, since in $D=5$ there is no $\U(1)$ factor in the $R$-symmetry group.

Also here the commutator of two supersymmetries contains a scalar-dependent gauge transformation
\begin{equation}
  \left[\delta (\epsilon _1),\delta (\epsilon _2)\right]= \ldots + \delta _G\left(\theta ^I=-\frac{\sqrt{6}}{4}\rmi\kappa ^{-1}h^I\bar \epsilon _2^i\epsilon _{1i}\right)\,.
 \label{QQtoGD5}
\end{equation}


\section{Final Remarks}
\label{ch:finalremarks}

We reviewed the constructions of the supergravity actions for ${\cal N}=2$ with vector and hypermultiplets. We restricted ourselves to actions at most quadratic in spacetime derivatives in the bosonic side, linear for the fermionic parts. We used the superconformal techniques to obtain these results.\footnote{There are many other approaches. We gave an overview in the introduction.} With the same techniques also theories with higher derivatives \index{higher derivatives} can be obtained. See e.g. the review~\cite{Mohaupt:2000mj}. See also \cite{Banerjee:2011ts,Bergshoeff:2012ax,Banerjee:2015uee} for various aspects of higher-derivative actions in this context and \cite{Bonetti:2018lfb} for an application to get Killing spinor identities. New possibilities with higher-derivative terms and supersymmetry breaking (related to nonlinear realizations of supersymmetry) have also been explored in \cite{Antoniadis:2019hbu}.

We made several further choices. For the Weyl multiplet, we used the `standard Weyl multiplet'.\index{standard Weyl multiplet} An alternative would be the dilaton Weyl multiplet \index{dilaton Weyl multiplet} \cite{Bergshoeff:1986mz,Siegel:1996px,Bergshoeff:2001hc,Butter:2017pbp}. As far as we know, this does not lead to other matter couplings when considering theories quadratic in spacetime derivatives.

In Sect. \ref{ss:pureN2} we discussed already the choices of the second compensating multiplet for $D=4$. A similar choice exists also for $D=5$ and $D=6$. We have chosen in the further treatment the hypermultiplet as second compensating multiplet because this exhibits the structure of quaternionic-K\"{a}hler manifolds as projective manifolds in the same way as special K\"{a}hler manifolds are obtained. A disadvantage is that this multiplet is on-shell from the start. When one uses the tensor multiplet  or the nonlinear multiplet  \cite{deWit:1981tn,deWit:1984xe}, one keeps the theory off-shell. See e.g. the construction for $D=6$ in \cite{Coomans:2011ih}.
In our treatment we have chosen to eliminate auxiliary fields at an early stage, i.e. before the gauge fixing to the Poincar\'{e} theory. It may be useful to eliminate these fields only at a later stage such that off-shell Poincar\'{e} supergravity is obtained. The techniques explained in the book can then still be used.

The ideas of localization, considering rigid supersymmetric theories on a curved background, are often based on off-shell supergravity. For ${\cal N}=2$ this has been considered in \cite{Klare:2013dka}.
In view of these ideas, Euclidean versions of supergravity are studied. Special geometry in Euclidean versions is constructed in \cite{Cortes:2003zd,Cortes:2005uq,Cortes:2009cs,Cortes:2015wca}.\footnote{Special geometry in other signatures was considered in \cite{Lledo:2006nr,Sabra:2017xvx}.} The off-shell $D=5$ Minkowski supergravity treated in this book can also be reduced over time to an Euclidean $D=4$ theory \cite{deWit:2017cle}.

The main methods that we have explained in this book are still applicable for all these extensions. We hope that the extensive explanations that we gave here will be useful for future developments of the theory and for exploring the physics of solutions to the field equations.\footnote{Apart from Sect. \ref{ss:emcharges}, we have refrained in this review to discuss solutions, although a lot of interesting results have been obtained, with or without supersymmetry breaking.}

%% file: appA_Notation.tex
\chapter{Notation}\label{appendix_notations}

We use the same conventions as in \cite{Freedman:2012zz}. This implies that the sign factors $s_1$ to $s_9$ in Appendix~A of that book, see also (\ref{sfactors}), are equal to $+1$.
In particular, we use the metric signature $(-+\ldots +)$.
If you prefer the opposite,
insert a minus sign for every upper index which you see, or for an
explicit metric $\eta_{ab}$ or $g_{\mu\nu}$. The gamma matrices $\gamma
_a$ should then be multiplied by an $\rmi$ to implement this change of
signature consistently.

We collect the indices that we use in Table~\ref{tbl:useindices}.
\begin{table}[t]
  \caption{\it Use of indices}\label{tbl:useindices}
\begin{center}$
\begin{array}[b]{|lll|}\hline
 \mu  & 0,\ldots ,3 & \mbox{Local spacetime}     \\
 a & 0,\ldots ,3 & \mbox{Tangent spacetime}     \\
  i  & 1,2 & \SU(2) \\ \hline
\multicolumn{3}{|l|}{\mbox{In Chap.~\ref{ss:ingredients}
and~\ref{ss:Weyl}.}}\\
 \alpha & 1,\ldots , 4 & \mbox{Spinor indices} \\
A & & \mbox{All the gauge transformations} \\
I & & \mbox{All gauge transformations excluding translations }\\ \hline
\multicolumn{3}{|l|}{\mbox{From Chap.~\ref{ss:matter} onwards.}}\\
 I & 0,\ldots ,n=n_V & \mbox{Vector multiplets }\\
 X & 1,\ldots ,4n_H\ & \mbox{Scalars in hypermultiplets}\\
 {\cal A} & 1,\ldots ,2n_H& \mbox{Spinors (or $\USp(2n_H)$ vector) in hypermultiplets}\\
\vec{V} & 1,2,3 &\mbox{Triplet of $\SU(2)$, see Sect. \ref{SU2vectorNotation}}\\
\hline
 \multicolumn{3}{|l|}{\mbox{From Chap.~\ref{ss:gaugefix} onwards, indices $I,X,{\cal A}$ are split and}}\\
 \alpha & 1,\ldots ,n=n_V & \mbox{Independent coordinates in special K\"{a}hler}\\
 X & 1,\ldots ,4(n_H+1)\ & \mbox{Scalars in hypermultiplets}=\{0,\ralpha,u\} \\
 {\cal A} & 1,\ldots ,2(n_H+1) & \mbox{Spinors (or $\USp(2(n_H+1))$ vector) in hypermultiplets}=\{i,A\}\\
 \ralpha & 1,2,3& \mbox{Compensating directions, part of $X$}\\
 u& 1,\ldots ,4n_H & \mbox{Independent coordinates in quaternionic-K\"{a}hler, part of $X$}\\
 A & 1,\ldots ,2n_H & \mbox{Spinors (or $\USp(2(n_H))$ vector) in hypermultiplets, part of ${\cal A}$}\\ \hline
\end{array}$
\end{center}
\end{table}

(Anti)symmetrization is done with weight one:
\begin{equation}
  A_{[ab]}=\ft12 \left( A_{ab}-A_{ba}\right) \qquad  \mbox{and}\qquad A_{(ab)}=\ft12 \left(
A_{ab}+A_{ba}\right)\,.
 \label{weight1}
\end{equation}
The antisymmetric tensors are often contracted with $\gamma $ matrices as
in $\gamma \cdot T\equiv \gamma ^{ab}T_{ab}$.

\section{Bosonic Part} \label{app:bosonicConv}

For the curvatures and connections, we repeat here the main formula from the conventions of \cite[Appendix A.1]{Freedman:2012zz} in order to be able to compare with other papers in the literature.
\begin{eqnarray}
R_{\mu\nu}{}^{ab}(e) &=& 2 \partial_{[\mu} \omega_{\nu]}{}^{ab}(e) + 2
\omega_{[\mu}{}^{ac}(e) \omega_{\nu]}{}_c{}^b(e)\, ,\nonumber\\
R^\mu{}_{\nu\rho\sigma}&=&R_{\rho\sigma}{}^{ab}e_a^\mu e_{\nu b}=
2\partial_{[\rho}\Gamma^\mu_{\sigma]\nu} +2\Gamma^\mu_{\tau[\rho}
\Gamma^\tau_{\sigma]\nu}\,,\nonumber\\
R_{\mu\nu}&=& R_{\rho\mu}{}^{ba}e_b{}^\rho e_{\nu a}=R^\rho {}_{\nu \rho
\mu }\,, \qquad R = g^{\mu\nu}
R_{\mu\nu}\, ,  \nonumber\\
  G_{\mu \nu }&=&e^{-1}\frac{\delta }{\delta g^{\mu \nu }}\int \rmd ^4x\, eR
  =R_{\mu \nu }-\ft12 g_{\mu \nu }R\,.
  \label{defR}
\end{eqnarray}
The formulations in terms of spin connection $\omega $ and in terms of
Levi-Civita connection \index{Levi-Civita connection}$\Gamma $ are equivalent by demanding
\begin{equation}
0=\nabla _\mu e_\nu{}^a = \partial_\mu e_\nu{}^a + \omega_\mu{}^{ab}(e)
e_{\nu b} - \Gamma^\rho_{\mu\nu} e_\rho{}^a\,, \qquad g_{\mu\nu} =
e_\mu{}^a \eta_{ab} e_\nu{}^b\,, \label{Nablae0}
\end{equation}
which leads to
\begin{eqnarray}
\omega_\mu{}^{ab}(e) &=& 2 e^{\nu[a} \partial_{[\mu} e_{\nu]}{}^{b]} -
e^{\nu[a}e^{b]\sigma} e_{\mu c} \partial_\nu e_\sigma{}^c\,,\nonumber\\
\Gamma^\rho_{\mu\nu} &=& \ft12 g^{\rho\lambda}\left(2\partial_{(\mu}
g_{\nu)\lambda} - \partial_\lambda g_{\mu\nu}\right)\,,\qquad \Gamma
_{\mu \nu }^\nu =\ft12\partial _\nu \ln g \,.\label{omegae}
\end{eqnarray}
Note that the Ricci tensor \index{Ricci tensor}and scalar curvature are of opposite sign as in several older papers in this field. The sign is now chosen such that Einstein
spaces with positive scalar curvature are compact.

The anticommuting Levi-Civita tensor  \index{Levi-Civita tensor} is real, and taken to be
\begin{equation}
\varepsilon_{0123}=1\,,\qquad \varepsilon^{0123}=-1\,,
\label{valueLeCi}
\end{equation}
the $-$ sign is related to the one timelike
direction.\footnote{Note that in many papers,
e.g.~\cite{VanNieuwenhuizen:1981ae}, one takes in 4
dimensions an imaginary Levi-Civita tensor to avoid factors of $\rmi$ in
definitions of duals (\ref{dual46}).\label{fn:epsIm}} For convenience, we give
below the formulae for an arbitrary dimension $D$ and number of timelike
directions $t$. The contraction identity for these tensors is ($p+n=D$)
\begin{equation}
\varepsilon_{a_1\ldots a_nb_1\ldots b_p} \varepsilon^{a_1\ldots
a_nc_1\ldots c_p}= (-)^t\, p!\, n!\, \delta_{[b_1}^{[c_1}\ldots
\delta_{b_p]}^{c_p]}\,. \label{contractepsilon}
\end{equation}
For the local case, we can still define constant tensors
\begin{equation}
\varepsilon_{\mu_1\ldots \mu_D}=e^{-1}e_{\mu_1}^{a_1}\ldots e_{\mu_D}^{a_D}
\varepsilon_{a_1\ldots a_ D}   \,,\qquad
\varepsilon^{\mu_1\ldots \mu_D}=e e^{\mu_1}_{a_1}\ldots e^{\mu_D}_{a_D}
\varepsilon^{a_1\ldots a_ D}   \,.\label{LeviCivLocal}
\end{equation}
They are thus \textit{not} obtained from each other by raising or
lowering indices with the metric. In some papers in the literature a difference is made between $\epsilon_{\mu_1\cdots \mu _D}$  and $\varepsilon _{\mu _1\cdots \mu _D}$, such that one of the two has the factor $e^{-1}$ in the definition (\ref{LeviCivLocal}) and the other one has not. We only use the symbol defined by (\ref{LeviCivLocal}).

\bexer
 Show that the tensors in (\ref{LeviCivLocal})
are indeed constants, i.e.\ that arbitrary variations of the frame field
cancel in the full expression. You can use the so-called
Schouten identities, \index{Schouten identities} which means that antisymmetrizing in more
indices than the range of the indices, gives zero. The constancy thus
implies that one can have (\ref{valueLeCi}) without specifying whether the
$0123$ are local or flat indices.
\eexer

The definition of dual tensor \index{dual tensor}is in 4 and 6 dimensions, respectively,
\begin{equation}
D=4\ :\  \tilde  F^{ab}\equiv -\ft12 \rmi\varepsilon ^{abcd}F_{cd}
\,,\qquad D=6\ :\  \tilde  F^{abc}\equiv -\ft1{3!} \varepsilon
^{abcdef}F_{def} \,.\label{dual46}
\end{equation}
The minus sign in the definition of the dual is convenient for historical
reasons. Indeed, when, as written in footnote~\ref{fn:epsIm}, this
$\varepsilon $ is $\rmi$ times the $\varepsilon $ in these earlier papers
then this agrees with the operation that was taken there (e.g.
in~\cite{VanNieuwenhuizen:1981ae}). In 4 dimensions the dual is an
imaginary operation, and the complex conjugate of a self-dual tensor is
its anti-self-dual partner, while in 6 dimensions the (anti-)self-dual
part of a real tensor is real.

The self-dual \index{self-dual} and anti-self-dual tensors are introduced in even
dimensions as
\begin{equation}
F_{a_1\ldots a_n}^\pm\equiv \ft12\left(F_{a_1\ldots a_n}\pm \tilde
F_{a_1\ldots a_n}\right)\,. \label{def+-dual}
\end{equation}

It is useful to observe relations between (anti)self-dual tensors. In 4
dimensions:
\begin{eqnarray}
&&G^{+ab}H^-{}_{ab}=0 \,,\qquad G^{\pm c(a}H^{\pm b)}{}_c  =   -\ft14 \eta ^{ab}G^{\pm cd}H^\pm {}_{cd}\,, \nonumber\\
&& G^+_{ c[a}H^-_{ b]}{}^c  =  0\,.
 \label{identDual4}
\end{eqnarray}
In 6 dimensions:
\begin{eqnarray}
&& G^{\pm abc}H^\pm {}_{abc}=0\,,\qquad G^{+cd(a}H^{-b)}{}_{cd}  =  \ft16 \eta ^{ab}G^{+ cde}H^- {}_{cde}\,,\nonumber\\
& &  G^\pm _{cd[a}H^\pm{}_{ b]}{}^{cd}  = 0\,,\qquad
 G^\pm _{a[bc}H^\pm _{cd]}{}^a=0\,.
 \label{identDual6}
\end{eqnarray}

\section{\texorpdfstring{$\SU(2)$}{SU(2)} Conventions}
\label{app:SU2Notation}
We use the Levi-Civita $\varepsilon^{ij} $ for which the
important property is that
\begin{equation}
 \varepsilon _{ij}\varepsilon ^{jk}=-\delta _i{}^k\,,
 \label{propLC2}
\end{equation}
where in principle $\varepsilon ^{ij}$ is the complex conjugate of
$\varepsilon _{ij}$, but we can use ($\varepsilon =\rmi \sigma _2$)
\begin{equation}
  \varepsilon _{12}=\varepsilon ^{12}=1\,.
\label{vareps12}
\end{equation}

\subsection{Raising and Lowering Indices}

There is an important difference on the use of the $i,j$ indices between the formulae for $D=4$ and $D=5,\,6$.
For $D=4$ these indices are raised and lowered by complex (or charge) conjugation (here $A^i$ is used for any doublet)
\begin{equation}
  D=4\ :\ (A^i)^C = A_i\,,\qquad (A_i)^C = A^i\,.
 \label{D4indexC}
\end{equation}
For $D=5,\,6$, these indices are raised or lowered using $\varepsilon ^{ij}$. We use
NorthWest--SouthEast (NW--SE) convention, \index{NW--SE convention} which means that this is the
direction in which contracted indices should appear to raise or lower
indices: see
\begin{equation}
D=5,\,6\ :\   A^i=\varepsilon ^{ij}A_j\,,\qquad  A_i=A^j\varepsilon _{ji}\,.
\label{NWSEconv}
\end{equation}
Note that this also implies that for these dimensions, interchanging the position of contracted indices leads to a minus sign:
\begin{equation}
  A^iB_i=-A_iB^i\,.
 \label{SwitchiIndices}
\end{equation}
Another useful relation is that for any antisymmetric tensor in $ij$:
\begin{equation}
   A^{[ij]}=-\ft12\varepsilon^{ij}A^k{}_k\,.
 \label{as=epstrace}
\end{equation}
 \bexer
 Check that we can consider $\varepsilon _{ij}$ as the tensor $\delta
 _i{}^j$ with the $j$ index lowered. For this it is important to
 write $\delta _i{}^j$ and not $\delta^j {}_i$. Also $\varepsilon ^{ij}$
 is the corresponding tensor with raised indices.
 \eexer
Implicit summation (in NW--SE
direction) is also used for bilinears of fermions, e.g.
\begin{equation}
  \bar \lambda \chi \equiv \bar \lambda ^i\chi _i\,.
 \label{implicitIndBil}
\end{equation}

\subsection{Triplets}
\label{SU2vectorNotation}

$\SU(2)$-triplets \index{$\SU(2)$-triplet} can be converted into $2\times 2$ matrices using the anti-hermitian $\vec{\tau}$, defined as $\vec{\tau}_i{}^j=\rmi \vec{\sigma}_i{}^j$, being $\vec{\sigma}_i{}^j$ standard Pauli matrices. We define the triplets as the traceless anti-hermitian matrices
\begin{align}
Y_i{}^j\equiv \vec{\tau}_i{}^j \cdot \vec{Y}, \qquad \vec{Y}=-\ft12 \vec{\tau}_i{}^j Y_j{}^i, \qquad \lambda_i{}^j Y_j{}^i=-2\vec{\lambda}\cdot \vec{Y}.
\label{tripletvecij}
\end{align}
For $D=5,6$ indices can be raised and lowered and thus
\begin{align}
D=5,6\ :\ \vec{\tau}^{ij}=\varepsilon^{ik}\vec{\tau}_k{}^j=\vec{\tau}^{ji}=(\rmi \sigma_3,-\unity,-\rmi \sigma_1)=(\vec{\tau}_{ij})^*.
\end{align}
The correspondence between real symmetric matrices and triplets is
\begin{align}
Y^{ij}=\vec{\tau}^{ij}\cdot \vec{Y}=(Y_{ij})^*= \varepsilon ^{ik}\varepsilon ^{j\ell}(Y^{k\ell})^*,\qquad \vec{Y}=\ft12 \vec{\tau}^{ij} Y_{ij}=\ft12 \vec{\tau}_{ij} Y^{ij}.
\label{Yijupcc}
\end{align}
The auxiliary field $Y^{ij}$ of vector multiplets in $D=4$ has the reality property (\ref{realYd4}), so that this is consistent with the introduction of a real vector $\vec{Y}$.

Useful formulae for the symmetric tensors are
\begin{align}
\vec{\tau}_{ij}\cdot \vec{\tau}^{kl}&=\delta_i^k\delta_j^l+\delta_j^k\delta_i^l,\nonumber\\
A_{ij}B^{jk}&=\delta_i^k \vec{A}\cdot \vec{B}+(\vec{A}\times \vec{B})\cdot \vec{\tau}_i{}^k.
\label{productSU2}
\end{align}
This has been written explicitly for the $\SU(2)$ generators in (\ref{propUaHtr})--(\ref{Udu56}).

\subsection{Transformations, Parameters and Gauge Fields}
As mentioned in (\ref{propUaHtr})--(\ref{Udu56}), the generators $U_i{}^j$ satisfy the same properties in $D=4,5,6$, but in $D=4$ the $U^i{}_j$ are defined as the complex conjugates, which implies (\ref{propU4}), while for $D=5,6$, the property that $U_{ij}=U_{ji}$ after raising and lowering of indices implies
\begin{equation}
  D=4\ : \ U_i{}^j = -U^j{}_i\,,\qquad D=5,6\ :\ U_i{}^j = U^j{}_i\,.
 \label{UdifferentD}
\end{equation}
The correspondence with the parameters is different with the two conventions (see (\ref{paramgenerators}))
\begin{equation}
  D=4\ :\ \delta_{\SU(2)} = \lambda_i{}^jU_j{}^i= -2\vec{\lambda}\cdot \vec{U} \,,\qquad D=5,6\ :\
\delta_{\SU(2)} =\lambda^{ij}U_{ij}=2\vec{\lambda}\cdot \vec{U}\,.
 \label{convULambda}
\end{equation}
Using (\ref{U123algebra}), this implies
\begin{equation}
  \left[\delta_{\SU(2)} (\vec{\lambda} _1),\, \delta_{\SU(2)} (\vec{\lambda} _2)\right]= \delta_{\SU(2)} (\vec{\lambda} _3)\,,\qquad \vec{\lambda} _3=\mp 2\, \vec{\lambda} _1\times\vec{\lambda} _2\,,
 \label{SU2commlambda}
\end{equation}
where the upper sign is for $D=4$ and the lower for $D=5,6$.
Despite the difference in properties in (\ref{UdifferentD}), which can also be used for the parameters, there is one form that can be used in general $\delta =\ldots -\lambda ^j{}_iU_j{}^i$.

The $\SU(2)$ transformation of a doublet, like $A^i$, can be written in different ways
\begin{eqnarray}
 D=4  \ : & \delta A^i = A^j\lambda _j{}^i= -\lambda ^i{}_jA^j \,,\qquad & \delta A_i = -\lambda _i{}^jA_j = A_j\lambda ^j{}_i\,,\nonumber\\
 D=5,6 \ : & \delta A^i = \lambda ^{ij}A_j=   -\lambda ^i{}_jA^j \,,\qquad  & \delta A_i =-\lambda_{ij}A^j = A_j\lambda ^j{}_i\,.
 \label{lambdatransfSU2}
\end{eqnarray}

\section{Gamma Matrices and Spinors} \label{app:gammaspinors}
A general treatment of gamma matrices and spinors is given in \cite[Sect.~3]{VanProeyen:1999ni}. In that review general spacetime signatures
are treated. Of course, that material is not original, and is rather a
convenient reformulation of earlier
works~\cite{Scherk:1978fh,Kugo:1983bn,VanNieuwenhuizen:1985be,Regge:1983uu}.
Another approach to the theory of spinors has been presented
in~\cite{D'Auria:2000ec}. For Lorentzian signatures, the conventions that we use are given in \cite[Chap. 3]{Freedman:2012zz}.

Coefficients $t_r=\pm 1$ appear in Majorana flip relations \index{Majorana flip relation} and relations between spinors and conjugated spinors
\begin{align}
  &\bar\lambda \gamma_{a_1\ldots a_r}\chi = t_r \bar\chi \gamma_{a_1\ldots a_r}\lambda\,,\nonumber\\
  & \chi _{a_1\ldots a_r}= \gamma_{a_1\ldots a_r}\lambda \ \rightarrow \ \bar \chi _{a_1\ldots a_r}=t_0t_r\bar\lambda  \gamma_{a_1\ldots a_r}\,,
\end{align}
and they are given by
\begin{equation}
\begin{array}{ccccccc}\noalign{\smallskip}\hline\noalign{\smallskip}
    & t_0 & t_1 & t_2  & t_3 & t_4 & t_5 \\ \hline\noalign{\smallskip}
  D=4 & + & - & - & + & + &   \\
  D=5 & + & + & - & -  & + & + \\
  D=6 & - & + & + & -  & - & +\\ \noalign{\smallskip}\hline\noalign{\smallskip}
\end{array}
 \label{trD456}
\end{equation}
However, take into account that often the $i,j$ indices are hidden, see (\ref{implicitIndBil}), such that e.g. in 5 dimensions
\begin{equation}
  D=5\ :\  \bar \lambda \chi \equiv \bar \lambda ^i\chi _i= \bar \chi _i\lambda ^i= -\bar \chi ^i\lambda _i= - \bar\chi \lambda\,.
 \label{D5hiddent}
\end{equation}

\subsection{\texorpdfstring{$D=4$}{D=4}} \label{app:gamma4}
One useful tool, which is explained in \cite[Chap. 3]{Freedman:2012zz}, is that complex conjugation can be performed by using charge conjugation,\index{charge conjugation} since for a Lagrangian the two are equal.
The most important rules are as follows\footnote{These properties follow from the definition that the charge conjugate of a spinor $\chi $ is $\chi ^C= \rmi C\gamma ^0\chi $ and $\bar\chi=\chi ^TC$.  }:
\begin{itemize}
\item  For scalars in spinor space, charge conjugation is equal to complex conjugation.
\item For matrices in $\SU(2)$ space (and scalars in spinor space), the choice of the position of the indices is chosen such that (complex and) charge conjugation changes the height of the index: e.g.\  $(M^{ij})^C=M_{ij}$, or $(M^i{}_j)^C=M_i{}^j$.
\item Majorana spinors in a Weyl basis satisfy $(\chi ^i) ^C=\chi _i$. The position of the $i$ index indicates the chirality, see
  Tables~\ref{tbl:countWeyl} and \ref{tbl:matterMult}, and thus charge conjugation changes the chirality.
  \item For a bispinor built from two spinors and a matrix $M$ in spinor (and possible $\SU(2)$) space):
\begin{equation}
  (\bar \lambda M \psi )^C =\overline{ \lambda^C} M^C \psi ^C \,,\qquad \mbox{e.g.\ }(\bar \lambda ^iM_i{}^j\psi _j)^C = \bar \lambda _iM^i{}_j\psi ^j\,,
 \label{D4Cbispinor}
\end{equation}
 and $(MN)^C = M^C N^C$.
  \item The action in spinor space can be derived from $\gamma _a^C=\gamma _a$, and $\gamma _*^C=-\gamma _*$
  \item For the
  fermions of hypermultiplets, the index $\Ae $ is by charge conjugation raised or lowered and at the same time replaced by its barred one, $\bAe$, e.g.
  the charge conjugate of $\zeta ^\Ae =P_L\zeta ^\Ae$ is $\zeta _\bAe= P_R\zeta _\bAe$.
\end{itemize}
For the frame fields of hypermultiplets in 4 dimensions:
\begin{equation}
(f^X{}_{i\Ae})^C=(f^X{}_{i\Ae})^*= f^{Xi\bAe}= \varepsilon ^{ij}\rho ^{\bAe\Be}f^X{}_{j\Be}\,,\qquad   f_{i\bAe X}=(f^{i\Ae}{}_X)^*\,.
 \label{fC}
\end{equation}
For the $\USp(2r)$ connection one has
\begin{equation}
  (\omega _{XA}{}^B)^C=(\omega _{XA}{}^B)^*=\bar \omega ^{\bAe}{}_\bBe= -\rho ^{\bAe\De}\omega _{X\De}{}^\Ee\rho
  _{\Ee\bBe}\,.
 \label{omegaCd4}
\end{equation}
The charge conjugate of the antisymmetric metric $C_{\Ae\Be}$ is $C^{\bAe\bBe}$:
\begin{equation}
  (C_{\Ae\Be})^* = C^{\bAe\bBe}= \rho ^{\bAe \Ce}\rho ^{\bBe\De}C_{\Ce\De}\,.
 \label{Ccc}
\end{equation}
These matrices are also used to raise and lower indices on bosonic quantities with the NW--SE convention.

\subsection{\texorpdfstring{$D=5$}{D=5}} \label{app:gamma5}
In odd dimensions the
antisymmetric product of all the matrices is proportional to the unit
matrix. We choose the sign as follows:
\begin{equation}
  \gamma ^{abcde}=\rmi\varepsilon^{abcde}\,.
\label{gammaepsilon}
\end{equation}
\bexer Check that this implies
\begin{eqnarray}
  \gamma ^{abcd}  & = & \rmi\varepsilon^{abcde}\gamma _e\,,   \nonumber\\
2  \gamma ^{abc}  & = & \rmi\varepsilon^{abcde}\gamma _{ed}\,,   \nonumber\\
3!  \gamma ^{ab}  & = & \rmi\varepsilon^{abcde}\gamma _{edc}\,,   \nonumber\\
4!  \gamma ^{a}  & = & \rmi\varepsilon^{abcde}\gamma _{edcb}\,,   \nonumber\\
5!    & = & \rmi\varepsilon^{abcde}\gamma _{edcba} \,,  \nonumber\\
 \rmi\varepsilon^{abcde}\gamma _{ef}  & = & 4 \gamma ^{[abc}\delta^{d]}_f
 \,, \nonumber\\
 \rmi\varepsilon^{abcde}\gamma _{efg}  & = & 12\gamma
 ^{[ab}\delta^{cd]}_{gf}\,.
 \label{auxgamma}
\end{eqnarray}
\eexer

 The charge conjugation $C$ and $C\gamma_a$ are antisymmetric in $D=5$.
We include a multiplication by $\varepsilon ^{ij}$ in the definition of charge conjugation\footnote{The choice of sign is motivated by the requirement that the translation parameter in the supersymmetry commutation relation reduces from $D=5$ to $D=4$ as in  (\ref{xia54}) below.}:
\begin{equation}
  (\lambda ^i)^C\equiv
  - \rmi\gamma ^0C^{-1}(\lambda^j)^*\varepsilon ^{ji}.
 \label{defSympld5}
\end{equation}
The elementary spinors are `symplectic Majorana',\index{symplectic Majorana} which means that they are invariant under this $C$ operation.

The practical rules are as follows:
\begin{itemize}
\item For bosons charge conjugation is equal to complex conjugation. However, a bispinor $\bar \lambda ^i\chi _i$ built from symplectic Majorana spinors is pure imaginary.
  \item For a bosonic matrix in $\SU(2)$ space, $M$, the charge conjugate is defined as $M^C\equiv \sigma _2
  M^* \sigma _2$, or explicitly $(M_i{}^j)^C=-\varepsilon _{ik}(M_k{}^\ell )^*\varepsilon ^{\ell
  j}$.
  \item For a bispinor that is also a scalar in $\SU(2)$ space: $(\bar \lambda M \psi )^C =-\overline{ \lambda^C} M^C \psi ^C $, where $M$ can be a matrix in spinor space and/or $\SU(2)$ space.
  \item Symplectic Majorana spinors are invariant under $C$.
  \item For matrices in spinor space: $\gamma _a^C=-\gamma _a$.
\end{itemize}
As an example, see that the expression
\begin{equation}
  \bar \lambda ^i \gamma ^\mu \partial _\mu \lambda _i\,,
\label{realexpr}
\end{equation}
is real for symplectic Majorana spinors.

For hypermultiplets the reality condition depends on a matrix
$\rho_{\Ae\bBe}$. The charge
conjugation under which the symplectic Majorana spinors are invariant is
then similar to (\ref{defSympld5}):
\begin{equation}
  (\zeta  ^\Ae)^C\equiv
  - \rmi\gamma ^0 C^{-1}(\zeta ^\Be)^*\rho ^{\bBe\Ae}=\zeta ^\Ae\,.
 \label{defSymplHyperd5}
\end{equation}

\subsection{\texorpdfstring{$D=6$}{D=6}} \label{app:gamma6}
Here there is again chirality, as for every even dimension, but moreover,
there are real self-dual tensors. Similarly, we define
\begin{equation}
\gamma _*=\gamma_0\ldots \gamma _5=-\gamma^0\ldots \gamma^ 5  \,,
\label{gamma*D6}
\end{equation}
without a factor $ \rmi$. The essential formula is
\begin{equation}
  \gamma_{abc}\gamma _*=-\tilde  \gamma_{abc}\,,
\label{gamma7dual}
\end{equation}
 \bexer
 Show that $\gamma _{abc}P_L=\gamma _{abc}^-$ as in 4 dimensions where
 $\gamma _{ab}P_L=\gamma _{ab}^-$.
 \eexer
Spinors can satisfy the (symplectic) Majorana condition and be chiral at
the same time, and we generally use such `symplectic Majorana--Weyl' \index{symplectic Majorana--Weyl} spinors. Thus complex conjugation does not change chirality, and we
can raise and lower $i$ indices as in (\ref{NWSEconv})
\bigskip

We choose the charge conjugation matrix to be symmetric.\footnote{That is
a choice in 6 dimensions, as we can also use the antisymmetric charge
conjugation matrix $C'=C\gamma _*$.}. We use the same definition for charge conjugation as in (\ref{defSympld5}), and thus this also applies for symplectic Majorana spinors.

The rules for charge conjugation are
\begin{itemize}
\item For bosons charge conjugation is equal to complex conjugation. Also a bispinor $\bar \lambda ^i\chi _i$ built from symplectic Majorana spinors is real.
  \item For a bosonic matrix in $\SU(2)$ space, $M$, the charge conjugate is defined as $M^C\equiv \sigma _2
  M^* \sigma _2$, or explicitly $(M_i{}^j)^C=-\varepsilon _{ik}(M_k{}^\ell )^*\varepsilon ^{\ell
  j}$.
  \item For a bispinor that is also a scalar in $\SU(2)$ space: $(\bar \lambda M \psi )^C =-\overline{ \lambda^C} M^C \psi ^C $, where $M$ can be a matrix in spinor space and/or $\SU(2)$ space.
  \item Symplectic Majorana spinors are invariant under $C$.
  \item For matrices in spinor space: $\gamma _a^C=\gamma _a$ and $\gamma _*^C=\gamma _*$.
\end{itemize}

\subsection{Products of \texorpdfstring{$\gamma $}{gamma} Matrices and Fierzing}

There are some useful identities
for calculations in arbitrary dimensions~\cite{vanHolten:1982mx}. For a
product of two antisymmetrized gamma matrices, one can use
\begin{eqnarray}
&&  \gamma _{a_1\ldots a_i}\gamma ^{b_1\ldots b_j}=\sum_{k=|i-j|}^{i+j}
  \frac{i!j!}{s!t!u!}\delta ^{[b_1}_{[a_i}\cdots \delta ^{b_s}_{a_{t+1}}
  \gamma _{a_1\ldots a_t]}{}^{b_{s+1}\ldots b_{j]}}\,,
 \label{GammaGamma} \\
&& s=\ft12(i+j-k)\,,\qquad t=\ft12(i-j+k)\,,\qquad
u=\ft12(-i+j+k)\nonumber\,.
\end{eqnarray}
The numeric factor can be understood as follows. Between the $i$ indices
of the first gamma factor, we select $s=i-t$ of them for the contraction. That
choice is a factor $\binom{i}{s}$. The same number of indices $s$ is
chosen between the $j$ indices of the second factor. That is a factor
$\binom{j}{s}$. Finally, we can contract these $s$ indices in $s!$ ways.
In~\cite{vanHolten:1982mx} a few extra rules are given and a diagrammatic
technique is explained that is based on the work of
Kennedy~\cite{Kennedy:1981kp}.

For contractions of repeated gamma matrices, one has the formula
\begin{eqnarray}
  &&\gamma_{b_1\ldots b_k} \gamma _{a_1\ldots a_\ell }\gamma^{b_1\ldots b_k}
  =c_{k,\ell }\gamma _{a_1\ldots a_\ell }
  \nonumber\\
  &&c_{k,\ell }= (-)^{k(k-1)/2}k!(-)^{k\ell }\sum_{i=0}^{\min (k,\ell )}
  \binom{\ell}{i}\binom{D-\ell}{k-i}(-)^i\,,
\label{sandwich}
\end{eqnarray}
for which tables were given in~\cite{vanHolten:1982mx} in dimensions 4,
10, 11 and 12. Useful examples in 4 dimensions are
\begin{equation}
  \gamma ^a\gamma ^{bc}\gamma _a=\gamma ^{ab}\gamma _c\gamma _{ab}=0,\qquad
  \gamma ^{ab}\gamma _{cd}\gamma _{ab}=4\gamma _{cd}.
 \label{gamgamgamd4}
\end{equation}

Further, there is the Fierz relation.\index{Fierz relation} We know that the gamma matrices are
matrices in dimension $\Delta =2^{{\rm Int ~}D/2}$, and that a basis of
$\Delta \times \Delta $ matrices is given by the set
\begin{equation}
\left\{\unity ,\gamma_a, \gamma_{a_1a_2}, \ldots ,\gamma ^{a_1\ldots
a_{[D]}}
 \right\}
\qquad \mbox{where}\qquad
  \left\{ \begin{array}{ll}
     [D]=D & \mbox{for even }D\,, \\{}
   [D]=(D-1)/2 & \mbox{for odd }D \,,
  \end{array}\right.
\label{basisFierz}
\end{equation}
of which only the first has non-zero trace. This is the basis of the
general Fierz formula for an arbitrary matrix $M$ in spinor space:
\begin{equation}
2^{\mathop{\rm Int}\nolimits (D/2)}  M_\alpha {}^\beta=
\sum_{k=0}^{[D]}(-)^{k(k-1)/2}\ft1{k!}
 \left(  \gamma _{a_1\ldots a_k}\right)  _\alpha {}^\beta \,\,\trace
  \left( \gamma ^{a_1\ldots a_k}M\right)\,.
  \label{Fierz}
\end{equation}
Further Fierz identities can be found in~\cite{Kennedy:1981kp}.
 \bexer
 Check that Fierz identities for chiral spinors in 4 dimensions lead to
\begin{eqnarray}
&&\bar\psi _L\phi _L\,\bar \chi _L\lambda _L=
 -\ft12\bar\psi _L\lambda _L\,\bar\chi _L\phi _L
 +\ft18\bar\psi _L\gamma ^{ab}\lambda _L\,\bar\chi _L\gamma _{ab}\phi _L\,,\nonumber\\
&&\bar\psi _L\phi _L\,\bar \chi _R\lambda _R=
 -\ft12\bar\psi _L\gamma ^a\lambda _R\,\bar\chi _R\gamma _a\phi _L\,.
 \label{Fierzchiral4}
\end{eqnarray}
An extra minus sign w.r.t. (\ref{Fierz}) appears here because fermions
$\lambda $ and $\phi $ are interchanged. In 5 dimensions the Fierz
equation is
\begin{equation}
  \bar \psi ^i\phi ^j\bar \chi ^k\lambda ^\ell =
  -\ft14\bar \psi ^i\lambda ^\ell\bar \chi ^k\phi ^j
  -\ft14\bar \psi ^i\gamma ^a\lambda ^\ell\bar \chi ^k\gamma _a\phi ^j
  +\ft12\bar \psi ^i\gamma ^{ab}\lambda ^\ell\bar \chi ^k\gamma _{ab}\phi
  ^j.
 \label{Fierz5}
\end{equation}
In 6 dimensions ($\bar \psi _R=\bar \psi P_L$) one has
\begin{eqnarray}
&&\bar\psi _R\phi _L\,\bar \chi _L\lambda _R=
 -\ft14\bar\psi _R\gamma ^a\lambda _R\,\bar\chi _L\gamma _a\phi _L
 +\ft1{48}\bar\psi _R\gamma ^{abc}\lambda _R
            \,\bar\chi _L\gamma _{abc}\phi  _L
 \,,\nonumber\\
&&\bar\psi _R\phi _L\,\bar \chi _R\lambda _L=
 -\ft14 \bar\psi _R\lambda _L\,\bar \chi _R\phi _L
 +\ft18 \bar\psi _R\gamma ^{ab}\lambda _L\,\bar \chi _R\gamma_{ab}\phi _L
 \,.
 \label{Fierzchiral6}
\end{eqnarray}
For those that want to go further, a more complicated identity for
doublet spinors in 4 dimensions is
\begin{equation}
  \varepsilon ^{jk }\gamma ^{ab}\lambda_i
  \bar \lambda _j\gamma _{ab}\lambda_k =8\varepsilon ^{jk }
  \lambda _k \bar \lambda _i\lambda _j.
 \label{Fierzdoubletquartetspinor}
\end{equation}
It uses the first equation of (\ref{Fierzchiral4}), symmetries of the
bilinears and manipulations between the $\varepsilon $ symbols.

  \eexer

\section{Spinors from 5 to 6 and 4 Dimensions}
\label{app:spinors5to4}

For some part of this text (especially for hypermultiplets) we started from explicit formulae for symplectic Majorana spinors in $D=5$ and these formulae are useful also in 4 and 6 dimensions. The spinors have different properties in 4 and 6 dimensions, but we can easily translate the formulae.

Let us temporarily \emph{write $\tilde \zeta ^A$ for the symplectic Majorana--Weyl spinors in $D=6$}, which are right-handed. We consider them now as 4-component spinors, and translate to the spinors $\zeta ^A$ of $D=5$ by
\begin{equation}
  \zeta ^A = \rmi\tilde\zeta ^A\,,\qquad \bar \zeta^A= -\rmi \overline {\tilde \zeta}{} ^A\,.
 \label{D65right}
\end{equation}
For spinors that are left-handed in 6 dimensions, e.g. the supersymmetry parameter $\epsilon ^i$, we have
\begin{equation}
  \epsilon ^i = \tilde \epsilon ^i\,,\qquad \bar \epsilon ^i =  \overline {\tilde \epsilon }{}^i\,.
 \label{D65left}
\end{equation}
Finally, the last $\gamma $-matrix from 6 dimensions, $\gamma _5$ gets translated to an imaginary unit: $\rmi $ on a left-handed spinor, and $-\rmi$ on a right-handed spinor. Combined with the above factors, we thus have
\begin{equation}
  \gamma _5\tilde \epsilon ^i = \rmi\epsilon ^i\,,\qquad \gamma _5\tilde \zeta ^A = - \zeta ^A\,.
 \label{gamma5D65}
\end{equation}

Reduction from 5 to 4 dimensions is a bit more involved \cite{Rosseel:2004fa}. We give here the rules and some illustrative examples.
\emph{Now the tilde indicates quantities for $D=5$.}

The rules are based on the fact that the $\gamma_4$ matrix in 5 dimensions is identified with $\gamma _*$ in 4 dimensions, and the charge conjugation matrix of 5 dimensions  is $\tilde C=C\gamma _*$ in 4 dimensions. An important difference is that in 5 dimensions $i,j$ indices are raised and lowered as in (\ref{NWSEconv}), while this is related to charge conjugation and change of chirality in 4 dimensions, see Sect. \ref{app:gamma4}. The symplectic Majorana condition of $D=5$,  (\ref{defSympld5}), can be written as
\begin{equation}
  (\tilde \lambda ^i)^* = \rmi \tilde C \gamma ^0 \tilde \lambda ^j\varepsilon _{ji}\,.
 \label{ccD5}
\end{equation}
This is now translated in terms of the 4-dimensional charge conjugation as
\begin{equation}
  (\tilde \lambda ^i)^C = -\gamma _*\tilde \lambda ^j\varepsilon _{ji}\,.
 \label{condsymplD4}
\end{equation}
Now translation rules depend on the choice in 4 dimensions whether $\lambda ^i$ is left- or right-handed:
\begin{equation}
  \lambda ^i = \pm\gamma _*\lambda ^i\,,\qquad \lambda _i = \mp\gamma _*\lambda _i\,.
 \label{lambdaLorR}
\end{equation}
In order to be consistent with (\ref{condsymplD4}) we can then write
\begin{equation}
  \tilde \lambda ^i = \lambda ^i\mp\lambda _j\varepsilon ^{ji}\,.
 \label{tillambdalambda}
\end{equation}
This then also implies
\begin{equation}
 \overline{ \tilde \lambda ^i} = (\tilde \lambda ^i)^T {\cal C}= \left(\lambda^i \mp\lambda _j\varepsilon ^{ji}\right)^T C\gamma ^*= \pm\bar \lambda ^i +\bar \lambda _j\varepsilon ^{ji}\,.
 \label{barredD54}
\end{equation}
Similar rules apply to the spinors of the hypermultiplets, replacing $\varepsilon _{ij}$ by $\rho _{\Ae\bBe}$. We took in $D=4$ the choice $\zeta ^\Ae = P_L\zeta ^\Ae$, hence we have
\begin{equation}
  \tilde \zeta ^\Ae = \zeta ^\Ae - \zeta _\bBe\rho ^{\bBe\Ae}\,,\qquad
 \overline {\tilde \zeta} ^\Ae=\bar \zeta ^\Ae +\bar \zeta _\bBe \rho^{\bBe\Ae}\,.
 \label{tilzeta54}
\end{equation}

As a first example, let us start from the translation parameter in terms of two supersymmetries, which are left-handed in $D=4$, as in (\ref{xiaeps12}):
\begin{eqnarray}
\tilde \xi ^a(\tilde \epsilon _1,\tilde \epsilon _2)&=&\ft12 \overline {\tilde \epsilon} _2^i\gamma^a\tilde \epsilon_1^j\varepsilon _{ji}\nonumber\\
&=&\ft12\left(\bar \epsilon_2 ^i +\bar \epsilon _{2k}\varepsilon ^{ki}\right)\gamma ^a \left(\epsilon _1 ^j -\epsilon _{1\ell}\varepsilon ^{\ell j}\right)\varepsilon _{ji}\nonumber\\
&=&\ft12\bar \epsilon_2 ^i\gamma ^a\epsilon _{1i} +\ft12\bar \epsilon _{2i} \gamma ^a\epsilon _1 ^i=\xi ^a\,,
 \label{xia54}
\end{eqnarray}
since the other terms disappear by chirality. The right-hand side is the correct expression in 4 dimensions. The 5-dimensional commutator contains another component:
\begin{eqnarray}
\tilde \xi ^4(\tilde \epsilon _1,\tilde \epsilon _2)&=&\ft12\overline {\tilde \epsilon} _2^i\gamma^4\tilde \epsilon_1^j\varepsilon _{ji}\nonumber\\
&=&\ft12\left(\bar \epsilon_2 ^i +\bar \epsilon _{2k}\varepsilon ^{ki}\right)\gamma_* \left(\epsilon _1 ^j -\epsilon _{1\ell}\varepsilon ^{\ell j}\right)\varepsilon _{ji}\nonumber\\
&=&-\ft12\bar \epsilon_2 ^i\epsilon _1^j\varepsilon _{ij} +\ft12\bar \epsilon _{2i} \epsilon _{1j}\varepsilon ^{ij}\,.
 \label{xi454}
\end{eqnarray}

As a second example we translate the non-closure relation obtained in (\ref{nonclosurezeta})
\begin{align}
 \left[\delta(\tilde \epsilon _1),\delta (\tilde \epsilon _2)\right]\tilde \zeta ^\Ae  & =  \tilde \xi ^\mu \partial  _\mu \tilde \zeta ^\Ae+\ft14\left[(\overline {\tilde  \epsilon} _2^i\tilde \epsilon _1^j)- \gamma ^\nu (\overline {\tilde \epsilon} _2^i\gamma _\nu \tilde \epsilon _1^j)\right]\varepsilon _{ji}\rmi\tilde \Gamma ^\Ae\,,\nonumber\\
   &   \rmi \tilde \Gamma^\Ae \equiv  \slashed{\nabla} \tilde \zeta^\Ae + \ft 12 W_{\Be\Ce\De} {}^\Ae
 \tilde \zeta^\Be\overline {\tilde{\zeta}}^\Ce \tilde \zeta^\De\,.
 \label{nonclosureD5tilde}
\end{align}
We define in 4 dimensions the $\Gamma^\Ae$ as right-handed, and for convenience redefine the $\Gamma ^A$ in $D=4$ from the rule with the lower sign in (\ref{tillambdalambda}) with a factor $\rmi$ (and thus $-\rmi$ for the left-handed component in order to define Majorana spinors in $D=4$), i.e.
\begin{equation}
  \rmi \tilde \Gamma ^\Ae =  -\Gamma ^\Ae + \Gamma _{\bBe}\rho ^{\bBe\Ae} \,,
 \label{deftildeGammaGamma}
\end{equation}
which leads to
\begin{equation}
 - \Gamma  ^\Ae=- P_R \Gamma  ^\Ae=  P_R\rmi\tilde \Gamma ^\Ae\,,  \qquad   \Gamma  _\bAe=P_L\Gamma  _\bAe=-P_L\rmi\tilde \Gamma  ^\Be\rho _{\Be\bAe}\,.
 \label{d54Gamma}
\end{equation}
We thus obtain the expression $\Gamma ^ \Ae$ from projecting to the right-handed part:
\begin{equation}
-  \Gamma  ^\Ae=\slashed{\nabla} \zeta^\Ae + \ft 12 W_{\Be\Ce\De} {}^\Ae
 (-\zeta_\bEe\rho ^{\bEe\Be}) {\bar{\zeta}}^\Ce  \zeta^\De\,,
 \label{calcGammaD4}
\end{equation}
since Fierz identities together with the symmetry of the $W$ tensor annihilate the part with 3 antichiral fermions.
We thus obtain:
\begin{equation}
  \Gamma ^A = -\slashed{\nabla} \zeta^\Ae +\ft12 W_{\Be\Ce}{}^{\bDe\Ae}
 \zeta_\bDe {\bar{\zeta}}^\Be  \zeta^\Ce\,, \qquad W_{\Be\Ce}{}^{\bDe\Ae}\equiv \rho ^{\bDe\Ee}W_{\Be\Ce\Ee}{}^\Ae\,.
 \label{nonclosureGammaD4}
\end{equation}
In the same way, the left-handed part of the commutator  (\ref{nonclosureD5tilde}) is
\begin{eqnarray}
 \left[\delta( \epsilon _1),\delta (\epsilon _2)\right] \zeta ^\Ae  & = &\xi ^\mu \partial  _\mu \zeta ^\Ae
 +\ft12 \varepsilon ^{ij}\bar \epsilon _{2i}\epsilon _{1j} \Gamma _\bBe\rho ^{\bBe\Ae}
 +\ft14 \gamma ^\nu \Gamma ^A \left(\bar \epsilon _2^i \gamma _\nu \epsilon _{1i}+ \bar \epsilon _{2i} \gamma _\nu \epsilon _1^i\right)\,.\nonumber\\
 \label{nonclosureD4}
\end{eqnarray}
Note that the reduction to 4 dimensions implies that the spacetime derivative $\partial _4$ vanishes. However the $\nu =*$ part in (\ref{nonclosureD5tilde}) contributes such that the left-handed parts cancel a similar contribution from $\overline {\tilde \epsilon} _2^i\,\tilde \epsilon _1^j$.

%% file: appB_Superalgebras.tex
\chapter{Superalgebras}\label{app:groups}

Lie superalgebras \index{Lie superalgebras}have been classified in \cite{Kac:1977qb}. We cannot go through the full classification mechanism of course,
but will consider the most important superalgebras, the `simple Lie superalgebras', \index{simple Lie superalgebra} which have no non-trivial invariant subalgebra. However,
one should know that for superalgebras there are more subtle issues, as
e.g.\ not any semi-simple superalgebra is the direct sum of simple
superalgebras. A good review is~\cite{Frappat:1996pb}. The fermionic
generators of such superalgebras are in representations of the bosonic part. If that `defining representation' of the bosonic algebra in the fermionic generators is completely reducible, the algebra is said to be `of classical type'. The others are `Cartan type superalgebras' $W(n)$, $S(n)$, $\tilde S(n)$ and $H(n)$, which we will further neglect. For
further reference, we give in Table~\ref{tbl:LieSA}
\begin{table}[!t]
\caption{\it Lie superalgebras of classical type. Note
that the superalgebras $\SU(m|m)$, which are indicated in this table with
bosonic subalgebra $\SU(m)\oplus\SU(m)$ are often indicated as $PSU(m|m)$, while then $\SU(m|m)$ refers to the (non-simple) algebra including the $\U(1)$ factor.}
\label{tbl:LieSA}
\begin{center}
\begin{tabular}{|l|l|l|l|l|}
\hline\noalign{\smallskip}
Name & Range  & Bosonic algebra  & Defining& Number of  \\
     &        &                  & repres. & generators\\ \hline \hline
$\SU(m|n)$&$m\geq 2$&$ \SU(m)\oplus \SU(n)$&$(m,\bar n)\oplus
$&$m^2+n^2-1,$\\
     & $ m\neq n $       & \ \ \ $\oplus \U(1)$&\ $(\bar m,n)$ &\ \ \  $2mn  $\\
& $ m= n $&\ \ \ no $\U(1)$   & 
&$2(m^2-1),2m^2$\\
     \hline
 \multicolumn{2}{c|}{
\begin{tabular}{|l}
$\Sl(m|n) $       \\
$\SU(m-p,p|n-q,q) $   \\
$\SU^*(2m|2n) $\\
$\Sl \,{}^\prime(n|n)$
\end{tabular}}
& \multicolumn{3}{l}{\begin{tabular}{l} $\begin{array}{l}
\Sl (m)\oplus \Sl (n)  \\
\SU (m-p,p)\oplus \SU(n-q,q)\\
\SU^*(2m)\oplus \SU^*(2n)
\end{array}  $
\hspace{-10pt} $\left.
\begin{array}{l}
\oplus \SO(1,1)   \\
\oplus \U(1)   \\
\oplus \SO(1,1)
\end{array} \hspace{-2pt}\right\}$\hspace{-5pt} \begin{tabular}{l}
if  \\
 $m\neq n$
\end{tabular}
\\ $\Sl (n,\Cbar)$\end{tabular}}
\\ \hline\hline
$\OSp(m|n)$&$ m\geq 1$ & $\SO(m)\oplus \Sp(n)$ & $(m,n)$ & $\ft12(m^2-m+$\\
          & $ n=2,4,..$ &                   &         &\ \ \  $n^2+n),mn$\\ \hline
 \multicolumn{2}{c|}{
\begin{tabular}{|l}
\phantom{$\SU(m-p,p|n-q,q) $}  \\[-4mm]
$\OSp(m-p,p|n)$  \\
$\OSp(m^*|n-q,q)$
\end{tabular} }&
 \multicolumn{2}{c}{
\begin{tabular}{l}
\phantom{$\SU (m-p,p)\oplus \SU(n-q,q)$}\hspace{-5pt}  \\[-4mm]
$ \SO(m-p,p)\oplus \Sp(n)$ \\
  $ \SO^*(m)\oplus \USp(n-q,q)$
\end{tabular} }&
\begin{tabular}{l} $n$ even
 \\ $m,n,q$ even
\end{tabular}
\\ \hline\hline
$D(2,1,\alpha)$ & $0<\alpha\leq 1$ & $\SO(4)\oplus S\ell (2)$&$(2,2,2)$&9, 8 \\
                \hline
 \multicolumn{2}{c|}{
\begin{tabular}{|l}
\phantom{$\SU(m-p,p|n-q,q) $}   \\[-4mm]
$D^p(2,1,\alpha)$
\end{tabular} }&
 \multicolumn{2}{c}{
\begin{tabular}{l}
\phantom{$\SU (m-p,p)\oplus \SU(n-q,q)$} \\[-4mm]
 $\SO(4-p,p) \oplus \Sl(2)$
\end{tabular} }&$p=0,1,2$ \\ \hline\hline
$F(4)$ & & $\overline {\SO(7)}\oplus \Sl(2)$ & $(8,2)$& 24, 16 \\
\hline
 \multicolumn{2}{c|}{
\begin{tabular}{|l}
\phantom{$\SU(m-p,p|n-q,q) $} \\[-4mm]
$F^p(4)$\\ $F^p(4)$
\end{tabular} }&
 \multicolumn{2}{l}{
\begin{tabular}{l}
\phantom{$\SU (m-p,p)\oplus \SU(n-q,q)$} \\[-4mm]
$\SO(7-p,p)\oplus \Sl(2)$\\ $\SO(7-p,p)\oplus \SU(2)$
\end{tabular} }&
\begin{tabular}{l}
$p=0,3$    \\ $p=1,2$
\end{tabular}
\\ \hline\hline
$G(3)$ & & $G_2\oplus \Sl (2)$ & $(7,2)$ & 17, 14\\  \hline
 \multicolumn{2}{c|}{
\begin{tabular}{|l}
\phantom{$SU(m-p,p|n-q,q) $} \\[-4mm]
$G_p(3)$
\end{tabular} }&
 \multicolumn{2}{c}{
\begin{tabular}{l}
\phantom{$\SU (m-p,p)\oplus \SU(n-q,q)$} \\[-4mm]
$G_{2,p}\oplus  \Sl(2)$
\end{tabular}} & $p=-14,2$\\ \hline\hline
$P(m-1)$&$m\geq 3$ & $\Sl (m)$ & $(m\otimes m)$&$m^2-1,m^2$ \\
\hline\hline $Q(m-1)$&$m\geq 3$ & $\SU(m)$ & Adjoint &$m^2-1,m^2-1$ \\
\hline
 \multicolumn{2}{c|}{
\begin{tabular}{|l}
\phantom{$\SU(m-p,p|n-q,q) $} \\[-4mm]
$Q(m-1)$ \\
$Q((m-1)^*)$  \\
$UQ(p,m-1-p)$
\end{tabular} }&
 \multicolumn{2}{c}{
\begin{tabular}{l}
\phantom{$\SU (m-p,p)\oplus \SU(n-q,q)$} \\[-4mm]
$\Sl (m)$\\
 $\SU^*(m)$ \\
 $\SU(p,m-p)$
\end{tabular}} & \\ \hline\hline
\end{tabular} 
\end{center}\end{table}
the list of the real forms of superalgebras `of classical
type'~\cite{Kac:1977em,Parker:1980af,Nahm:1978tg}. In this table `defining representation' gives the  fermionic generators as a representation of the bosonic subalgebra. The `number of generators' gives the numbers of (bosonic,fermionic) generators in the superalgebra.
We mention first the algebra as an algebra over $\Cbar$, and then give
different real forms of these algebras. With this information, you can
reconstruct all properties of these algebras, up to a few exceptions. The
names which we use for the real forms is for some algebras different from
those in the mathematical literature
\cite{Kac:1977em,Parker:1980af,Nahm:1978tg} and chosen such that it is
most suggestive of its bosonic content. There are isomorphisms as
$\SU(2|1)=\OSp(2^*|2,0)$, and $\SU(1,1|1)=\Sl(2|1)=\OSp(2|2)$. In the
algebra $D(2,1,\alpha)$ the three $\Sl(2)$ factors of the bosonic group
in the anticommutator of the fermionic generators appear with relative
weights $1$, $\alpha$ and $-1-\alpha$. The real forms contain,
respectively, $\SO(4)=\SU(2)\times \SU(2)$, $\SO(3,1)=\Sl(2,\Cbar)$ and
$\SO(2,2)=\Sl(2) {\times} \Sl(2)$. In the first and last case $\alpha$ should be
real, while $\alpha= 1 +\rmi a$ with real $a$ for $p=1$. In the limit
$\alpha=1$ one has the isomorphisms $D^p(2,1,1)=\OSp(4-p,p|2)$.

For the real forms of $\SU(m|m)$, the one-dimensional subalgebra of the
bosonic algebra is not part of the irreducible algebra. In some papers these are called P$\SU(m|m)$, keeping the name $\SU(m|m)$ for the algebra including the abelian factor. Furthermore, in that case there are subalgebras obtained from projections of those mentioned here with only
one factor $\SU(n)$, $\Sl(n)$, $\SU^*(n)$ or $\SU(n-p,p)$ as bosonic
algebra.

%% file: appC_Comparison.tex
\chapter{Comparison of Notations}
\label{ss:compareN2d4conv}
Unfortunately the normalization of $F$ and various other functions vary
in the ${\cal N}=2$ literature. A lot of standard notations can be summarized in factors $s_1,\ldots , s_9$. The first three of these were already identified in the book of Misner--Thorne--Wheeler~\cite{MisnerThorneWheeler}. More explanations are given in \cite[Appendix A]{Freedman:2012zz}. We summarize here a few relevant formulae
\begin{align}
   & \eta _{ab}= s_1{\mathop{\rm diag}\nolimits}(-+++)\,, \nonumber\\
&R_{\mu\nu}{}^\rho{}_\sigma= s_2\left(\partial_\mu \Gamma^\rho_{\nu\sigma}
-\partial_\nu \Gamma^\rho_{\mu\sigma} +
\Gamma^\rho_{\mu\tau}\Gamma^\tau_{\nu\sigma} -
\Gamma^\rho_{\nu\tau}\Gamma^\tau_{\mu\sigma} \right)\,,\nonumber\\
  &s_2s_3  R_{\mu \nu }=  R^\rho {}_{\nu \rho \mu }\,,\nonumber\\
&D_\mu \psi =  \partial _\mu +s_2s_4\ft14 \omega_\mu{}^{ab}\gamma _{ab}\,,\qquad D_\mu V ^a =\partial _\mu V^a + s_2s_4\omega_\mu{}^{ab}V_b\,,\nonumber\\
&\varepsilon_{0123}=s_5\,,\qquad \varepsilon^{0123}=-s_5\,,\nonumber\\
&  \gamma _\mu \gamma _\nu +\gamma _\nu \gamma _\mu =2s_6g_{\mu \nu }\,,\nonumber\\
& \gamma _*=  \gamma _5 = s_7 \rmi \gamma _0\gamma _1\gamma _2\gamma _3\,,\qquad   \ft12\varepsilon _{abcd}\gamma ^{cd} =-\rmi \frac{s_5}{s_7} \rmi \gamma_* \gamma _{ab}\,,\nonumber\\
&   \left(\bar \psi \lambda\right) ^*=s_8(-t_0t_1)\bar \psi
  \lambda\,, \nonumber\\
& \left[\delta (\epsilon _1),\delta (\epsilon _2)\right]\phi  = s_9 \ft12 \bar \epsilon
  _2 \gamma ^\mu\epsilon _1 \partial _\mu \phi\,.
\label{sfactors}
\end{align}
Note that the factor $(-t_0t_1)=1$ for $D=4$ and $D=6$, see (\ref{trD456}).

 In Table~\ref{tbl:compnot},
we compare notations between this book and some of the original papers in the development of ${\cal N}=2$ matter-coupled supergravity in $D=4$.
\begin{enumerate}
\item[Here]: means the notation used in this book, and for most part also used in\footnote{The exception is the
Ricci sign (sign of $R$) that has been chosen negative in these papers.}
\cite{deWit:1995tf,VanProeyen:1995sw,deWit:1996jd}.
  \item[I]: The second column refers to many of the original papers on $D=4$ ${\cal N}=2$ superconformal  tensor calculus and special geometry.
These are e.g.
\cite{deWit:1979pq,deWit:1980ug,deRoo:1980mm,deWit:1980gt,deWit:1981tn,
VanProeyen:1983wk,deWit:1984xe,deWit:1984rz,%
deWit:1984pk,deWit:1985px,Cremmer:1985hj,Cecotti:1989qn,deWit:1992cr,%
deWit:1992nm,deWit:1993wf,Claus:1998fk}.
  \item[II]: The third column contains the notation of the
the matter-coupled
Lagrangians of \cite{Andrianopoli:1996vr,Andrianopoli:1997cm}.
\end{enumerate}

\begin{table}[!t]
\caption{\it Comparison of notations}
\label{tbl:compnot}\begin{center}\begin{tabular}{|ccc|}\hline
Here& I &II \\
\hline
$s_i=1$&&\\
 $s_{1,2,3,4,5}=$ &$(+1,-1,-1,1,\rmi)$ & $(-1,-1,-2,1,-1)$ \\
$s_{6,7,8,9}=$ &$(+1,-1,+1,+4)$ &$(+1,+1,+1,2\rmi)$ \\
$\gamma _\mu $& $\gamma _\mu $&$\rmi\gamma _\mu $  \\[1mm]
 $g_{\alpha\bar \beta}$&$-g_{A\bar B}$&$g_{ij^*}$ \\
${\cal K}$&  $ -K $& ${\cal K}$\\
$F$ & $-\ft \rmi 4 F$&    \\[1mm]
$X^I$ & $ X^I$ &$L^\Lambda$ \\
$Z^I$ &   $Z^I$ &$X^\Lambda$  \\
$F_I$ & $-\ft \rmi{4}F_I$&$M_\Lambda$ \\[2mm]
$N_{IJ}$ & $-N_{IJ}$&$- N_{\Lambda\Sigma}$ \\[2mm]
${\cal N}_{IJ}$&$\rmi{\cal N}_{IJ}$&${\cal N}_{\Lambda\Sigma}$\\[2mm]
$C_{\alpha\beta\gamma}$& $e^{-K}Q_{ABC}$&$\rmi C_{ijk}$\\[1mm]
$F_{\mu\nu}^I$&$F_{\mu\nu}^I$&$-2{\cal F}_{\mu\nu}^\Lambda$  \\[1mm]
$f_{IJ}{}^K$   & $- g f_{IJ}{}^K$  & $f_{\Lambda \Sigma }^\Delta $ \\[1mm]
${\cal P}_I $ &&$-{\cal P}_\Lambda $\\[1mm]
$\epsilon $ & $2\epsilon $&$\sqrt{2}\epsilon $ \\[1mm]
$\psi _\mu $ & $\psi _\mu $&$\sqrt{2}\psi _\mu $  \\[1mm]
$\Omega ^I$ & $\Omega ^I$&  \\[1mm]
$\eta _i$& $\eta _i$ &  \\[1mm]
$\phi _\mu $&$\frac{1}{2}\phi _\mu $ &  \\[1mm]
$f_\mu {}^a$&$\ft12f_\mu {}^a$&   \\[1mm]
$\lambda _{\rm K}$&$\ft12\Lambda _{\rm K} $ & \\[1mm]
$V_\mu {}^i{}_j$& $\ft12{\cal V}_\mu {}^i{}_j$ &   \\[1mm]
$\chi ^i$ & $\chi ^i$&   \\[1mm]
$T^-_{ab}\varepsilon ^{ij}$ & $ T^{-ij}_{ab}$ & \\[1mm]
\hline
\end{tabular}\end{center}\end{table}
A few comments are in order. In the original papers often the Pauli signature was used, which means that there is an index $a=4$ or $\mu =4$, we translated for vectors to $V_4=\rmi V_0$ in order to put them under the general scheme. The factor $s_2s_3$ determines the normalization of the Ricci scalar $R=g^{\mu \nu }R_{\mu \nu }$.

Note also that in the older papers an antisymmetrization $[ab]$ was
$(ab-ba)$ without the factor as in~(\ref{weight1}). Also one used
\begin{equation}
  \sigma _{ab}= \ft12\gamma _{ab}\,.
 \label{sigmaab}
\end{equation}

Finally, we want to point out that there are also a few changes in notation between this book and the notation the papers \cite{Bergshoeff:2002qk,Bergshoeff:2004kh}. To use the latter papers in the present notation, one has to make the following substitutions
\begin{align}
&k_I{}^X\leftarrow\ -k_I{}^X \,,\qquad A_\mu ^I\leftarrow\ W_\mu ^I\,,\qquad {\cal D}_\mu \leftarrow\ \hat{D}_\mu \,,\nonumber\\
& t_{IA}{}^B\leftarrow\ -t_{IA}{}^B,\,\qquad \vec{P}_I\leftarrow\ \ft12\vec{P}_I\,,\qquad \vec{\sigma}_i{}^j\leftarrow \ \rmi\vec{\tau }_i{}^j \,.
  \label{translationsoldD5}
\end{align}